\documentclass{egpubl}
\usepackage{eg2024}
 
\ConferencePaper        
\CGFccby

\usepackage[T1]{fontenc}
\usepackage{dfadobe}  

\usepackage{cite}  
\BibtexOrBiblatex
\electronicVersion
\PrintedOrElectronic

\ifpdf \usepackage[pdftex]{graphicx} \pdfcompresslevel=9
\else \usepackage[dvips]{graphicx} \fi

\usepackage{egweblnk} 

\usepackage{bbm}
\usepackage{dsfont}
\newif\ifdraft

\ifdraft
\newcommand{\rgc}[1]{{\color{orange}[\textbf{RG:} #1]}}
\newcommand{\abc}[1]{{\color{olive}[\textbf{AB:} #1]}}
\newcommand{\OSH}[1]{{\color{magenta}[\textbf{OSH:} #1]}}
\newcommand{\dcc}[1]{{\color{red}[\textbf{DC:} #1]}}
\newcommand{\ahc}[1]{{\color{purple}[\textbf{AH:} #1]}}

\newcommand{\todo}[1]{{\color{red}[\textbf{TODO:} #1]}}

\newcommand{\new}[1]{{\color{red}#1}}

\else
\newcommand{\rgc}[1]{}
\newcommand{\abc}[1]{}
\newcommand{\ahc}[1]{}
\newcommand{\OSH}[1]{}
\newcommand{\dcc}[1]{}

\newcommand{\todo}[1]{}
\newcommand{\new}[1]{{\color{black}#1}}

\fi

\newcommand{\ourmethod}{SENS}

\newcommand{\figref}[1]{Fig.\ \ref{#1}}
\newcommand{\secref}[1]{Sec.\ \ref{#1}}

\newcommand{\tableref}[1]{Table \ref{#1}}

\usepackage[percent]{overpic}

\usepackage{pifont}
\newcommand{\cmark}{\ding{51}}%
\newcommand{\xmark}{\ding{55}}%
\usepackage{makecell}
\usepackage{multirow}

\title{\ourmethod{}: \new{Part-Aware} Sketch-based Implicit Neural Shape Modeling}

\author[A.\ Binninger, A.\ Hertz, O.\ Sorkine-Hornung, D.\ Cohen-Or, R.\ Giryes]
{\parbox{\textwidth}{\centering Alexandre Binninger$^{1}$\orcid{0000-0002-9833-4126},
        Amir Hertz$^{2}$\orcid{0000-0003-3037-3556},
        Olga Sorkine-Hornung$^{1}$\orcid{0000-0002-8089-3974},
        Daniel Cohen-Or$^{2}$\orcid{0000-0001-6777-7445},
        Raja Giryes$^{2}$\orcid{0000-0002-2830-0297}
        }
        \\
{\parbox{\textwidth}{\centering $^1$ETH Zurich, Switzerland $\quad \quad \quad$
         $^2$Tel Aviv University, Israel
       }
}
}

\usepackage{amsmath}
\usepackage{amssymb}
\usepackage{amsfonts}
\usepackage{enumitem}

\usepackage[ruled]{algorithm2e}

\usepackage{microtype}
\usepackage{units}
\usepackage{gensymb}
\usepackage{threeparttable}
\usepackage{array}
\usepackage{booktabs}
\usepackage[dvipsnames, table]{xcolor}
\usepackage{overpic}
\usepackage{pgfplots}
\usetikzlibrary{pgfplots.groupplots}
\pgfplotsset{compat=1.7}
\usepackage{tikz,tikz-3dplot} 
\usepackage{tkz-euclide}

\definecolor{lightBronze}{HTML}{e5ae87}

\usetikzlibrary{3d, arrows}

\usepackage{dsfont}
\usepackage{wrapfig}

\usepackage{mdframed}
\mdfsetup{skipabove=\topskip,skipbelow=\topskip}
\newmdenv[	linewidth=0.75pt,
			leftline=false,
			rightline=false,
   			innerleftmargin=0pt,
			innerrightmargin=0pt]{topbot}

\usepackage{balance}

\begin{document}

\teaser{
  \includegraphics[trim=10 10 10 10, clip, width=\textwidth]{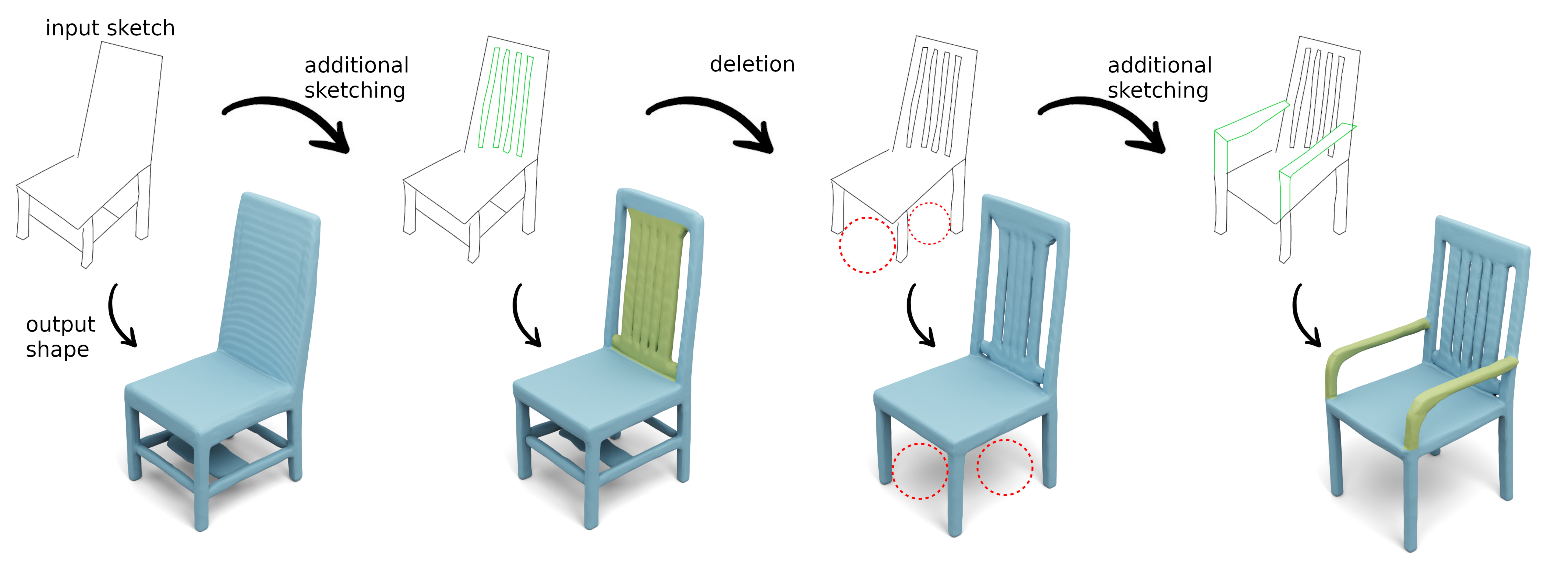}
  \caption{\ourmethod{} generates shapes and enables ongoing edits via sketching. Adding details or removing parts from the sketch is reflected in the output shape.}
  \label{fig:teaser}
}
\maketitle

\begin{abstract}
  We present \ourmethod{}, a novel method for generating and editing 3D models from hand-drawn sketches, including those of abstract nature. Our method allows users to quickly and easily sketch a shape, and then maps the sketch into the latent space of a part-aware neural implicit shape architecture. \ourmethod{} analyzes the sketch and encodes its parts into ViT patch encoding, subsequently feeding them into a transformer decoder that converts them to shape embeddings suitable for editing 3D neural implicit shapes. \ourmethod{} provides intuitive sketch-based generation and editing, and also succeeds in capturing the intent of the user's sketch to generate a variety of novel and expressive 3D shapes, even from abstract and imprecise sketches. Additionally, \ourmethod{} supports refinement via part reconstruction, allowing for nuanced adjustments and artifact removal. It also offers part-based modeling capabilities, enabling the combination of features from multiple sketches to create more complex and customized 3D shapes. We demonstrate the effectiveness of our model compared to the state-of-the-art using objective metric evaluation criteria and a user study, both indicating strong performance on sketches with a medium level of abstraction. Furthermore, we showcase our method's intuitive sketch-based shape editing capabilities, and validate it through a usability study.

\end{abstract}

\section{Introduction}

Data-driven techniques have become the de facto state-of-the-art for recovering a shape from a partial representation in computer graphics. Training neural networks can leverage prior domain knowledge of the data to deal with the innate ambiguity of the input. Neural implicit fields are currently widely used as a generative model because of their ability to represent arbitrary shapes at arbitrary resolutions  \cite{chen2018,DeepSDFPark2019,Atzmon2019,OrEl2022StyleSDF,Tewari2022Advances}.
However, generative models either allow one to randomly sample from the latent space or interpolate between known latent representations, and hence offer only very limited control over the output shape, which hinders creativity. Thus, editing implicit representations for creative processes is not straightforward 
\cite{hertz2022spaghetti,DualSDFHao2020}.

In this paper, we approach the generation and editing of neural implicit shapes based on free-form sketching. Sketching is an intuitive and effective way to visually communicate shape information. Moreover, sketch-based modeling and editing can be particularly impactful in fields such as architecture, game development and product design, where 3D models are an essential part of the workflow.
Despite vigorous efforts in sketch-based 3D modeling, it remains a challenging problem: First, the reconstruction of a 3D shape from an image is inherently ill-posed, since a raw image without annotation is generally a representation of a 3D object merely from a single viewpoint. Second, sketches can vary significantly in style and abstraction level, ranging from fast, casual or even sloppy styles to professional, rigorous sketches. In this paper, we define \emph{abstract sketches} as hand-drawn representations that may lack geometric accuracy and focus more on capturing the essence or key features of the intended 3D shape rather than its exact specifications.
When assuming near-perfect correspondence between the sketched silhouettes or other shape features and the output shape, high quality results can be achieved, see e.g.\ \cite{Lun2017ShapeMVD, li2018robust, MonsterMash:2020, zhou2023gasketching}. Similarly, exceptional 3D results can be extracted from high quality input technical drawings that include 3D clues, such as hidden lines \cite{Li:2020:Sketch2CAD} or symmetric strokes \cite{hahnlein2022}.
However, designing a sketch-based 3D modeling system that is agnostic to the level of sketch abstraction of the input and the personal style of the user, accommodating inexact or unskilled drawings, is challenging.

Aside from using sketches to retrieve scenes for modeling \cite{eitz2012sbsr}, data-driven generating techniques have always been susceptible to being mere retrievals of the datasets \cite{tatarchenko2019singleview,forgery2022somepalli}. Providing guarantees that shape-generating systems create novel shapes is thus imperative.
We therefore approach the problem using a part-aware generative model to avoid this retrieval pitfall.
Part-aware modeling can mitigate the issue, since the generation first detaches the different parts, before assembling the whole shape coherently.  This motivates us to use SPAGHETTI \cite{hertz2022spaghetti}, a part-aware neural implicit shape representation model, as our backbone.

We present \ourmethod{}, a method that leverages part-aware neural implicit representation to output novel shapes out of an input sketch. Our framework decomposes the input sketch into patches that are fed into a Vision Transformer \cite{ViTDosovitskiy2020}. A transformer decoder then outputs the latent code into the latent space used by SPAGHETTI \cite{hertz2022spaghetti}. Using this space, editing can be applied to specific isolated parts of the shapes. For example, the user can manually select a part of the generated shape, such as the back of a chair, and redraw it by restricting the modification to the selected part only. Furthermore, our method offers the ability to systematically replace selected parts of a generated shape, providing an effective means of refining the model and removing any undesired artifacts. \ourmethod{} also offers the possibility to outline the obtained shape while modeling, granting the user the possibility to modify the sketch directly and lowering the sketching skill cap. 

We compare \ourmethod{} with state-of-the-art sketch-to-shape techniques, \new{encompassing both empirical and quantitative analyses. To illustrate that our method goes beyond simple shape retrieval, we present the top-4 shapes retrieved from the shapes generated by our approach.} We further validate the quality of \ourmethod{}'s generation ability via a comparative perceptual user study. We also showcase the editing possibilities of our method in an interactive environment. Our key contributions are:
\begin{itemize}
    \item Sketch-based modeling based on single-view sketches of diverse levels of abstraction.
    \item State-of-the-art results for shape generation with limited retrieval.
    \item New editing capabilities that allow for part-based shape refinement and localized sketch-based reshaping and combinations.
\end{itemize}

\section{Related work} \label{sec:related_work}

\begin{table}[t]
    \centering
    \caption{Comparison of sketch-based shape generation methods.}
    \label{tab:method_comparison}
    \small
\setlength{\tabcolsep}{2pt}
    \begin{tabular}{lccc}
        \toprule
        & Single view & Editing & Abstract sketches\\ \midrule
        ShapeMVD \cite{Lun2017ShapeMVD} & \xmark & \xmark & \xmark \\
        Pixel2Mesh \cite{pixel2mesh} & \cmark & \xmark & \xmark \\
        ProSketch \cite{zhong2021prosketch} & \xmark & \xmark & \xmark \\
        Sketch2Mesh \cite{Sketch2Mesh} & \cmark & \cmark & \xmark \\
        DeepSketch \cite{10.1016/j.cag.2022.06.005} & \cmark & \xmark & \xmark \\
        Ours & \cmark & \cmark & \cmark \\
        \bottomrule
    \end{tabular}
\end{table}

\begin{figure*}[t]
    \center
    \newcommand{\under}{-1.5}
    \begin{overpic}[trim= 2 2 2 2, clip, width=1.\linewidth]{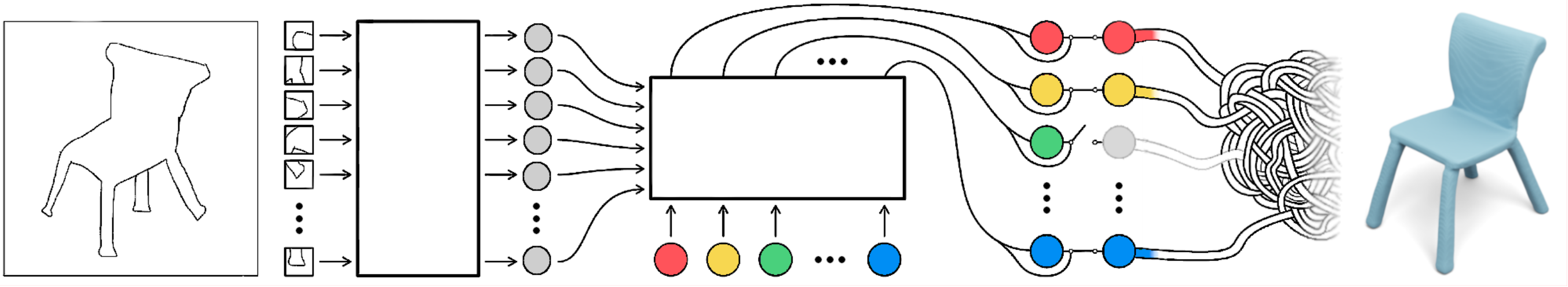}
    \put(2.5,  19){\textcolor{black}{Input sketch}}
    \put(18.1,  19){\textcolor{black}{Sketch encoding}}
    \put(37,  19){\textcolor{black}{Sketch to shape decoder}}
    \put(65,  19){\textcolor{black}{SPAGHETTI -- Shape decoder}}
    \put(18.5,  8){\textcolor{black}{ViT}}
    \put(34.3,  9){\textcolor{black}{transformer decoder}}
    \put(4,  \under){\textcolor{black}{\small Patch projection}}
    \put(16.8,  \under){\textcolor{black}{\small Visual embeddings}}
    \put(31.3,  \under){\textcolor{black}{\small Learned part queries}}
    \put(49,  \under){\textcolor{black}{\small Masked part embeddings}}
    \put(75,  \under){\textcolor{black}{\small Output shape}}
    
    \end{overpic}
    \caption{\ourmethod{} takes as input a $256\times 256$ normalized grayscale sketch. It is partitioned into $16 \times 16$ patches, and then passed through a Vision Transformer. A transformer decoder is then used to generate the latent variable $z \in \mathds{R}^{m \times d_{\text{model}}}$, which is a \iffalse GMM-based\fi part-aware latent space with $m$ parts represented by latent vectors of dimension $d_{\text{model}}$ that conditions the neural implicit representation given by SPAGHETTI, which is used to generate the output shape. By the part-aware latent space we get a mapping between sketch and shape parts.}
    \label{fig:methodSummary}
    \vspace{-0.8cm}
\end{figure*}


    

\textit{Sketch-based modeling.} Sketch-based modeling was extensively researched before the recent burst of data-driven techniques. As we focus on the latter, we only present a fragment of this domain and refer the reader to \cite{cani:hal-00336304, Bonnici2019} for a more complete survey. Teddy \cite{Igarashi1999} was one of the first modeling systems introduced for casual modeling, and has inspired many works since \cite{Tai2004, FiberMesh:2007, schmidt2008, bernhardt:inria-00336688, gingold2009, MonsterMash:2020, zhang2022creatureshop}. Some methods offer sketch-based creation by targeting a specific class of shapes, such as garment modeling systems \cite{turquin:inria-00510171, fondevilla:hal-03280215}. Virtual reality provides an environment in which sketches are three-dimensional, resolving partial ambiguities for shape modeling \cite{Verhoeven:RodMesh:2019, CassieYASBS21, yu:hal-03691151}. Using inputs with additional information such as concept sketches \cite{LiftingGHLSB20, hahnlein2022} or manual annotations \cite{xu2014} can facilitate reconstruction but requires higher sketching skills.

\textit{Neural networks shape representation types.}
The rise of deep learning for 3D geometry inspired the use of many shape representations. 
Explicit representations are popular for their expressiveness and editing possibilities. However, mesh representations require using graph neural networks \cite{meshcnnhanocka2018,pixel2mesh, feng2019meshnet}, which are computationally harder to process due to the inherent lack of regularity. Parametric representations offer mathematical accuracy but are hard to acquire and often rely on other representations for learning, such as meshes \cite{despoina2019}, point clouds \cite{gopal2020} or distance fields \cite{smirnov2020dps}.
Voxel representations leverage the regularity of the grid to ease the design of effective networks \cite{zhang2018, wu2018}, but they are resolution dependent and lead to poor representations of details.
Point clouds are easy to acquire and process but do not embed geometrical structures 
\cite{fan2016, Yin2018, yang2019}. We refer the reader to \cite{mirbauer2022} for a comprehensive survey on neural shape representations.

\textit{Neural implicit shape generation and modeling.} Neural implicit representations emerged as an alternative representation. DeepSDF learns the truncated sign distance function \cite{DeepSDFPark2019}, while other methods are based on a binary outside/inside classification \cite{Mescheder2019OccupancyNetworks, chen2018}. \new{In the realm of occupancy networks, some approaches have been developed to learn latent codes on regular grids \cite{peng2020convolutional}. However, recent advancements propose employing irregular grids for latent vector distribution \cite{zhang20223dilg}, or even utilizing sets of latent vectors \cite{3DShape2VecSetzhang2023}.} Since implicit representations are level-sets of a function, they are often restricted to closed meshes, though this can be avoided by learning unsigned distance functions \cite{chibane2020ndf,guillard2022udf}. They are also restricted to watertight surfaces and are difficult to modify directly. To solve these issues, mixed representations have also emerged. Deepcurrents handles boundaries using an explicit representation \cite{DeepcurrentPalmer2021}. Since implicit representations are hard to edit, DualSDF \cite{DualSDFHao2020} proposes a combined explicit representation that the user can edit. SPAGHETTI is a part-aware generative network \cite{hertz2022spaghetti} which relies on Gaussian mixture models to represent each part of the shape and provides editing via an affine transform of each Gaussian cluster. Part-aware representations can also help avoid the caveat of falling into mere retrieval \cite{componet2018, forgery2022somepalli}, which is a property we use in this work.

\begin{figure}[t]
\centering
	\scriptsize
	\setlength{\tabcolsep}{1pt}
	\begin{tabular}{ccccc}
    volumetric
    & outline
    & partial outline 
    & abstract sketch
    & freehand sketch \\
    tracing
    & rendering
    & rendering
    & \cite{vinker2022clipasso}
    &  \cite{zhong2021prosketch} \\
    \includegraphics[width=0.19\linewidth]{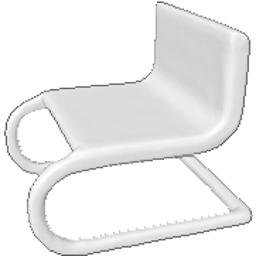}\ & \includegraphics[width=0.19\linewidth]{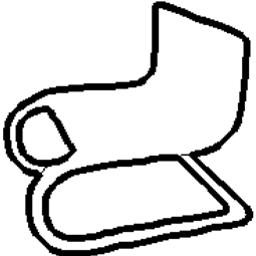}\ &
	\includegraphics[width=0.19\linewidth]{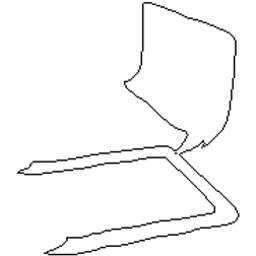}\ &
	\includegraphics[width=0.19\linewidth]{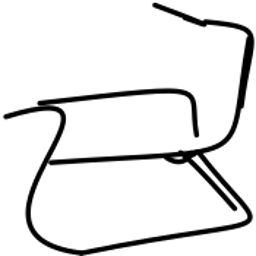}\ & \includegraphics[width=0.19\linewidth]{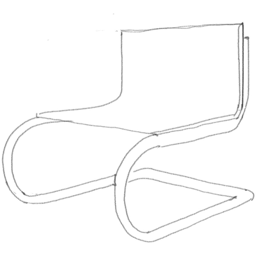} \\ 
 (i) & (ii) & (iii) & (iv) & (v)
	\end{tabular}
  \caption{We used a variety of sketch styles as inputs to our method. The target shape is an implicit shape rendered via volume rendering, outline rendering and partial outline rendering. The abstract sketch is produced using CLIPasso \cite{vinker2022clipasso} on the volume rendering. Expert freehand sketches come from ProSketch \cite{zhong2021prosketch}.}
  \label{fig:sketches}
  \vspace{-0.4cm}
\end{figure}

\textit{Neural sketch-to-mesh methods.} Using neural networks for sketch-based modeling is an active area of research in computer graphics, and there has been notable progress in recent years in developing neural network-based approaches for generating 3D models from 2D sketches. 
ShapeMVD \cite{Lun2017ShapeMVD} and SketchCNN \cite{li2018robust} reconstruct shapes from 2D sketches using a convolutional neural network, but require multiple views and do not support abstract sketches. ProSketch \cite{zhong2021prosketch} and DeepSketch \cite{10.1016/j.cag.2022.06.005} are trained on a mix of synthetic and professional sketches. Some view-aware modeling systems exist: Sketch2Mesh \cite{Sketch2Mesh} proposes an encoder/decoder architecture to reconstruct 3D shapes that can be refined via a user interface; Garment Ideation \cite{GarmentIdeation} is a feature aggregation-based iterative method targeted towards garment ideation that predicts a winding number to generate 3D shapes; concurrently to our work, LAS-Diffusion \cite{zheng2023lasdiffusion} proposes a multi-class diffusion method based on an attention mechanism and GA-Sketching \cite{zhou2023gasketching} proposes a multi-view method with modeling options via iterative refinement, but they fall short in effectively processing abstract sketches. Edit3D \cite{Cheng2022-edit3d} employs a unified latent space to generate 3D shapes, sketches, and RGB images, thereby establishing a correspondence between these three types of representations that enables shape and color editing.
Delanoy et al. \cite{delanoy2017} propose a method to recover a volumetric shape from an input sketch.
Parametric representations are also used for sketches \cite{smirnov2019}. Sketch2CAD \cite{Li:2020:Sketch2CAD} is based on the generation of primitives, Free2CAD \cite{Li:2022:Free2CAD} decomposes an input sketch into a sequence of strokes that are mapped to a sequence of CAD instructions, and GeoCode  \cite{GeoCode2022} offers sketch-based modeling of parametric shapes with additional part-aware control of the relevant parameters.
Note that neural methods can also recover shapes from non-sketch images. Pixel2Mesh \cite{pixel2mesh} recovers a mesh from an image while 3D-R2N2 \cite{3D-R2N2choy2016} and NeRFs \cite{mildenhall2020nerf} can reconstruct a shape from multiple views. SKED \cite{mikaeili2023sked} is a NeRF-based method which provides a sketch-guided text-based shape editing method.
\tableref{tab:method_comparison} presents the strengths and weaknesses of the works most related to ours.

\section{Method} \label{sec:method}

\ourmethod{} generates a neural implicit shape from a \emph{single-view} input sketch. More specifically, it associates to a sketch a latent code that can be interpreted by a neural implicit shape decoder. To this end, we design a neural network that learns to match a sketch to its corresponding shape's latent code in the latent space of SPAGHETTI \cite{hertz2022spaghetti}. \new{SPAGHETTI is designed to convert a latent vector into a collection of $m$ Gaussians, where each Gaussian represents a part of the object. Subsequently, each part goes through a ``mixing network'', a transformer encoder that ensures global consistency across the shape. An ``occupancy network'' follows for decoding the final shape, as it returns the signed distance function from a query point. The main property of SPAGHETTI that we use lies in the fact that it} is a \emph{part-aware} implicit shape decoder, which means that its latent space is divided in several parts, and each part of the latent space encodes for a corresponding part in the resulting shape. This feature enables to train our network on partial inputs to mitigate shape retrieval, to train a refinement network that can regenerate selected latent parts, and to restrict the shape generation to specific parts during the modeling process.

\subsection{Data generation and input normalization}

To improve our network robustness with respect to the style and the level of abstraction of the input, we use a dataset with a variation of designs. Our dataset is based on a subset of the ShapeNet dataset \cite{shapenet2015chang}: the chair dataset with 6755 shapes, the lamp dataset with 833 shapes, and the airplane dataset with 1775 shapes. Each was rendered with six different views in three different manners: (i) volumetric rendering that relies on ray marching; (ii) outline rendering based on the depth map; and (iii) partial outline rendering, which are renders of SPAGHETTI's shapes after masking out parts of their latent code. In addition, (iv) abstract sketches of eight strokes were computed based on each view of the volumetric renderings by using CLIPasso \cite{vinker2022clipasso} with 2000 iterations. For chairs, we used an additional dataset, (v) ProSketch \cite{zhong2021prosketch}, to add freehand sketches drawn by experts. We display examples from our sketch dataset in \figref{fig:sketches}.
The data is augmented with random perspective transformation and horizontal symmetry. Using the fact that CLIPasso provides vector graphics outputs, we applied data augmentation to its abstract sketches by modifying the stroke width before rendering it as an image. We normalize the input by centering the sketch, cropping the empty borders and resizing it to a $256\times256$ image. Partial outline renderings are normalized and cropped in alignment with their respective full renders.

\subsection{Sketch-to-latent representation}   

Our network maps a sketch to the latent representation of a neural implicit shape generator, namely SPAGHETTI \cite{hertz2022spaghetti}. 
SPAGHETTI receives as input a latent representation that is mapped to a collection of $m$ vectors of dimension $d_{\text{model}}$ that represents a Gaussian mixture model (GMM), i.e. each of these $m$ vectors corresponds to a 3D Gaussian. SPAGHETTI outputs a 3D implicit shape by mapping each Gaussian to a part of the represented shape, and mixes these parts to produce a globally coherent shape.
In this work, we make use of this intermediate GMM-based latent space and map the input sketch directly to it. For each given shape, we precompute its collection of latent vectors $\{z_i\}_{i=1}^m$ using shape inversion \cite{hertz2022spaghetti}.
An overview of our network architecture is displayed in \figref{fig:methodSummary}.
Inspired by the DETR object detection model \cite{carion2020end}, our network is composed of an image encoder that takes an input sketch and outputs visual embeddings. A transformer decoder maps a set of learned part queries together with these visual embeddings to SPAGHETTI's multi-part latent space.   
The image encoder (\figref{fig:methodSummary} left) is a Vision Transformer network \cite{ViTDosovitskiy2020}. It divides $256\times 256$ sketch images into $16\times16$ patches. Each patch is mapped to a single visual embedding via a transformer encoder.
The transformer decoder (\figref{fig:methodSummary} middle) takes as input these visual embeddings and a set of $m$ part queries, and processes them using its self-attention and cross-attention layers. The part queries are \emph{learnable} vectors, \textit{i.e.} they are optimized at the same time as the network.
Finally, each output vector of the decoder is mapped to a latent part vector $\{\Tilde{z}_i\}_{i=1}^m$ of the neural implicit shape decoder, SPAGHETTI, which uses them to generate the output shape (\figref{fig:methodSummary} right).
The training loss we use is
%
$$\mathcal{L}_{\text{full}} = \dfrac{1}{m}\sum_{i=1}^{m} \Vert \Tilde{z}_{i} - z_i \Vert_{1},$$
where $z_i$ is the ground truth $i$th part vector of the 3D shape that corresponds to the input sketch, and $\tilde{z}_i$ is the prediction of \ourmethod{}.

\subsection{Partial shape}
\label{par:partialshape}

\ourmethod{} is trained to perform reconstruction also by additional outline renders of \emph{partial} 3D shapes. The goal is to reinforce the uncoupling between parts of the restored shape as demonstrated in Sec. \ref{sec:ablations}. 

The \emph{partial outline rendering} supervision for this task is obtained by randomly selecting a subset of part vectors $\{z_i[c_i]\}_{i=1}^{m}$ where the binary assignment $c_i$ indicates the presence of part $i$ in the subset. Then the subset of vectors is given to SPAGHETTI which generates the corresponding partial 3D implicit shape. Finally, we render the partial shape. See \figref{fig:sketches}(iii) for an example.

When feeding partial outline renders into \ourmethod{}, we use different loss functions. In this case, the output of the transformer decoder is passed through an MLP to an additional classification score $\Tilde{c_i} \in [0, 1]$ which indicates the presence of part $i$ in the input outline render. We optimize it by the binary cross entropy loss,
%
$$\mathcal{L}_{\text{cls}} = \dfrac{1}{m}\sum_{i=1}^{m} BCE(\Tilde{c_i}, c_i),$$
where $c_i$ is the ground truth indicator of part $i$ in the input render.
Moreover, the loss for the latent vector prediction of our network is
%
$$\mathcal{L}_{\text{part}} =  \dfrac{1}{\Vert \textbf{c} \Vert_{0}}\sum_{i=1}^{m} c_{i} \Vert \Tilde{z_i} - z_i\Vert_{1},$$
where $c_{i}$ are used to ignore latents of parts not present at the input and the normalizing factor $\Vert \textbf{c} \Vert_{0}$ counts the number of non-zero entries in $\textbf{c} = \left[ c_1, ..., c_m\right]$.

\begin{figure}[t]
 	\centering
 	\small
 	\setlength{\tabcolsep}{1pt}
 	\begin{tabular}{cccc}
     \includegraphics[width=0.24\linewidth]{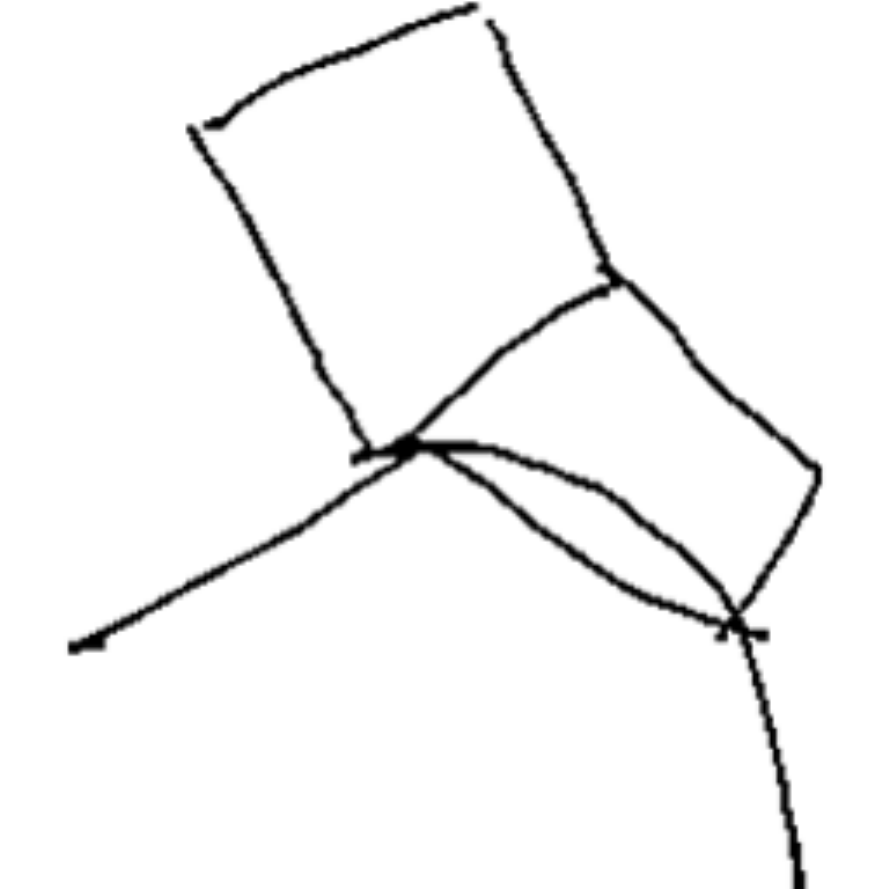}\
     & \includegraphics[trim= 120 40 120 150, clip, width=0.24\linewidth]{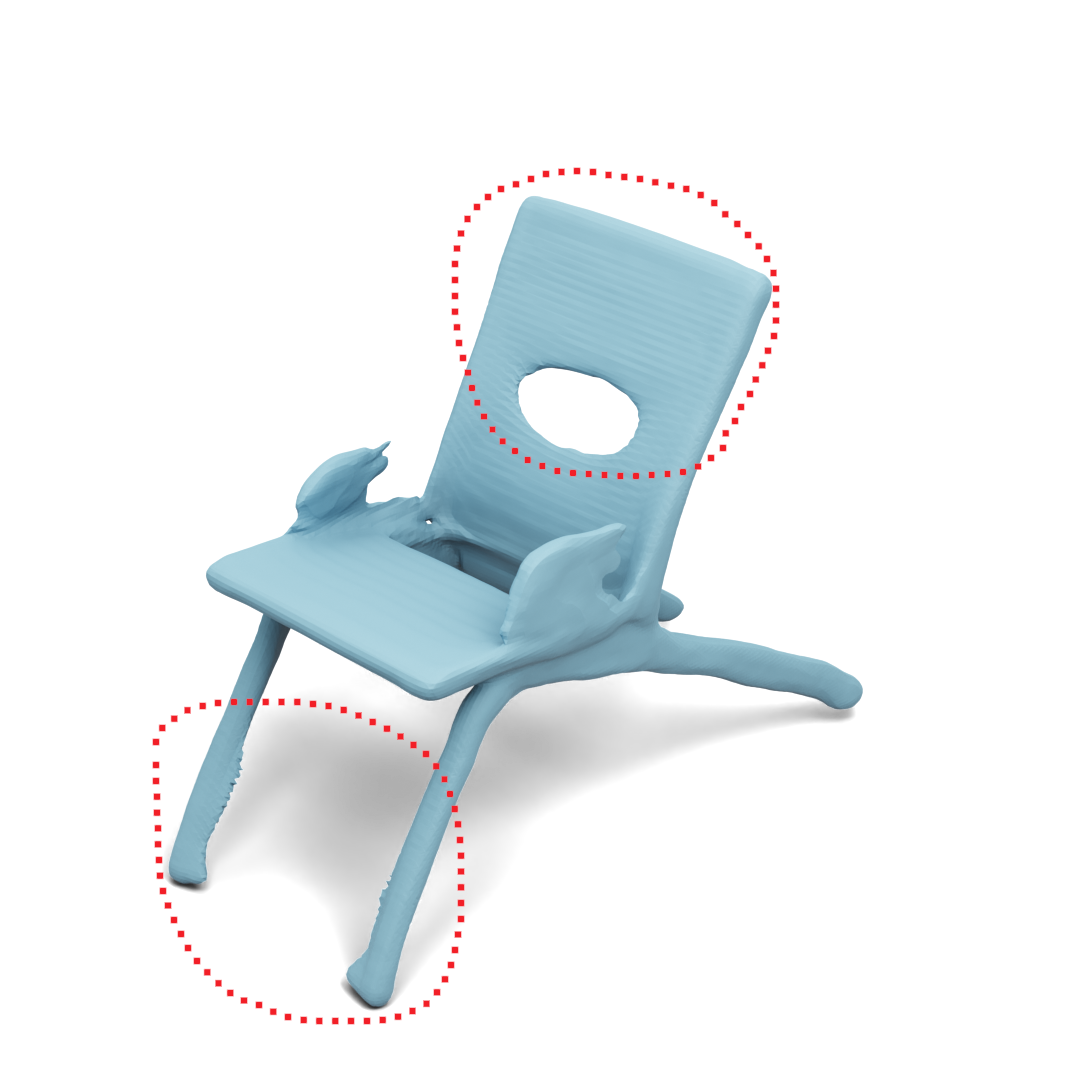}\
     & \includegraphics[trim= 120 40 120 150, clip, width=0.24\linewidth]{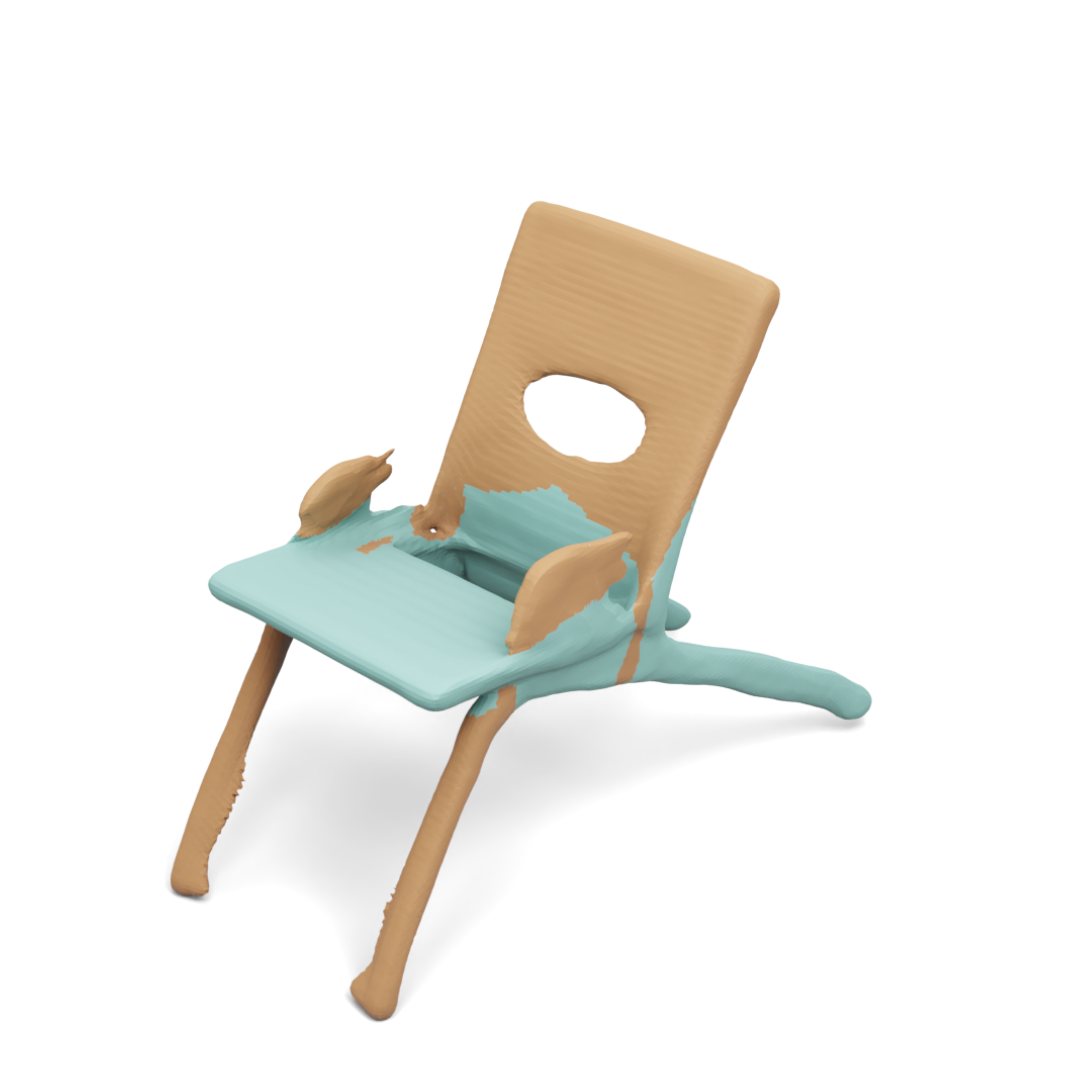}\
     & \includegraphics[trim= 120 40 120 150, clip, width=0.24\linewidth]{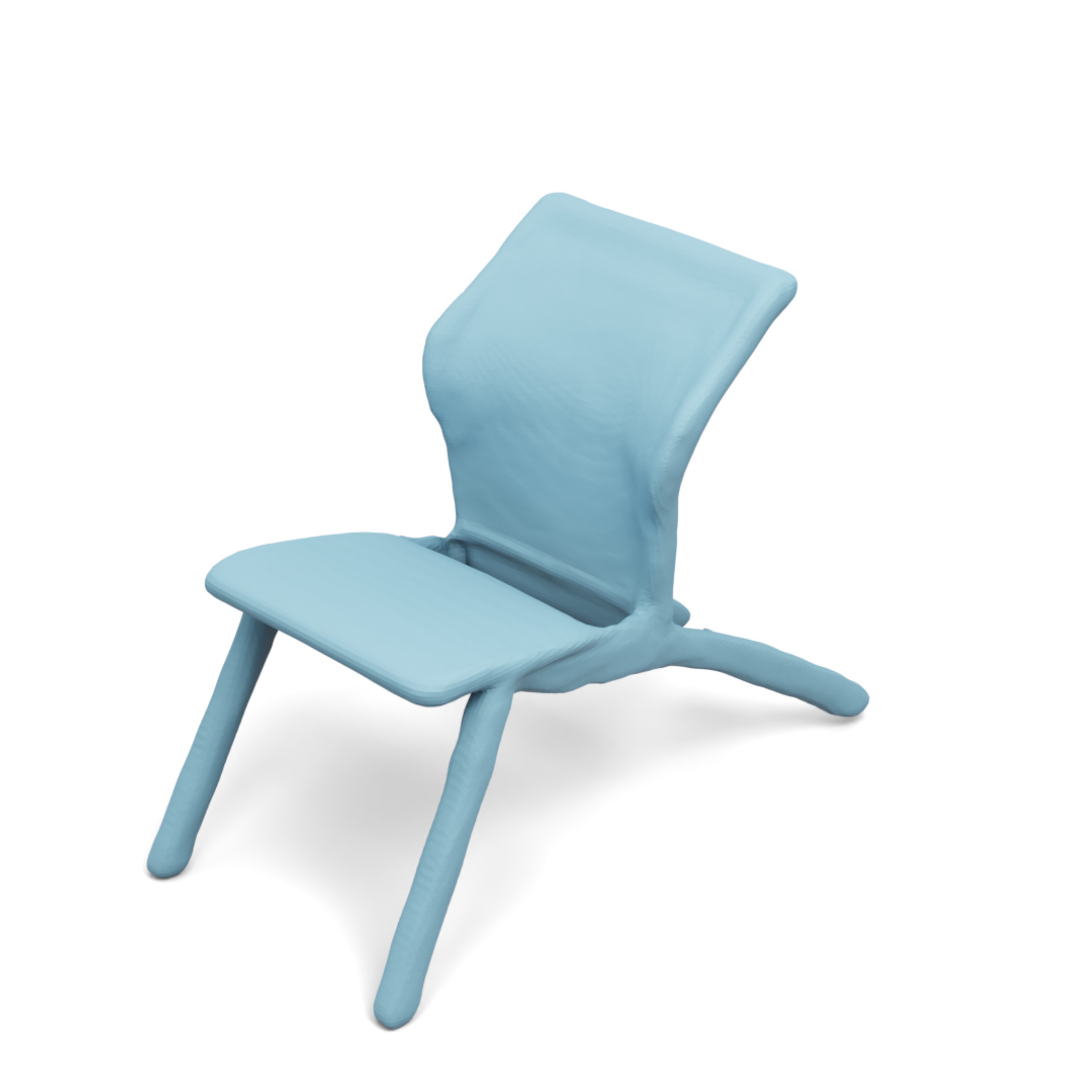}\
     \\
     (i) & (ii) & (iii) & (iv)
 	\end{tabular}
  \vspace{-0.2cm}
 	\caption{A sketch of poor quality (i) may yield inadequate results (ii). Users can select unsatisfactory parts of the output (ii, lasso selection on shape in red; iii, selected parts in orange). Our refinement network can predict a refined shape (iv) by regenerating the selected parts of the latent space based on the non-selected parts.}
 	\label{fig:refinement}
  \vspace{-0.4cm}
 \end{figure}

\subsection{Refinement network}
\label{sec:refinementNetwork}

The refinement network allows to regenerate parts of a given shape, and is illustrated in \figref{fig:refinement}. In some cases, poor quality or ambiguities in the input sketch (i) may lead to artifacts in the generated shape (ii). The user can select unsatisfactory parts from the output shape (iii, marked in orange). A selected part on the shape has a corresponding latent vector part in the GMM latent space.
Our refinement network, which is conditioned on the latent vector parts of the non-selected parts, outputs a set of vectors parts that replace the selected ones. Finally, the shape decoder regenerates the refined shape using the new latent vector parts (iv).

The refinement network is a bidirectional transformer encoder network that receives the set of latent vectors $\Tilde{z} \in \mathds{R}^{m\times d_{\text{model}}}$ such that the corresponding vectors of the selected parts are masked (i.e., zeroed).
It outputs ${\hat{z}} \in \mathds{R}^{m\times d_{\text{model}}}$, which contains the refined vectors in the entries corresponding to the selected parts.

The network uses a masking objective \cite{devlin2018bert}, where $5-40\%$ of the input vectors are masked, and the network has to predict their content based on the unmasked context. The loss is
$$\mathcal{L}_{\text{refine}} = \dfrac{1}{\Vert \mathds{1} \Vert_1}\sum_{i=1}^{m} \mathds{1}_i \Vert \hat{z}_{i} - z_i \Vert_{1},$$
where the indicator $\mathds{1}_i$ equals one if and only if the input vector $ \tilde{z}_i$ was masked and $\Vert \mathds{1} \Vert_1 = \sum_{i=1}^{m}{\mathds{1}_i}$ is a normalizing factor.

\section{Results} \label{sec:results}

\begin{figure}[t]
	\centering
	\small
	\setlength{\tabcolsep}{1pt}
	\begin{tabular}{cccc}
    \includegraphics[width=0.25\linewidth]{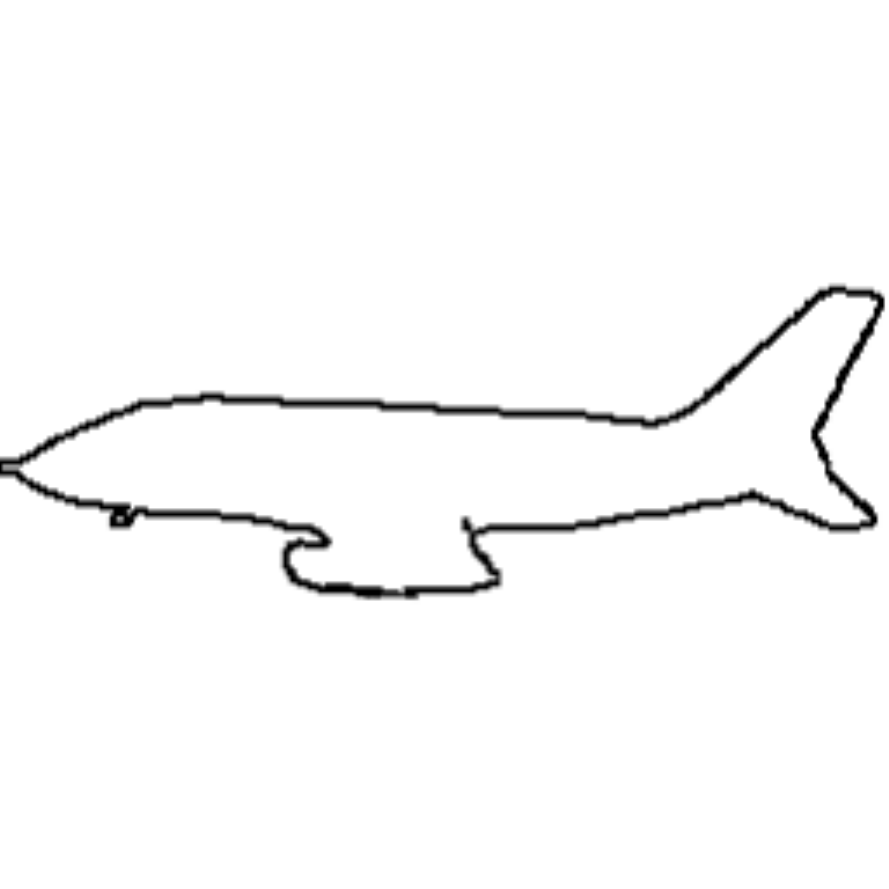}\ &
    \includegraphics[trim=60 60 60 60, clip, width=0.25\linewidth]{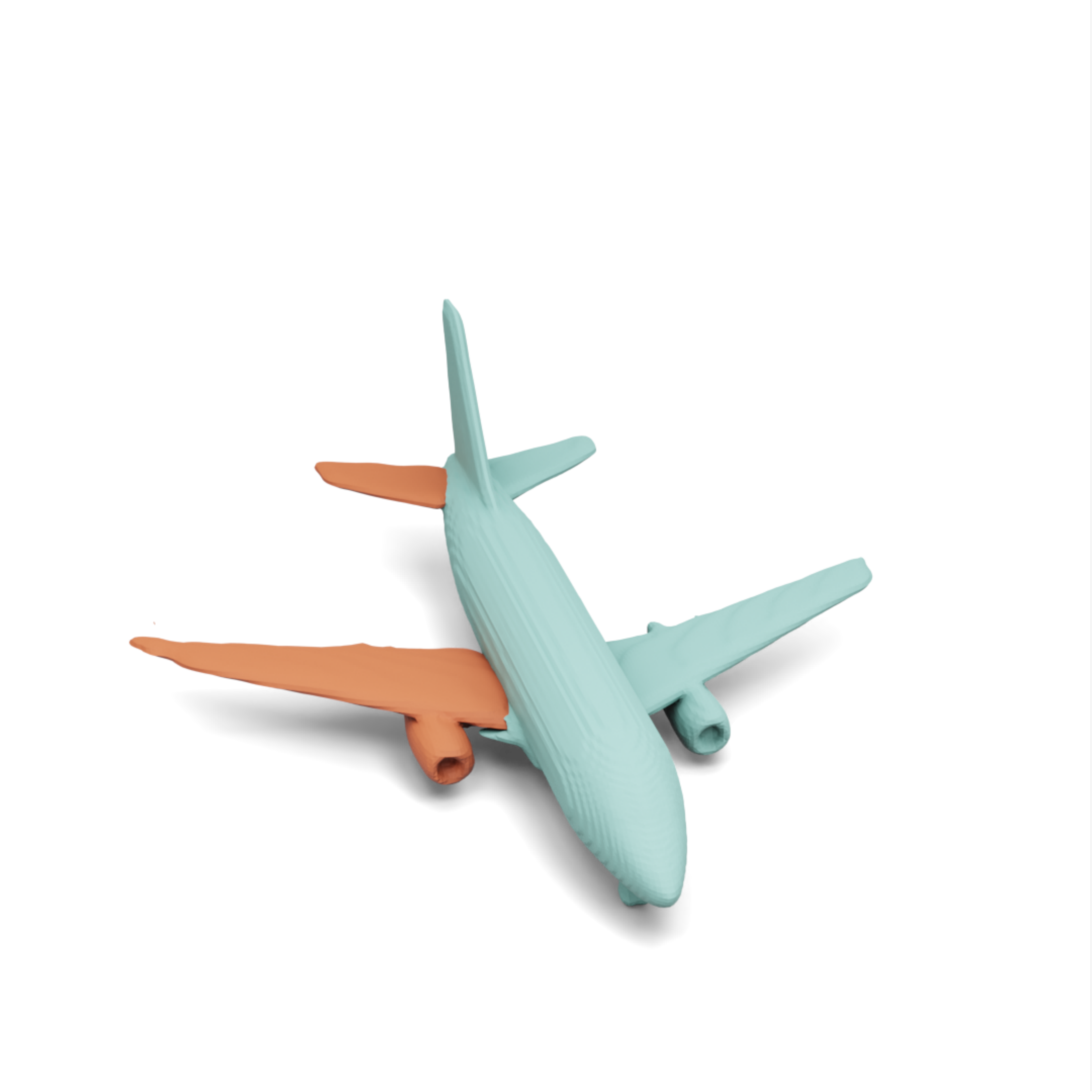}\
    & \includegraphics[width=0.25\linewidth]{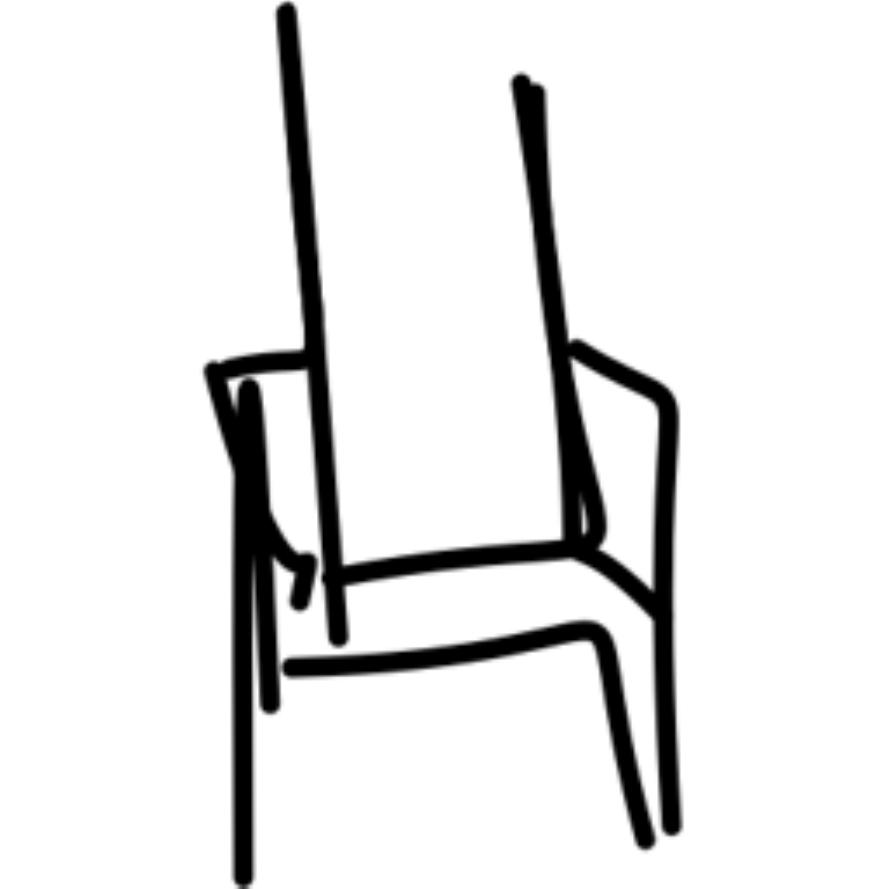}\
    & \includegraphics[trim=10 20 10 100, clip, width=0.25\linewidth]{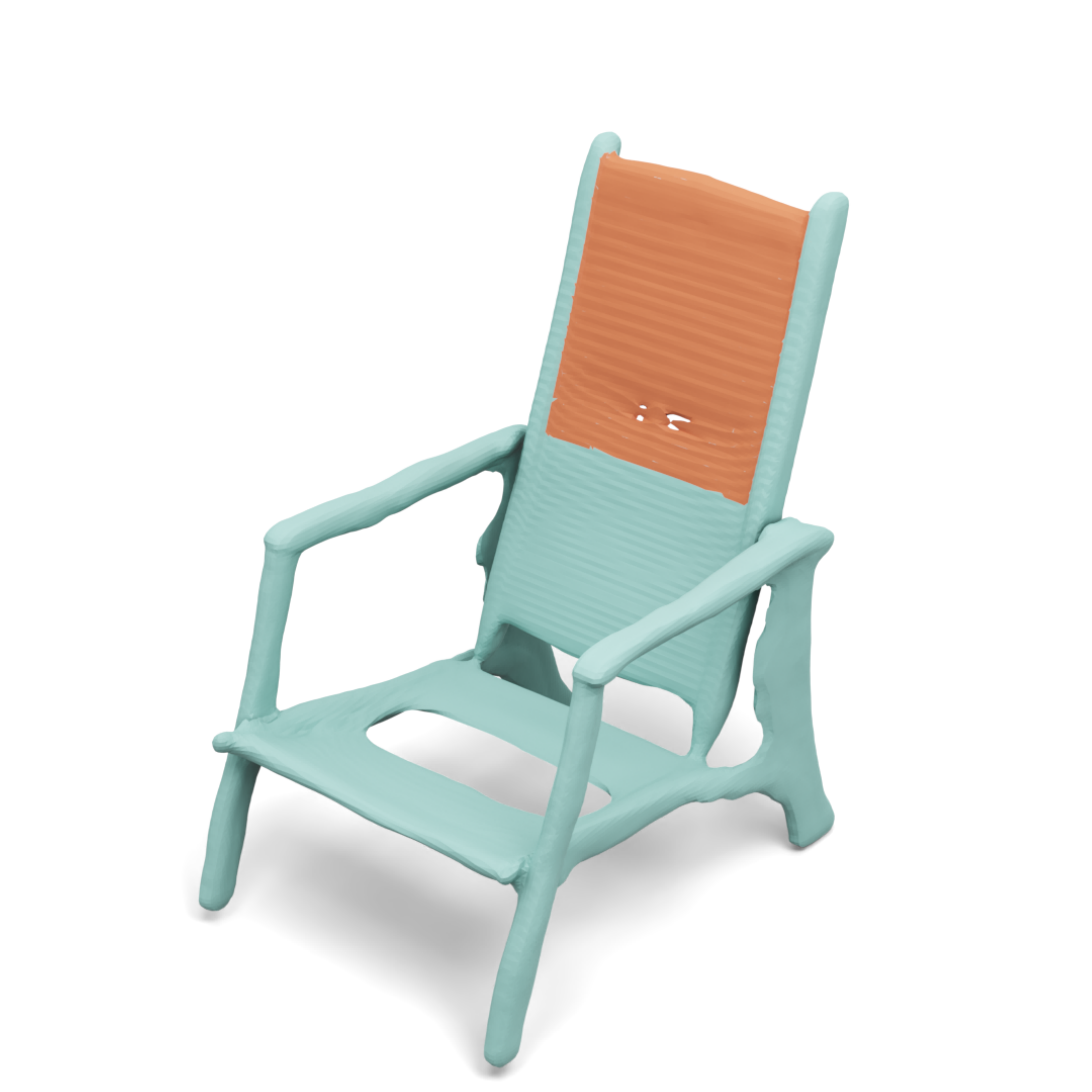}
	\end{tabular}
	\caption{We exemplify how our network performs shape completion from single-view sketches. If input sketch does not display the full shape, the network is still able to reconstruct it, notably taking advantage of the symmetry of the class of shapes in the dataset.}
	\label{fig:completion}
 \vspace{-0.5cm}
\end{figure}

\begin{figure*}[ht!]
	\centering
	\small
	\setlength{\tabcolsep}{1pt}
	\begin{tabular}{ccccc}
	\toprule
    Input sketch
    & Pixel2Mesh
    & Sketch2Mesh
    & DeepSketch
    & Ours
    \\
    
    & \cite{pixel2mesh}
    & \cite{Sketch2Mesh}
    & \cite{10.1016/j.cag.2022.06.005}
    & 
    \\
	\includegraphics[width=0.16\linewidth]{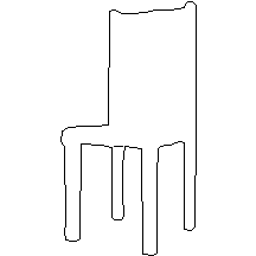}\ & \includegraphics[trim=5 5 5 5, clip, width=0.16\linewidth]{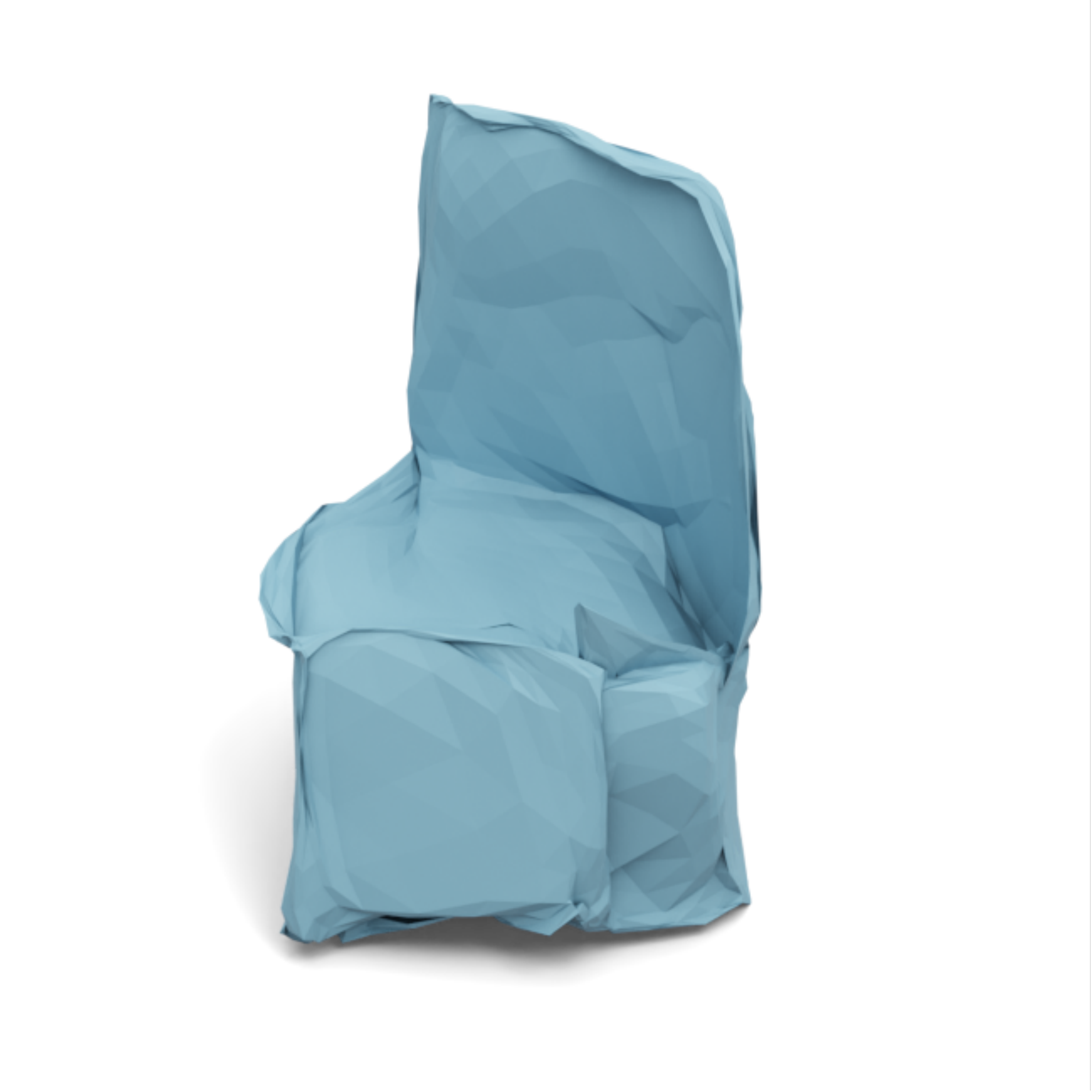}\ &
	\includegraphics[trim=5 5 5 5, clip, width=0.16\linewidth]{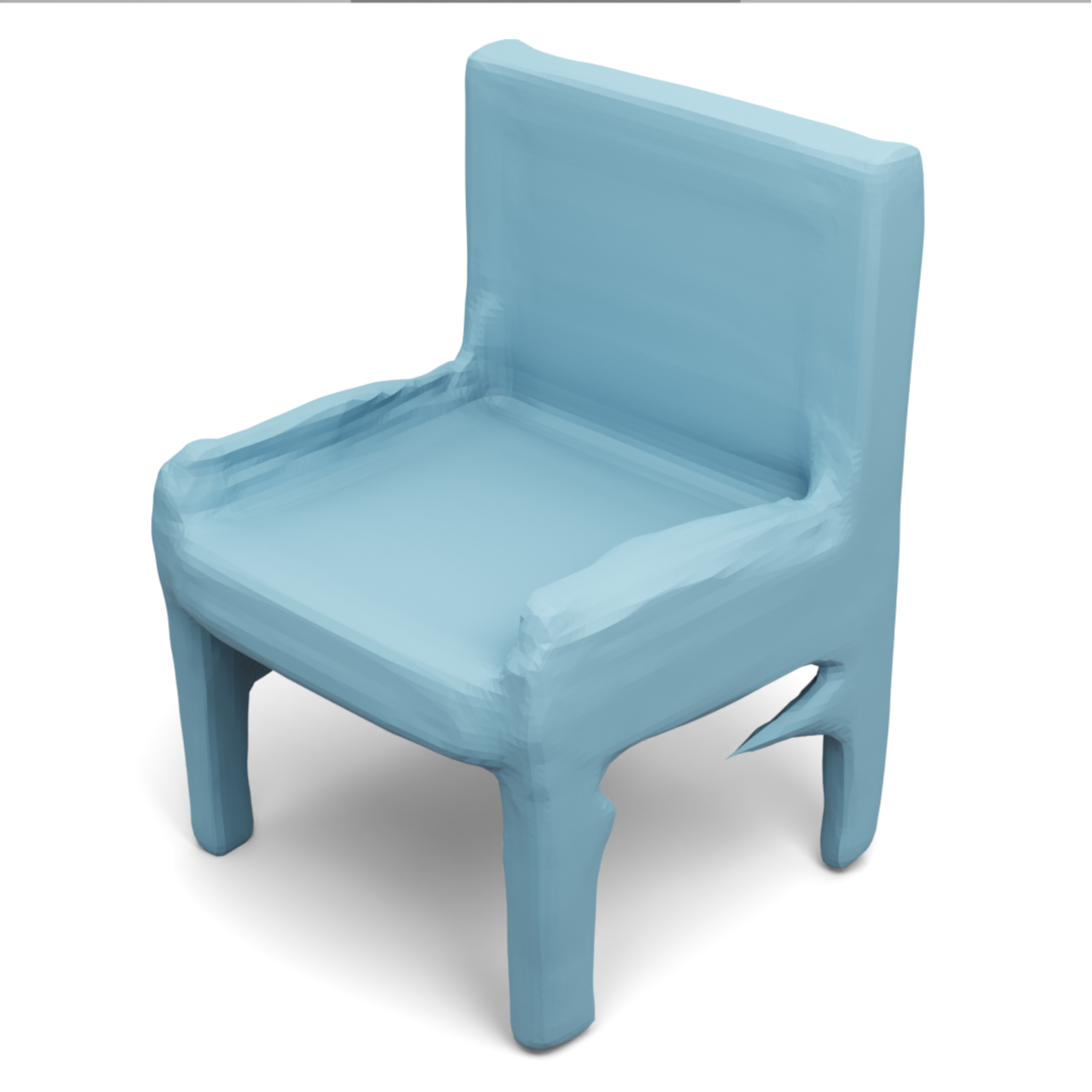}\ &
	\includegraphics[trim=160 50 160 250, clip, width=0.16\linewidth]{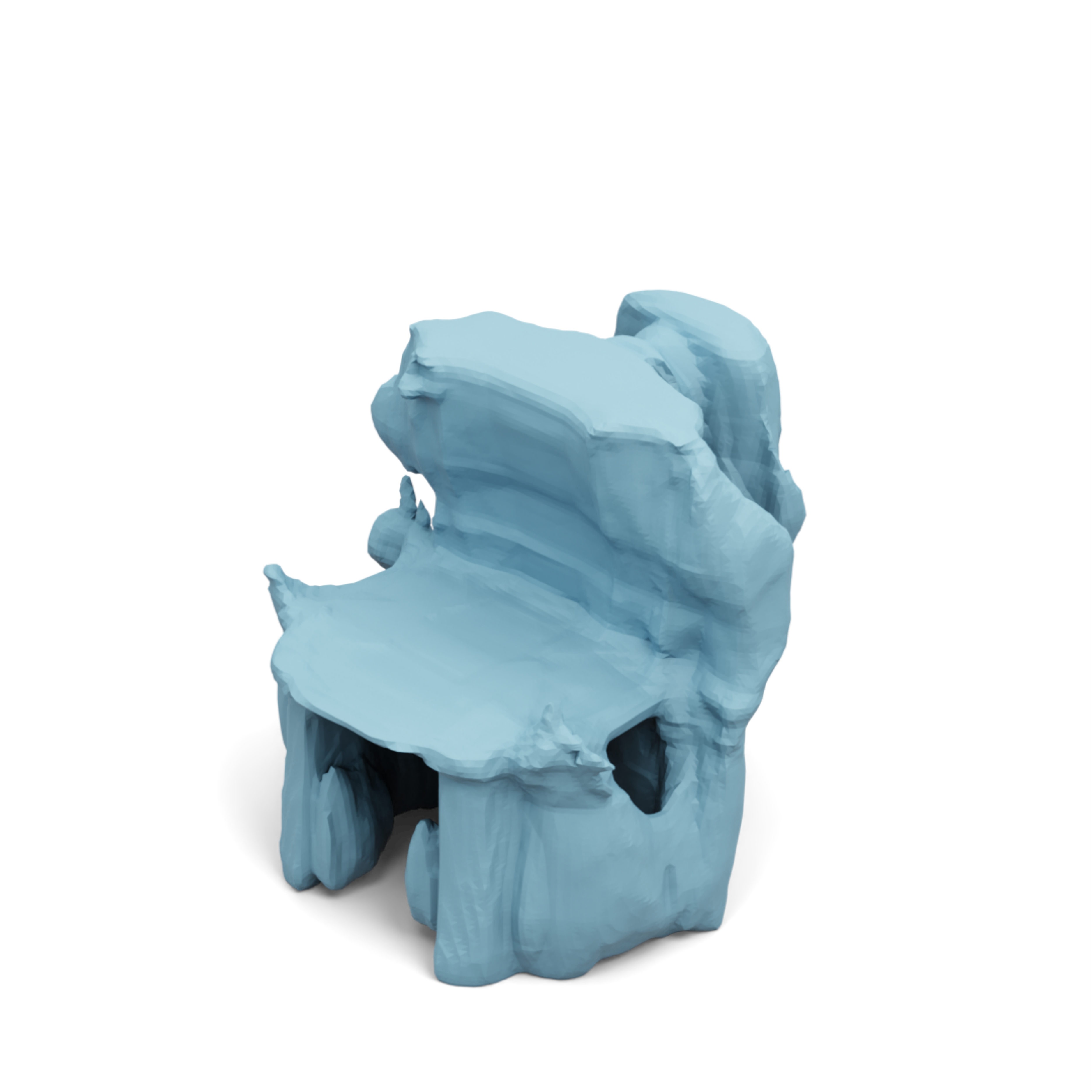}\ & \includegraphics[trim=5 5 5 5, clip, width=0.16\linewidth]{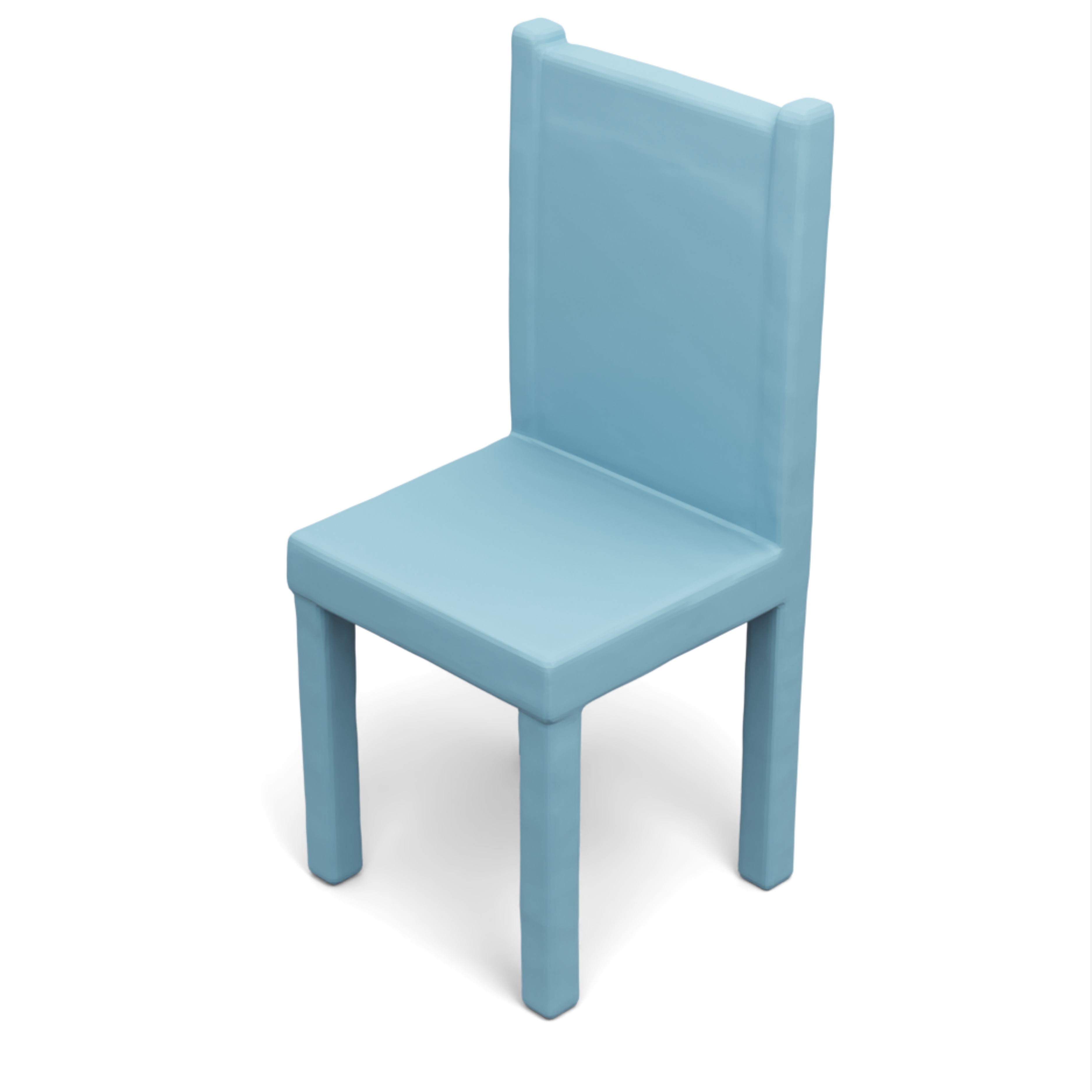}\\
 \midrule
	\includegraphics[width=0.16\linewidth]{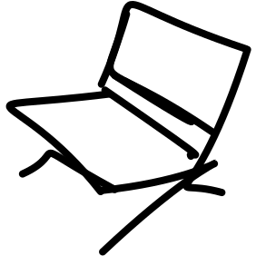}\ & \includegraphics[trim=5 5 5 5, clip, width=0.16\linewidth]{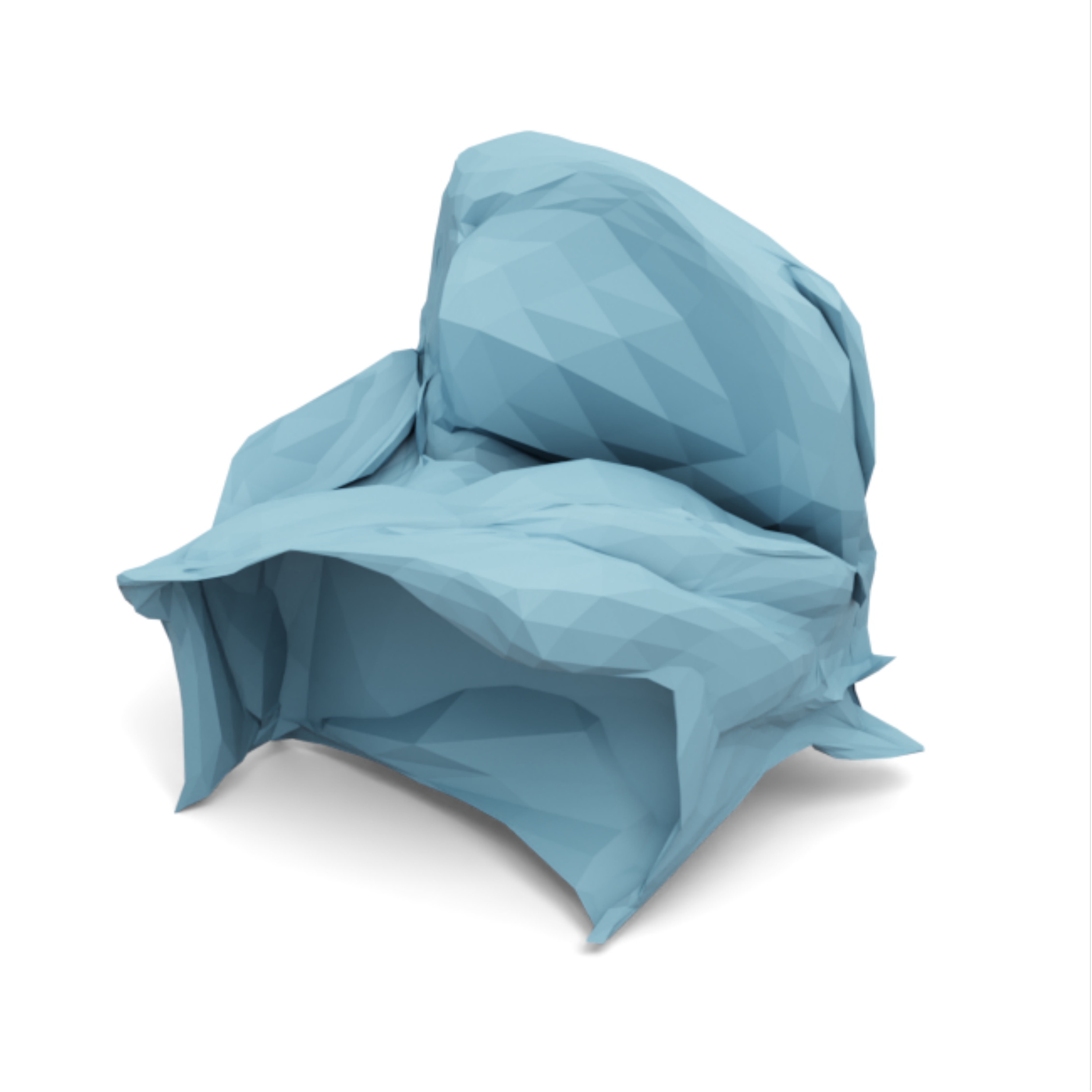}\ &
	\includegraphics[trim=5 5 5 5, clip, width=0.16\linewidth]{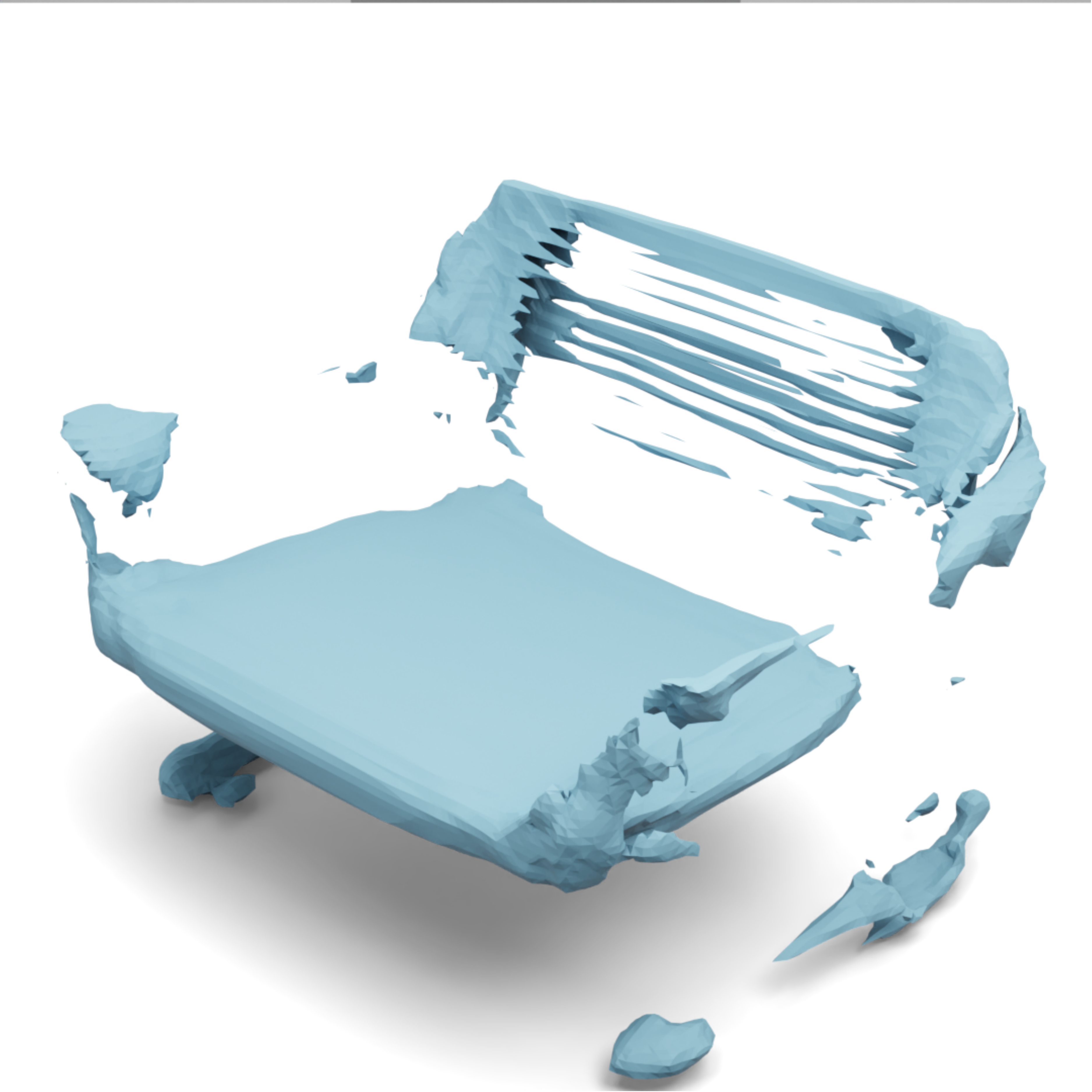}\ &
	\includegraphics[trim=160 50 160 250, clip, width=0.16\linewidth]{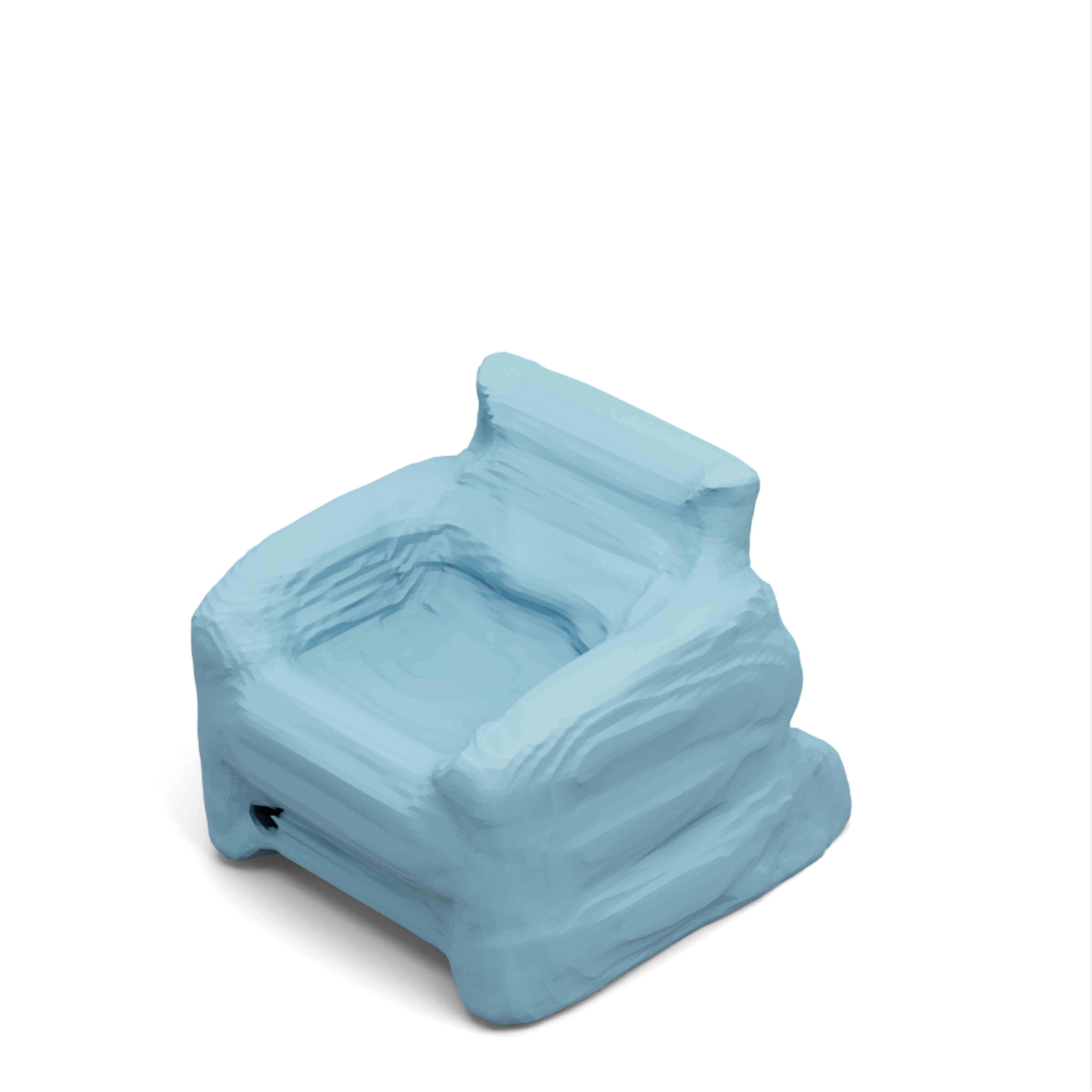}\ & \includegraphics[trim=5 5 5 5, clip, width=0.16\linewidth]{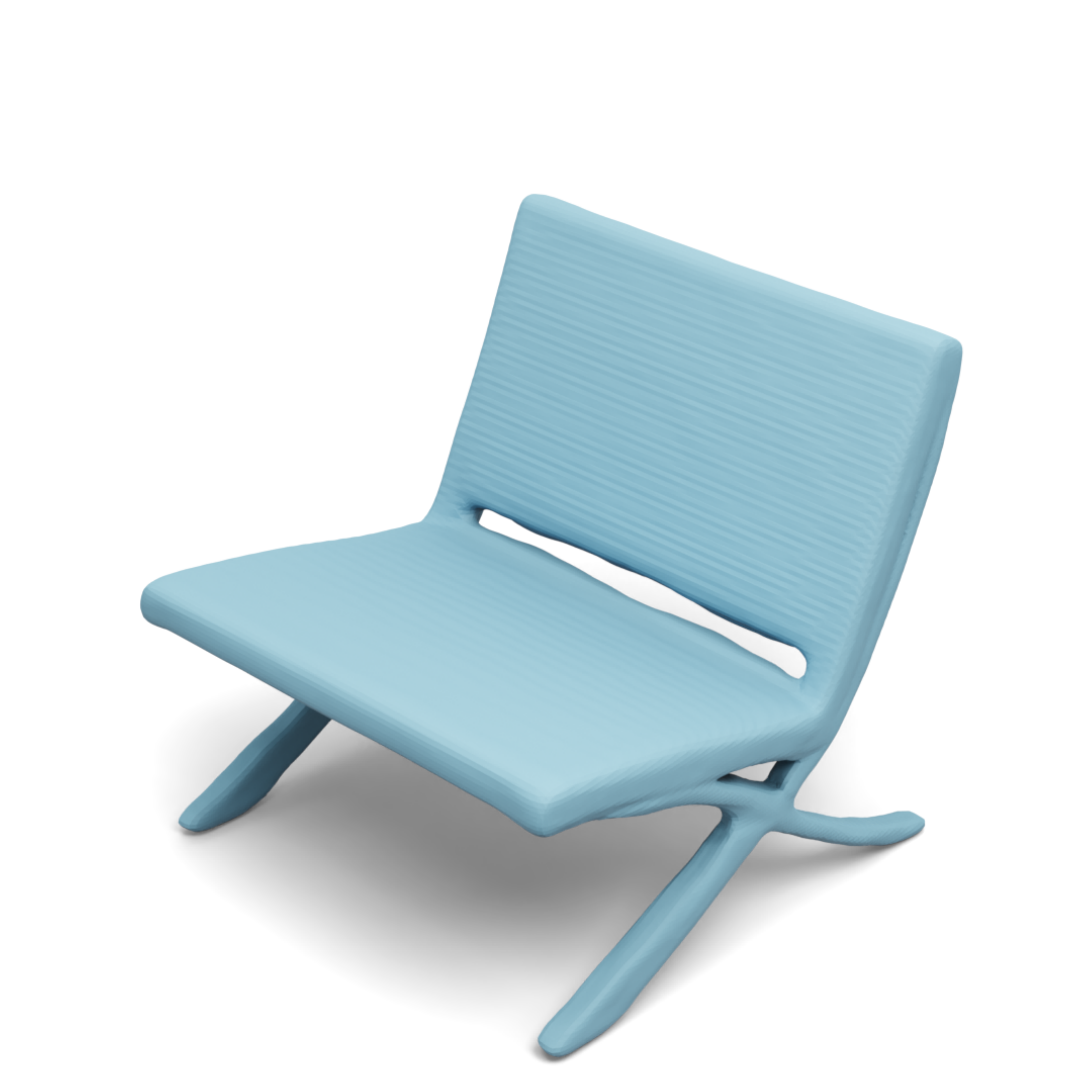}\\
 \midrule
	\includegraphics[width=0.16\linewidth]{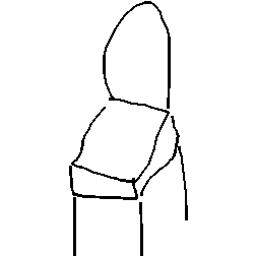}\ & \includegraphics[trim=5 5 5 5, clip, width=0.16\linewidth]{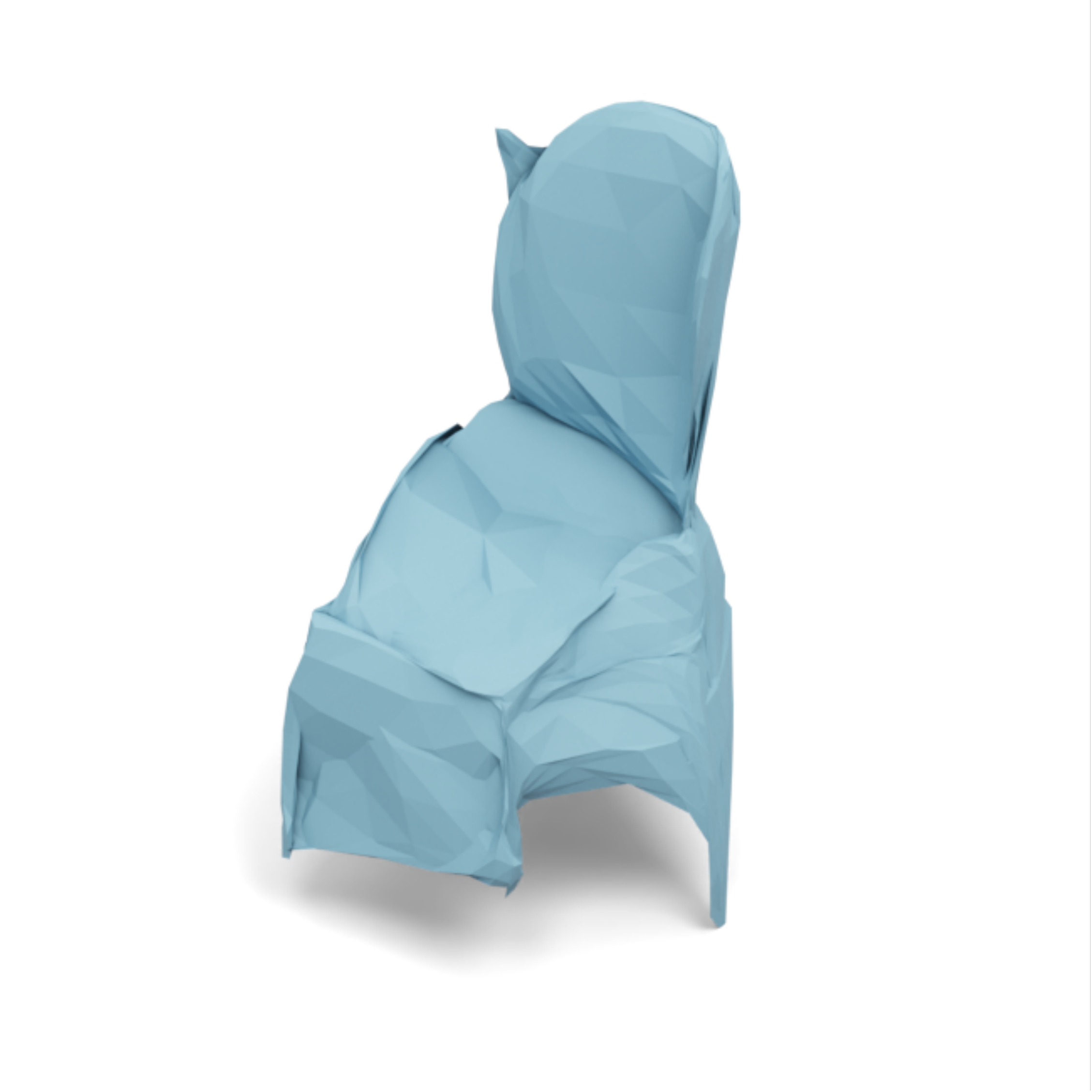}\ &
	\includegraphics[trim=5 5 5 5, clip, width=0.16\linewidth]{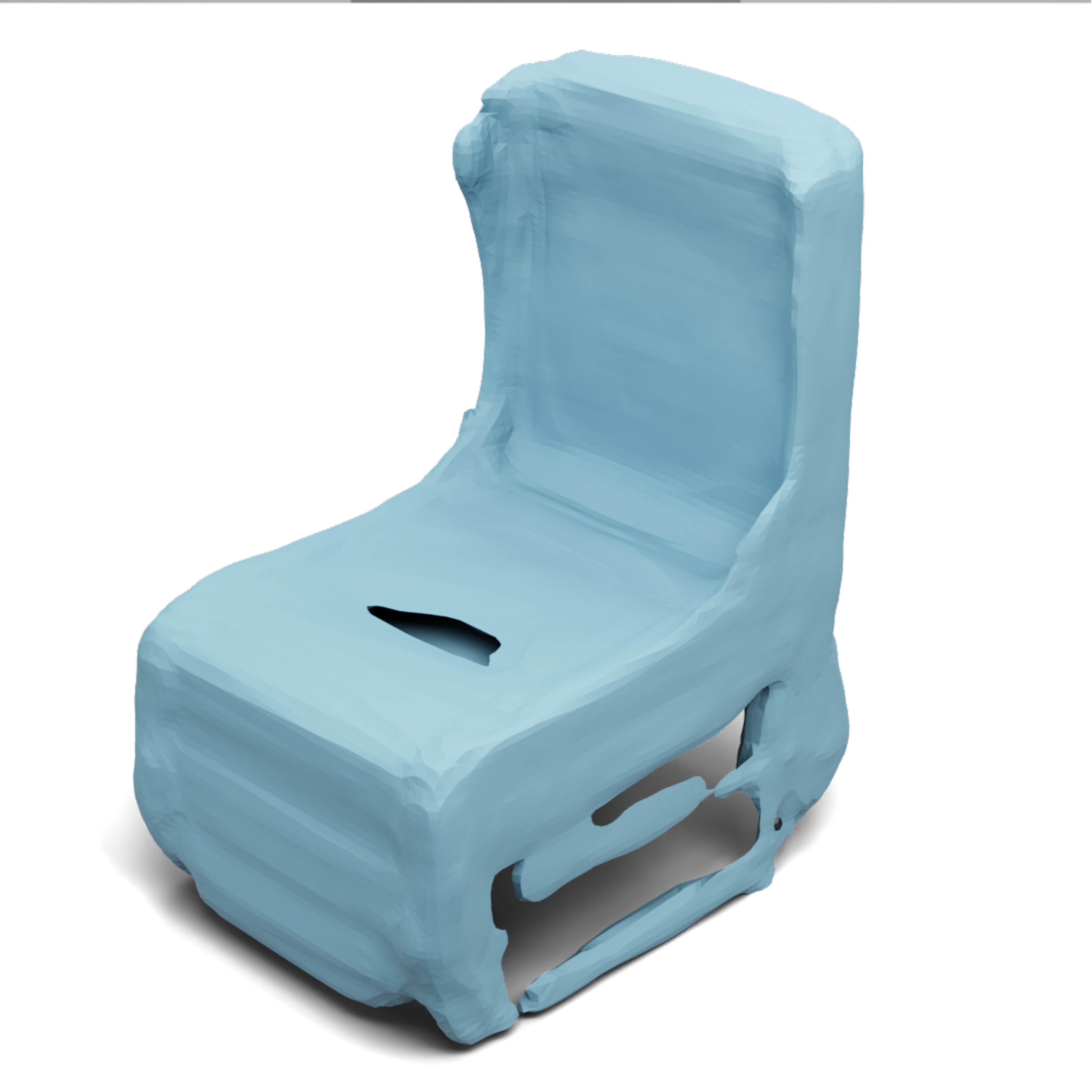}\ &
	\includegraphics[trim=160 50 160 250, clip, width=0.16\linewidth]{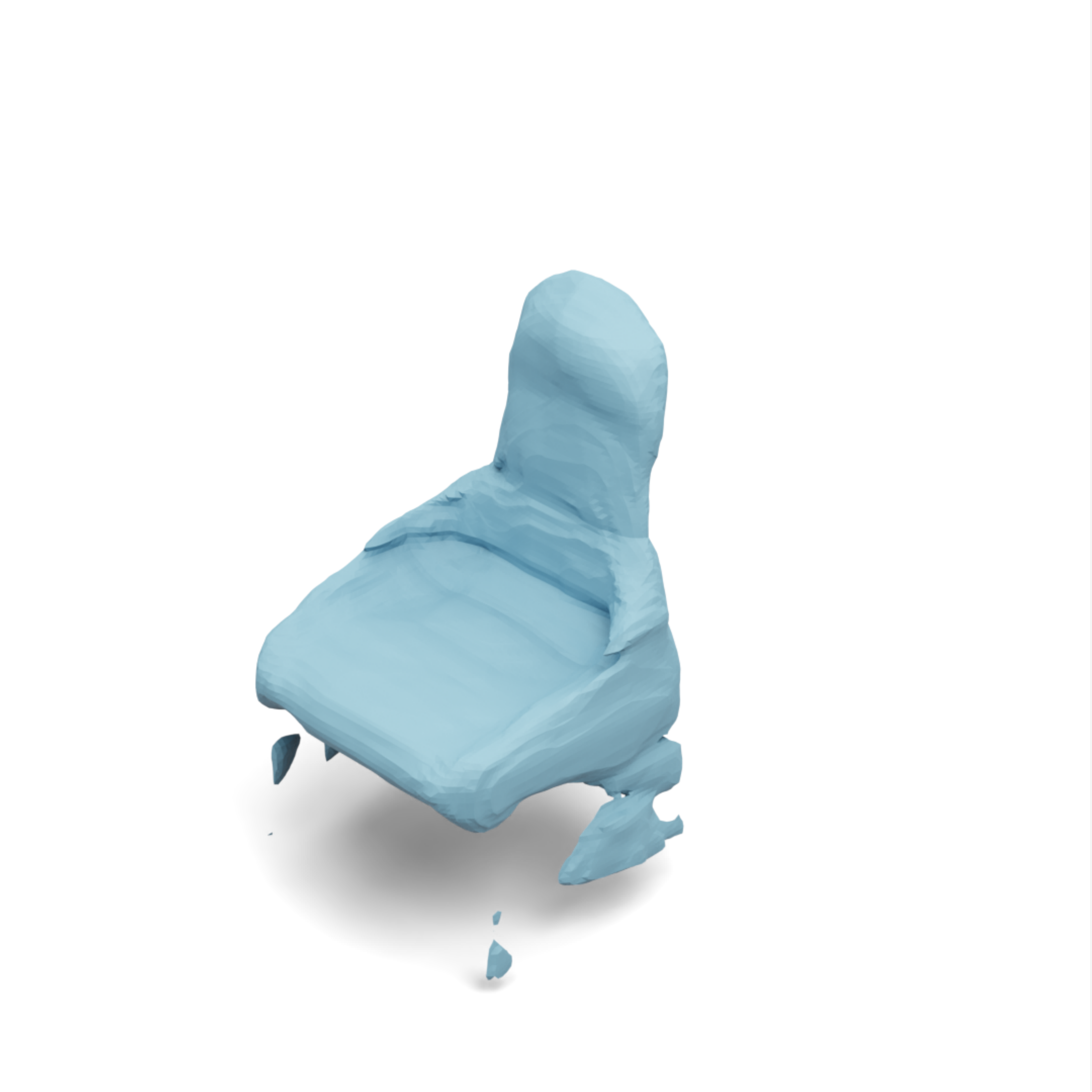}\ & \includegraphics[trim=5 5 5 5, clip, width=0.16\linewidth]{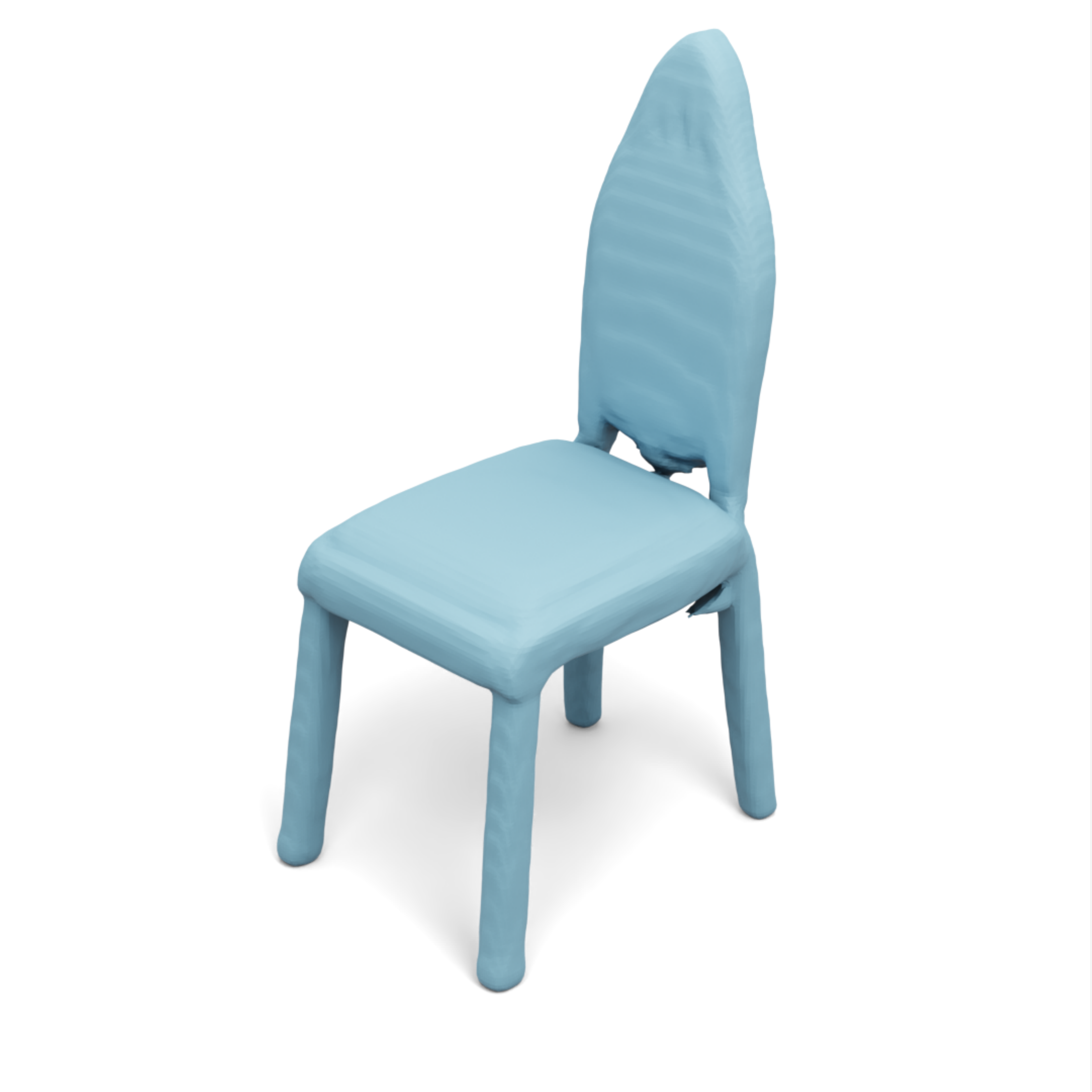} \\ 
 \midrule
	\includegraphics[width=0.16\linewidth]{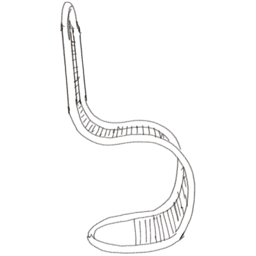}\ & \includegraphics[trim=5 45 5 5, clip, width=0.16\linewidth]{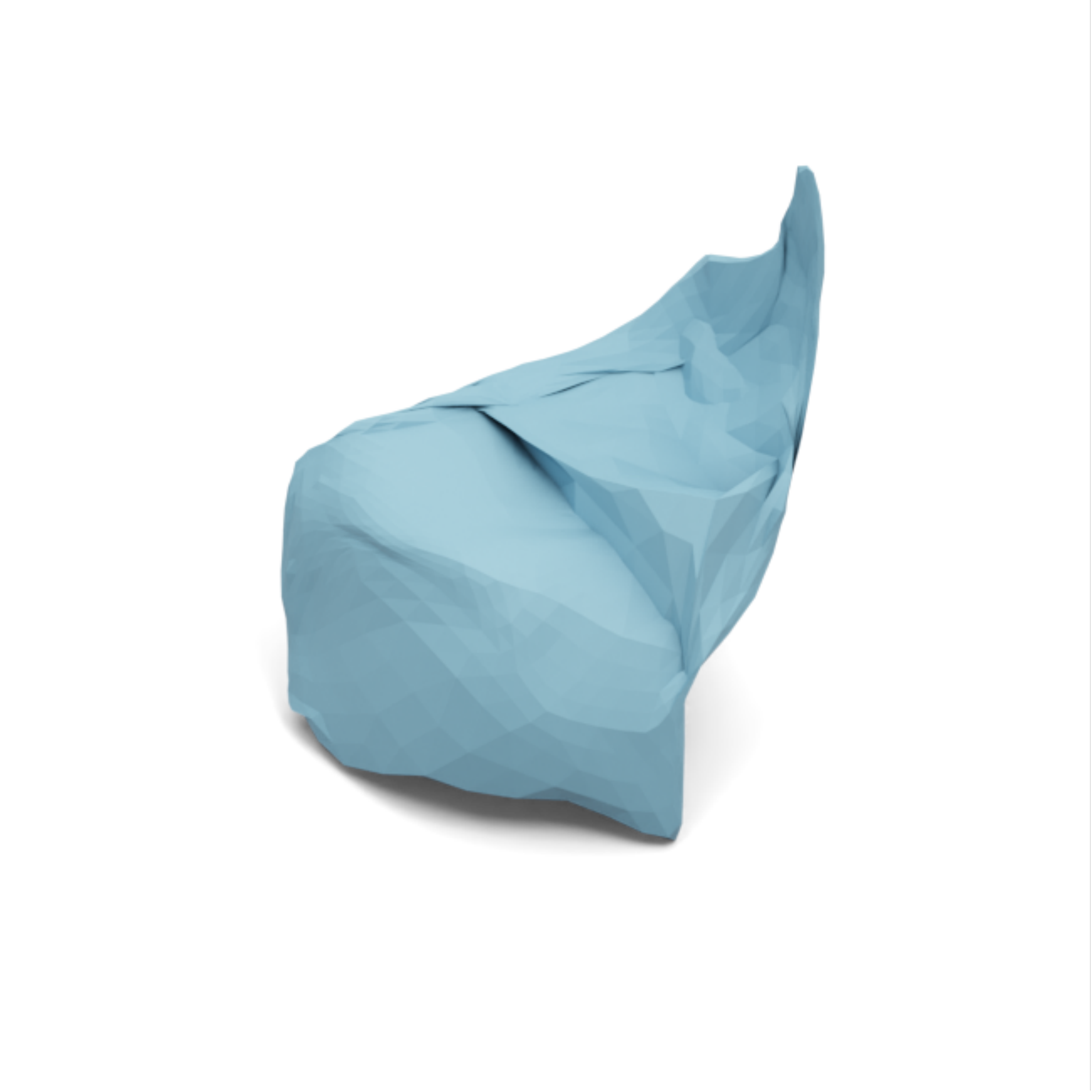}\ &
	\includegraphics[trim=5 5 5 5, clip, width=0.16\linewidth]{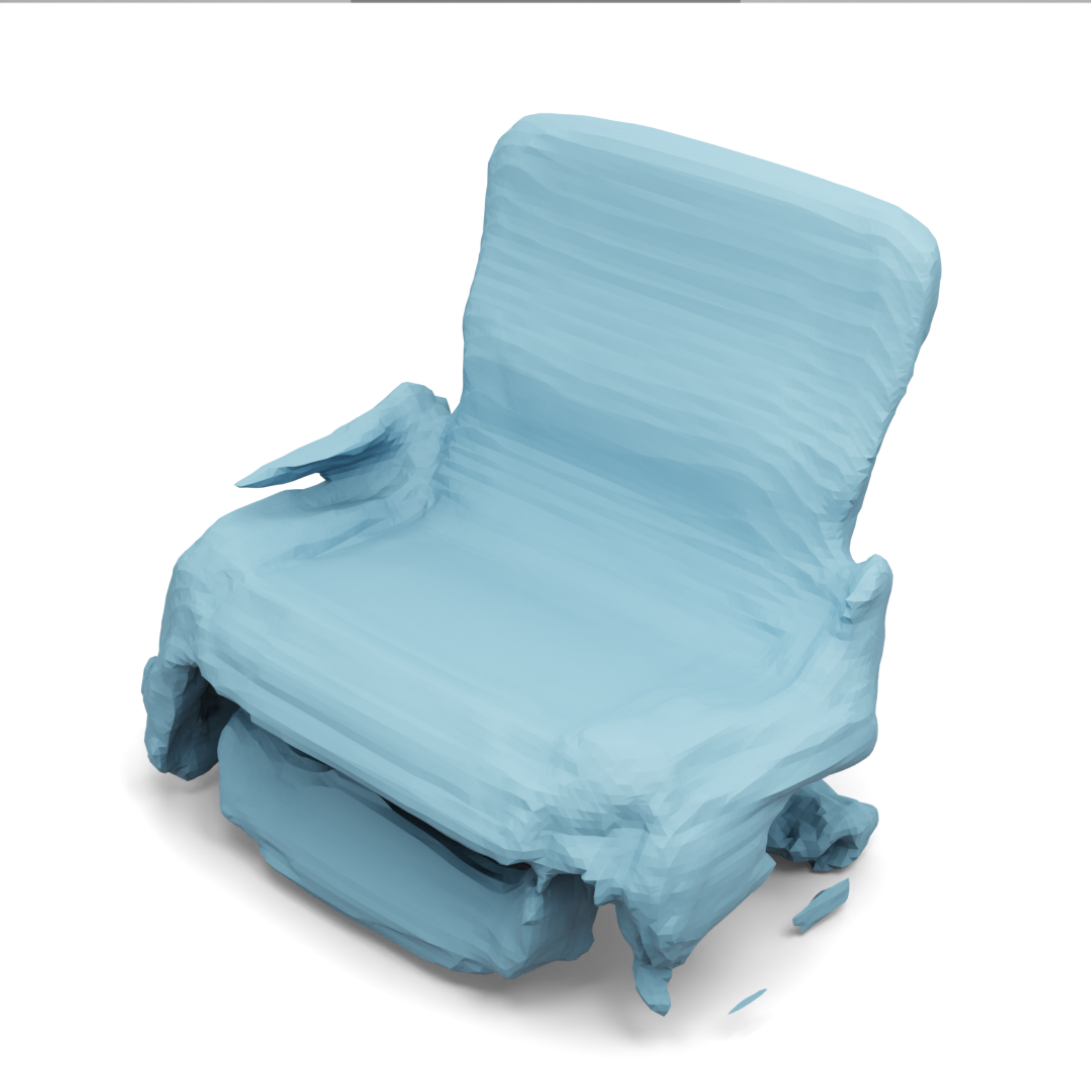}\ &
	\includegraphics[trim=160 50 160 250, clip, width=0.16\linewidth]{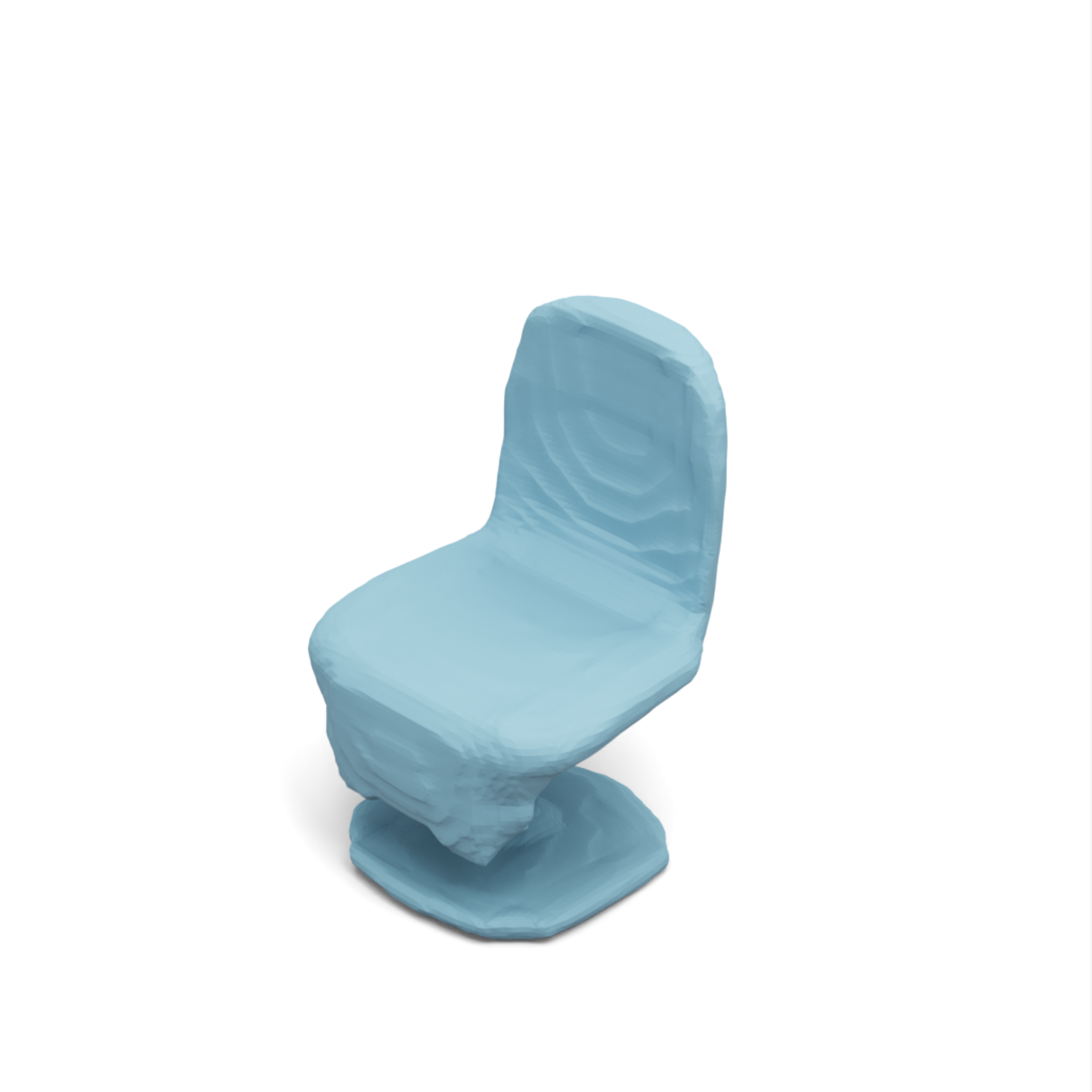}\ & \includegraphics[trim=5 5 5 5, clip, width=0.16\linewidth]{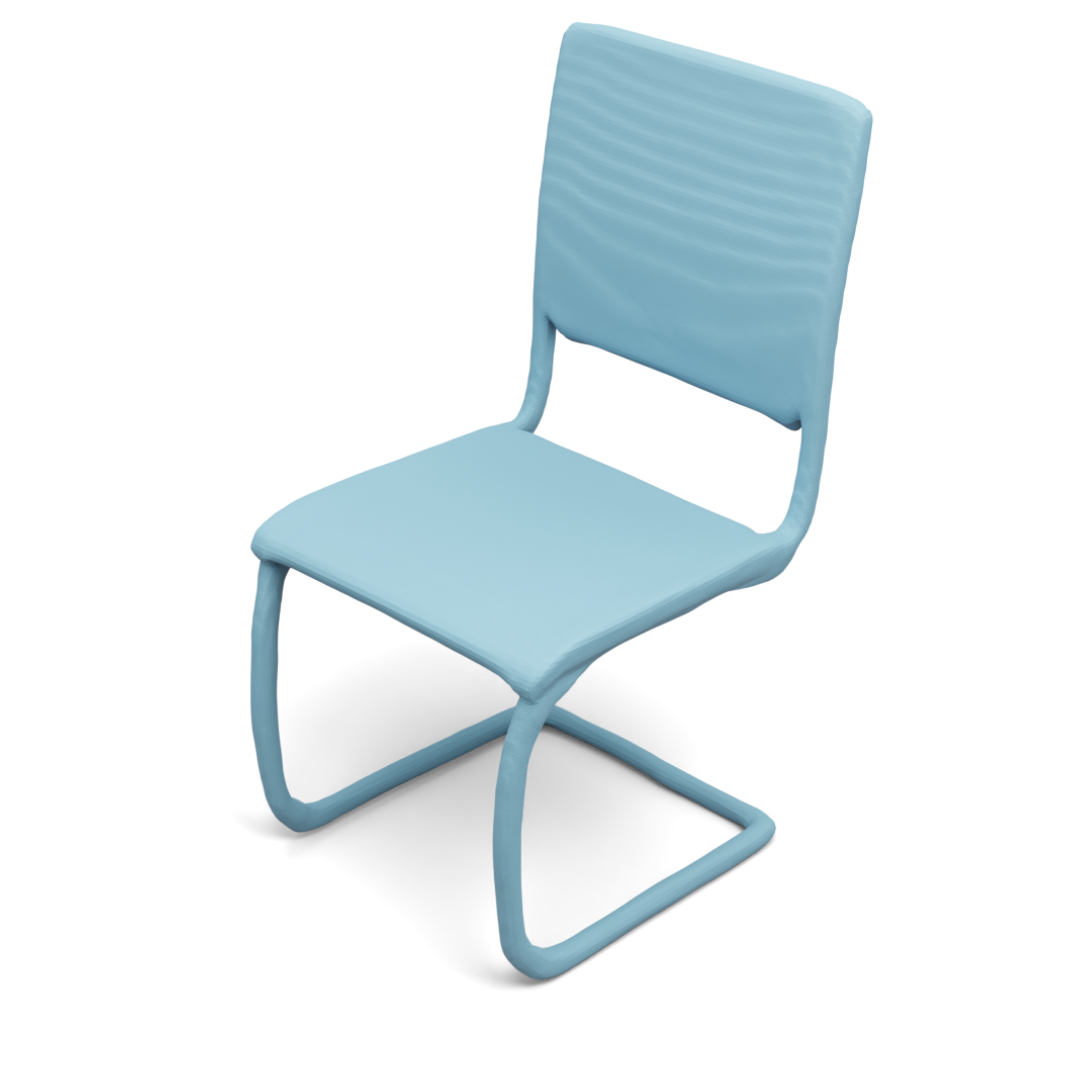}\\
	\bottomrule
	\end{tabular}\begin{tabular}{cc}
	\toprule
  \multicolumn{2}{c}{top-4 retrieval} \\
  \multicolumn{2}{c}{\,} \\
    \includegraphics[trim=10 10 10 10, clip, width=.0791925\linewidth]{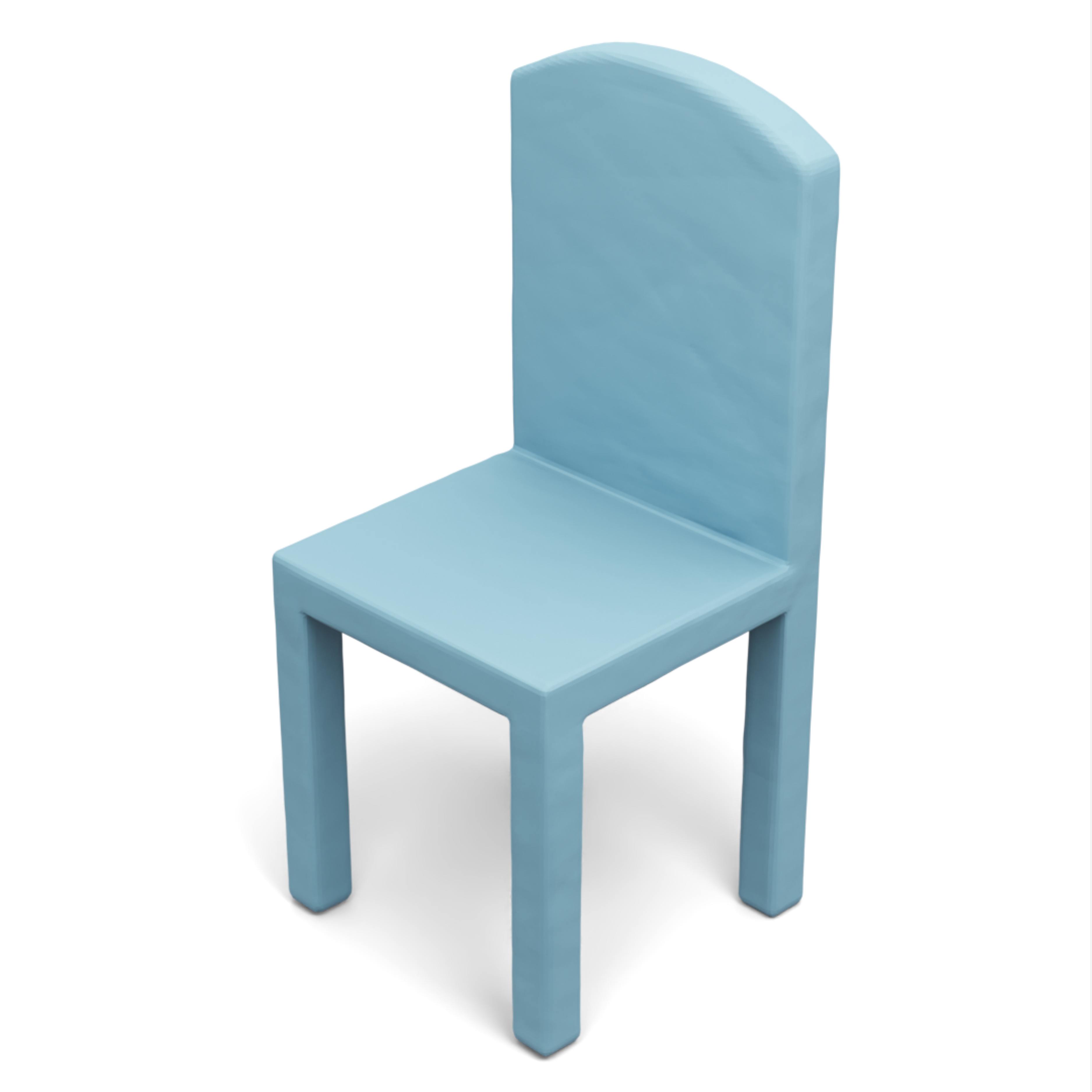} & 
    \includegraphics[trim=10 10 10 10, clip, width=.0791925\linewidth]{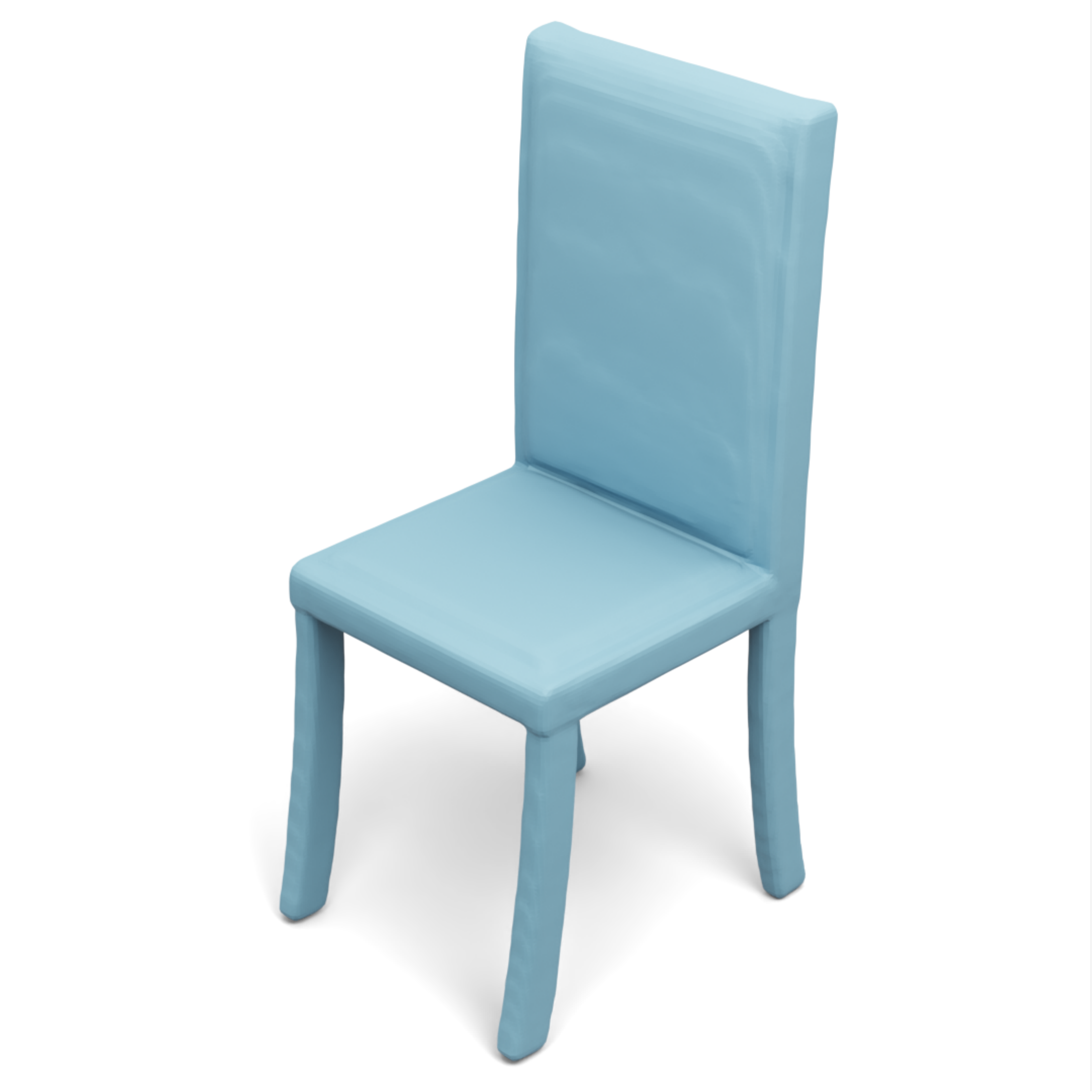} \\
    \includegraphics[trim=10 10 10 10, clip, width=.0791925\linewidth]{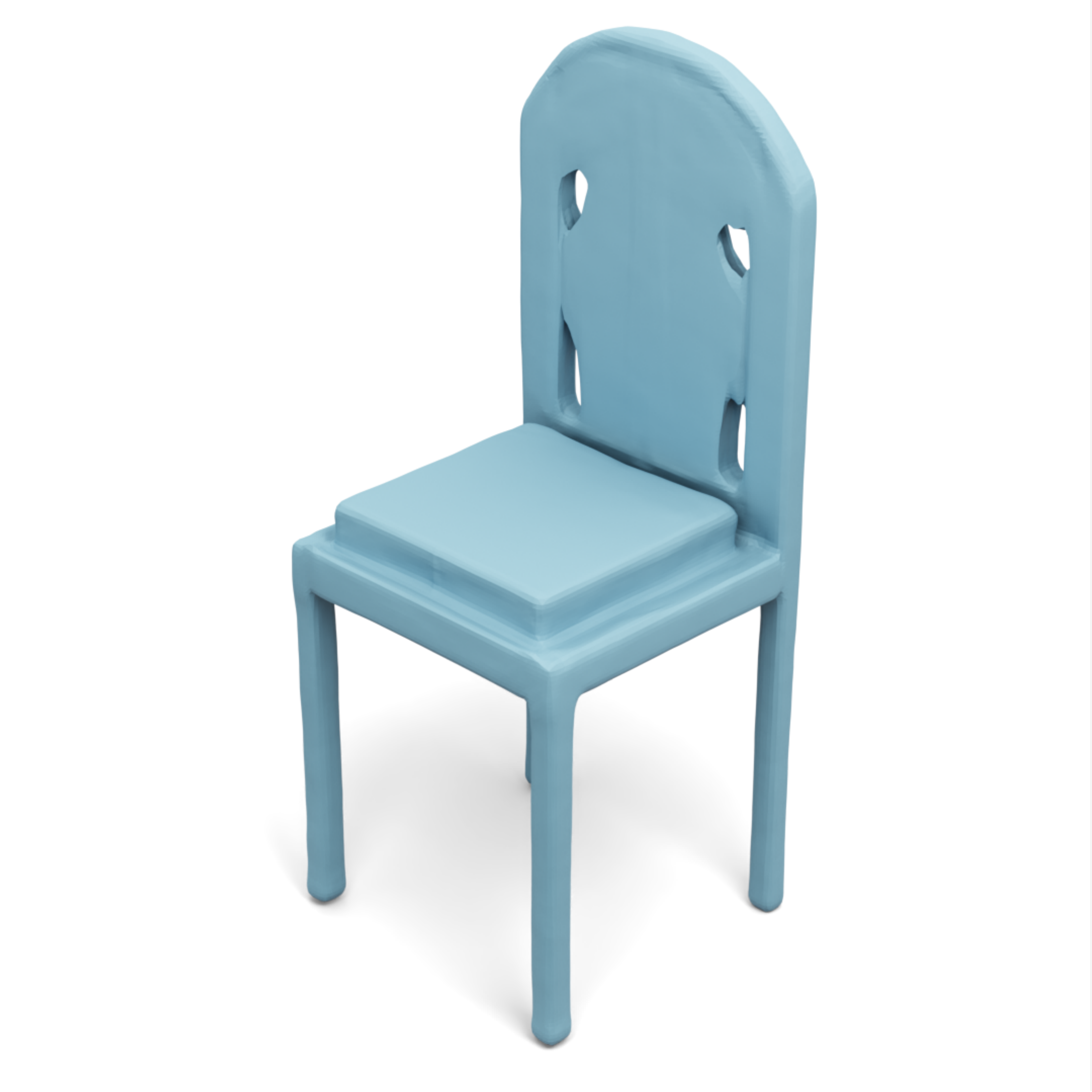} &
    \includegraphics[trim=10 10 10 10, clip, width=.0791925\linewidth]{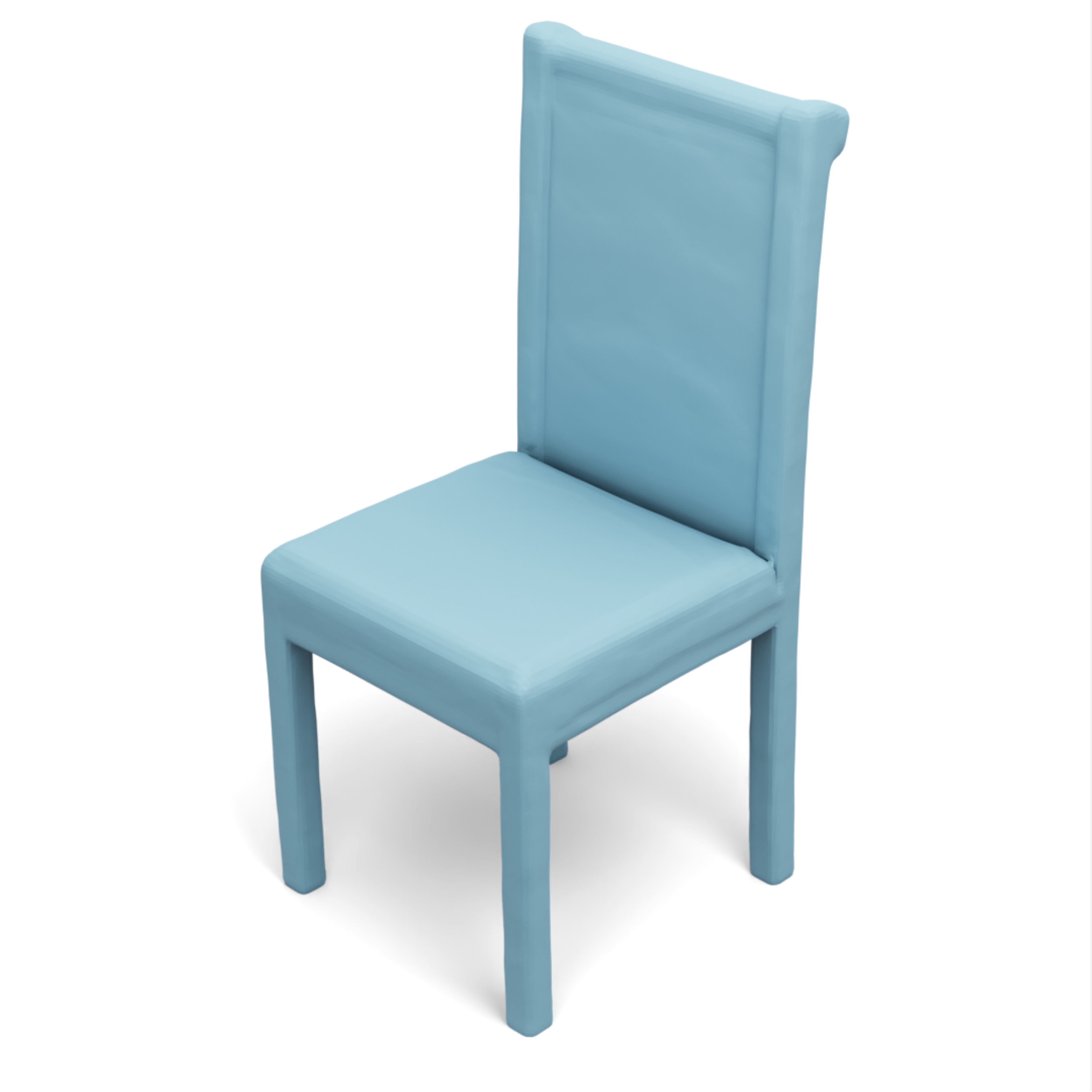}\\
    \midrule
    \includegraphics[trim=10 10 10 10, clip, width=.0791925\linewidth]{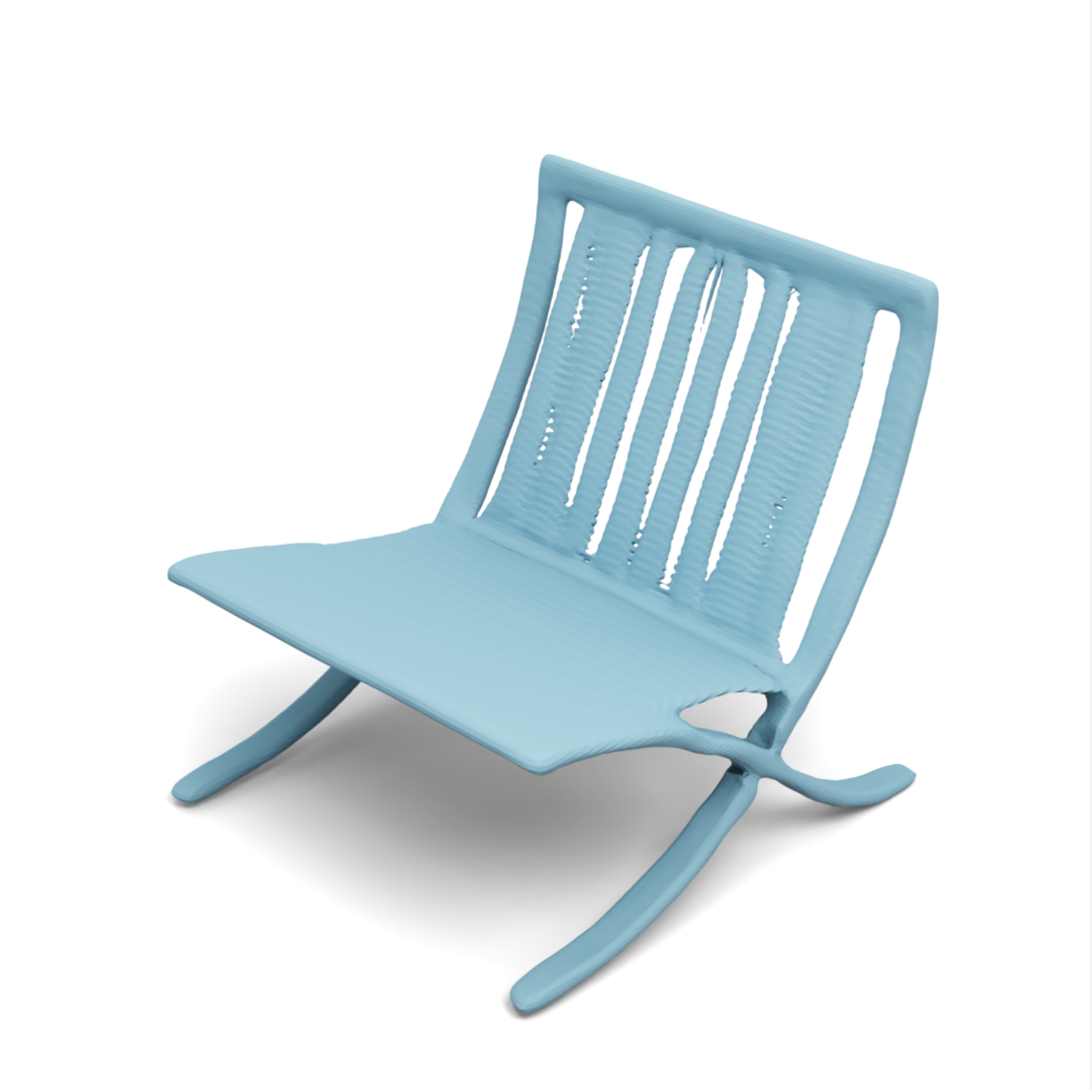} & 
    \includegraphics[trim=10 10 10 10, clip, width=.0791925\linewidth]{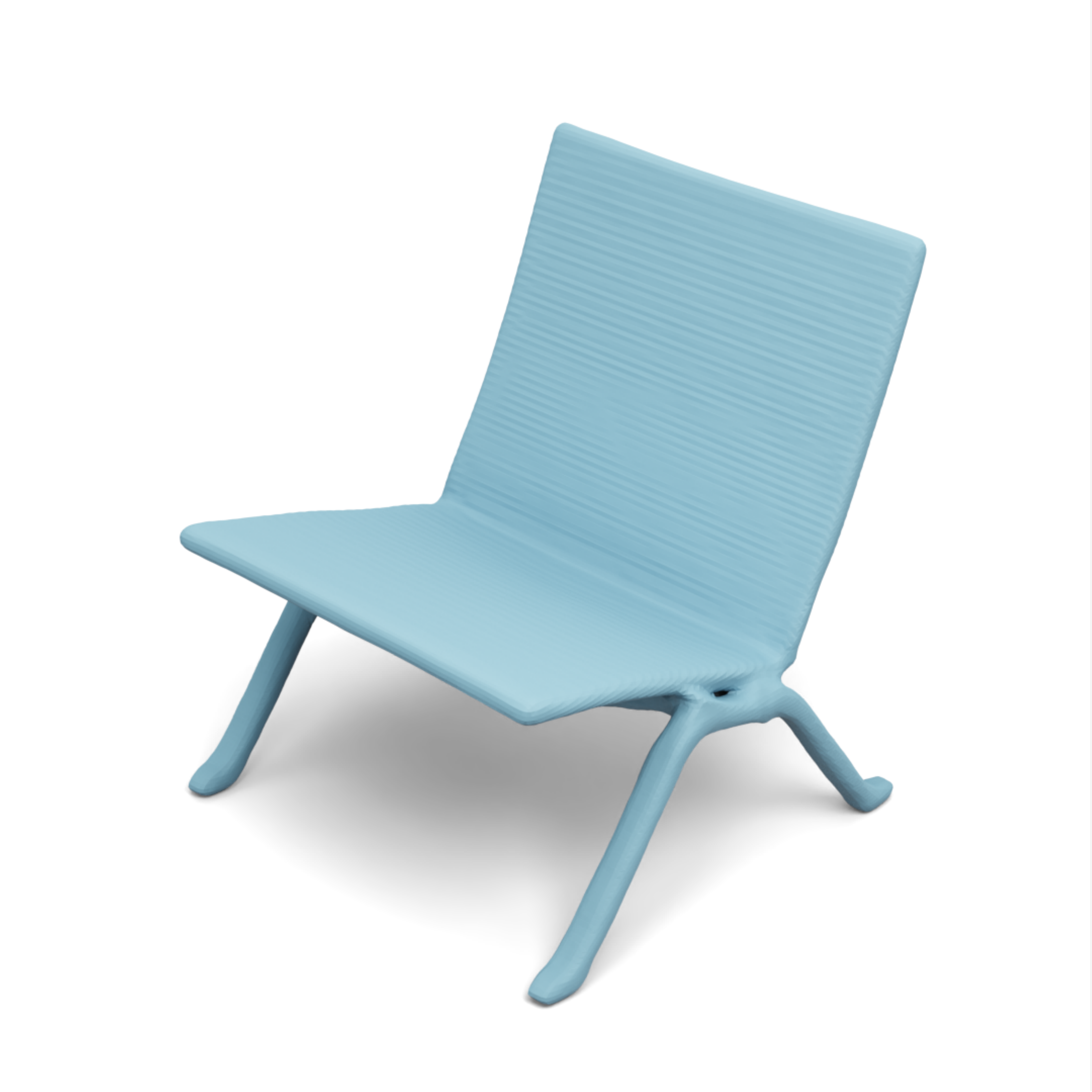} \\
    \includegraphics[trim=10 10 10 10, clip, width=.0791925\linewidth]{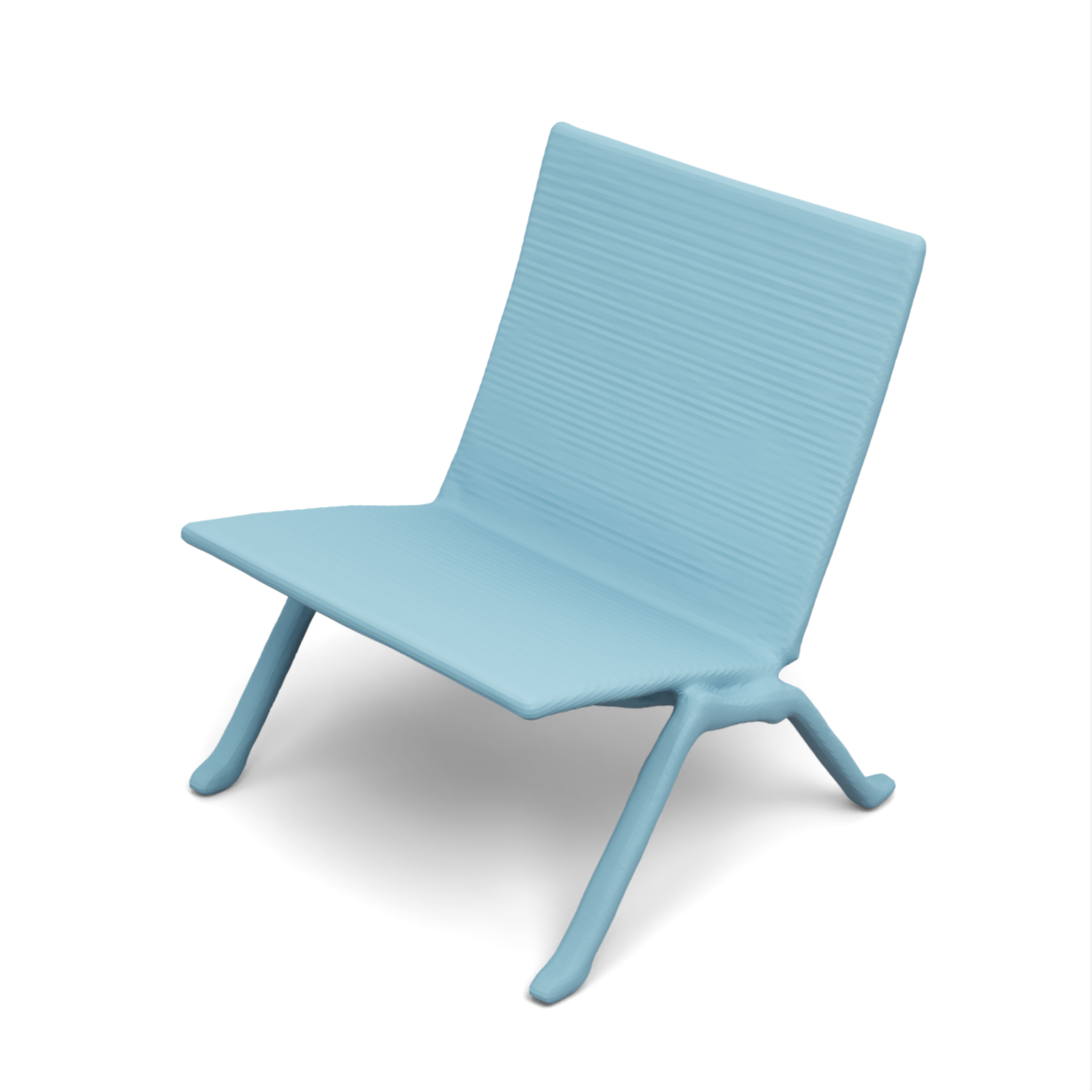} &
    \includegraphics[trim=10 10 10 10, clip, width=.0791925\linewidth]{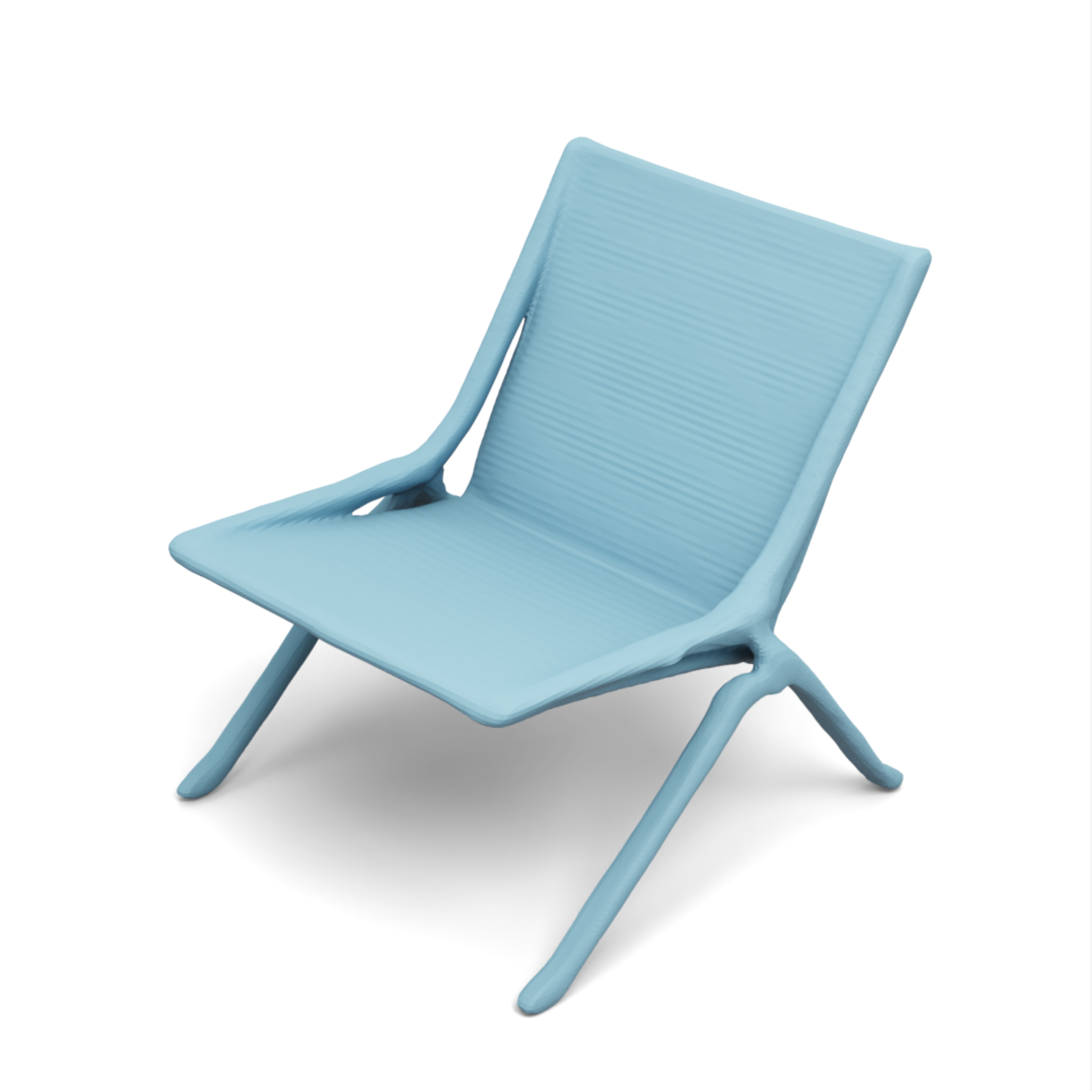}\\
    \midrule
    \includegraphics[trim=10 10 10 10, clip, width=.0791925\linewidth]{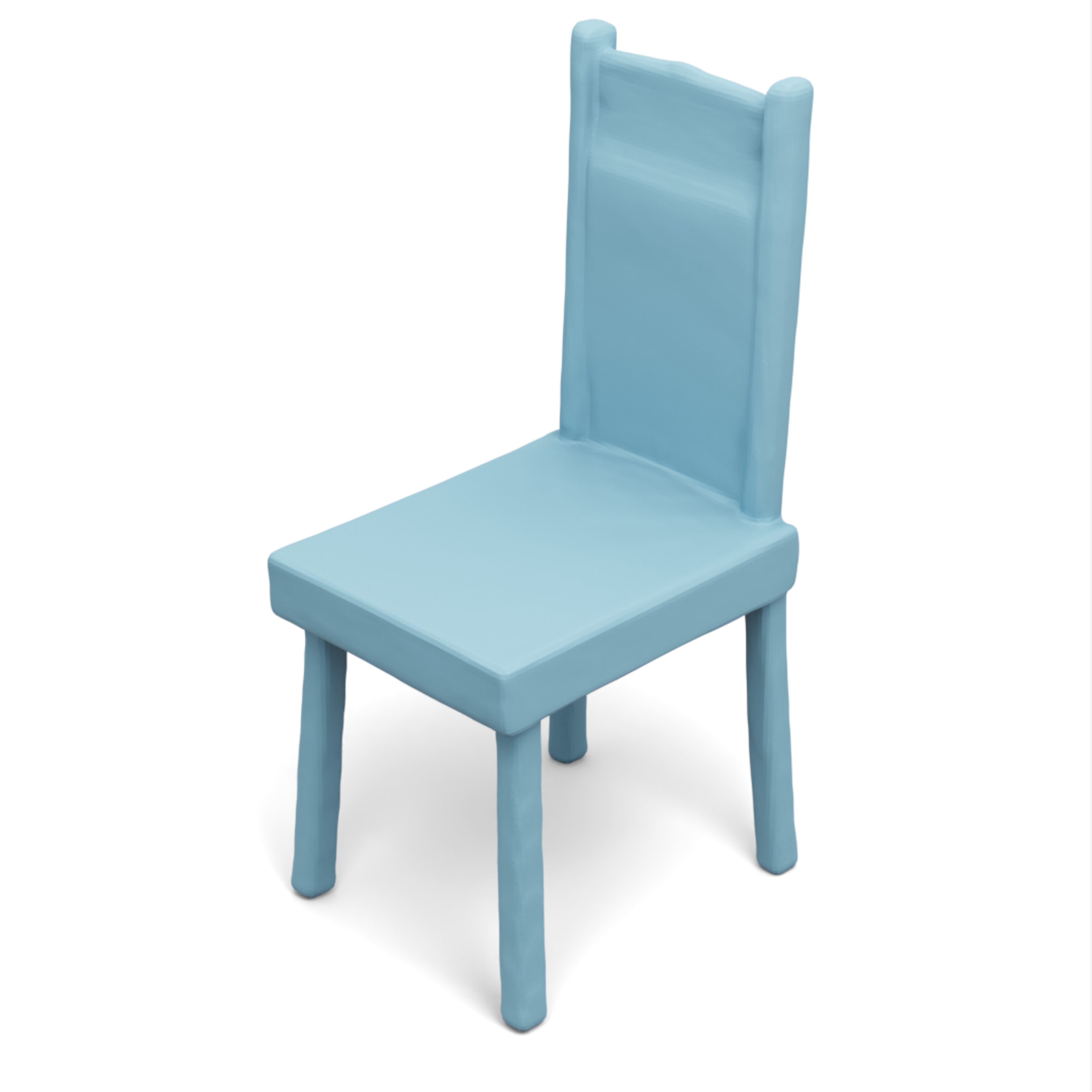} & 
    \includegraphics[trim=10 10 10 10, clip, width=.0791925\linewidth]{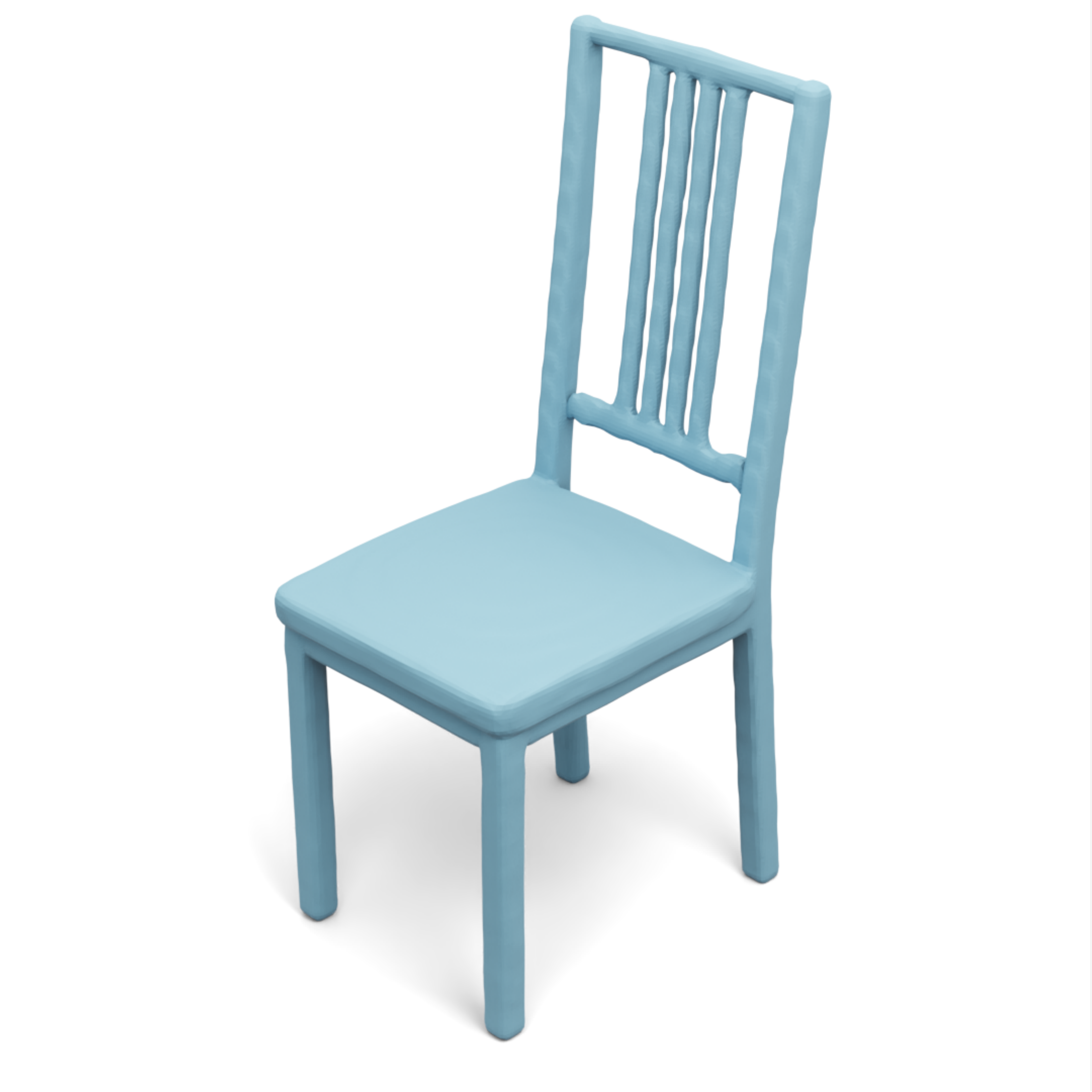} \\
    \includegraphics[trim=10 10 10 10, clip, width=.0791925\linewidth]{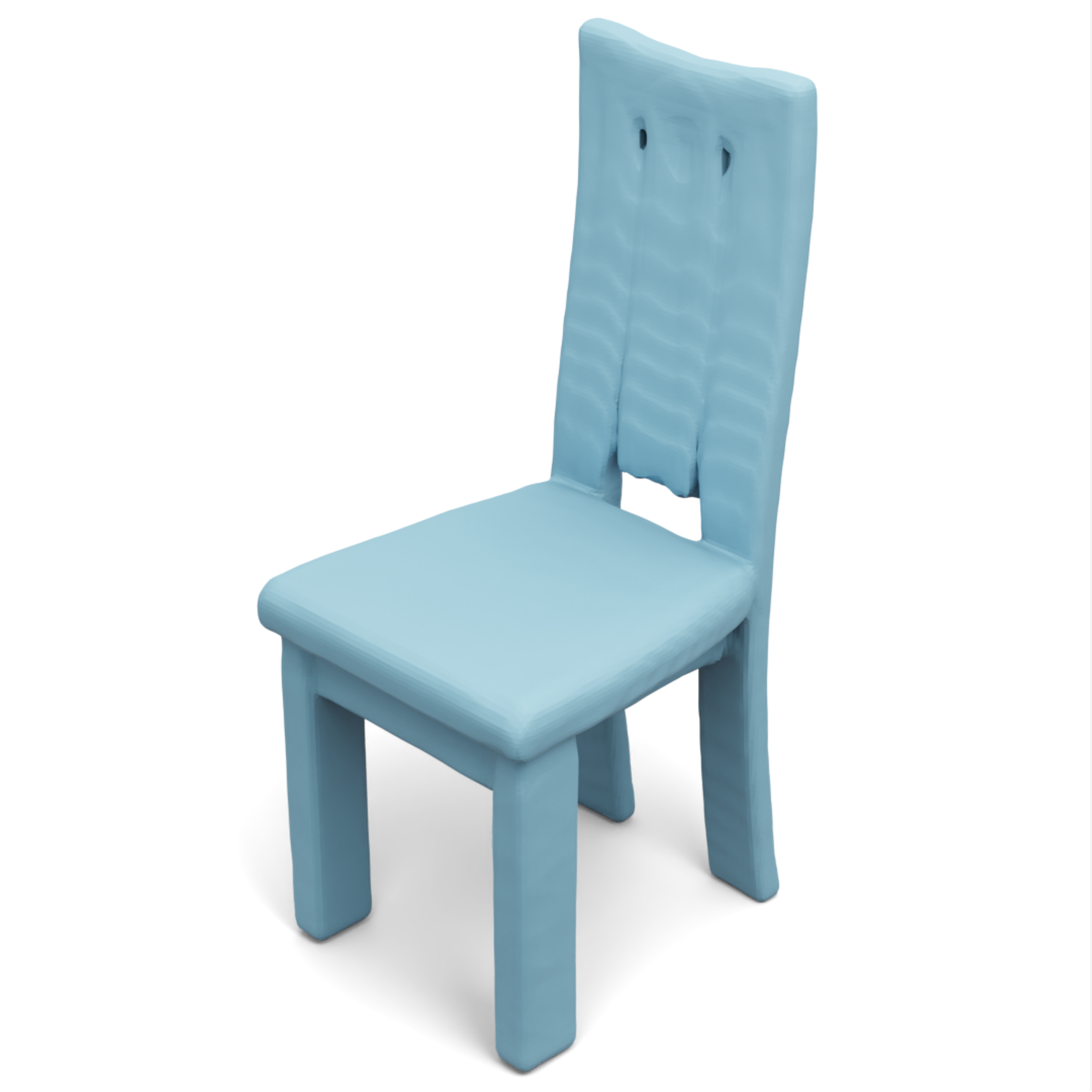} &
    \includegraphics[trim=10 10 10 10, clip, width=.0791925\linewidth]{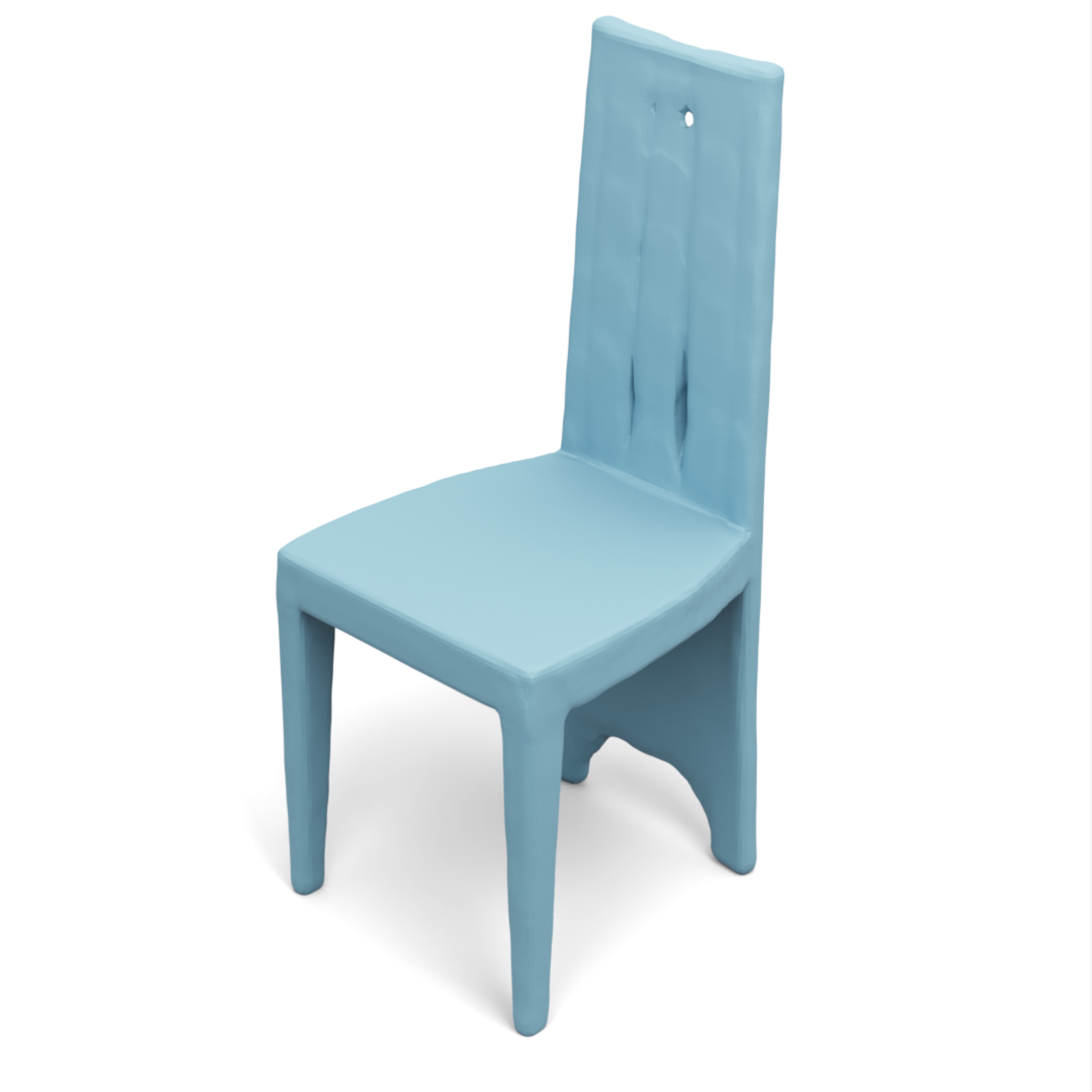}\\
    \midrule
    \includegraphics[trim=10 10 10 10, clip, width=.0791925\linewidth]{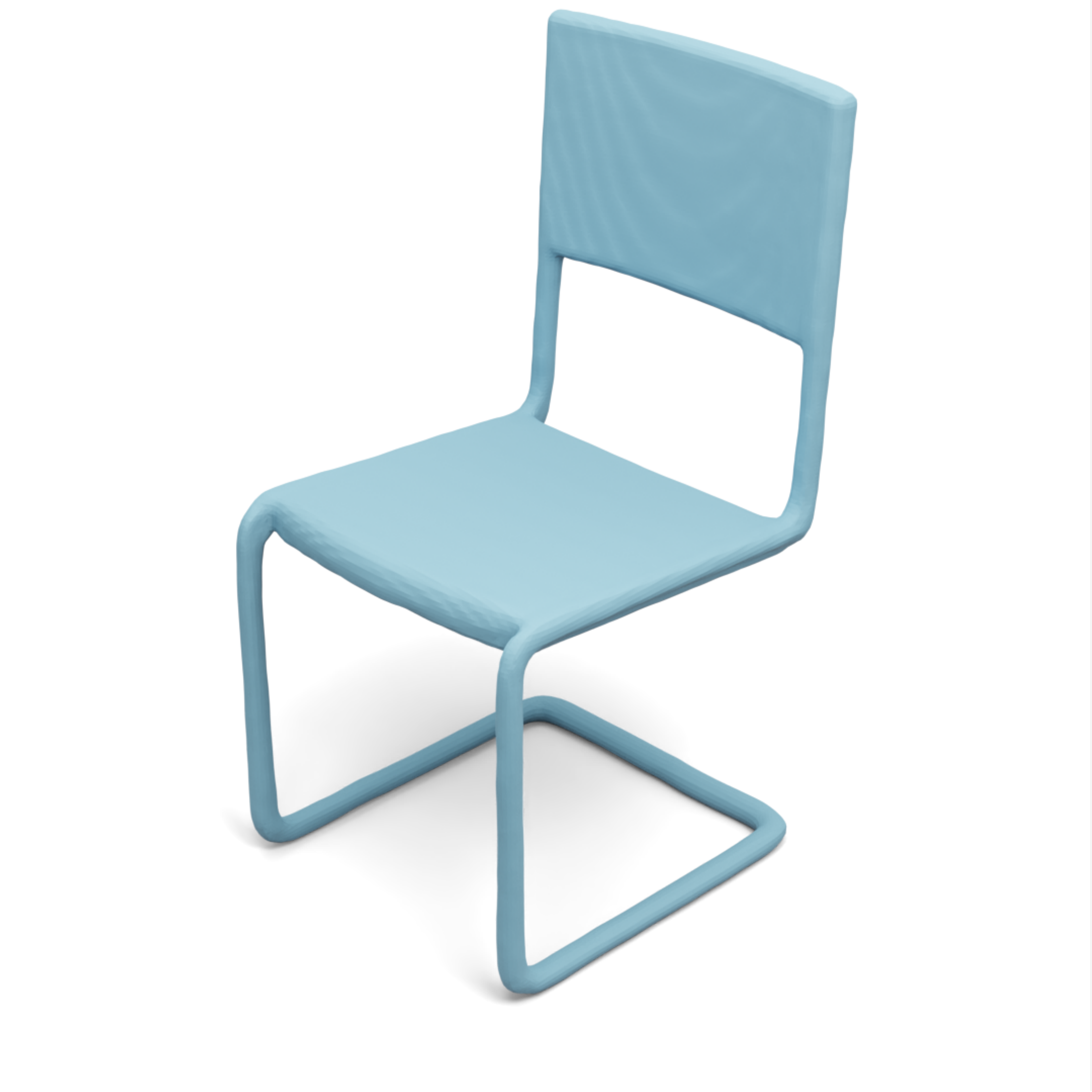} & 
    \includegraphics[trim=10 10 10 10, clip, width=.0791925\linewidth]{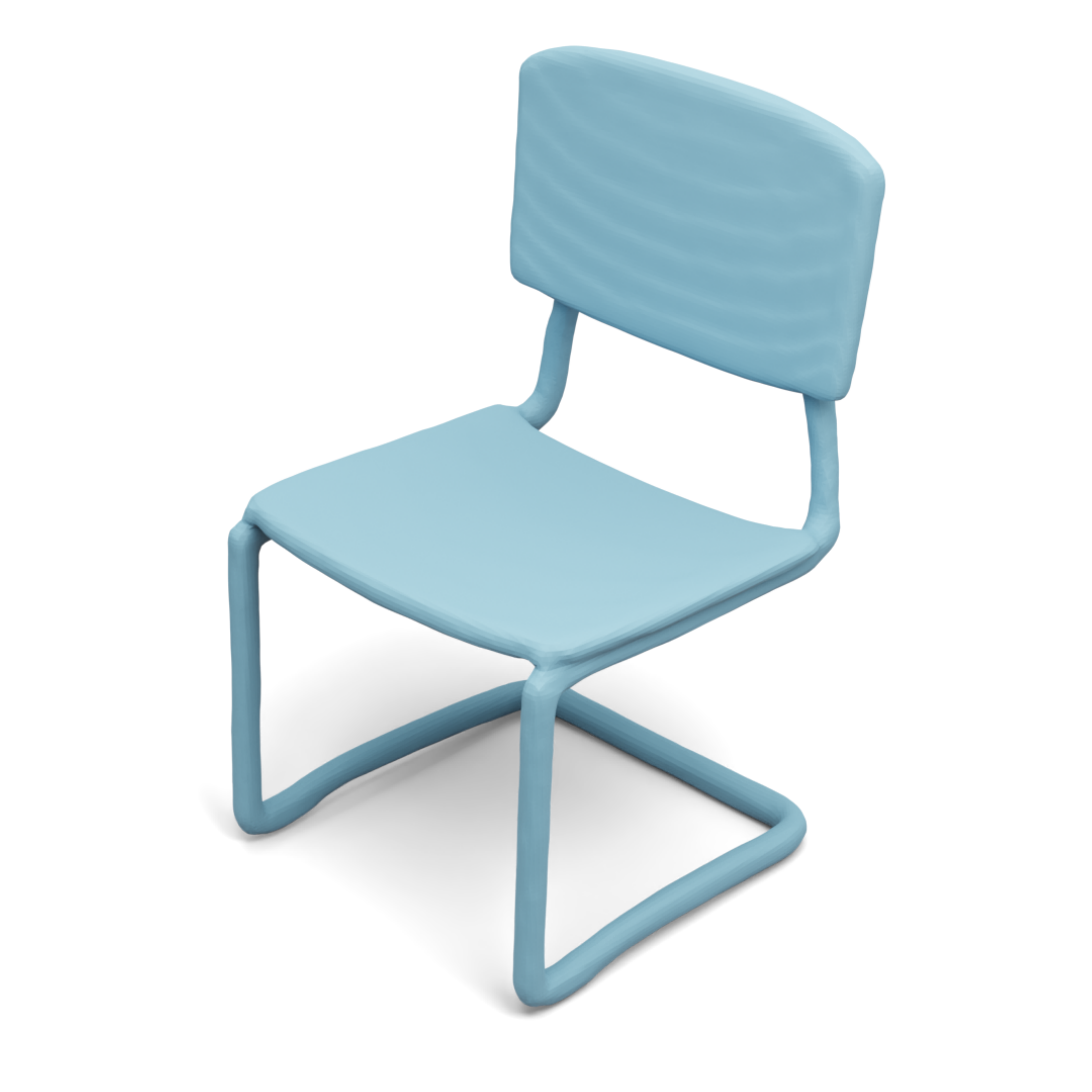} \\
    \includegraphics[trim=10 10 10 10, clip, width=.0791925\linewidth]{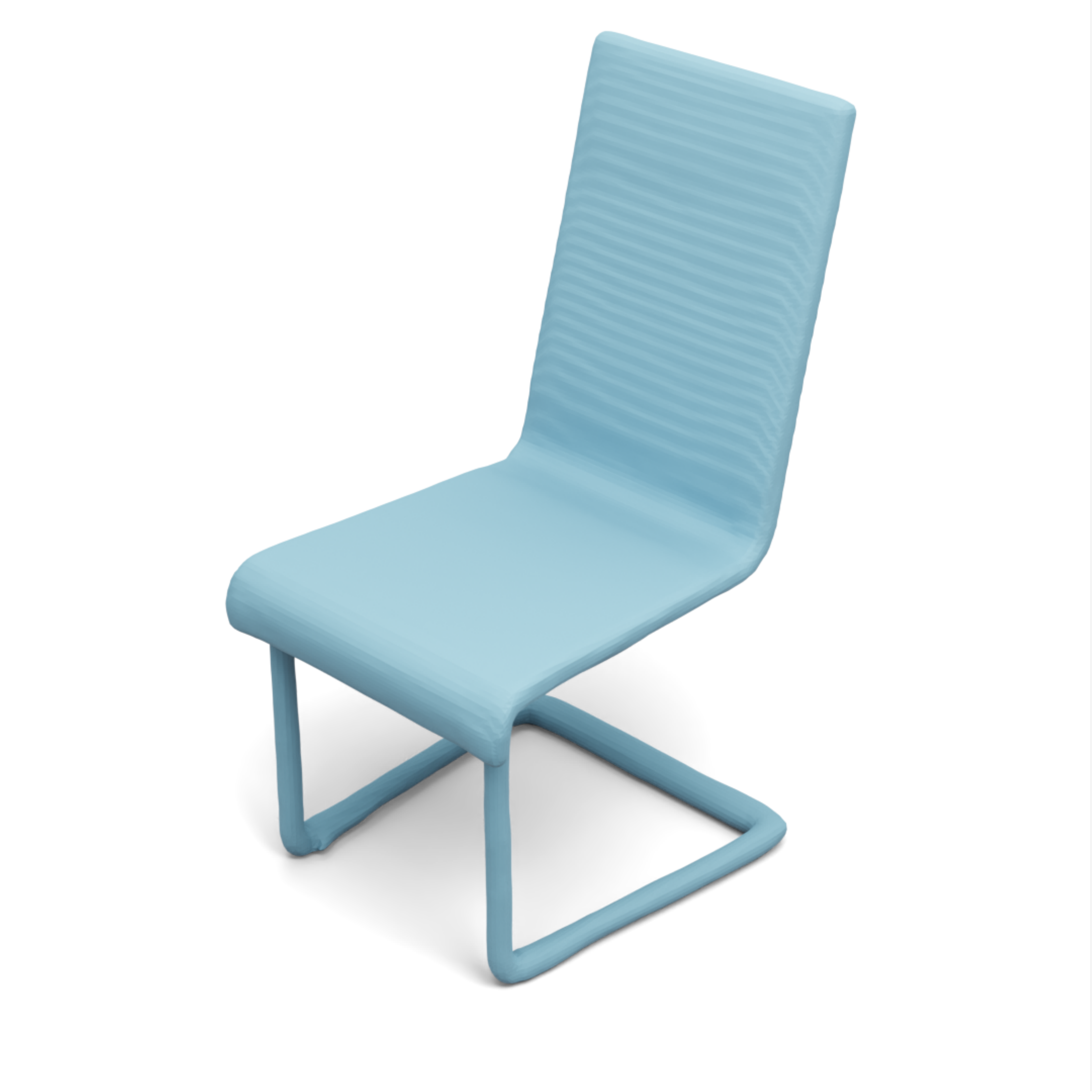} &
    \includegraphics[trim=10 10 10 10, clip, width=.0791925\linewidth]{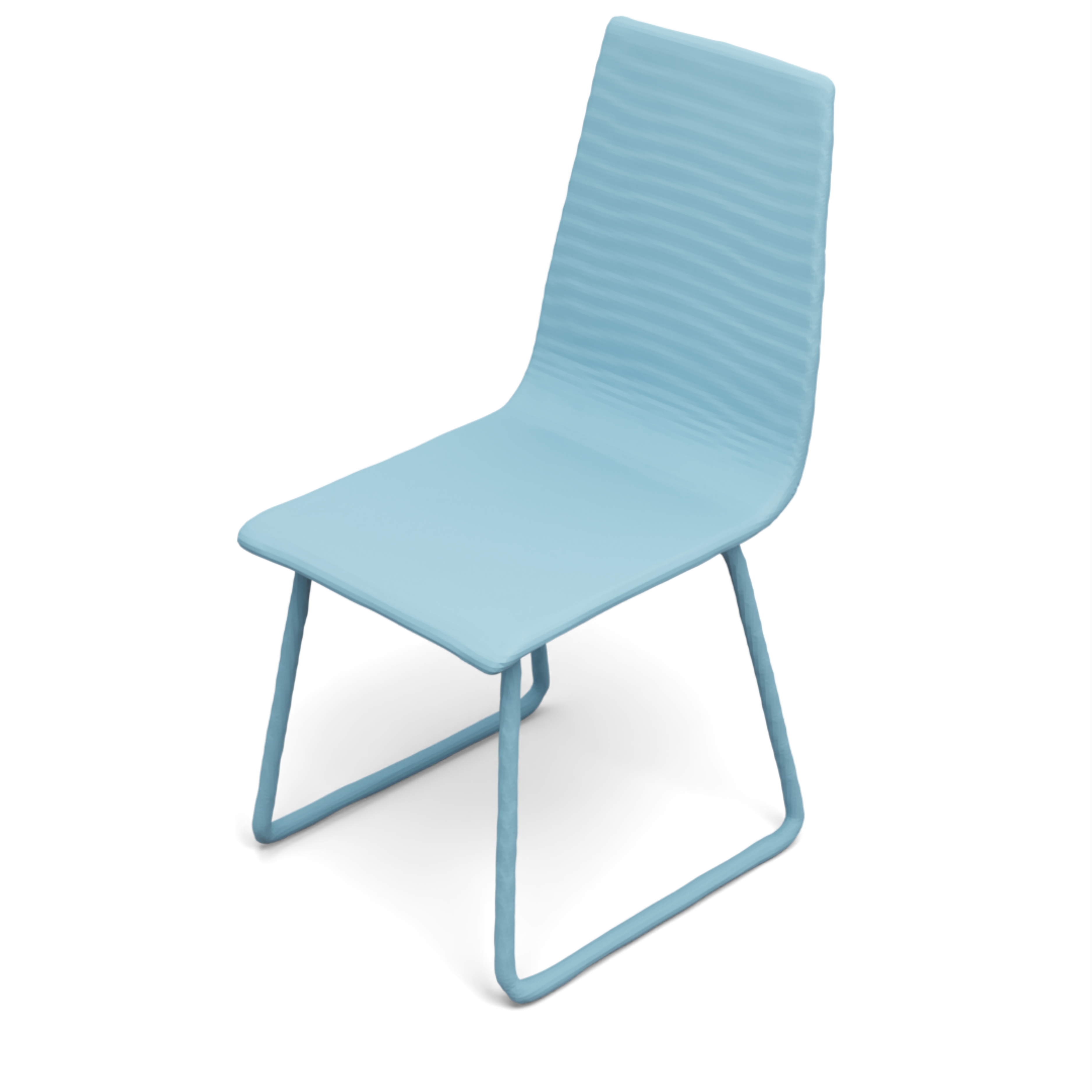}\\
    \bottomrule
  \end{tabular}   
	\caption{We compare our method with state-of-the-art single-view reconstruction methods on sketches of various styles such as an outline, an abstract sketch, a non-expert handmade sketch and an expert freehand sketch from the ProSketch dataset \cite{zhong2021prosketch}. Note that these sketches were not part of the dataset. We also show the top-4 retrieval: we first remesh the output shapes of SPAGHETTI \cite{huang2018} that were used for training \ourmethod{}, and compute the Chamfer Distance over 100,000 sampled points over the surface. We display the output shape of SPAGHETTI. The order is left to right, top to bottom.}
	\label{fig:comparison}
    \vspace{-0.5cm}
\end{figure*}

We show our shape generation and editing results, with quantitative evaluation and insights into retrieval, completion, ablation, and limitations. 
We trained two single-class \ourmethod{} networks over chairs and airplanes and a multi-class network that was trained jointly over chairs, airplanes and lamps.
The latent space of the pre-trained SPAGHETTI model consists of $m=16$ and $m=32$ parts with dimensions $d_{\text{model}}=512$ and $d_{\text{model}}=768$ for the single and multi-class networks respectively. We will publish our sketches dataset, code, pre-trained models and user interface upon acceptance.

\subsection{Generation comparison}

\begin{figure}[t]
    \centering
    \small
    \setlength{\tabcolsep}{1pt}
    \begin{tabular}{cccccc}
    \multicolumn{2}{c}{Input sketches} & \multicolumn{2}{c}{ShapeMVD} & \multicolumn{2}{c}{Ours} \\ 
    \includegraphics[width=0.195\linewidth]{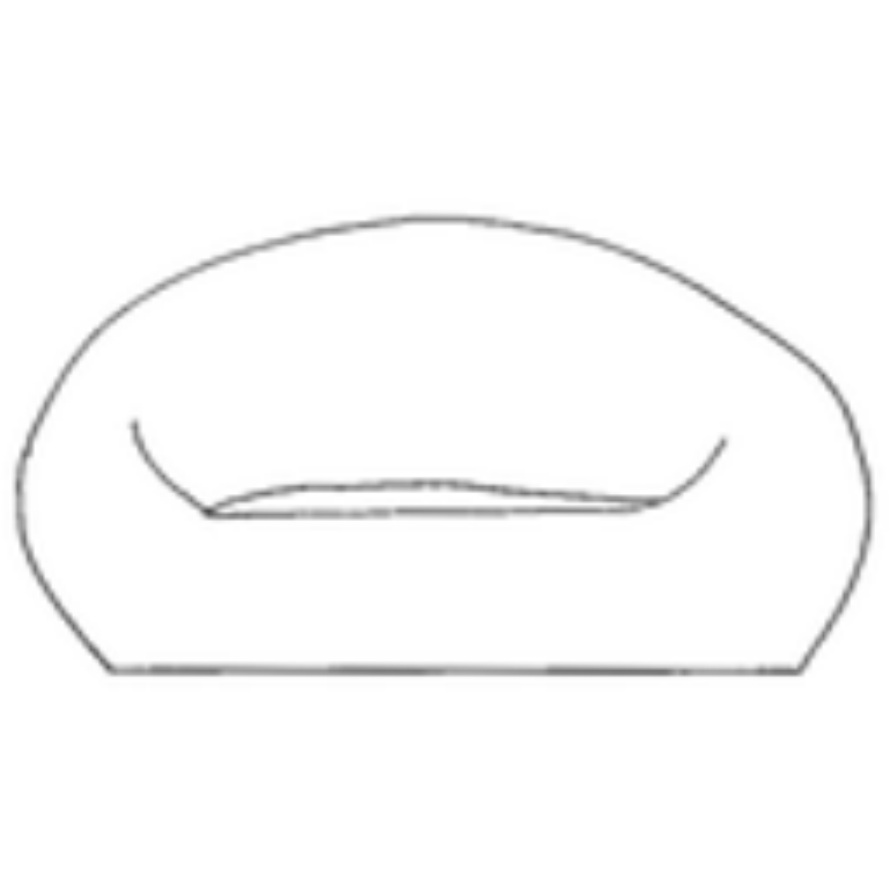} & \includegraphics[width=0.195\linewidth]{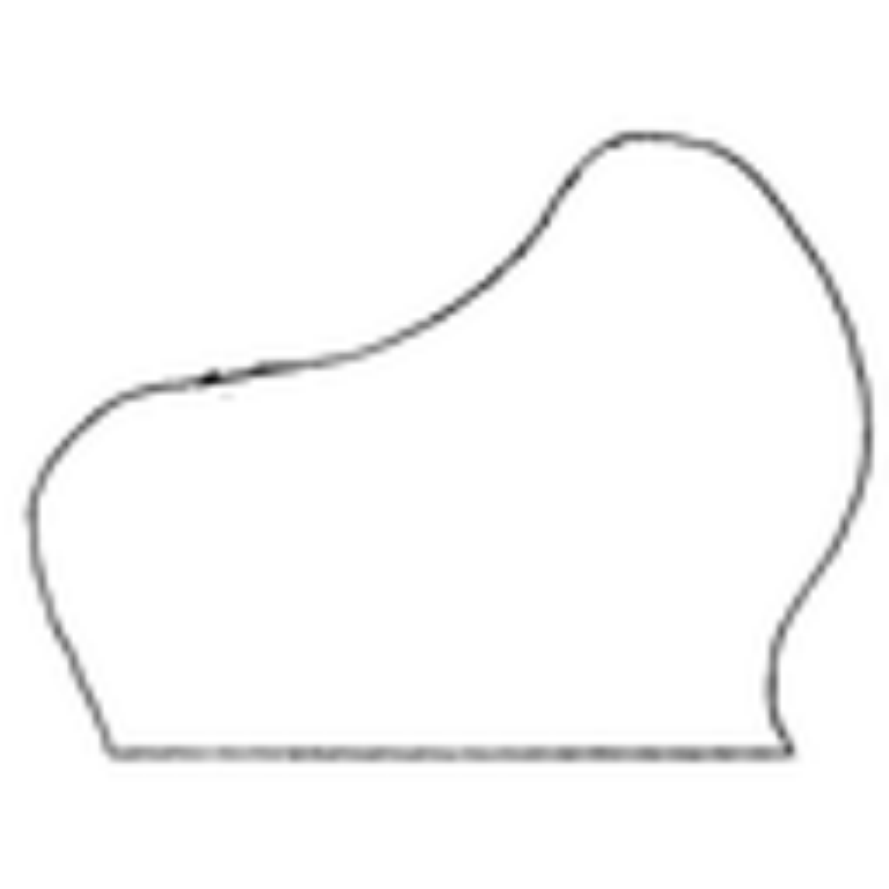} & \multicolumn{2}{c}{\includegraphics[trim =30 30 30 30, clip, width=0.195\linewidth]{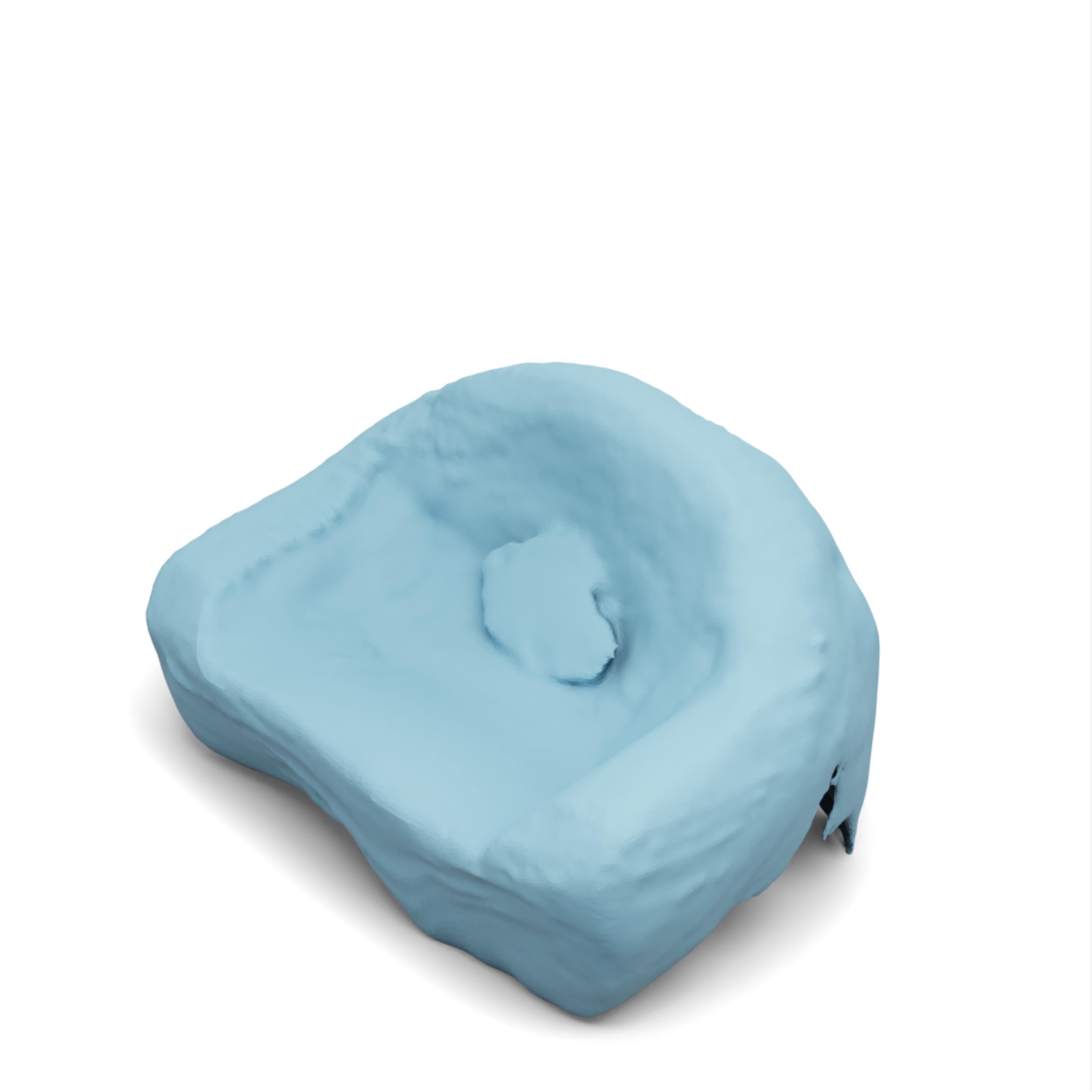}} & \includegraphics[trim = 30 30 30 30, clip, width=0.195\linewidth]{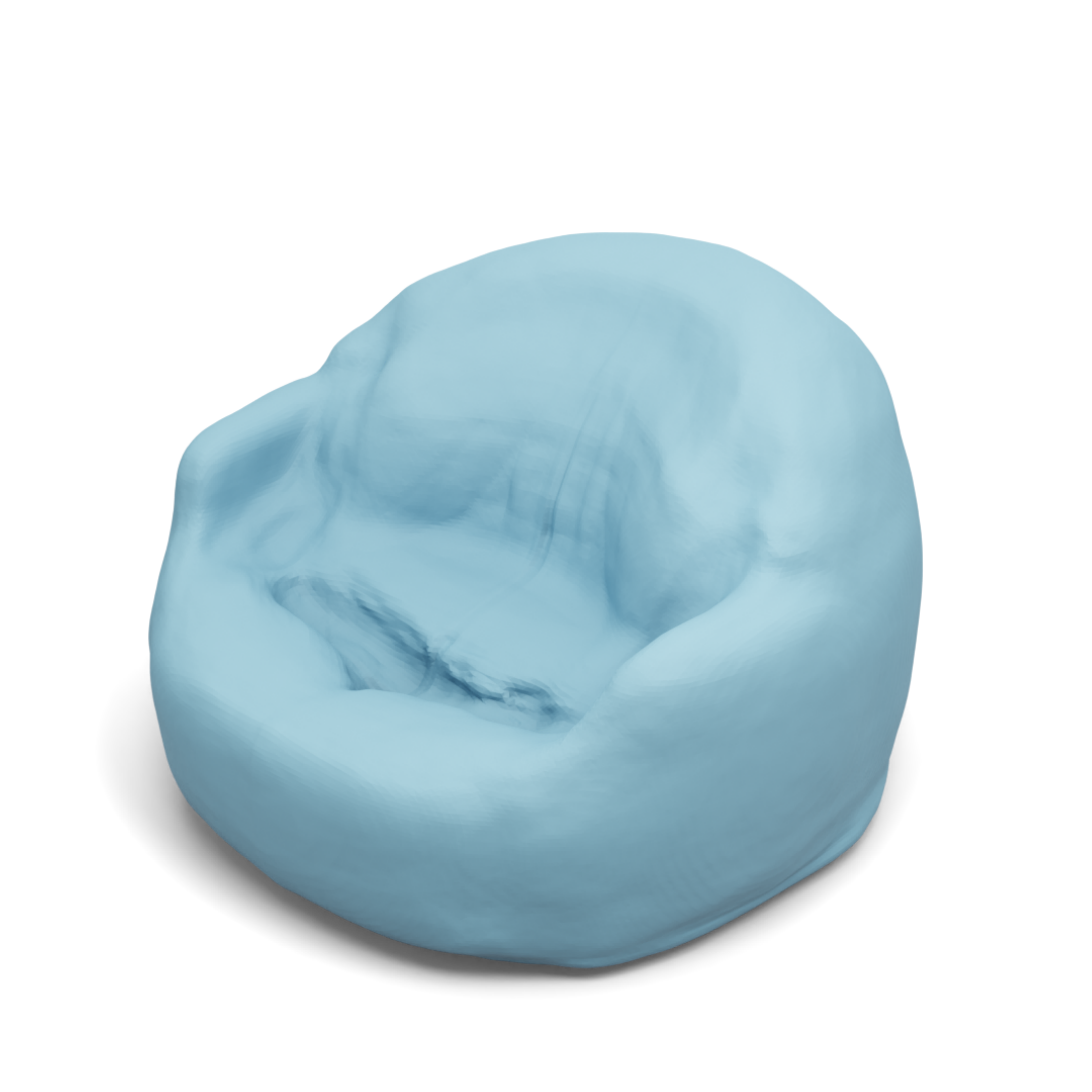} & \includegraphics[trim = 30 30 30 30, clip, width=0.195\linewidth]{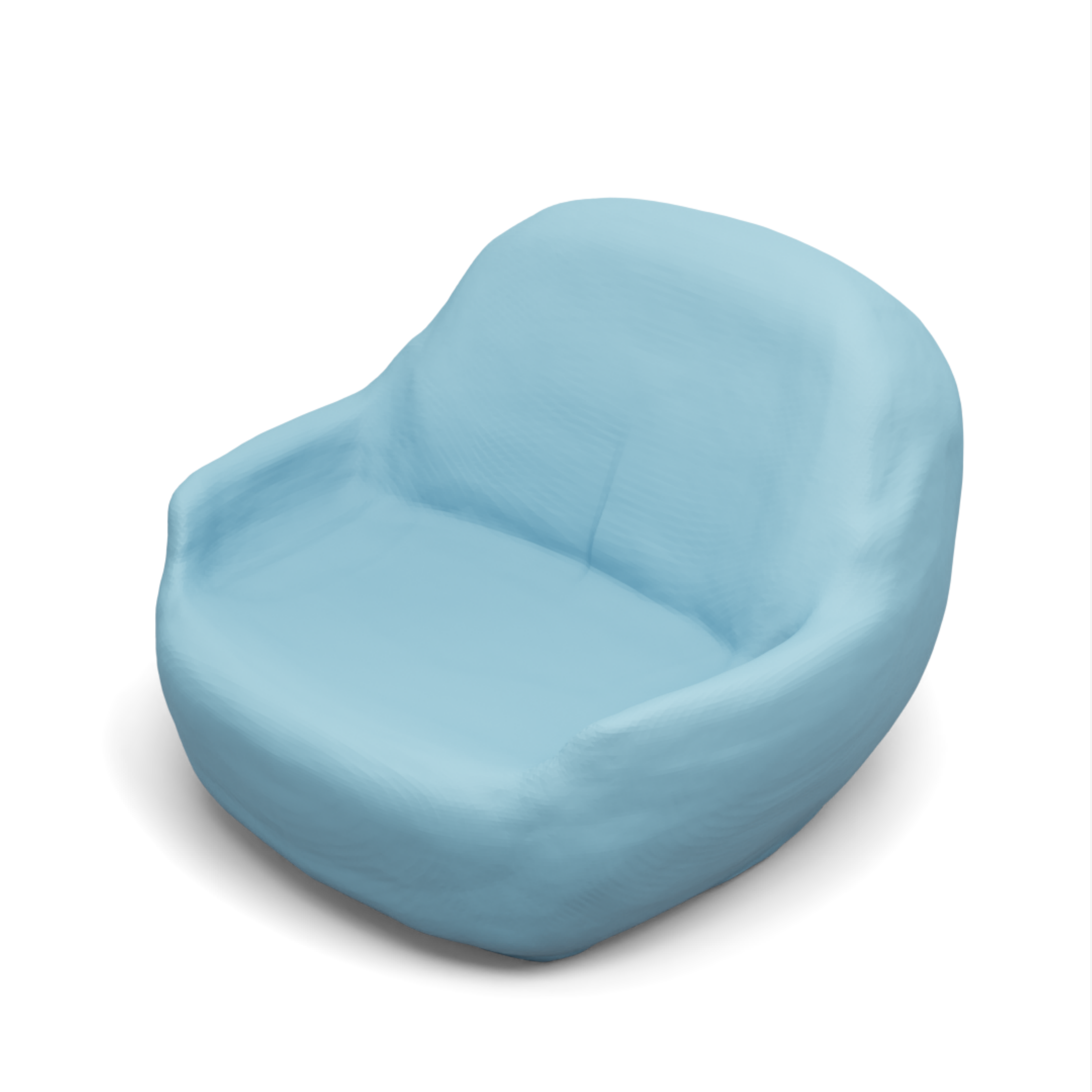} \\ \includegraphics[width=0.195\linewidth]{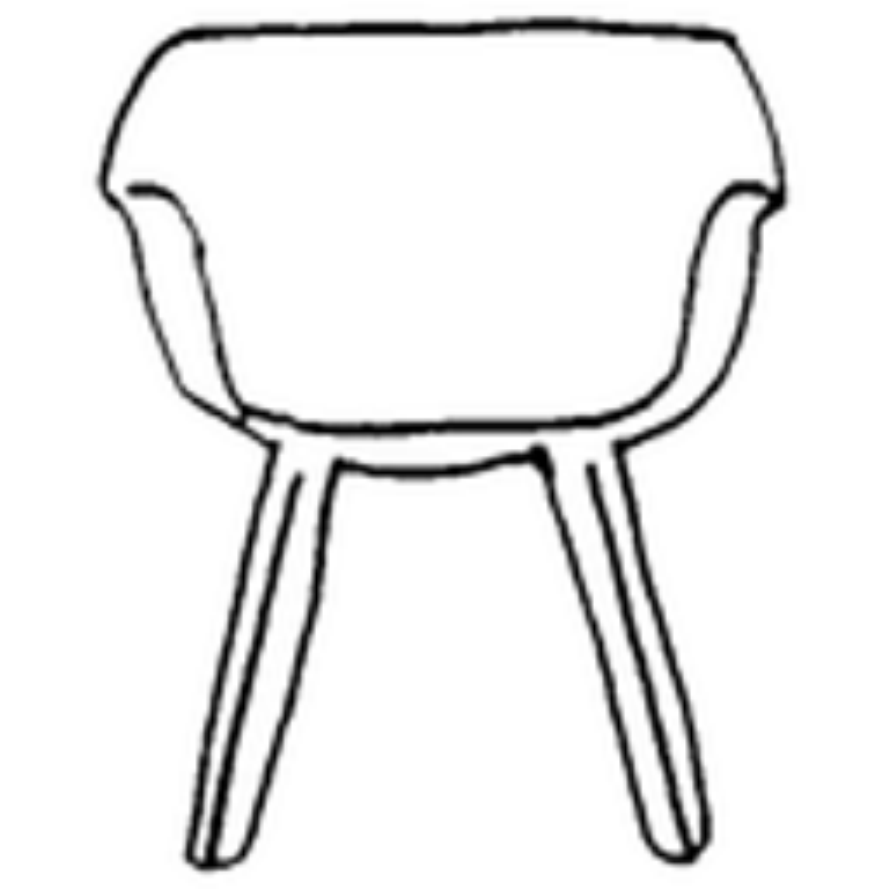} & \includegraphics[width=0.195\linewidth]{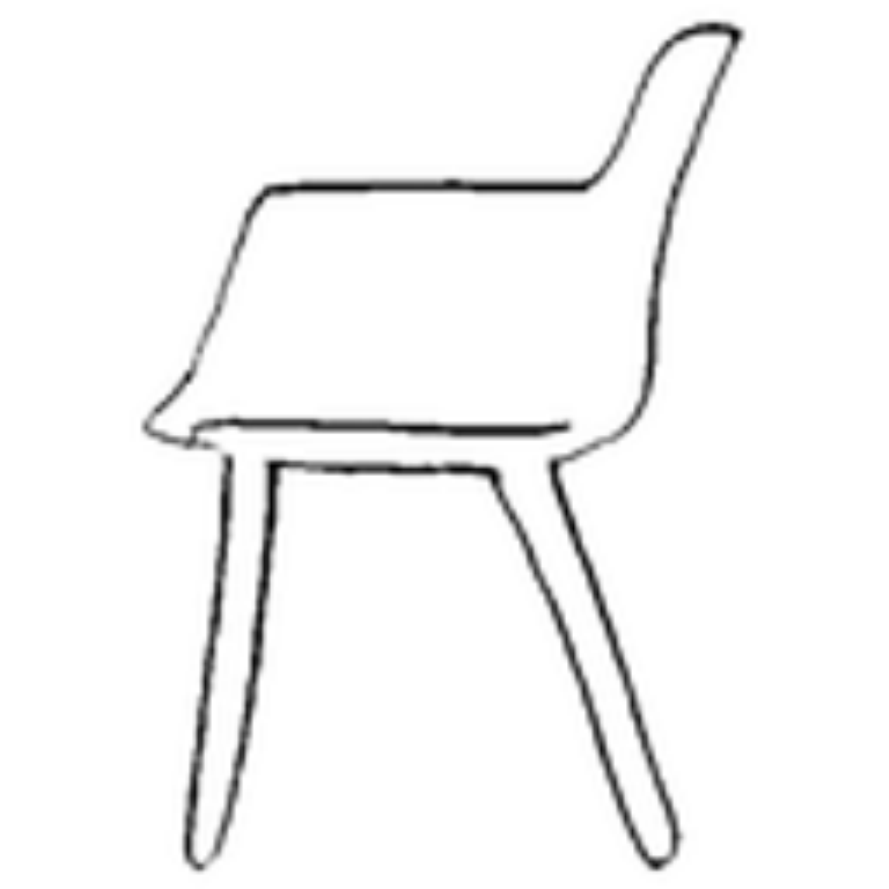} & \multicolumn{2}{c}{\includegraphics[trim =30 30 30 30, clip, width=0.195\linewidth]{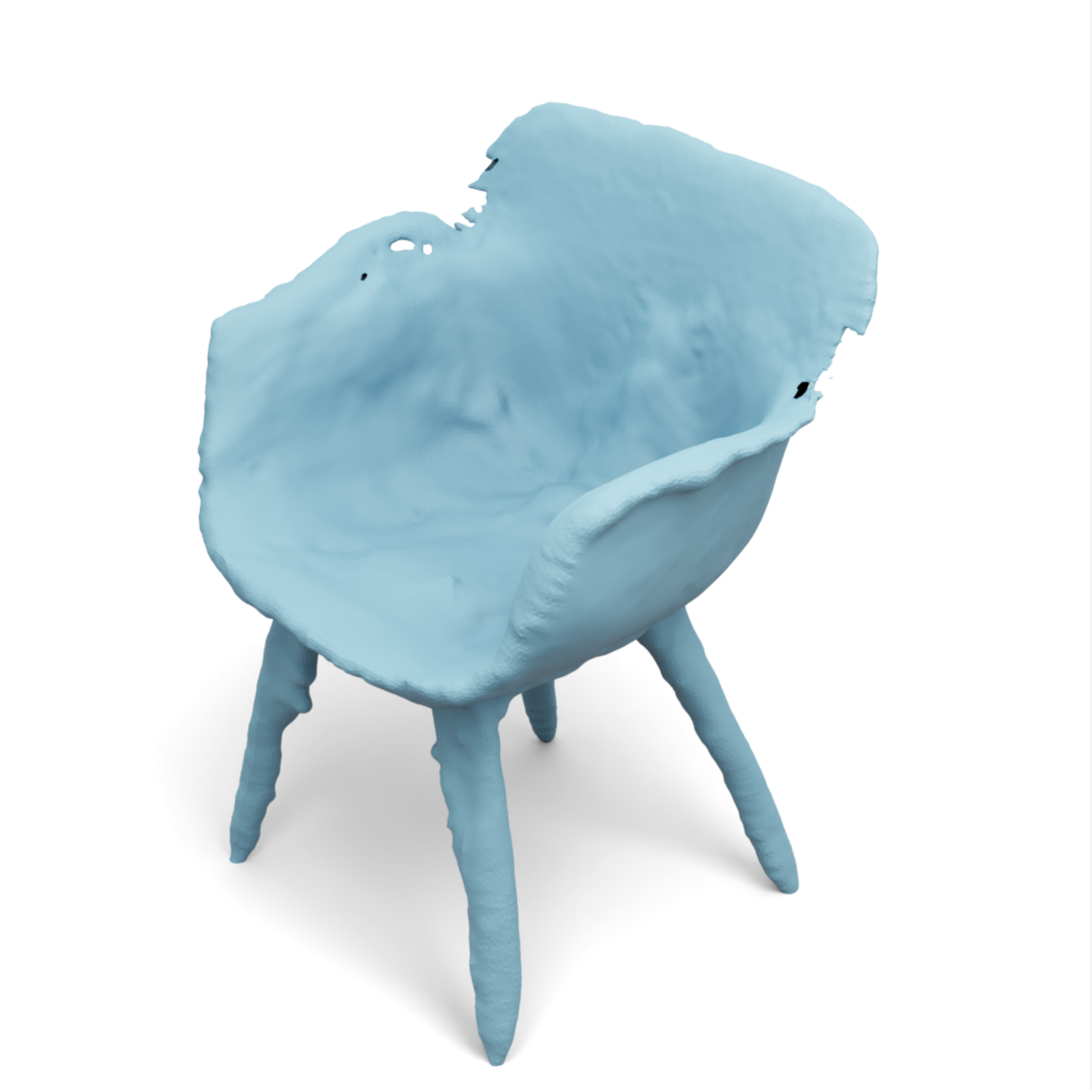}} &  \includegraphics[trim = 30 30 30 30, clip, width=0.195\linewidth]{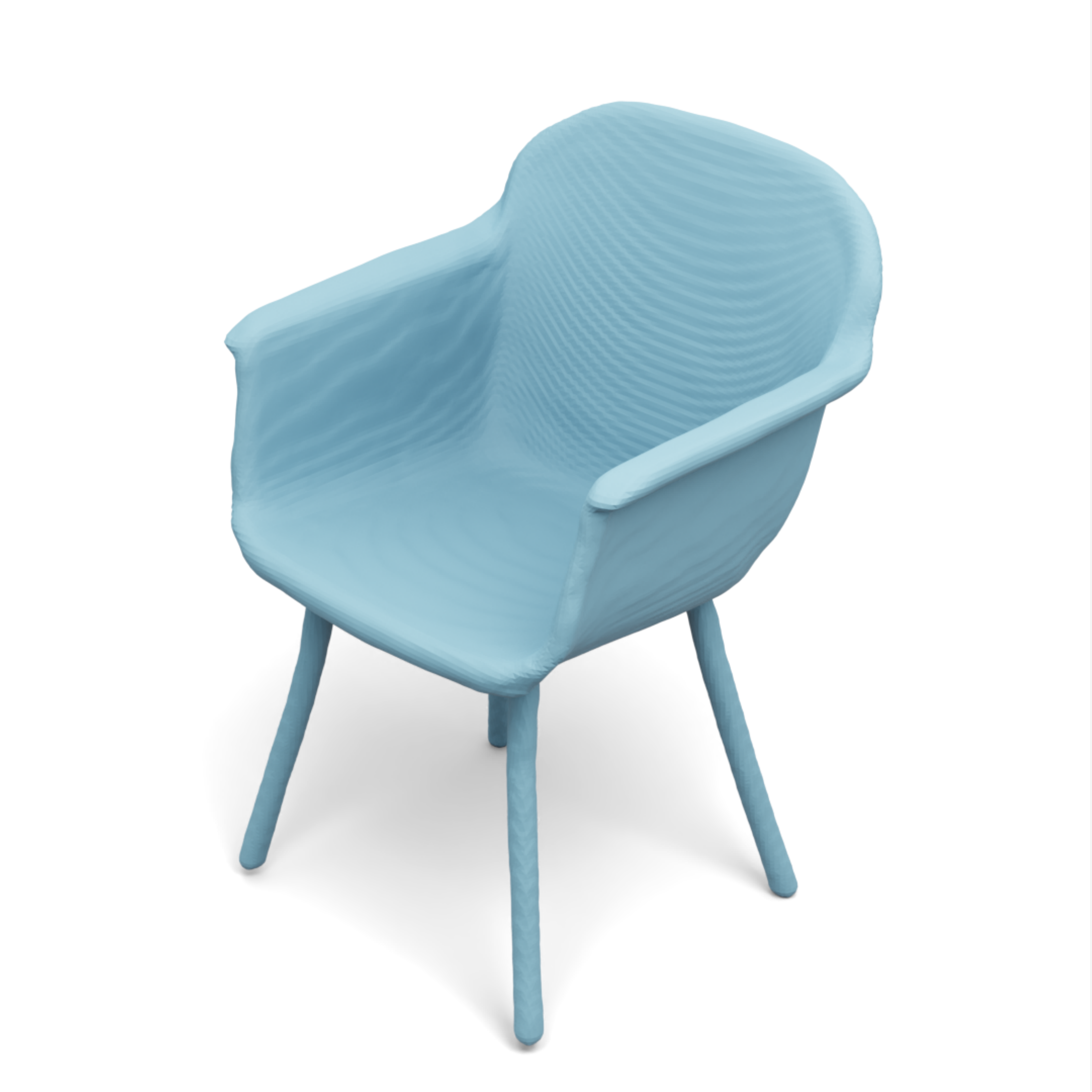} & \includegraphics[trim = 30 30 30 30, clip, width=0.195\linewidth]{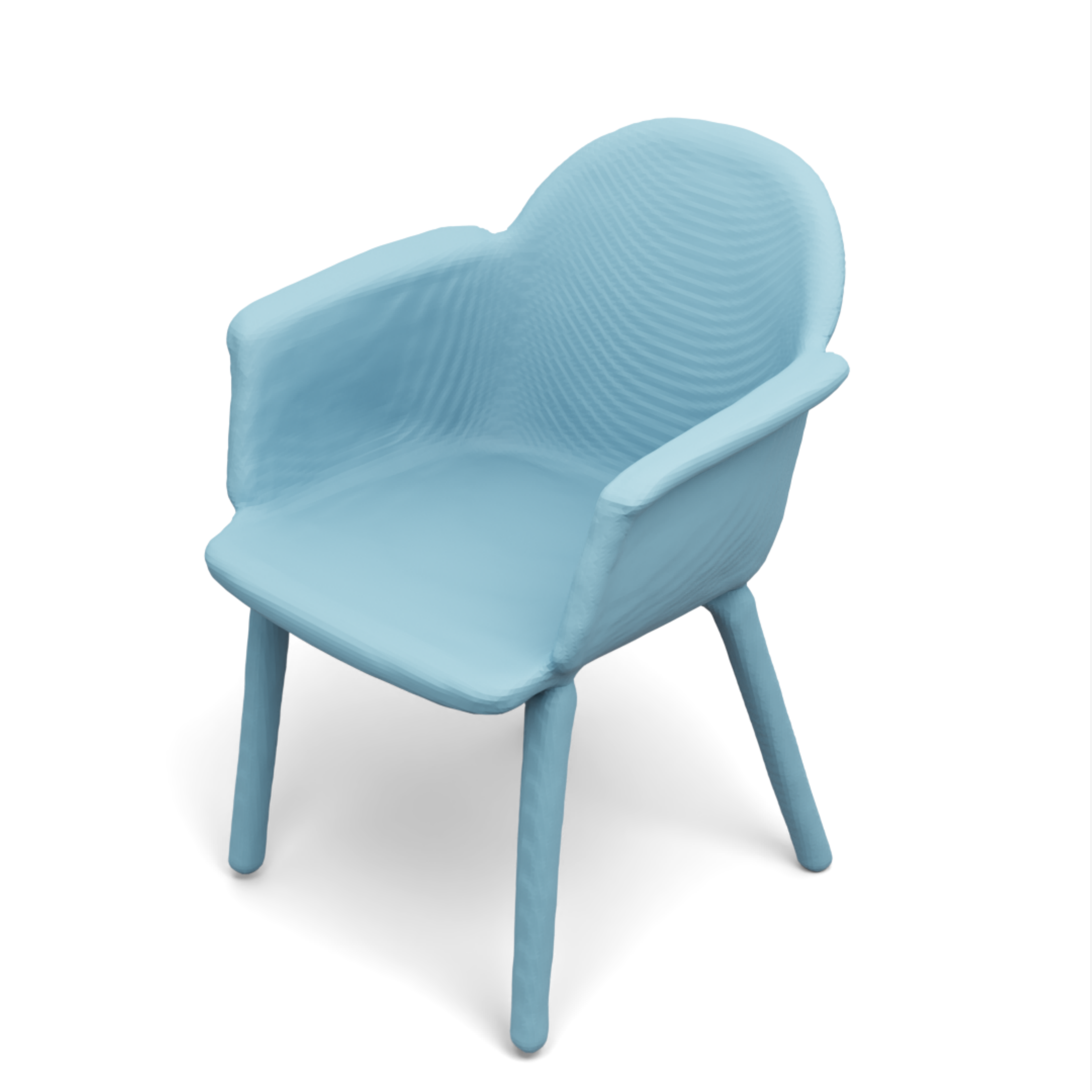} \\ \includegraphics[width=0.195\linewidth]{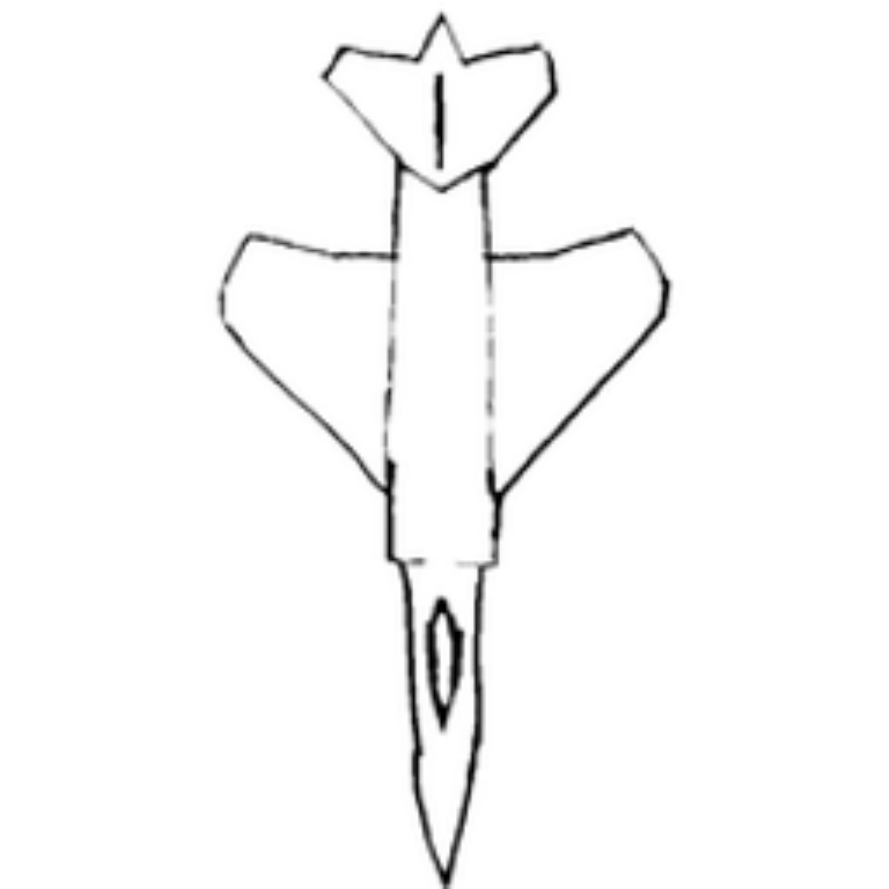} & \includegraphics[width=0.195\linewidth]{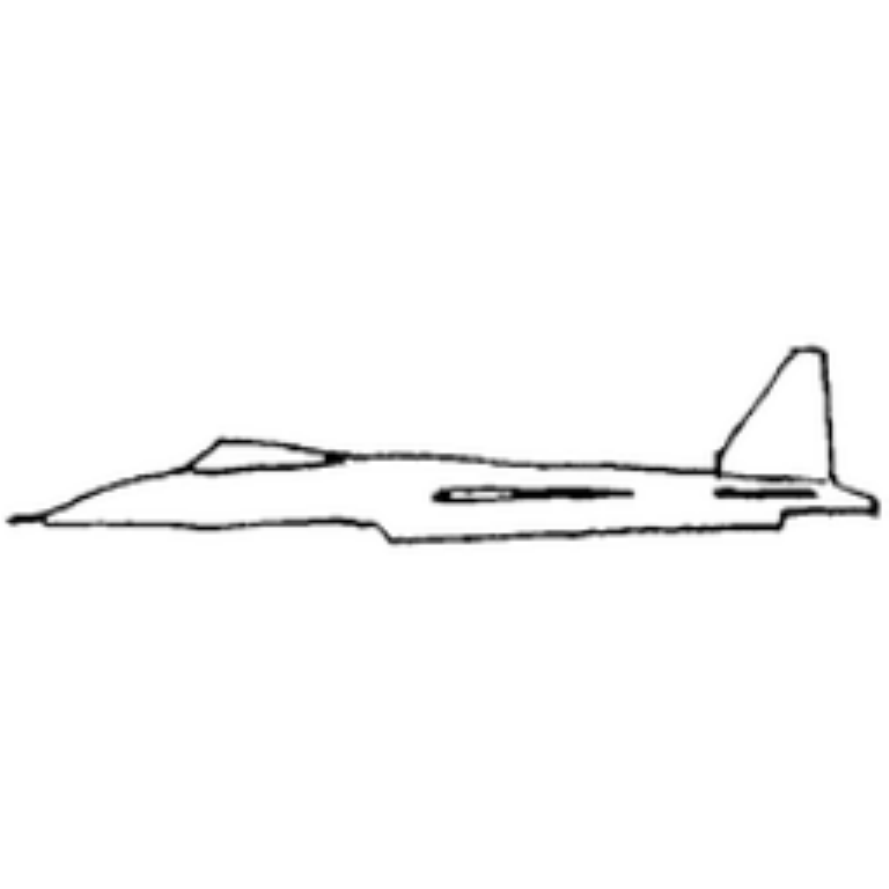} & \multicolumn{2}{c}{\includegraphics[trim =30 5 30 70, clip, width=0.195\linewidth]{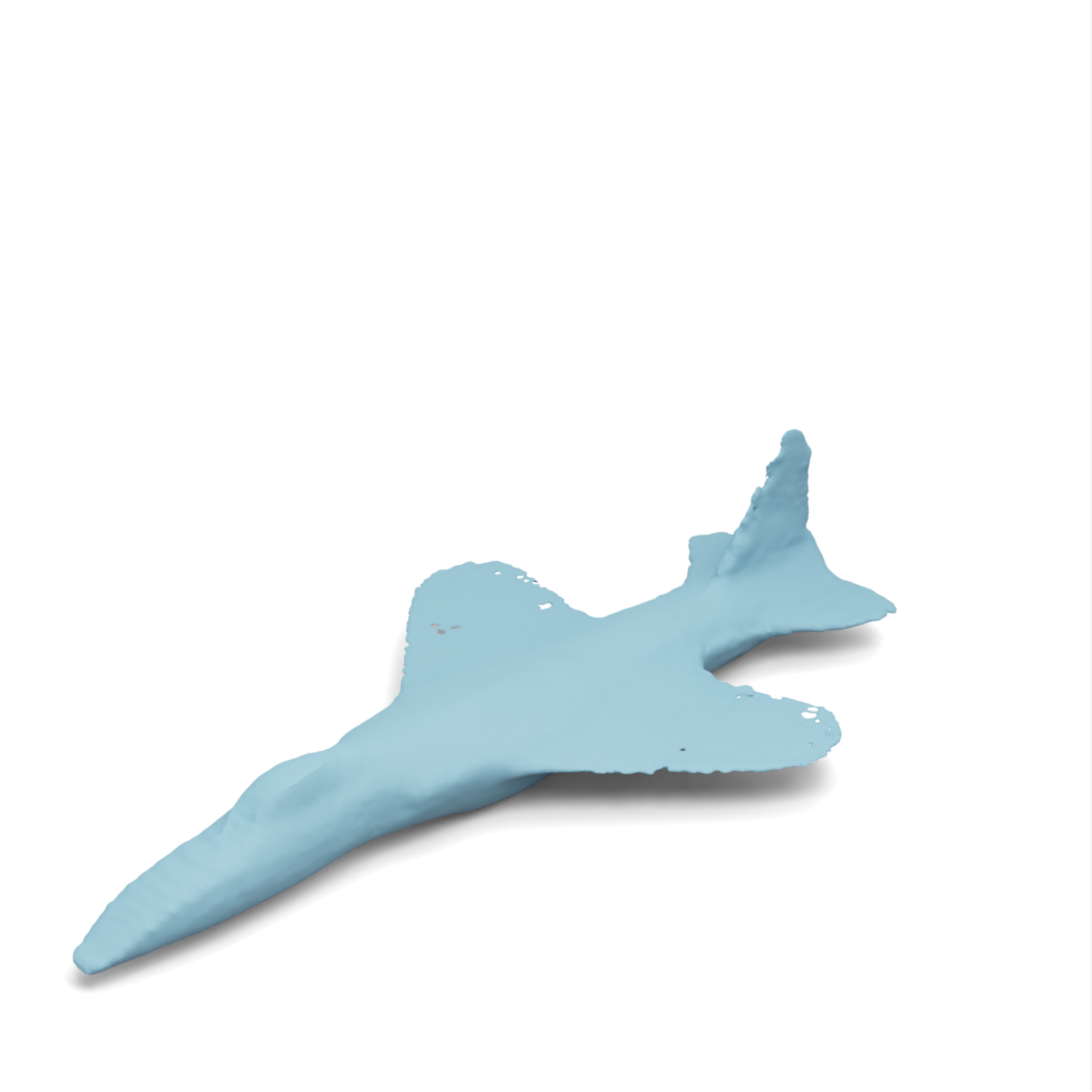}} &  \includegraphics[trim = 30 160 30 30, clip, width=0.195\linewidth]{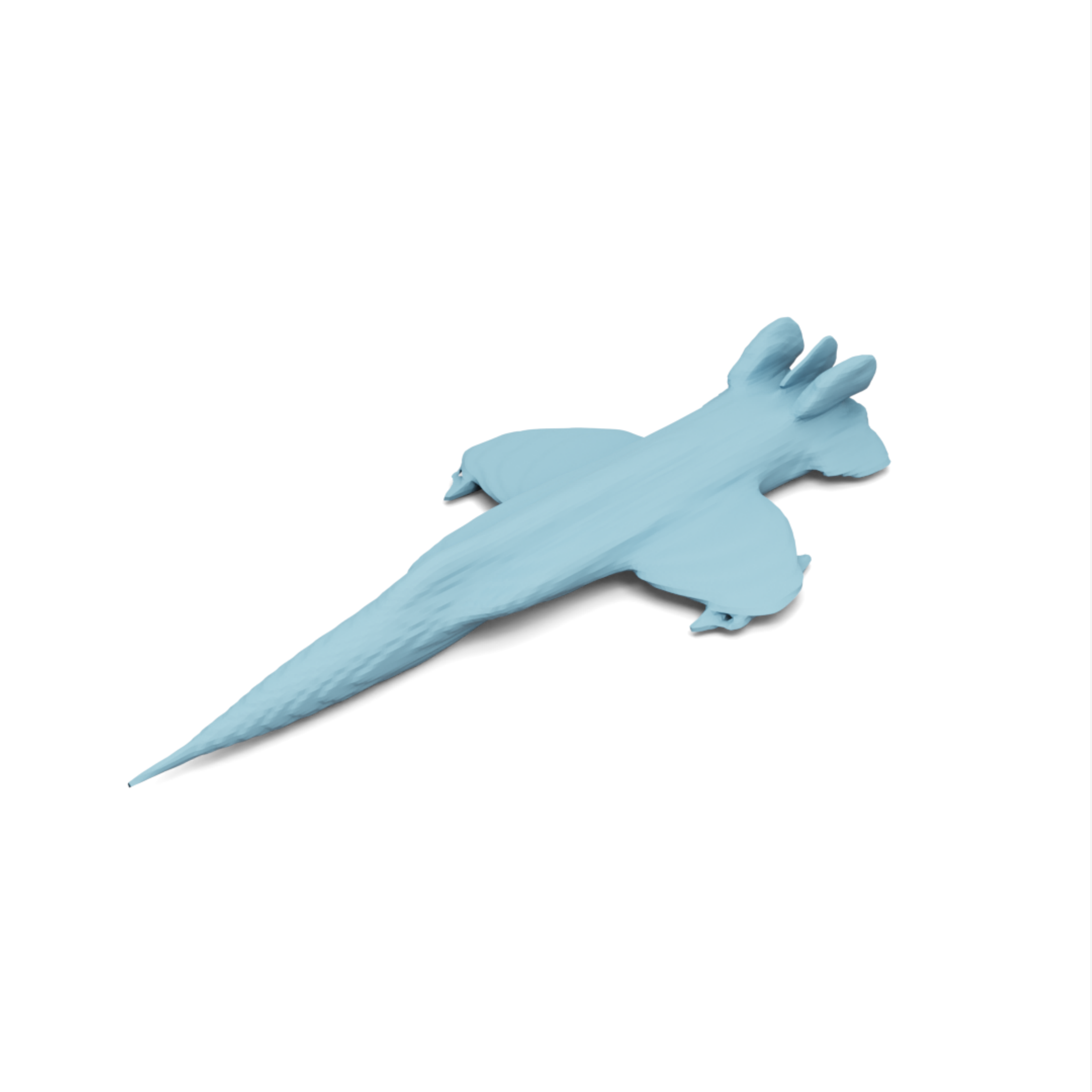} & \includegraphics[trim = 30 160 30 30, clip, width=0.195\linewidth]{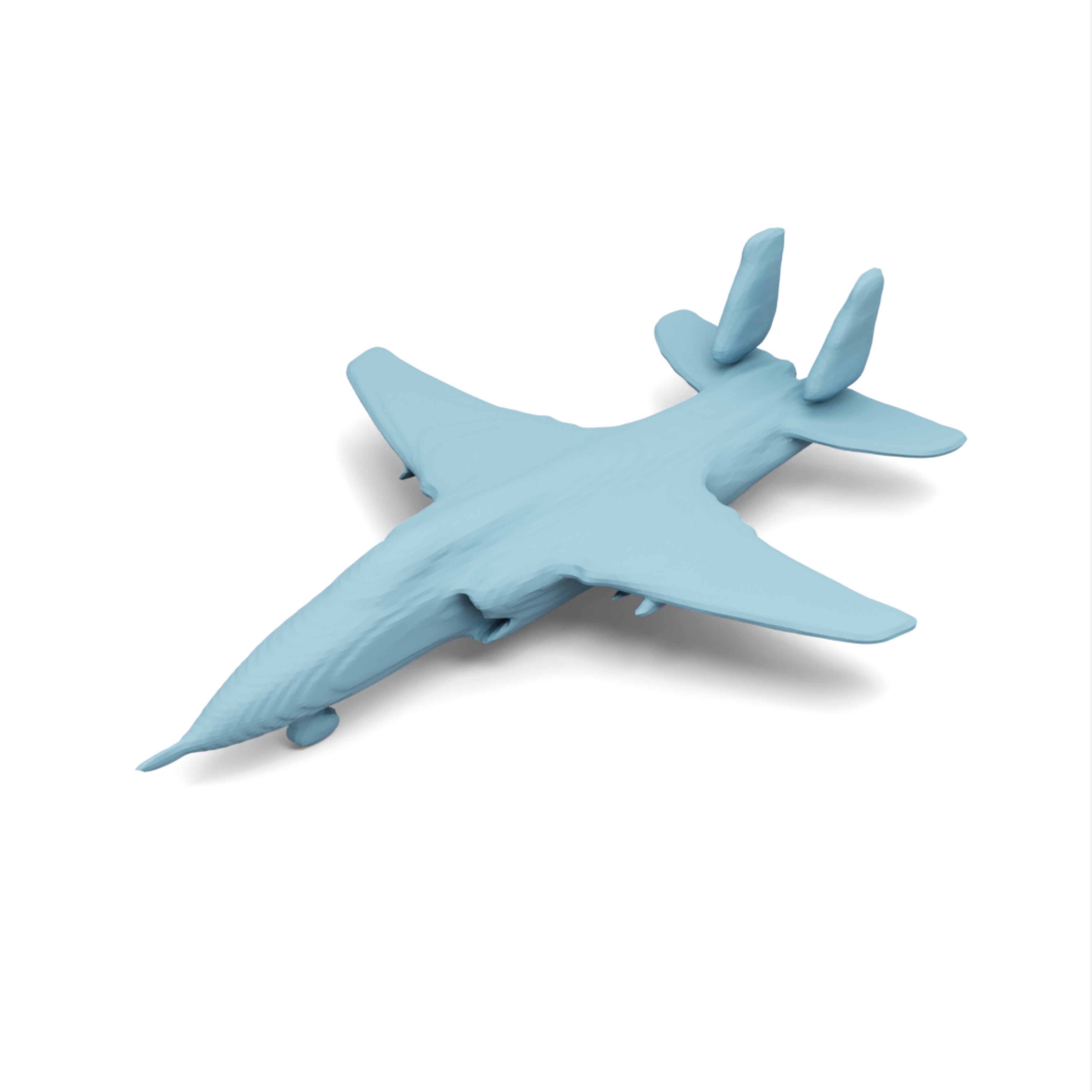}
    \end{tabular}
    \caption{We compare \ourmethod{} with ShapeMVD \cite{Lun2017ShapeMVD}, a sketch-to-shape method requiring multi-view input sketches.
    The pairs of input sketches belong to ShapeMVD test set. Since \ourmethod{} 
    relies on a single-view input, we show the results for both input sketches.}
    \label{fig:comparisondoubleview}
 \vspace{-0.3cm}
\end{figure}

\ourmethod{} can generate a shape from a single input sketch. As we trained our neural network on a combination of outline renderings, abstract sketches and expert freehand sketches (\figref{fig:sketches}), we are able to produce sensible outputs from sketches of diverse styles. \figref{fig:bigfigure}, \figref{fig:bigfiguremulti}, and our supplementary material show some examples of our sketch-based generation.

We compare \ourmethod{} in \figref{fig:comparison} with three single-view image to shape methods, namely Pixel2Mesh \cite{pixel2mesh}, Sketch2Mesh \cite{Sketch2Mesh} and DeepSketch \cite{10.1016/j.cag.2022.06.005}. Pixel2Mesh is a generic, non-sketch-specific, image-to-shape method. Though able to reconstruct a shape that maps the outline of the input, the result is less aesthetically pleasing. While DeepSketch and Sketch2Mesh are targeted towards sketch-to-shape applications, their methods struggle to produce reasonable output from abstract sketches. Deepsketch is trained on synthetic shapes \cite{zhong2020} and expert freehand sketches \cite{zhong2021prosketch}, and even though Sketch2Mesh is trained on several sketch styles, the external contours of the input sketches remain the same. We do not use its refinement because it requires additional camera view parameters.

We also compare with ShapeMVD \cite{Lun2017ShapeMVD}, a multi-view reconstruction method in \figref{fig:comparisondoubleview}, using input sketches from their own test dataset. The inputs to ShapeMVD are two orthogonal views that are precisely aligned. Their method predicts the depth map and normal map to output a point cloud from which a mesh is extracted using screened Poisson Surface Reconstruction \cite{kazhdan2013screenedpoisson}. Because the additional view reduces the ambiguity, their method is able to generate shapes that are more accurate to the input, but which seem to be more prone to artefacts. We noticed that ShapeMVD failed at shape generation from abstract sketches, thereby raising the level of skill required to use it.

\subsection{Evaluation}

For an objective evaluation, we ran Pixel2Mesh, Sketch2Mesh, DeepSketch and \ourmethod{} on the AmateurSketch dataset \cite{Qi2021AmateurSketch}, which contains 3000 freehand sketches of ShapeNet chairs of medium abstraction level. We then computed the chamfer distance (CD), the Earth Mover's distance (EMD) and the shading-image-based Fréchet Inception distance (FID) \cite{heusel2018gansfid,parmar2021cleanfid,zheng2022sdfstylegan}. Our results are reported in \tableref{tab:distances}, and we refer the reader to our supplementary material for more details about the used metrics. Note that \ourmethod{} performs better in all the metrics referenced here.

As an additional perceptual evaluation, we conducted a user study. We randomly sampled 24 sketches from the AmateurSketch \cite{Qi2021AmateurSketch} dataset on which we applied the methods we compare with. Users were asked to rank the four chairs for how realistic and how similar to the input sketch they are. \tableref{tab:userstudy} shows the results for both questions in separate columns.
54 people took part in our user study. Note that \ourmethod{} consistently ranks highest both in terms of realism and similarity. More details are to be found in the supplementary material.

\begin{table}[htb]
\small
\caption{Performance comparison of shape reconstruction methods on the AmateurSketch dataset \cite{Qi2021AmateurSketch} using chamfer distance (CD), earth mover's distance (EMD), and Fréchet inception distance (FID). Lower values indicate better performance. Comparison is done with Pixel2Mesh \cite{pixel2mesh}, Sketch2Mesh \cite{Sketch2Mesh}, and DeepSketch \cite{10.1016/j.cag.2022.06.005}. The supplementary material contains additional comparisons.}
\label{tab:distances}
\centering
\setlength{\tabcolsep}{10pt}
\begin{tabular}{lccc}
\toprule
    Method  & CD$\downarrow$ & EMD$\downarrow$ & FID$\downarrow$ \\
\midrule
Pixel2Mesh & 0.2191 & 0.1658 & 401.7 \\
Sketch2Mesh & 0.2113 & 0.1573 & 368.4 \\
DeepSketch & 0.1520 & 0.1142 &  292.2 \\
SENS & \textbf{0.1186} & \textbf{0.0946} & \textbf{171.3}\\
\bottomrule
\end{tabular}
\end{table}

\begin{table}[htb]
\small
\caption{Perceptual evaluation through a user study, highlighting the performance of our method in comparison to Pixel2Mesh \cite{pixel2mesh}, Sketch2Mesh \cite{Sketch2Mesh}, and retrained DeepSketch \cite{10.1016/j.cag.2022.06.005} in terms of realism and similarity to input sketches (1 is best rank).}
\label{tab:userstudy}
\centering
\setlength{\tabcolsep}{3pt}
\begin{tabular}{lccccccccc}
\toprule
    Question  & \multicolumn{4}{c}{Realistic} & & \multicolumn{4}{c}{Similar} \\
    Rank  & $1$ & $2$ & $3$ &$4$ & & $1$ & $2$ & $3$ &$4$ \\
\midrule
Pixel2Mesh  &  0.1  &  1.1  &  13.1  &  \textbf{85.7}  & &  0.6  &  16.0  &  24.5  &  \textbf{59.0}  \\
Sketch2Mesh  &  10.5  &  \textbf{52.0}  &  34.1  &  3.4  & &  1.8  &  28.6  &  \textbf{44.8}  &  24.8  \\
DeepSketch  &  2.0  &  37.4  &  \textbf{49.9}  &  10.6  & &  4.1  &  \textbf{49.5}  &  30.3  &  16.1  \\
SENS  &  \textbf{87.4}  &  9.5  &  2.9  &  0.2  & &  \textbf{93.5}  &  5.9  &  0.5  &  0.1  \\
\bottomrule
\end{tabular}
\end{table}

\subsection{Shape completion}

As explained in \secref{par:partialshape}, our network predicts latent codes $\Tilde{\mathbf{z}} \in \mathds{R}^{m\times d_{\text{model}}}$ and a continuous score $\Tilde{\mathbf{c}} \in \left[0, 1\right]^m$, where $\Tilde{c}_i$ indicates the probability that the $i$th component of $\Tilde{\mathbf{z}}$ is represented in the sketch. While the use of partial outline rendering allows our training to disentangle the different parts of the input sketch, the prediction of the mask $\Tilde{\mathbf{c}}$ is useful to determine the confidence of the network in the reconstruction of each part. Because \ourmethod{} reconstructs a shape from a single viewpoint, it often has to reconstruct parts of the shape that are not depicted in the input sketch. We show in \figref{fig:completion} several examples of completion. The part $i$ of a shape is said to be \emph{completed} if the mask probability $c_i$ is below a certain threshold, here set to $0.01$. Completed parts are displayed in orange.

\subsection{Shape retrieval}

It is crucial for shape-generation techniques to address the retrieval problem. This means that a method should be able to generate a desired shape based on a given sketch, and not just retrieve a shape from the training dataset that approximates a reasonable result.
In \figref{fig:comparison}, we provide evidence that \ourmethod{} does not merely retrieve shapes. The main enabler for this is the part-aware property of \ourmethod{} as it is trained to produce disentangled part vectors that are combined to generate the whole shape. For instance, while the first and second output shapes share similar legs as their respective first retrievals, they exhibit significant differences in the back area. The rounded back of the third chair is not present in the top-4 shape retrieval results. While the fourth shape has an identical structure to its top retrieval, the back, seat, and legs' lengths vary. 

\subsection{Editing}

The ability to generate 3D shapes from sketches can simplify 3D modeling. 
Yet, a user may desire to edit the generated shape, which is a complex task. One major advantage of \ourmethod{} is the ability to easily edit shapes through sketching (\figref{fig:teaser}). We implemented a user interface using the Visualization Toolkit (VTK) \cite{schroeder2006visualization} featuring a drawing canvas and a viewer that displayed the generated shape after its conversion to a mesh via marching cube \cite{lorensen87marchingcubes}. We present a live demonstration of the editing possibilities in a video attached to the supplementary material.

\subsubsection{Outline rendering}
\begin{figure}[t]
	\centering
	\small
	\setlength{\tabcolsep}{1pt}
	\begin{tabular}{cccccc}
    \includegraphics[width=0.16\linewidth]{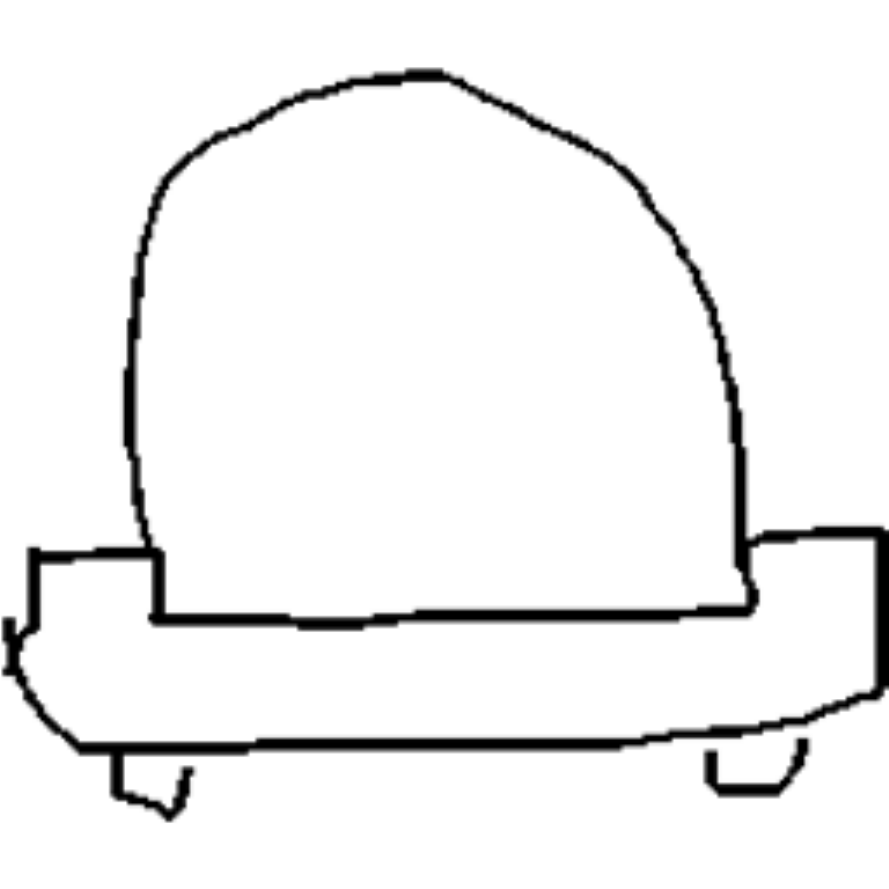}
    &
    \includegraphics[width=0.16\linewidth]{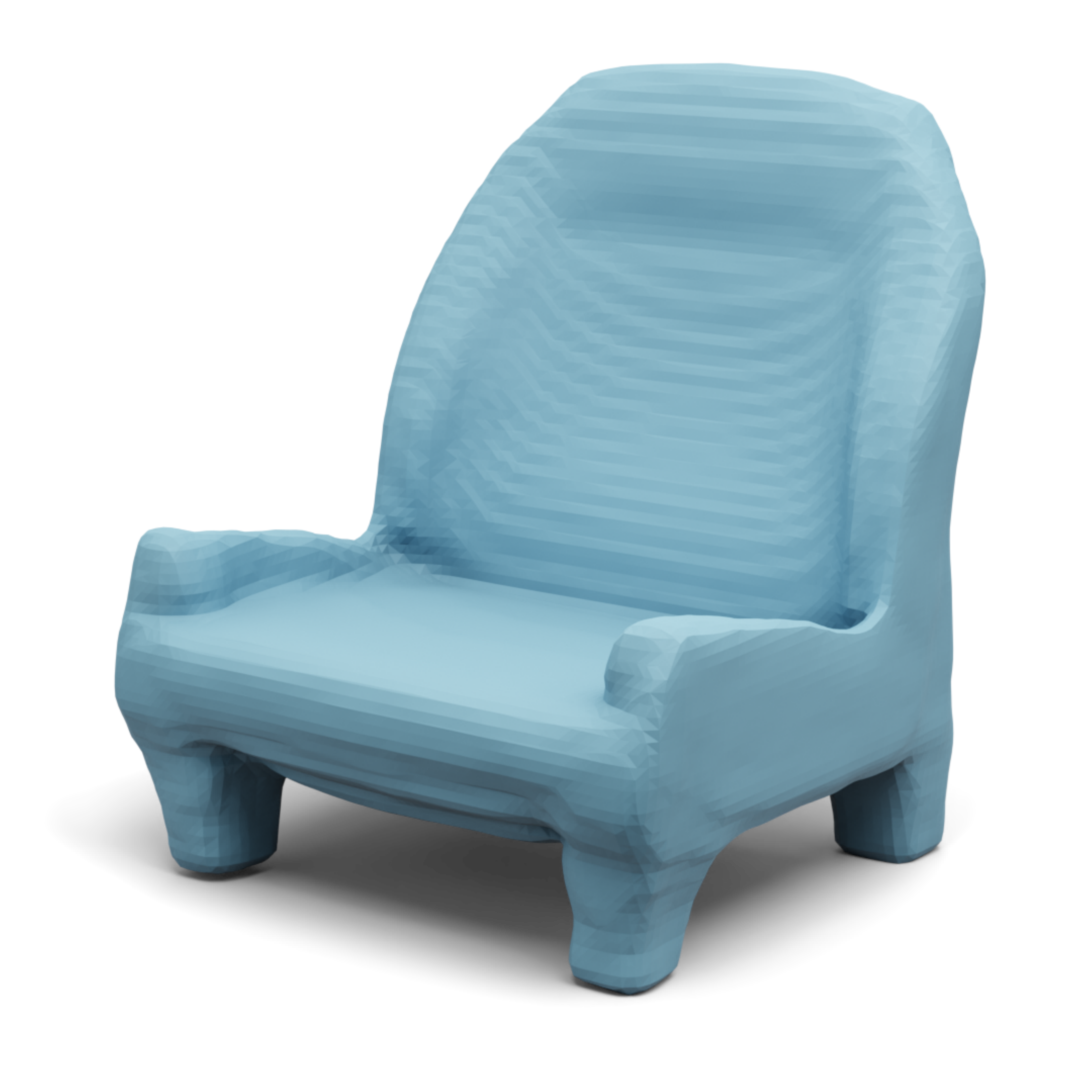}
    &
    \includegraphics[width=0.16\linewidth]{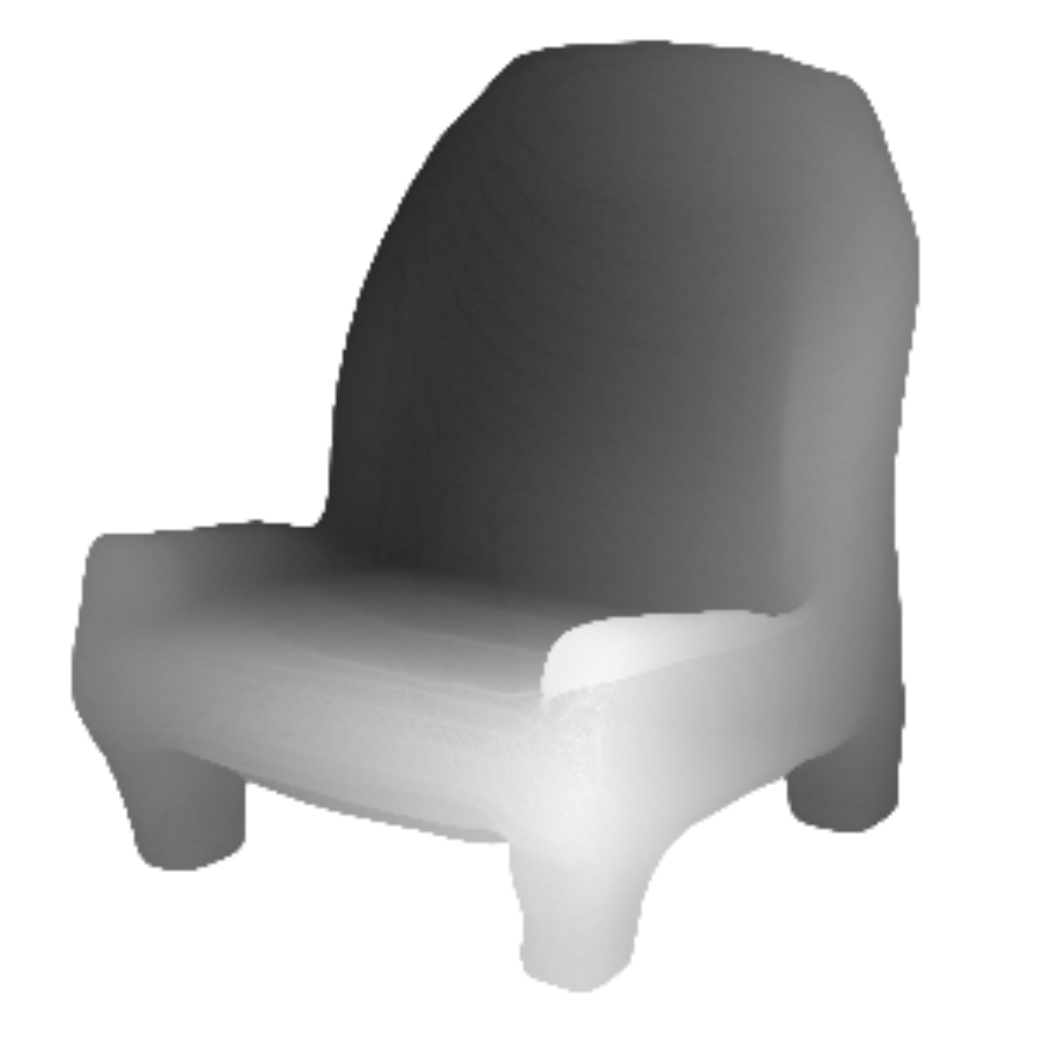}
    &
    \includegraphics[width=0.16\linewidth]{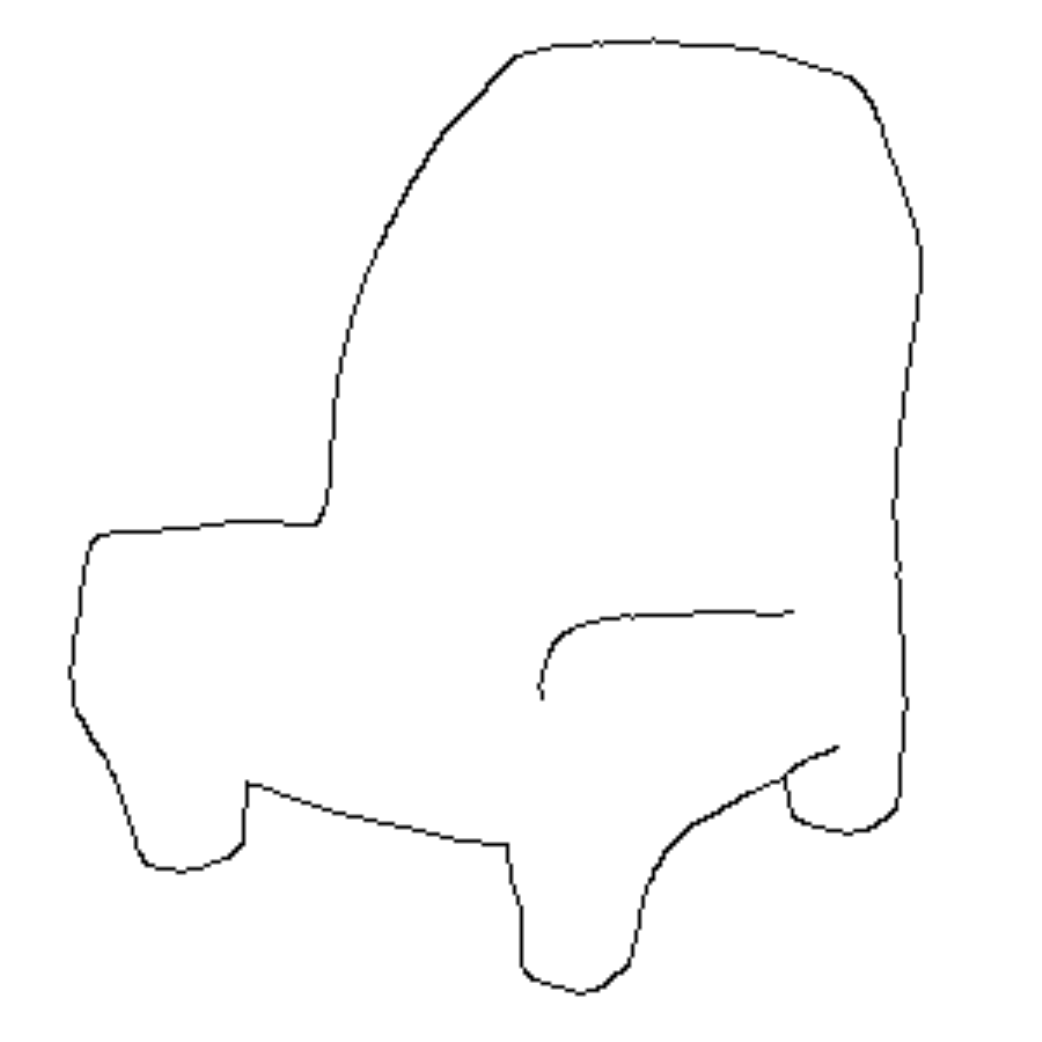}
    &
    \includegraphics[width=0.16\linewidth]{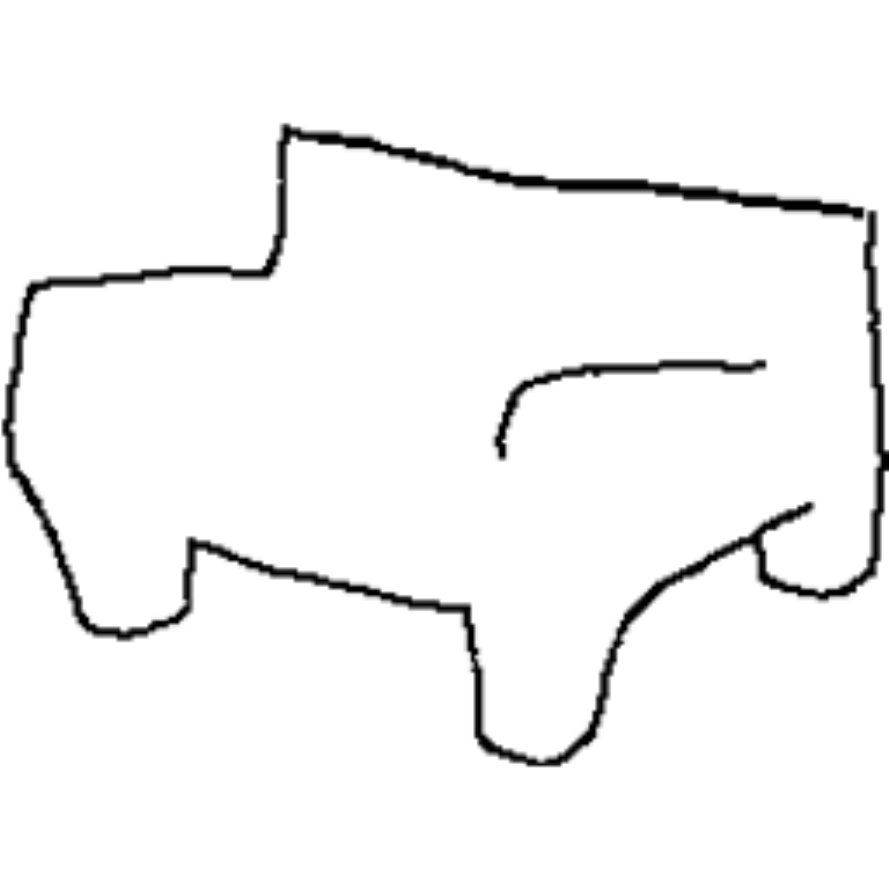}
    &
    \includegraphics[width=0.16\linewidth]{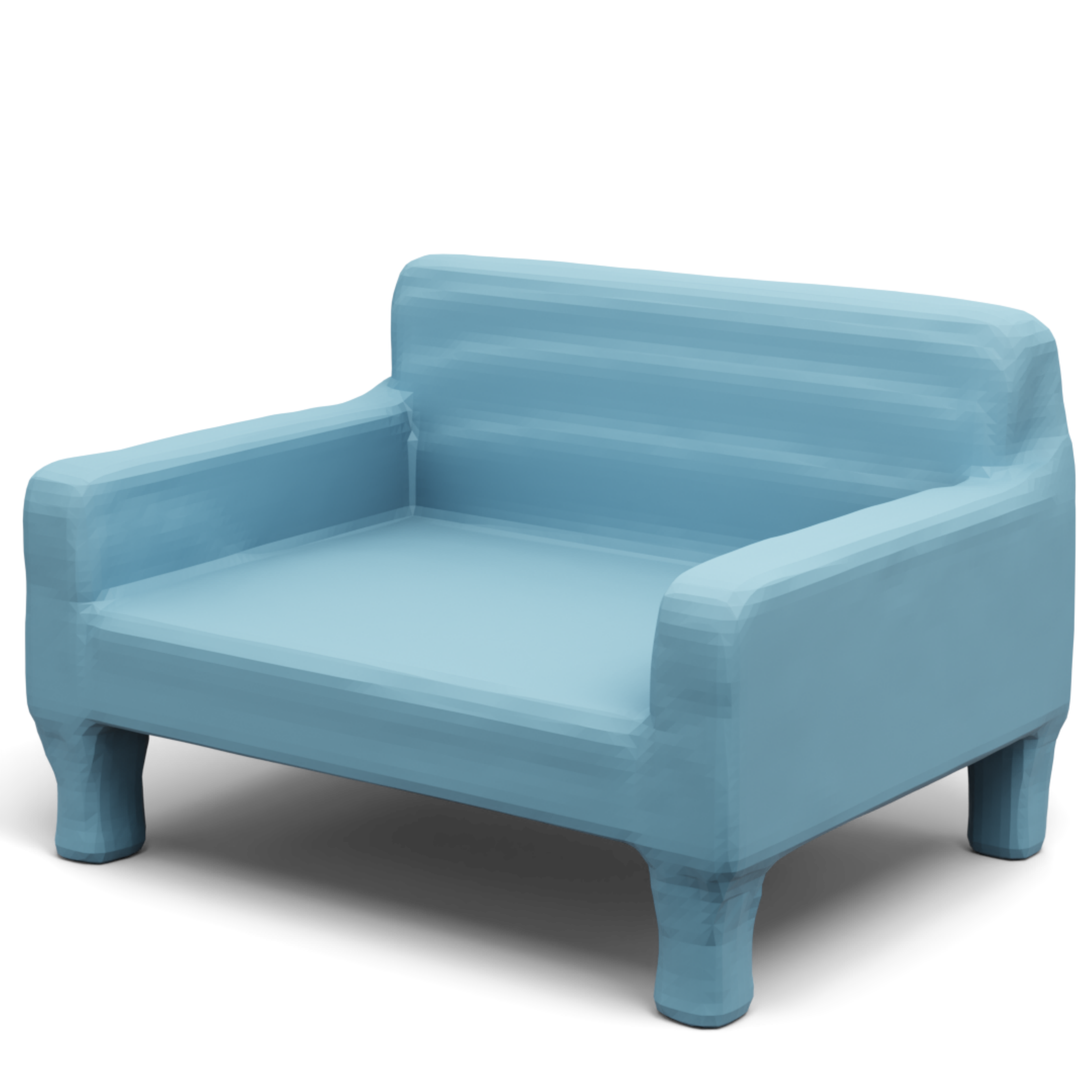} \\
    (i) & (ii) & (iii) & (iv) & (v) & (vi)
	\end{tabular}
	\caption{Our outline rendering pipeline. An initial drawing (i) serves for shape generation (ii). We render its depth map (iii), which is in turn used for edge extraction (iv). The outline can be modified (v) and used as an input for further shape generation (vi).}
	\label{fig:outline}
 \vspace{-0.5cm}
\end{figure}

Our interface proposes an outline rendering method of the displayed shape, enabling users to perform direct modifications on the drawing canvas. The pipeline is illustrated in \figref{fig:outline}: after an initial drawing (i) generates a starting shape (ii), the shape is rendered as a depth map (iii), which is then smoothed via a Gaussian filter. Edges are then extracted using the Canny edge detection method \cite{Canny86}.
Consequently, the outline aligns with the shape's orientation on the screen (iv). As a result, our interface allows users to first create an abstract sketch of a chair, generate its outline, and then directly edit the outline (v) for further shape generation (vi). This simplification of the 3D modeling process greatly reduces the demand for advanced sketching skills.

\subsubsection{Refinement via part reconstruction}

Because SPAGHETTI is a part-aware shape decoder, it is possible to select parts of the latent code and use a refinement network to regenerate them based on the unselected parts, as described in \secref{sec:refinementNetwork}.
The selection is illustrated in \figref{fig:refinement} and operates as follows: first, the user employs a freehand lasso selection on the screen (ii). Then, our interface detects which faces of the mesh are picked by the lasso selection. The parts of the latent code that encode for the generation of these picked faces are then labeled as "selected". Once a part is selected, we display in orange all the faces that are generated by this part (iii), not only the originally picked faces. Our interface will mask the selected parts and feeds the latent code to the refinement network, which generates new parts of the latent code to replace the selected one (iv).
While the refinement network was initially trained to reconstruct $5\%-40\%$ of masked latent vectors, there are no practical constraints on the number of vector components that can be masked for refinement. The refinement strategy can be particularly useful for removing artifacts from the generated shape, as exemplified in \figref{fig:refinement} and in the supplementary video.

\subsubsection{Part-based modeling}
\begin{figure}[t]
	\centering
	\small
	\setlength{\tabcolsep}{1pt}
	\begin{tabular}{ccc}
    \includegraphics[width=0.33\linewidth]{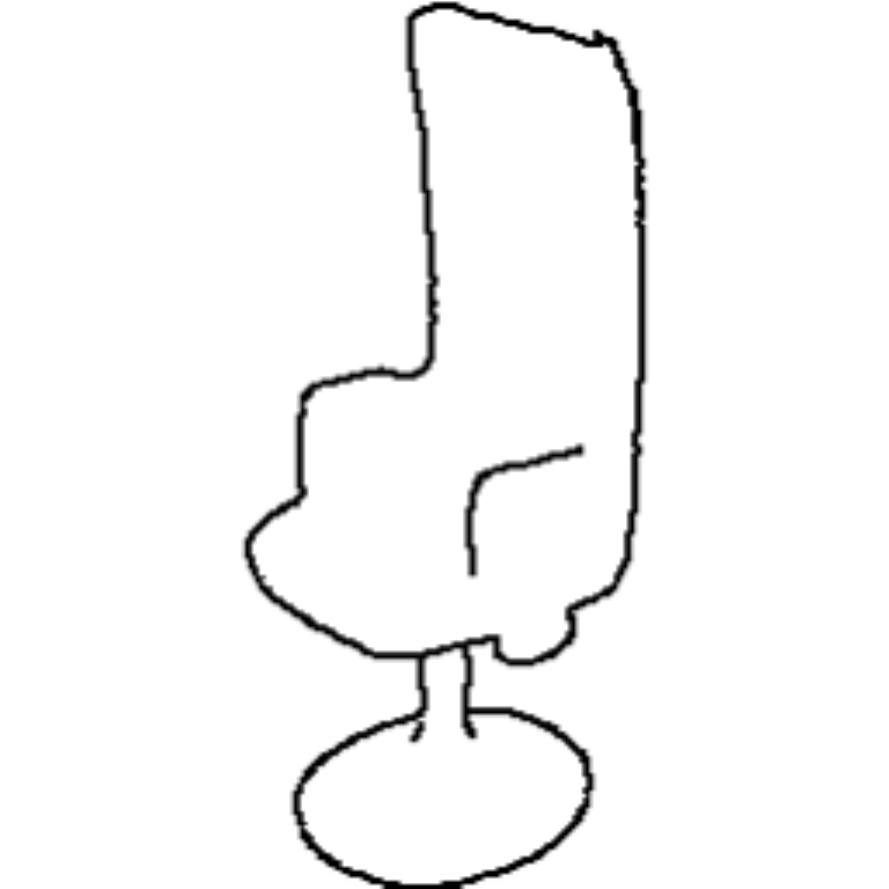}
    &
    \includegraphics[width=0.33\linewidth]{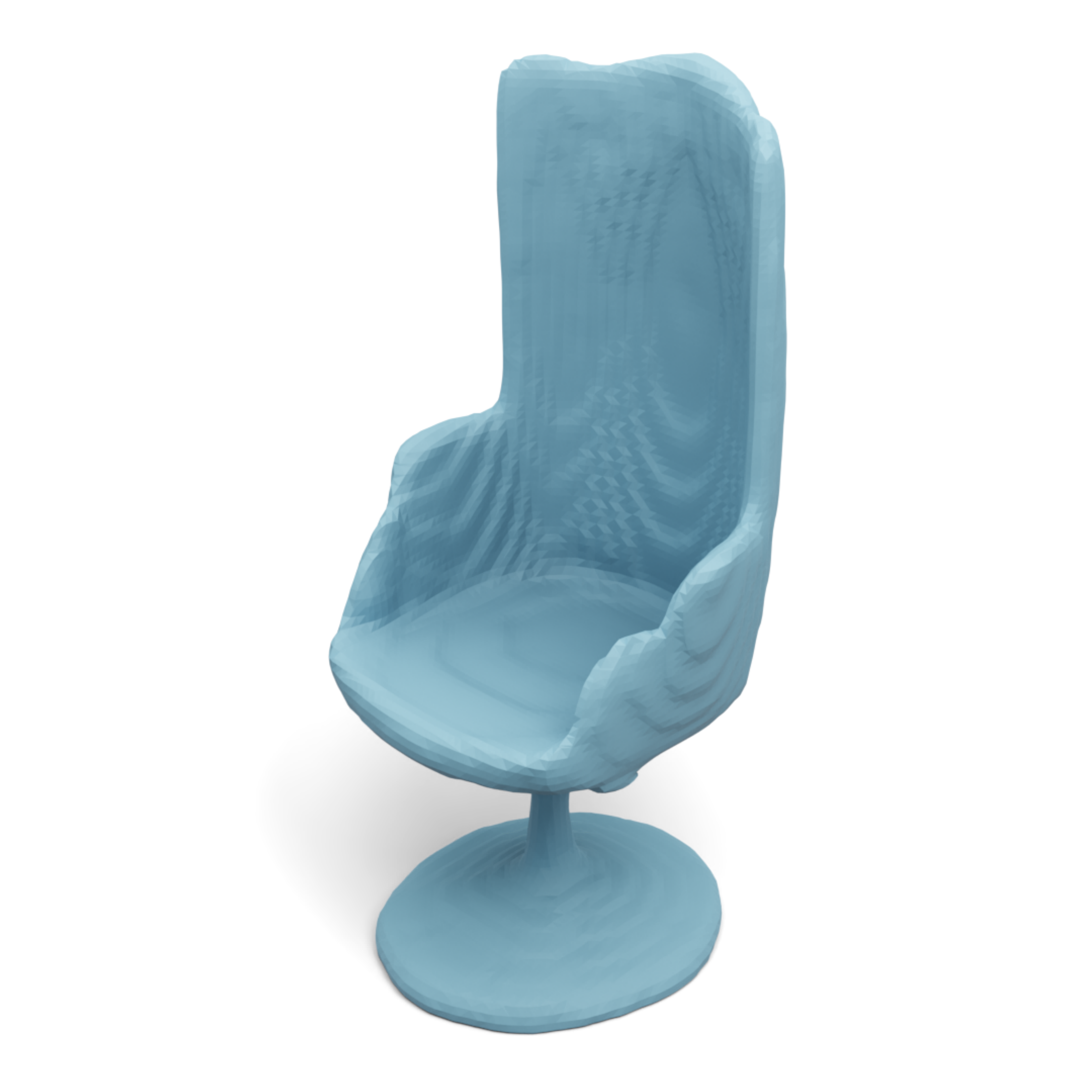}
    &
    \includegraphics[width=0.33\linewidth]{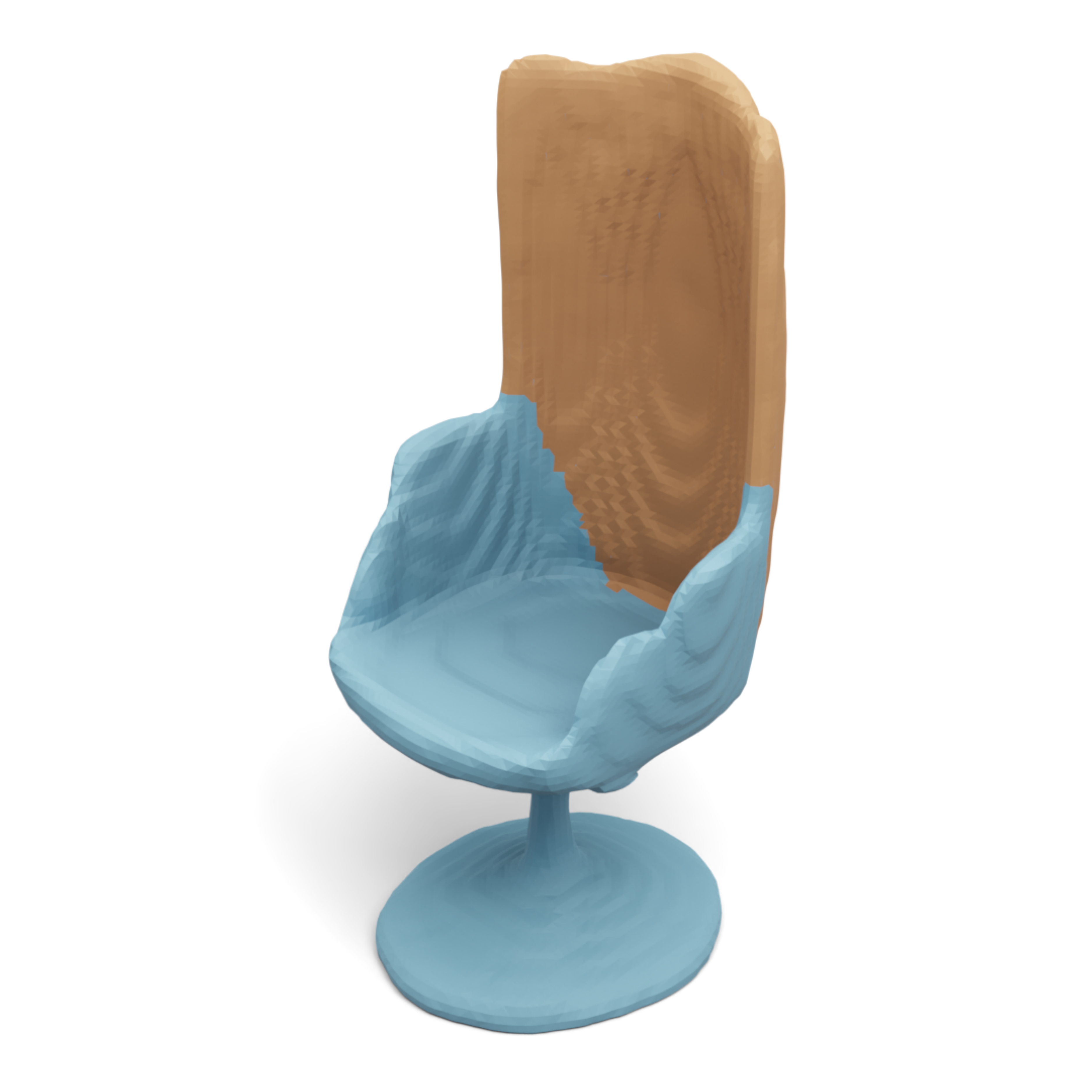}
    \\
    (i)& (ii)& (iii)
    \\
    \includegraphics[width=0.33\linewidth]{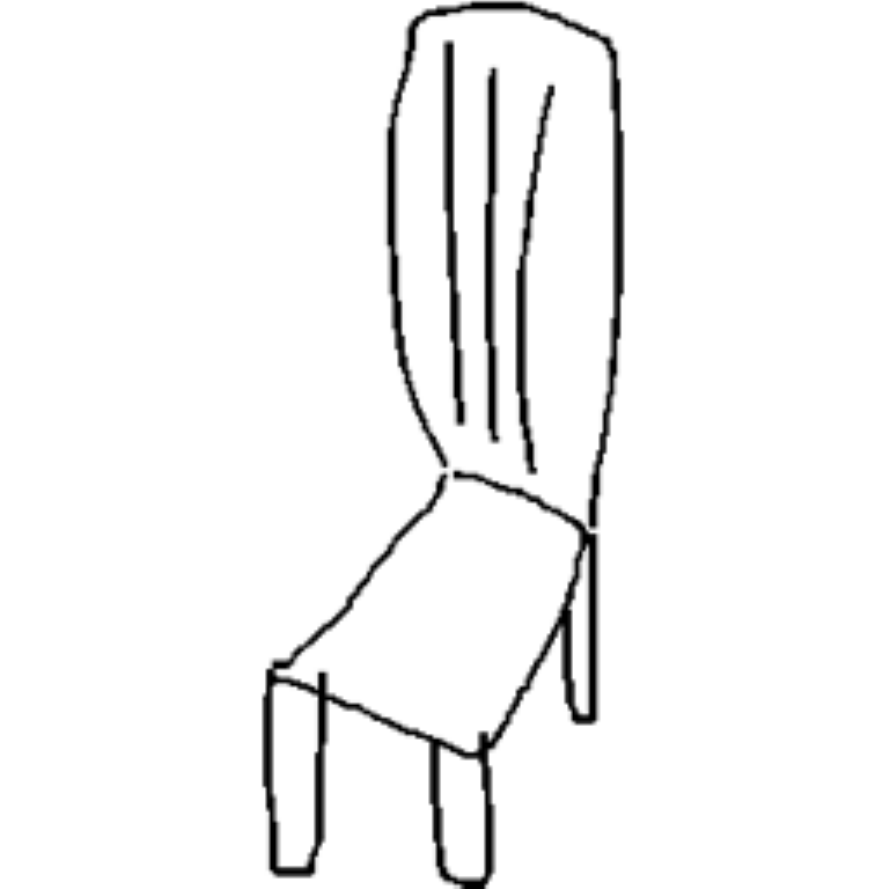}
    &
    \includegraphics[width=0.33\linewidth]{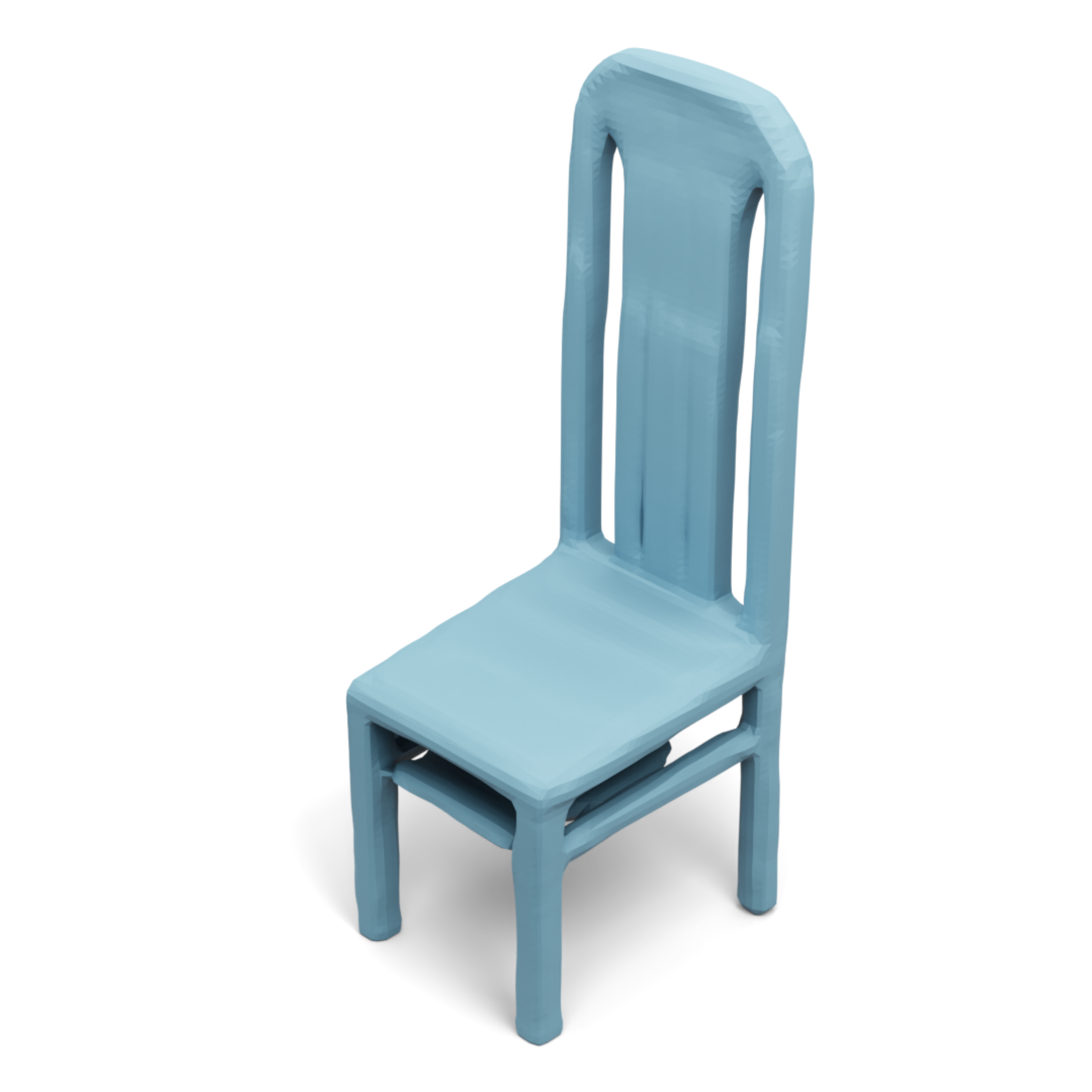}
    &
    \includegraphics[width=0.33\linewidth]{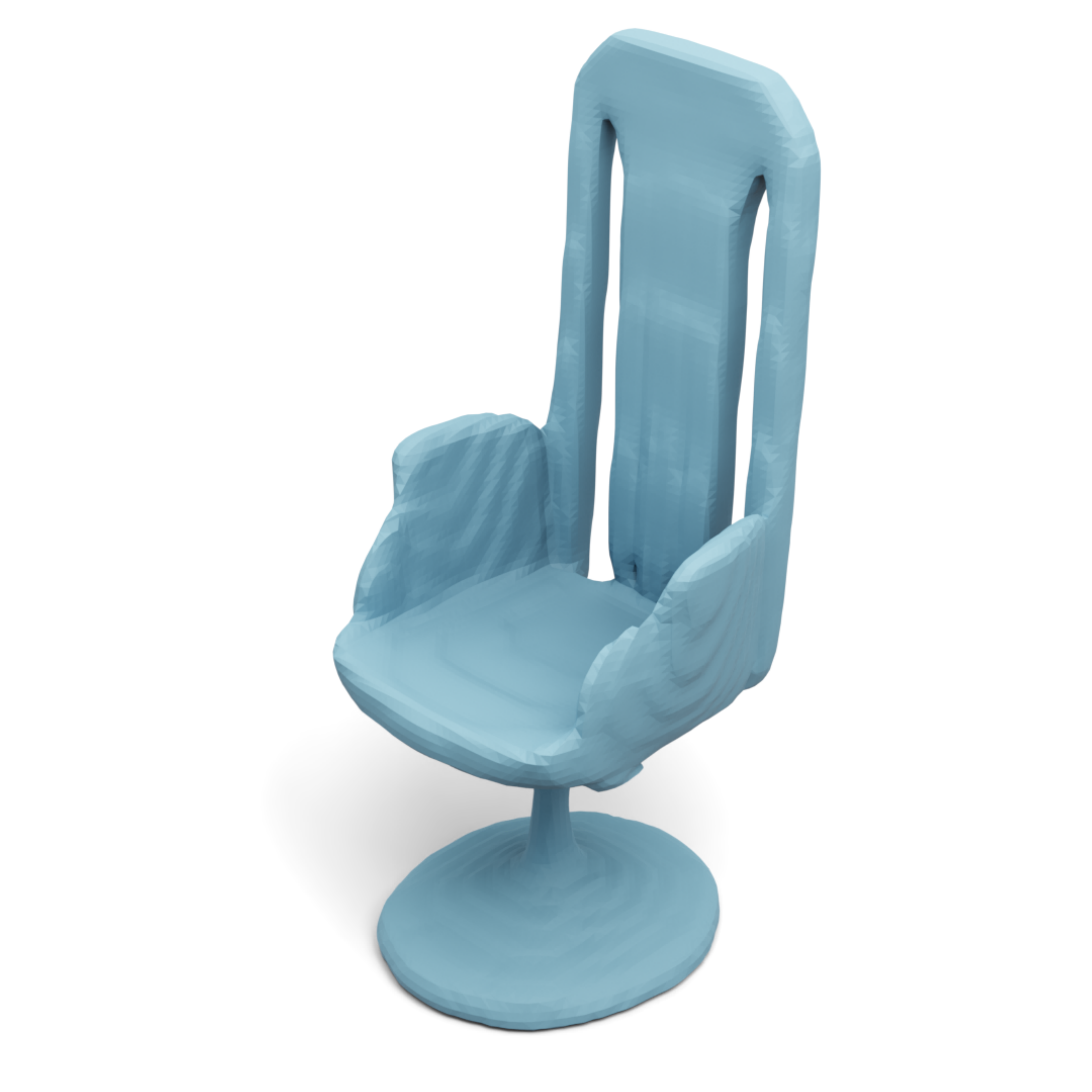}
    \\
    (iv) & (v) & (vi)
	\end{tabular}
	\caption{Part-based modeling example. The input sketch (i) is fed to our network to generate a shape (ii). The user can select parts of the resulting shape (iii). Given another sketch (iv), \ourmethod{} would generate a completely different shape (v). But using part-based modeling, our interface will only replace the selected parts (vi).}
	\label{fig:partbasedModeling}
 \vspace{-0.5cm}
\end{figure}

The use of a part-aware shape decoder also enables local modifications to the generated shapes. Indeed, \ourmethod{} accepts a sketch as input and produces a corresponding latent code that can be broken down into several parts. However, these latent parts can originate from different input sketches, hence allowing the fusion of features from distinct shapes. We provide an illustration of part-based modeling in \figref{fig:partbasedModeling}. The initial drawing (i) generates a latent code that, when decoded, yields a shape (ii). The user can select parts of the latent code, illustrated in orange on the output shape (iii). Drawing another sketch (iv) generates a new latent code, that if decoded by SPAGHETTI, would yield a completely different shape (v). Instead of replacing the entire latent code, only the selected parts are replaced, hence producing a new shape that blends features from both original shapes (vi). In this example, the resulting chair combines the base of the first chair with the backrest of the second chair. This technique represents a substantial improvement over traditional sketch-to-shape methods in significantly extending the modeling flexibility and generation capabilities, going beyond the dataset's inherent limitations. Note that our part-based modeling method can be used with sketches of different abstraction levels, which strengthens its flexibility.

\subsubsection{Evaluation}

To evaluate the usability of our method's editing capabilities, we carried out a user study with 8 participants from diverse backgrounds, possessing varying levels of modeling and sketching expertise. During this session, participants were tasked with two assignments: firstly, creating any chair design, ensuring they utilized all available editing tools to familiarize themselves with our system; and secondly, modeling three distinct shapes based on provided images. After the modeling session, participants completed two questionnaires to gauge the system usability and the efforts required to use it. We show the results in our supplementary material, where we detail the questionnaire outcomes and showcase a range of shapes crafted during the study. Feedback from participants was largely positive; they found the system intuitive and user-friendly, expressing satisfaction with their outputs.

\begin{figure}[t]
	\centering
	\small
	\setlength{\tabcolsep}{3pt}
	\begin{tabular}{cccc}
    Input sketch & Ablation partial loss & Ablation dataset & Ours (full) \\
    \includegraphics[width=0.2\linewidth]{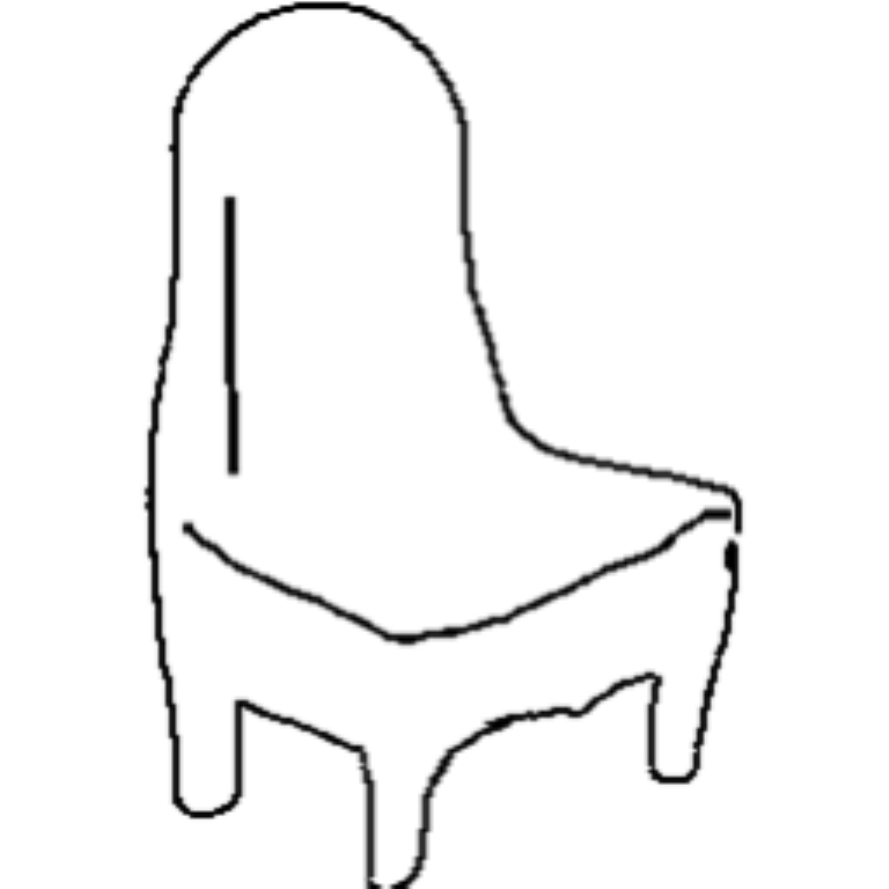}\
    & \includegraphics[trim=60 40 60 10, clip, width=0.2\linewidth]{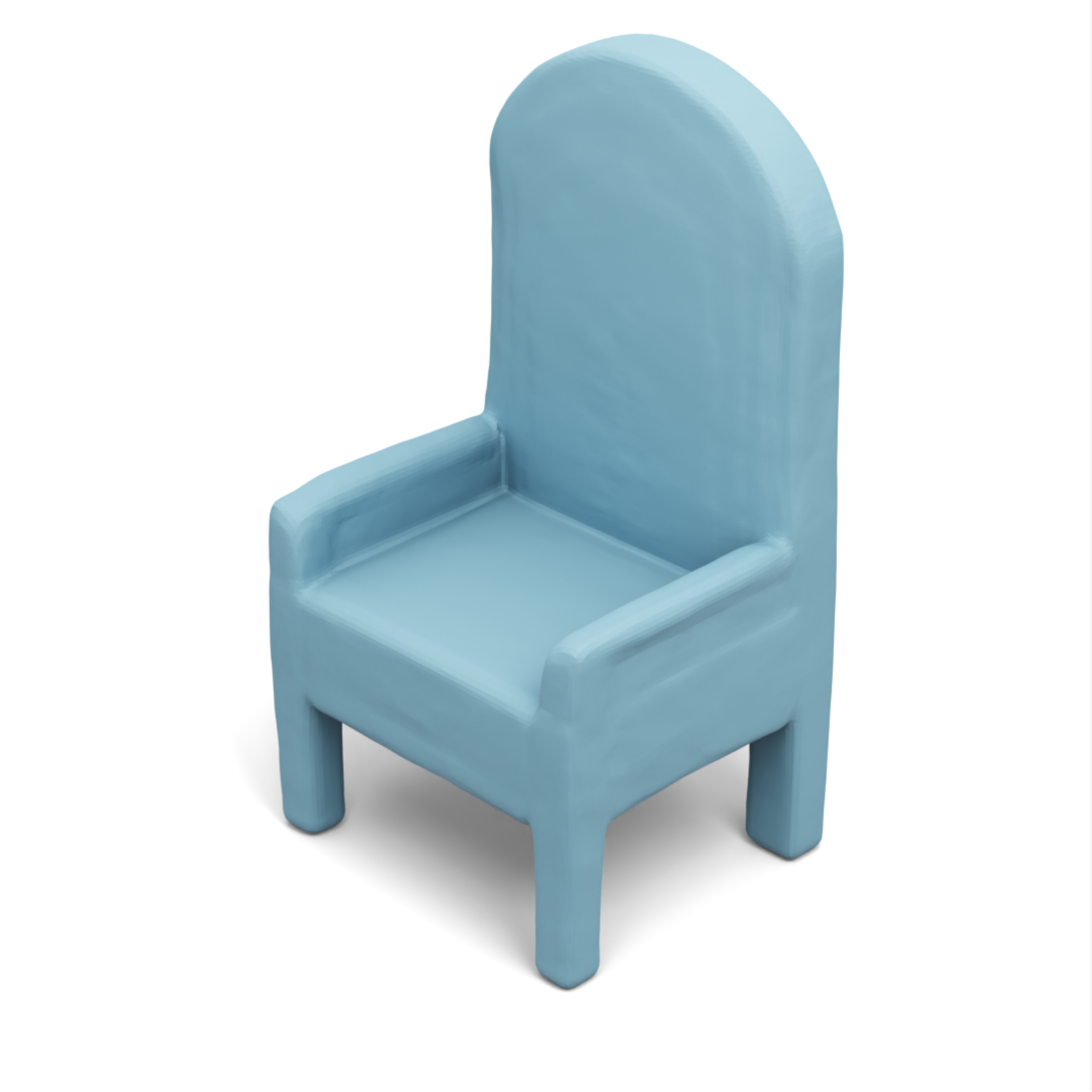}\
    & \includegraphics[trim=60 40 60 10, clip, width=0.2\linewidth]{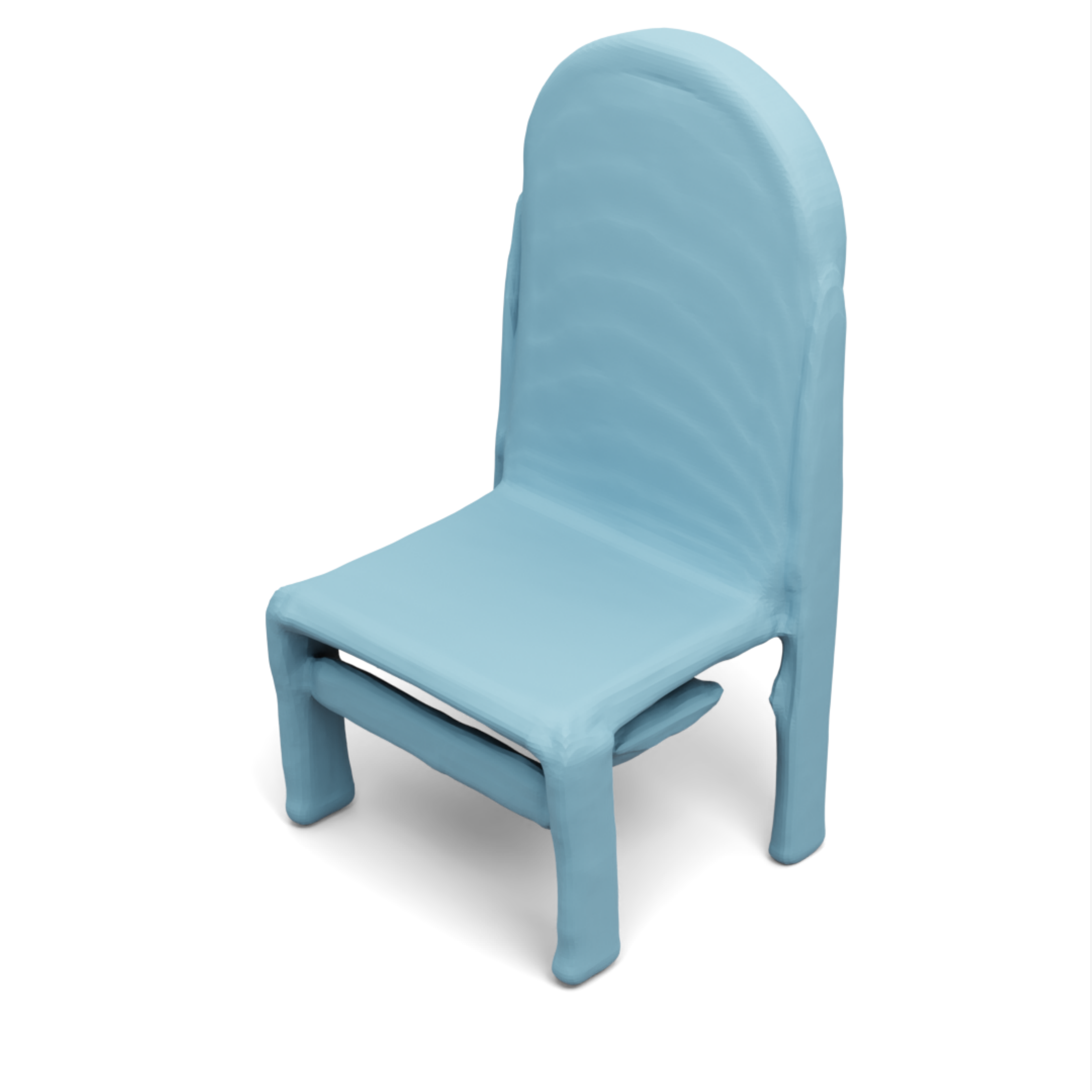}\
    & \includegraphics[trim=60 40 60 10, clip, width=0.2\linewidth]{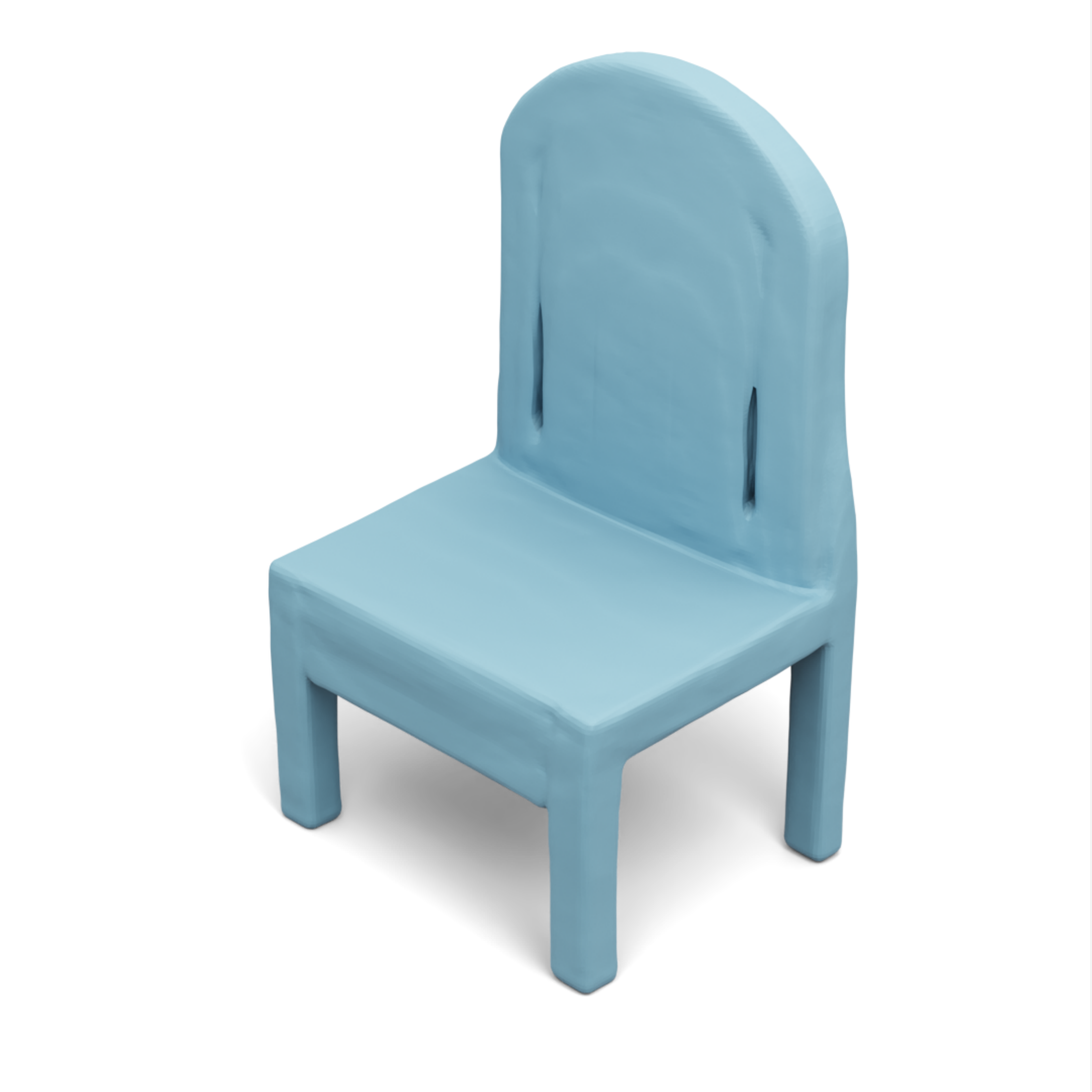}
    \\
    \includegraphics[width=0.2\linewidth]{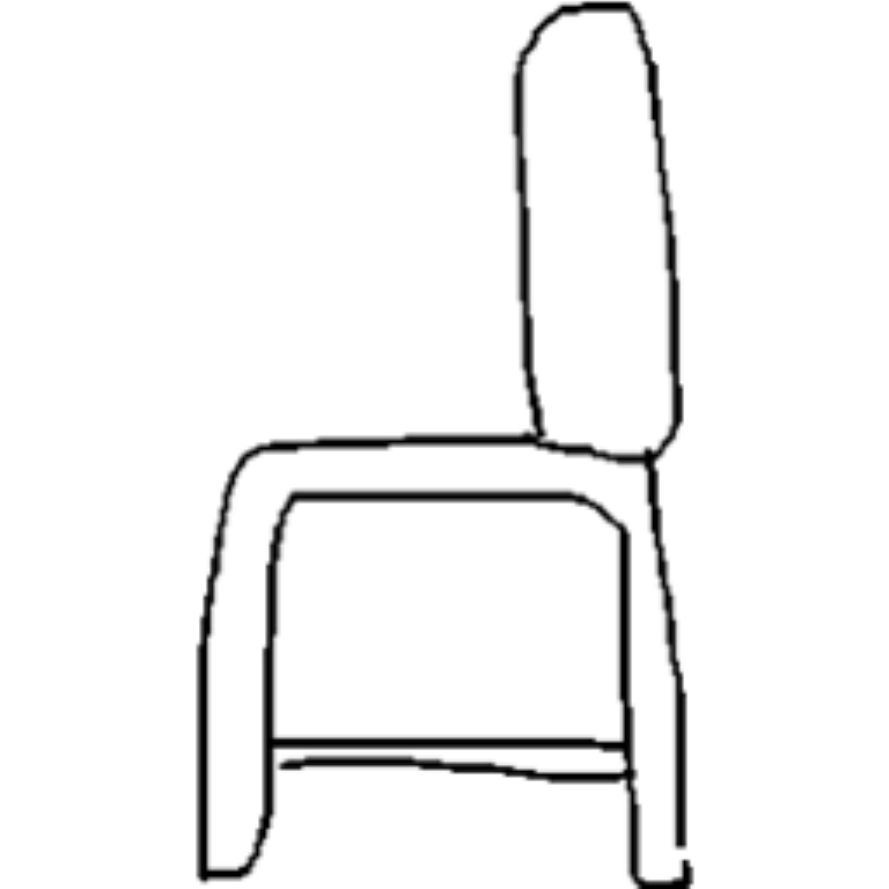}\ 
    & \includegraphics[trim=60 40 60 10, clip, width=0.2\linewidth]{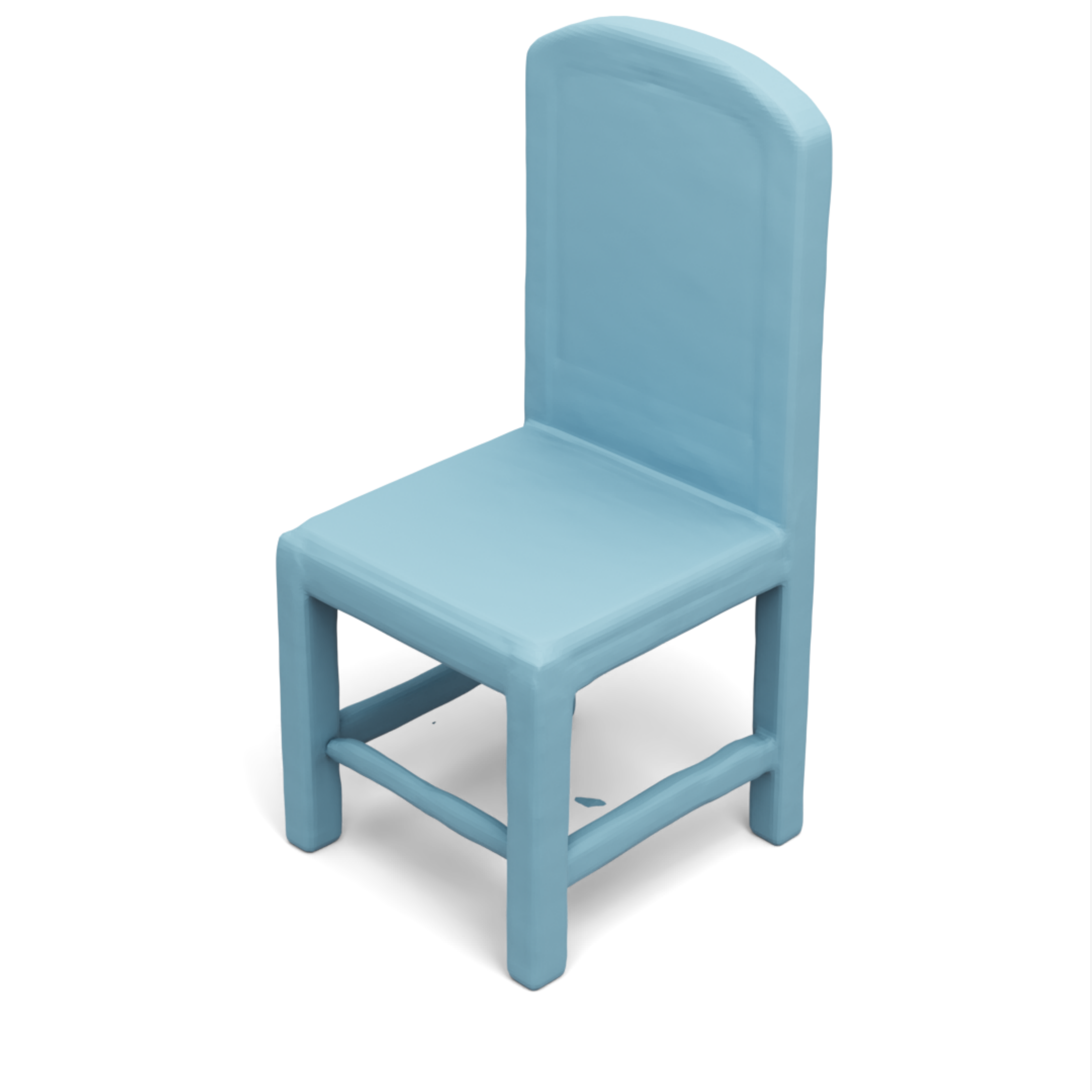}\
    & \includegraphics[trim=60 40 60 10, clip, width=0.2\linewidth]{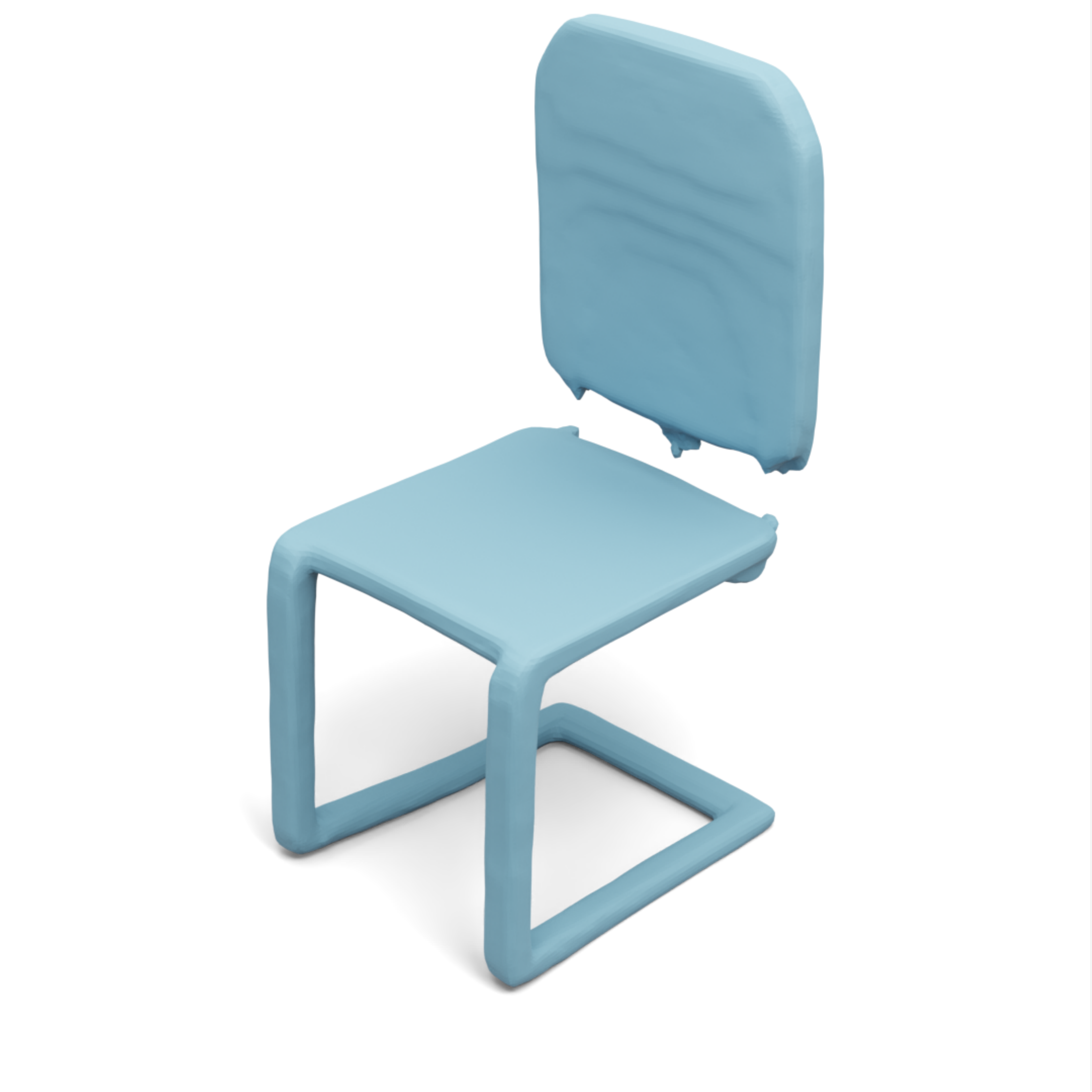}\
    & \includegraphics[trim=60 40 60 10, clip, width=0.2\linewidth]{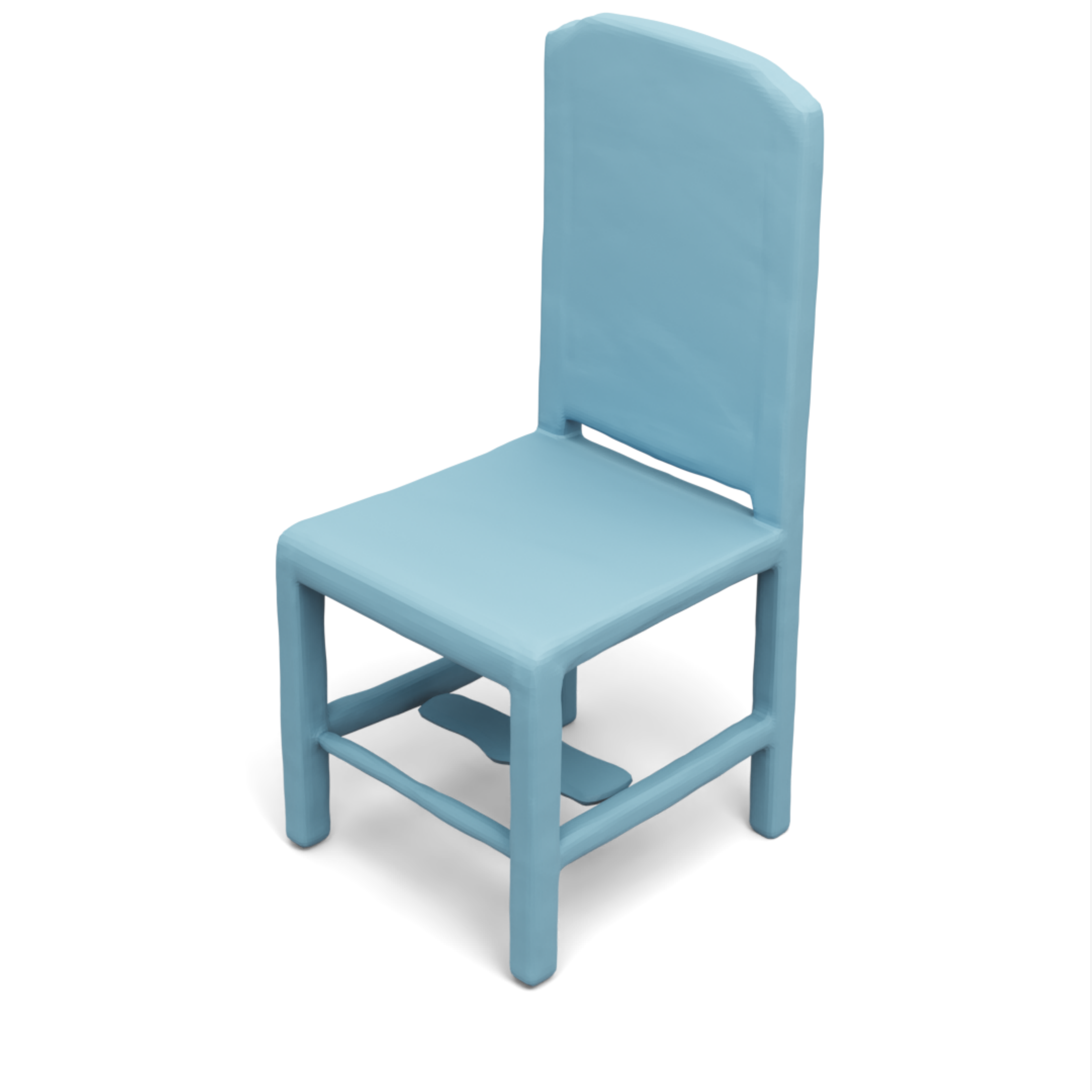}
    \\
    \includegraphics[width=0.2\linewidth]{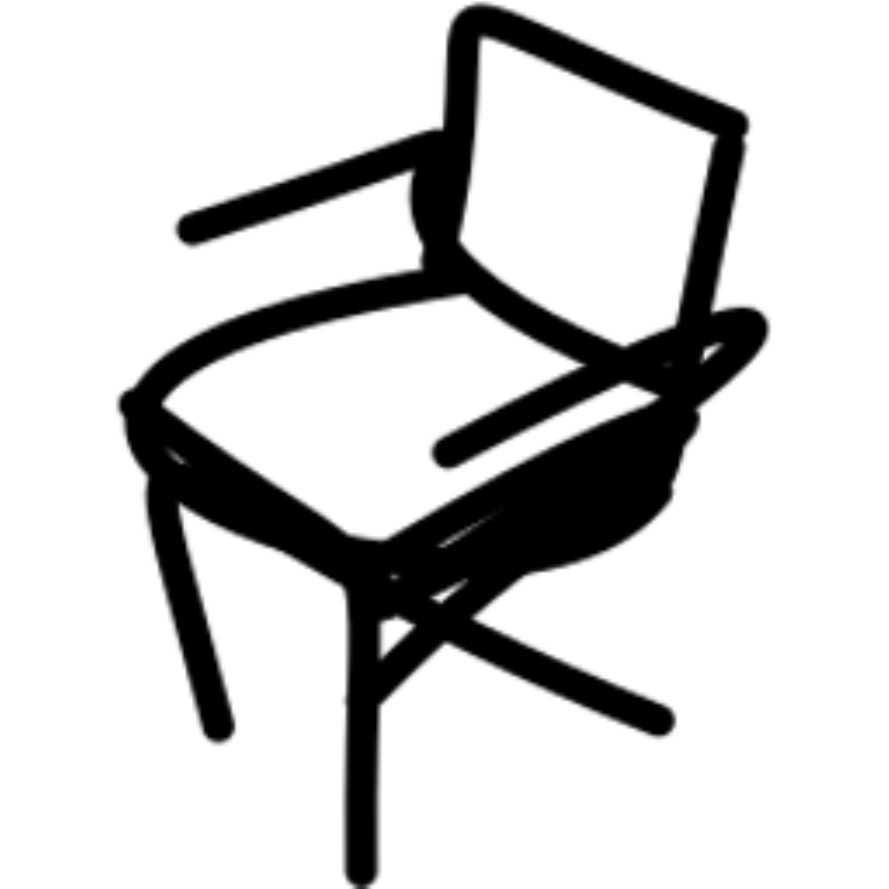}\
    & \includegraphics[trim=60 40 60 10, clip, width=0.2\linewidth]{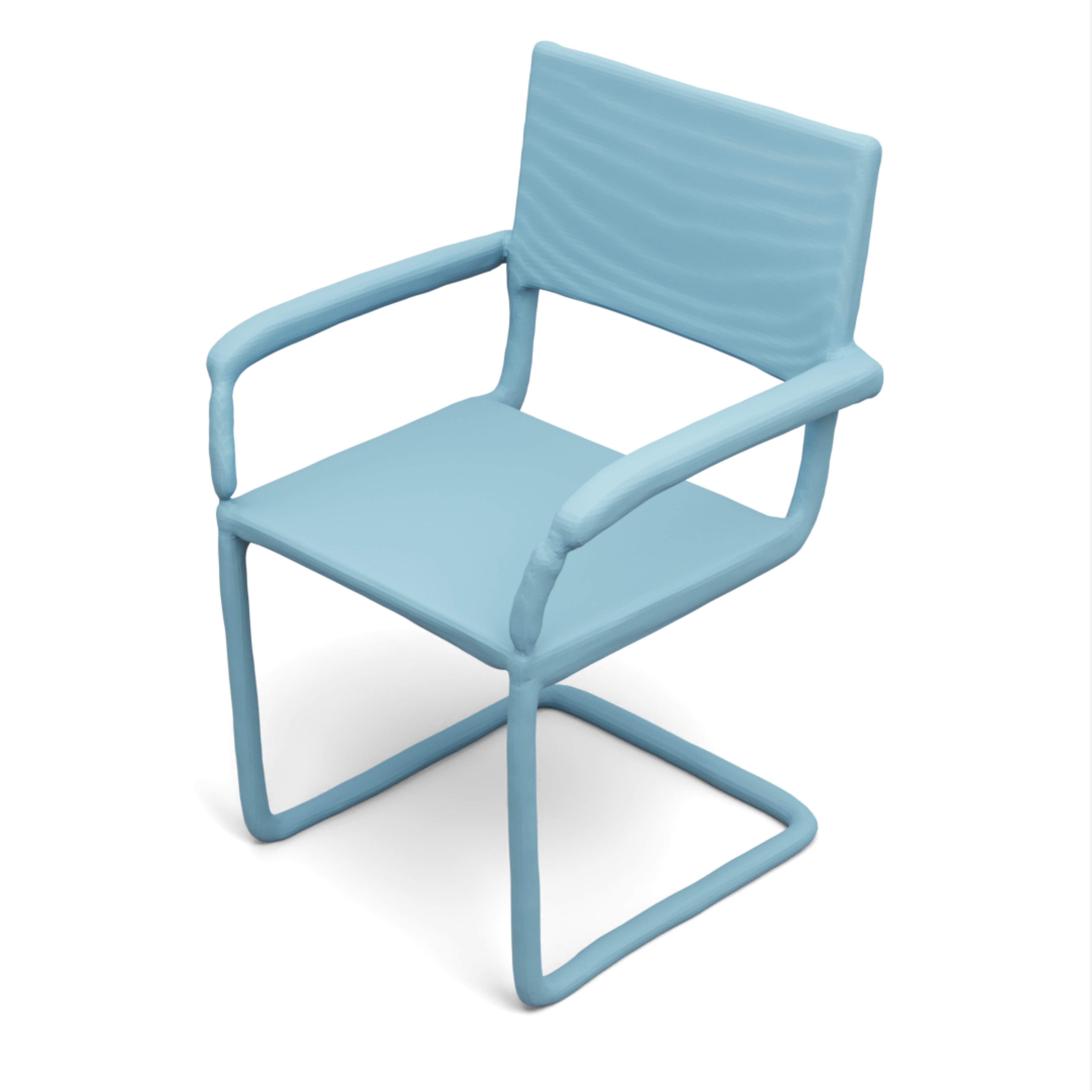}\
    & \includegraphics[trim=60 40 60 10, clip, width=0.2\linewidth]{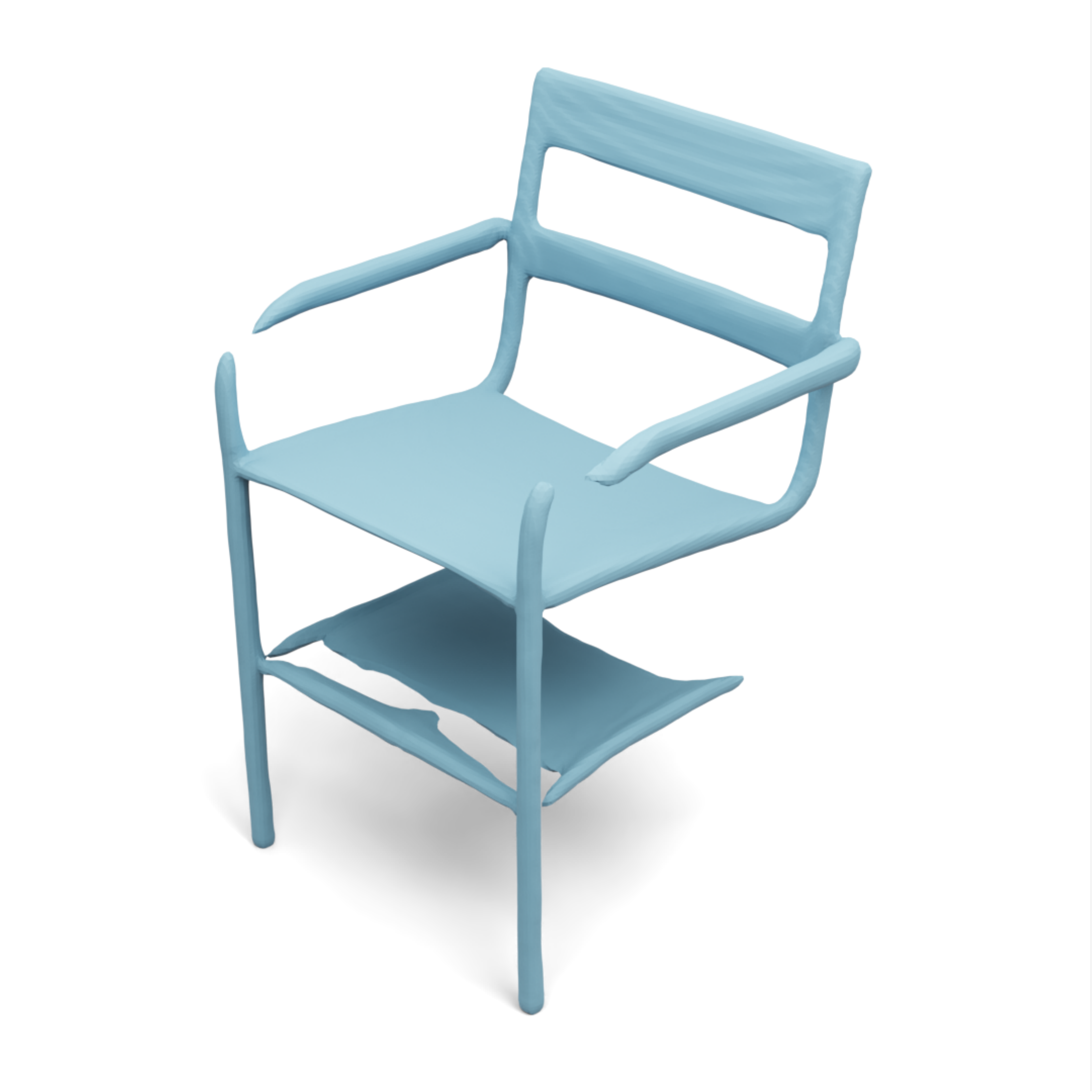}\
    & \includegraphics[trim=60 40 60 10, clip, width=0.2\linewidth]{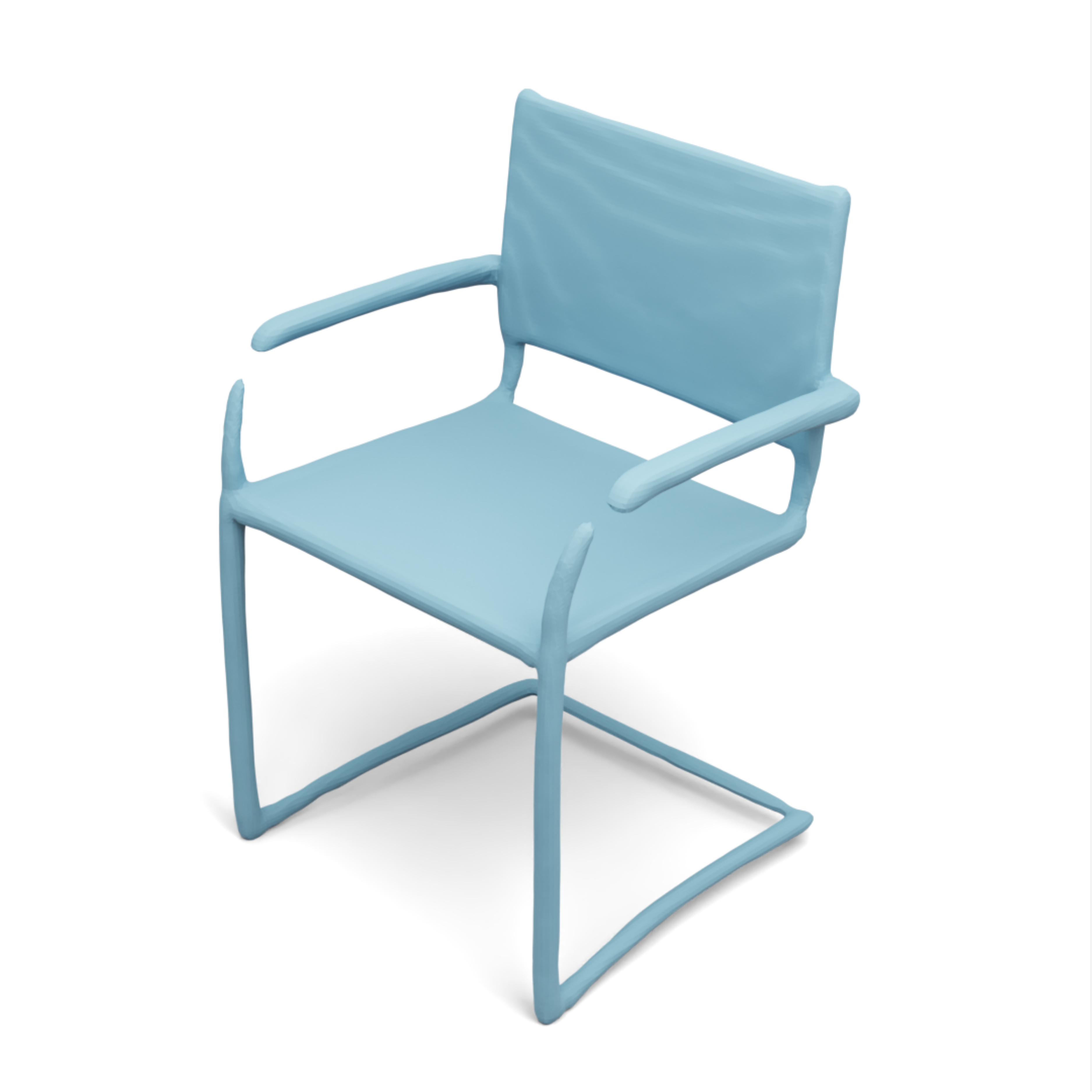}
    \\
	\end{tabular}
	\caption{We present our ablation study on three different input styles, namely a shape outline, a drawing and an abstract sketch. ``ablation partial loss'' means that the network did not train with partial loss; ``ablation dataset'' means that the network did not train with abstract sketches.}
	\label{fig:ablation}
   \vspace{-0.4cm}
\end{figure}

\subsection{Ablation studies}
\label{sec:ablations}

To analyze the relevance of different components of \ourmethod{}, we provide an ablation study. \new{Visual results} are displayed in \figref{fig:ablation} on inputs presented with increasing levels of abstraction from top to bottom. \new{For quantitative evaluations, refer to \tableref{tab:ablationQuantitativeComparison}}. To provide a fair comparison, no model in our ablation study has been trained on the ProSketch dataset, and all networks were trained for 40 hours. First, we trained the same network by removing the mask loss $\mathcal{L}_{\text{cls}}$ and the partial loss $\mathcal{L}_{\text{part}}$, both explained in \secref{par:partialshape} and referred to as ``ablation partial loss''. We claim that these losses improve the part disentanglement, hence allowing \ourmethod{} to produce shapes that are less prone to mere shape retrieval. This is particularly visible in the chairs' handles that are not present or not connected to the seat in the original drawing. Yet, they are visible in the output shape. \new{The quantitative comparison supports our analysis. The metrics indicate that eliminating the partial loss significantly decreases the distance between the shapes in the dataset and those generated, indicating a tendency towards retrieval.} Second, we trained \ourmethod{} without using abstract sketches, referred to as ``ablation dataset''. It clearly appears that the more abstract the input sketch, the more the obtained result decreases in quality, notably with some parts being absent from the output shape. \new{The quantitative metrics further demonstrate that incorporating sketches of varying abstraction levels enhances our method's adaptability to different input sketch styles. This is evidenced by the weaker performance on our metrics by the version of our method with dataset ablation.}

\begin{table}[b]
\centering
\caption{\new{Performance comparison of ablated methods on the AmateurSketch dataset \cite{Qi2021AmateurSketch} using chamfer distance (CD), earth mover's distance (EMD), and Fréchet inception distance (FID). The metrics indicate that our mask loss and partial loss enhance our method's ability to resist shape retrieval issues, and the use of abstract sketches increases its resilience to sketch abstraction.}}
\label{tab:ablationQuantitativeComparison}
\begin{tabular}{lccc}
\toprule
    Method  & CD & EMD & FID \\
\midrule
Ablation partial loss & 0.117 & 0.0940 & 170.6 \\
Ablation dataset & 0.1300 & 0.1023 & 181.9 \\ Ours (full) & 0.1235 & 0.0981 & 174.3 \\ 
\bottomrule
\end{tabular}
\end{table}

 \subsection{Multi-class reconstruction}

\begin{figure}[t]
\centering
\small
\setlength{\tabcolsep}{1pt}
\begin{tabular}{cccccc}
\rotatebox{90}{\footnotesize Input Sketch} & \includegraphics[width=0.18\linewidth]{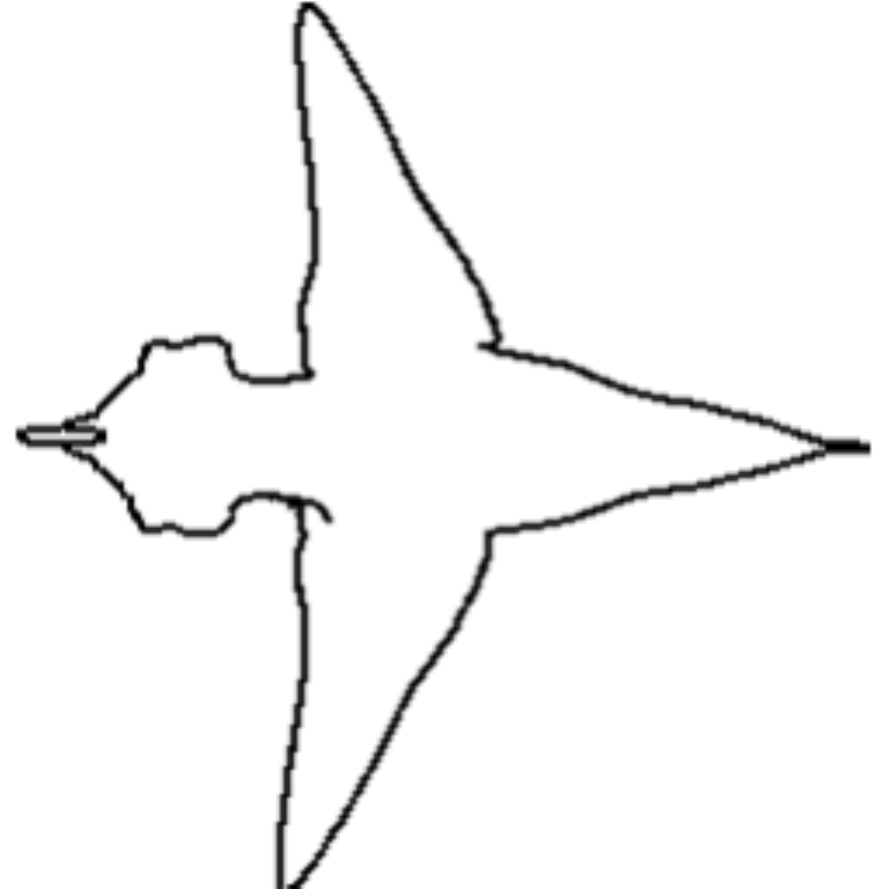}\ & \includegraphics[width=0.18\linewidth]{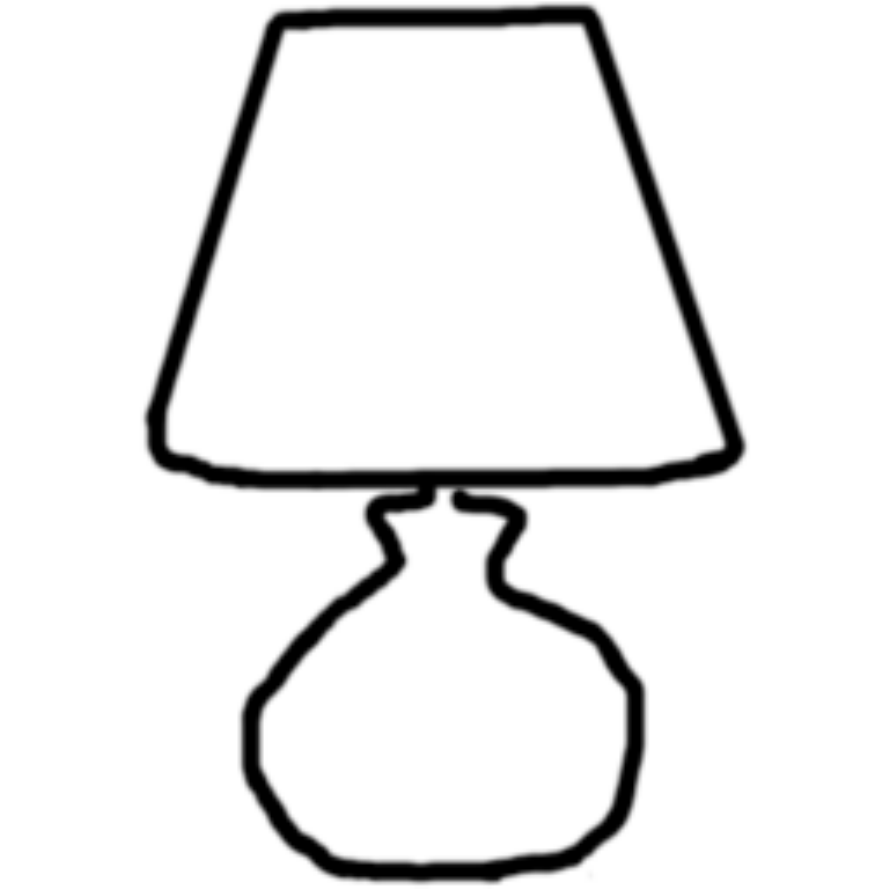}\
     & \includegraphics[width=0.18\linewidth]{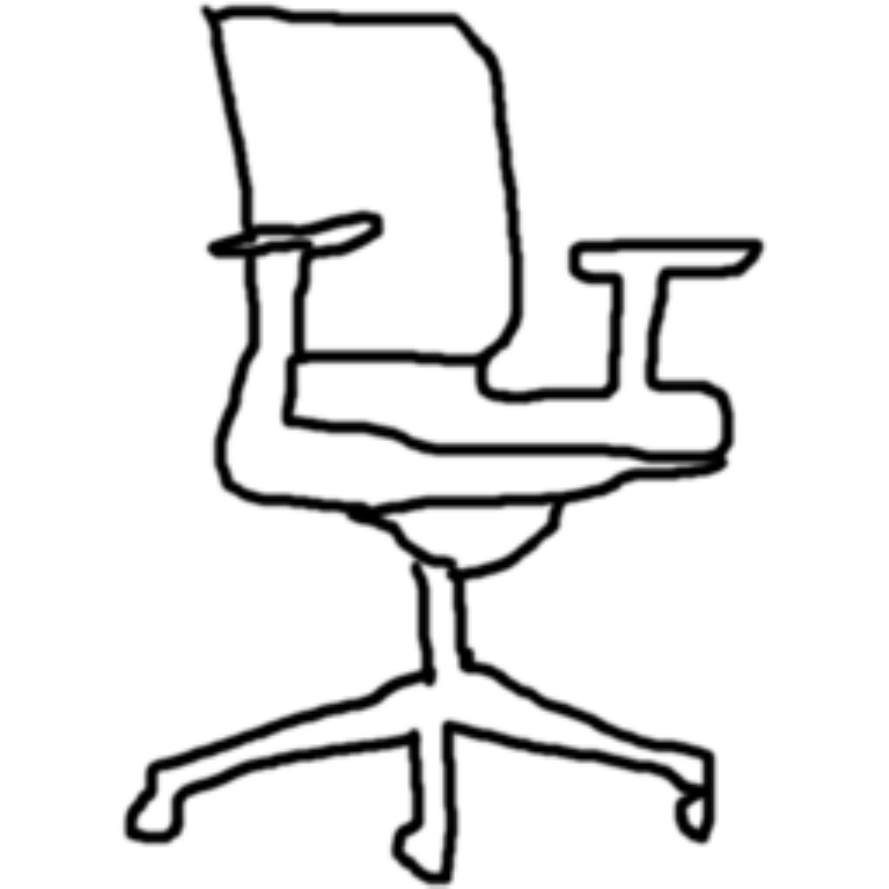} \
     & \includegraphics[width=0.18\linewidth]{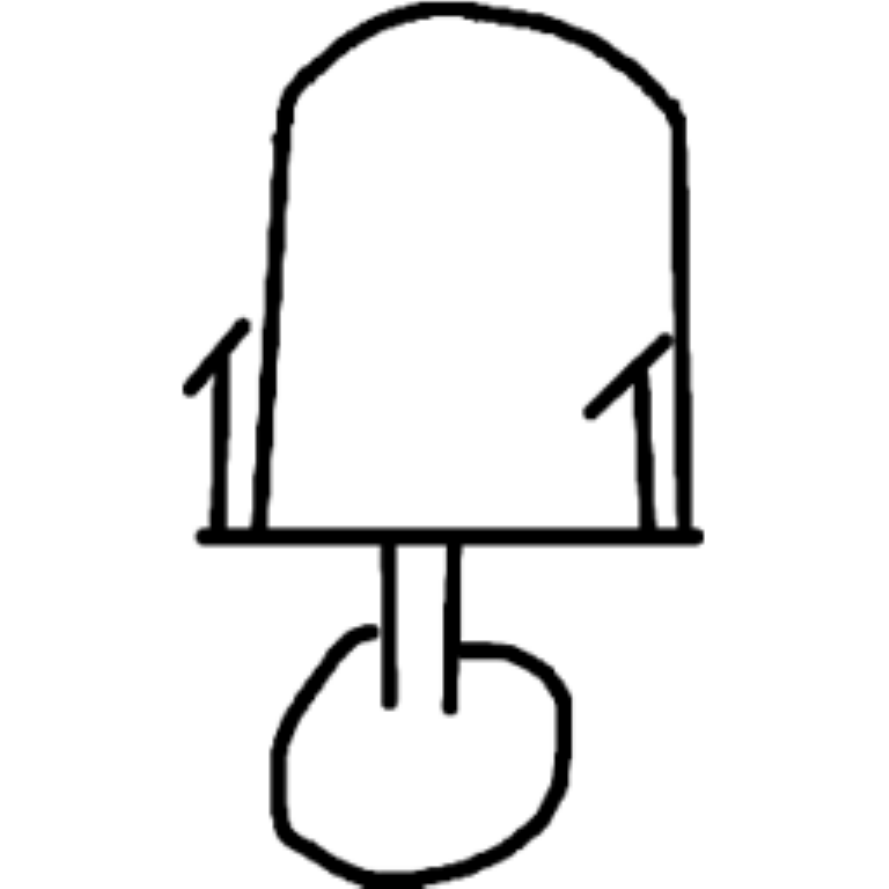} \
     & \includegraphics[width=0.18\linewidth]{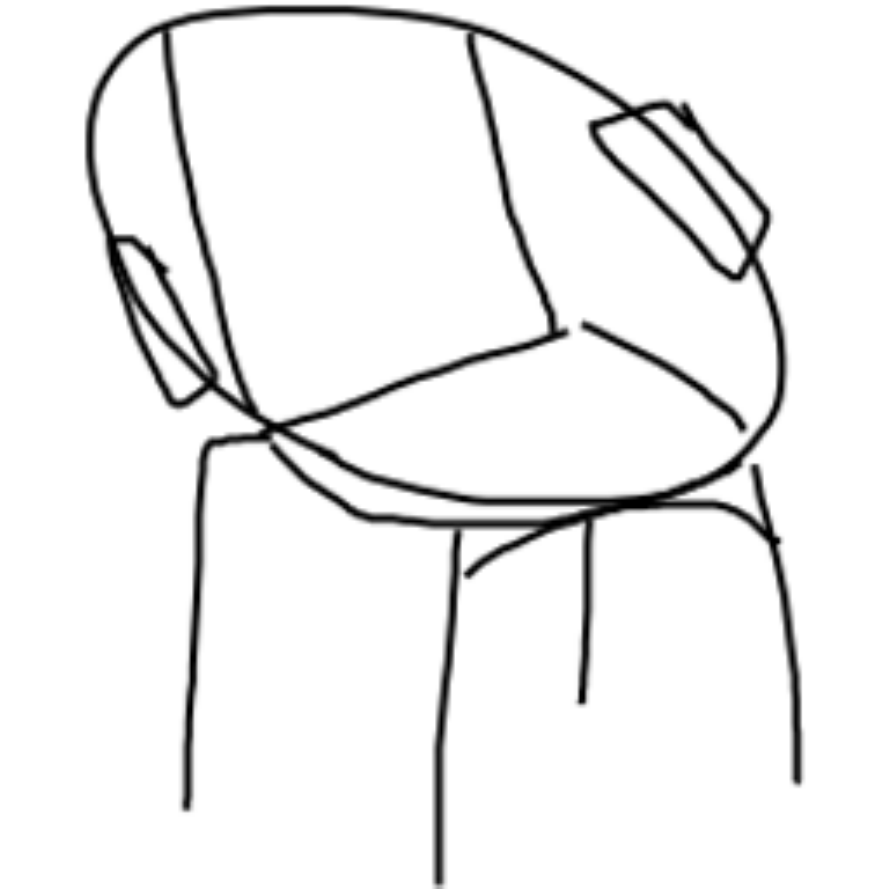}\\
     \rotatebox{90}{\footnotesize Multi-class} & \includegraphics[trim=20 20 20 20, clip, width=0.18\linewidth]{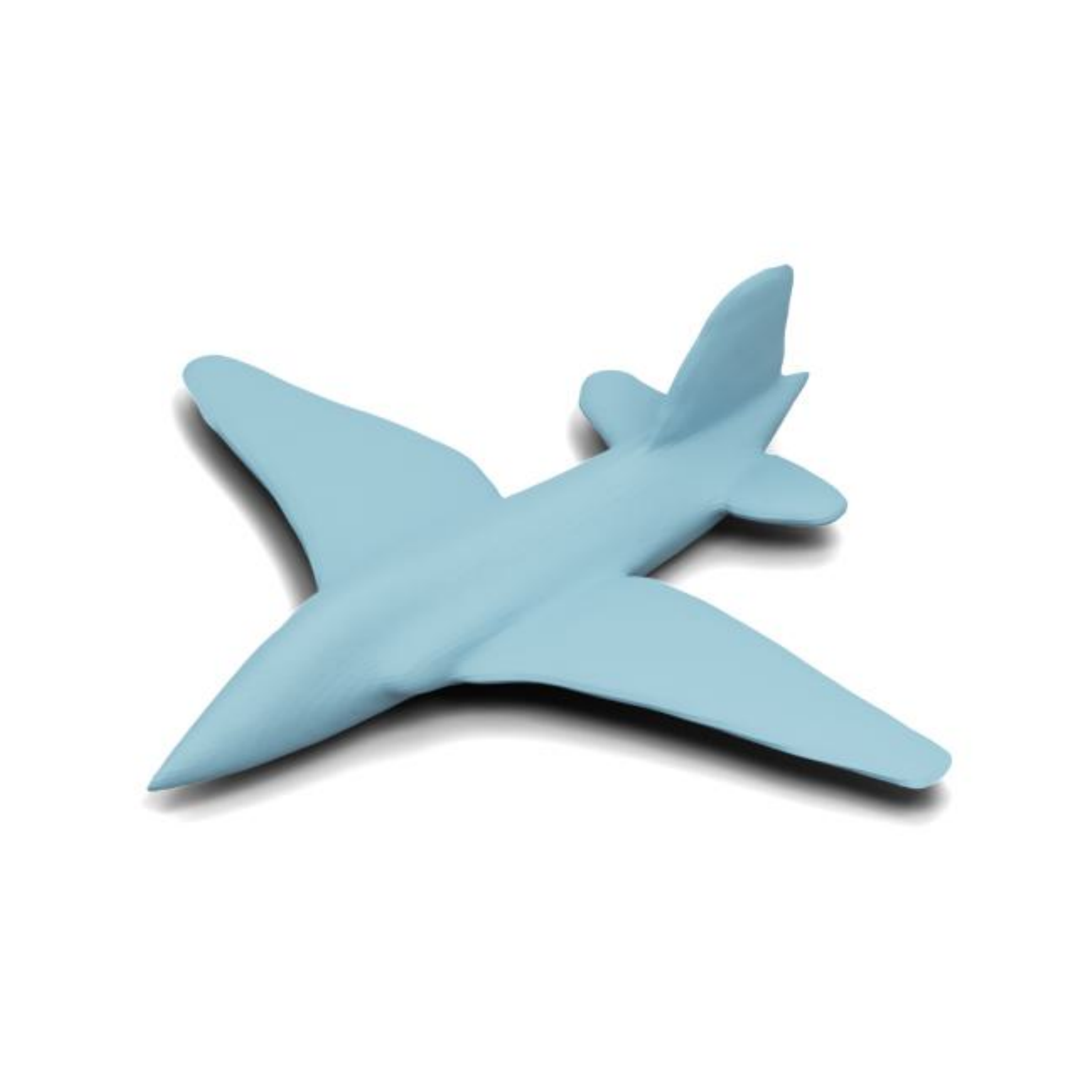} & \includegraphics[trim=20 20 20 20, clip, width=0.18\linewidth]{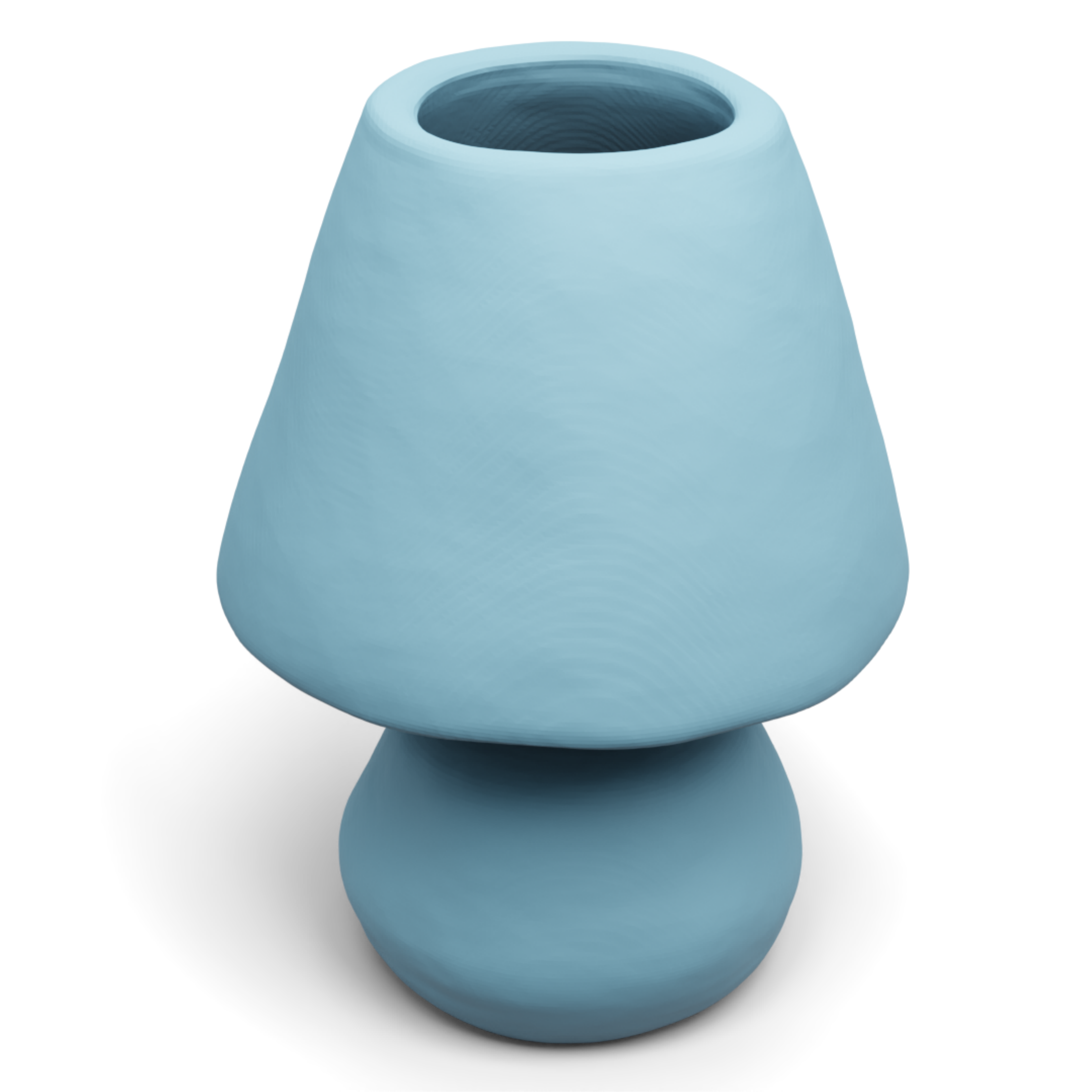} & \includegraphics[trim=20 20 20 20, clip, width=0.18\linewidth]{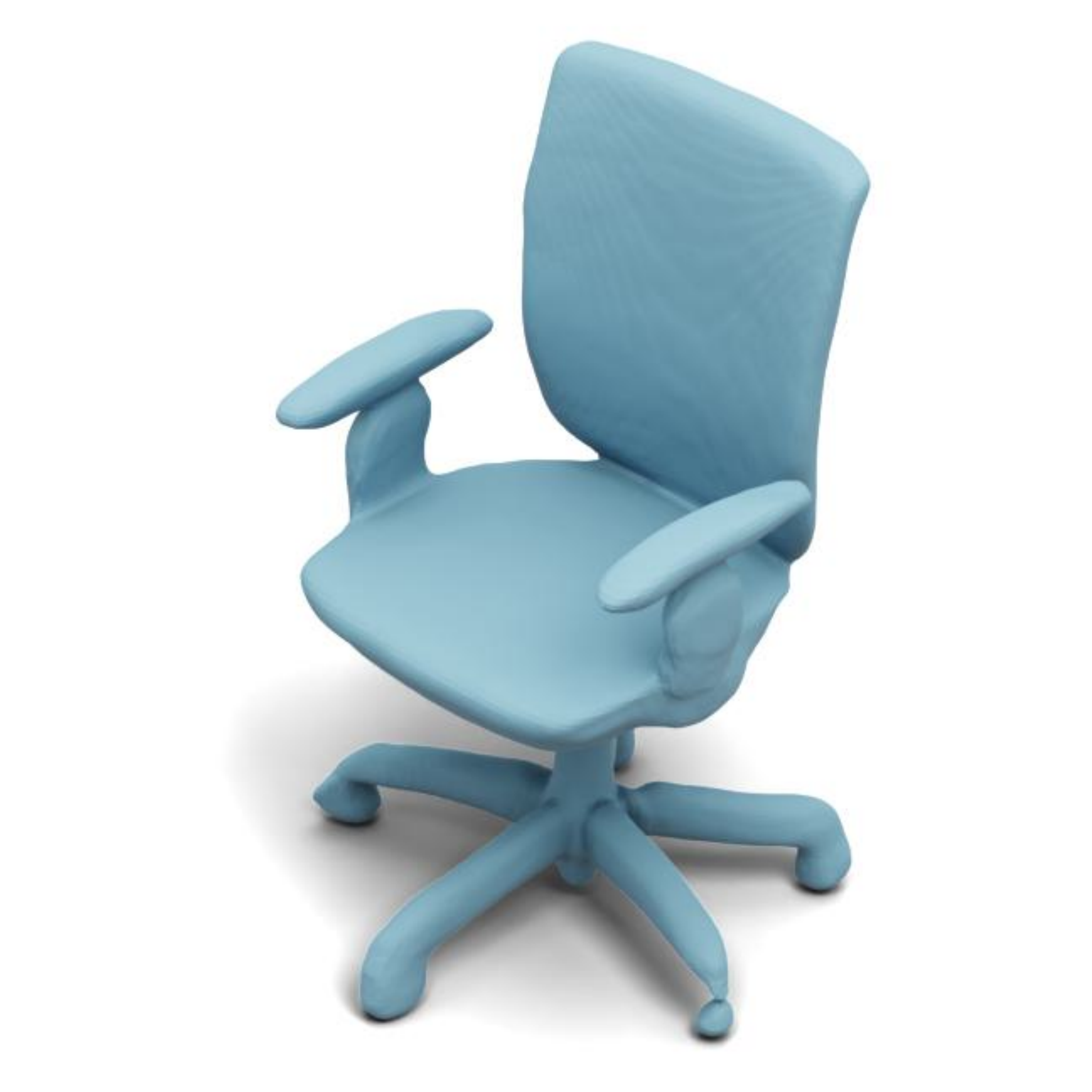} & \includegraphics[trim=20 20 20 20, clip, width=0.18\linewidth]{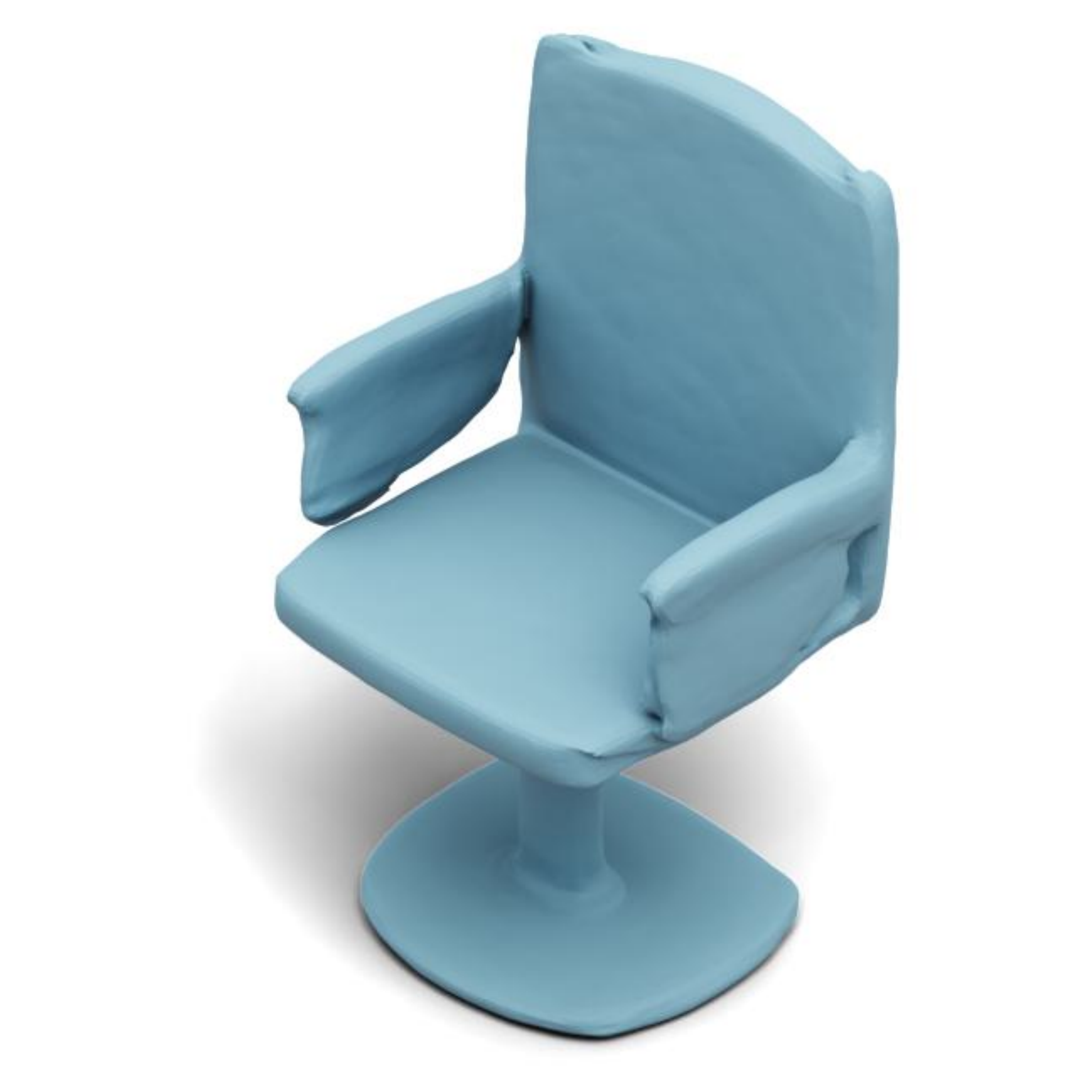}
     & \includegraphics[trim=30 30 30 30, clip, width=0.18\linewidth]{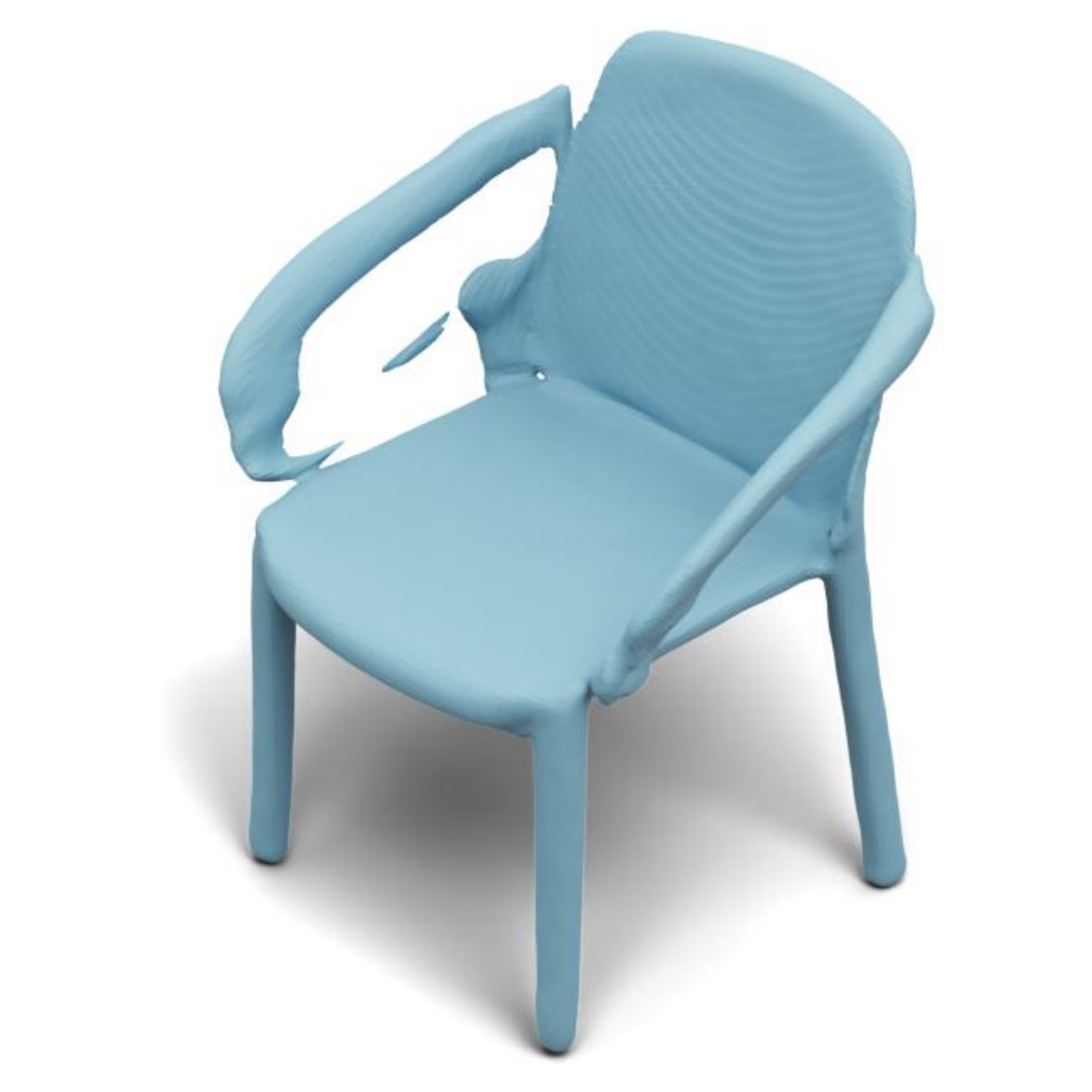}\\
     \rotatebox{90}{\footnotesize Single-class} & \includegraphics[trim=20 20 20 80, clip, width=0.18\linewidth]{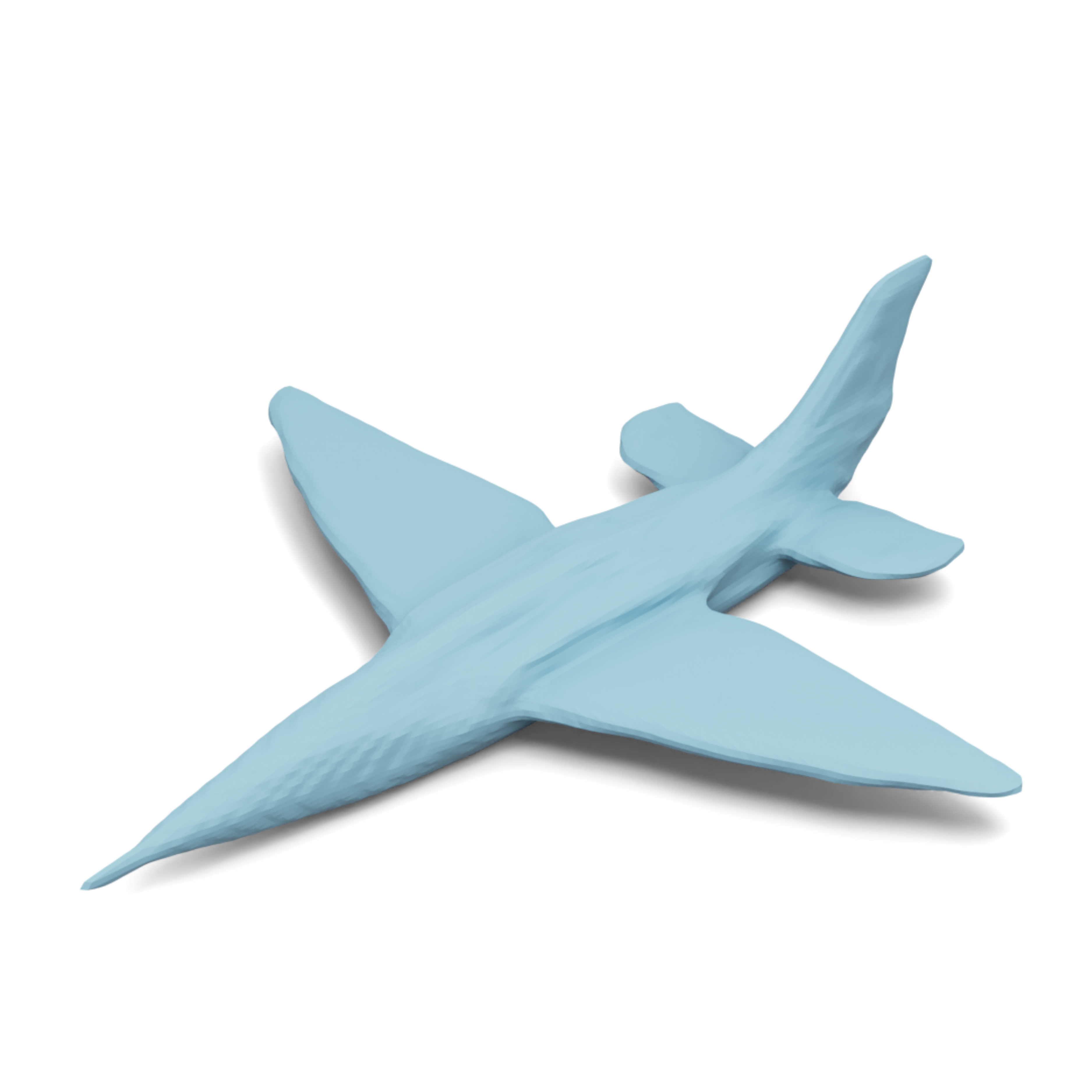} & \includegraphics[trim=20 20 20 20, clip, width=0.18\linewidth]{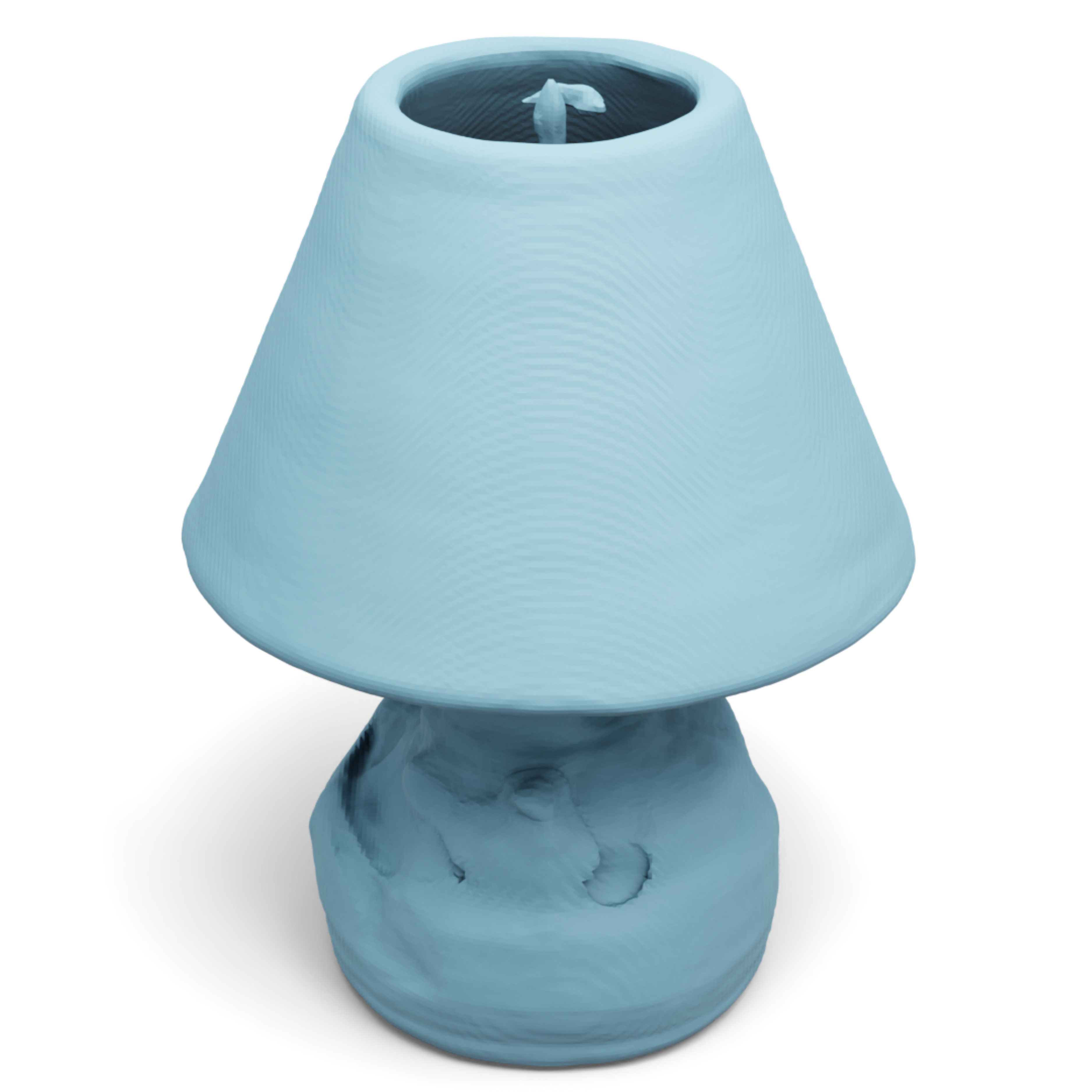} & \includegraphics[trim=20 20 20 20, clip, width=0.18\linewidth]{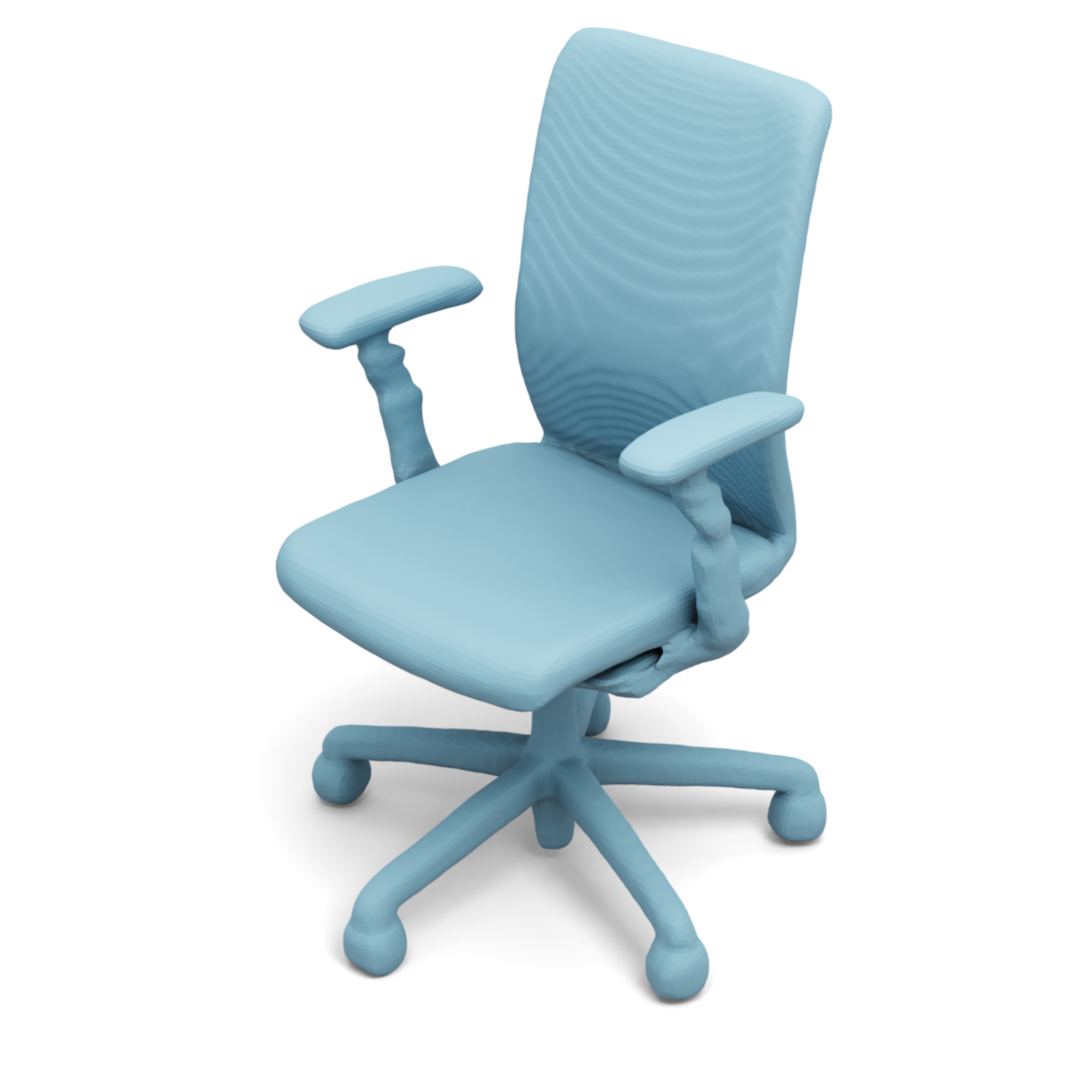} & \includegraphics[trim=20 20 20 20, clip, width=0.18\linewidth]{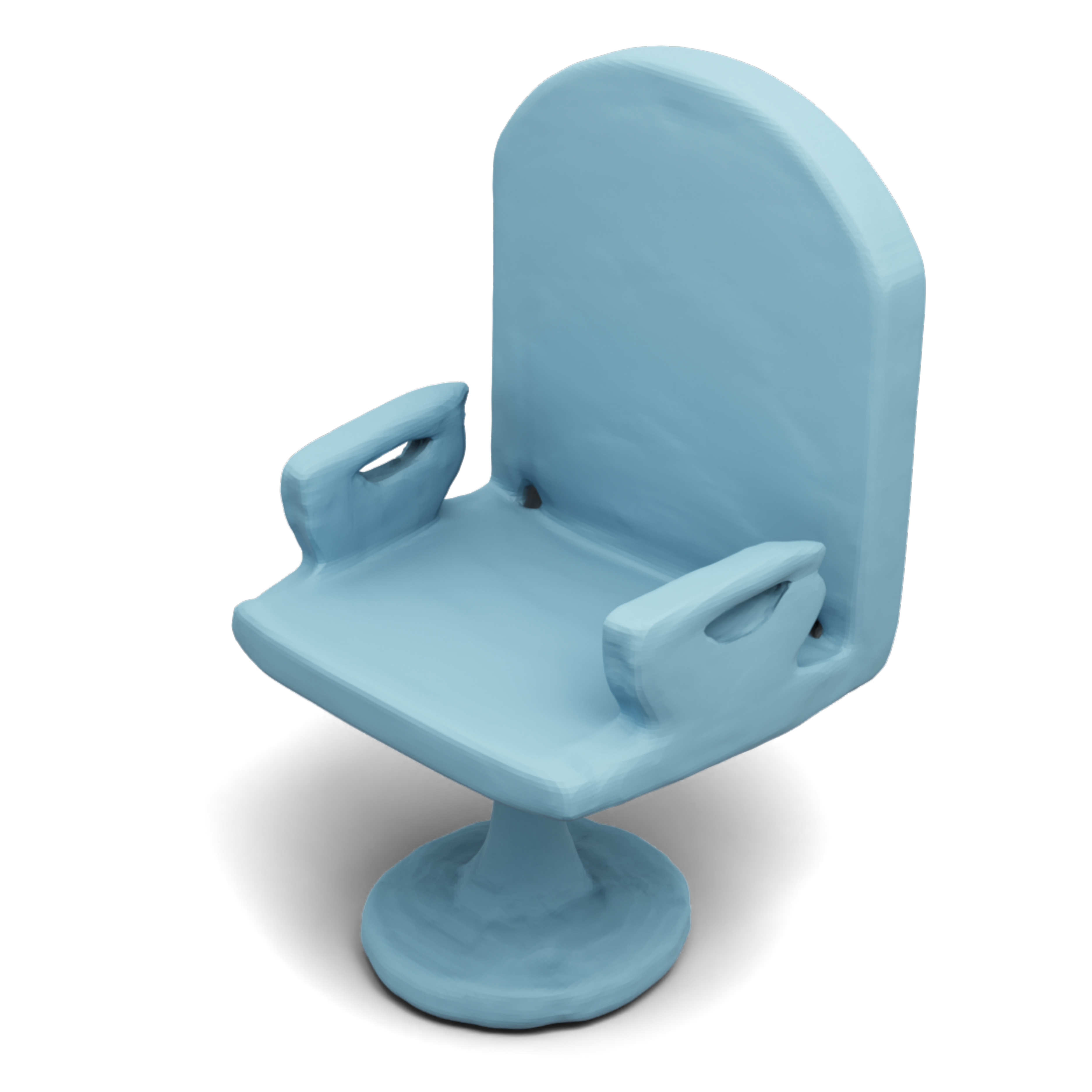} & \includegraphics[trim=20 20 20 20, clip, width=0.18\linewidth]{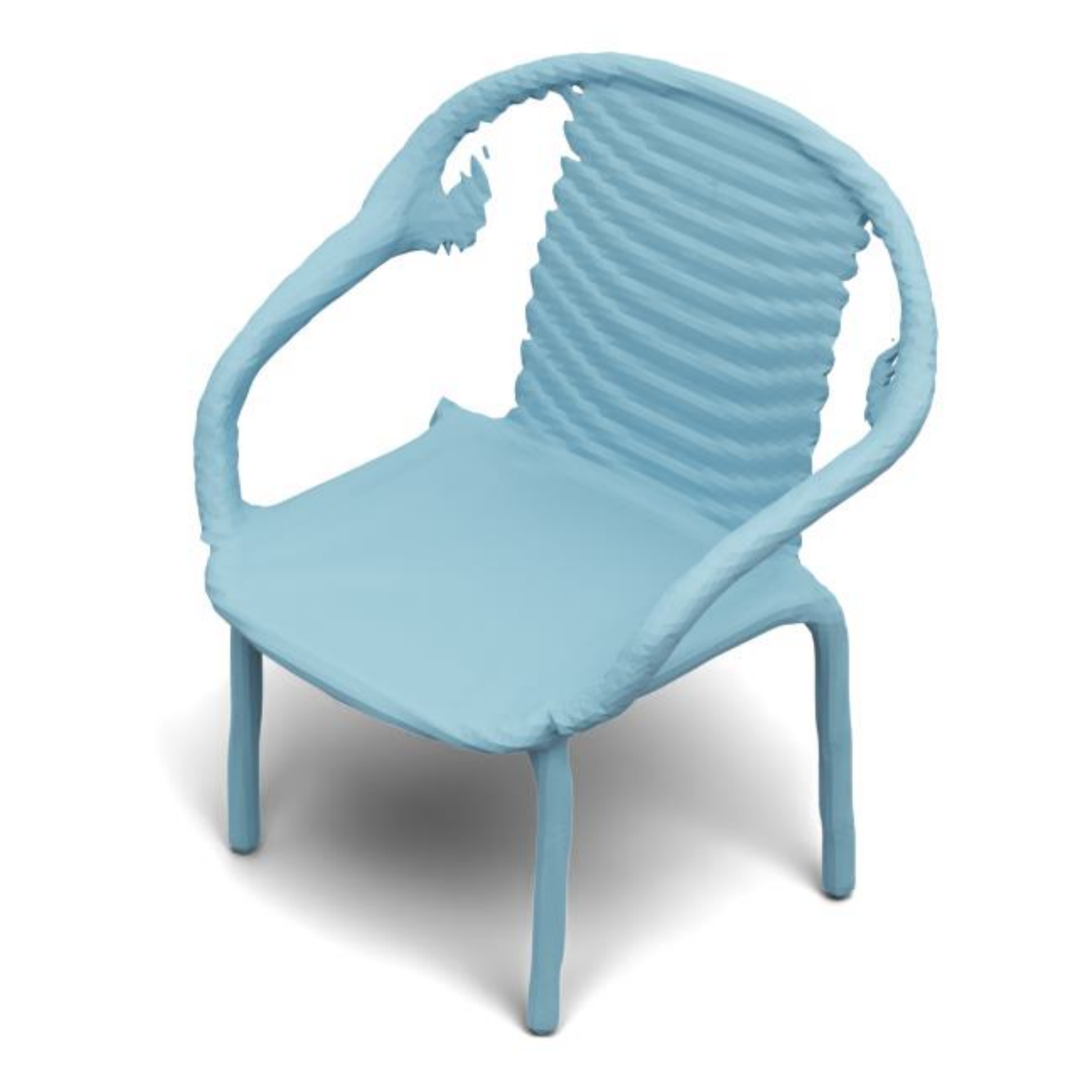}\\
\end{tabular}
\caption{We compare our multi-class and single-class sketch-to-shape models. The input of the last column comes from AmateurSketch. The other sketches are produced by us.}
\label{fig:multiclass}
\end{figure}

Until now, we had conditioned \ourmethod{} on a specific class of shapes. We demonstrate here that it is possible to condition \ourmethod{} on multiple classes at the same time. To account for the greater shape diversity, our multi-class shape generator relies on a higher number of Gaussians and the latent representation has a higher dimension. \figref{fig:multiclass} compares the results of the multi-class network to the single-class network of the respective category. We observe that the multi-class version produces successfully shapes that correspond to the right category. Compared to the single-class version, the output shapes are slightly less accurate, especially with sharp features. This can be observed in the chair and airplane sketches. Note though that for lamps, we get better results with the multi-class network. As lamps have a smaller training set, the multi-class network exhibits better generalization than the single class as it has access to more data.

\subsection{Limitations}

\begin{figure}[t]
	\centering
	\small
	\setlength{\tabcolsep}{1pt}
	\begin{tabular}{cccccc}
    \includegraphics[width=0.16\linewidth]{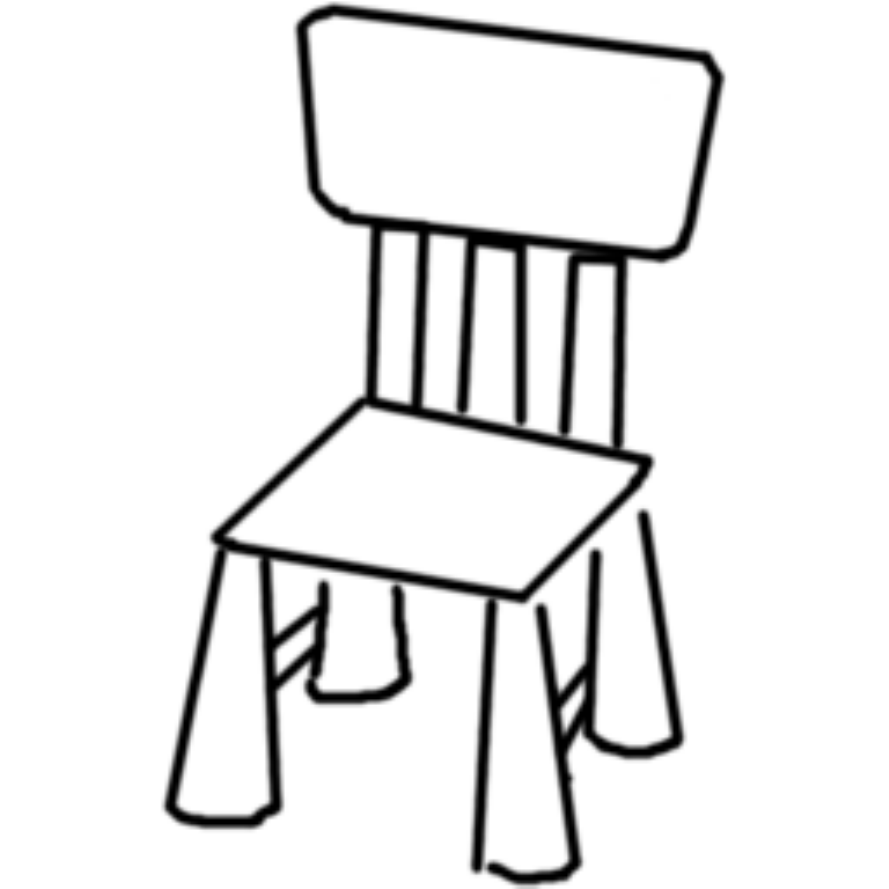}\
    &
    \includegraphics[trim=10 10 10 10, clip, width=0.16\linewidth]{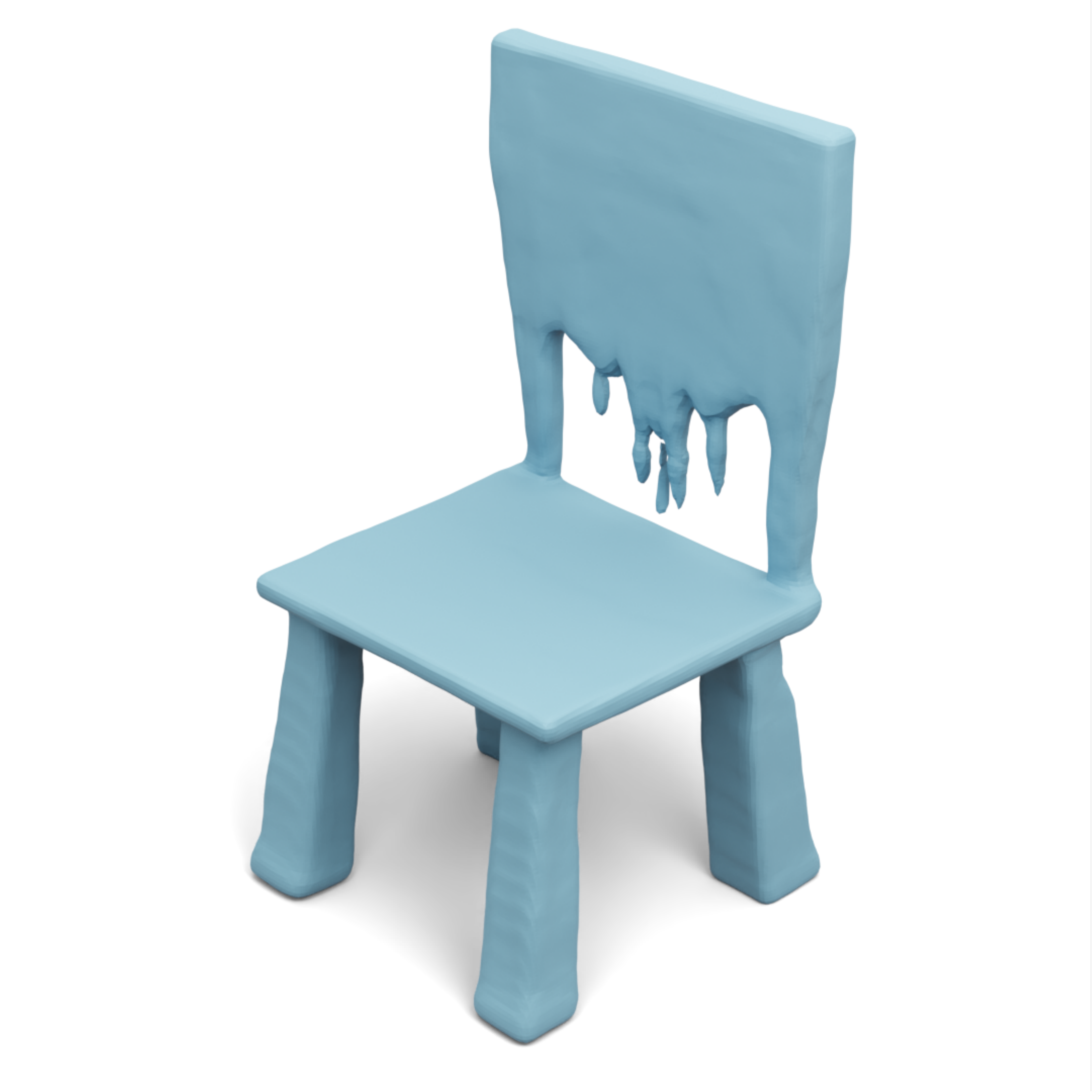}\
    & \includegraphics[width=0.16\linewidth]{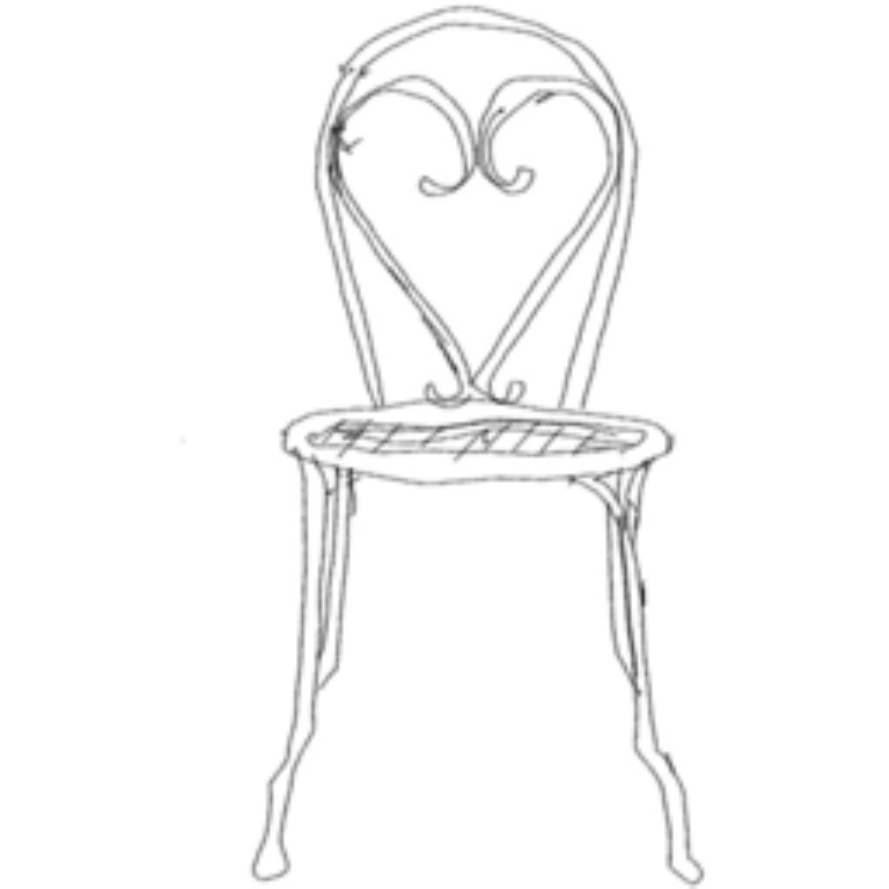}\
    & \includegraphics[trim=10 10 10 10, clip, width=0.16\linewidth]{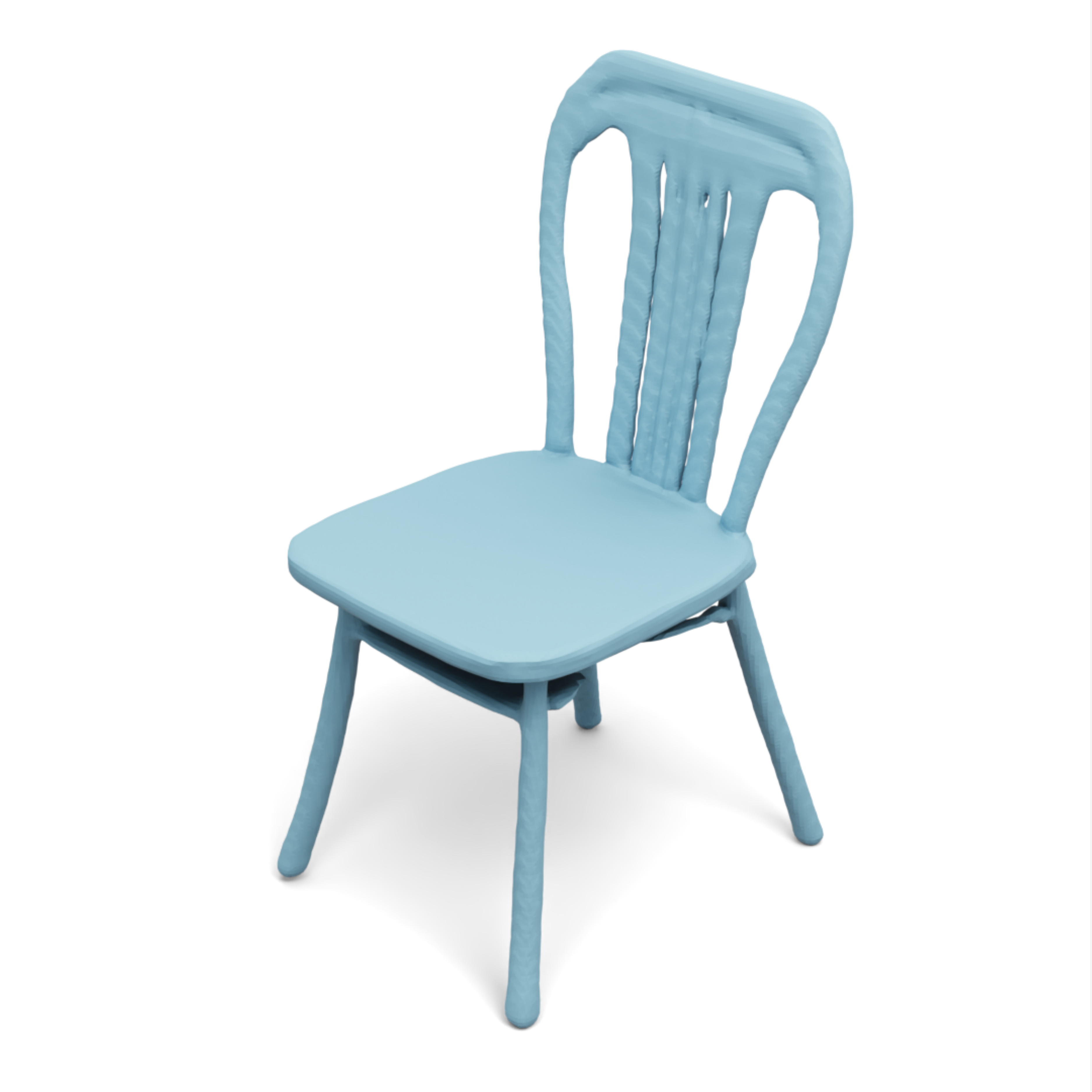}\
    & \includegraphics[width=0.16\linewidth]{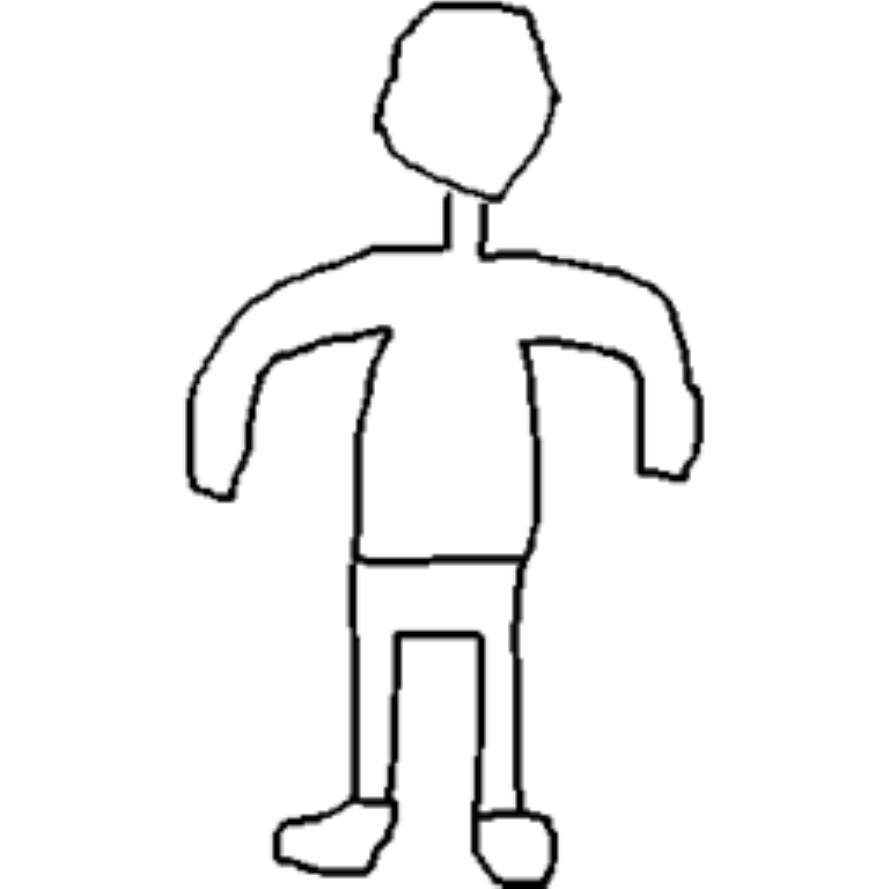}\
    & \includegraphics[trim=10 10 10 10, clip, width=0.16\linewidth]{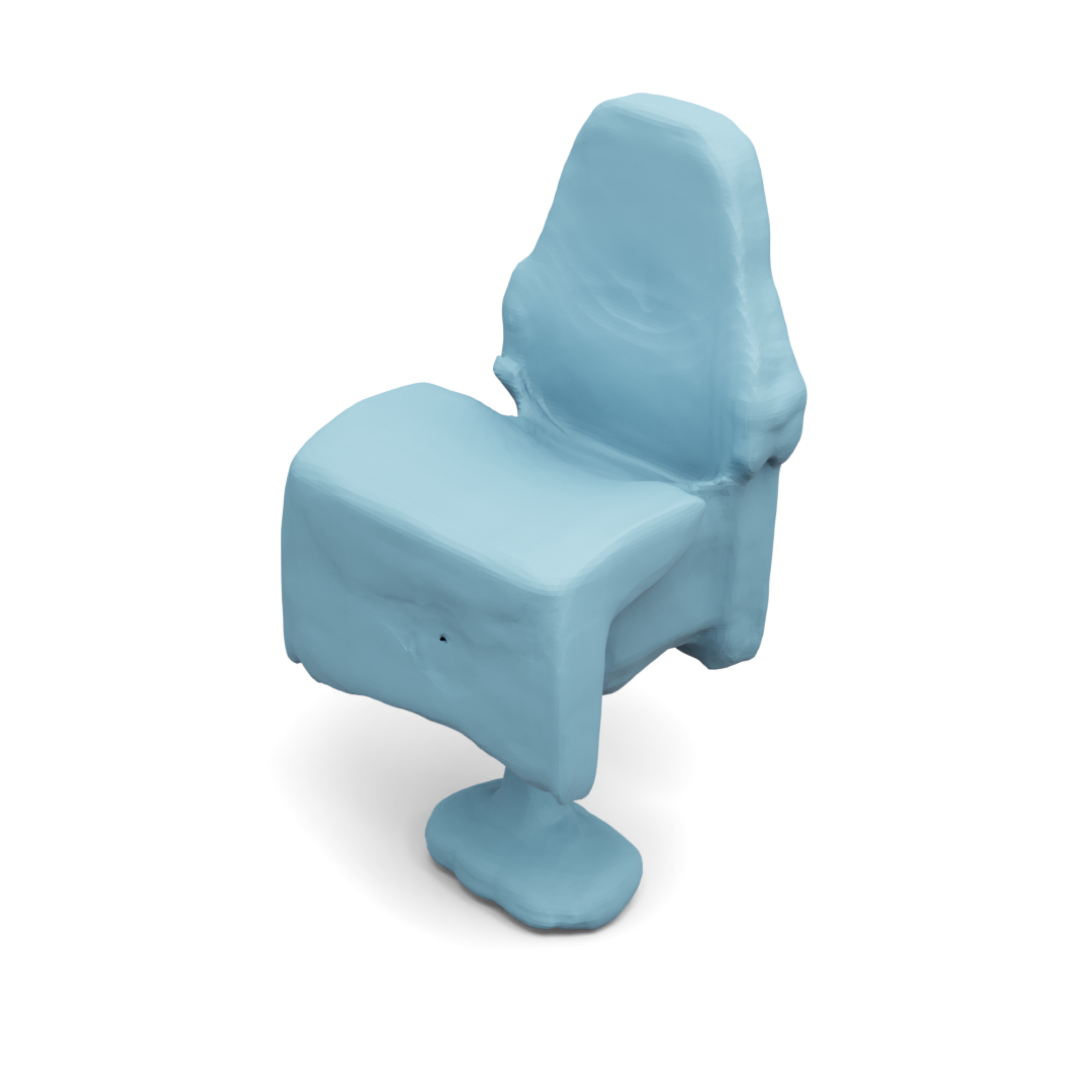}\\
    \multicolumn{2}{c}{(i)} & \multicolumn{2}{c}{(ii)} & \multicolumn{2}{c}{(iii)} 
	\end{tabular}
\vspace{-0.3cm}
 \caption{While our method quickly allows to obtain a shape from a drawing, it struggles in certain cases. (ii) comes from ProSketch \cite{zhong2021prosketch} but was not included in the training data.}
	\label{fig:limitations}
\vspace{-0.45cm}
\end{figure}

\begin{figure*}[t]
	\centering
	\small
	\setlength{\tabcolsep}{1pt}
	\begin{tabular}{cccccc}
    \includegraphics[width=0.16\linewidth]{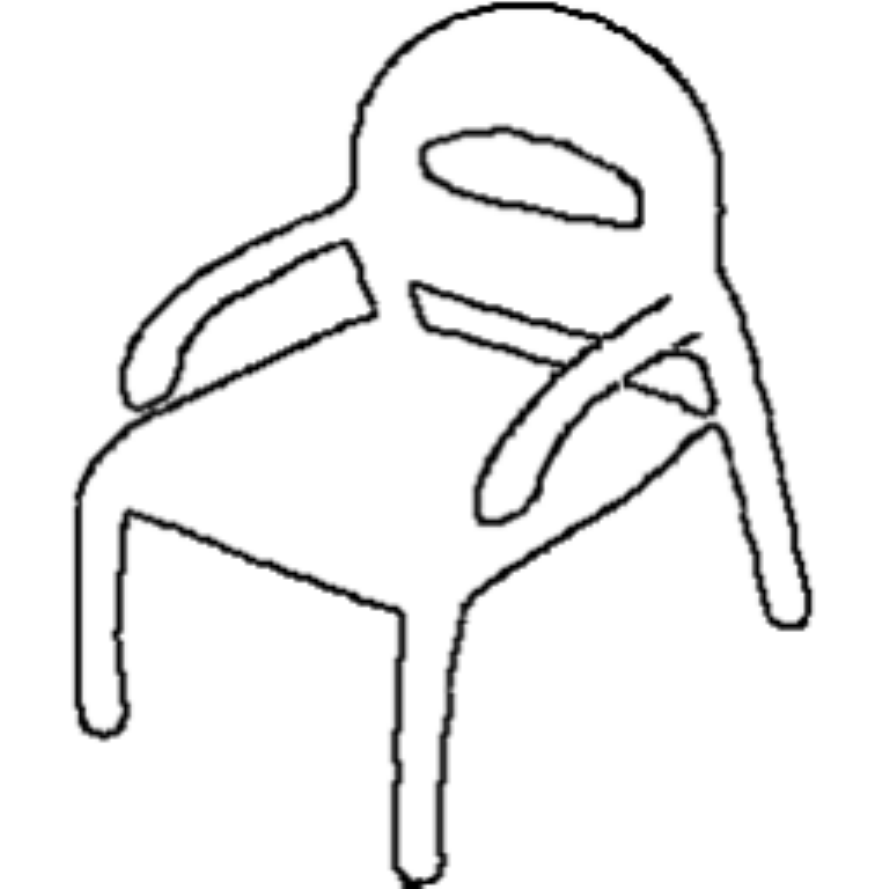}\
    & \includegraphics[width=0.16\linewidth]{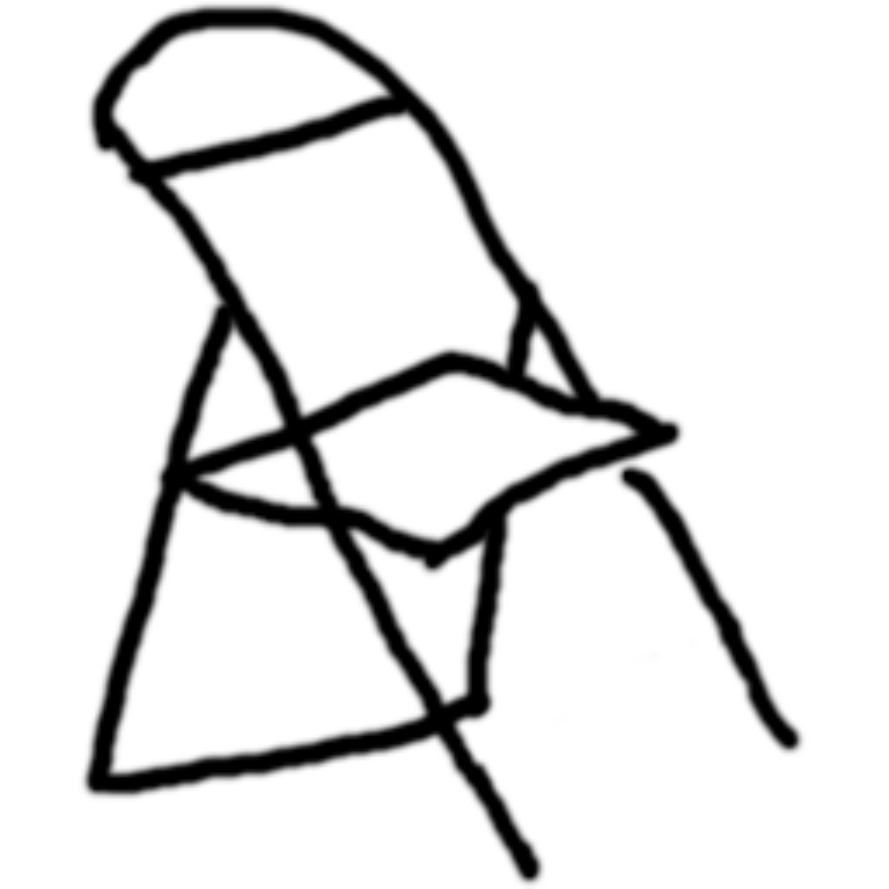}\
    & \includegraphics[width=0.16\linewidth]{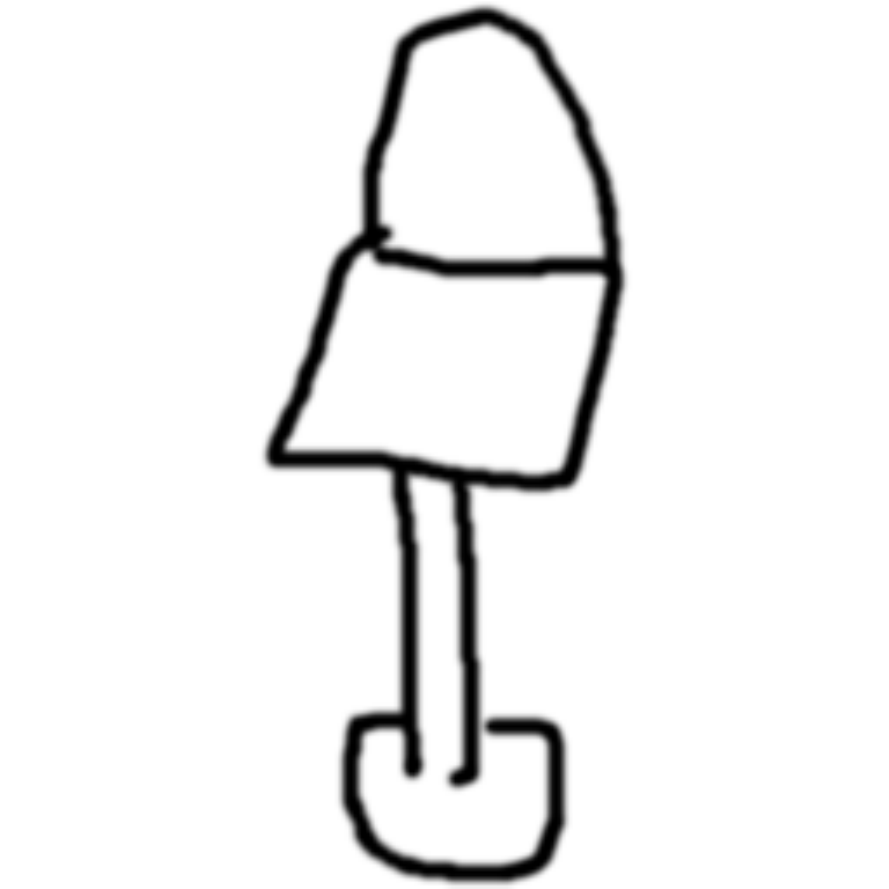}\
    & \includegraphics[width=0.16\linewidth]{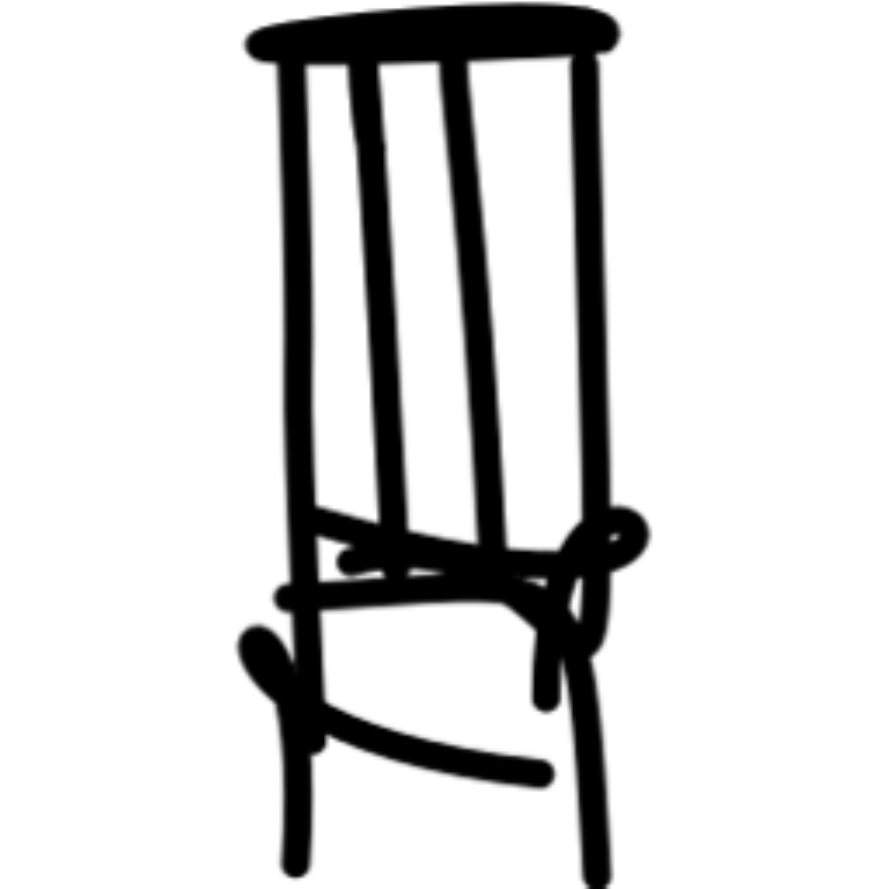}\
    & \includegraphics[width=0.16\linewidth]{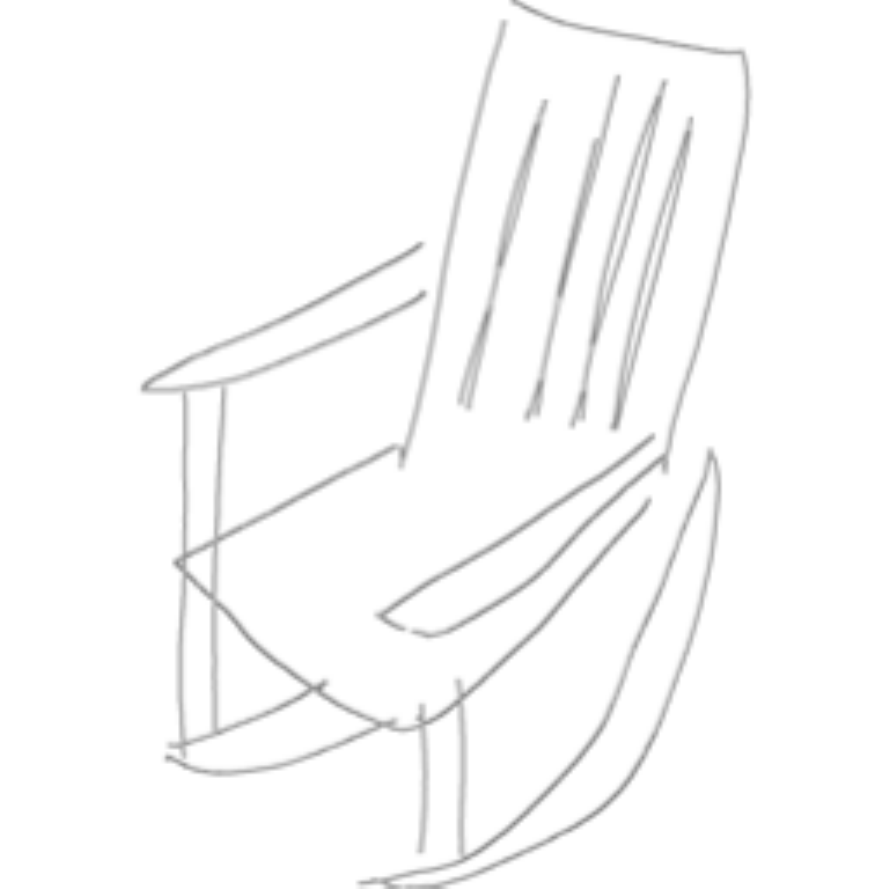}\
    & \includegraphics[width=0.16\linewidth]{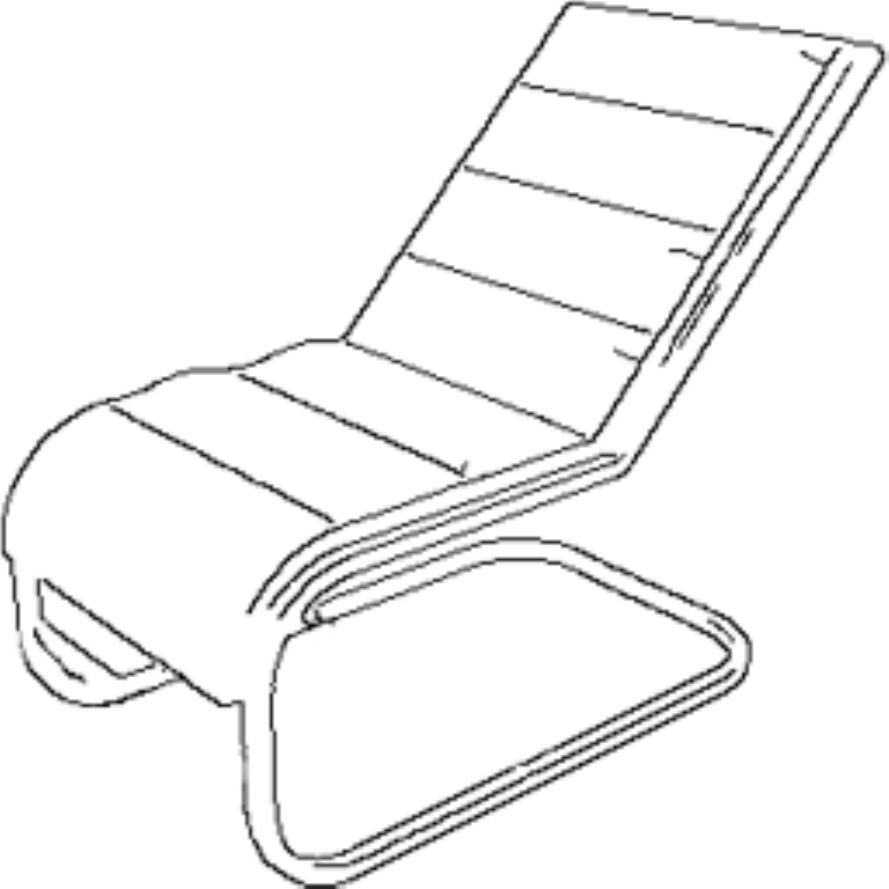} \\


\includegraphics[trim = 1 1 1 1, clip, width=0.16\linewidth]{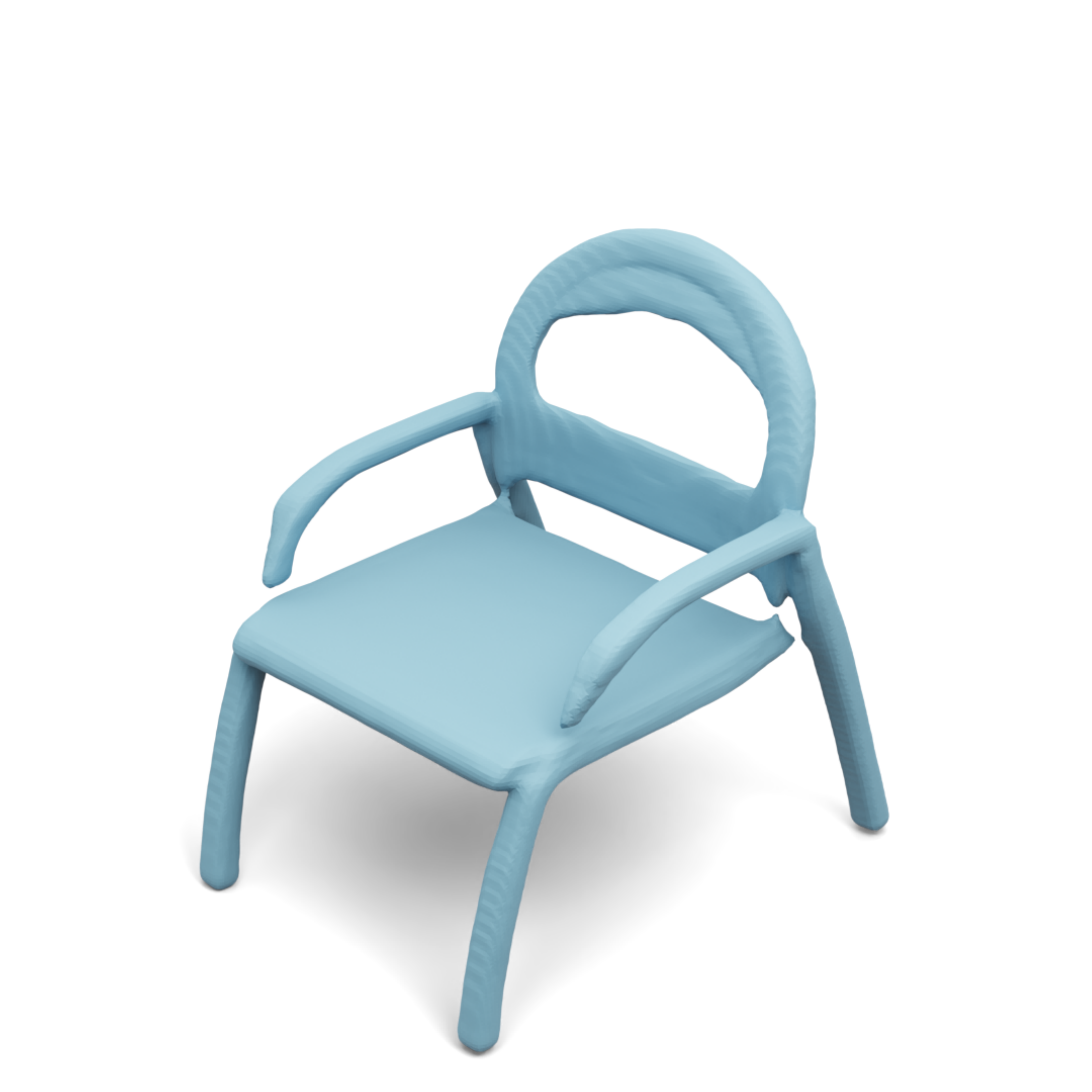}\
    & \includegraphics[trim = 1 1 1 1, clip, width=0.16\linewidth]{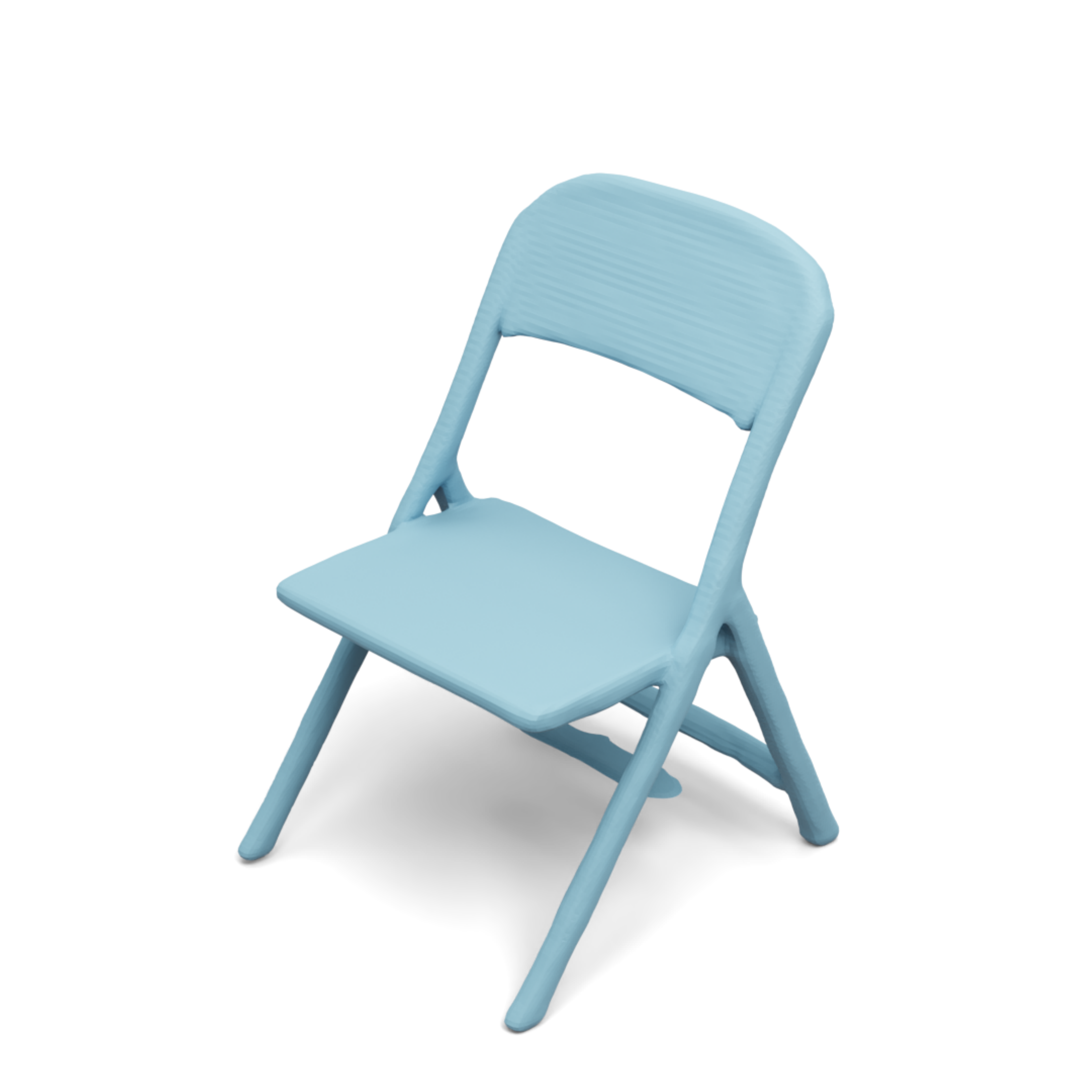}\
    & \includegraphics[trim = 1 1 1 1, clip, width=0.16\linewidth]{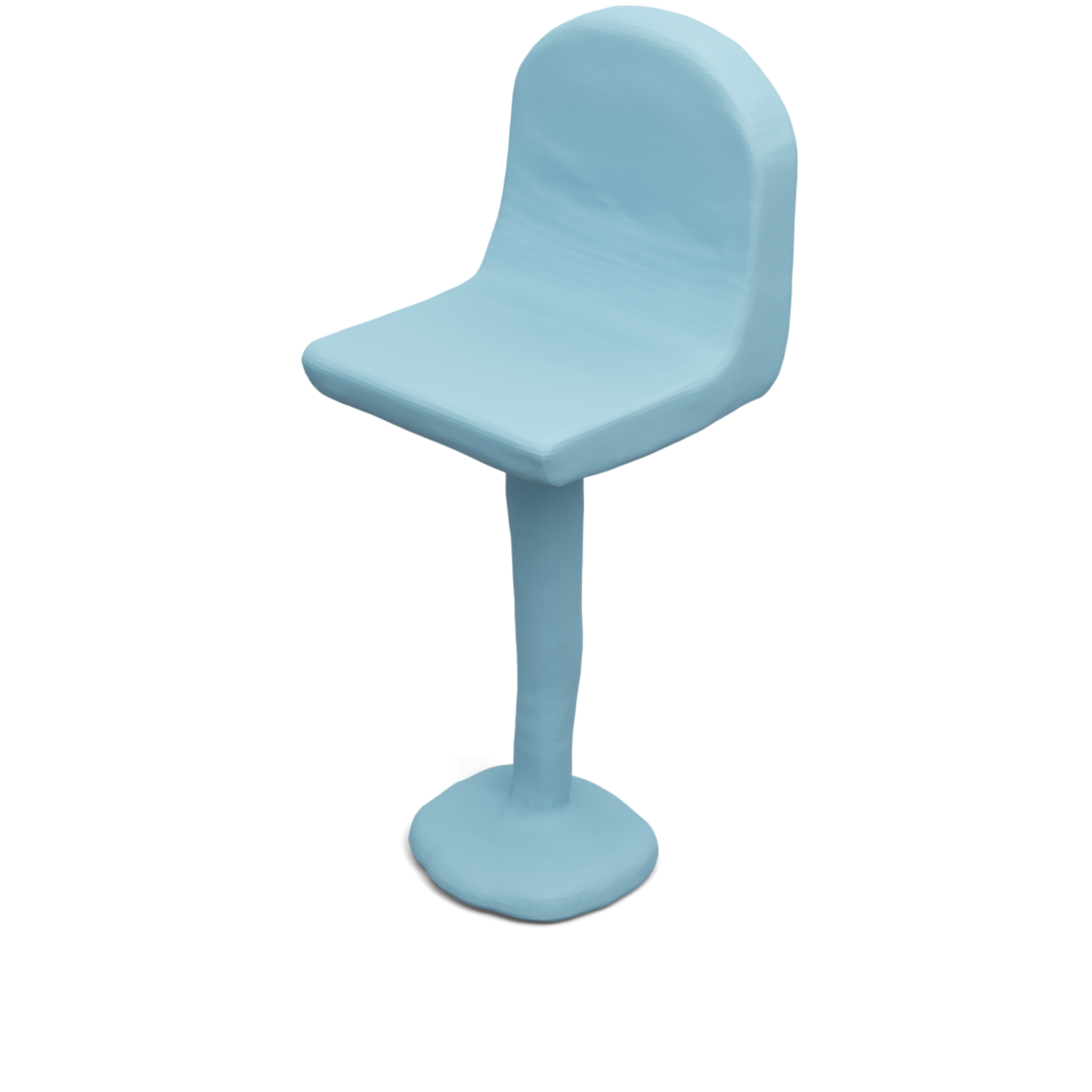}\
    & \includegraphics[trim = 1 1 1 1, clip, width=0.16\linewidth]{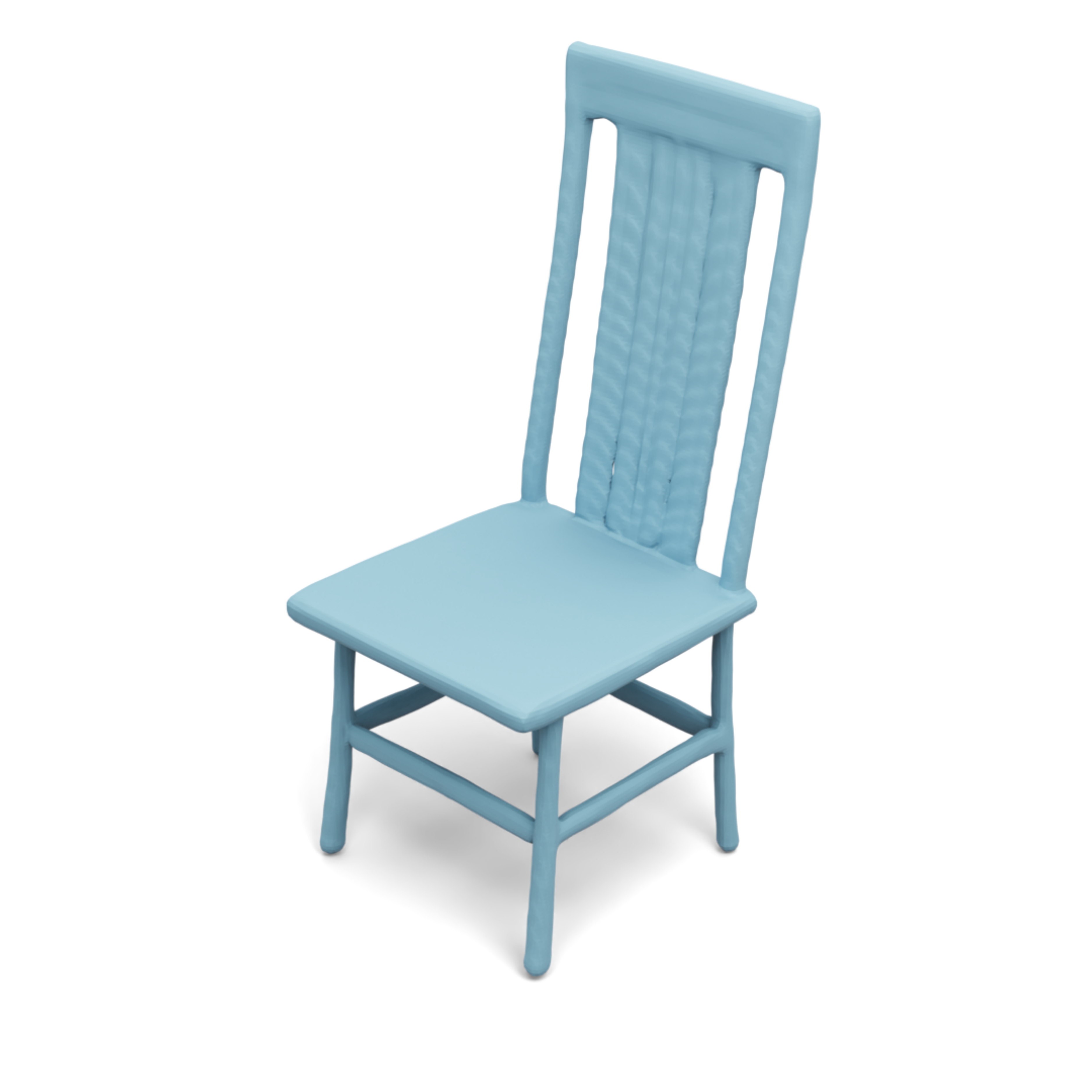}\
    & \includegraphics[trim = 1 1 1 1, clip, width=0.16\linewidth]{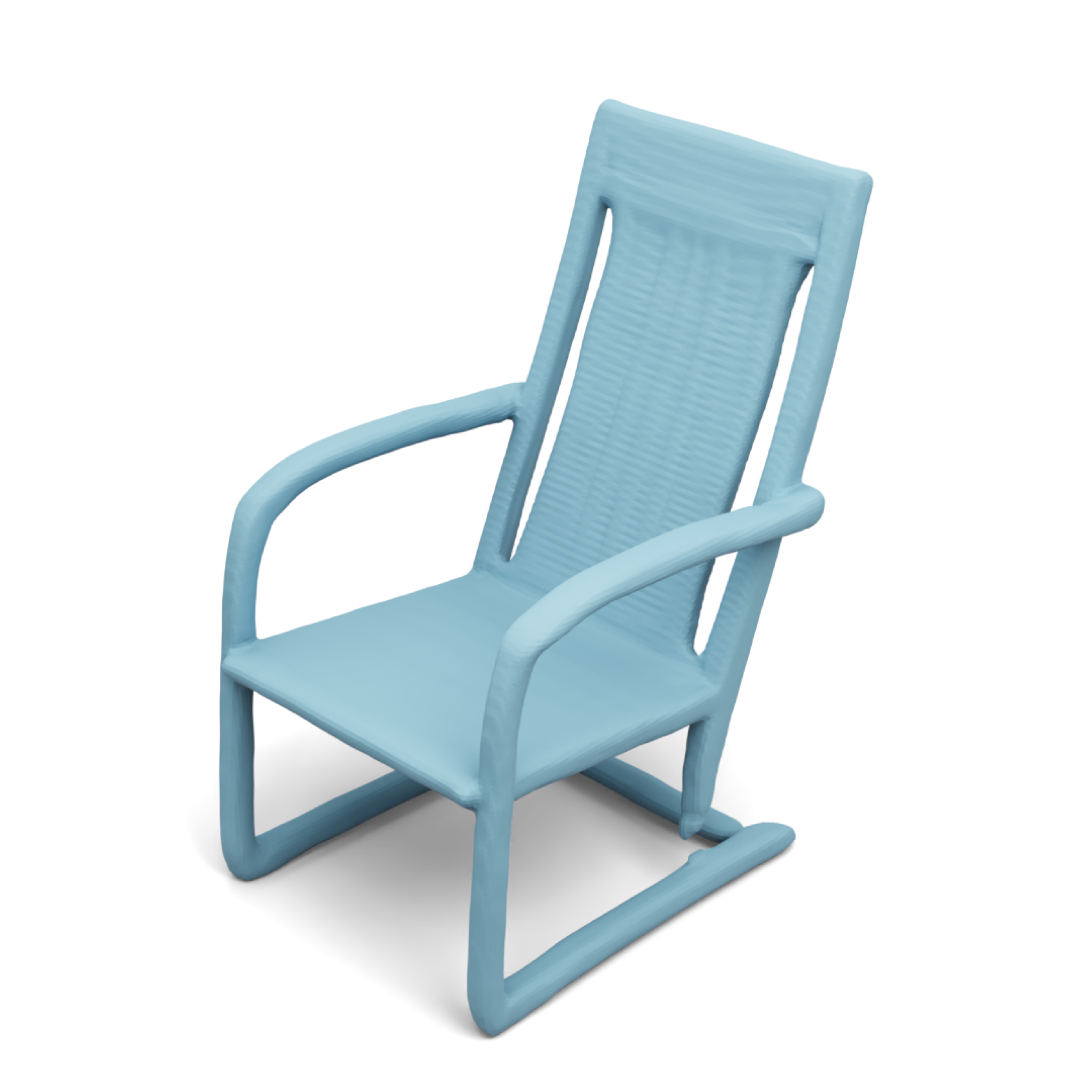}\
    & \includegraphics[trim = 1 1 1 1, clip, width=0.16\linewidth]{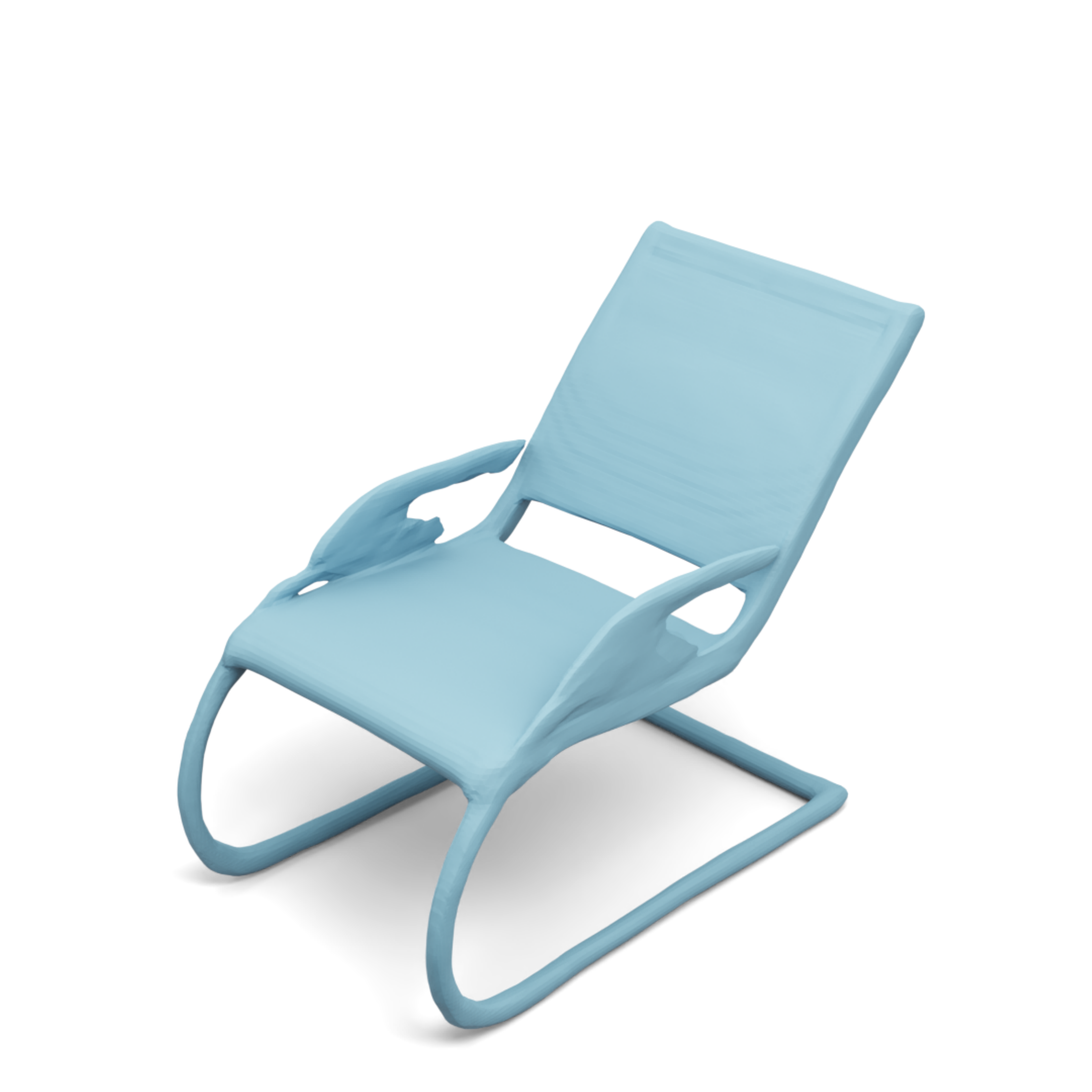} \\

    \includegraphics[width=0.16\linewidth]{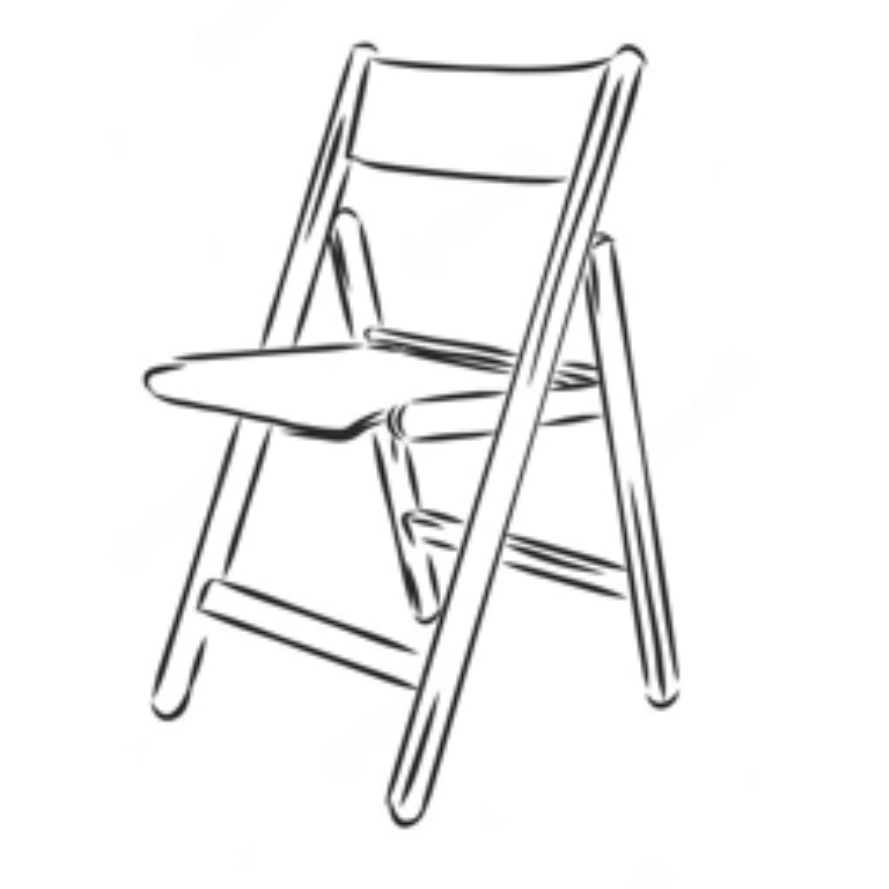}\
    & \includegraphics[width=0.16\linewidth]{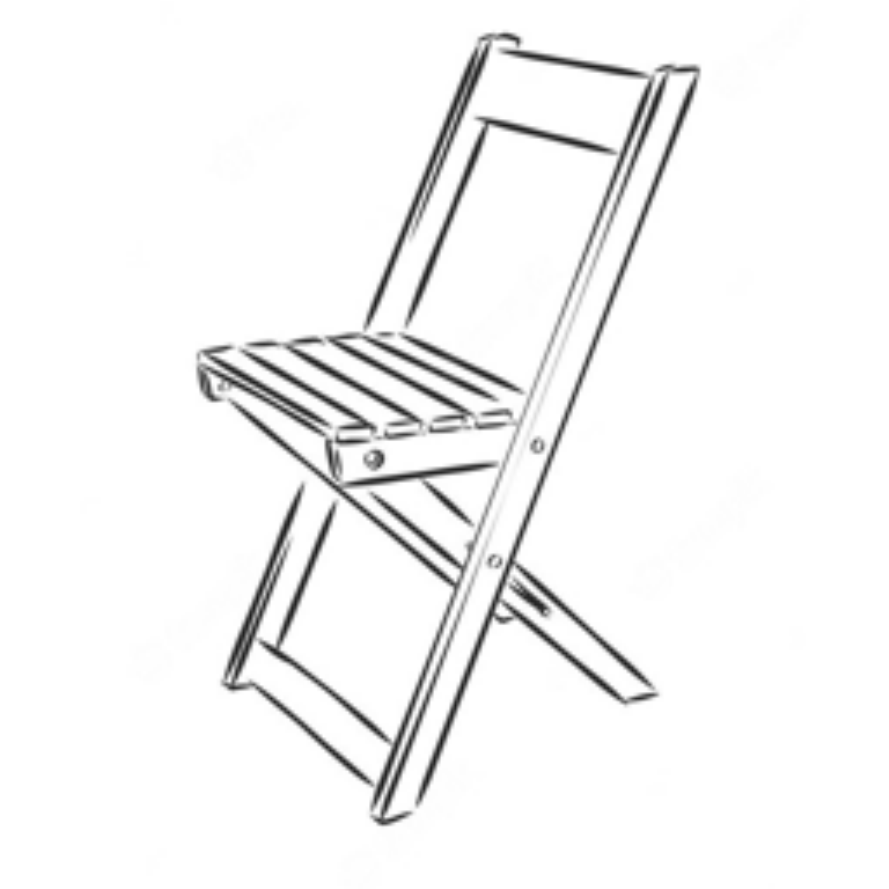}\
    & \includegraphics[width=0.16\linewidth]{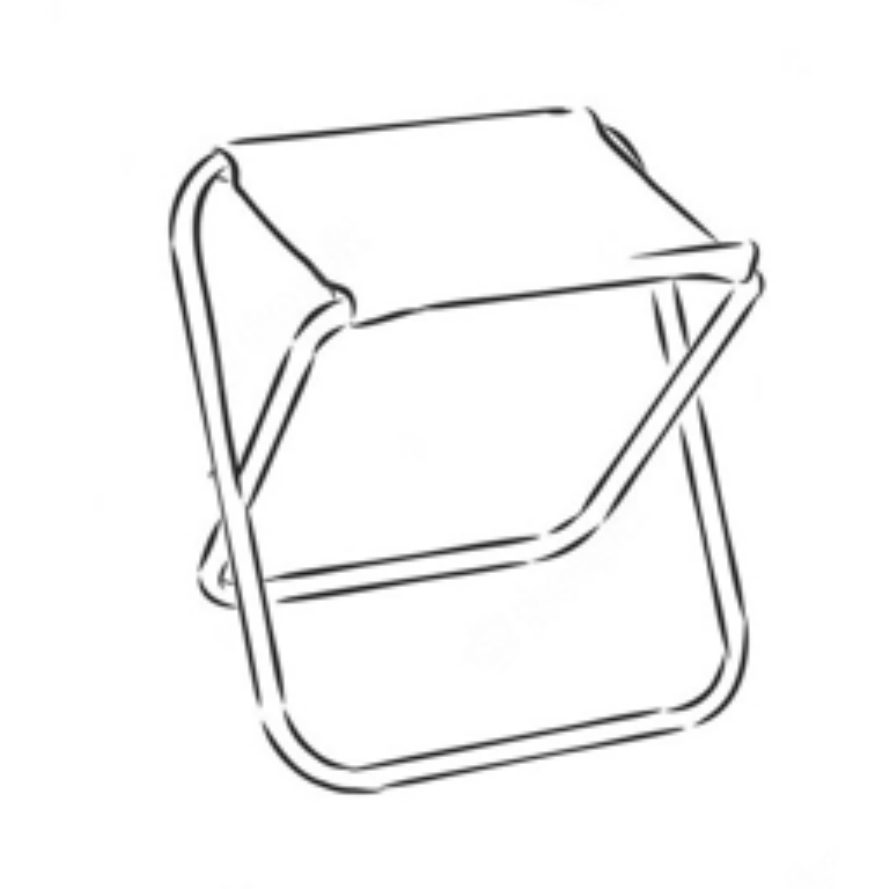}\
    & \includegraphics[width=0.16\linewidth]{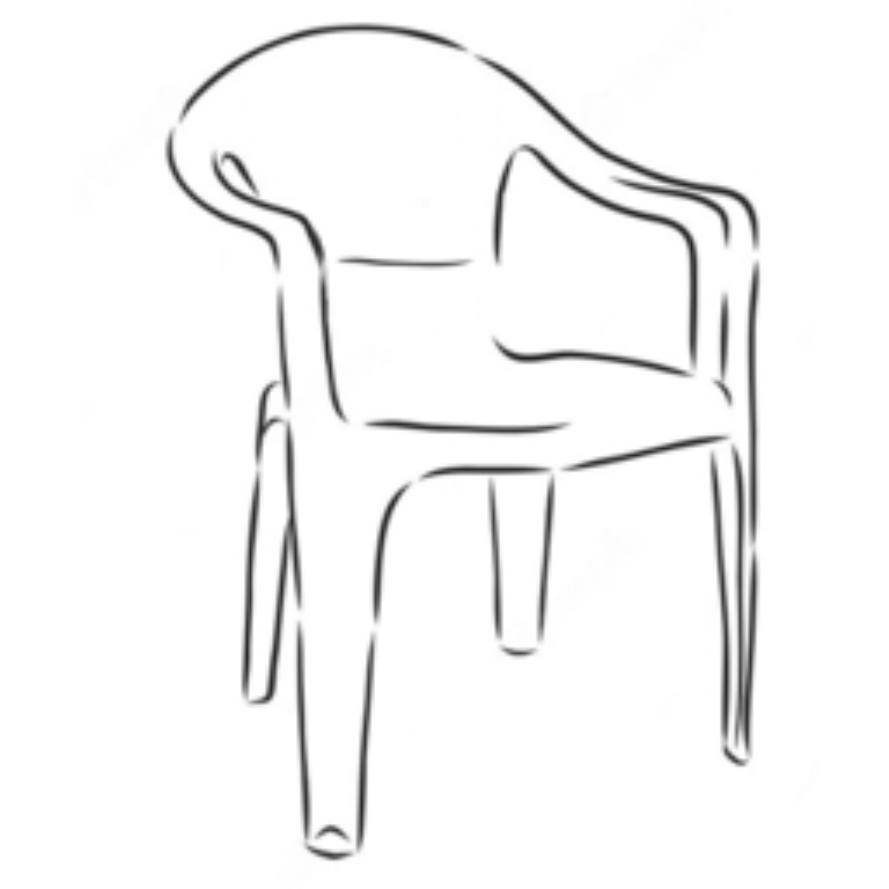}\
    & \includegraphics[width=0.16\linewidth]{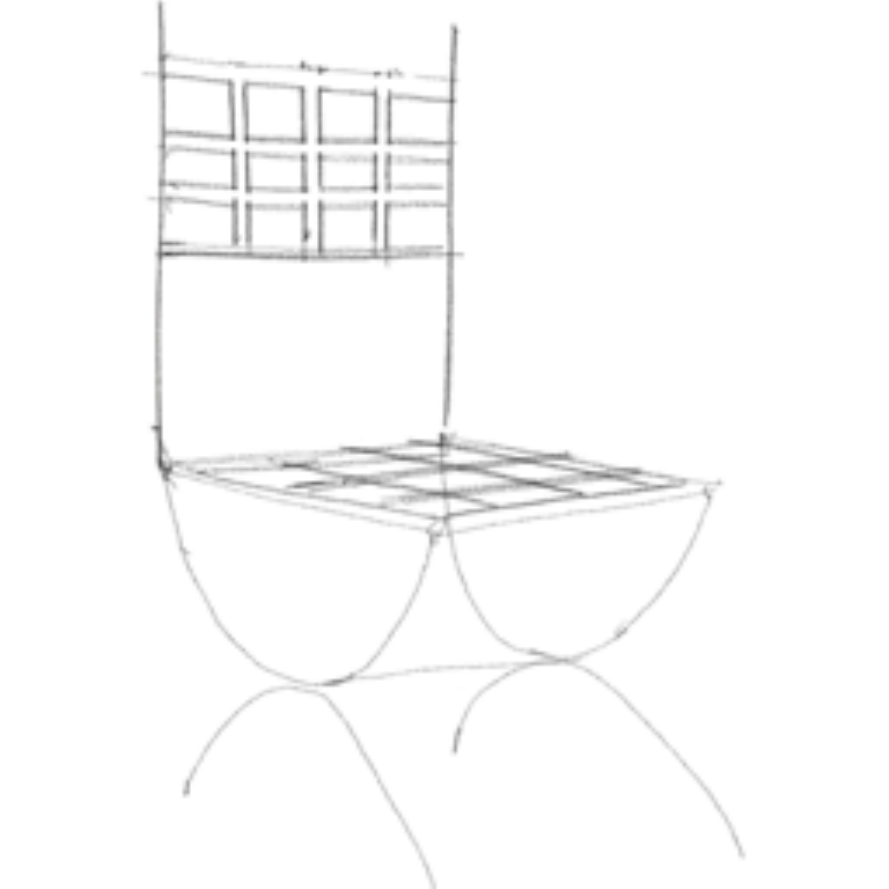}\
    & \includegraphics[width=0.16\linewidth]{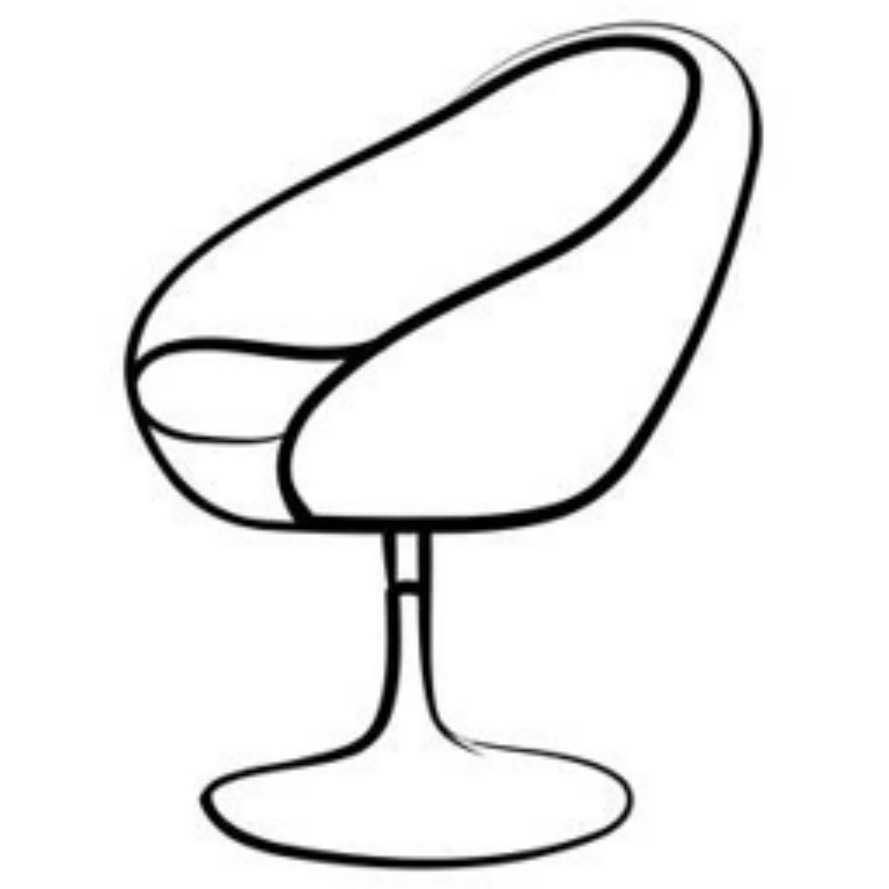} \\


\includegraphics[trim = 1 1 1 1, clip, width=0.16\linewidth]{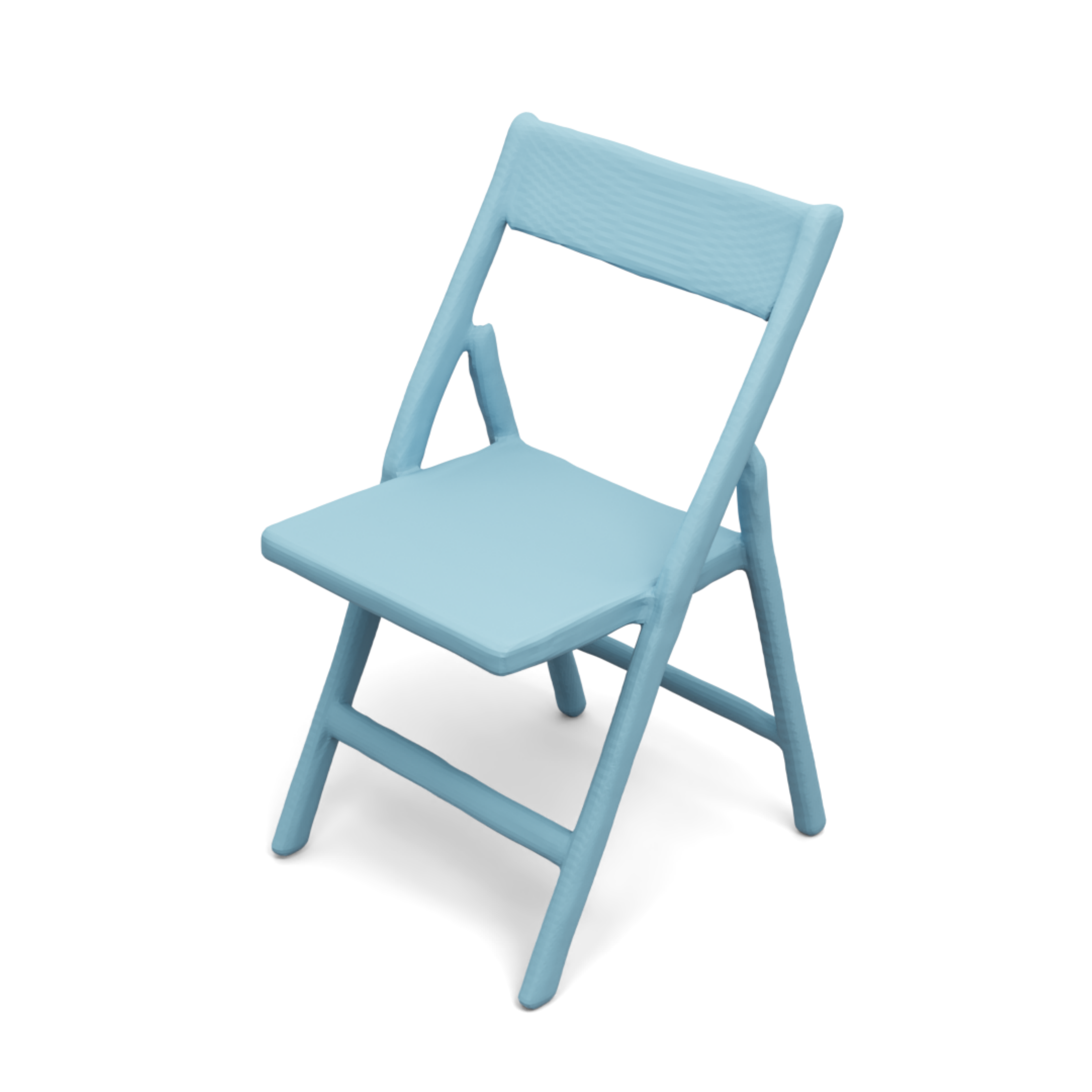}\
    & \includegraphics[trim = 1 1 1 1, clip, width=0.16\linewidth]{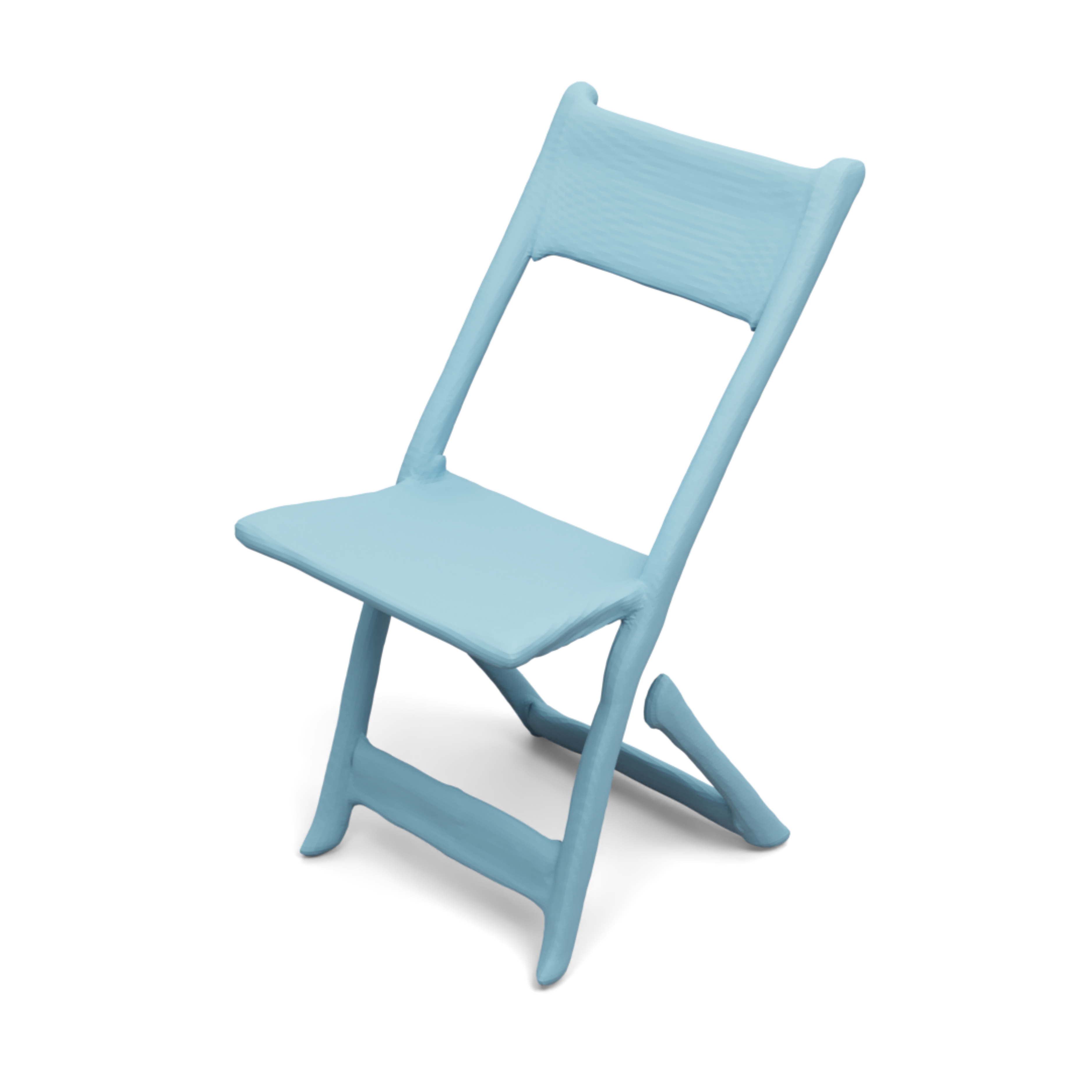}\
    & \includegraphics[trim = 1 1 1 1, clip, width=0.16\linewidth]{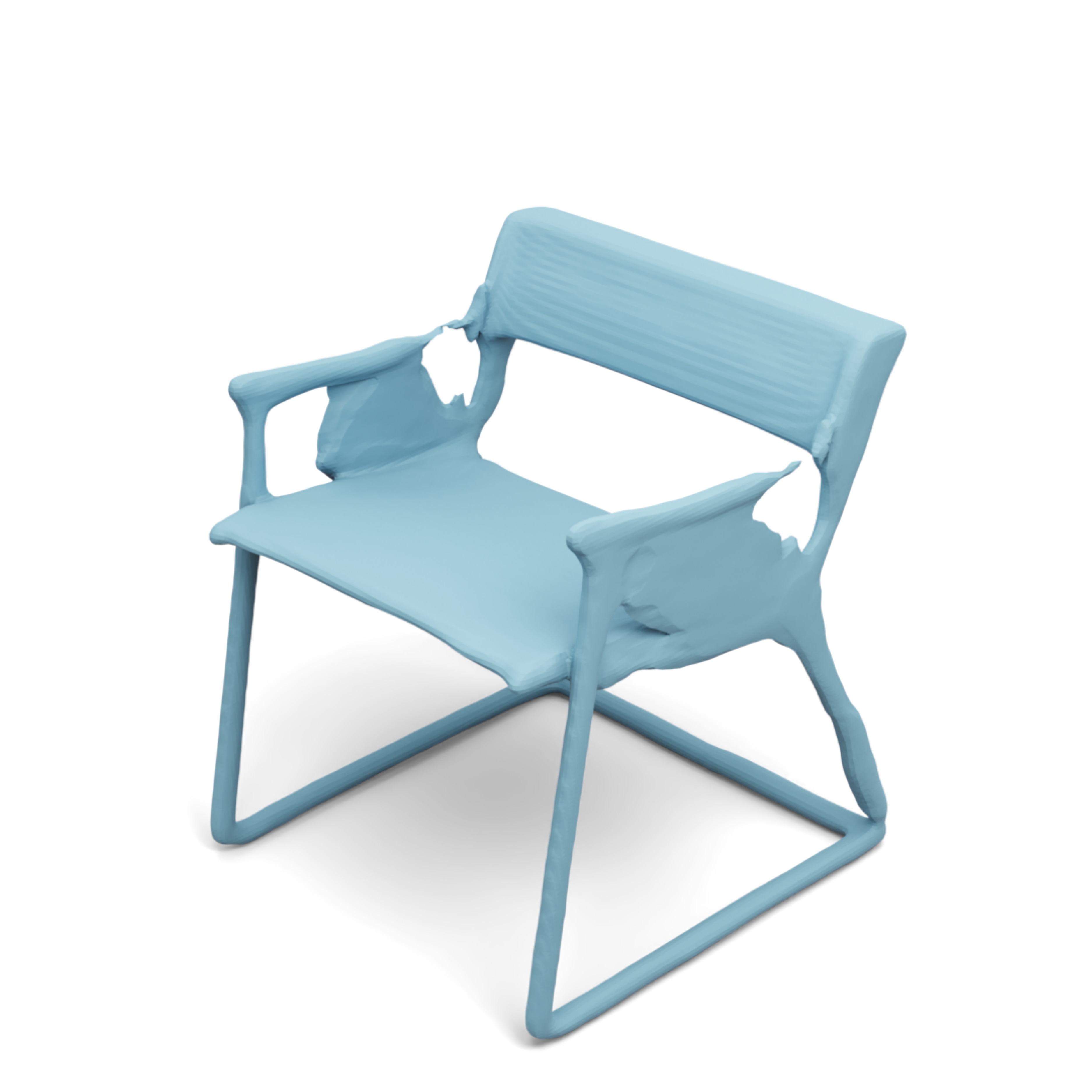}\
    & \includegraphics[trim = 1 1 1 1, clip, width=0.16\linewidth]{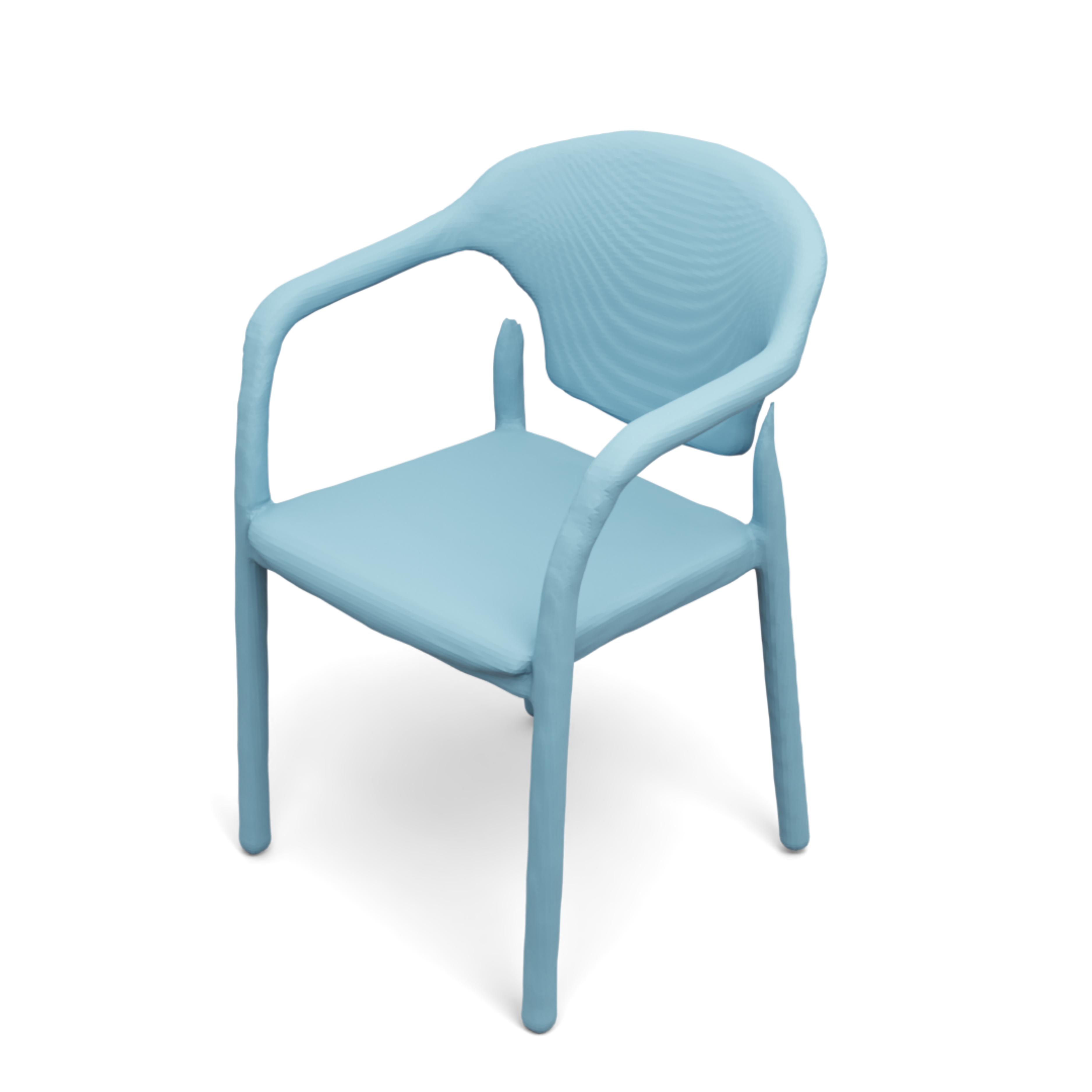}\
    & \includegraphics[trim = 1 1 1 1, clip, width=0.16\linewidth]{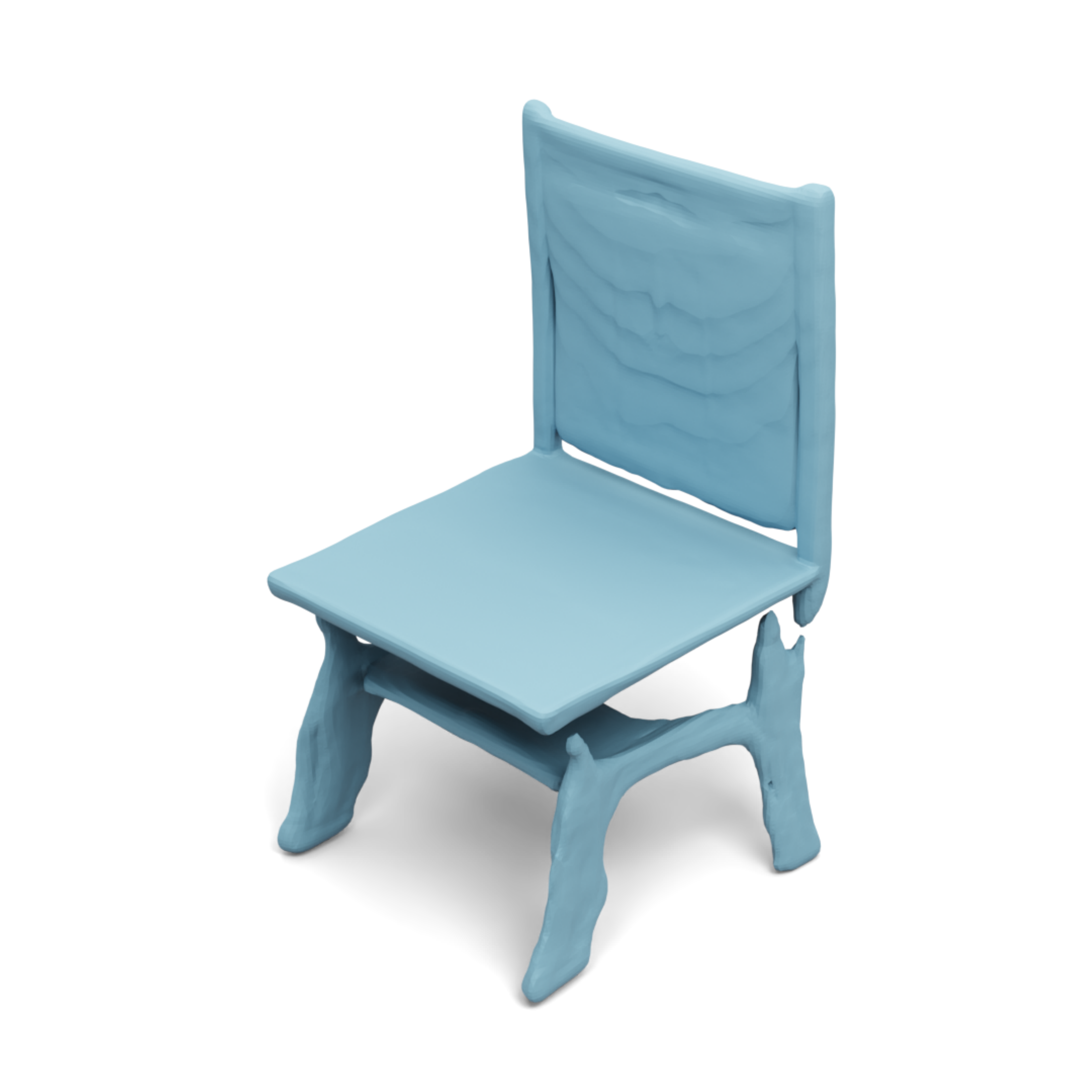}\
    & \includegraphics[trim = 1 1 1 1, clip, width=0.16\linewidth]{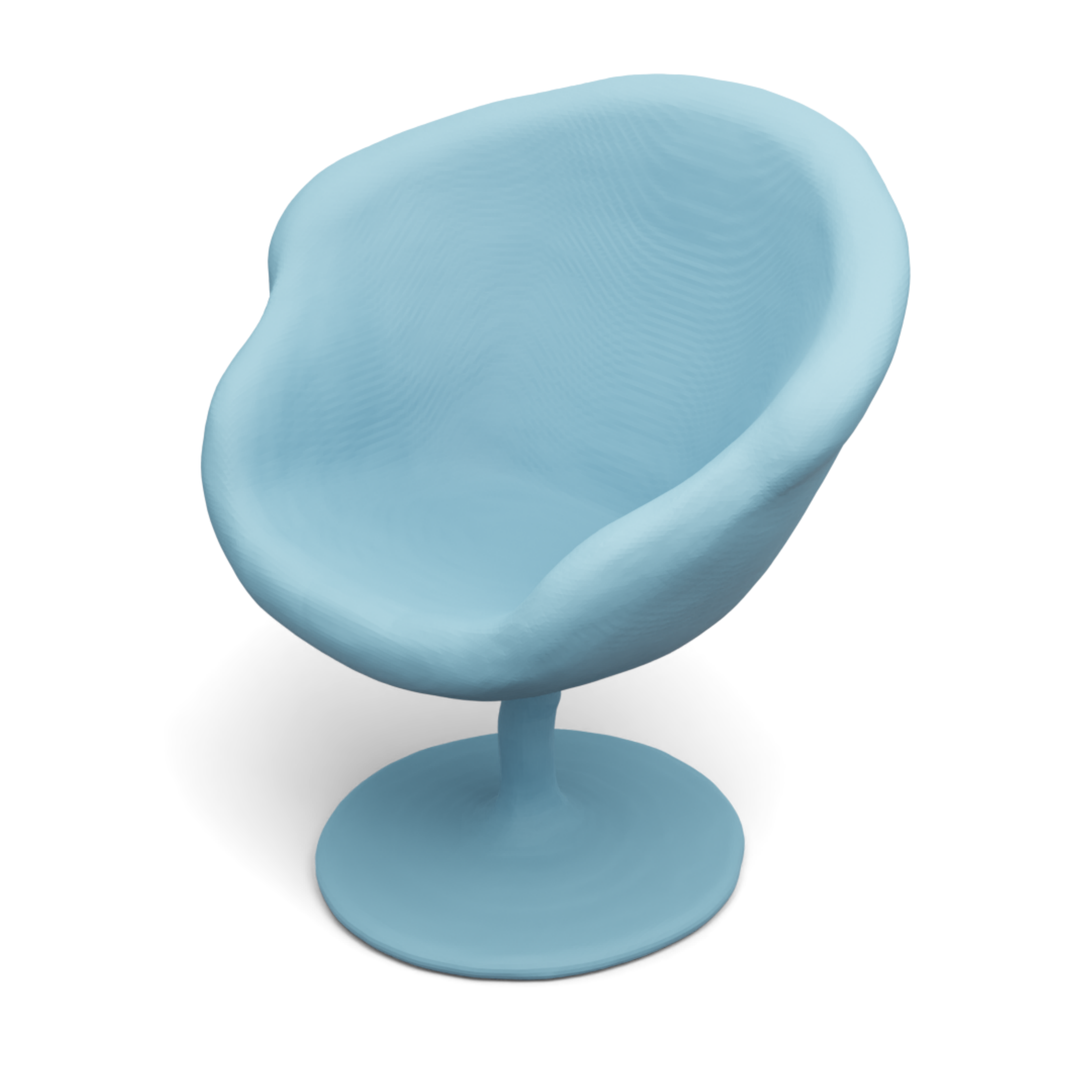} \\

    \includegraphics[width=0.16\linewidth]{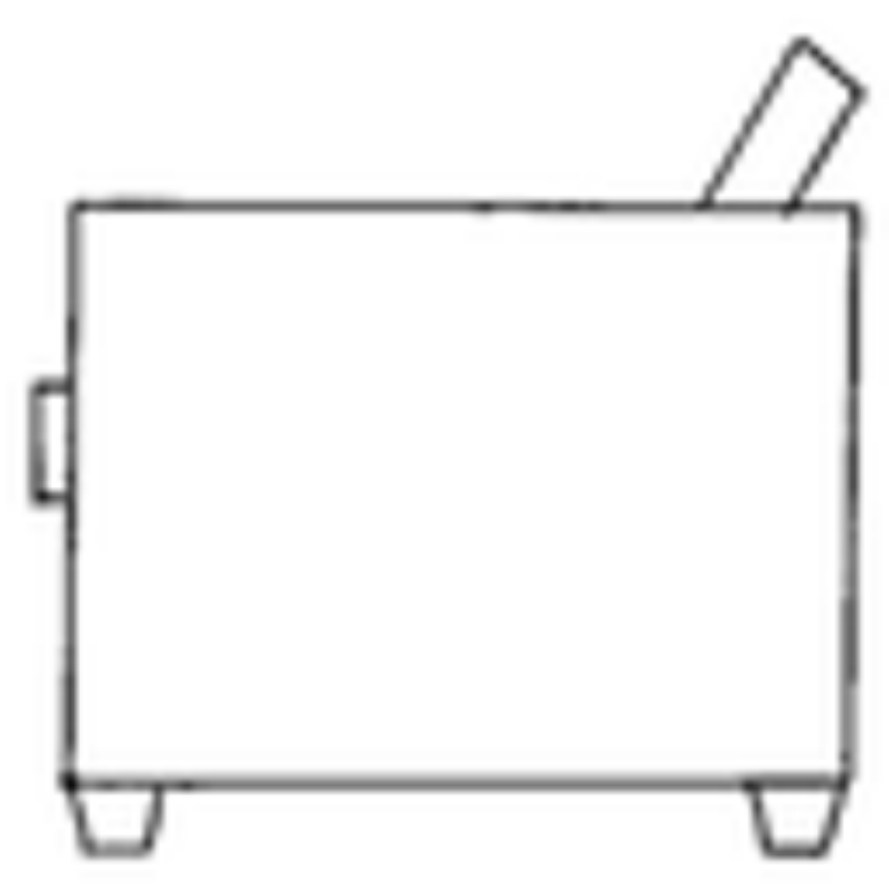}\
    & \includegraphics[width=0.16\linewidth]{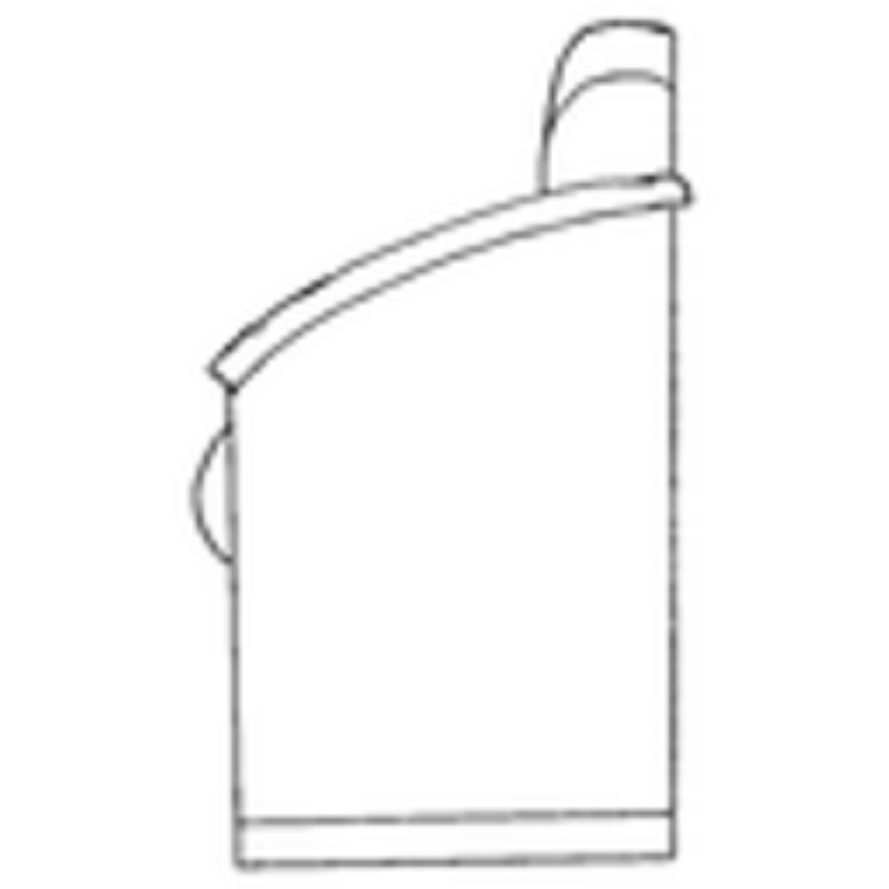}\
    & \includegraphics[width=0.16\linewidth]{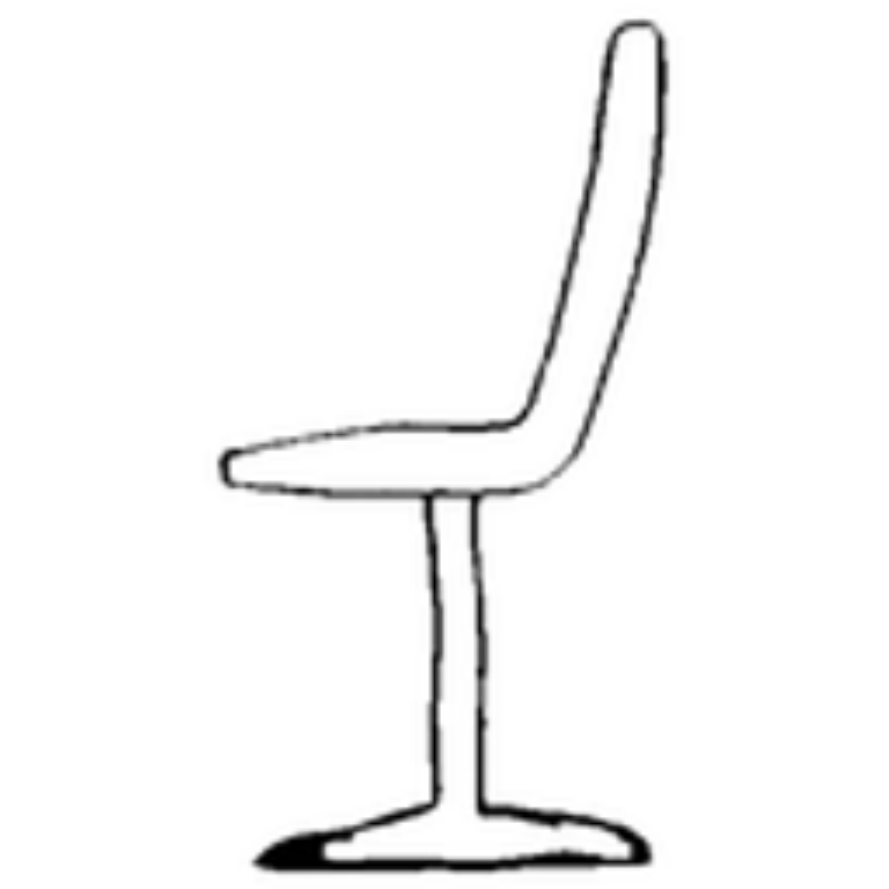}\
    & \includegraphics[width=0.16\linewidth]{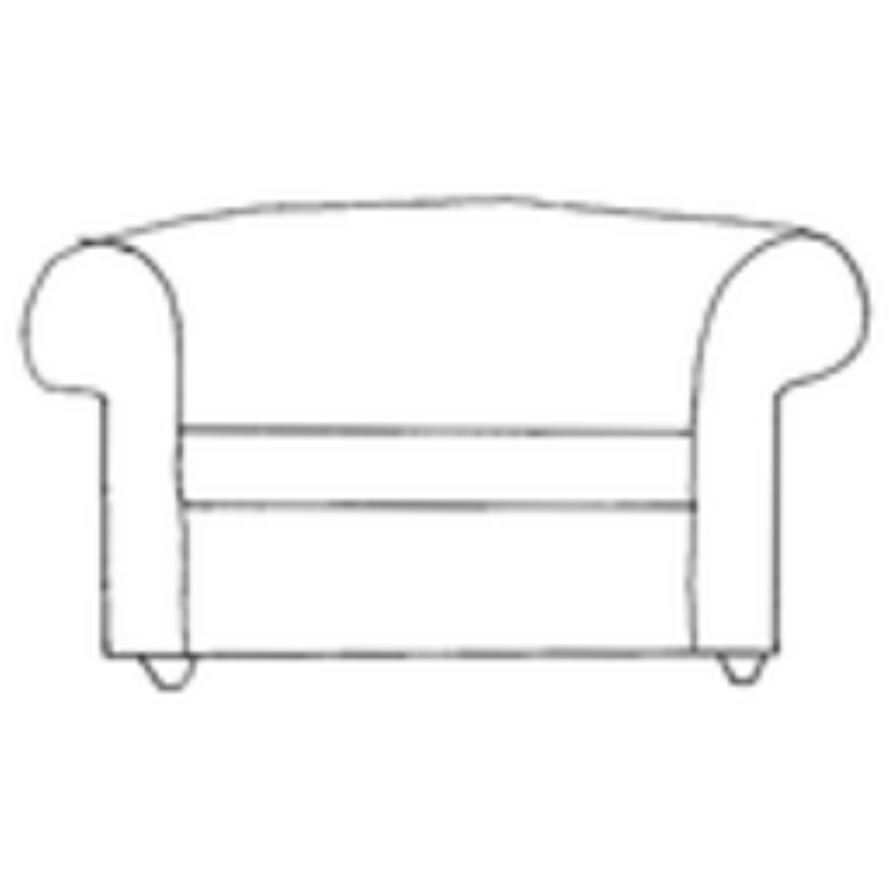}\
    & \includegraphics[width=0.16\linewidth]{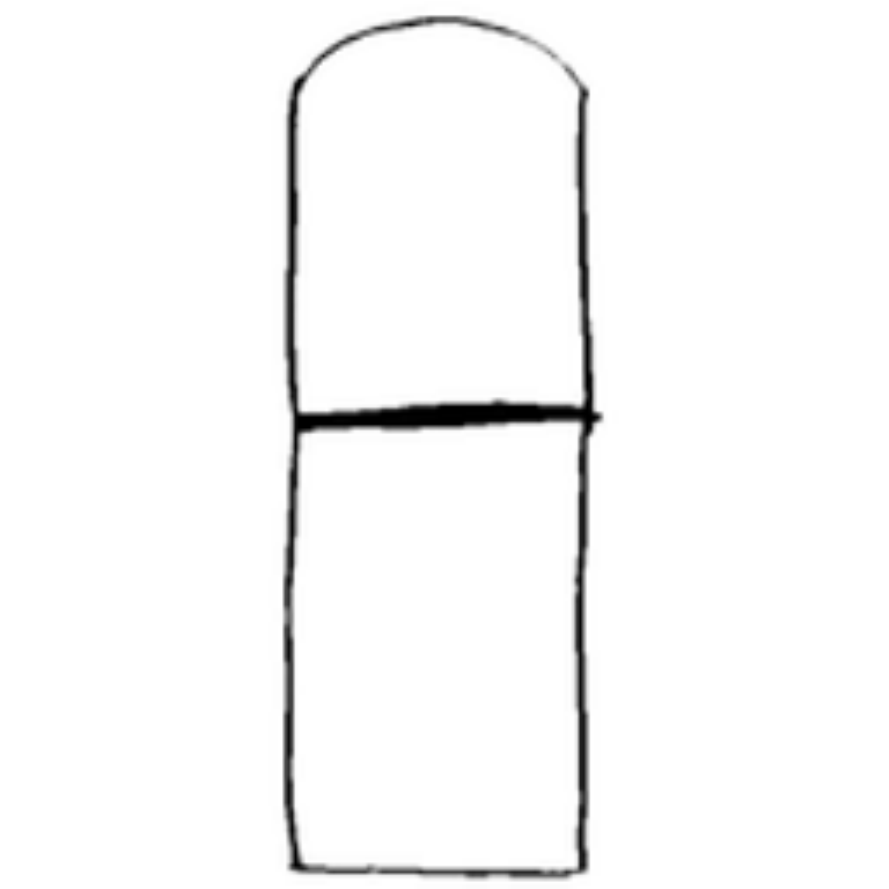}\
    & \includegraphics[width=0.16\linewidth]{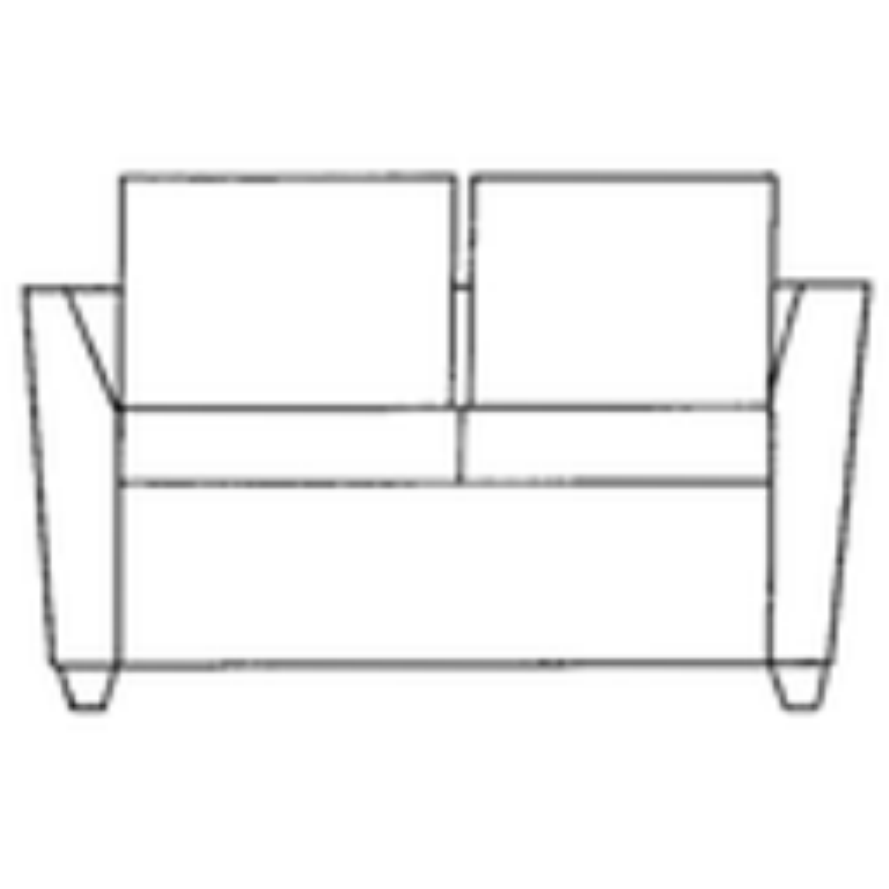} \\


\includegraphics[trim = 1 1 1 1, clip, width=0.16\linewidth]{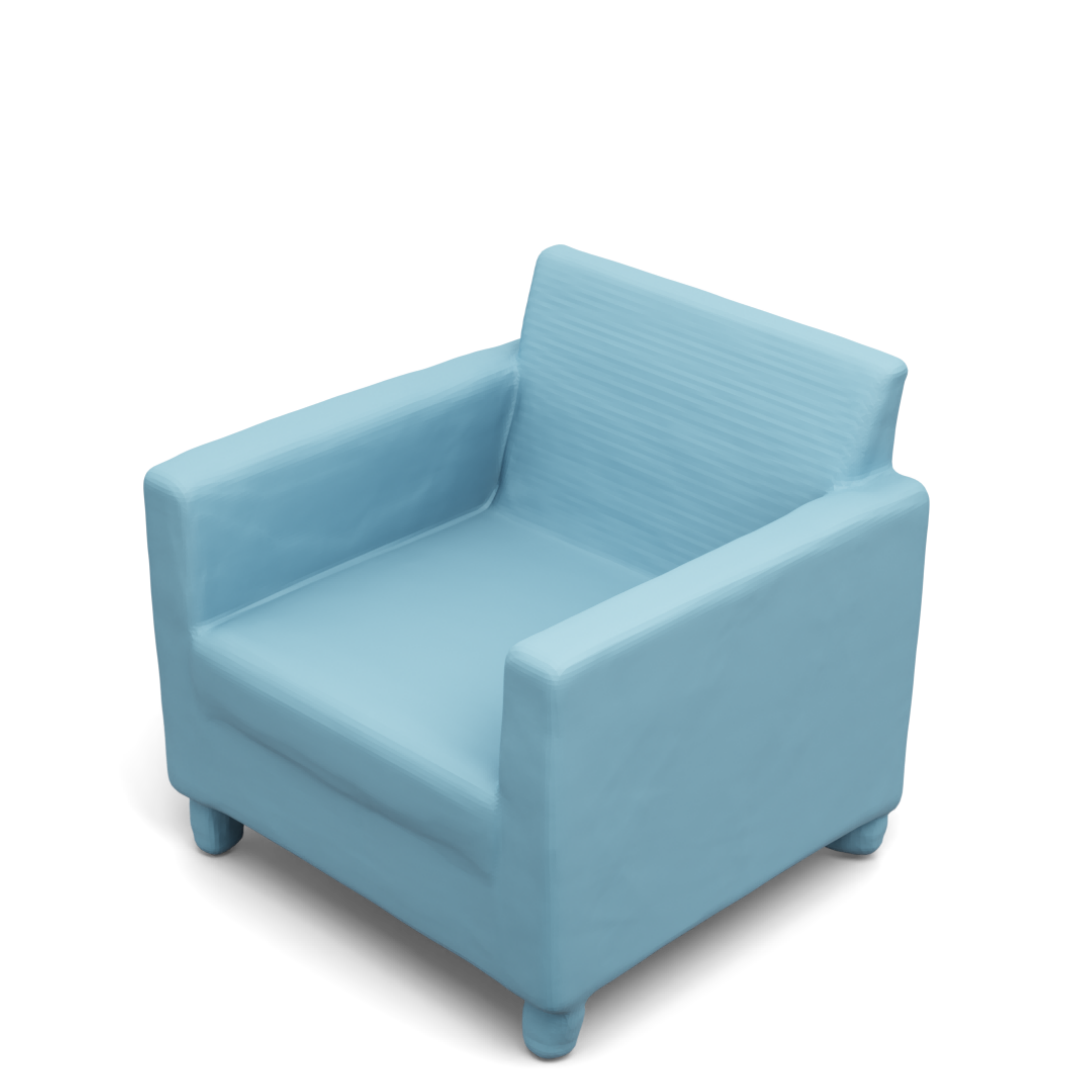}\
    & \includegraphics[trim = 1 1 1 1, clip, width=0.16\linewidth]{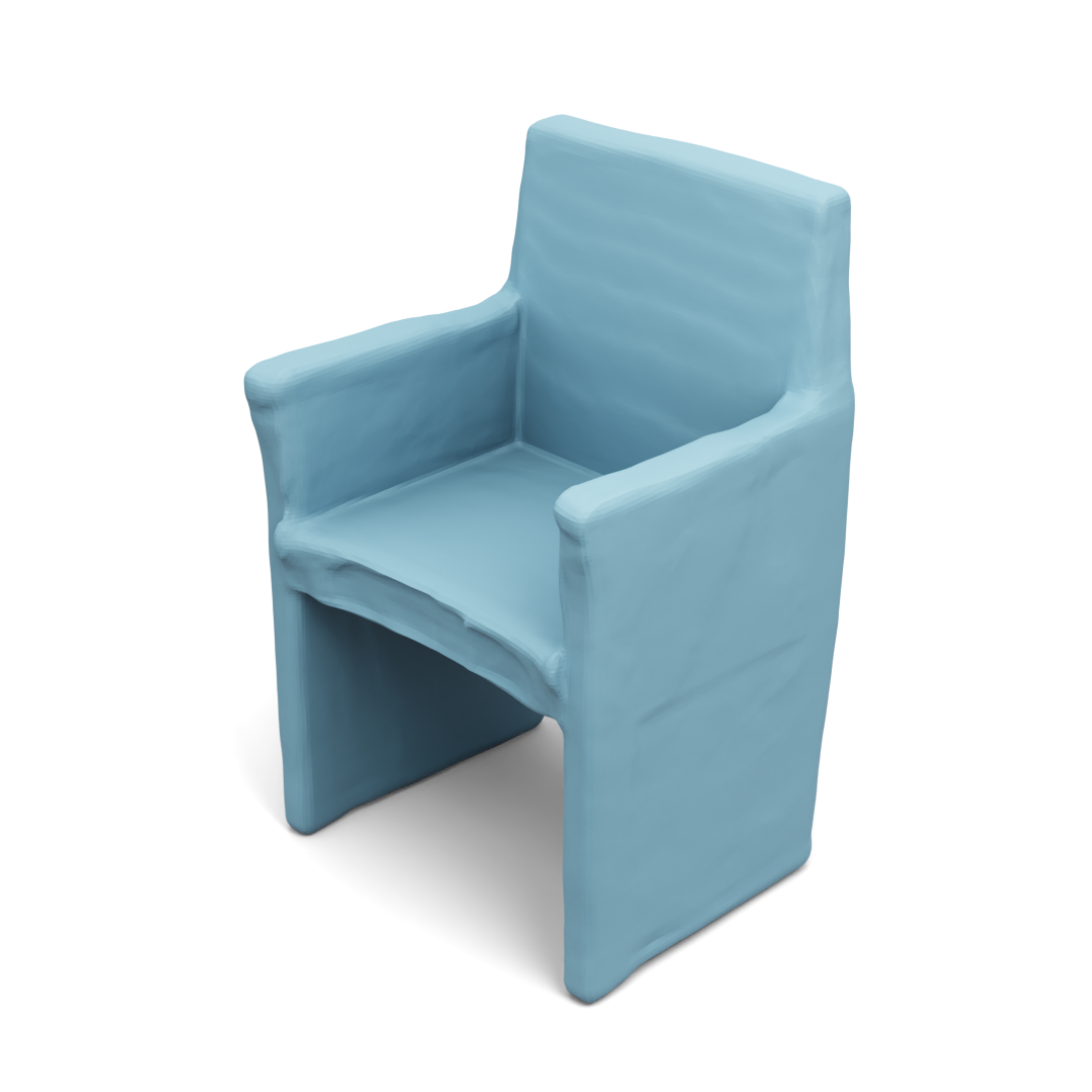}\
    & \includegraphics[trim = 1 1 1 1, clip, width=0.16\linewidth]{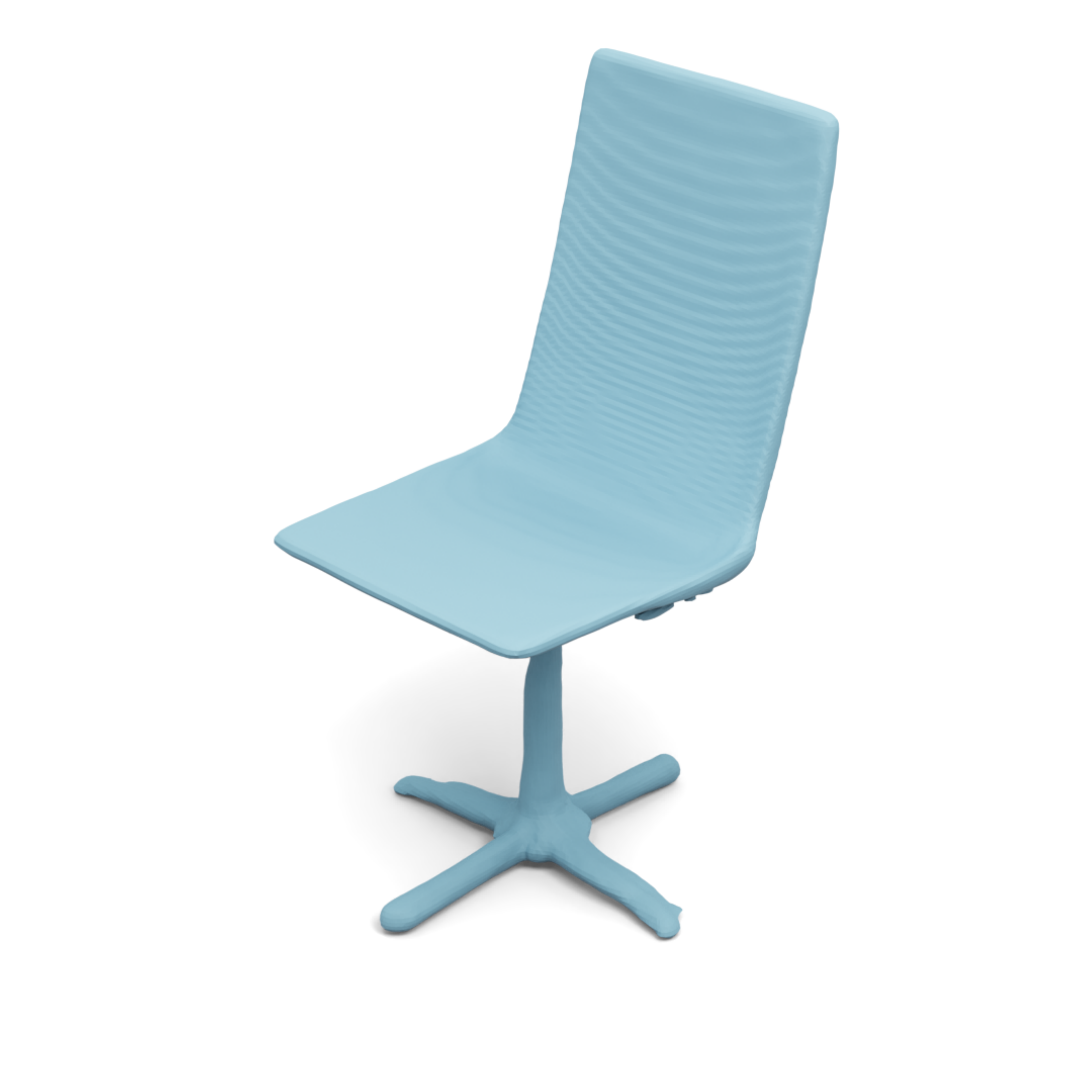}\
    & \includegraphics[trim = 1 1 1 1, clip, width=0.16\linewidth]{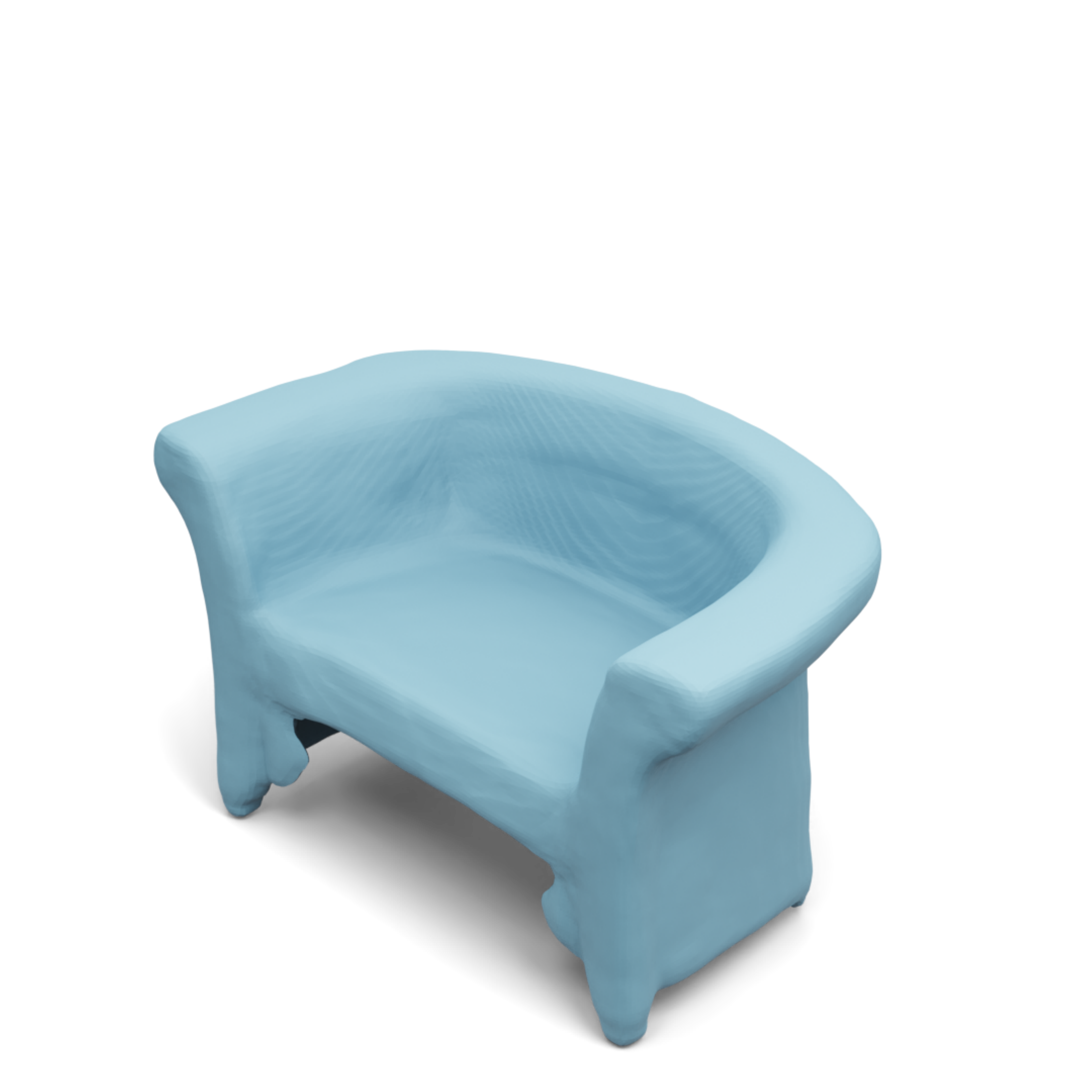}\
    & \includegraphics[trim = 1 1 1 1, clip, width=0.16\linewidth]{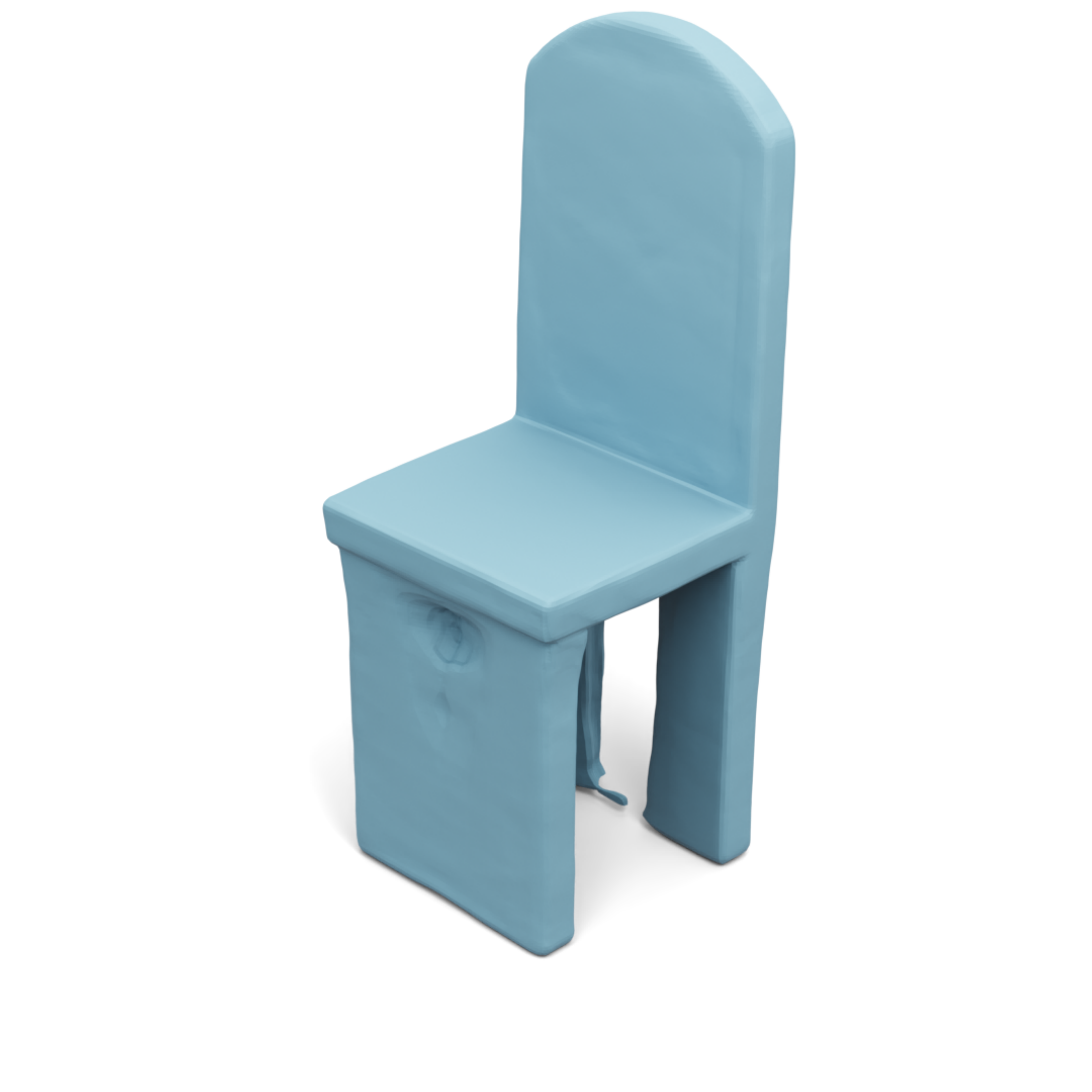}\
    & \includegraphics[trim = 1 1 1 1, clip, width=0.16\linewidth]{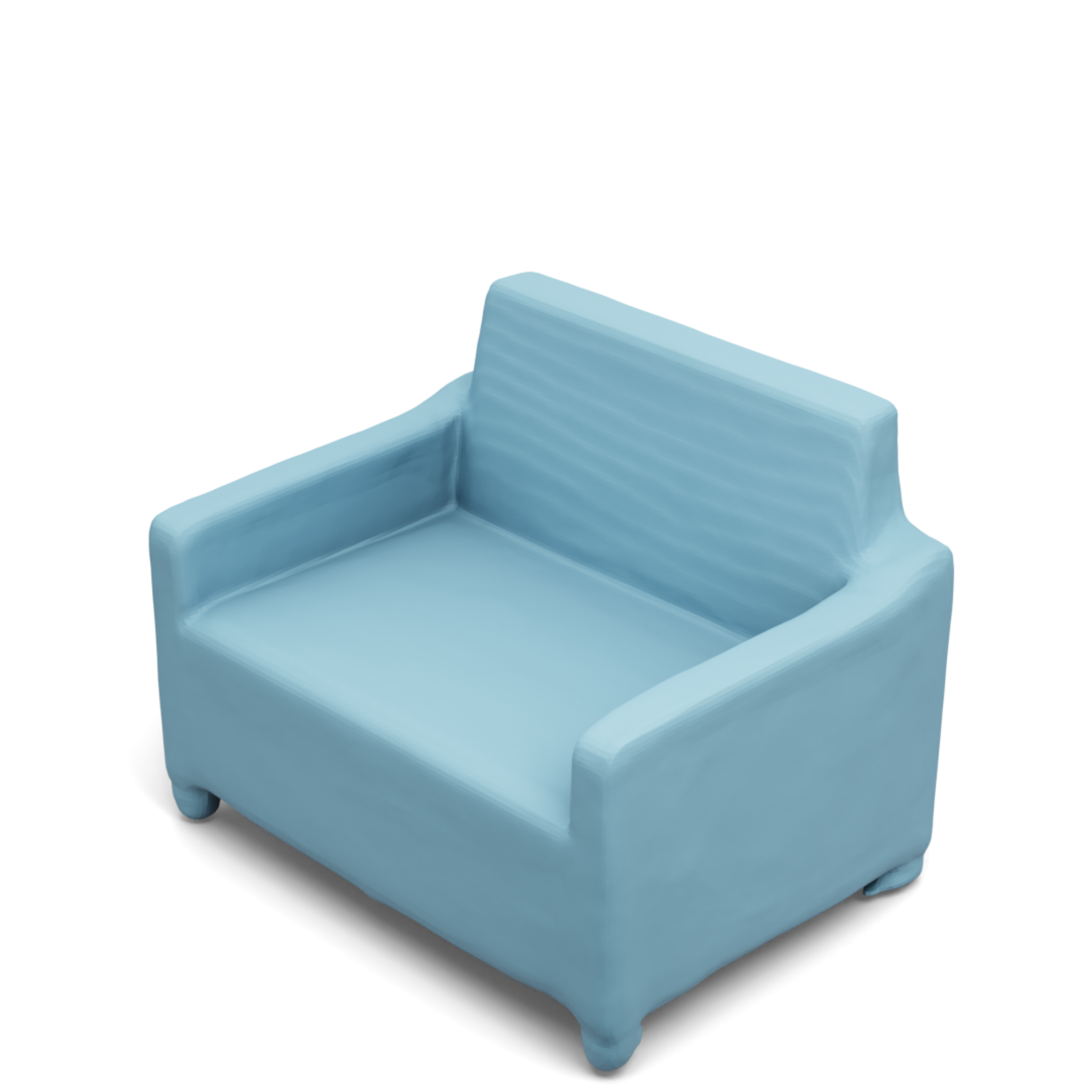} \\
	\end{tabular}
	\caption{We showcase our method using sketches of various styles and levels of abstraction. Chairs in the first row are casually drawn or produced via image processing techniques. The second row shows that our method works with sketches drawn by professionals. Images from the last row are front and side views of chairs originating from ShapeMVD \cite{Lun2017ShapeMVD}. We include them here to facilitate comparisons with further works.}
	\label{fig:bigfigure}
\end{figure*}

\begin{figure*}[t]
	\centering
	\small
	\setlength{\tabcolsep}{1pt}
	\begin{tabular}{cccccc}
    \includegraphics[width=0.16\linewidth]{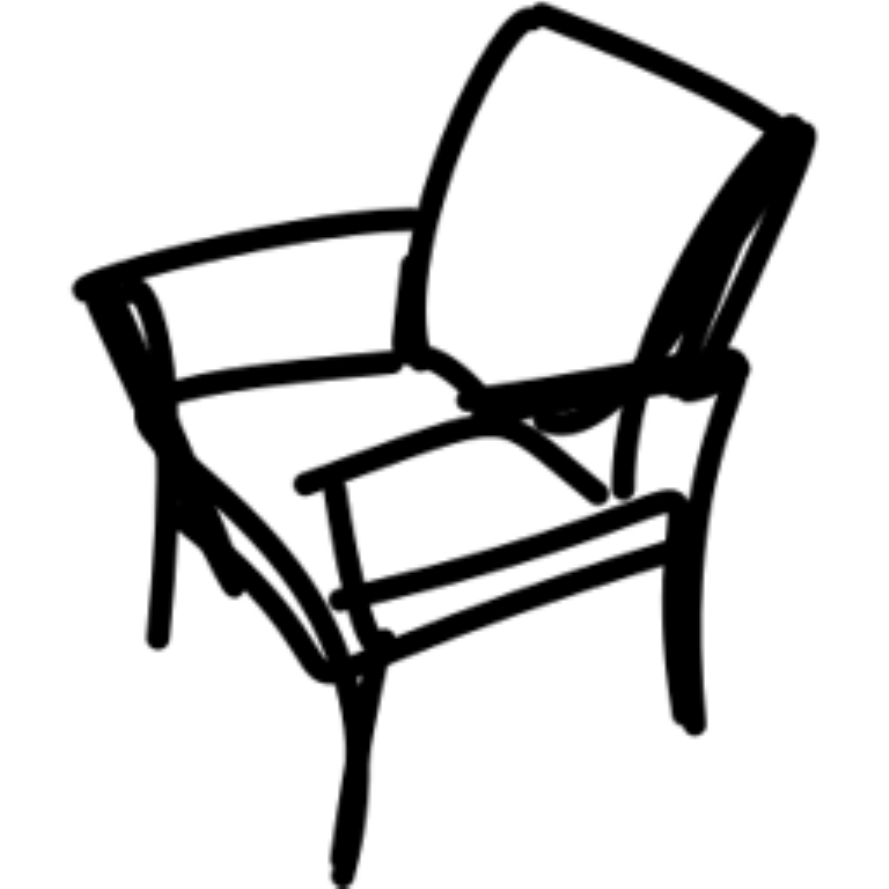}\
    & \includegraphics[width=0.16\linewidth]{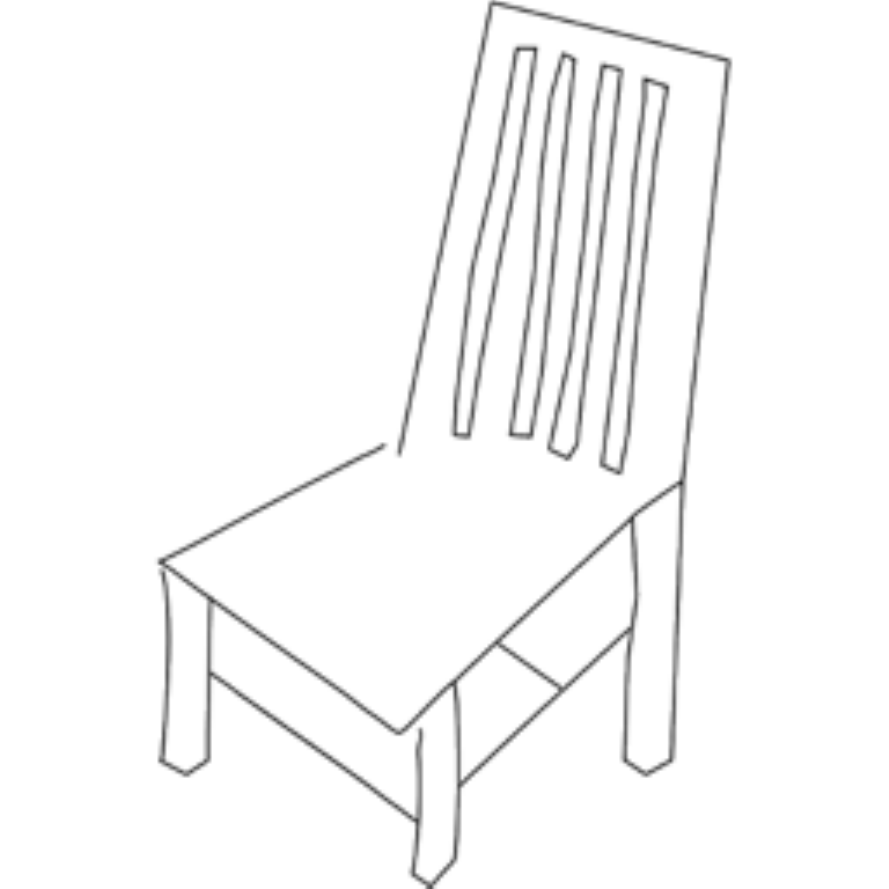}\
    & \includegraphics[width=0.16\linewidth]{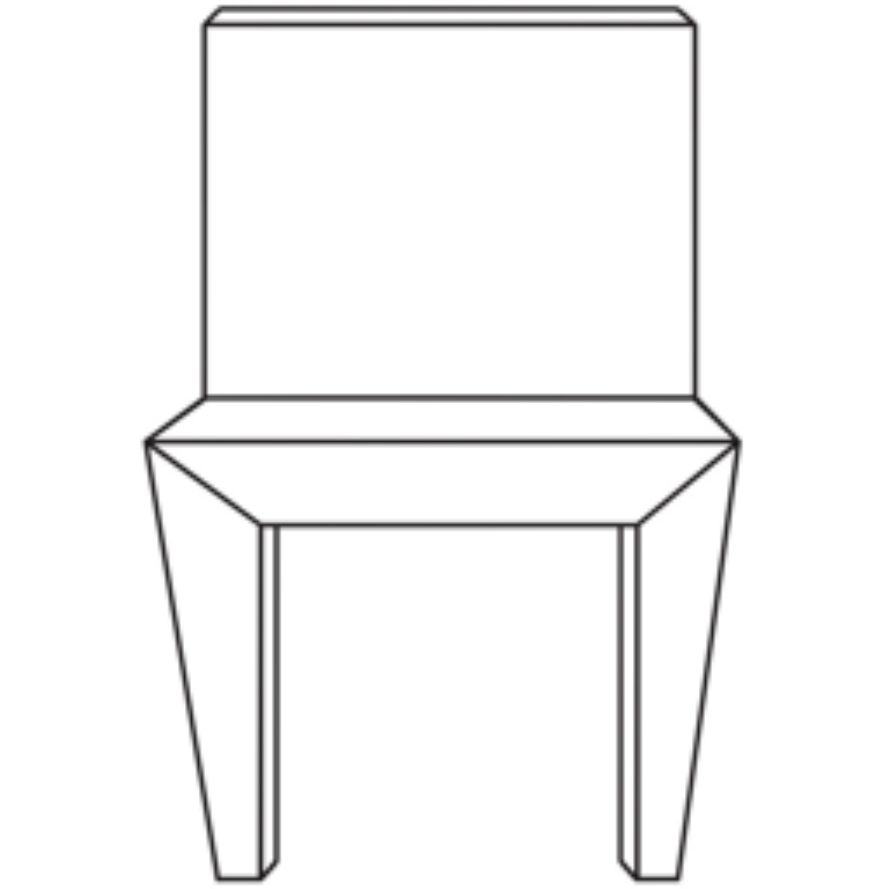}\
    & \includegraphics[width=0.16\linewidth]{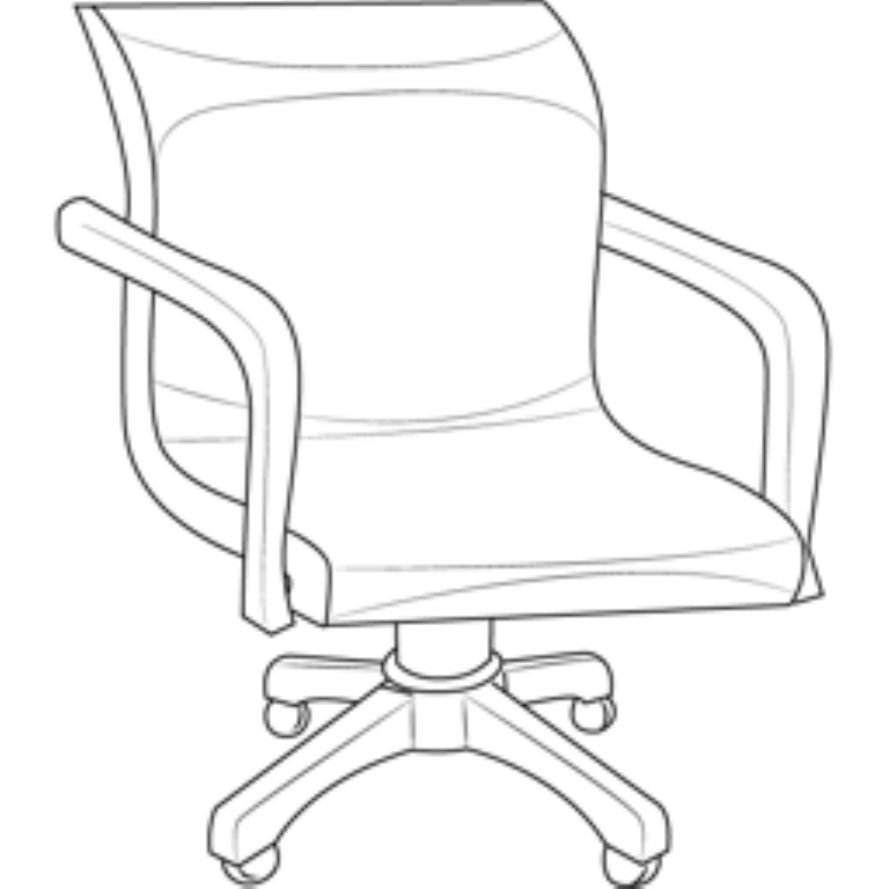}\
    & \includegraphics[width=0.16\linewidth]{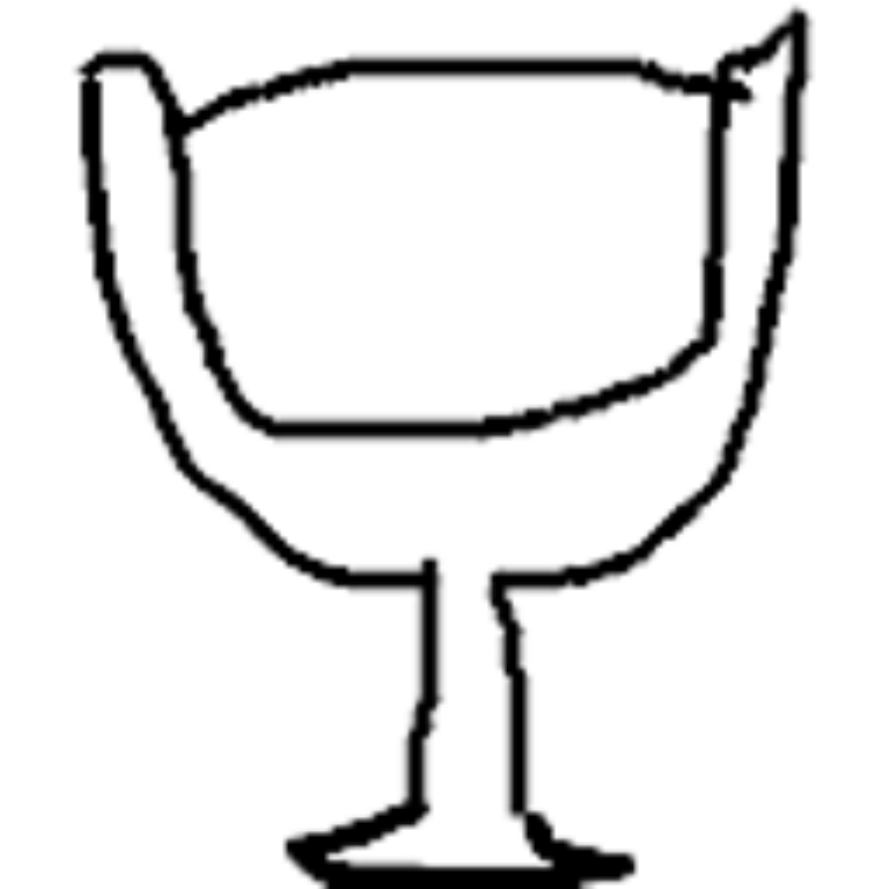}\
    & \includegraphics[width=0.16\linewidth]{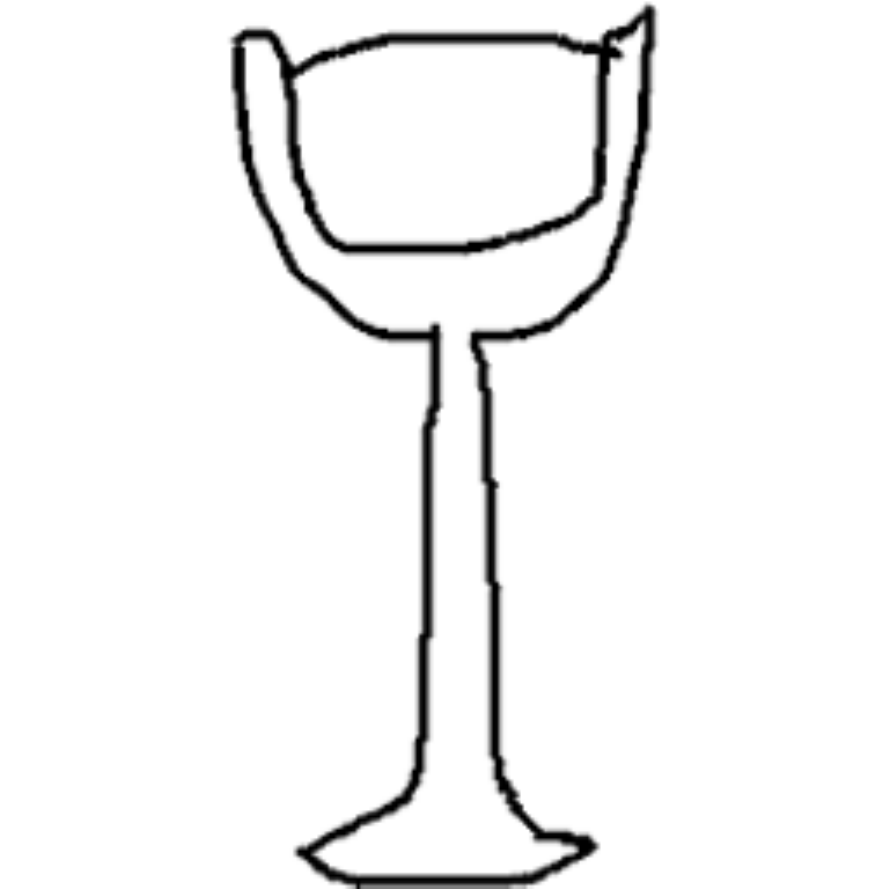} \\


\includegraphics[trim = 1 1 1 1, clip, width=0.16\linewidth]{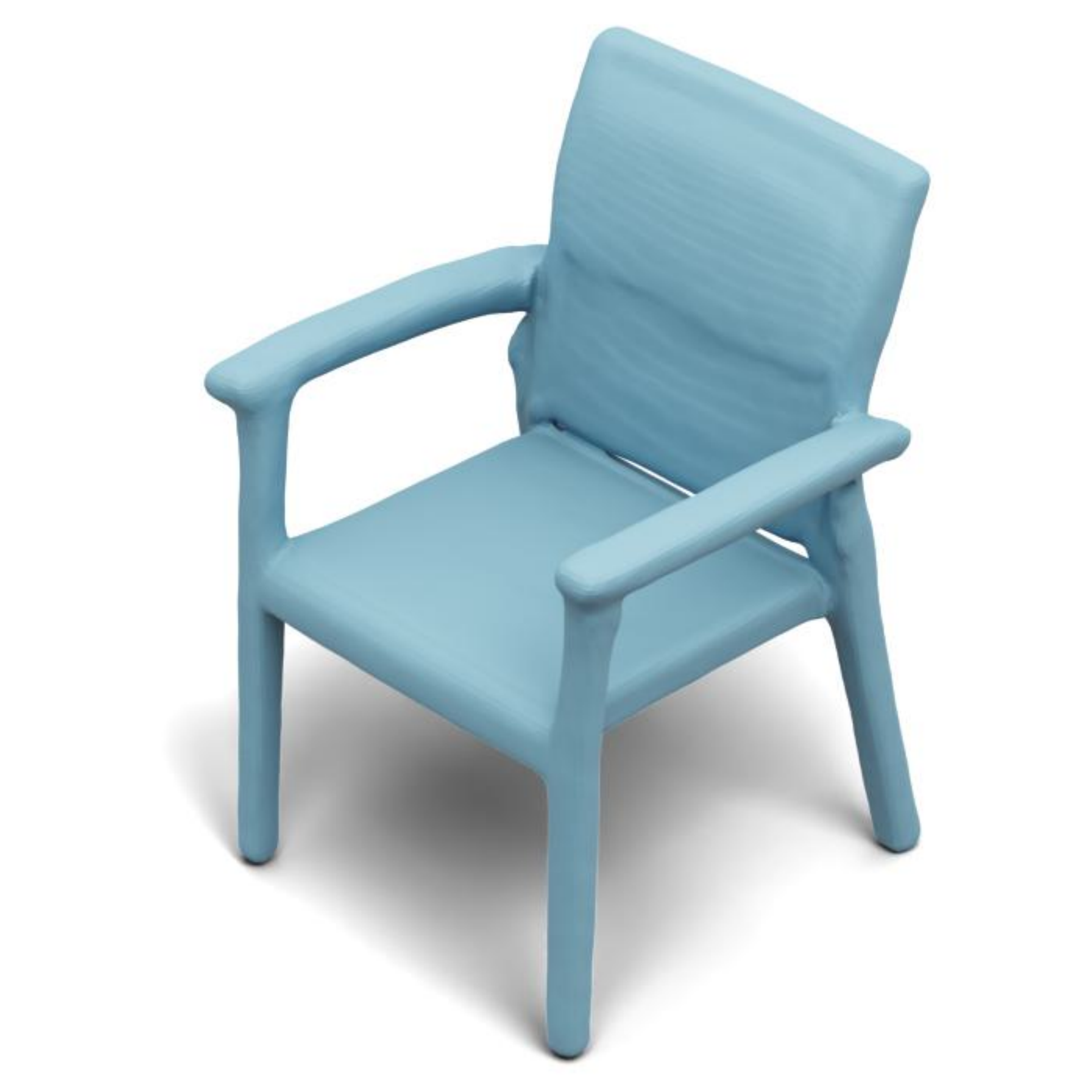}\
    & \includegraphics[trim = 1 1 1 1, clip, width=0.16\linewidth]{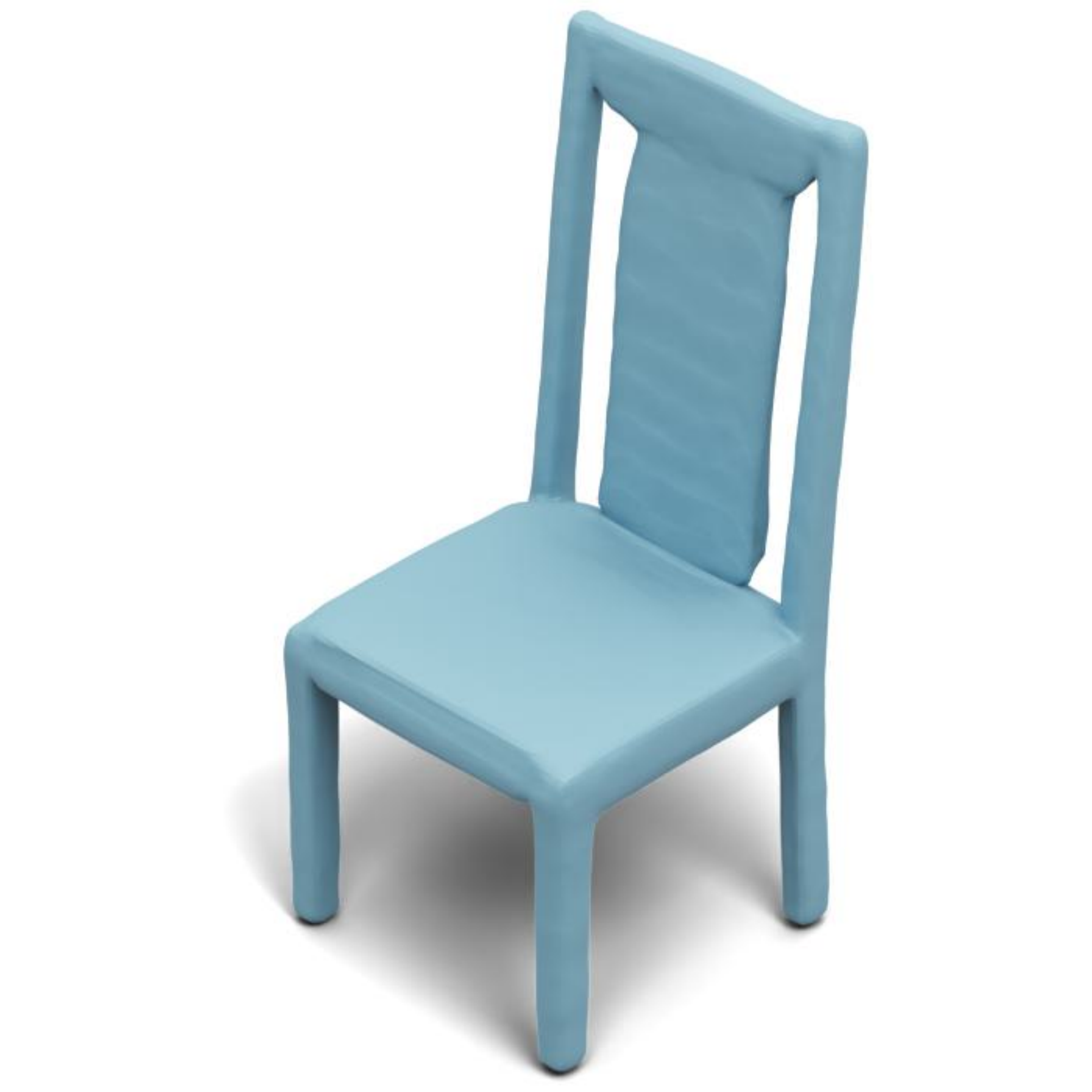}\
    & \includegraphics[trim = 1 1 1 1, clip, width=0.16\linewidth]{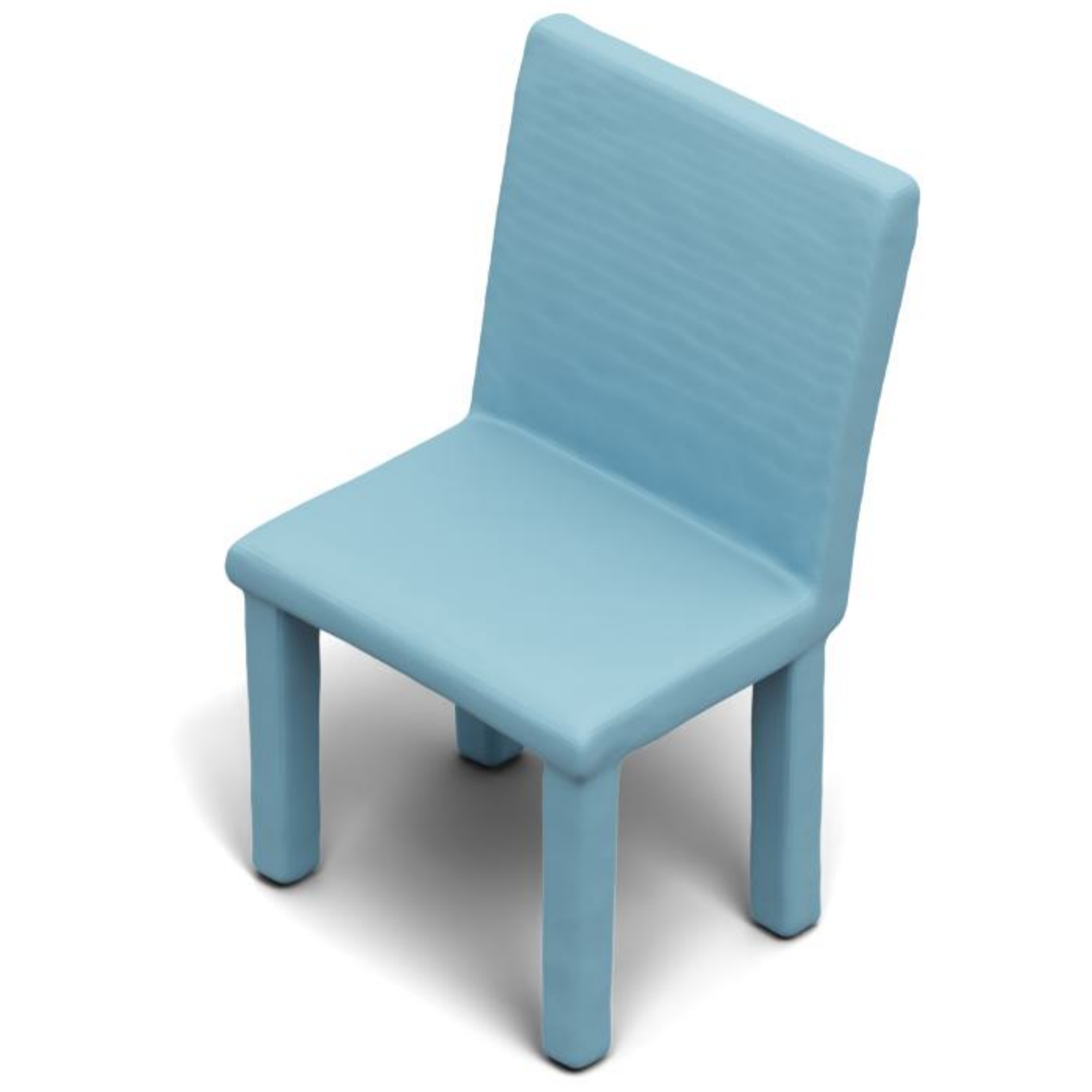}\
    & \includegraphics[trim = 1 1 1 1, clip, width=0.16\linewidth]{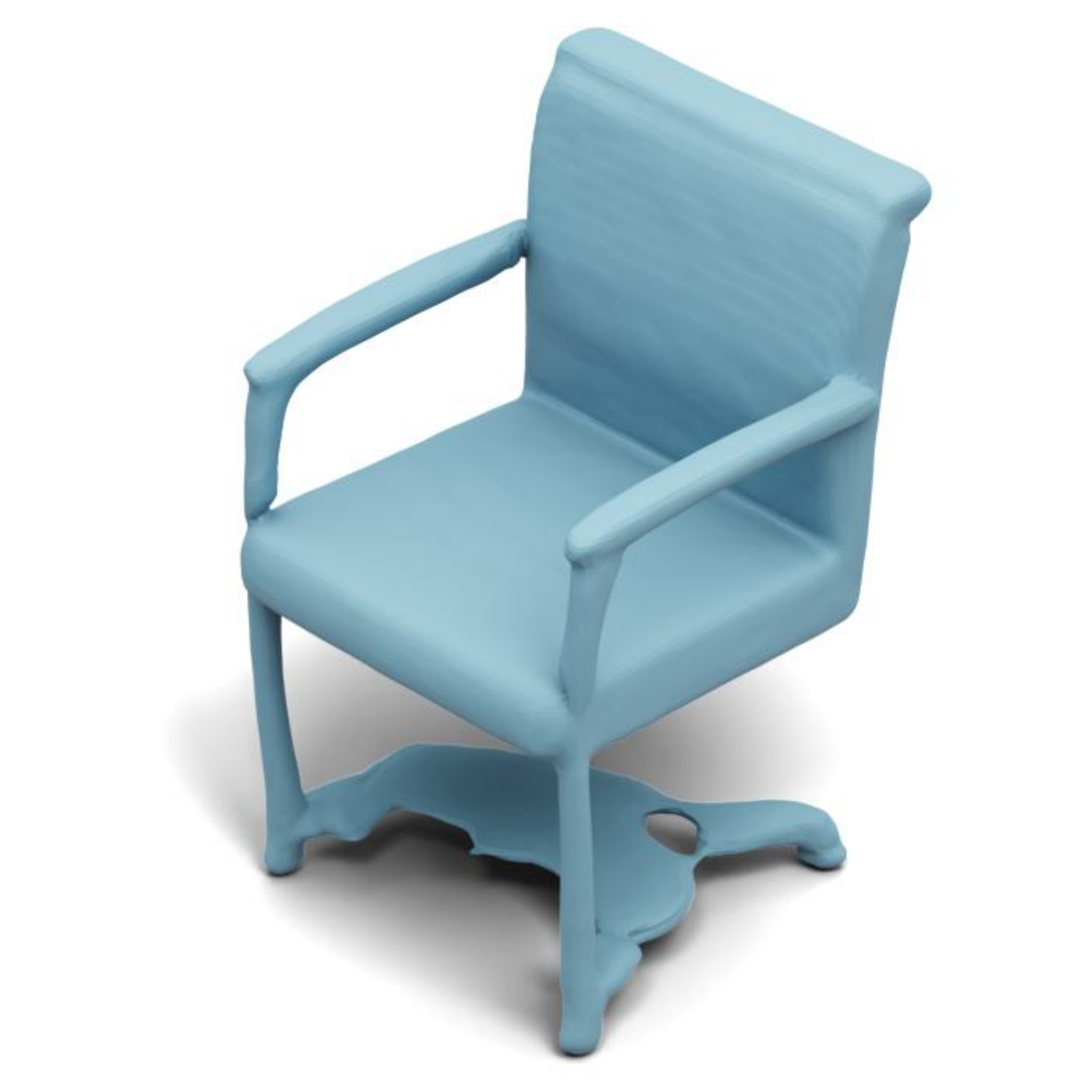}\
    & \includegraphics[trim = 1 1 1 1, clip, width=0.16\linewidth]{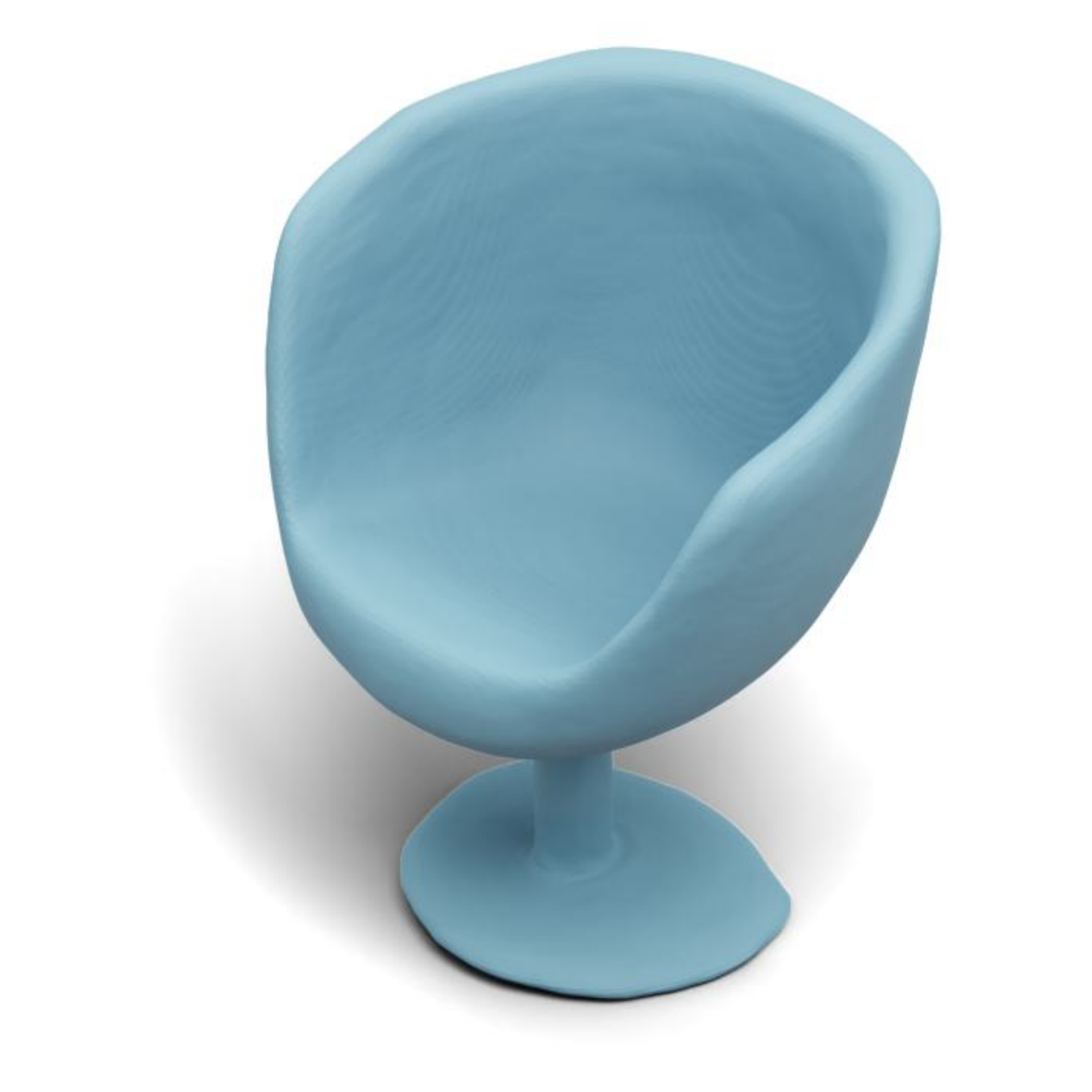}\
    & \includegraphics[trim = 1 1 1 1, clip, width=0.16\linewidth]{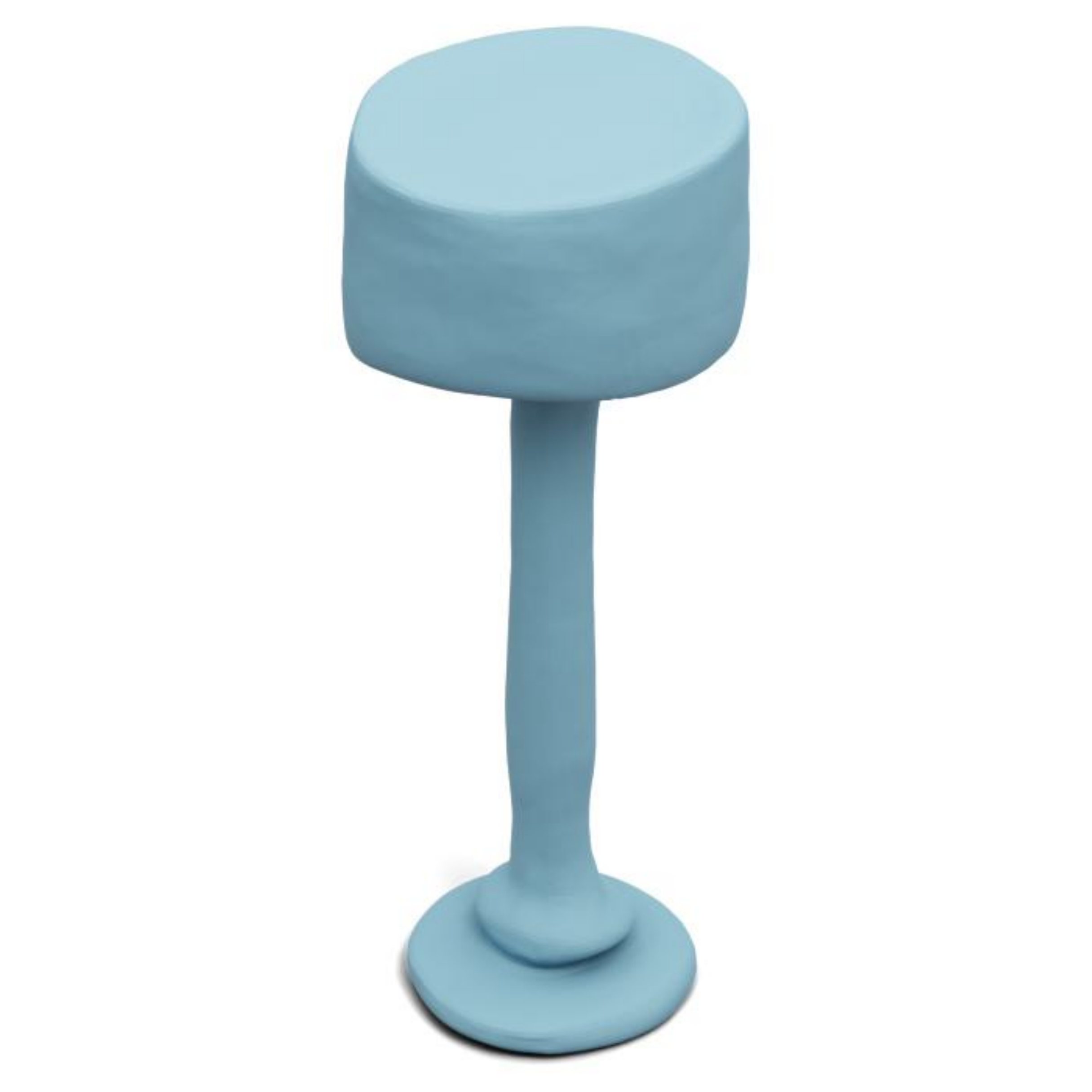} \\

    \includegraphics[width=0.16\linewidth]{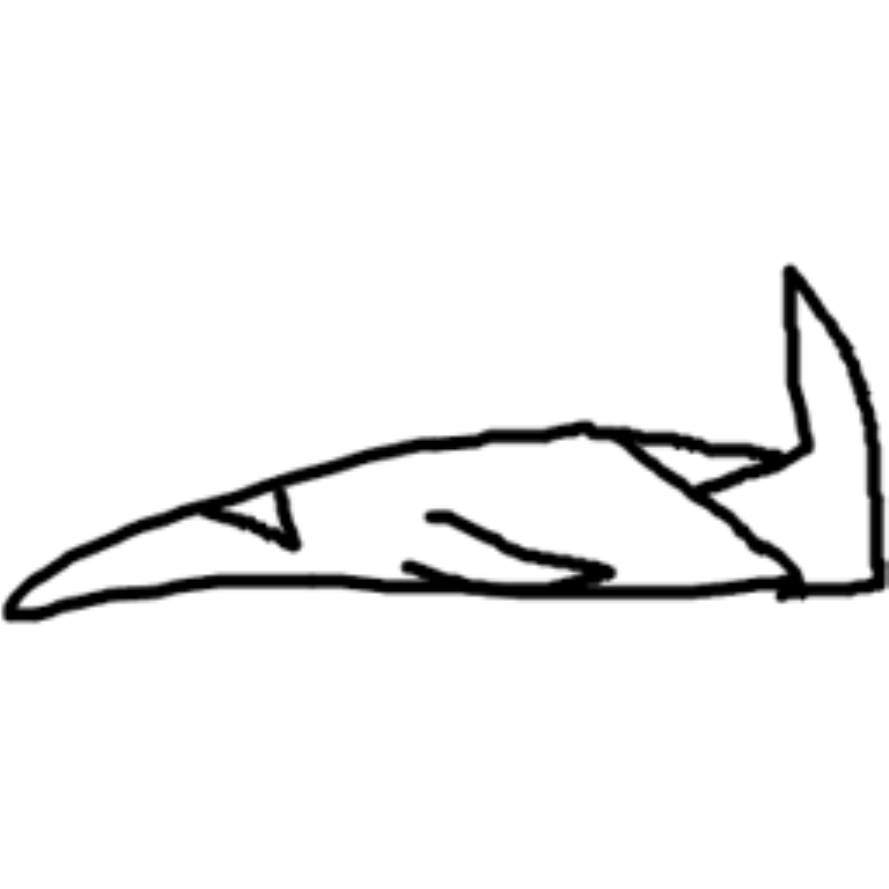}\
    & \includegraphics[width=0.16\linewidth]{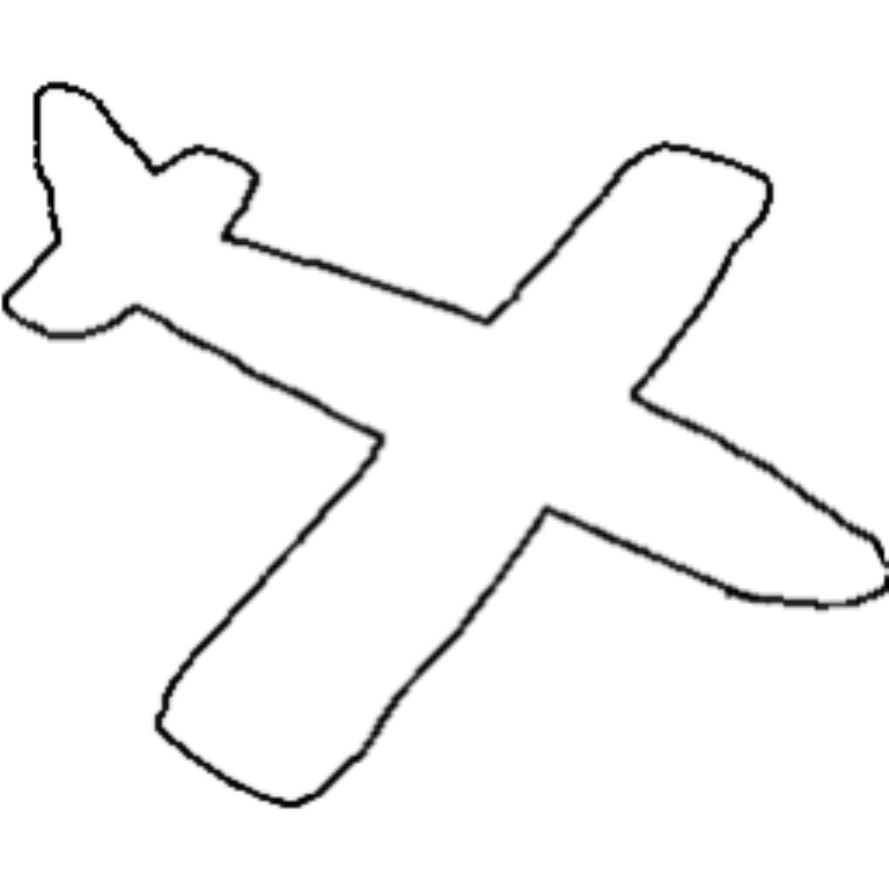}\
    & \includegraphics[width=0.16\linewidth]{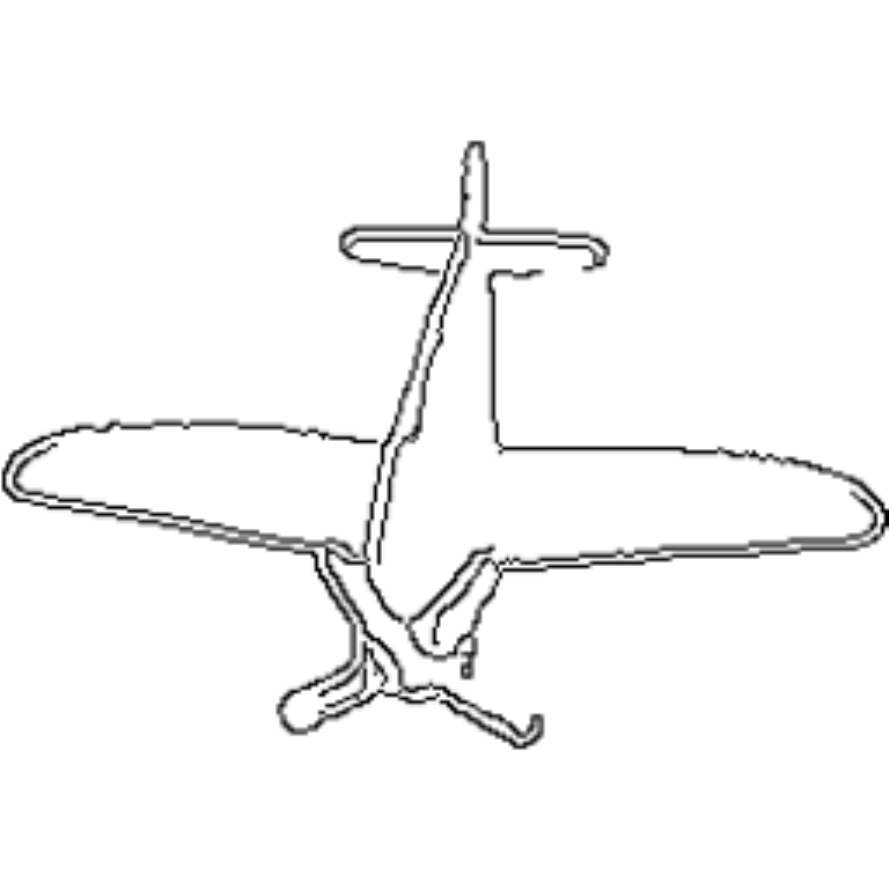}\  
    & \includegraphics[width=0.16\linewidth]{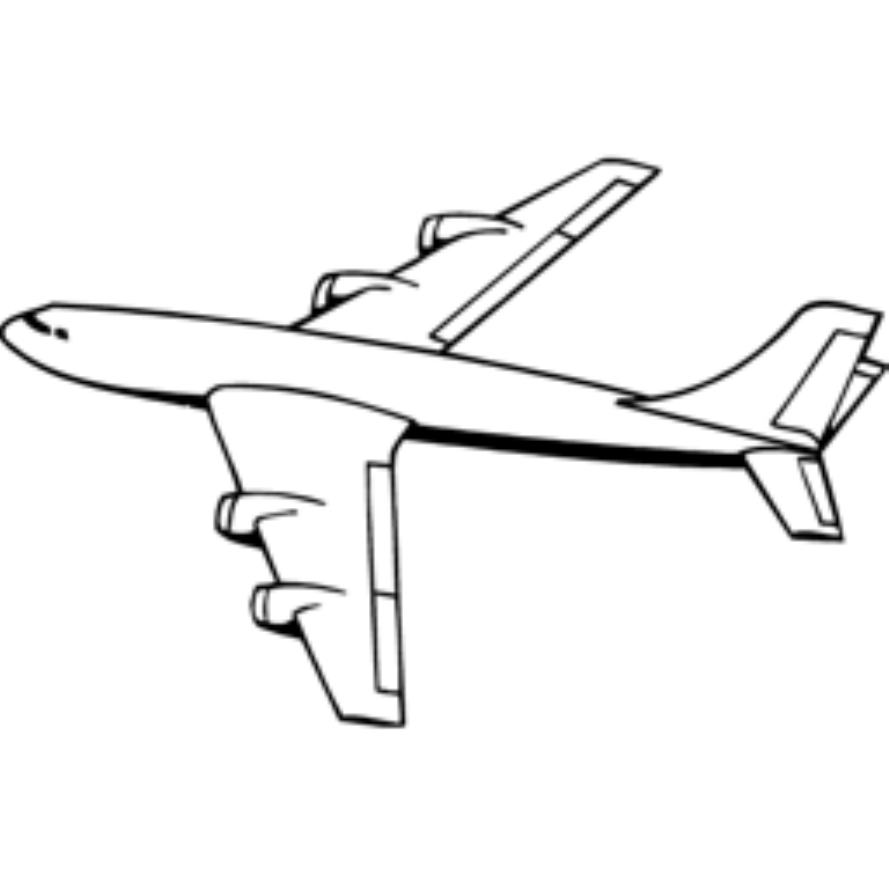}\
    & \includegraphics[width=0.16\linewidth]{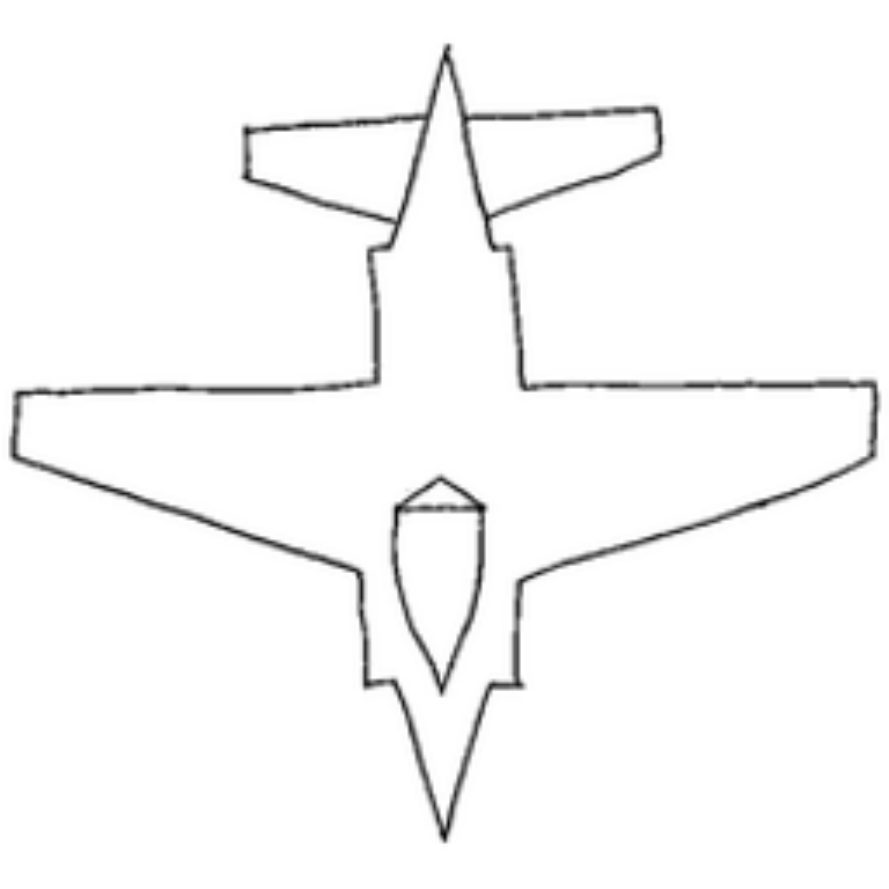}\
    & \includegraphics[width=0.16\linewidth]{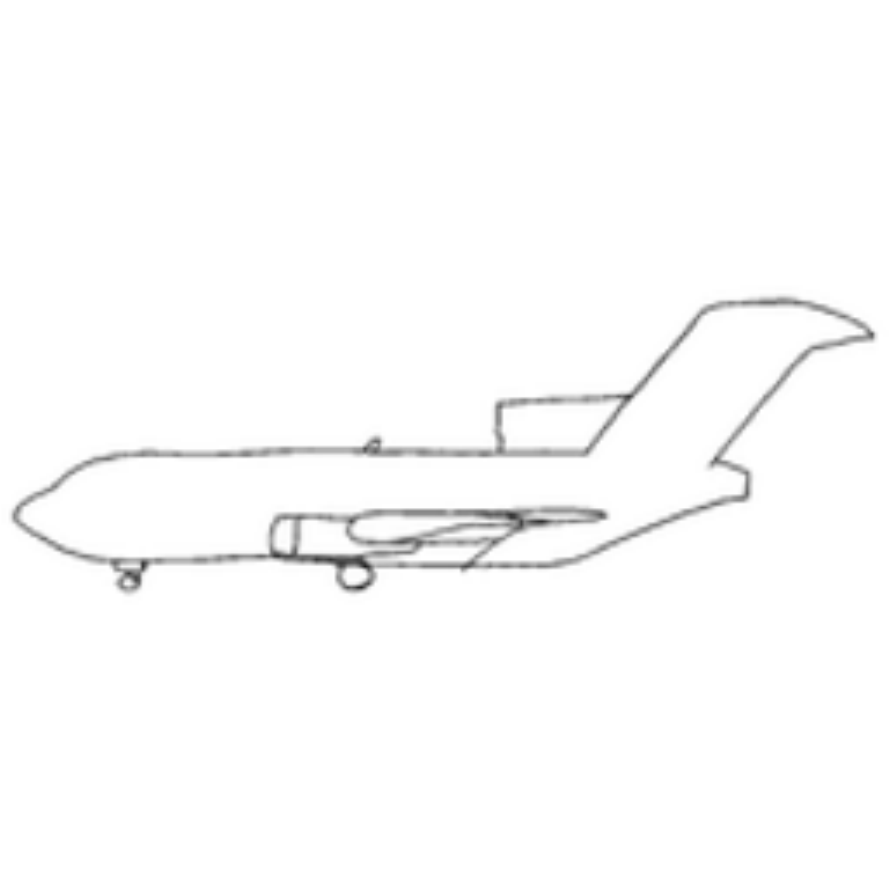} \\


\includegraphics[trim = 1 1 1 1, clip, width=0.16\linewidth]{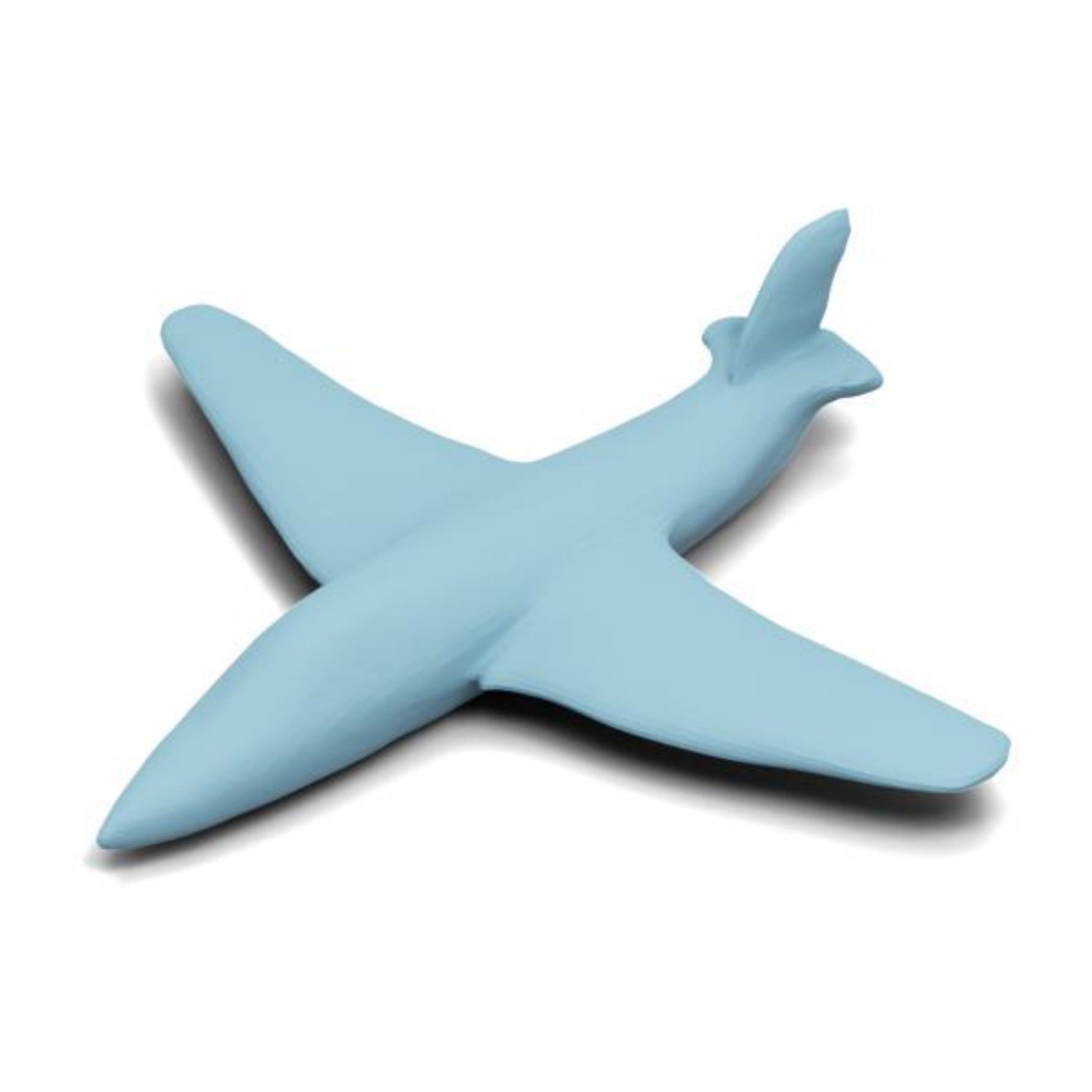}\
    & \includegraphics[trim = 1 1 1 1, clip, width=0.16\linewidth]{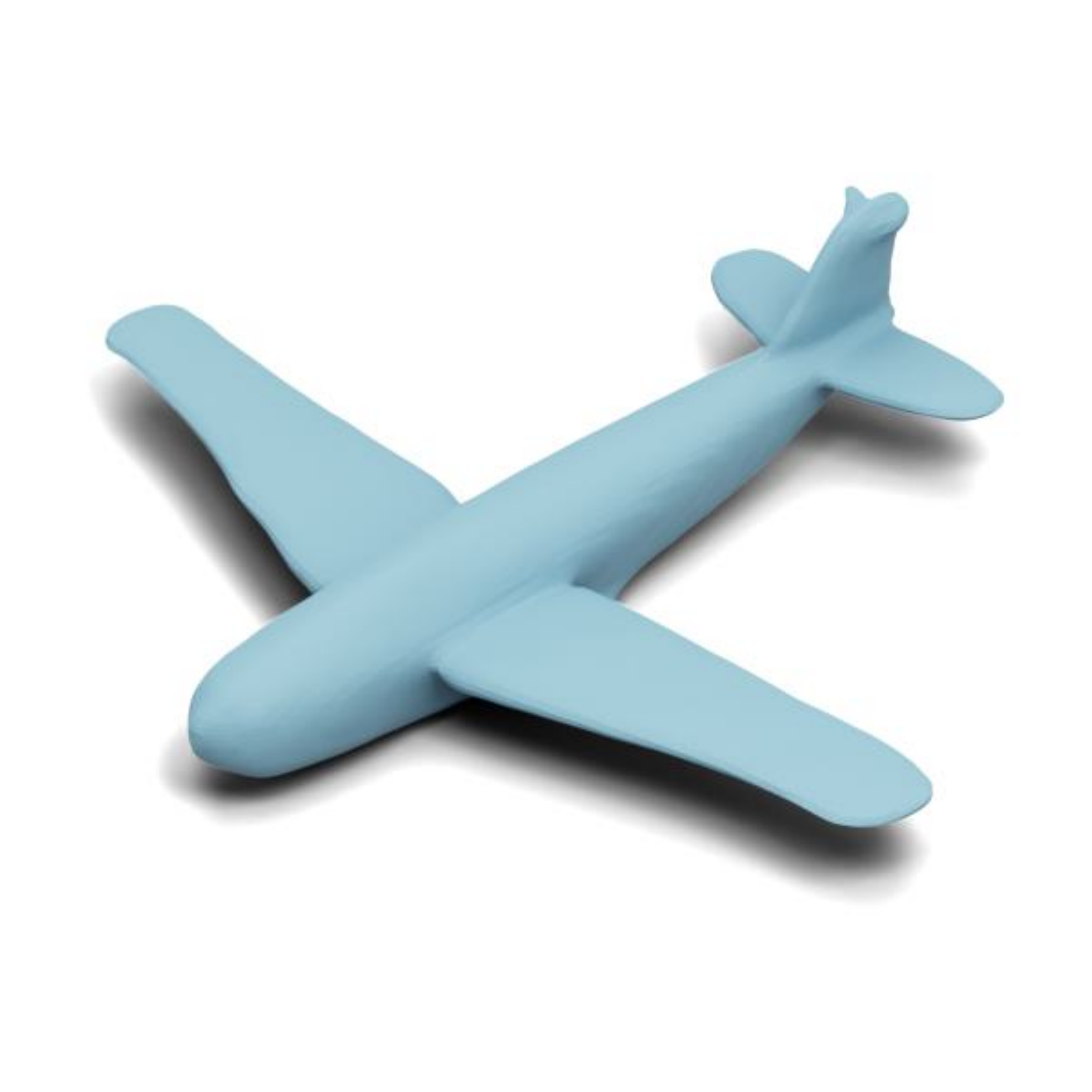}\
    & \includegraphics[trim = 1 1 1 1, clip, width=0.16\linewidth]{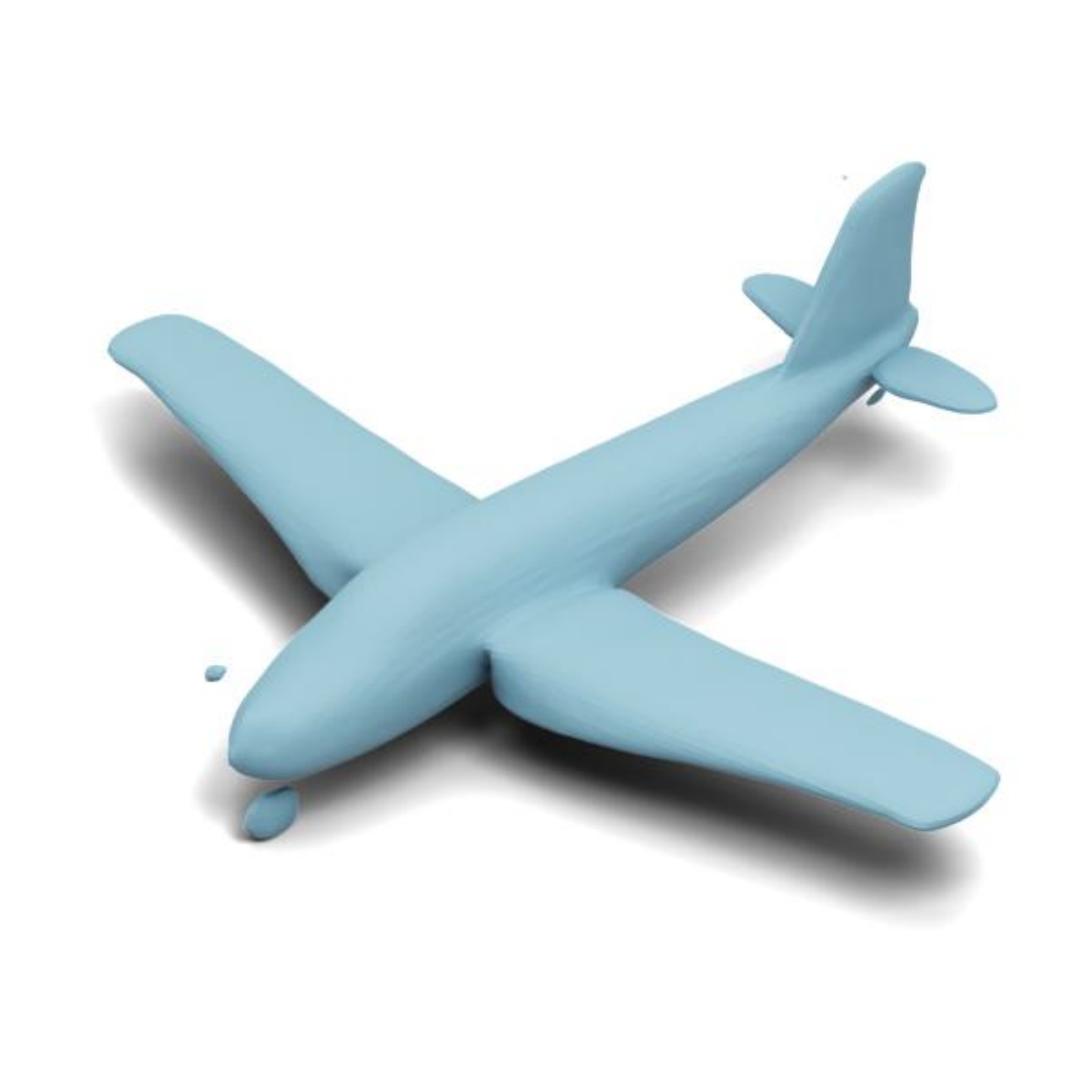}\
    & \includegraphics[trim = 1 1 1 1, clip, width=0.16\linewidth]{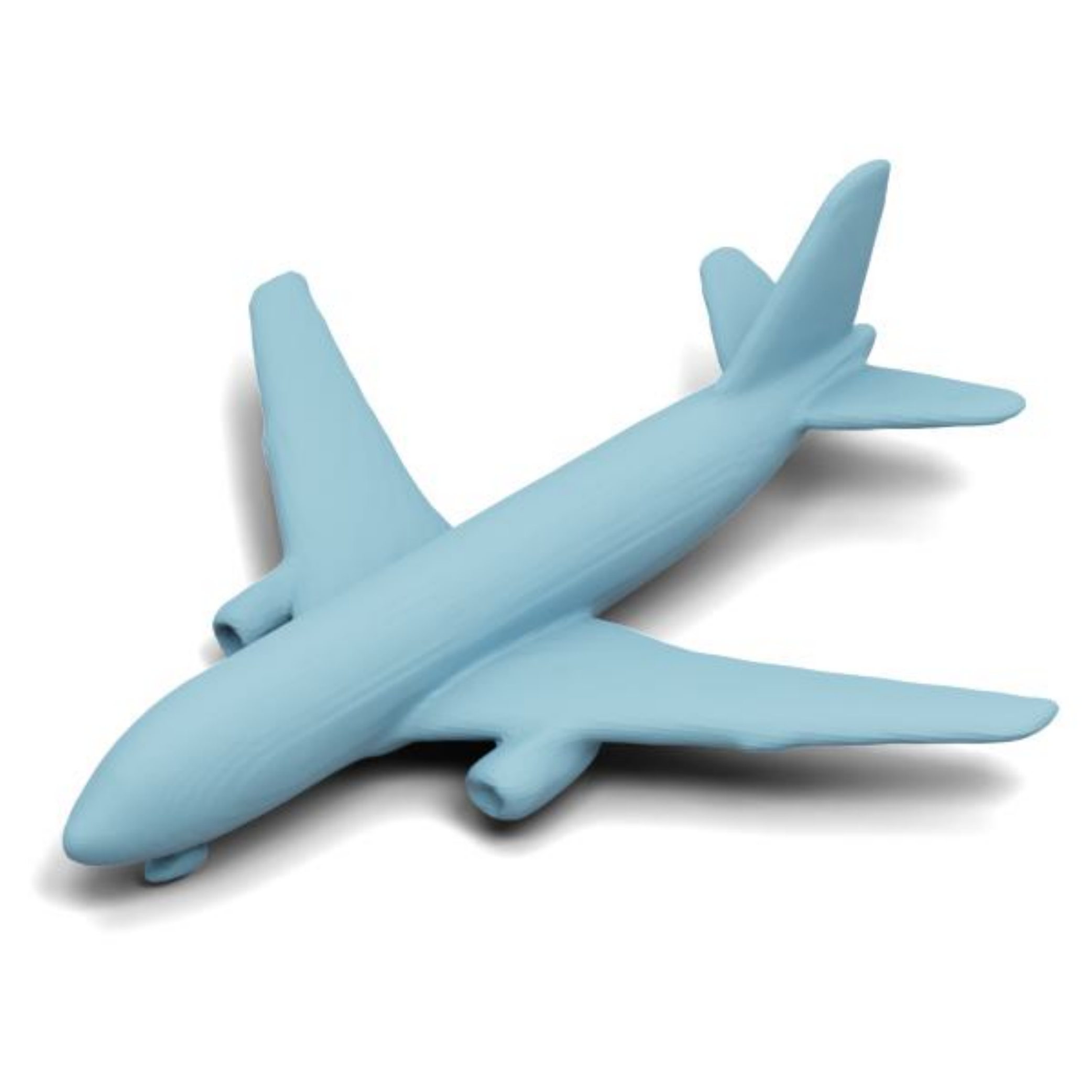}\
    & \includegraphics[trim = 1 1 1 1, clip, width=0.16\linewidth]{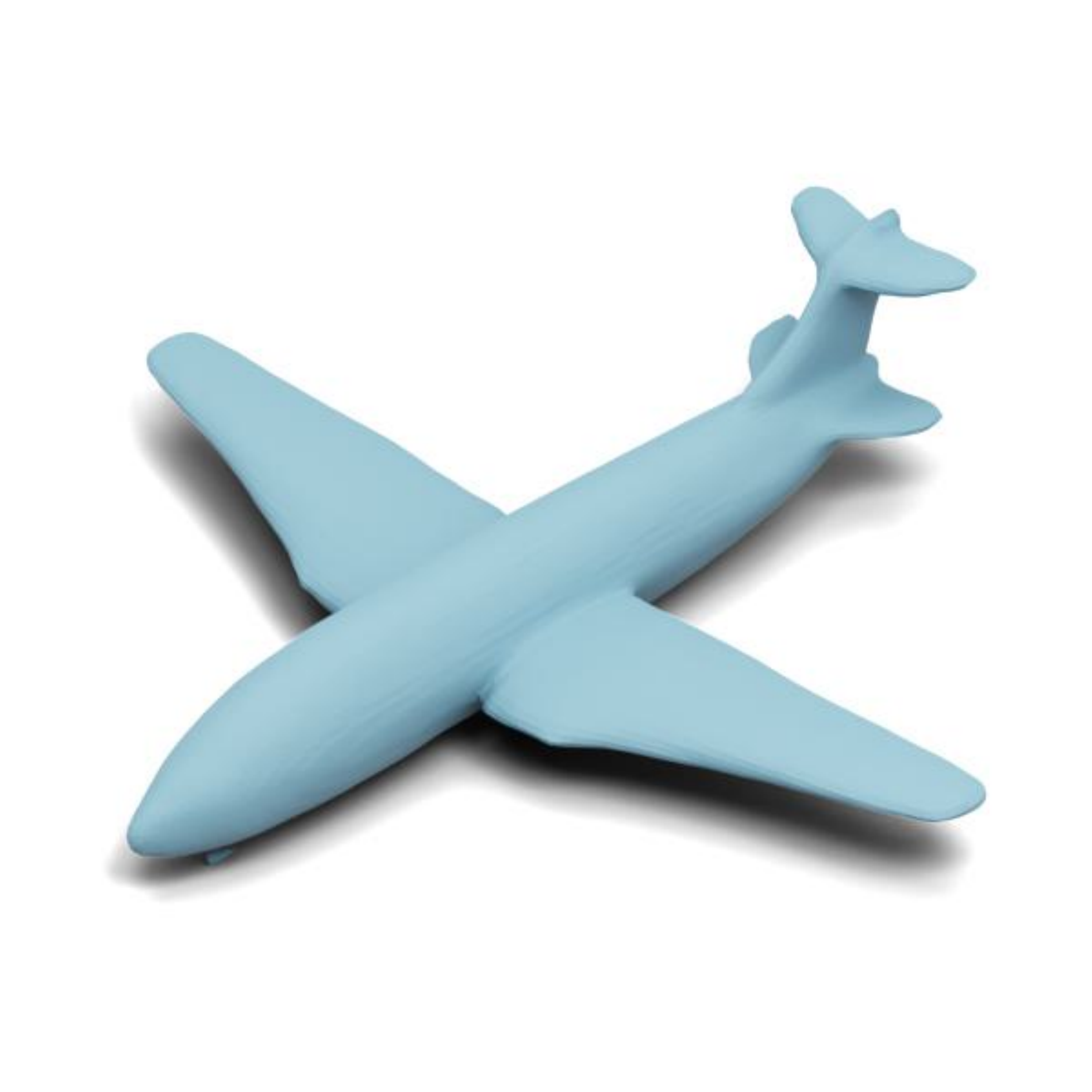}\
    & \includegraphics[trim = 1 1 1 1, clip, width=0.16\linewidth]{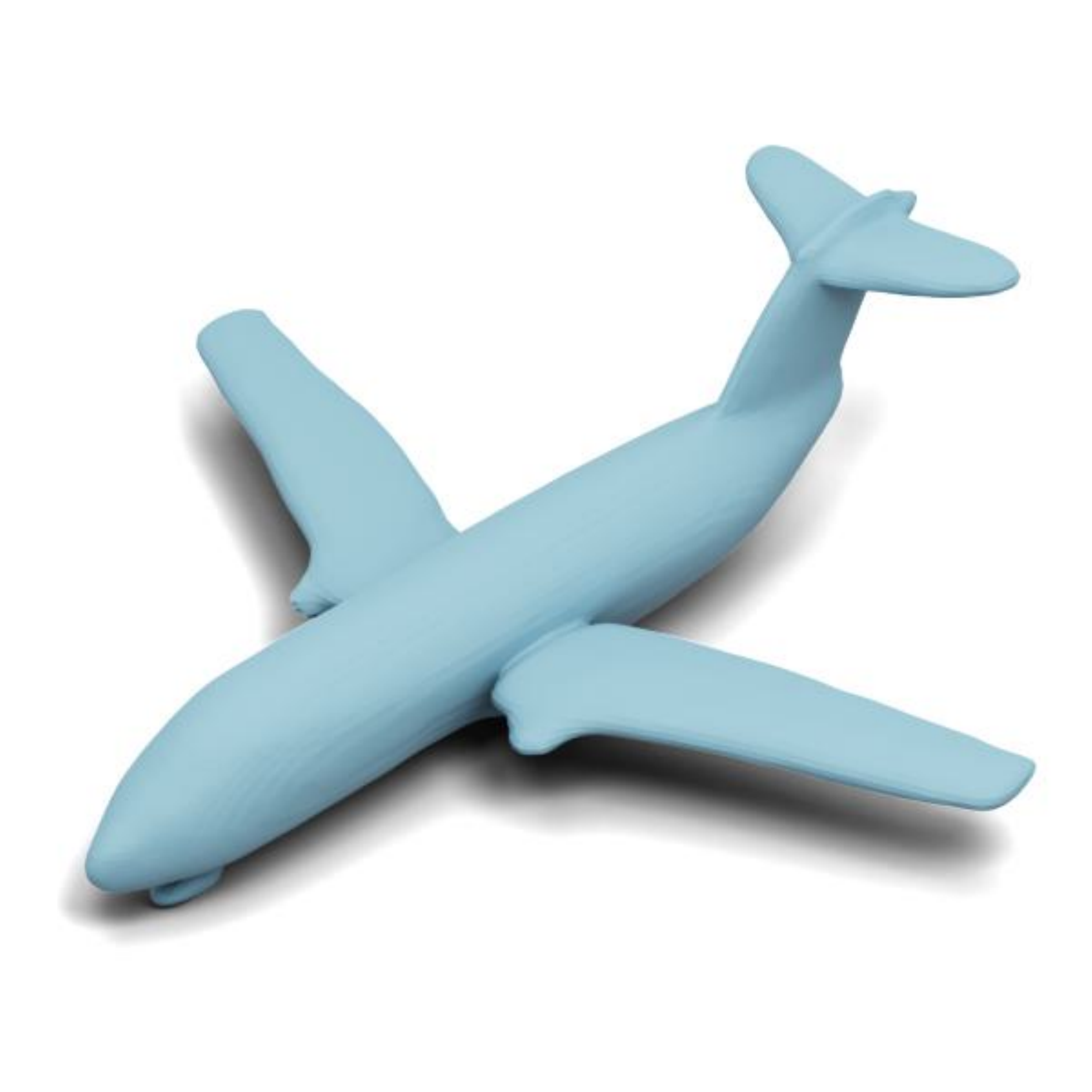} \\

    \includegraphics[width=0.16\linewidth]{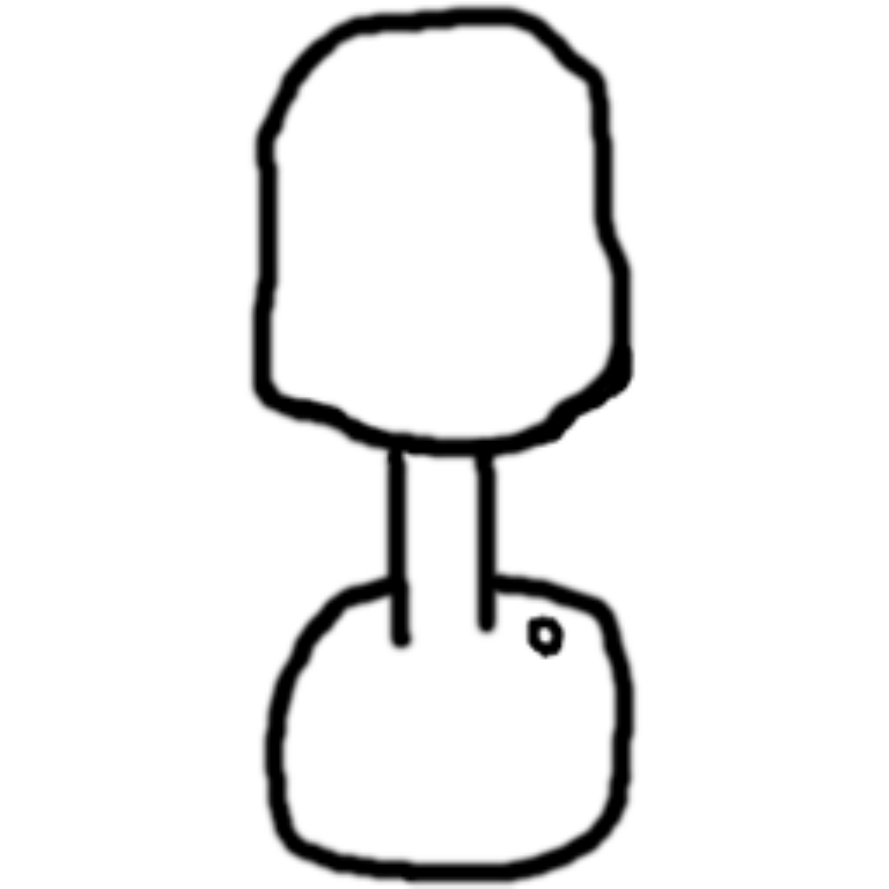}\
    & \includegraphics[width=0.16\linewidth]{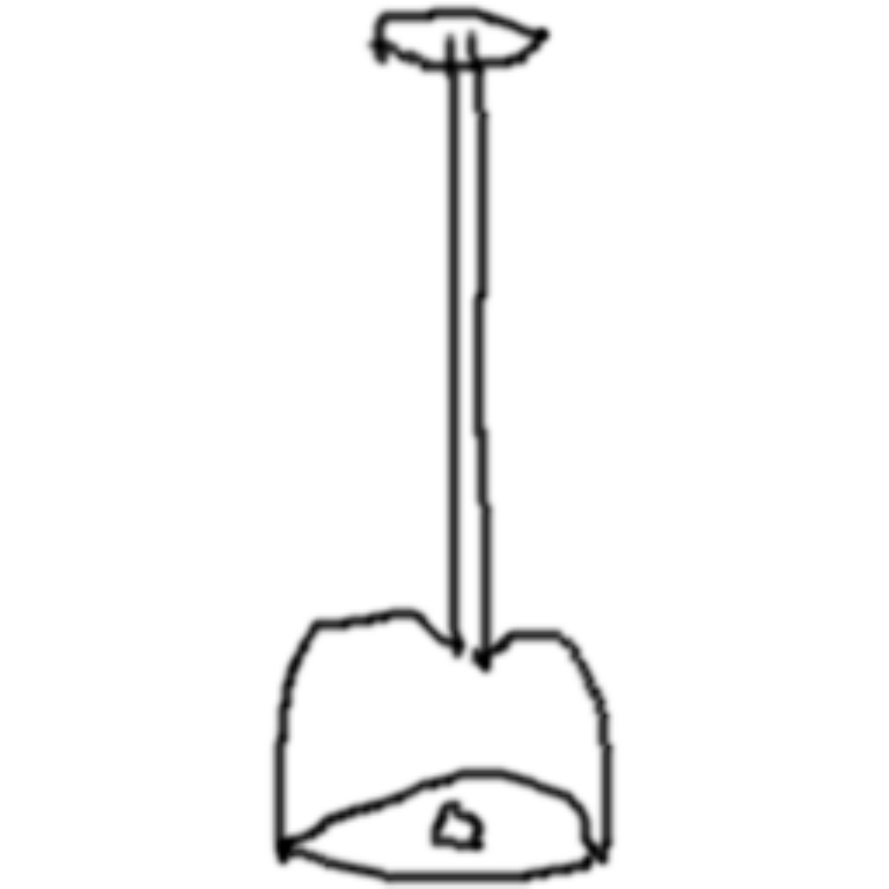}\
    & \includegraphics[width=0.16\linewidth]{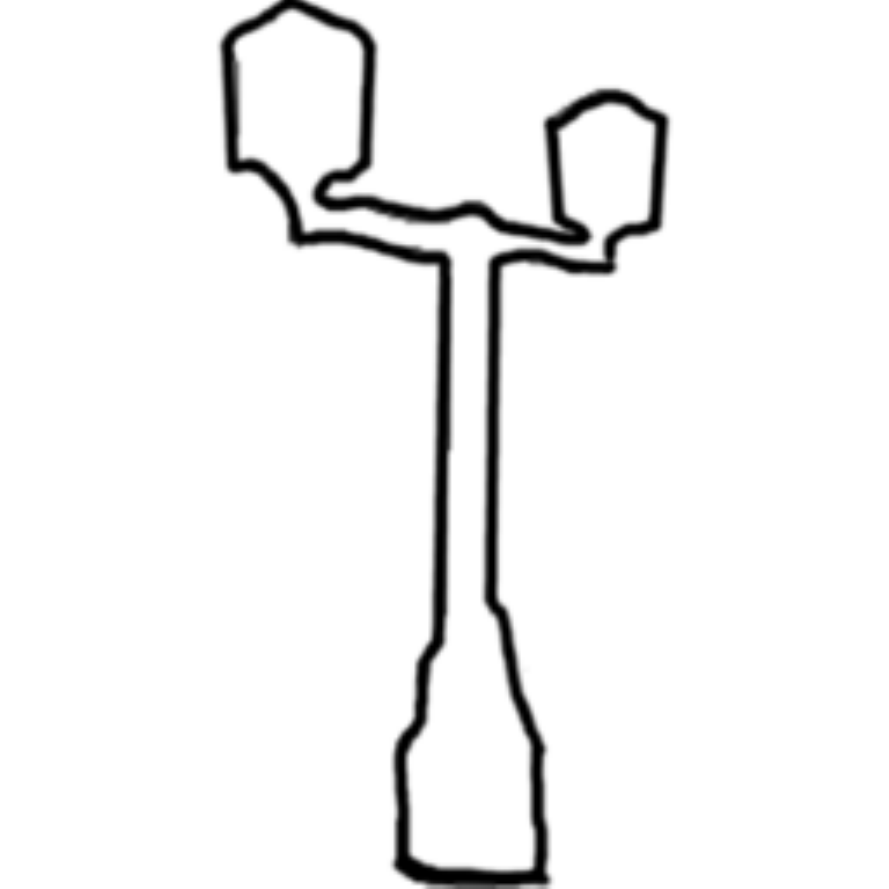}\
    & \includegraphics[width=0.16\linewidth]{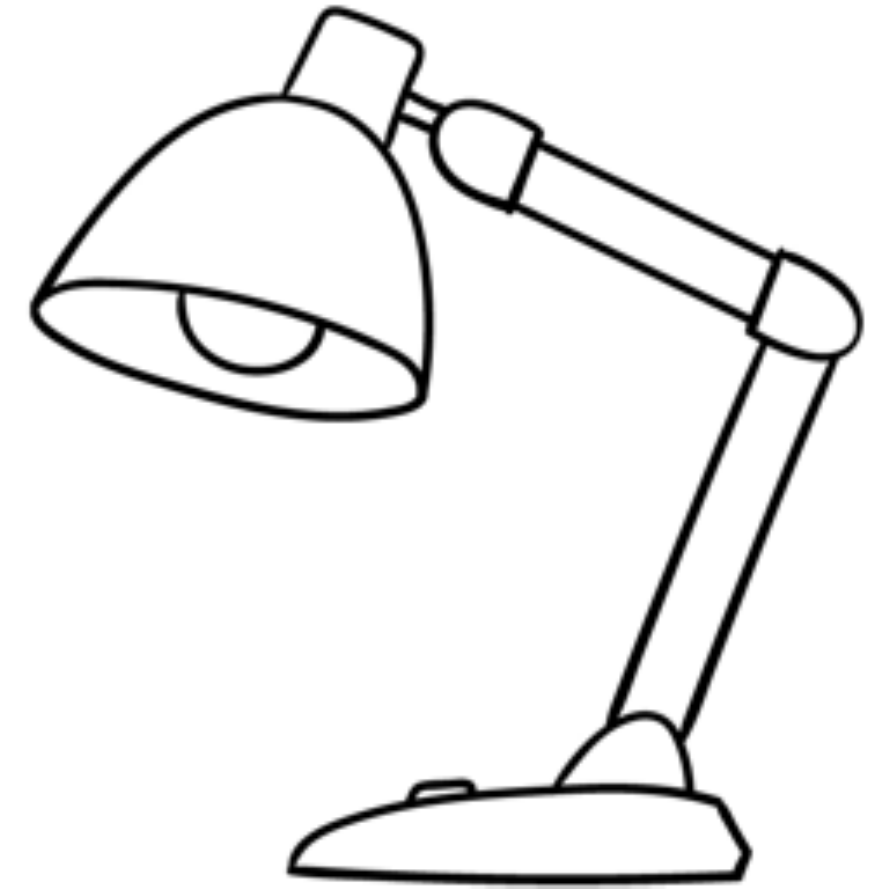}\
    & \includegraphics[width=0.16\linewidth]{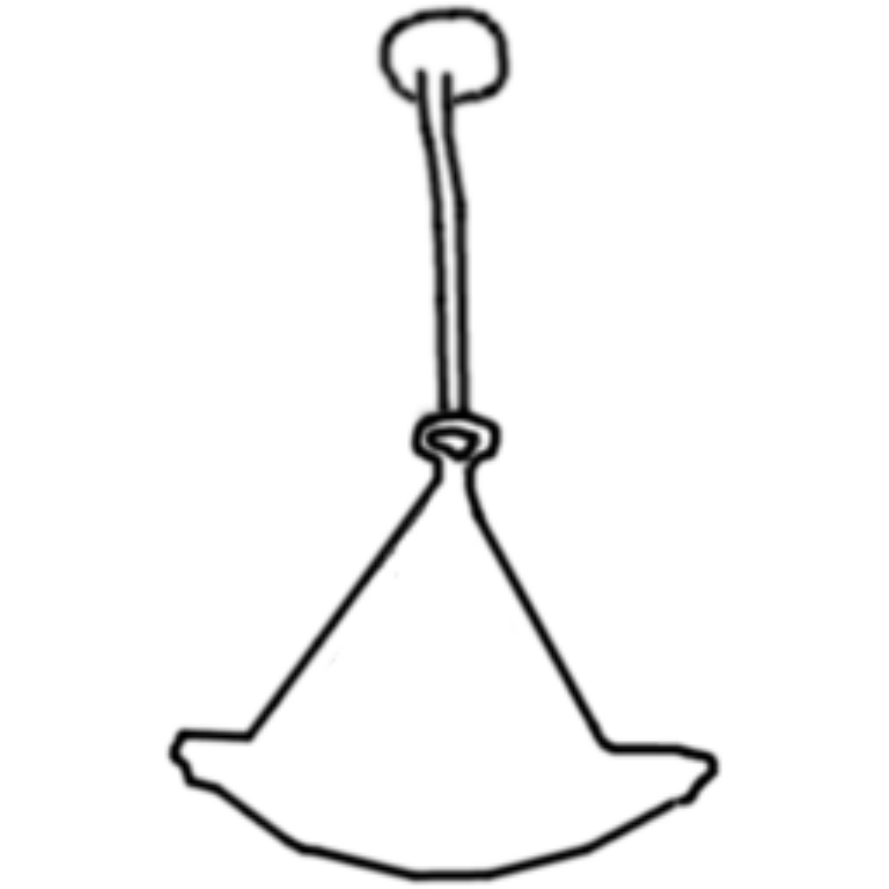}\
    & \includegraphics[width=0.16\linewidth]{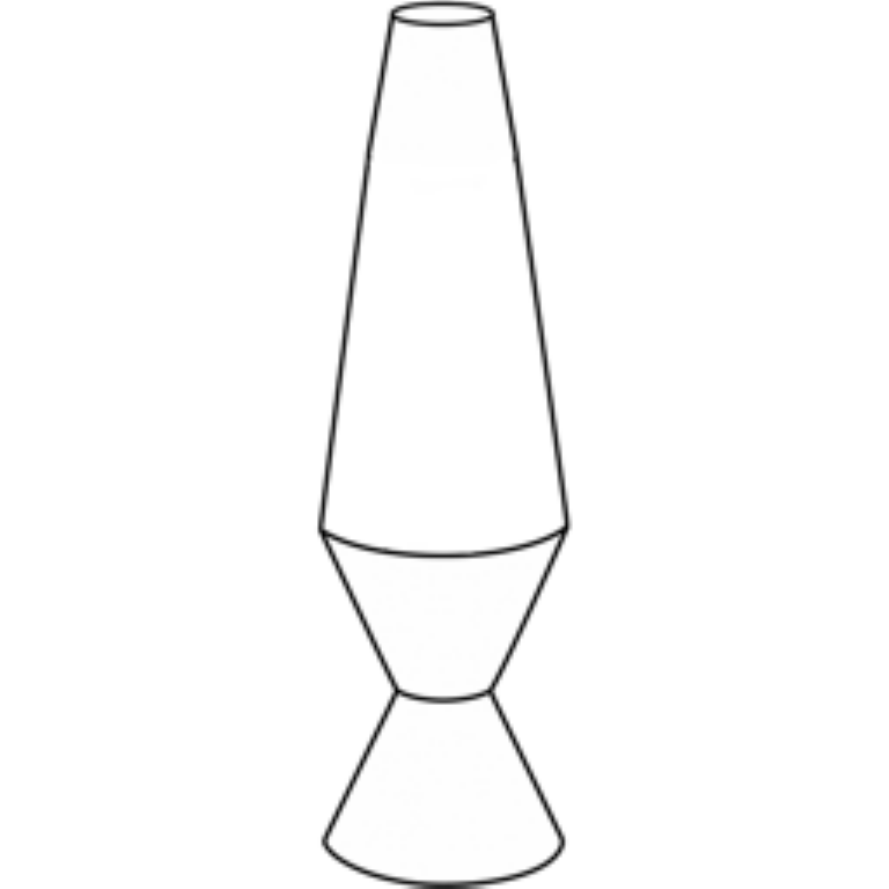} \\


\includegraphics[trim = 1 1 1 1, clip, width=0.16\linewidth]{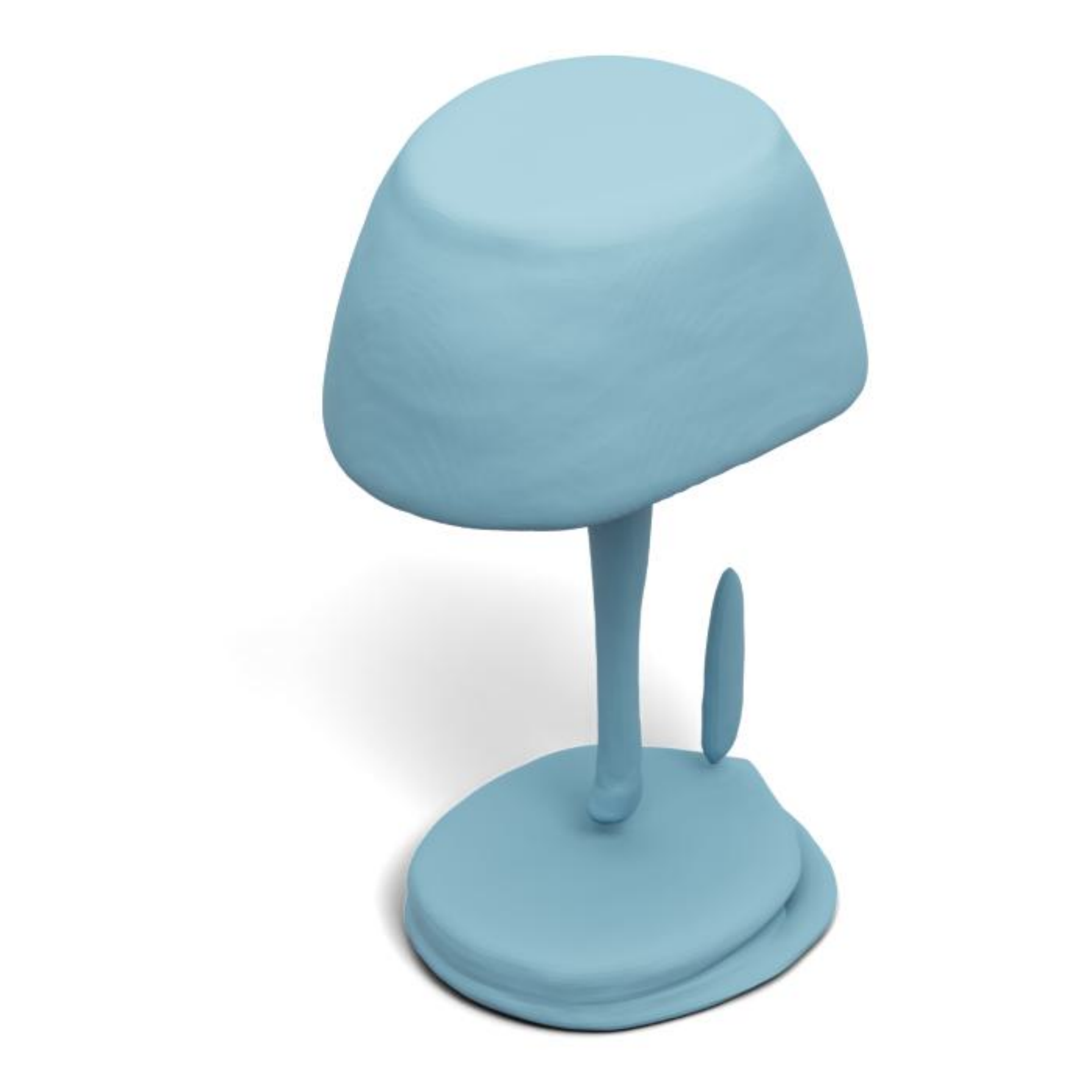}\
    & \includegraphics[trim = 1 1 1 1, clip, width=0.16\linewidth]{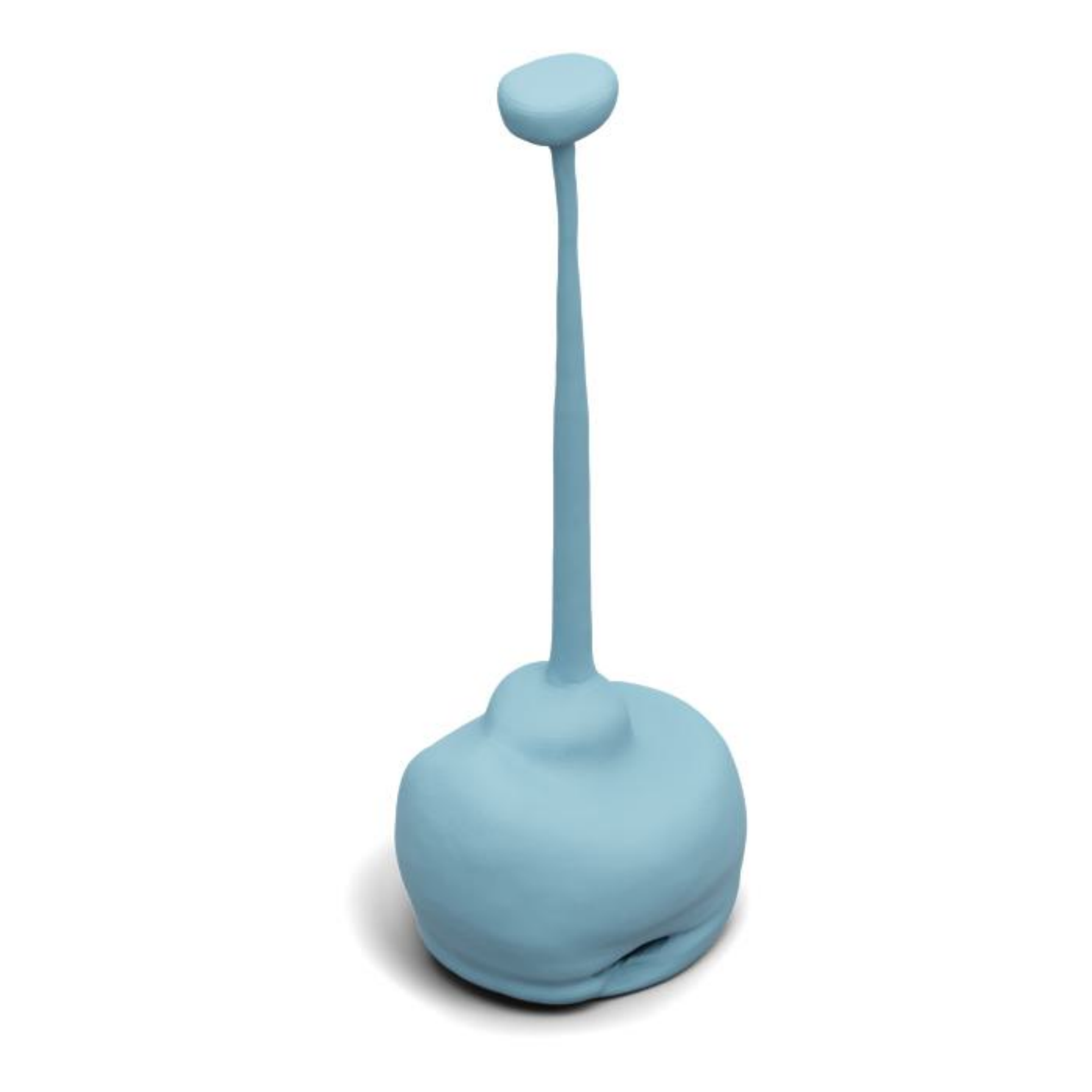}\
    & \includegraphics[trim = 1 1 1 1, clip, width=0.16\linewidth]{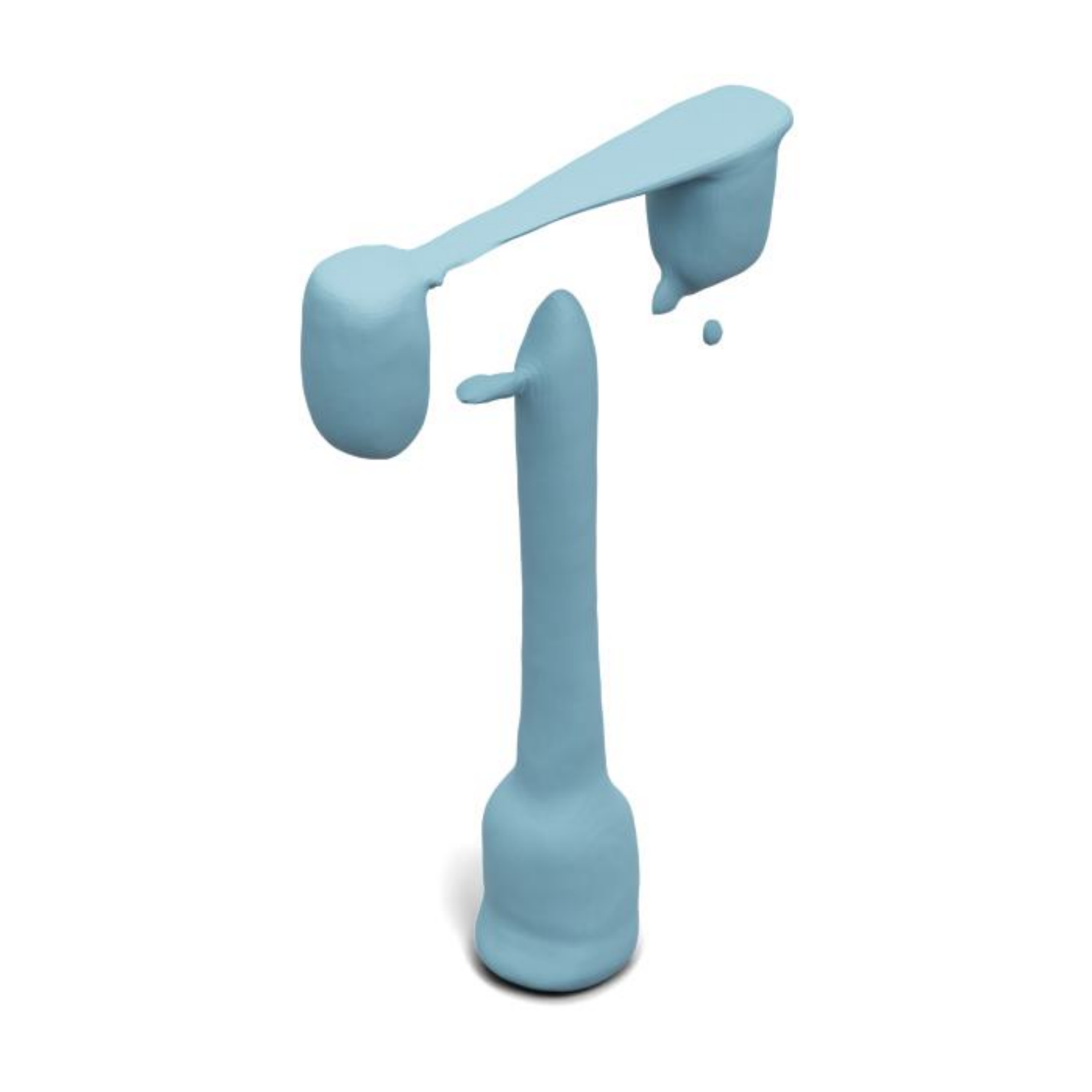}\
    & \includegraphics[trim = 1 1 1 1, clip, width=0.16\linewidth]{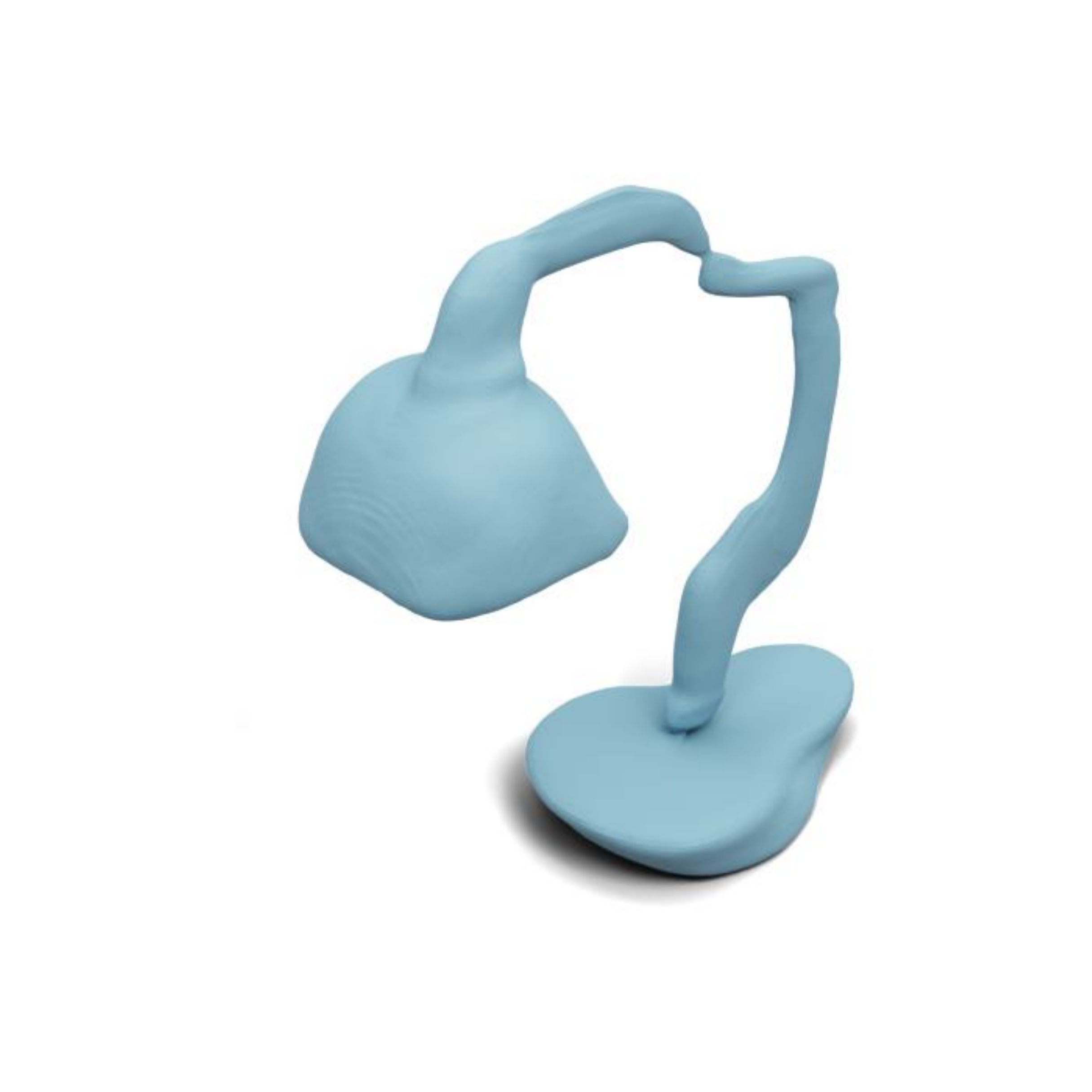}\
    & \includegraphics[trim = 1 1 1 1, clip, width=0.16\linewidth]{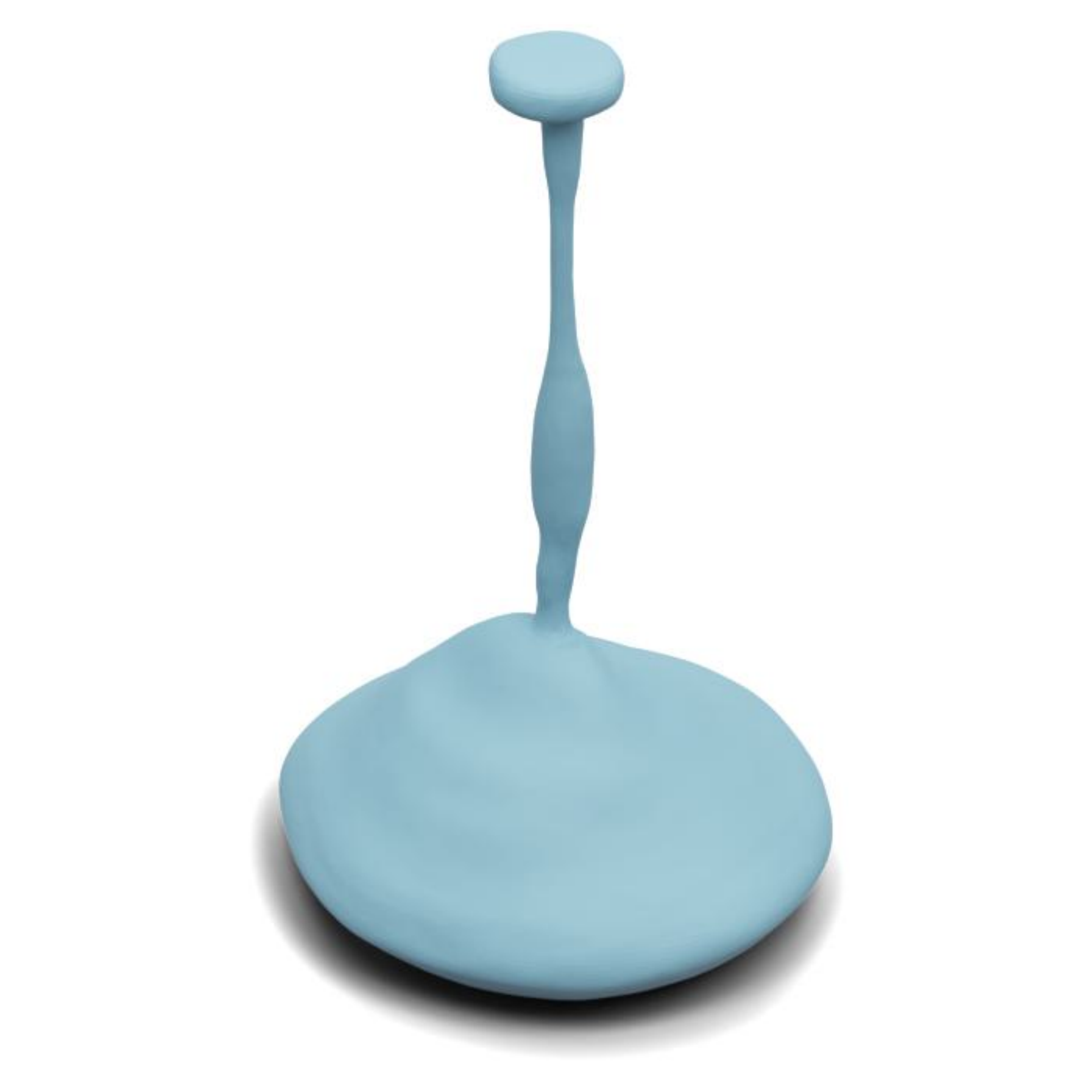}\
    & \includegraphics[trim = 1 1 1 1, clip, width=0.16\linewidth]{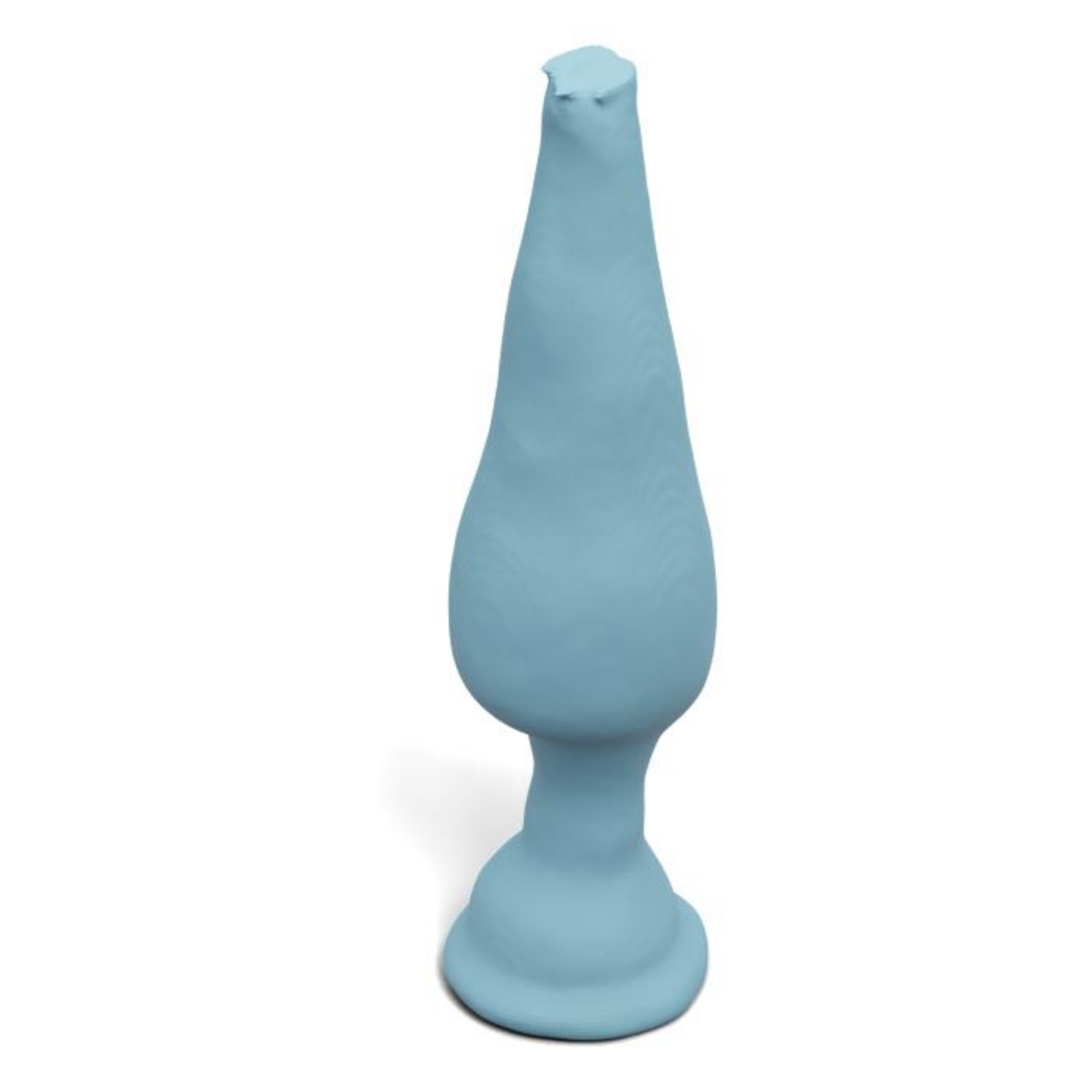} \\
    
	\end{tabular}
	\caption{\textbf{Multi-class \ourmethod{}} can produce chairs, planes and lamps out of sketches at diverse abstraction levels. Note that we do not indicate to the network at inference the kind of object we draw. In some cases, this can lead \ourmethod{} to misinterpret the class of the drawn shape (see top-right).}
	\label{fig:bigfiguremulti}
\end{figure*}

Single-view sketch-to-shape reconstruction is a challenging problem as it requires overcoming necessary ambiguities. \ourmethod{} tackles this by conditioning the network on a limited number of classes. Yet, it might struggle to produce a shape that corresponds to the input sketch if it cannot resolve these ambiguities. \figref{fig:limitations} shows such limitations. \figref{fig:limitations}(i) exhibits that \ourmethod{} might omit, deform, or add additional details that were not required by the user. This problem also appears in the airplane's tail in \figref{fig:comparisondoubleview}. Yet, \figref{fig:completion} demonstrates that tackling this ambiguity can benefit the consistency of the result. This often comes down to a trade-off between being close to the input or producing a coherent shape.
\figref{fig:limitations}(ii) shows that although the sketch may be drawn with precision, the final shape may not include the high-frequency details or patterns depicted in the sketch. \new{Such challenges can be attributed to our method's handling of sketches with varying abstraction levels, inherent limitations in the SPAGHETTI shape decoder's detail rendering capabilities, and the absence of view parameters to guide the generation process; factors that collectively impact the method's ability to deal with intricate details.}
As we condition \ourmethod{} to limited classes of shapes, the output is restricted to an object of such a class, even when the input sketch is unrelated. \figref{fig:limitations}(iii) presents such direct example. Also, the stool sketch in the middle row of \figref{fig:bigfigure} is not correctly mapped. Multi-class \ourmethod{} is subject to misinterpretation of the shape category, as exemplified in the top right corner of \figref{fig:bigfiguremulti}.
Finally, \ourmethod{} inherits some of SPAGHETTI's limitations, such as the necessity of training on a limited number of shape classes with similar structures, similar artifacts and lack of fine detail in the generated shapes, and potential under or over-clustering of parts within the same Gaussian, which restricts the desired level of control permitted by our selection tool for refinement or part-based modeling.

\section{Conclusion} \label{sec:conclusion}

In this paper, we present \ourmethod{}, a method for generating neural implicit shapes through sketching. 
The key concept of our approach is mapping different parts of the input sketch to a part-aware latent space.
Each latent code's part is consistently mapped to a different part of the generated shape.
Our part-aware reconstruction approach allows the network to integrate the relationships between different parts of the object, resulting in 3D models that are less prone to mere shape retrieval from the training dataset. In addition, we also offer part-based shape modeling, where users can select a part of a shape and redraw its corresponding sketch. This allows for even more precise model editing, and enables users to combine features from different shapes, thus expanding the scope of what can be modeled beyond the dataset's inherent limitations. Another implication of a part-aware latent space is the possibility to refine specific parts of the shape, hence allowing systematic artifacts removal in the final model. \new{Recent developments in generative diffusion-based models have shown promise for sketch-to-shape modeling, as highlighted in works like \cite{zheng2023lasdiffusion}. These models, when combined with part-aware shape decoders \cite{bandyopadhyay2023doodle}, offer new potential for advancing the field. This integration not only enhances current methodologies but also paves the way for innovative research directions in sketch-based shape generation.}

Among the key contributions of our method also lies the ability to generate shapes via a single sketch at various levels of abstraction. Moreover, we can edit their outline directly through sketching, reducing the need for advanced artistic skills in the modeling process.
We have shown through our experiments and comparisons with prior shape generation methods that \ourmethod{} generates models with a higher level of detail and realism while requiring less drawing expertise. We believe that our method provides a powerful tool for creating 3D models, offering both ease of use and high-quality results.

\section*{Acknowledgements}

\new{We thank the reviewers for their insightful and constructive comments. We use Silvia Sellán's Blender template for rendering. This work was supported in part by the European Research Council (ERC) under the European Union’s Horizon 2020 research and innovation program (grant agreement No.\ 101003104, ERC CoG MYCLOTH).}

\bibliographystyle{eg-alpha-doi} 
\bibliography{references}    
\newpage

\clearpage
\newpage

\end{document}


\maketitle

\begin{abstract}
  We provide more details related to data preparation, implementation, training and evaluation of our method.
\end{abstract}

\section{Network Architecture}

The network is composed of three parts: a Vision Transformer encoder, a Transformer decoder and an implicit shape decoder (SPAGHETTI). The Vision Transformer encoder consists in a "sketch to visual embeddings" Transformer encoder. It takes as input a $256\times 256$ grayscale image, decomposes it into $256$ patches of size $16 \times 16$, uses a learnable position encoding, and maps each patch to a visual embedding of dimension $h_d = 512$. The Vision Transformer itself consists in $8$ layers intertwining multi-head attention layers and feed-forward networks with layer normalization \cite{ViTDosovitskiy2020}. Then, we use a \emph{Transformer decoder} as our "visual embedding to shape latent code" network. It maps the $256$ visual embeddings to latent space code. The latent space code is composed of $m$ vectors of dimensions $d_{\text{model}}$. Single-class \ourmethod{} uses $m = 16$ and $d_{\text{model}} = 512$, while multi-class \ourmethod{} uses $m = 32$ and $d_{\text{model}} = 768$. The Transformer decoder also takes as input $m$ \emph{learnable} part queries of dimension $1.5h_d$ that are optimized simultaneously with the weights of the network. It is composed of $12$ cross-attention layers and feed-forward networks with layer normalization. The output of the Transformer decoder is then mapped to the latent code $z_h$ of the shape decoder latent space via an MLP with ReLU activation.
\section{Training}

\new{Single-class m}odels are trained on an \new{Nvidia} RTX 3090 GPU for 850 epochs. We use a gradual warmup scheduler \cite{GradualWarmupScheduler2017} to linearly increase the learning rate at each epoch. The learning rate starts at $10^{-7}$ and linearly increases to $10^{-6}$. 
\new{Our approach to training the multi-class model was based on a combined dataset from various classes, namely chairs, planes, and lamps. We include ShapeNet outline and partial outline renderings, as well as CLIPasso \cite{vinker2022clipasso} abstract sketches, and ProSketch chair sketches \cite{zhong2021prosketch}. The training was based on 630 epochs, and the training duration for the multi-class model was 96 hours, which is longer than the 60 hours required for the single-class model due to the increased amount of data per epoch. The same learning rate and scheduler were used.}
\section{Evaluation}

Our evaluation is performed on the AmateurSketch dataset \cite{Qi2021AmateurSketch}, which contains 3000 freehand sketches of ShapeNet shapes \cite{shapenet2015chang} of medium abstraction level. We only compare with the chair class, because this is the only class ubiquitously supported by all the methods we compare with.

\begin{table}[htbp]
\caption{Performance comparison of shape reconstruction methods on the AmateurSketch dataset \cite{Qi2021AmateurSketch} using chamfer distance (CD), earth mover's distance (EMD), and Fréchet inception distance (FID). Lower values indicate better performance. Comparison is done with Pixel2Mesh \cite{pixel2mesh}, Sketch2Mesh \cite{Sketch2Mesh}, and DeepSketch \cite{10.1016/j.cag.2022.06.005}. The notions ``cropped'' and ``padded'' refer to the differences in input normalization. DeepSketch results are shown with the network trained with their default training data and re-trained with our training data.}
\label{tab:distances}
\centering
\begin{tabular}{lccc}
\toprule
    Method  & CD$\downarrow$ & EMD$\downarrow$ & FID$\downarrow$ \\
\midrule
Pixel2Mesh & 0.2191 & 0.1658 & 401.7 \\
Sketch2Mesh (padded input) & 0.2113 & 0.1573 & 368.4 \\
Sketch2Mesh (cropped input) & 0.2325 & 0.1635 & 305.8 \\
DeepSketch (default dataset) & 0.1520 & 0.1142 &  292.2 \\
DeepSketch (our dataset) & 0.1920 & 0.1417 & 317.4 \\
SENS & \textbf{0.1186} & \textbf{0.0946} & \textbf{171.3}\\
\bottomrule
\end{tabular}
\end{table}

\subsection{Objective evaluation}

Our quantitative evaluation is based on several metrics. We compare our results with different methods: Pixel2Mesh \cite{pixel2mesh}, Sketch2Mesh \cite{Sketch2Mesh} and DeepSketch \cite{10.1016/j.cag.2022.06.005}. The comparison results are shown in \tableref{tab:distances}.

\subsubsection{Chamfer distance (CD)} The chamfer distance calculates the average distance between each point in one set to its closest point in the other set and is an intuitive way to quantify the dissimilarity between two point clouds. It is thus widely used for geometric comparison. The chamfer distance between two point sets $A$ and $B$ can be defined as follows:
$$d_{\text{chamfer}}(A, B) = \frac{1}{|A|} \sum_{a \in A} \min_{b \in B} \|a - b\|^2 + \frac{1}{|B|} \sum_{b \in B} \min_{a \in A} \|a - b\|^2.
$$

For each sketch in the AmateurSketch dataset, we extract a mesh from the implicit shape produced by our network. Then, we sample $100,000$ points on the surface of our output and on the reference mesh, and compute the chamfer distance between the two produced point clouds using the Point Cloud Utils library \cite{point-cloud-utils}.

\subsubsection{Earth mover's distance (EMD)} The earth mover's distance is a measure of dissimilarity between two probability distributions or point sets, and is often described as the minimum cost to transform one distribution into the other. 
The EMD between two point sets $A = \{a_i \in \mathds{R}^3\}_{i=1}^{n}$ and $B = \{b_j \in \mathds{R}^3\}_{j=1}^{m}$ can be formally defined as:
$$\text{EMD}(A, B) = \min_{\pi \in \Pi(A, B)} \sum_{i=1}^{n}{\sum_{j=1}^{m}{\pi_{i, j} \|a_i - b_j\| }},$$
where $\pi$ is a correspondence between $A$ and $B$, \textit{i.e.} $\Pi(A, B)$ is the set of $n \times m$ matrices, where rows and columns sum to one and $\pi_{i,j} \in  \left[0, 1\right]$ is the coefficient indicating how much points $a_i$ and $b_j$ correspond to each other. Due to the computational complexity of the EMD, we sample 1000 points on both meshes. We also use Point Cloud Utils library \cite{point-cloud-utils} for the computation of the EMD.

\subsubsection{Fréchet inception distance (FID)} To take visual perception into consideration, we use the Fréchet inception distance \cite{heusel2018gansfid}. FID evaluates the similarity between two sets of images, generated and real, by computing the Fréchet distance between the Gaussian distributions of their respective features. A lower FID value signifies a greater resemblance between the two image sets. The shading image based FID has been described in SDF-StyleGAN \cite{zheng2022sdfstylegan}, for which the authors report that it yields relevant results for measuring the plausibility and similarity of two shapes. We sample 20 views and render the shape $S_{\text{out}}$ produced by \ourmethod{} and the reference shape $S_{\text{ref}}$. The features are then extracted from these image via the Inception-V3 network \cite{szegedy2015rethinking}, an architecture trained over ImageNet \cite{deng2009}, which maps an image to a probability distribution over 1000 classes. From this probability distribution, we can extract the mean $\mu_i$ and the covariance matrix $\Sigma_i$ for each image $i$. The formula used to compute the FID is given by:
$$\text{FID} = \frac{1}{20} \sum_{i=1}^{20} \left( \lVert \mu_{i}^{\text{out}} - \mu_{i}^{\text{ref}} \rVert^{2} + \operatorname{Tr}\left(\Sigma_{i}^{\text{out}} + \Sigma_{i}^{\text{ref}} - 2\sqrt{\Sigma_{i}^{\text{ref}}\Sigma_{i}^{\text{out}}}\right) \right).$$
To compute the FID, we use the cleanFID library \cite{parmar2021cleanfid}.

\subsubsection{Interpretation} We report the results of our objective evaluation in \tableref{tab:distances}. First, we note that Sketch2Mesh \cite{Sketch2Mesh} fails to produce a shape in 112 cases when the input was cropped, and to provide a fair comparison we could not use their refinement because the camera view parameters are not an input of our method. We report the results for both cropped and padded input sketches, observing that the optimal method varies depending on the used metric.
Because the training procedure is available for DeepSketch\cite{10.1016/j.cag.2022.06.005}, we train this method for our evaluation in two ways: (1) using their default dataset, which includes their synthetic renders and ProSketch \cite{zhong2021prosketch}, and (2) using our training dataset which consists of our full outline rendering, ProSketch, and abstract CLIPasso \cite{vinker2022clipasso} renders. We indicate results for both training procedures. The evaluation on the default DeepSketch is done on padded input. Because cropped inputs are used for retraining DeepSketch on our dataset, we crop and center the AmateurSketch input sketches for its evaluation. Pixel2Mesh \cite{pixel2mesh} and our method are evaluated with cropped input sketches.

For both geometric and perceptual metrics, \ourmethod{} performs substantially better than the state of the art. This indicates that \ourmethod{} is particularly suitable for sketches with different levels of abstraction, and therefore is a relevant approach to allow people of various drawing skills to attempt sketch-based modeling. Since training DeepSketch on our dataset does not show any improvement on the metrics, this additionally indicates that the dataset is not the sole factor that explains the difference of performance between \ourmethod{} and the state of the art.

\begin{table}[htbp]
\caption{Performance comparison of multi-class shape reconstruction methods on the AmateurSketch dataset \cite{Qi2021AmateurSketch} using chamfer distance (CD), earth mover's distance (EMD), and Fréchet inception distance (FID). Lower values indicate better performance. Comparison is done with LAS-diffusion \cite{zheng2023lasdiffusion}.}
\label{tab:distancesMulticlasses}
\centering
\begin{tabular}{lccc}
\toprule
    Method  & CD$\downarrow$ & EMD$\downarrow$ & FID$\downarrow$ \\
\midrule
LAS-diffusion & 0.2112 & 0.1585 & 209.2 \\
SENS multi-class & 0.1171 & 0.0940 & 171.0 \\
\bottomrule
\end{tabular}
\end{table}

\subsubsection{Multi-class reconstruction} While LAS-Diffusion \cite{zheng2023lasdiffusion} is targeted toward a view-aware setting, this sketch-to-shape method can run without camera parameters. Since the authors provide the multi-class pretrained network for this task, we compare multi-class \ourmethod{} with LAS-Diffusion using the same evaluation metrics as for the single-class comparison. The results are reported in \tableref{tab:distancesMulticlasses}. We can see that our method performs better than LAS-diffusion on the AmateurSketch dataset. However, we emphasize that the multi-class LAS-diffusion has been trained on all the ShapeNet classes, while our method training was focused on only 3 classes. Moreover, while it is possible to run LAS-diffusion without input view information, the authors state in their ablation study that using a view-agnostic network tends to yield additional or wrong geometry. Therefore, no definitive conclusion can be drawn from this comparison.

Additionally, when comparing single-class and multi-class SENS, we notice that the metrics give very similar results. This shows that our multi-class setup has good generalization abilities.
\subsection{Subjective evaluation (user study)}

\begin{figure*}
    \centering
    \includegraphics[width=0.49\linewidth]{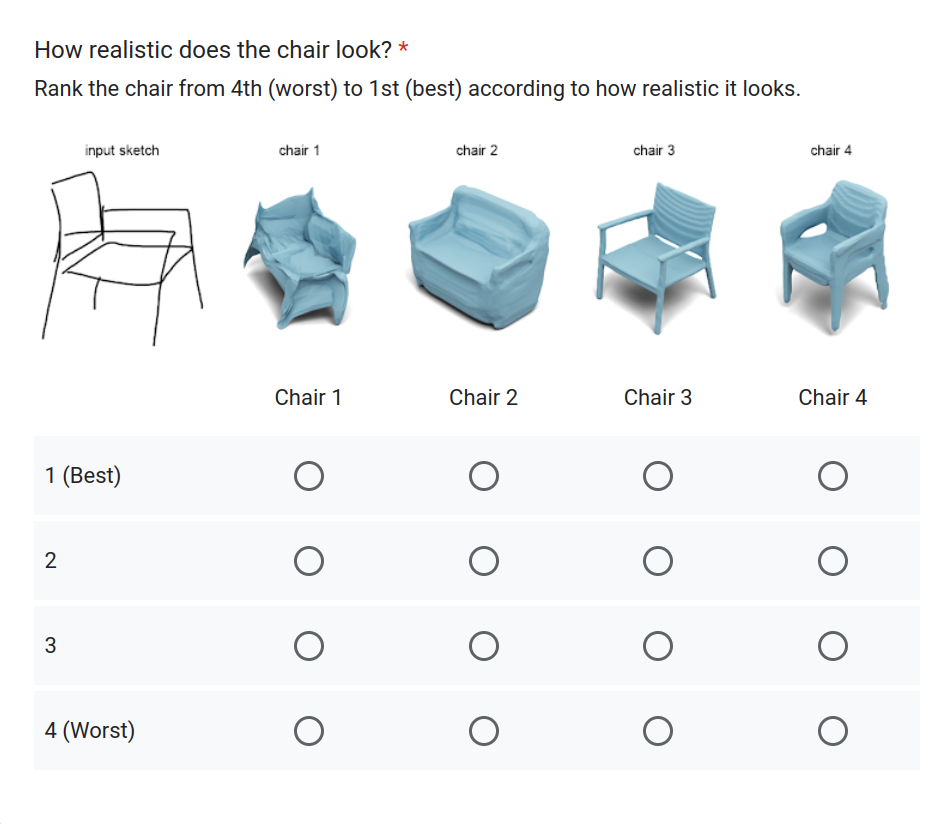}
    \includegraphics[width=0.49\linewidth]{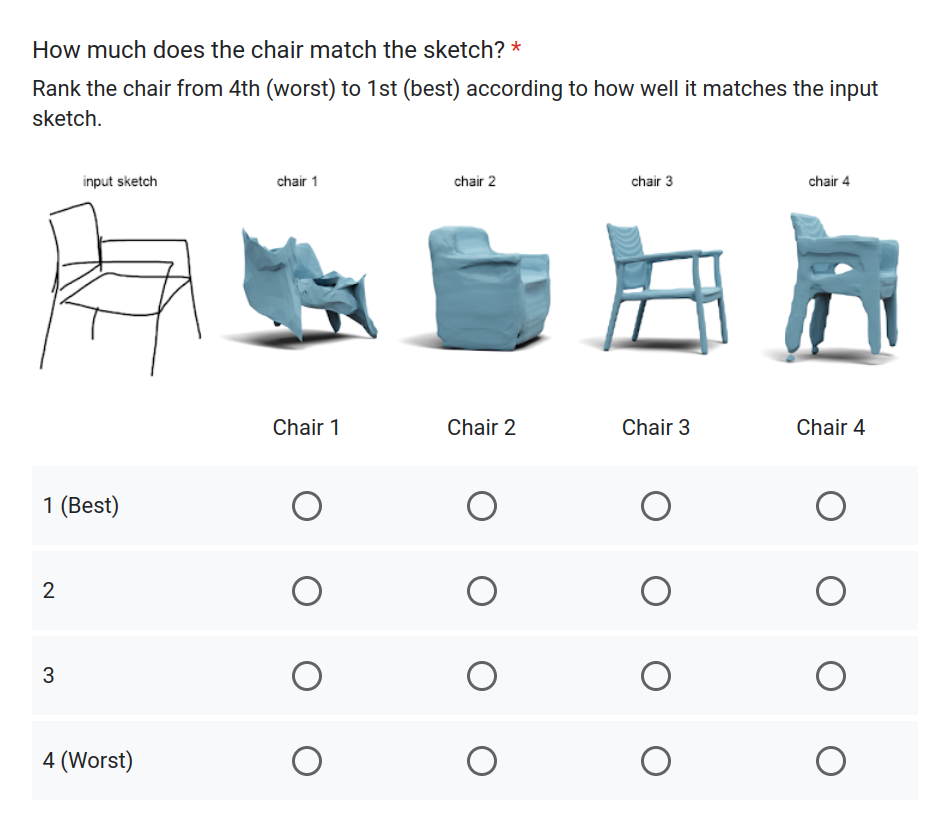}
    \caption{The two types of questions asked in our user study. When asking for how realistic the shape looks, the same view is applied for rendering the shapes. When asking for similarity with the input sketch, shapes are rendered with the same azimuth angle as the input sketch. The azimuth angle is provided by the AmateurSketch dataset.}
    \label{fig:userstudyQuestions}
\end{figure*}

\begin{figure*}[th]
    \centering
    \includegraphics[trim = 50 0 50 0, clip, width=0.49\linewidth]{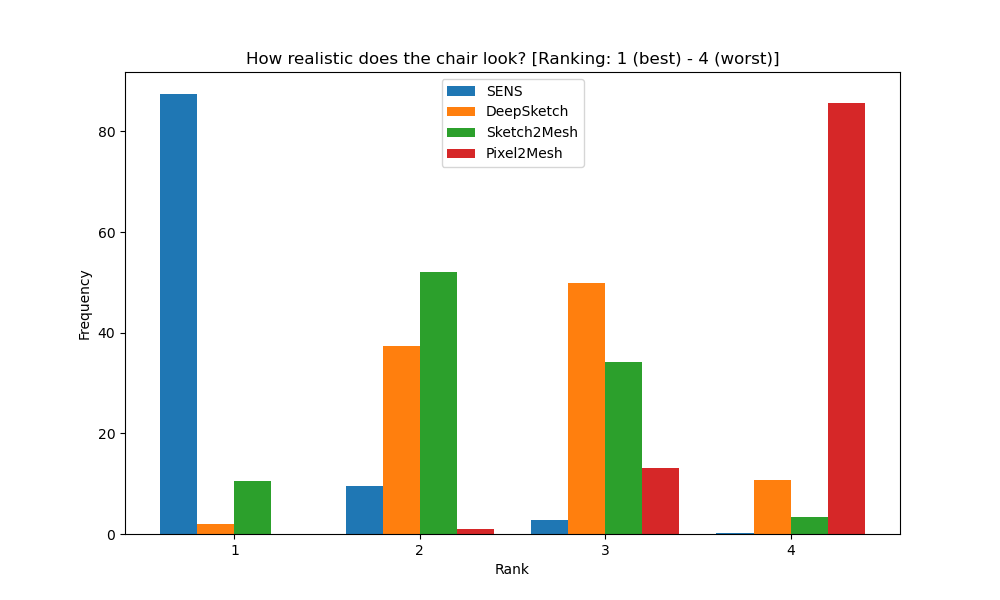}
    \includegraphics[trim = 50 0 50 0, clip, width=0.49\linewidth]{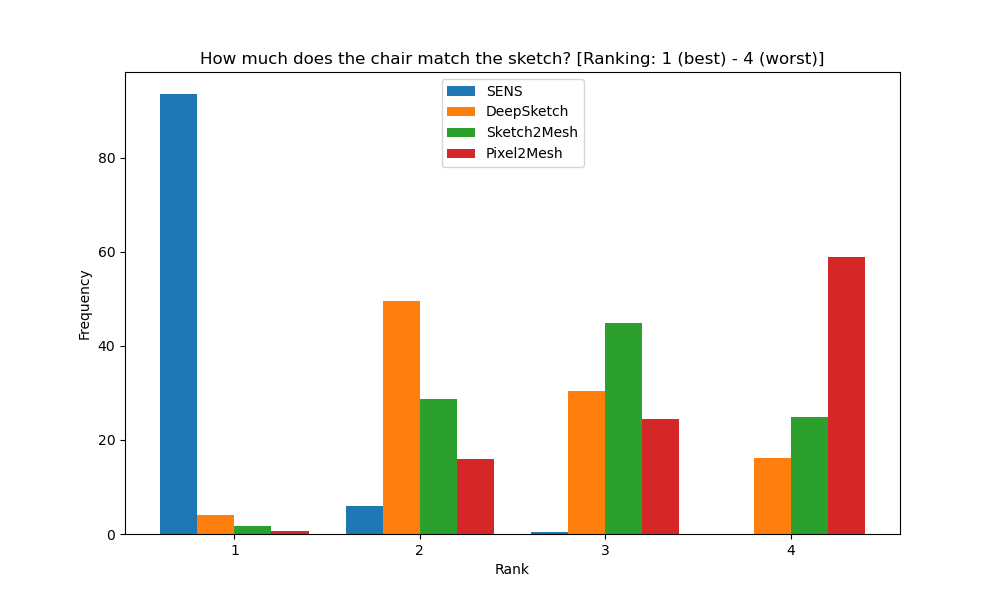}
    \caption{Results of our user study, displayed as an histogram. The results highlight the performance of our method in comparison to Pixel2Mesh \cite{pixel2mesh}, Sketch2Mesh \cite{Sketch2Mesh}, and retrained DeepSketch \cite{10.1016/j.cag.2022.06.005} in terms of realism and similarity to input sketches.}
    \label{fig:userstudyHistograms}
\end{figure*}

\begin{table}[t]
\caption{Perceptual evaluation through a user study, highlighting the performance of our method in comparison to Pixel2Mesh \cite{pixel2mesh}, Sketch2Mesh \cite{Sketch2Mesh} and retrained DeepSketch \cite{10.1016/j.cag.2022.06.005} in terms of realism and similarity to input sketches. The ranking in each question is from 1 (best) to 4 (worst).}
\label{tab:userstudy}
\small
\centering
\setlength{\tabcolsep}{3pt}
\begin{tabular}{lccccccccc}
\toprule
    Question \phantom{AAAA} & \multicolumn{4}{c}{Realistic} & \phantom{A} & \multicolumn{4}{c}{Similar to sketch} \\ \midrule
    Rank  & \cellcolor{green!25} $1$ & $2$ & $3$ &$4$ & & \cellcolor{green!25}$1$ & $2$ & $3$ &$4$ \\
\midrule
Pixel2Mesh  &  0.1  &  1.1  &  12.8  &  \textbf{86.0}  & &  0.4  &  15.8  &  24.8  &  \textbf{59.0}  \\
Sketch2Mesh  &  10.3  &  \textbf{53.1}  &  33.1  &  3.4  & &  1.6  &  28.5  &  \textbf{45.1}  &  24.8  \\
DeepSketch  &  1.7  &  36.6  &  \textbf{51.3}  &  10.3  & &  4.0  &  \textbf{50.0}  &  29.8  &  16.2  \\
SENS  &  \cellcolor{green!25}\textbf{87.9}  &  9.1  &  2.7  &  0.3  & &  \cellcolor{green!25}\textbf{94.0}  &  5.7  &  0.3  &  0.0  \\
\bottomrule
\end{tabular}
\end{table}

\begin{table}[t]
\caption{Median and interquartile range (IQR) of the results of our user study, for both realism and similarity to input sketches.}
\label{tab:userstudyMedian}
\centering
\setlength{\tabcolsep}{3pt}
\begin{tabular}{lccccc}
\toprule
Method & \multicolumn{2}{c}{Realistic} & & \multicolumn{2}{c}{Similar} \\
& Median & IQR & & Median & IQR\\ \midrule
Pixel2Mesh & 4.0 & 0.0 & & 4.0 & 1.0  \\
Sketch2Mesh & 2.0 & 1.0 & & 3.0 & 1.0  \\ 
DeepSketch & 3.0 & 1.0 & & 2.0 & 1.0 \\
SENS & 1.0 & 0.0 & & 1.0 & 0.0 \\ 
\bottomrule
\end{tabular}
\end{table}

To perform a perceptual evaluation of our work, we conduct a user study. We \emph{randomly} sample 24 sketches from the AmateurSketch dataset and render the output of \ourmethod{}, Pixel2Mesh \cite{pixel2mesh}, Sketch2Mesh \cite{Sketch2Mesh} (cropped input), and retrained DeepSketch \cite{10.1016/j.cag.2022.06.005}.
We show in \figref{fig:userstudyQuestions} the exact format used for the user study. For each sketch, we ask participants to rank the four methods' output in two questions: how realistic and how close to the input sketch the resulting chair looks. For the second question, we align the rendering view of the shape with the same azimuth angle as given by the AmateurSketch dataset. The order of the methods is randomized across the sketches, but the same order is used for both questions for each sketch. We recruit 54 individuals of diverse backgrounds and ages to partake in the user study, including 15 women and 39 men.

The results are reported in \tableref{tab:userstudy} and \figref{fig:userstudyHistograms}. According to this study, \ourmethod{} provides the most realistic shape in 87.9\% of the cases and the most similar to the input sketch in 94\% of the cases. Pixel2Mesh is often deemed to perform the worst, especially in terms of realism. Sketch2Mesh and DeepSketch both seem to perform equally well for both questions and rank second and third with nearly equal scores, as shown by the interquartile range in \tableref{tab:userstudyMedian}. Therefore, our user study is aligned with our objective evaluation. 

\subsection{Usability study}

\begin{figure*}[t]
	\centering
	\small
	\setlength{\tabcolsep}{0pt}
	\begin{tabular}{cccccccc}
    \includegraphics[width=0.125\linewidth]{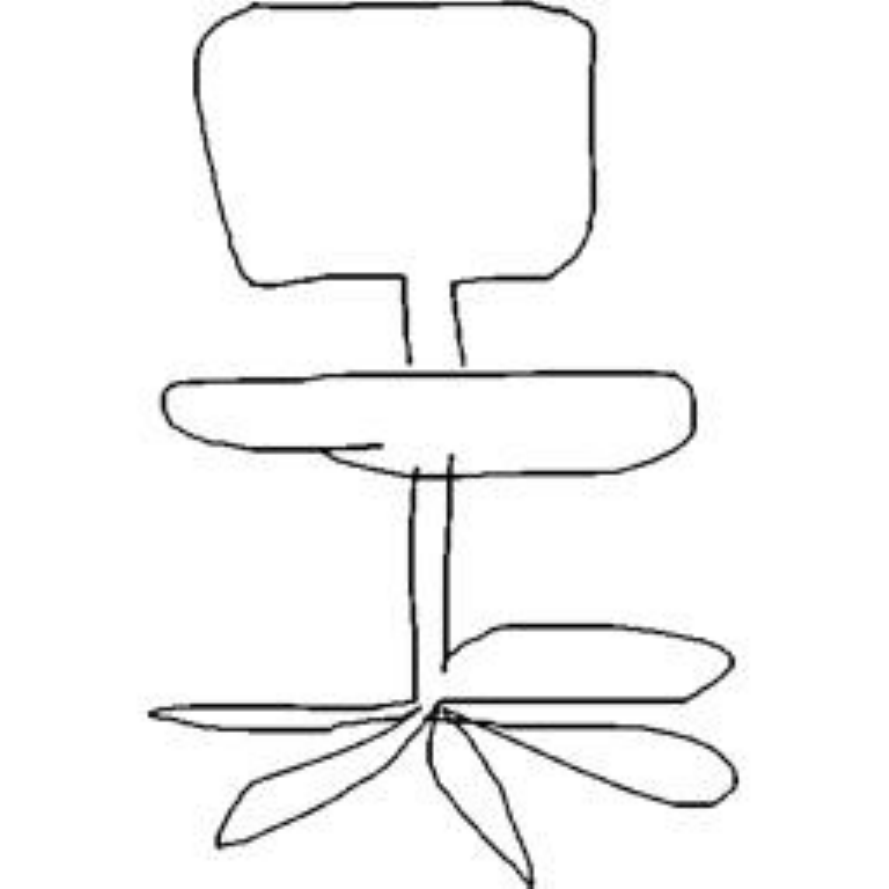}
    & \includegraphics[trim = 10 10 10 10, clip, width=0.125\linewidth]{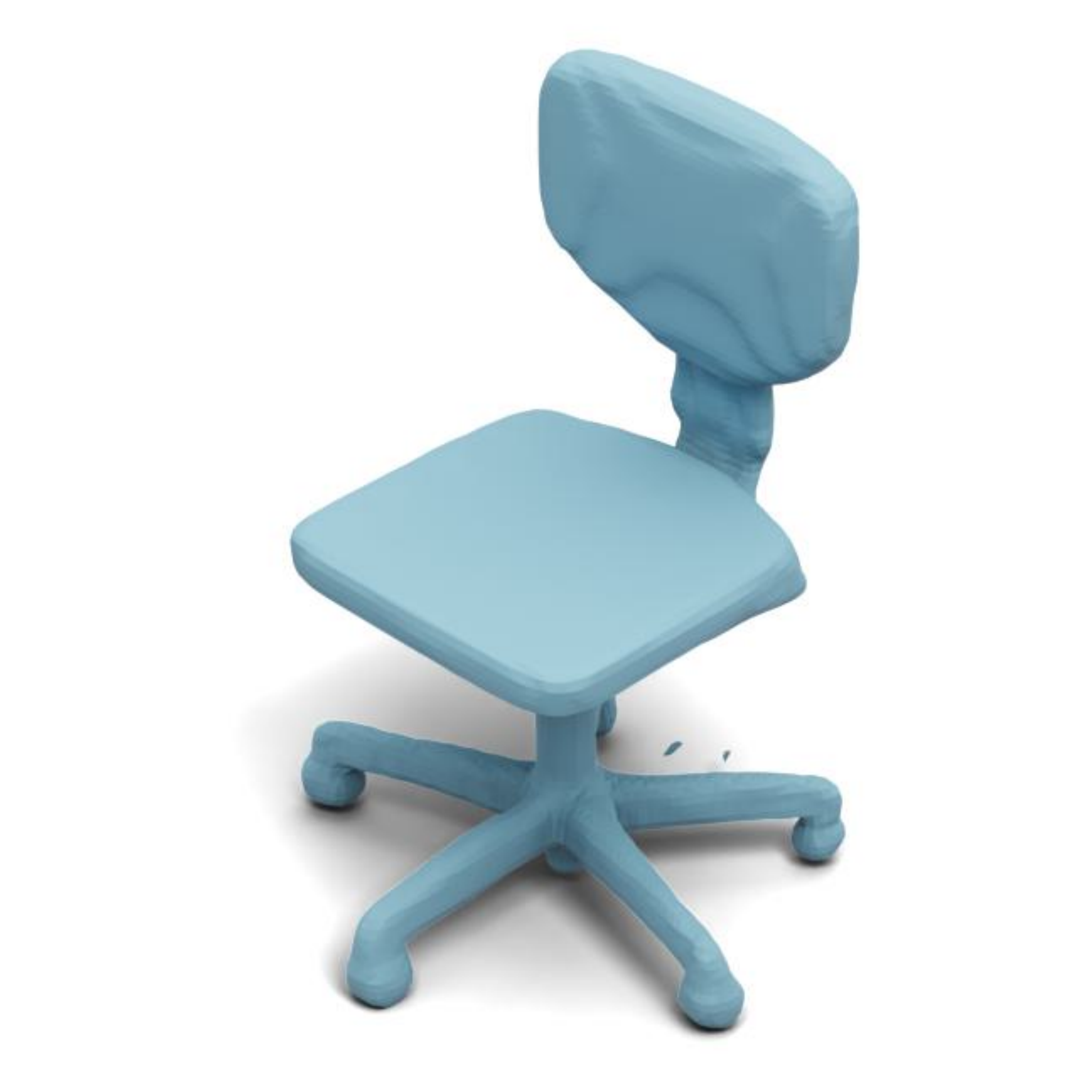}
    & \includegraphics[width=0.125\linewidth]{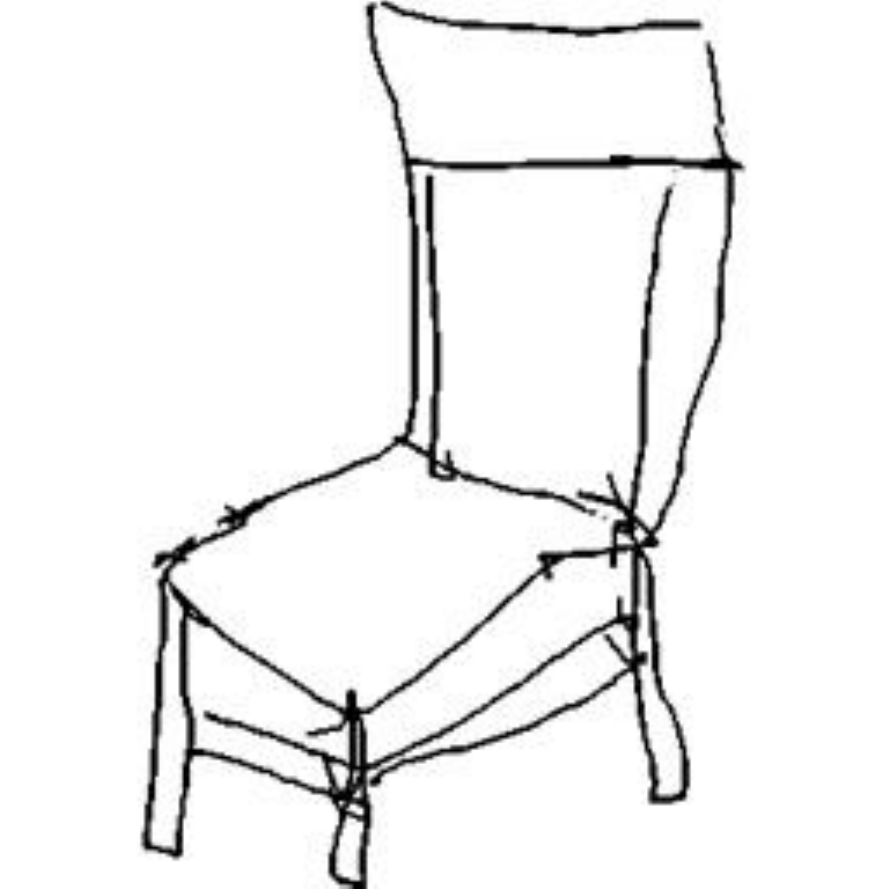}
    & \includegraphics[trim = 10 10 10 10, clip, width=0.125\linewidth]{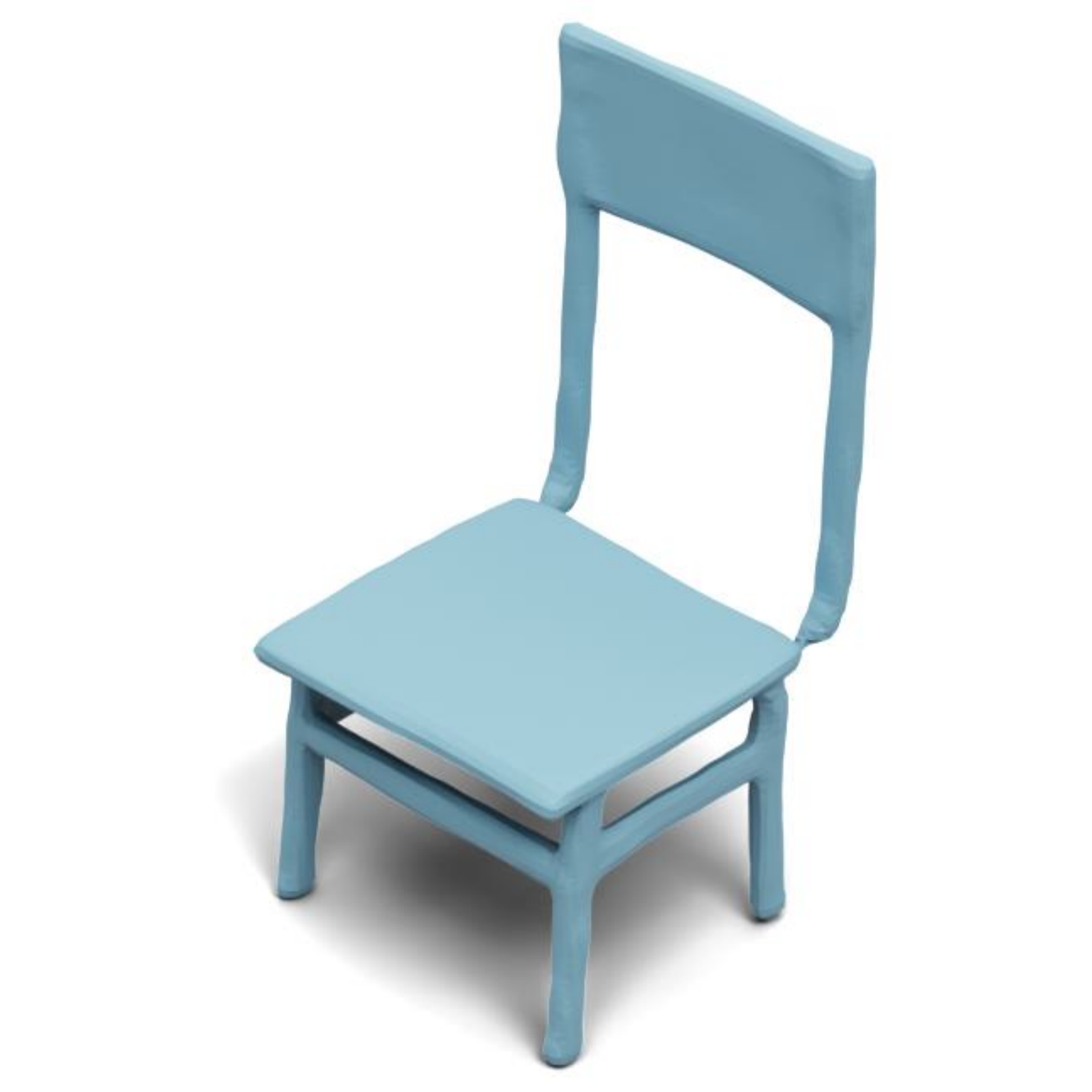}
    & \includegraphics[width=0.125\linewidth]{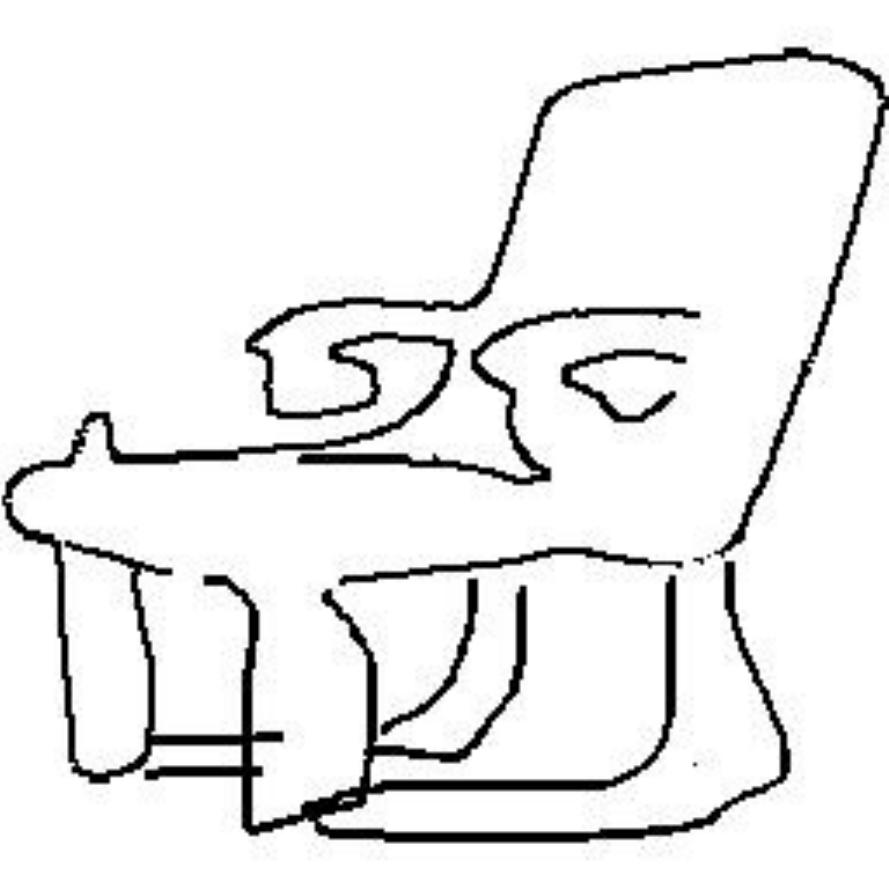}
    & \includegraphics[trim = 10 10 10 10, clip, width=0.125\linewidth]{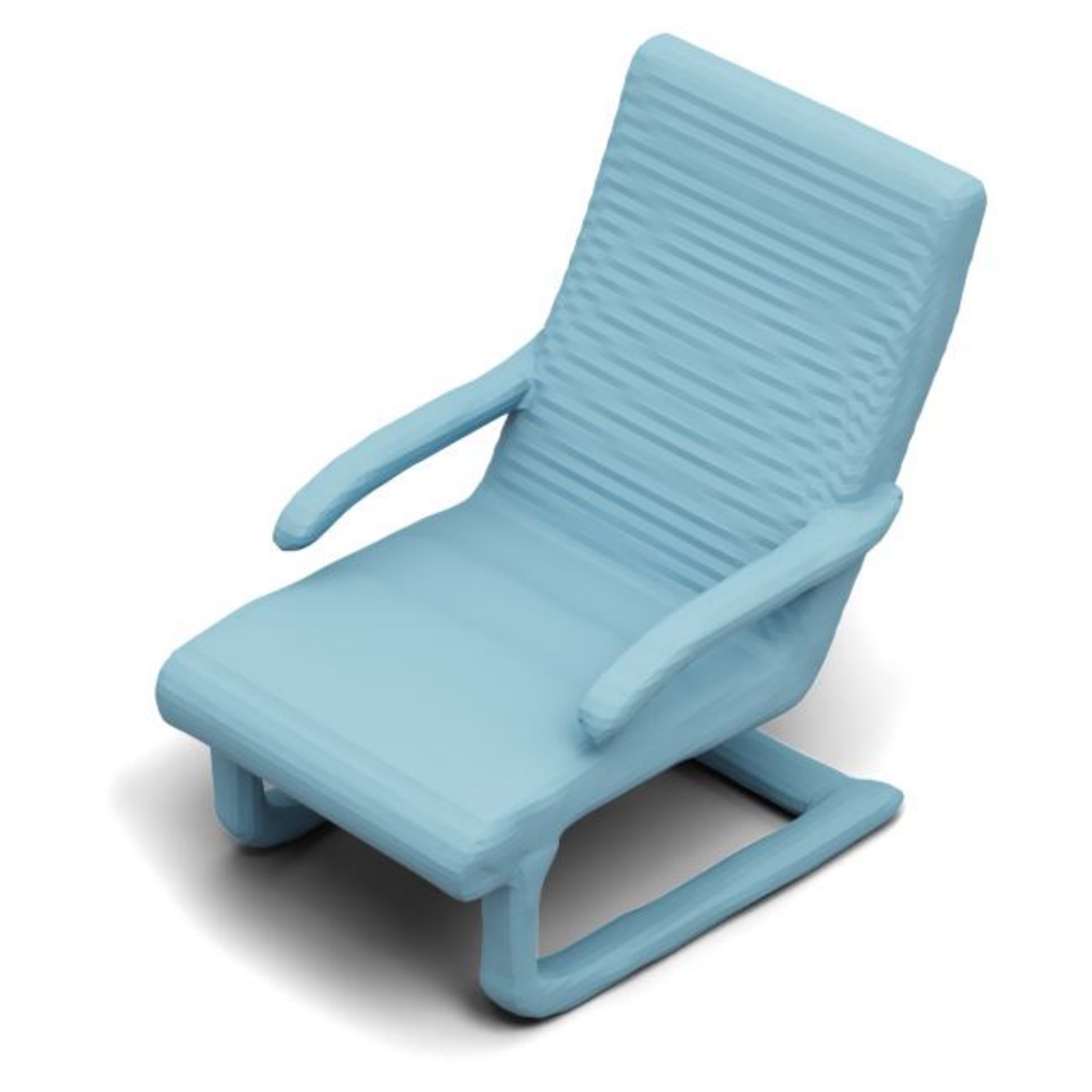}
    & \includegraphics[width=0.125\linewidth]{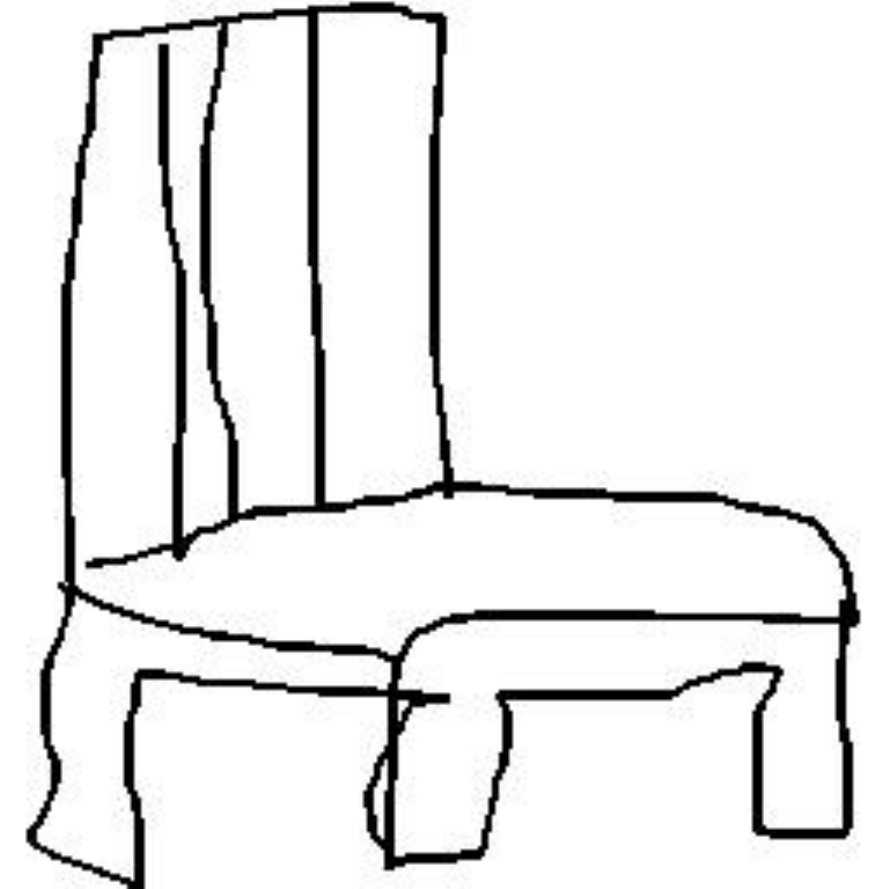} 
    & \includegraphics[trim = 10 10 10 10, clip, width=0.125\linewidth]{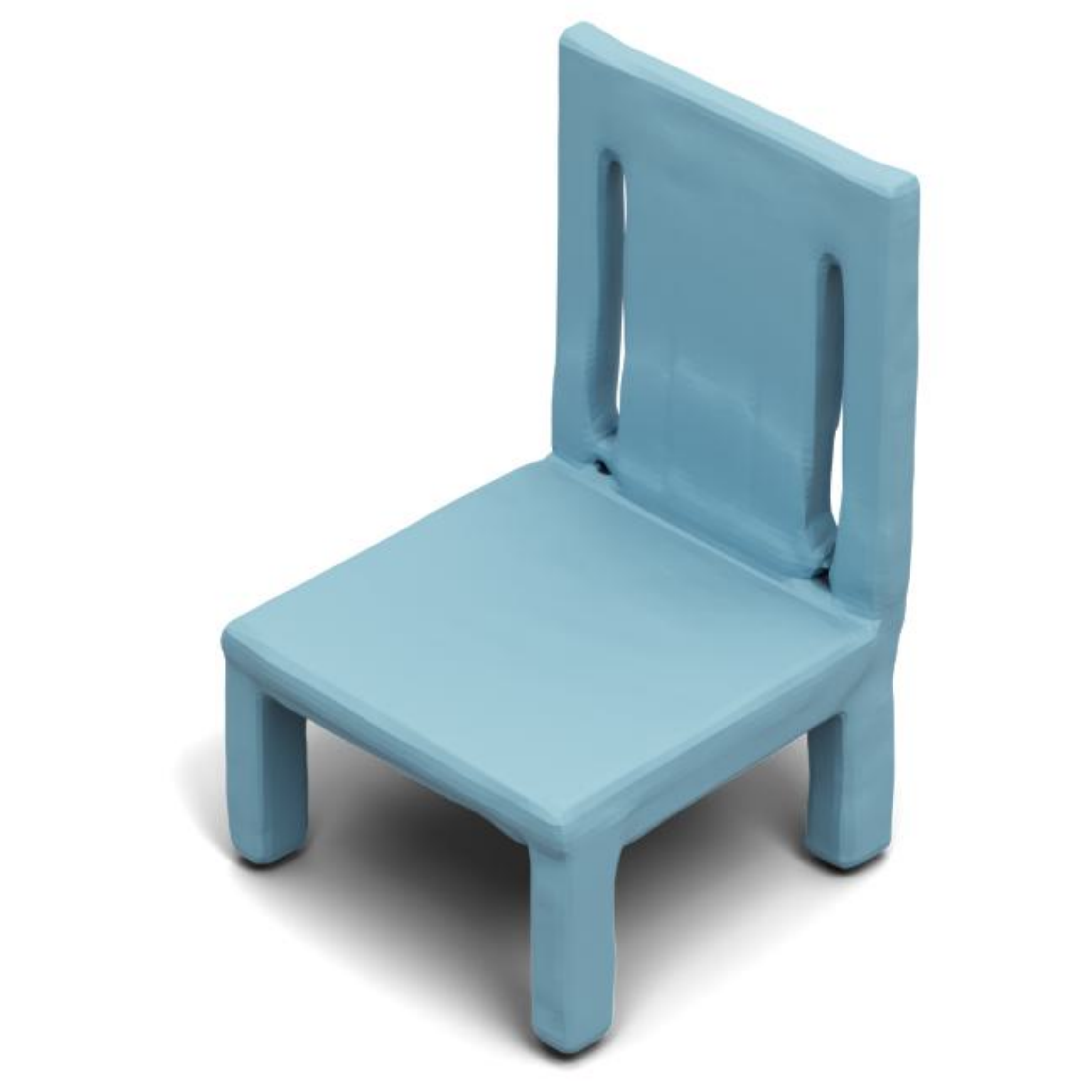}
    \\
    \includegraphics[width=0.125\linewidth]{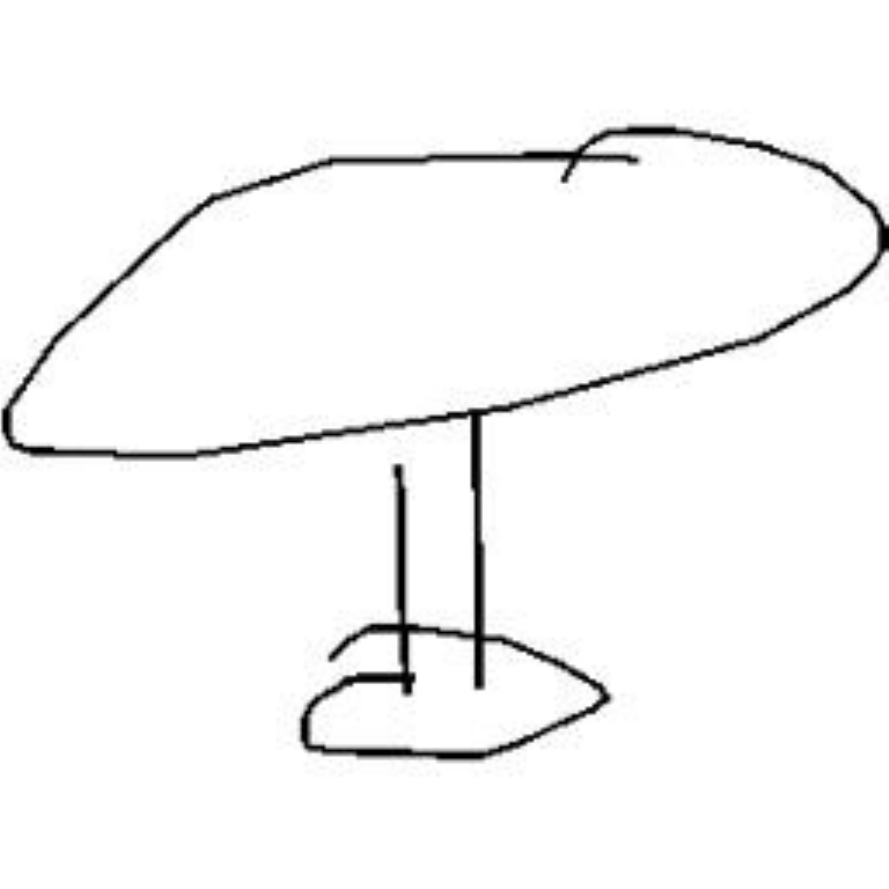}
    & \includegraphics[trim = 10 10 10 10, clip, width=0.125\linewidth]{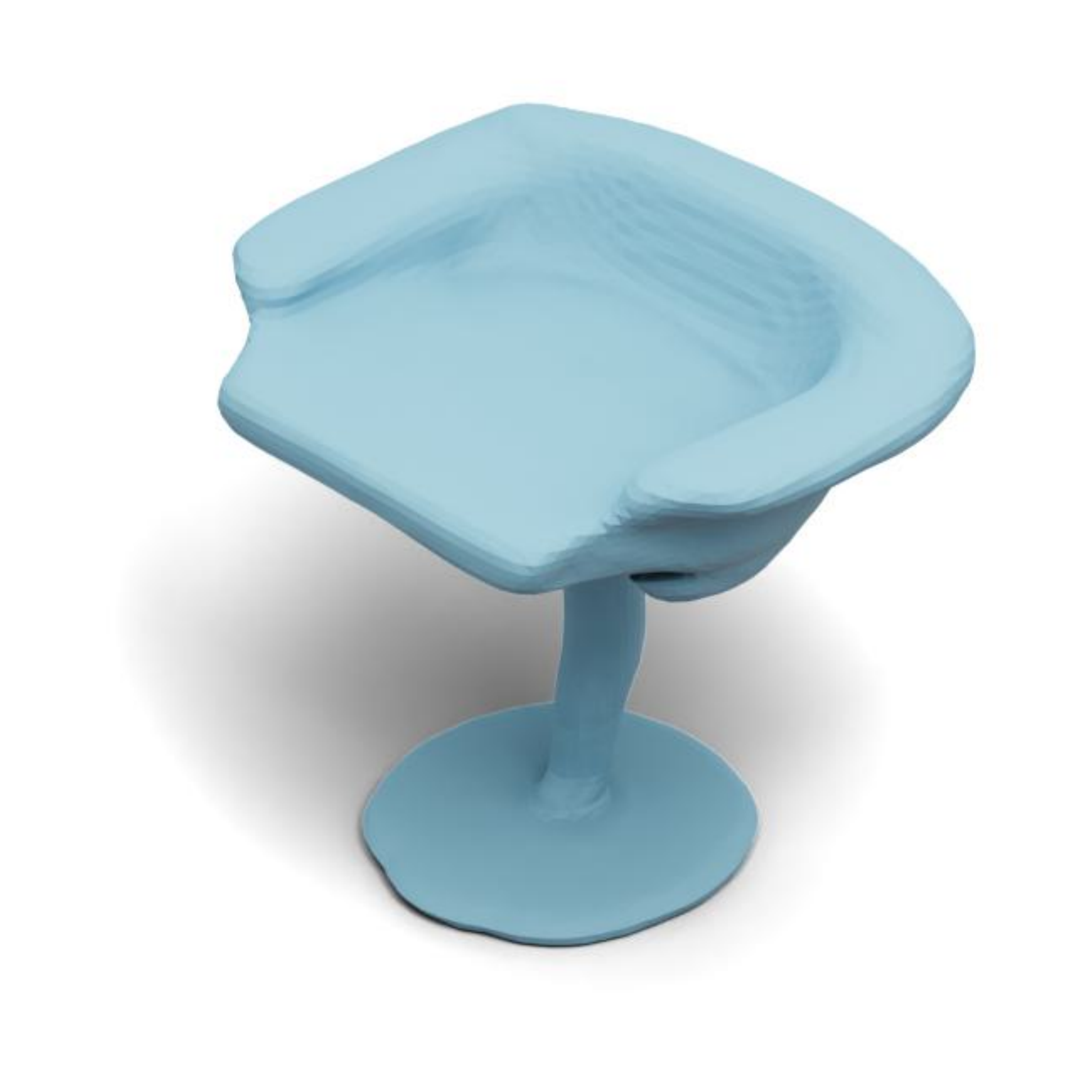}
    & \includegraphics[width=0.125\linewidth]{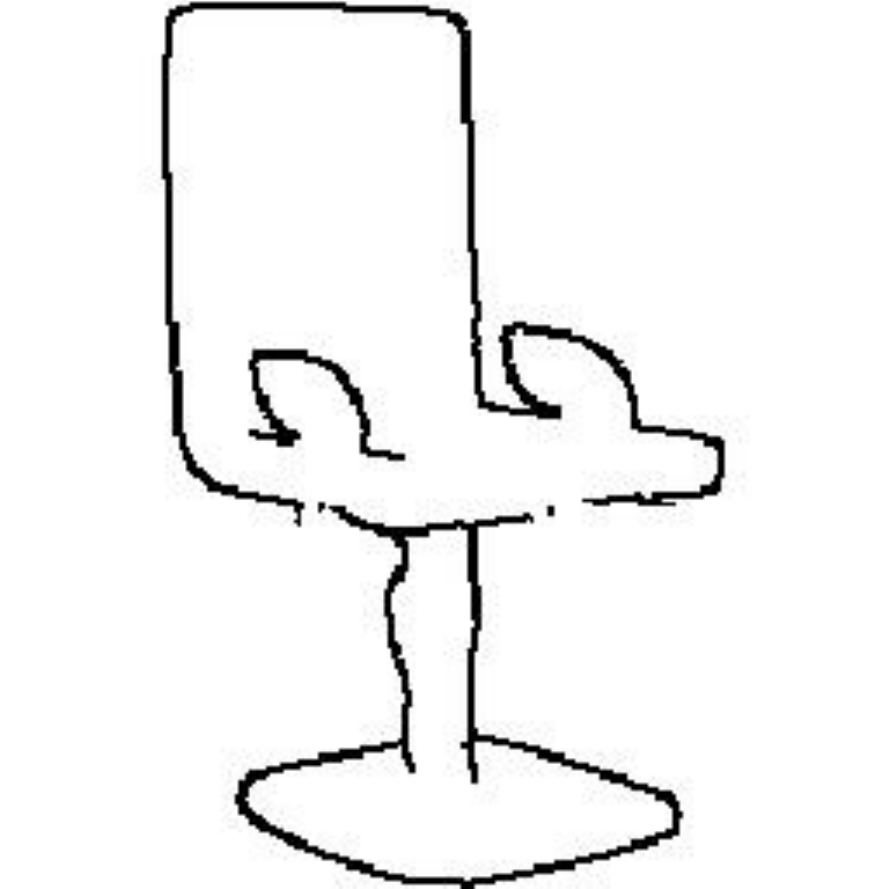}
    & \includegraphics[trim = 10 10 10 10, clip, width=0.125\linewidth]{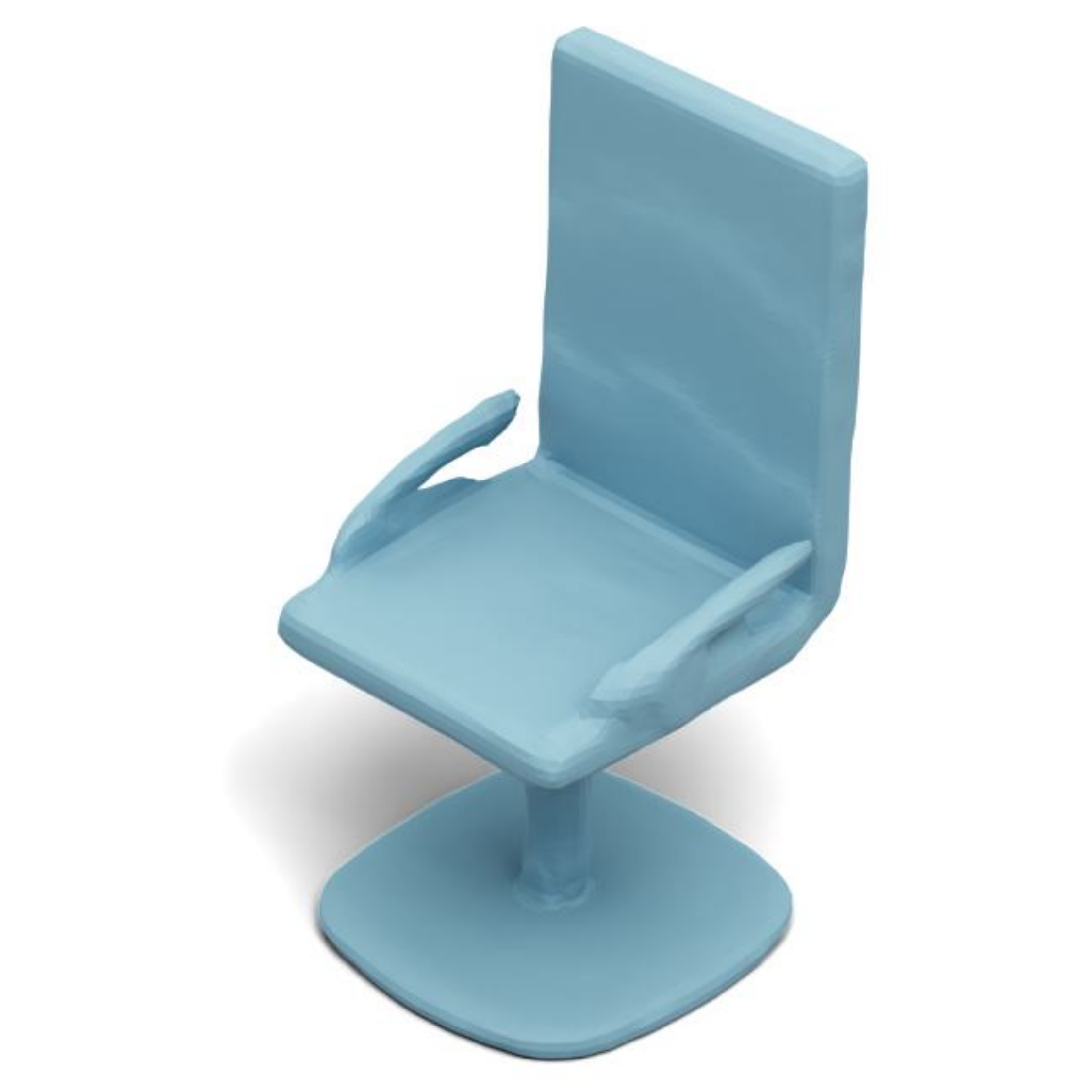}
    & \includegraphics[width=0.125\linewidth]{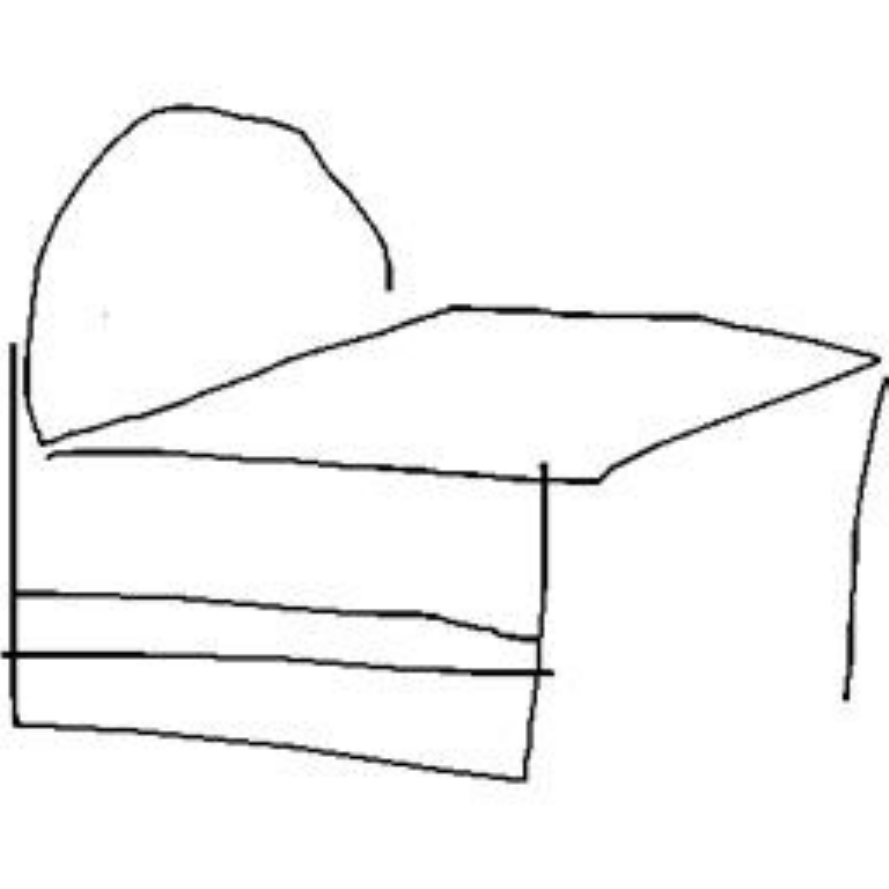}
    & \includegraphics[trim = 10 10 10 10, clip, width=0.125\linewidth]{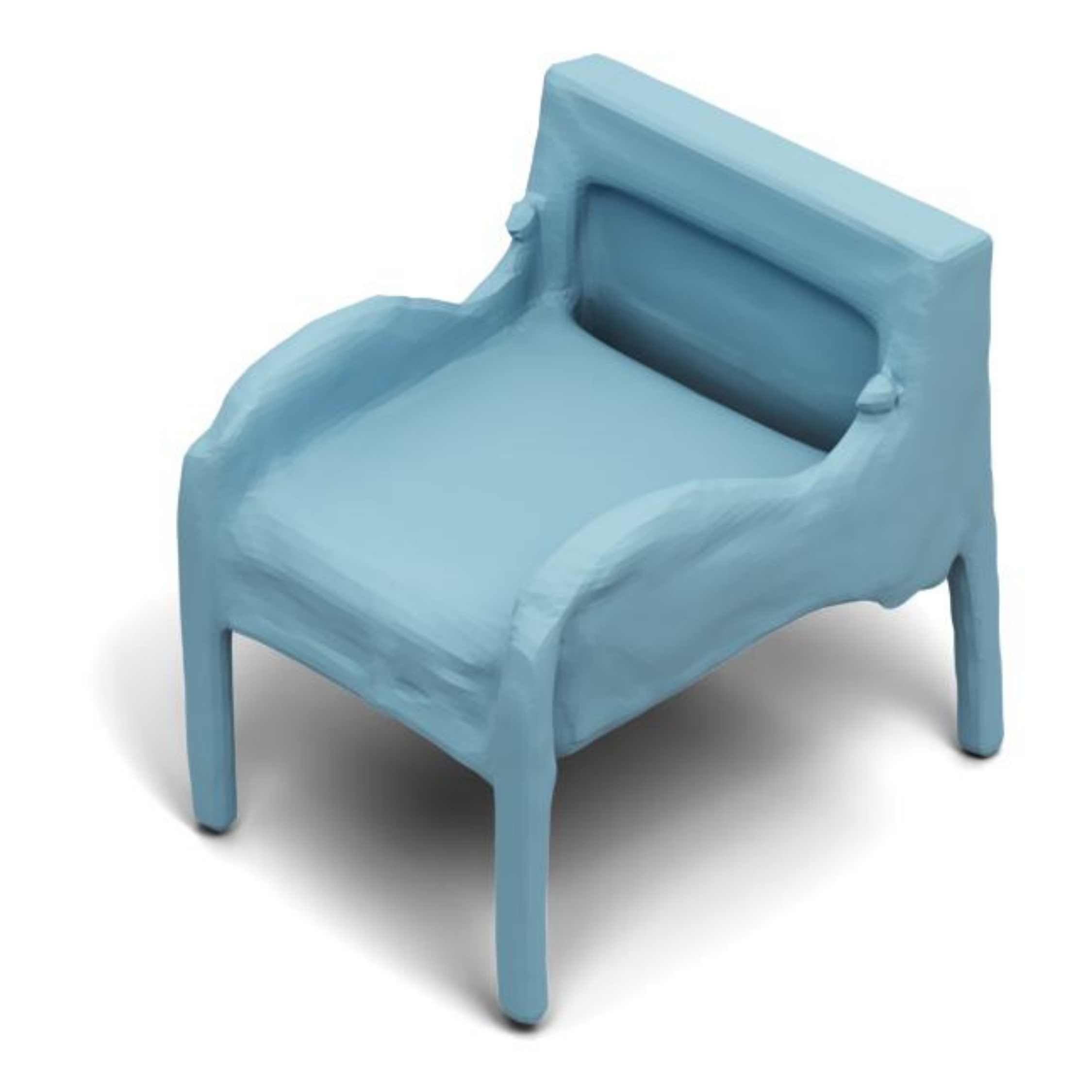}
    & \includegraphics[width=0.125\linewidth]{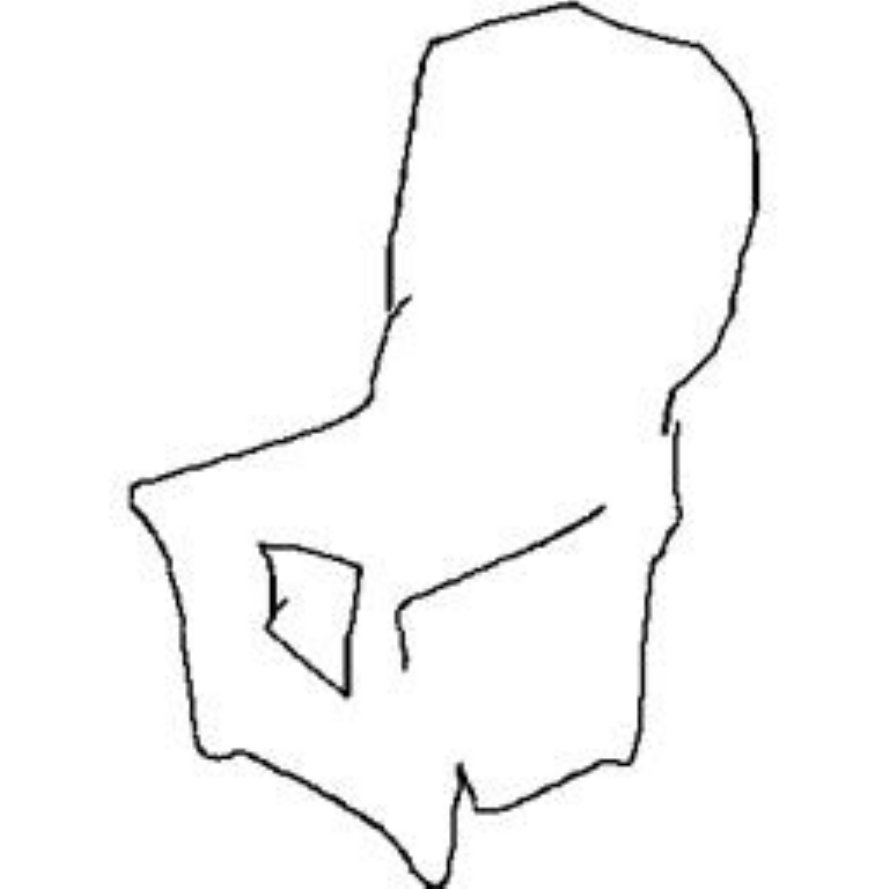}
    & \includegraphics[trim = 10 10 10 10, clip, width=0.125\linewidth]{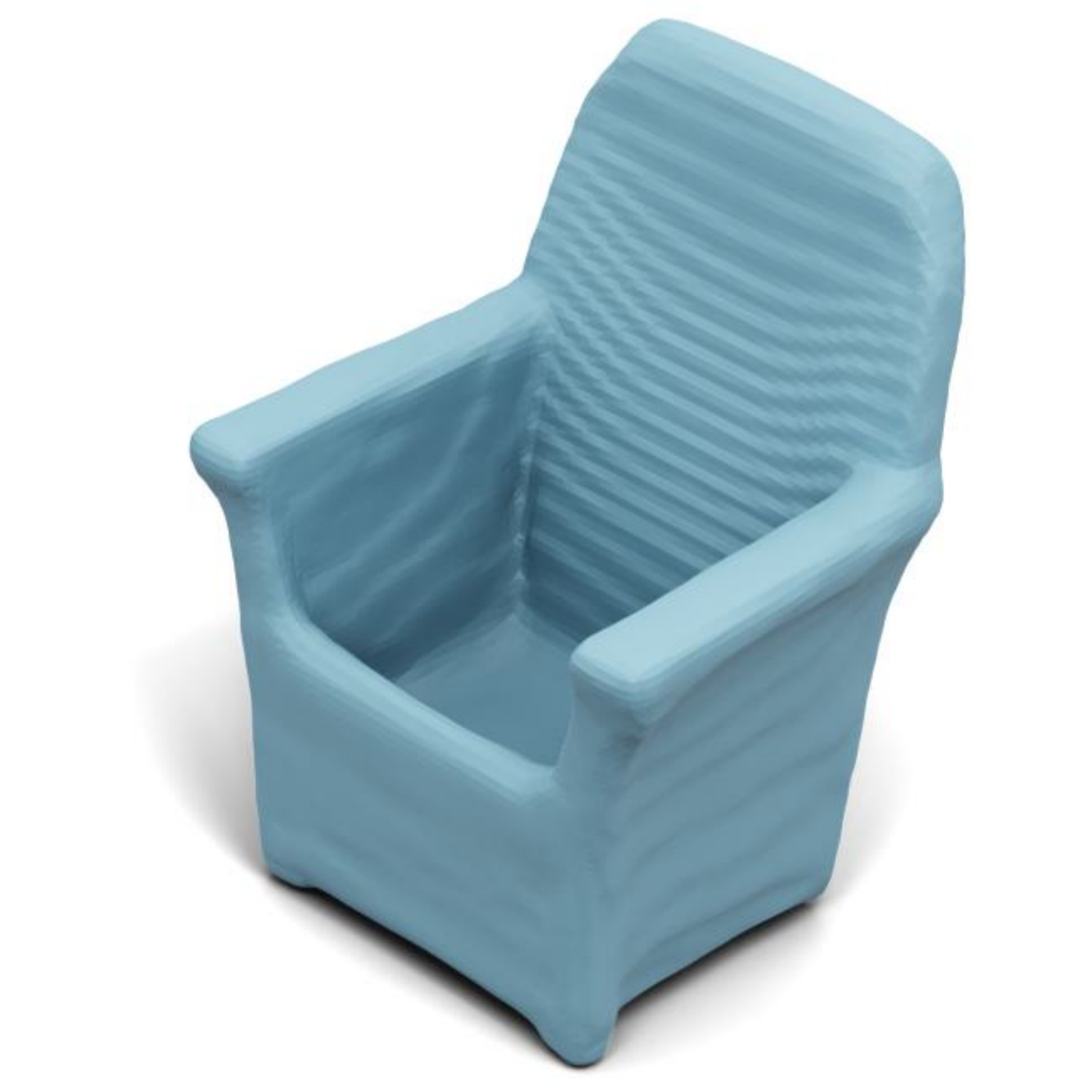}
  \end{tabular}
	\caption{Some sketches and shapes from the Task 1 of the usability study. The results come from each user (P1 to P8, ordered from left to right, top to bottom). Some sketches (P3, P6, and P8) are edited versions of the outline rendering from previously generated shapes. The displayed shapes are not solely generated by the input sketches, but might have been refined via part reconstruction or part-based modeling.}
	\label{fig:task1}
 \vspace{-0.3cm}
\end{figure*}

\begin{figure}[t]
	\centering
	\setlength{\tabcolsep}{1pt}
	\begin{tabular}{ccccc}
    \includegraphics[width=.2\linewidth]{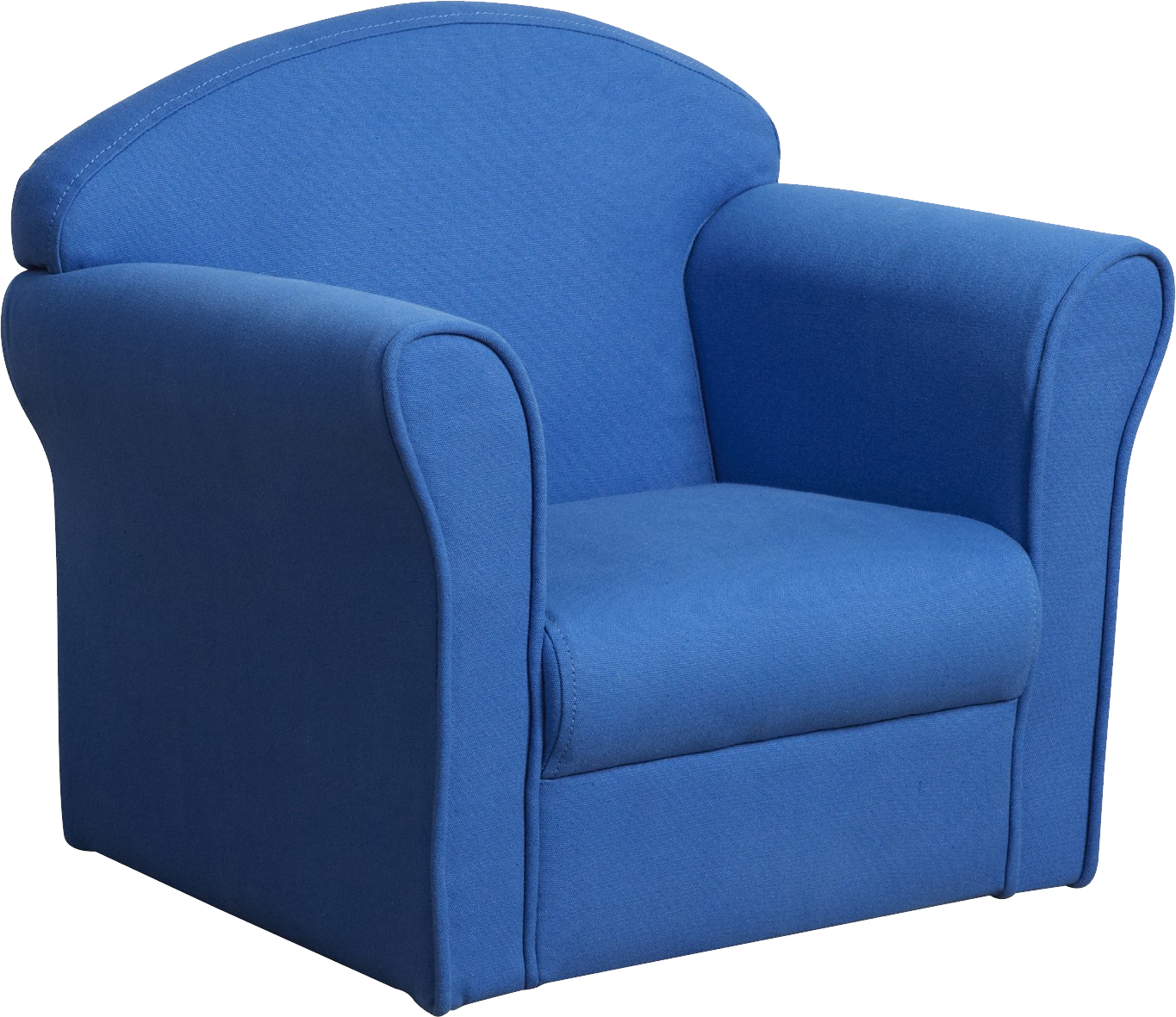}&
    \includegraphics[trim = 20 20 20 20, clip, width=.2\linewidth]{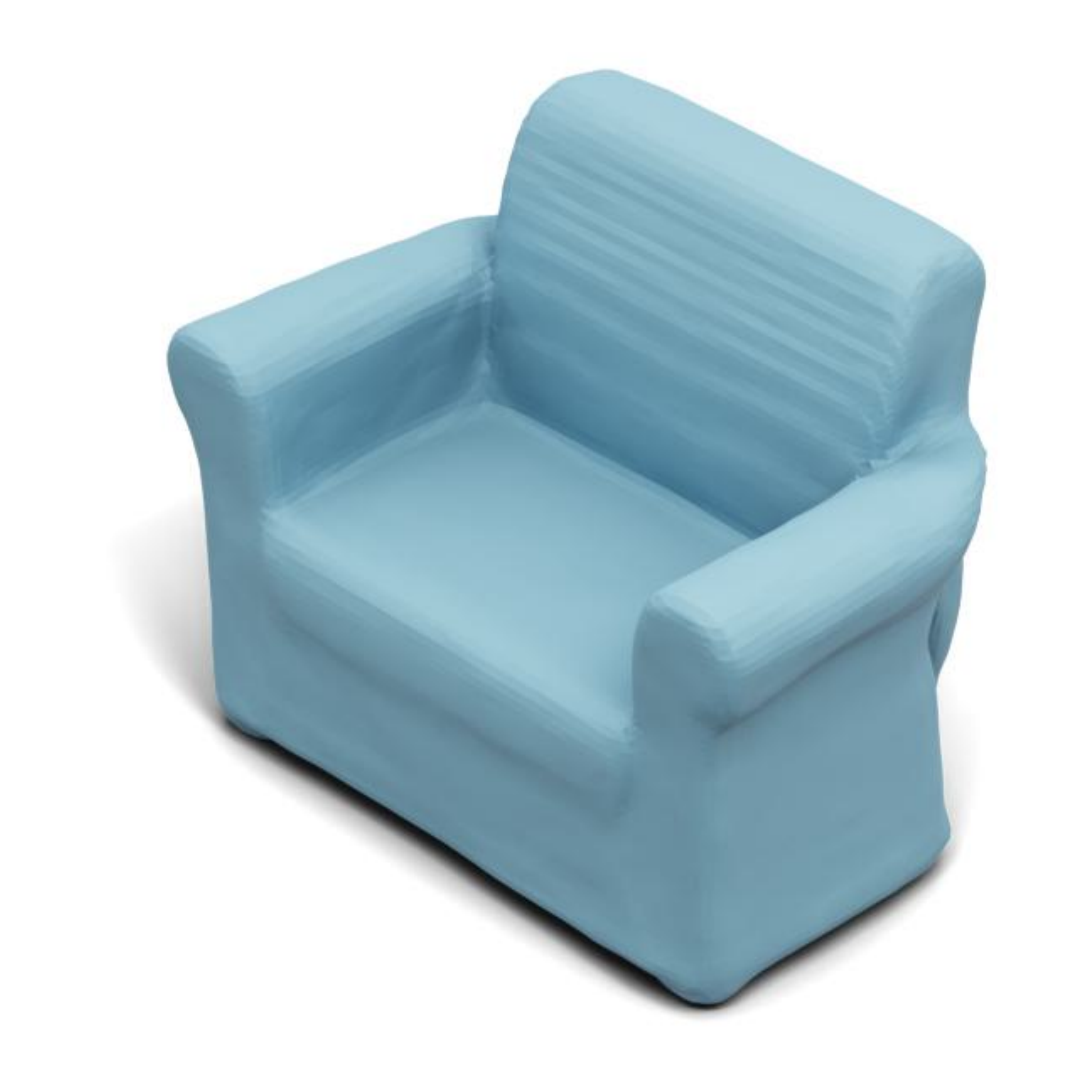}&
    \includegraphics[width=.2\linewidth]{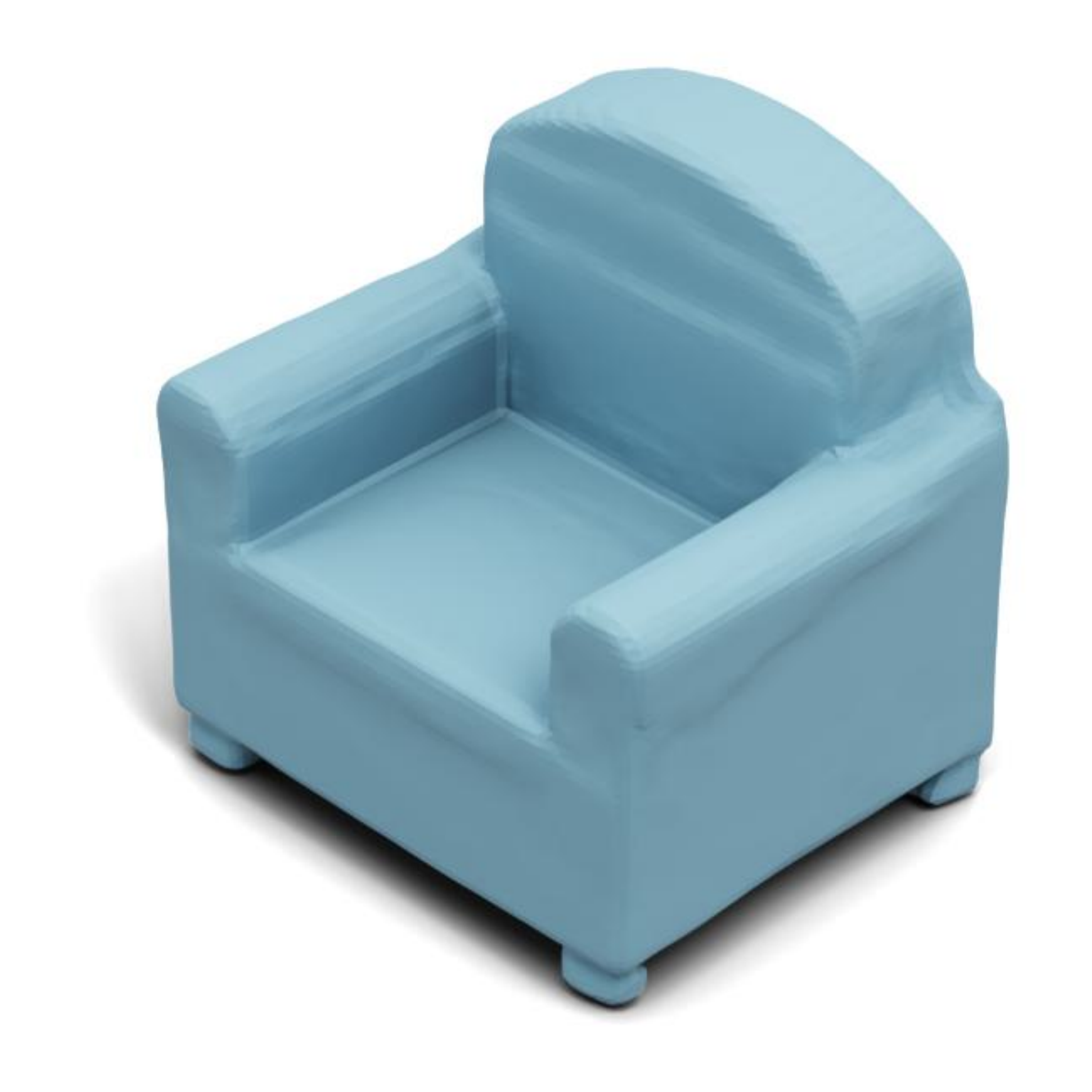}&
    \includegraphics[width=.2\linewidth]{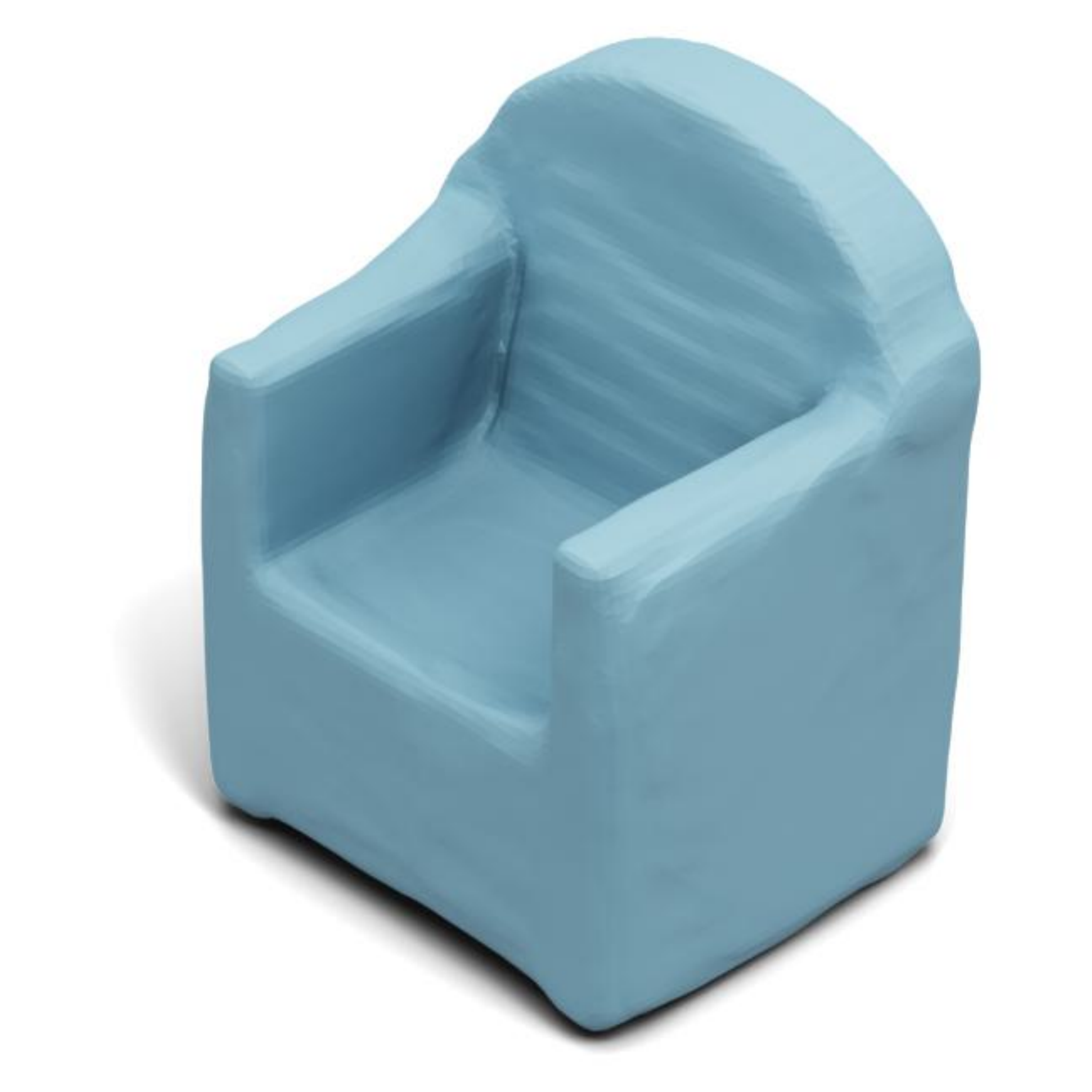}&
    \includegraphics[width=.2\linewidth]{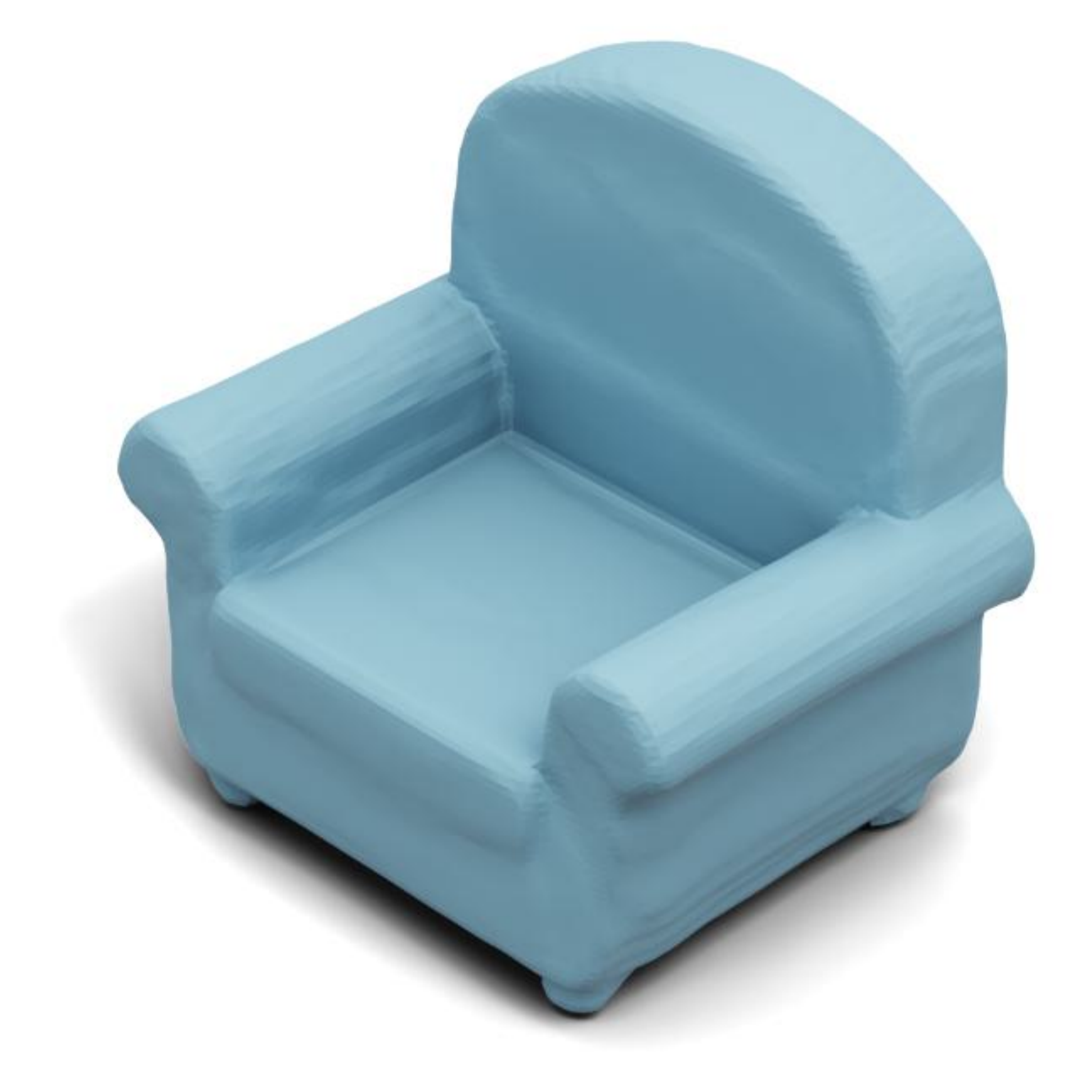}\\
    \includegraphics[trim = 30 15 30 25, clip, width=.125\linewidth]{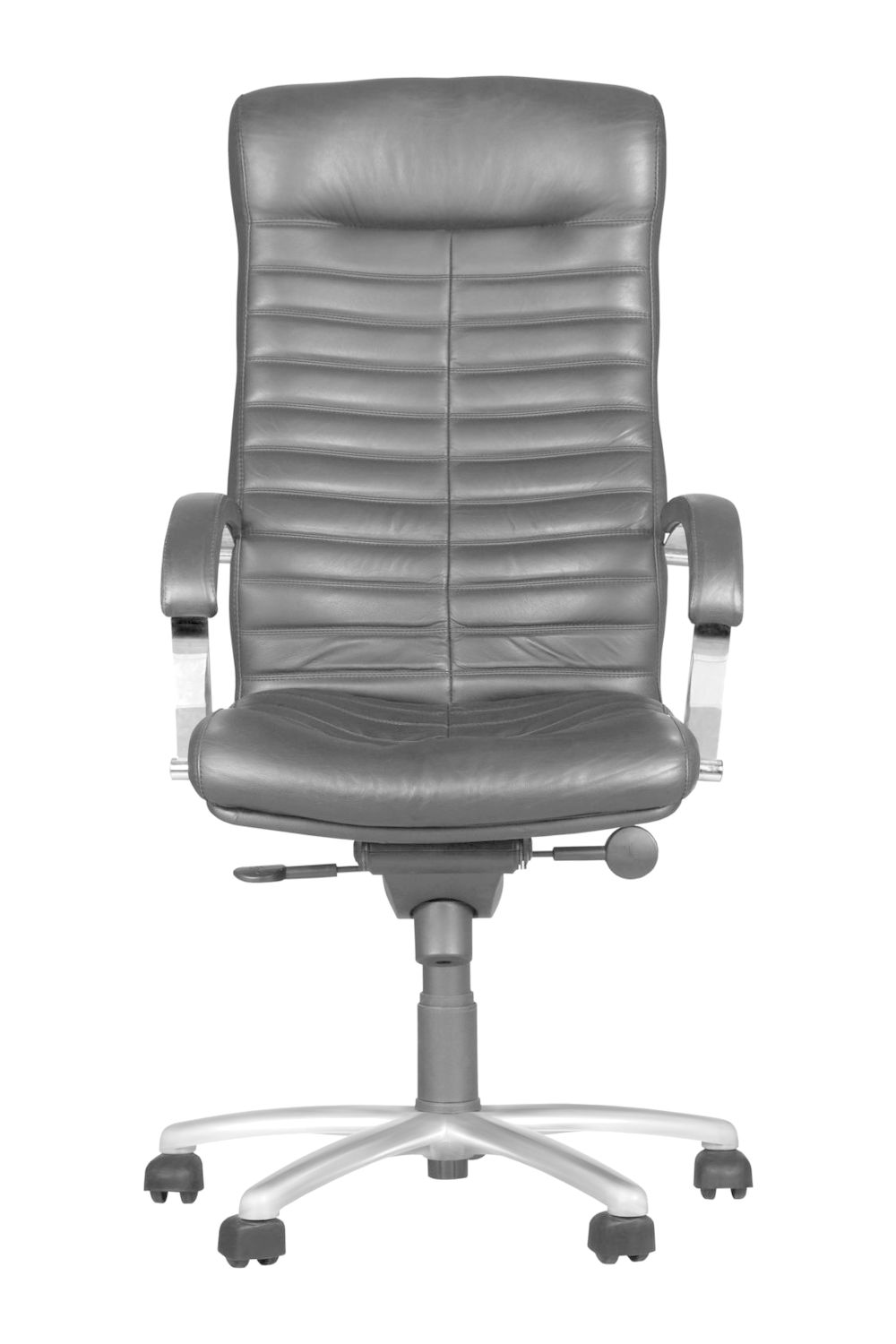}
    & \includegraphics[trim = 50 0 50 0, clip, width=.2\linewidth]{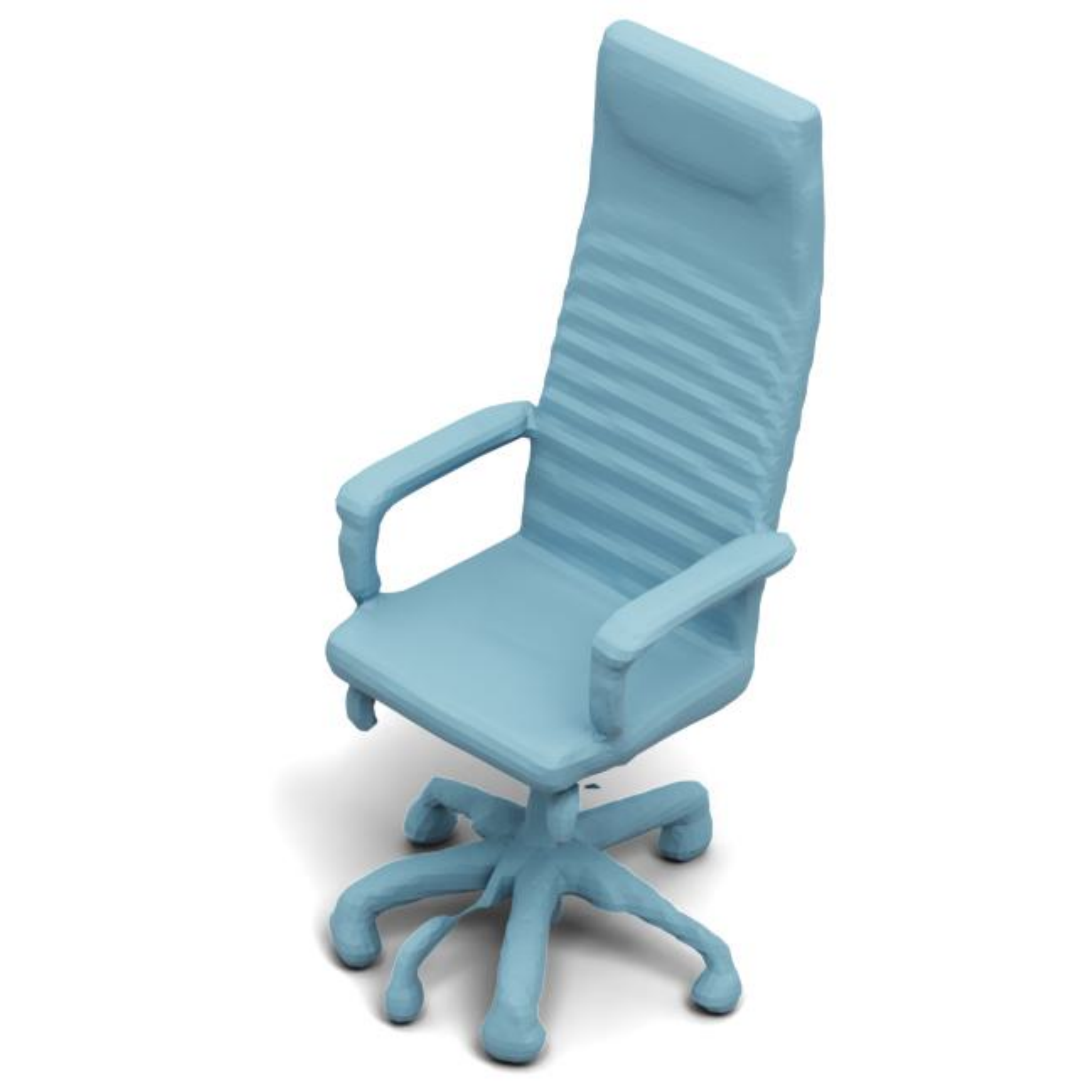}
    & \includegraphics[trim = 50 0 50 0, clip, width=.2\linewidth]{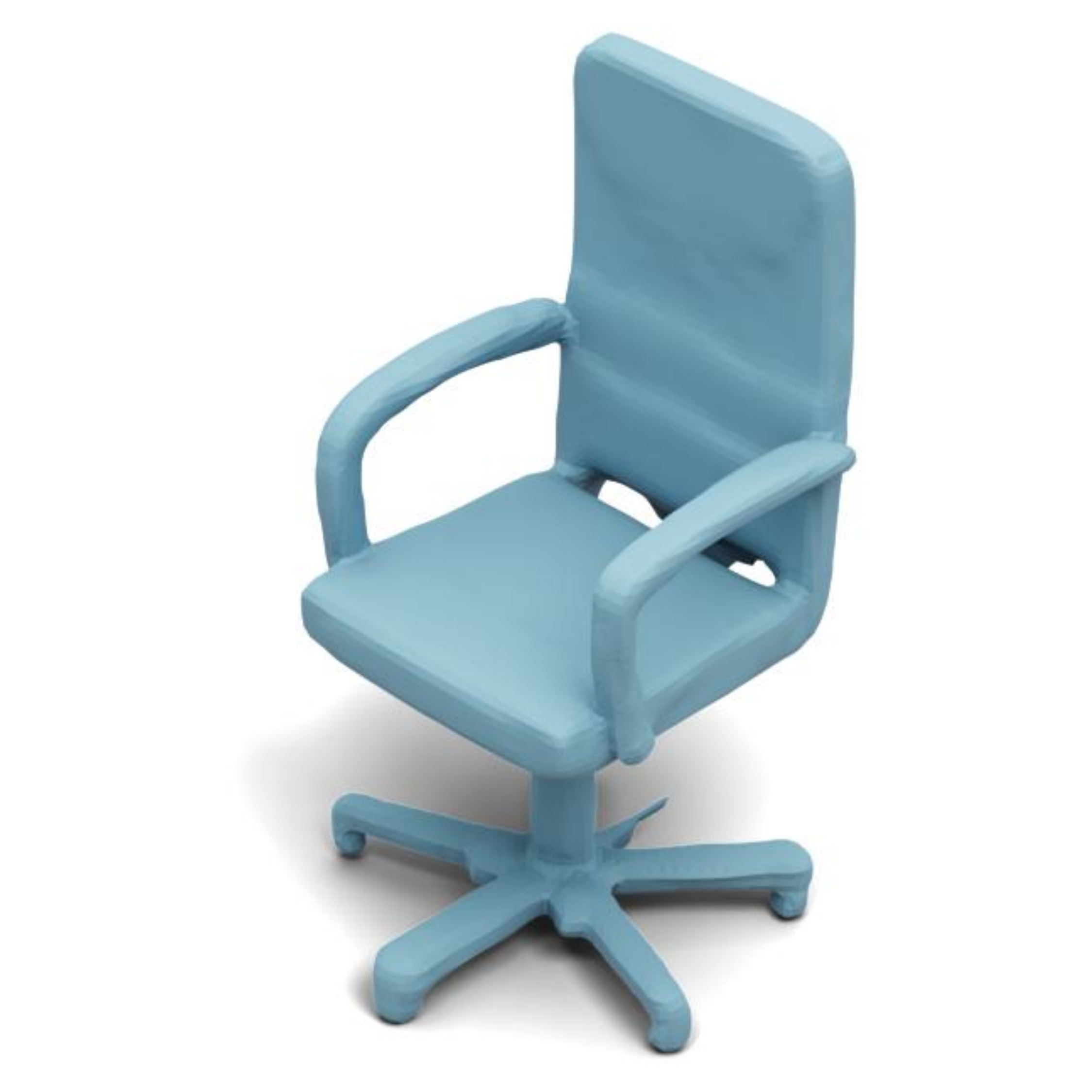}
    & \includegraphics[trim = 50 0 50 0, clip, width=.2\linewidth]{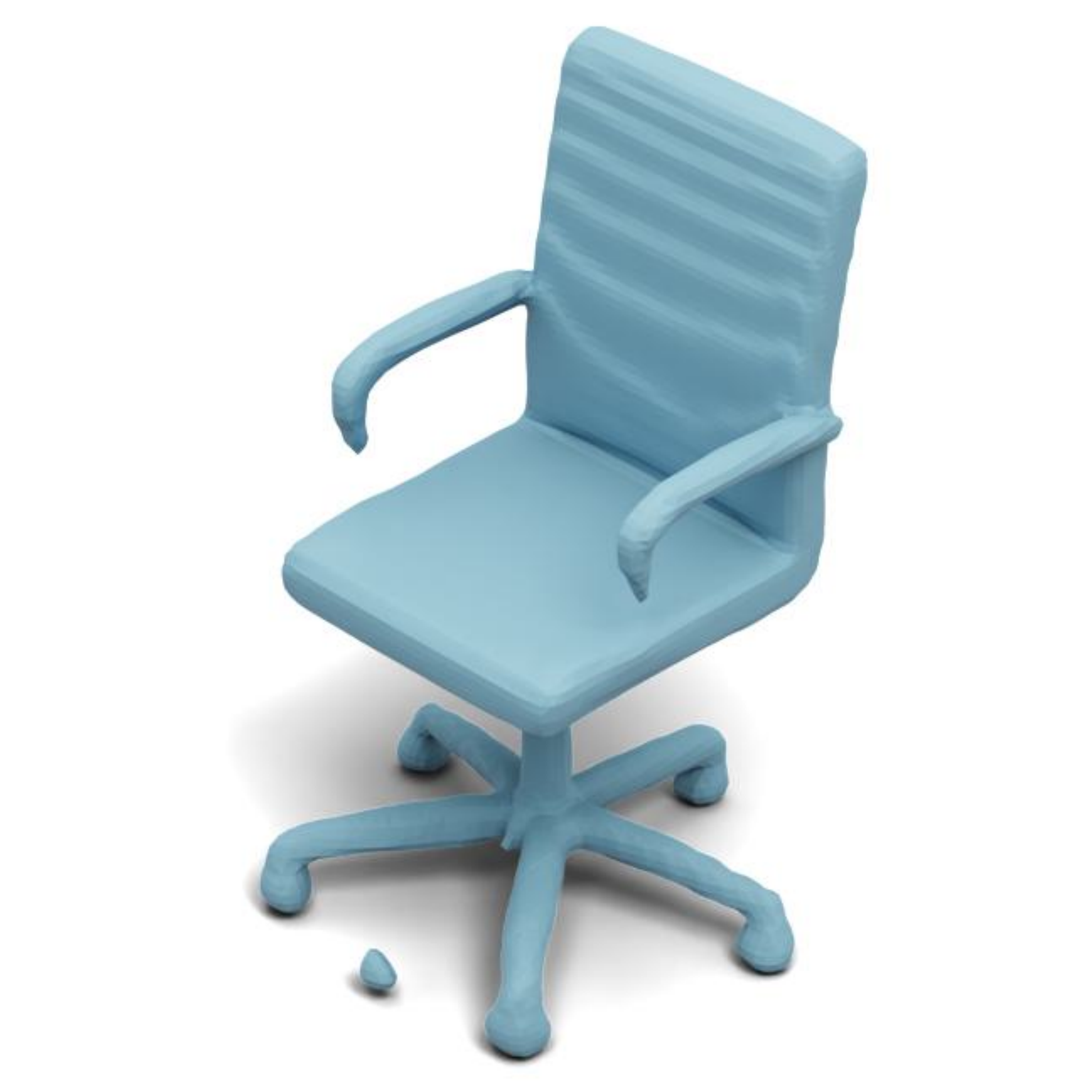}
    & \includegraphics[trim = 50 0 50 0, clip, width=.2\linewidth]{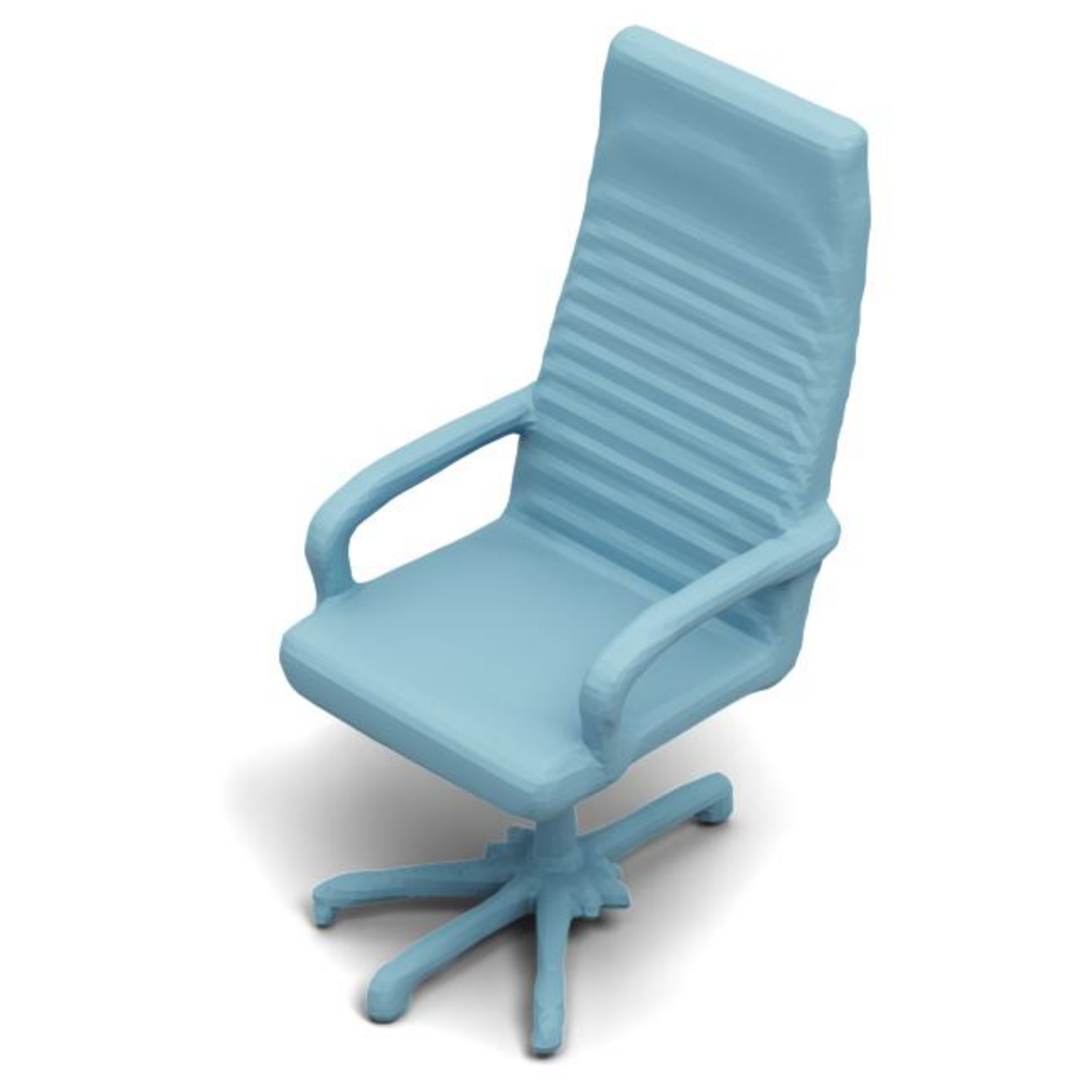}\\
    \includegraphics[width=.13\linewidth]{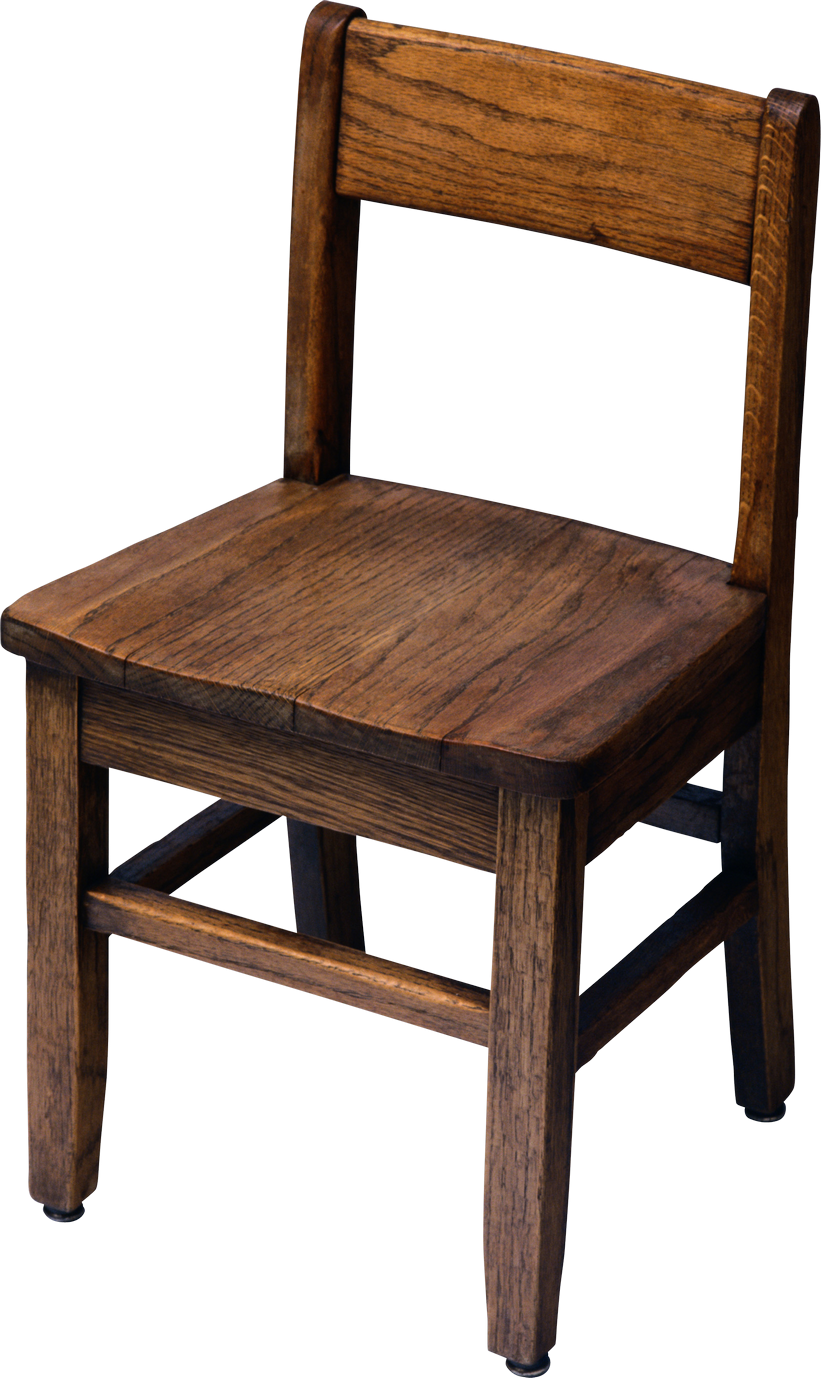}& 
    \includegraphics[trim = 50 0 50 0, clip, width=.2\linewidth]{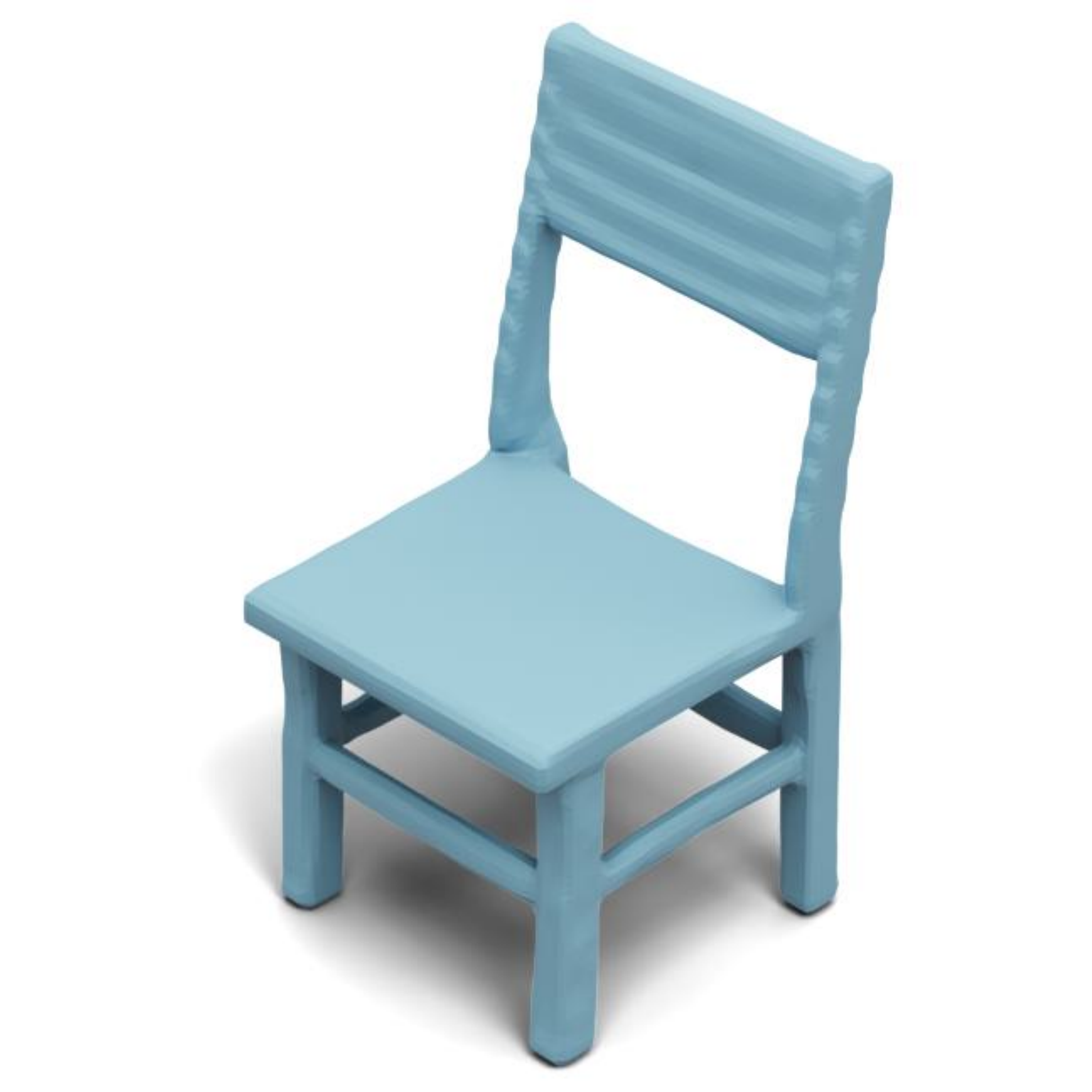}
    & \includegraphics[trim = 50 0 50 0, clip, width=.2\linewidth]{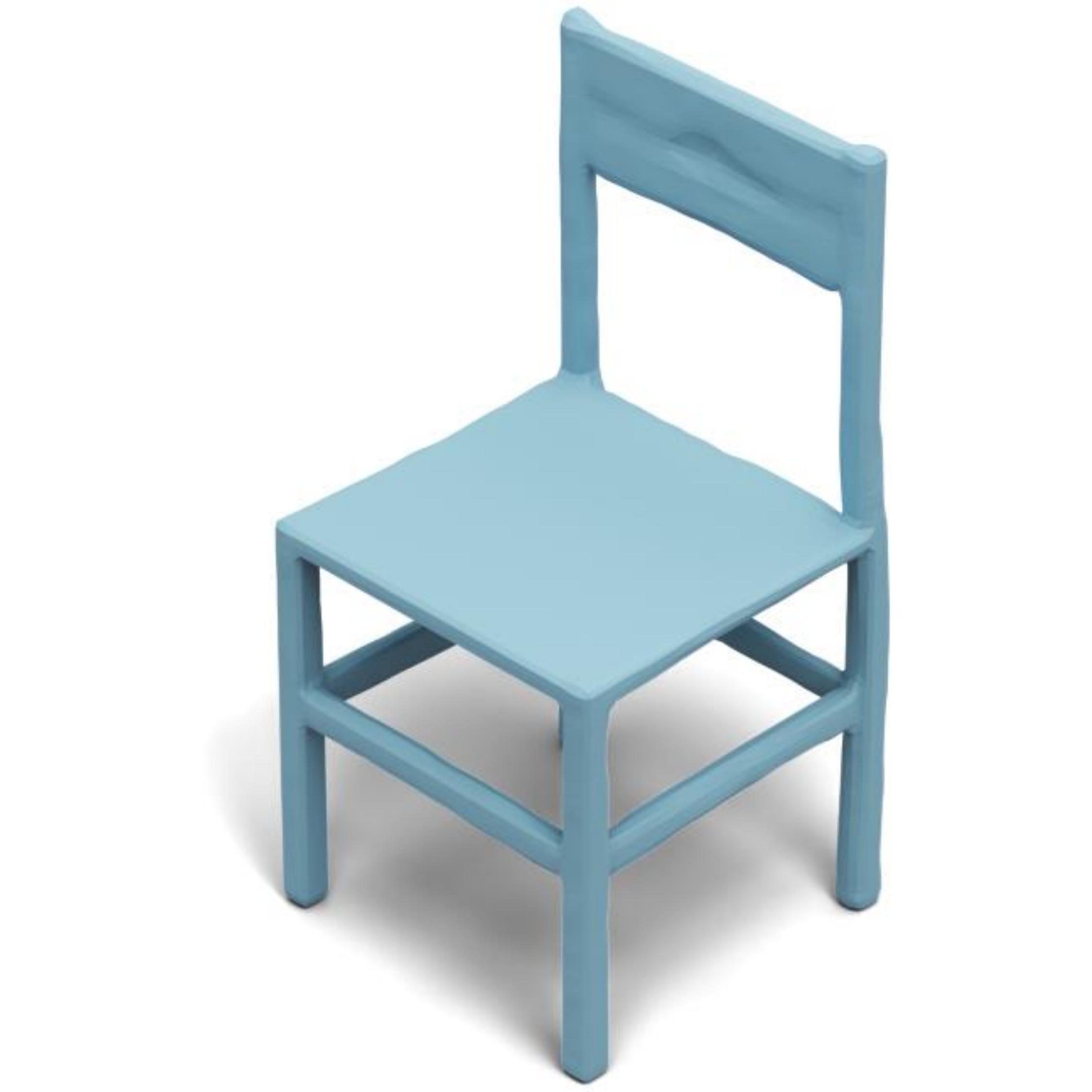}
    & \includegraphics[trim = 50 0 50 0, clip, width=.2\linewidth]{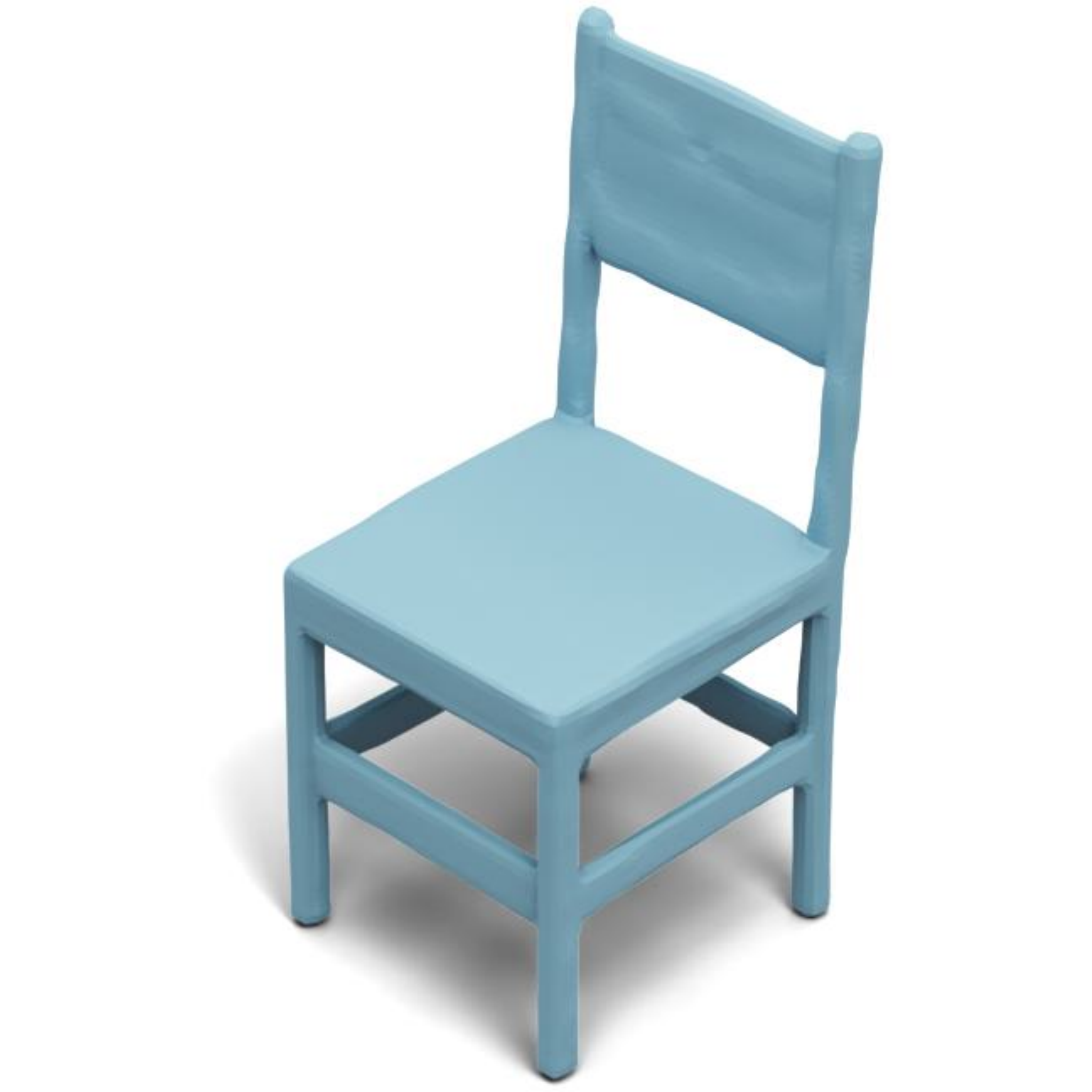}
    & \includegraphics[trim = 50 0 50 0, clip, width=.2\linewidth]{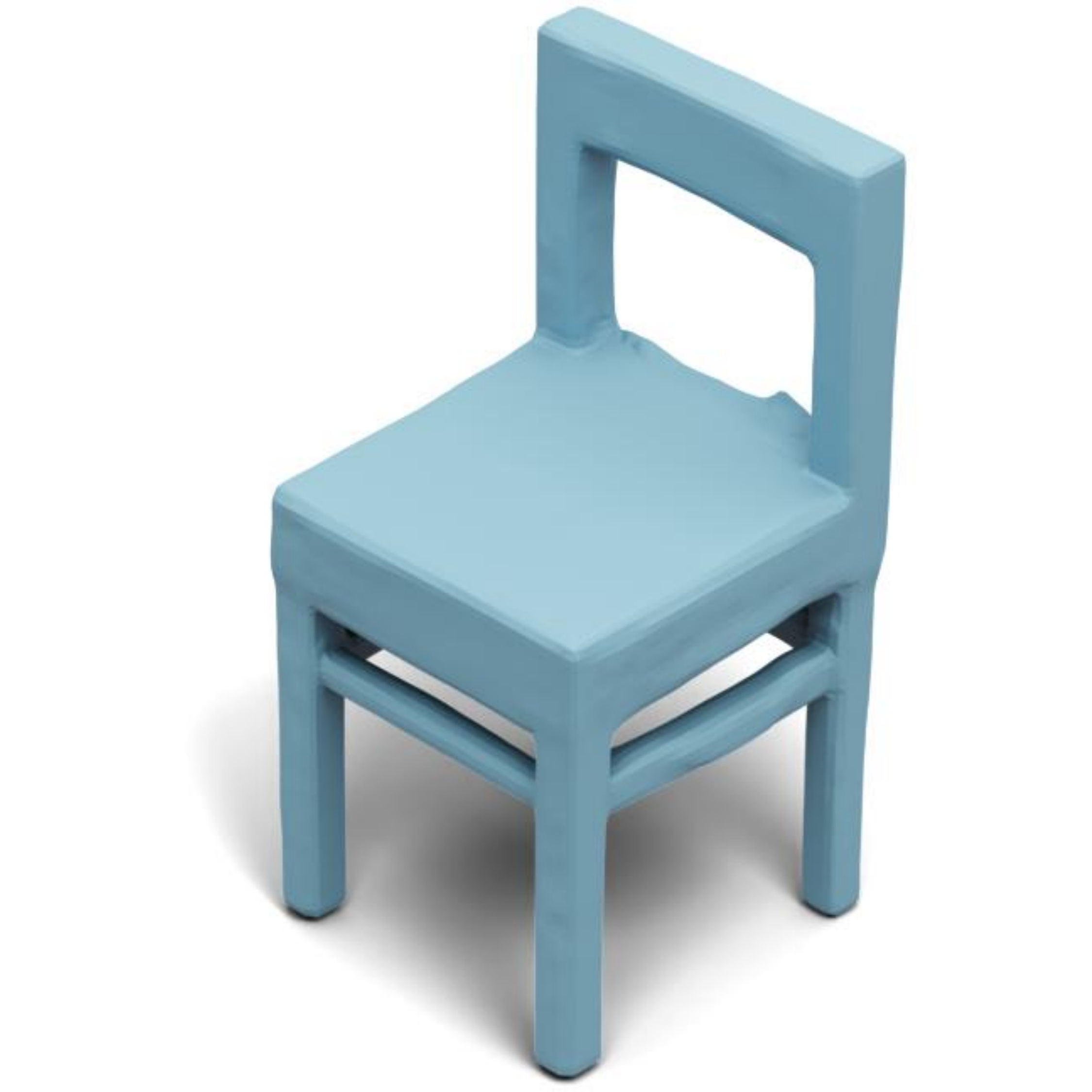}\\
  \end{tabular}   
	\caption{The three target shape images are displayed in the first column, with four attempts to model them during Task 2 of the usability study. The target shapes are sourced from the public domain.}
	\label{fig:task2}
 \vspace{-0.3cm}
\end{figure}

To evaluate the usability of our sketch-to-shape generation and editing methods, we carried out a usability study, drawing inspiration from the study presented in GA-Sketching \cite{zhou2023gasketching}. Eight participants from diverse backgrounds participated in the study. Among them, half were aged between 20 and 30, while the rest were above 30. The gender distribution was balanced, with 50\% women and 50\% men. In terms of 3D modeling experience, 25\% reported having no experience, 50\% had limited experience, and 25\% identified as hobbyists. When it came to 2D sketching or drawing, half the participants had no experience, 25\% reported limited experience, and 25\% described themselves as hobbyists. Notably, none of the participants were professional 2D illustrators or 3D artists.
The modeling session was divided into two phases. Initially, participants were introduced to the software's operation and its various functionalities, which included sketch-to-shape generation, outline rendering, part-based modeling, and part refinement. Subsequently, participants undertook two tasks. In Task 1, they had the freedom to sketch any chair design; however, they were required to use each of the software's functionalities at least once during the session, ensuring they became familiar with all available options. Task 2 involved modeling three specific shapes provided as reference images. While their sketches did not need to align with the image's perspective, the resulting shapes should closely resemble the target. The outcomes from both tasks are depicted in \figref{fig:task1} and \figref{fig:task2}. The outcomes of Task 1 underscore the system's resilience and adaptability. Even when participants, some of whom lacked advanced drawing skills, sketched rudimentary or imprecise chair designs, the algorithm consistently produced coherent 3D shapes. Often, only a few additional intuitive modeling steps were needed to refine the shape. Task 2 further demonstrates the system's ability to convert target ideas into concrete 3D models. Participants were able to transform target images into 3D chairs, even when the sketched perspectives differed from the reference images. This ease of transformation from a 2D reference image to a realistic 3D chair model accentuates the system's ability in bringing users' visions to realization.

\begin{figure}[t]
	\centering
	\includegraphics[trim = 80 30 140 30, clip, width=.85\linewidth]{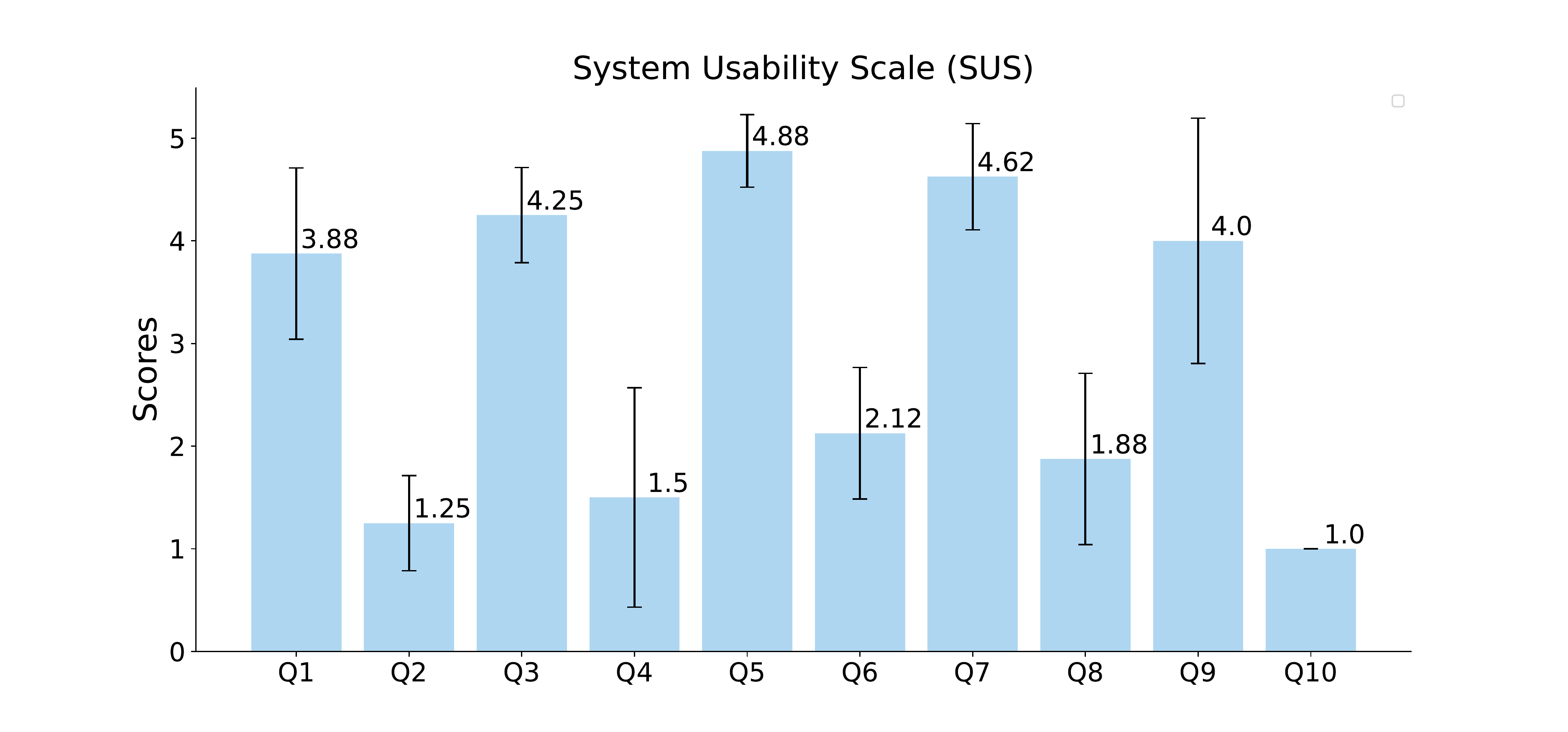}
	\caption{The mean of SUS scores. The whiskers represent the standard deviation. For questions with odd index, higher scores indicate better performance; for even-numbered questions, lower scores are preferable.}
	\label{fig:SUS}
 \vspace{-0.3cm}
\end{figure}

\begin{figure}[t]
	\centering
	\includegraphics[trim = 80 30 140 30, clip, width=0.85\linewidth]{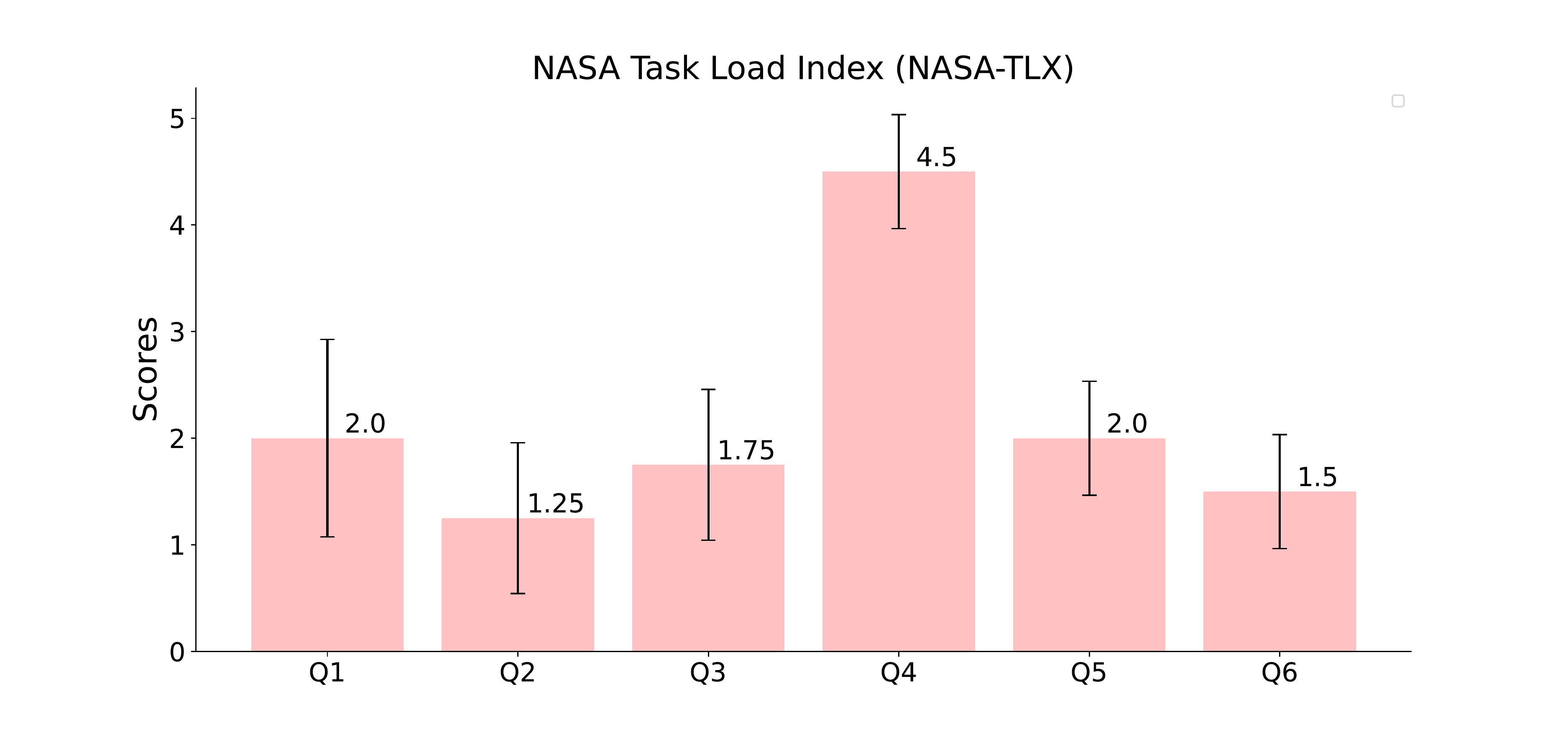}
	\caption{The mean of the NASA-TLX scores, which asks the participant to rate their experience according to six criteria to assess the intensity of the effort. The whiskers represent the standard deviation. The lower the better, except for Q4.}
	\label{fig:NASA-TLX}
 \vspace{-0.5cm}
\end{figure}

After completing the modeling session, participants were invited to complete a feedback form including both the System Usability Scale (SUS) questionnaire \cite{SUSJohn} and the NASA Task Load Index (NASA-TLX) questionnaire \cite{HART1988139}. The SUS questionnaire contains ten questions which evaluate the system's usability, and gauge its usefulness, ease of use, and consistency. The NASA-TLX questionnaire is designed to measure task-related effort intensities, such as mental (Q1), physical (Q2), and temporal (Q3) demands, as well as performance (Q4), effort (Q5), and frustration levels (Q6). The results are shown in \figref{fig:SUS} and \figref{fig:NASA-TLX}. 
Notably, the exceptionally low SUS scores for Q2 and Q4, combined with elevated scores for Q5 and Q7, and notably the unanimous score of $1$ for Q10, suggest a high intuitiveness with the editing options. This observation is further corroborated by the low scores reflected in the NASA-TLX. The marginally subpar scores for Q6 and Q9 appear to align with the absence of very high-frequency details from sketches to the resulting shape, a limitation we acknowledge in the main paper. However, it is worth noting the significant elevation in the NASA-TLX Q4 score, implying participants' satisfaction with their performance. Participants could readily conceptualize an initial rudimentary shape, even from the most abstract sketches and for those with very limited experience.
\vspace{-0.3cm}

\subsection{Additional visual results}

\begin{figure*}[t]
	\centering
	\small
	\setlength{\tabcolsep}{1pt}
 \begin{tabular}{cccccccc}
\includegraphics[width=0.125\linewidth]{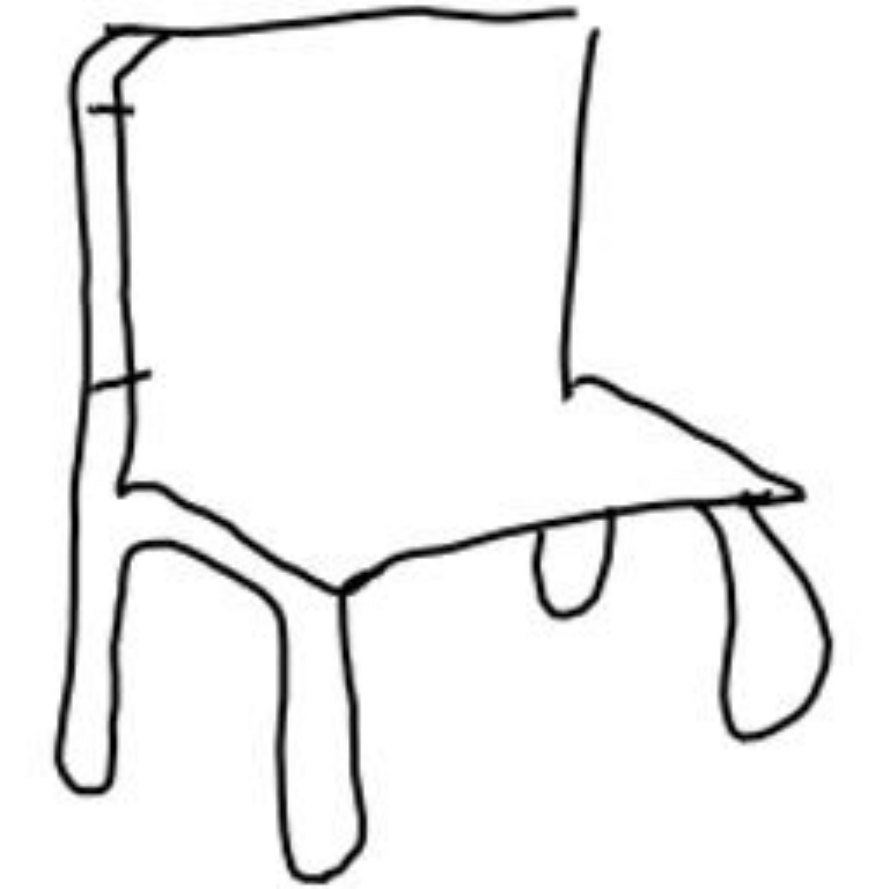}
&\includegraphics[width=0.125\linewidth]{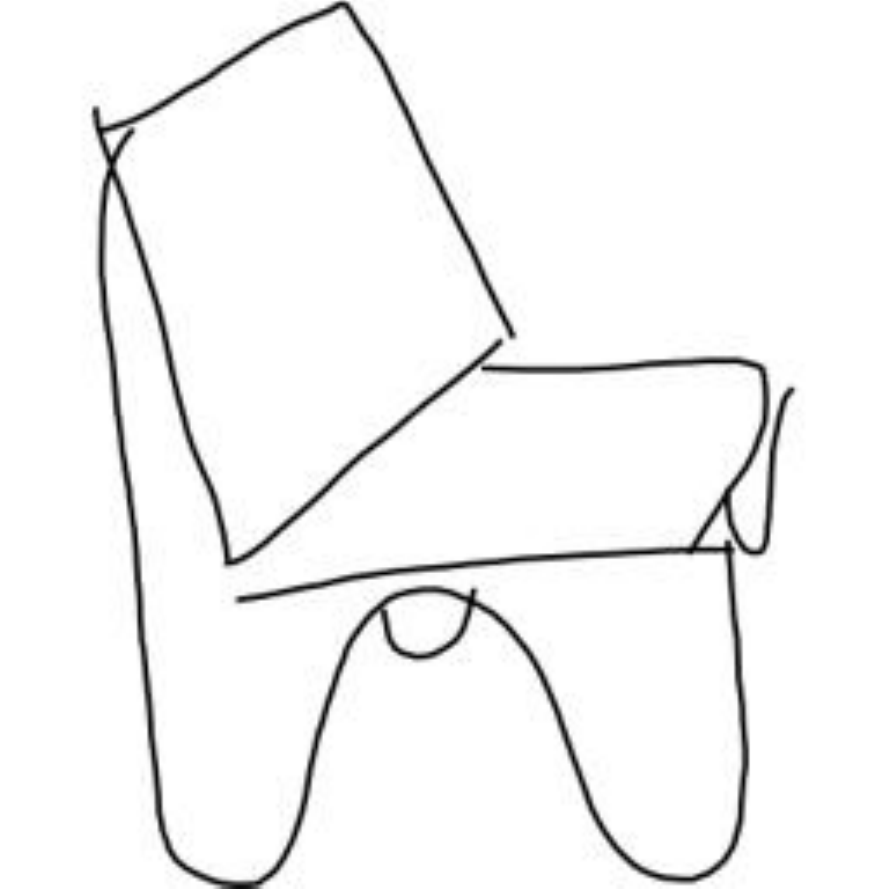}
&\includegraphics[width=0.125\linewidth]{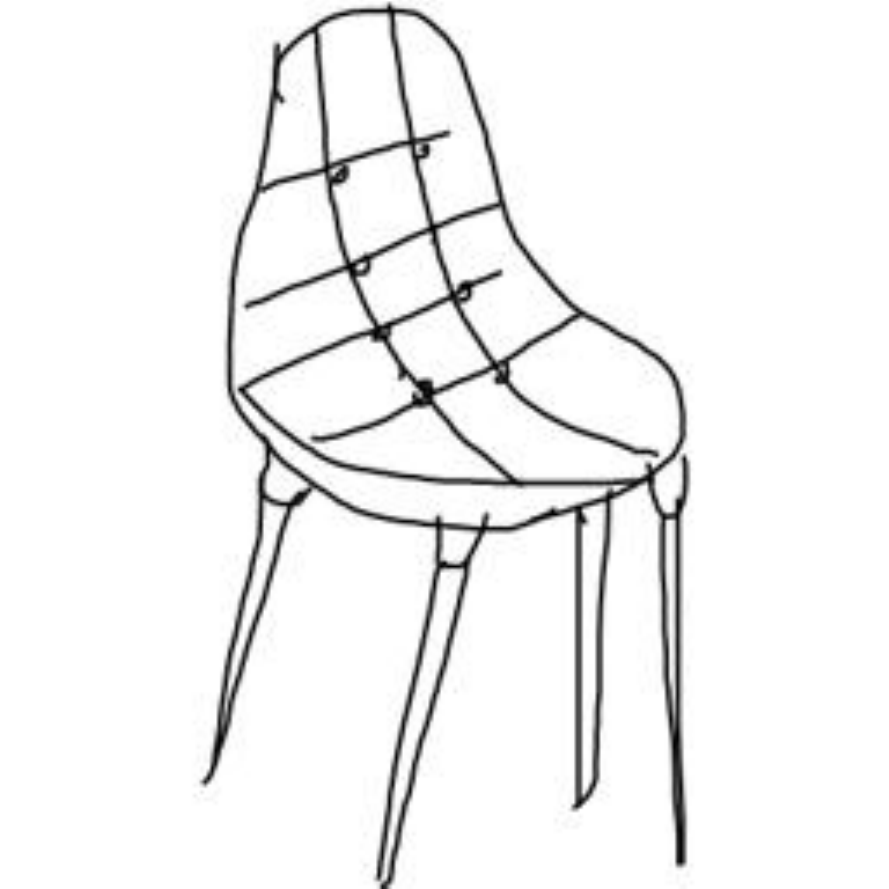}
&\includegraphics[width=0.125\linewidth]{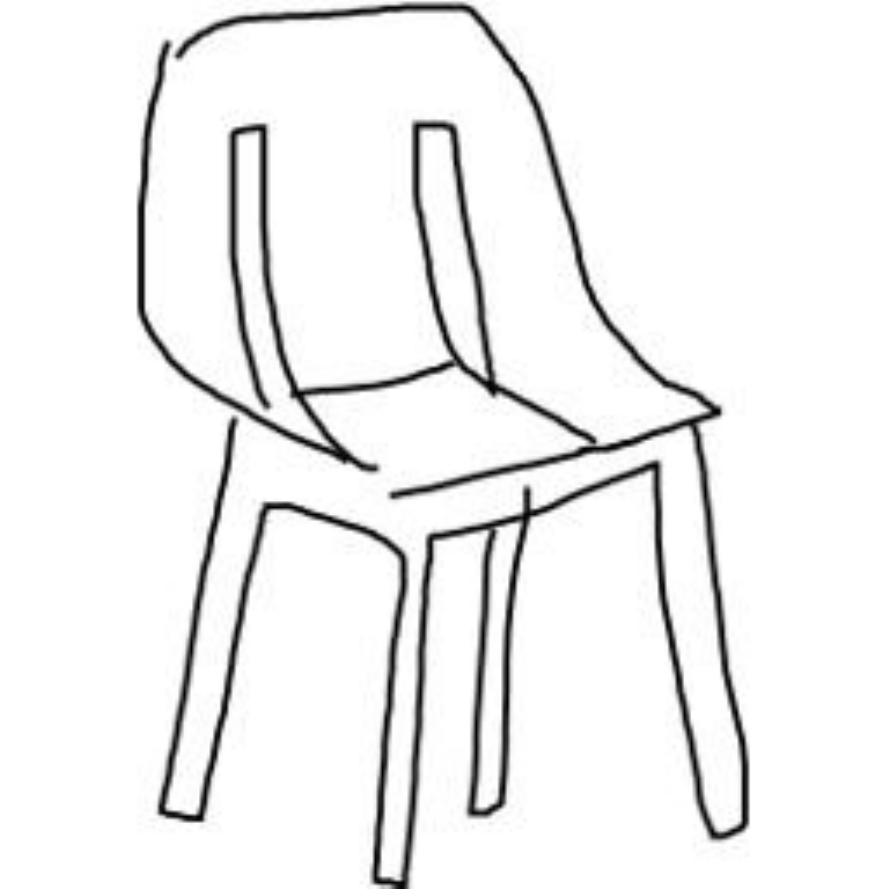}
&\includegraphics[width=0.125\linewidth]{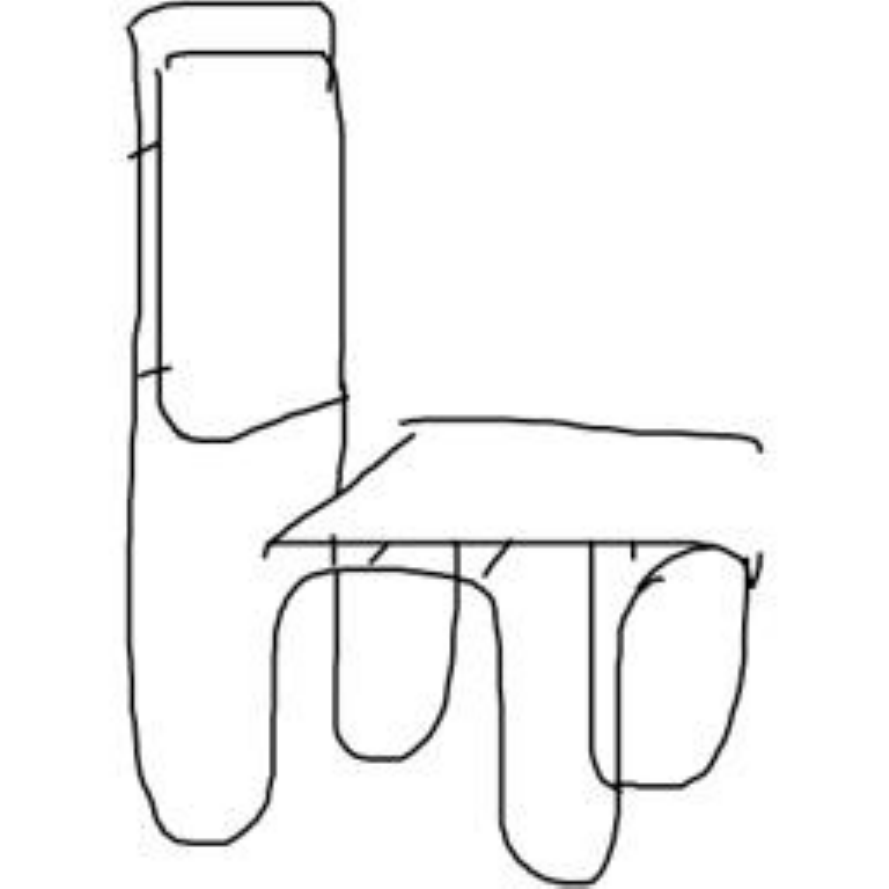}
&\includegraphics[width=0.125\linewidth]{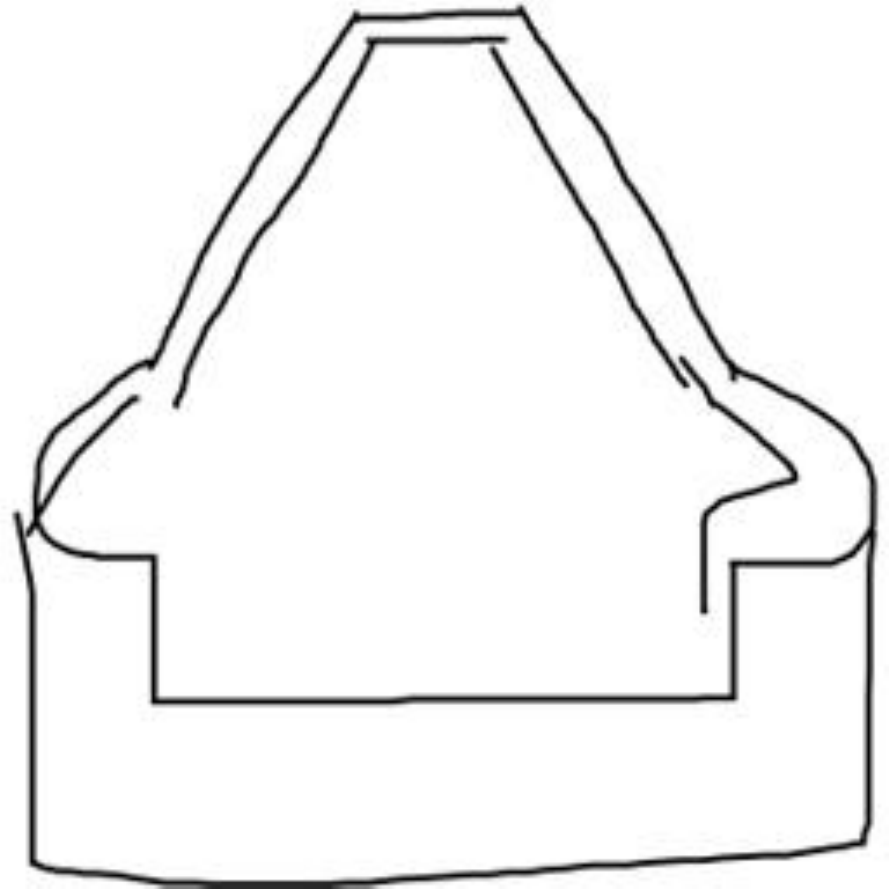}
&\includegraphics[width=0.125\linewidth]{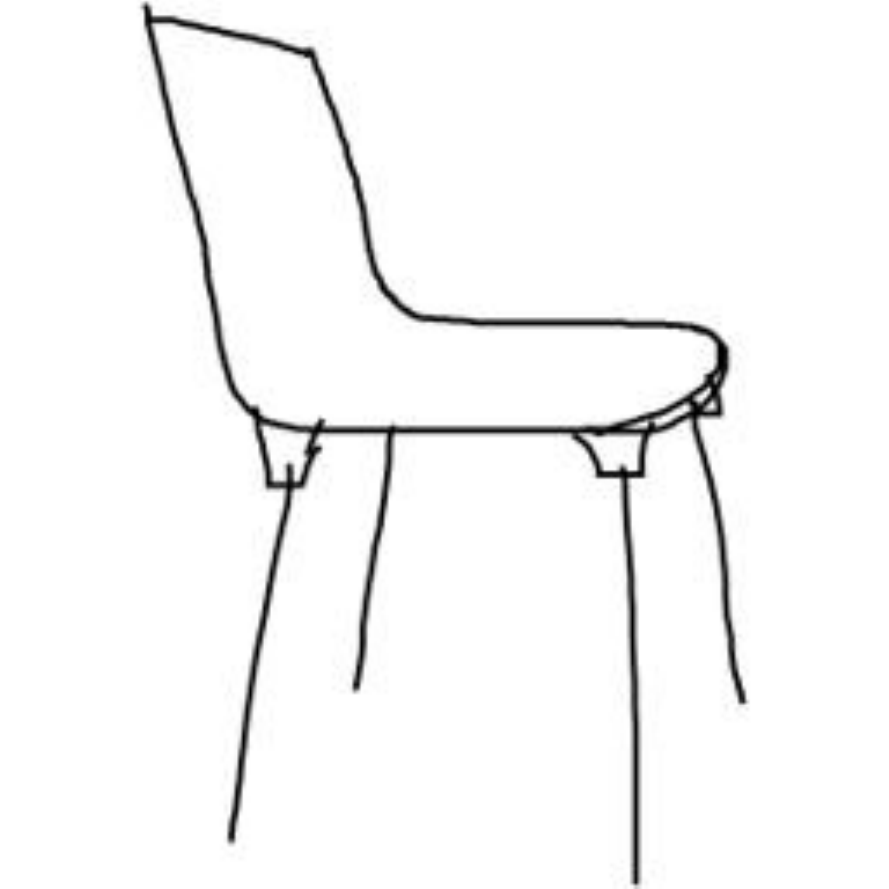}
&\includegraphics[width=0.125\linewidth]{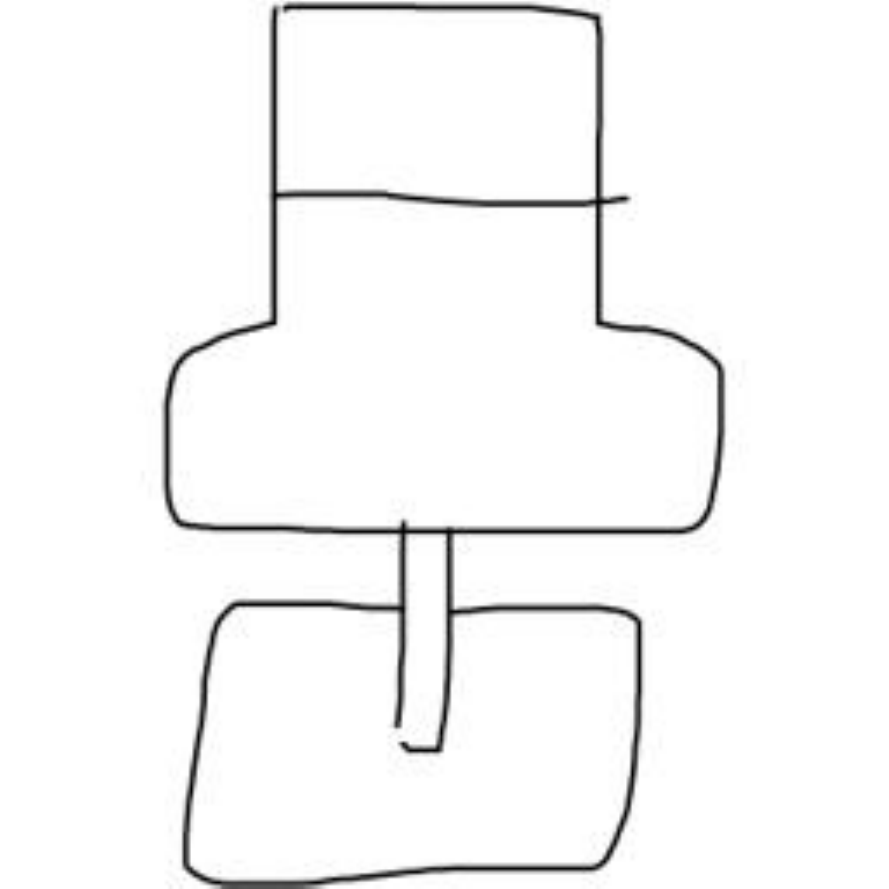}
\\
\includegraphics[trim = 1 1 1 1, clip, width=0.125\linewidth]{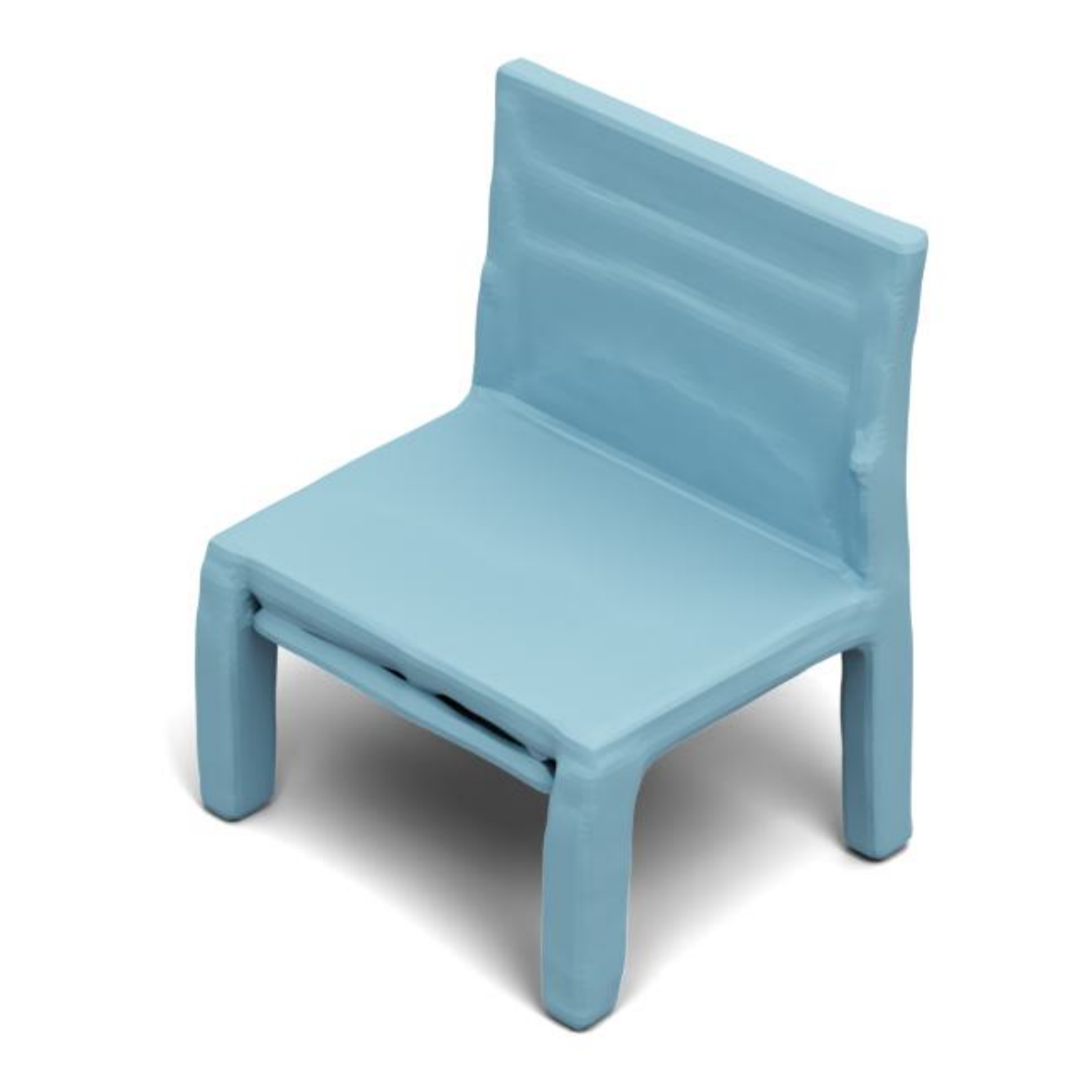}
&\includegraphics[trim = 1 1 1 1, clip, width=0.125\linewidth]{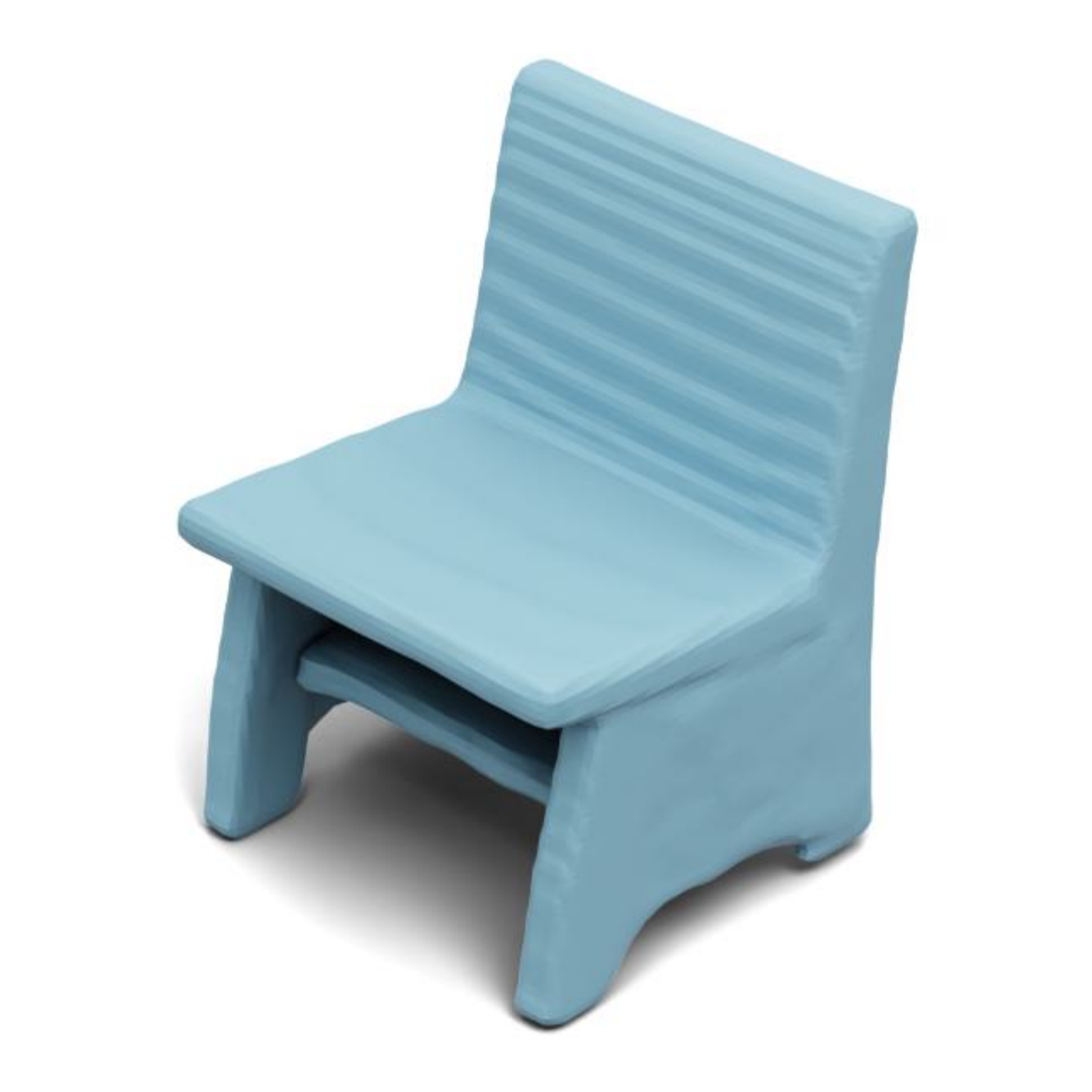}
&\includegraphics[trim = 1 1 1 1, clip, width=0.125\linewidth]{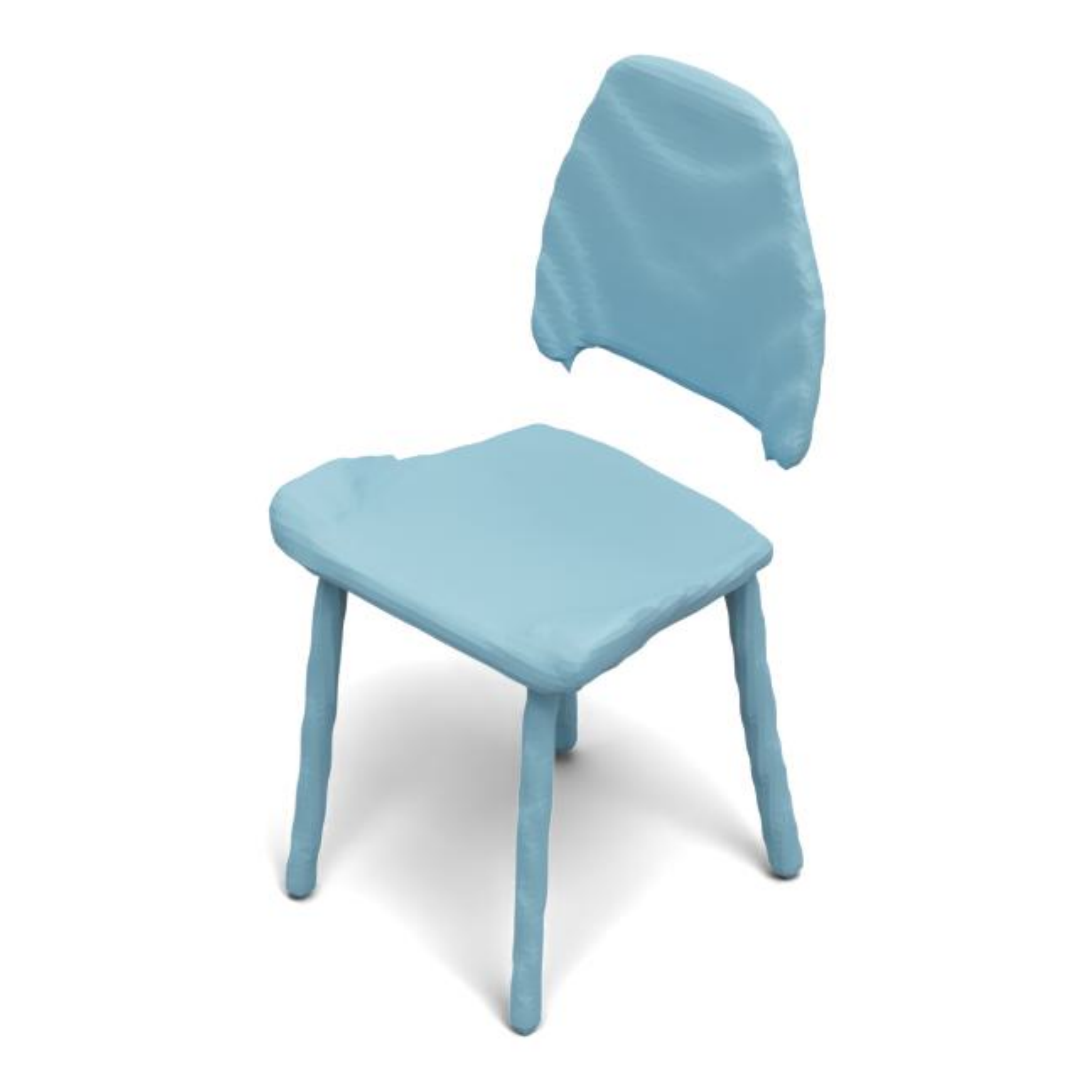}
&\includegraphics[trim = 1 1 1 1, clip, width=0.125\linewidth]{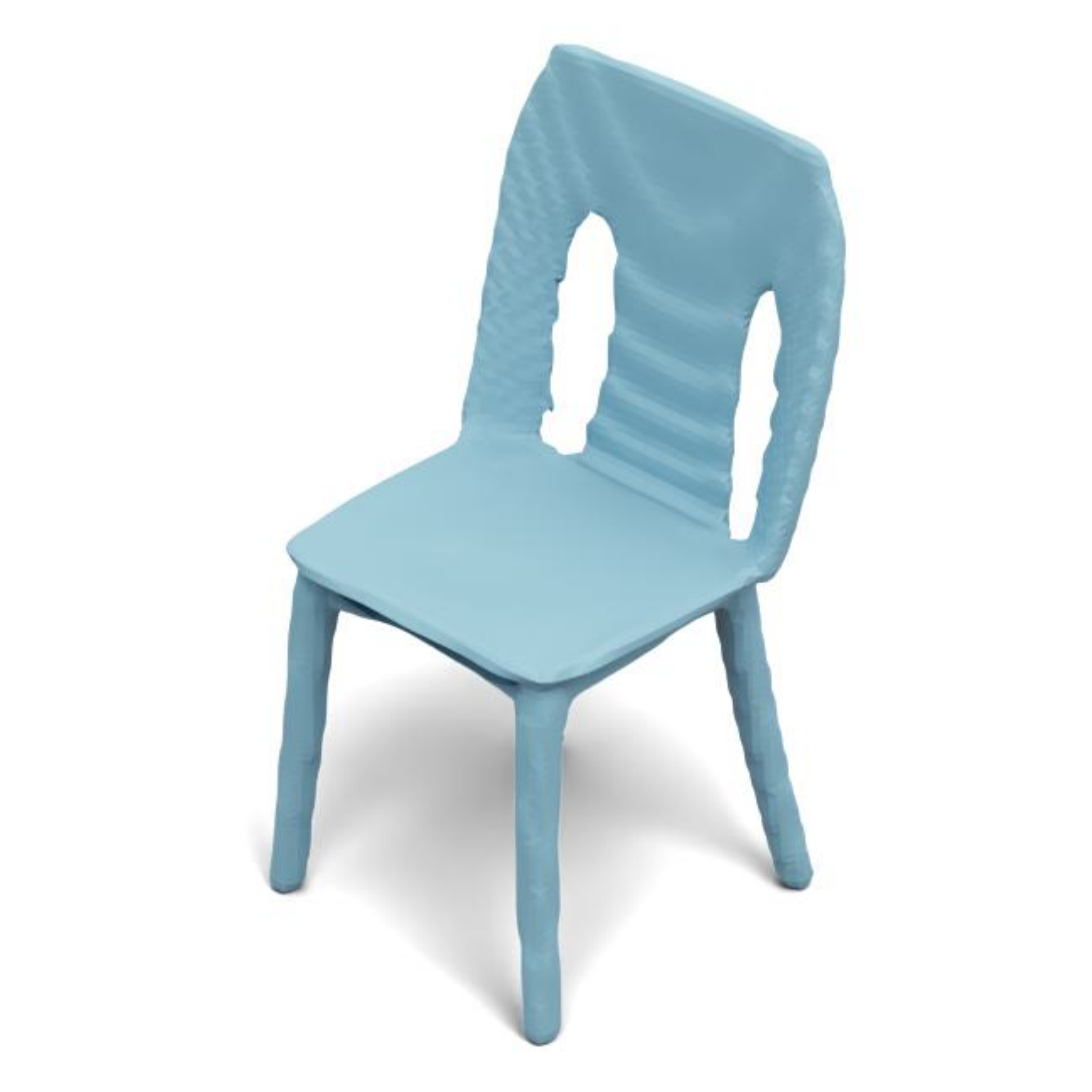}
&\includegraphics[trim = 1 1 1 1, clip, width=0.125\linewidth]{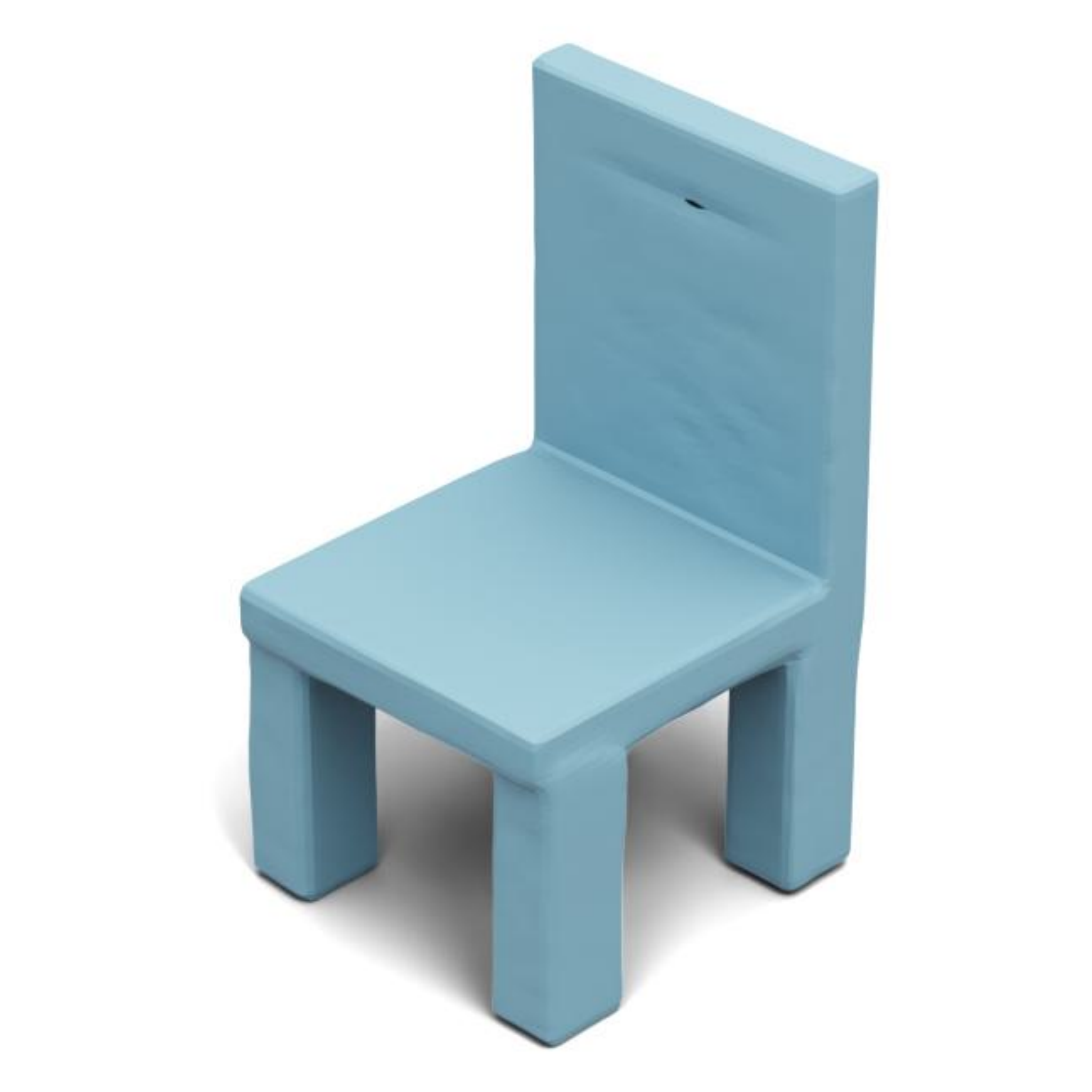}
&\includegraphics[trim = 1 1 1 1, clip, width=0.125\linewidth]{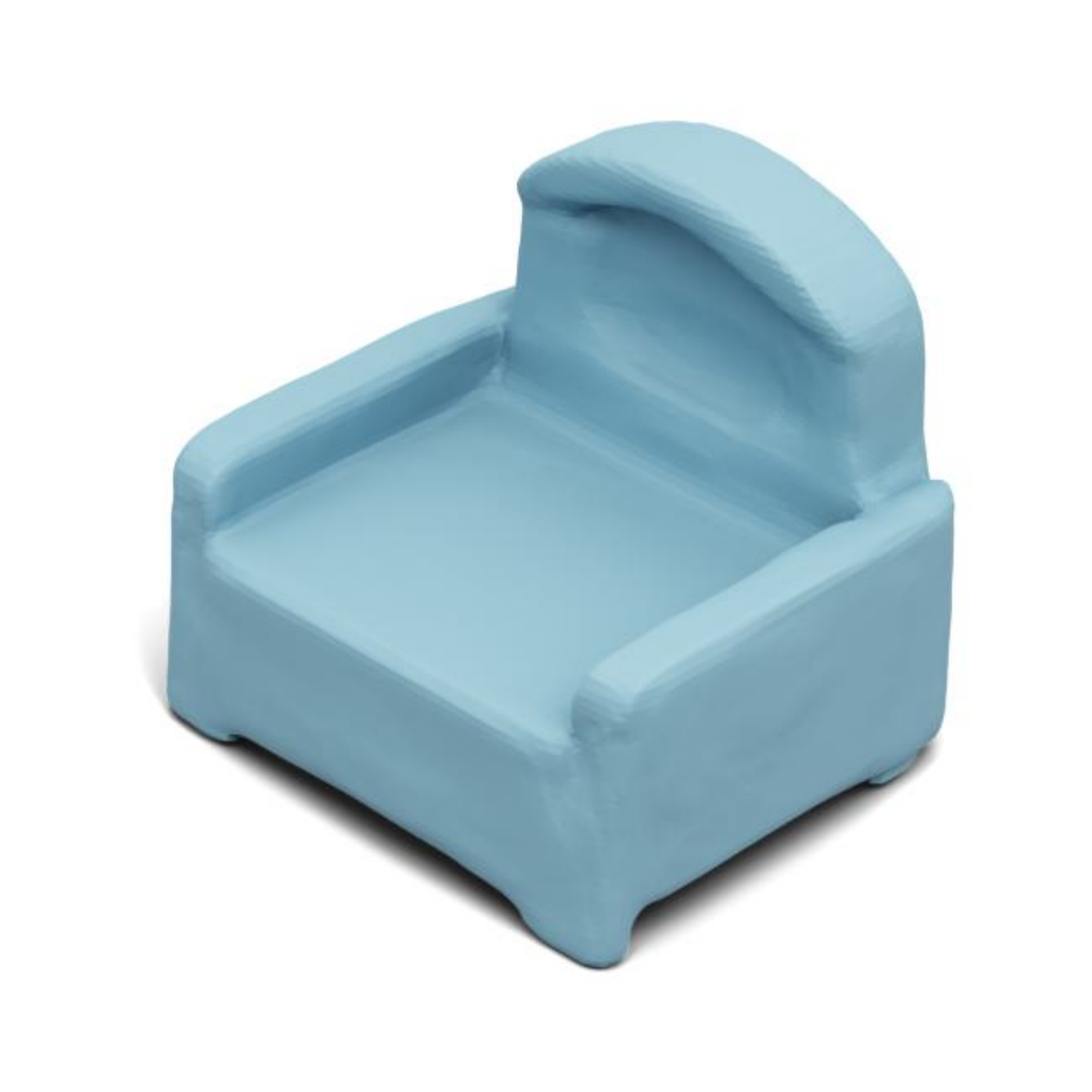}
&\includegraphics[trim = 1 1 1 1, clip, width=0.125\linewidth]{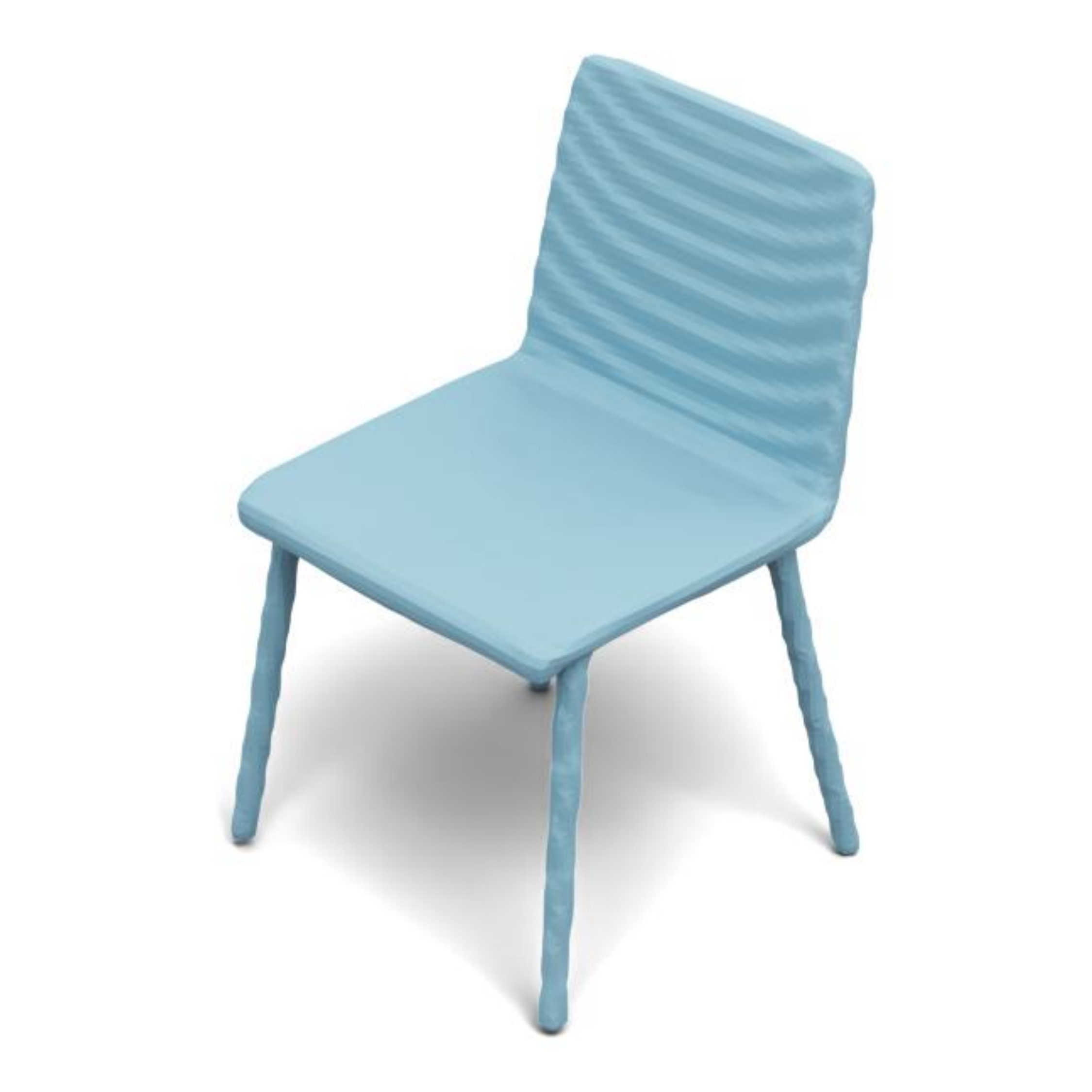}
&\includegraphics[trim = 1 1 1 1, clip, width=0.125\linewidth]{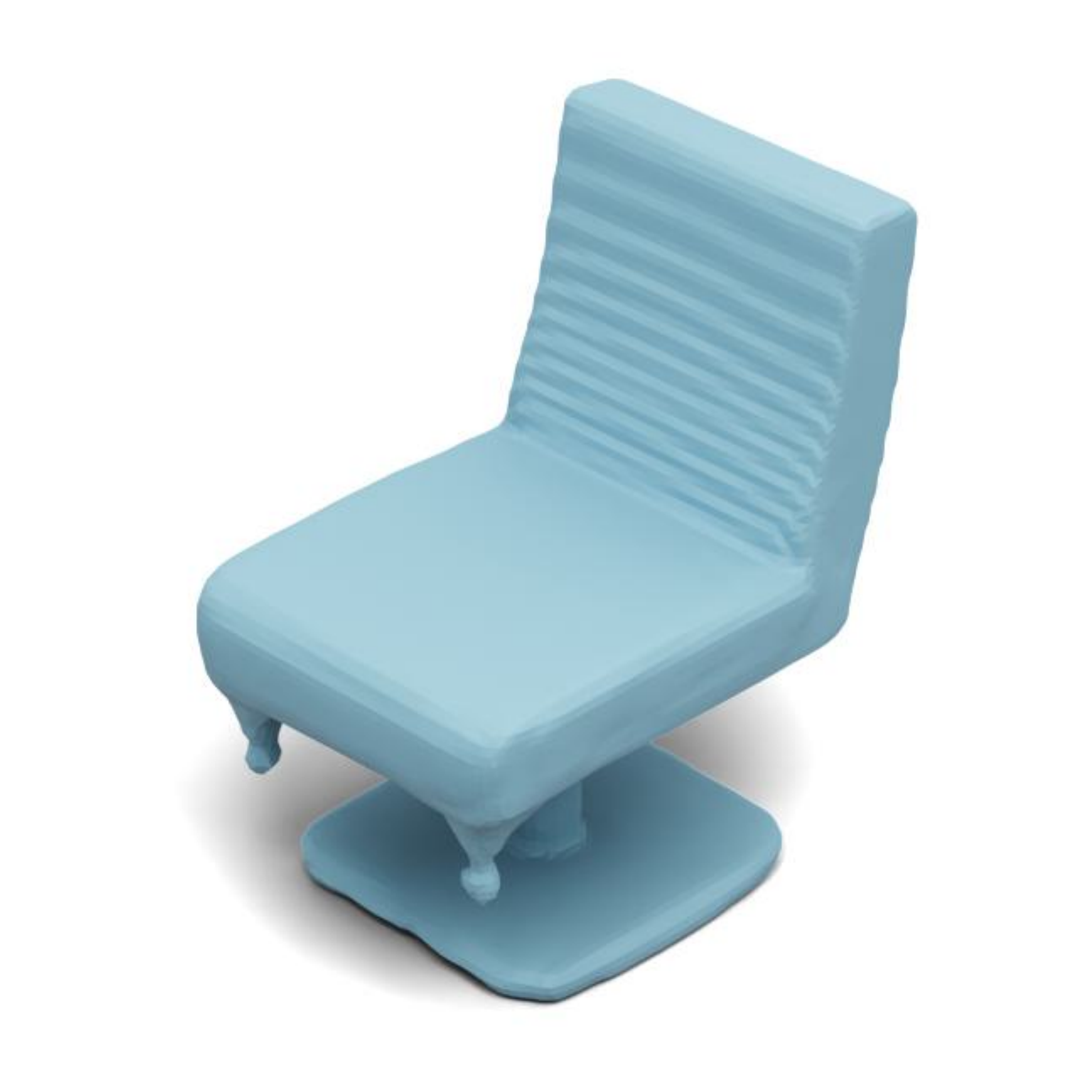}
\\
\includegraphics[width=0.125\linewidth]{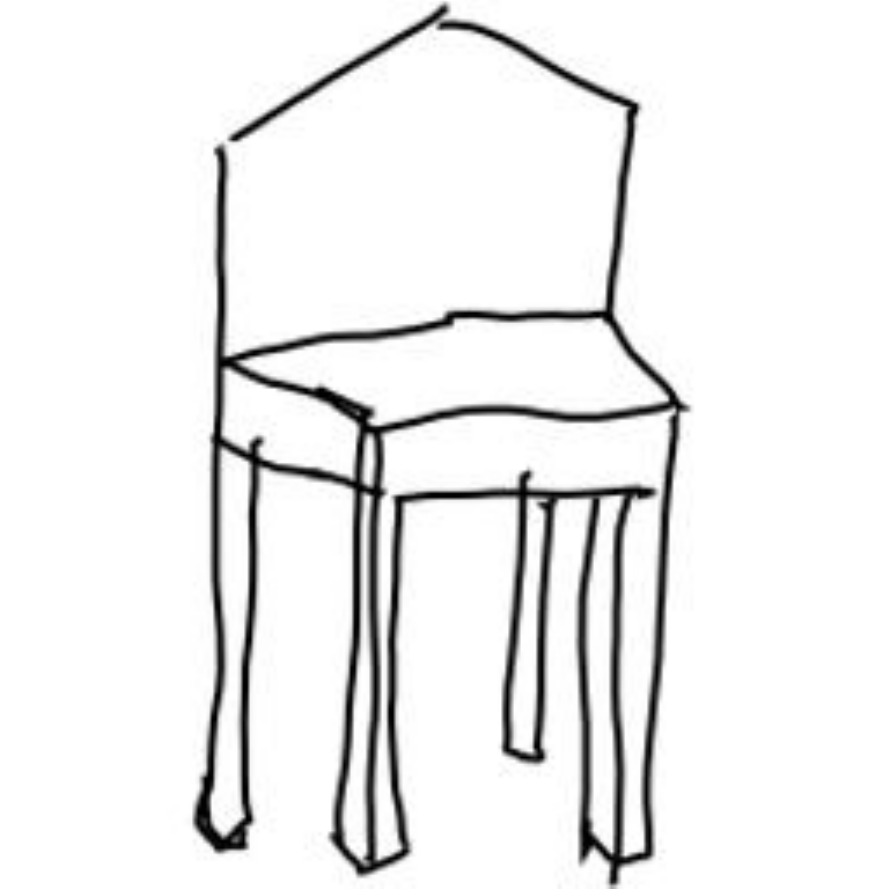}
&\includegraphics[width=0.125\linewidth]{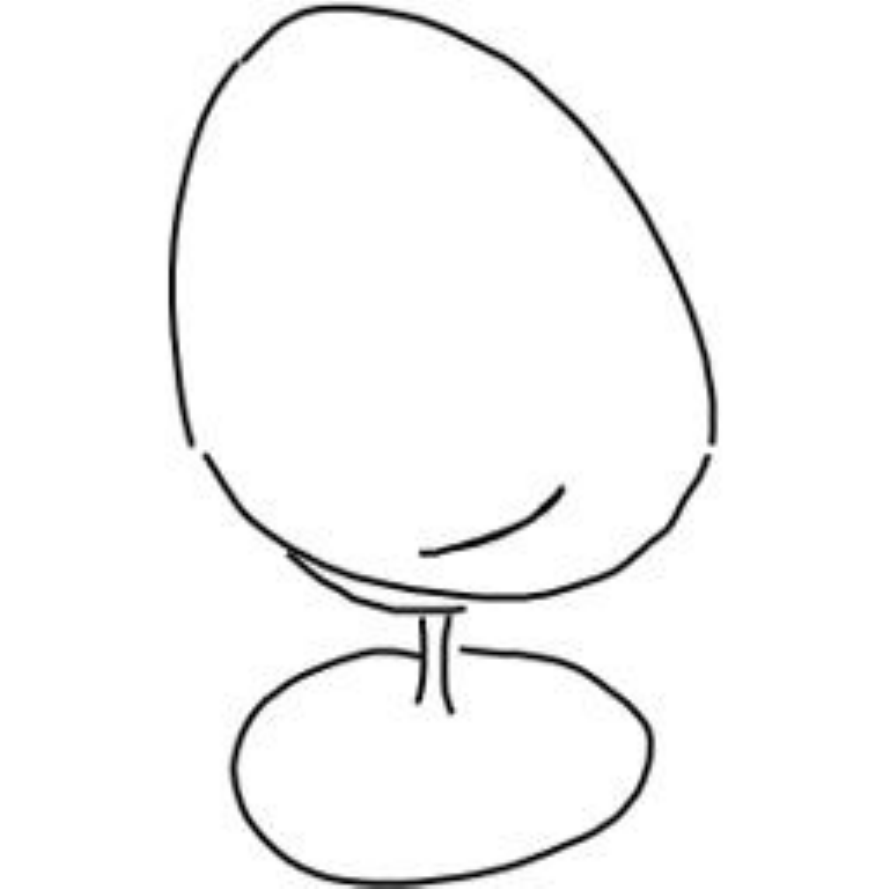}
&\includegraphics[width=0.125\linewidth]{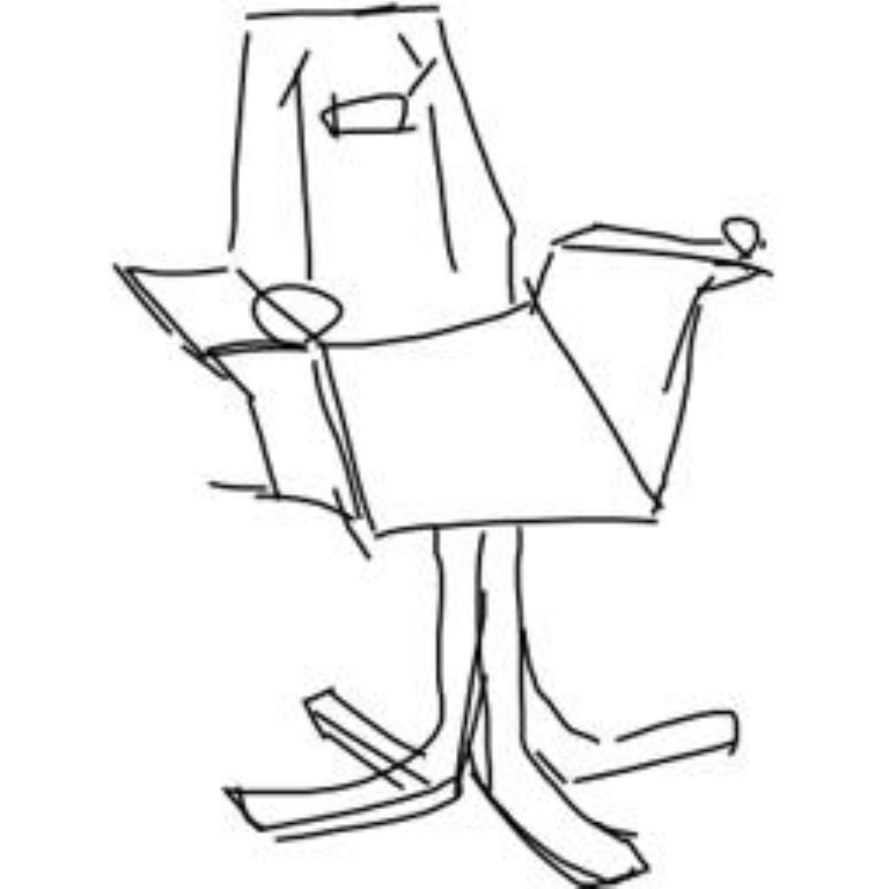}
&\includegraphics[width=0.125\linewidth]{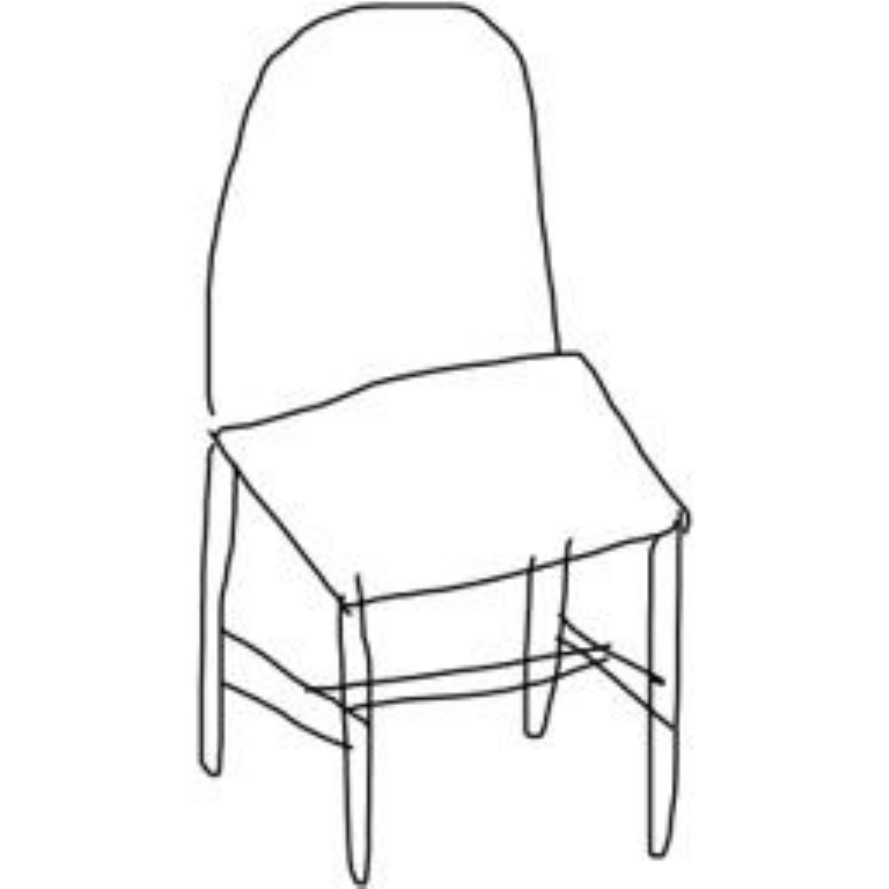}
&\includegraphics[width=0.125\linewidth]{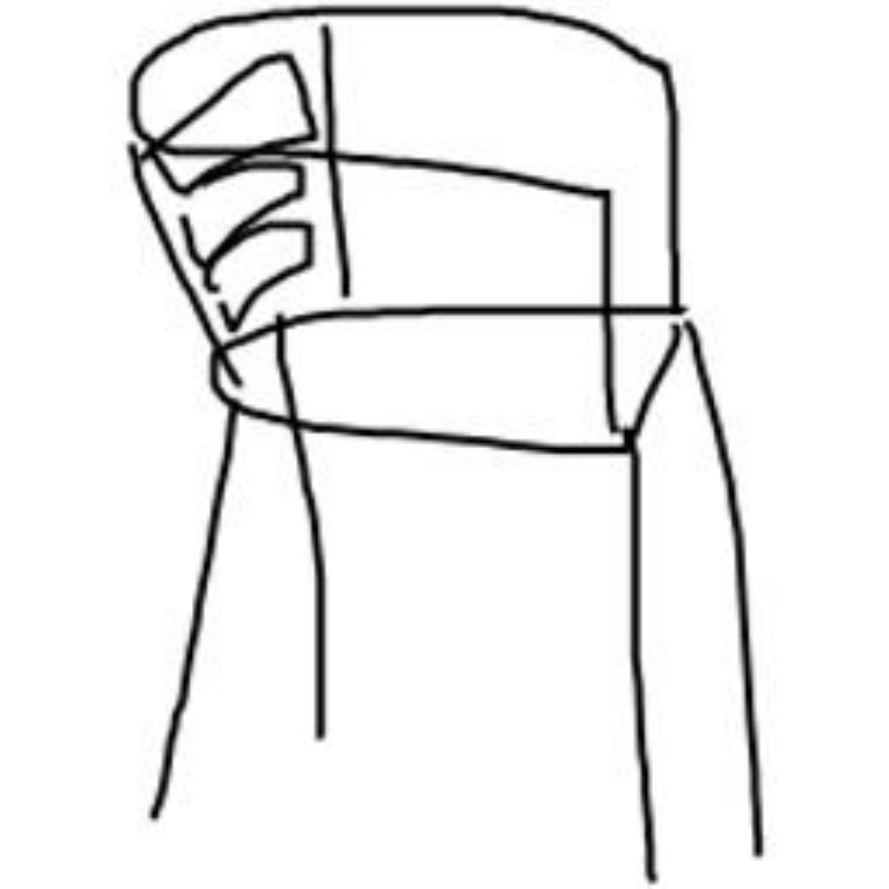}
&\includegraphics[width=0.125\linewidth]{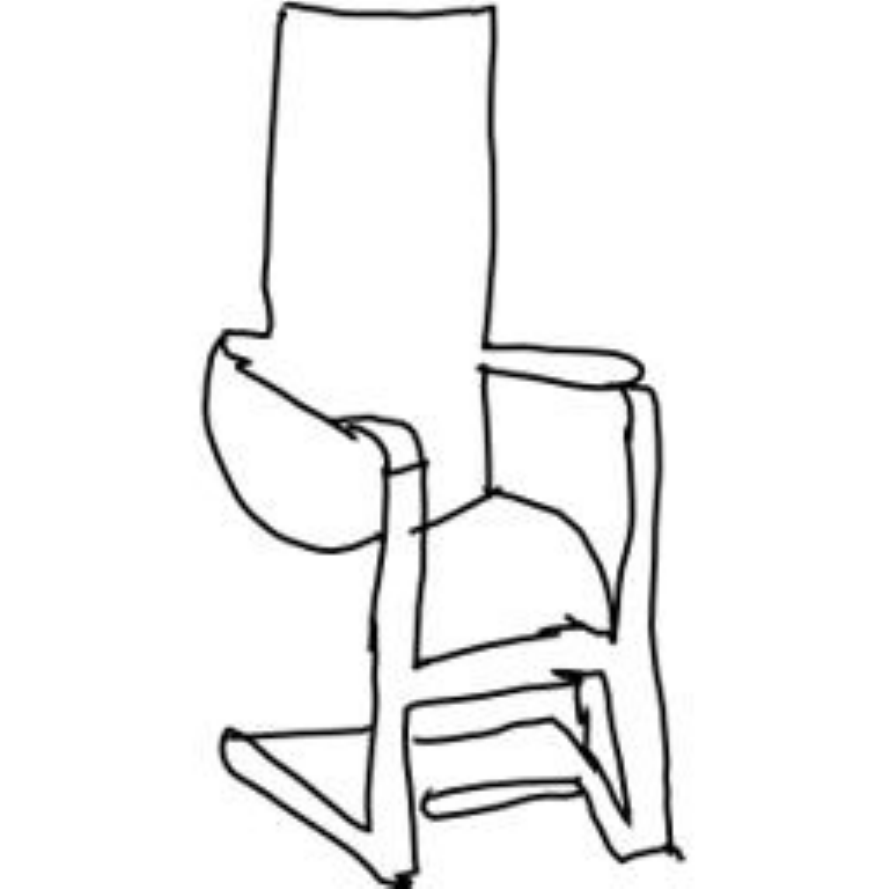}
&\includegraphics[width=0.125\linewidth]{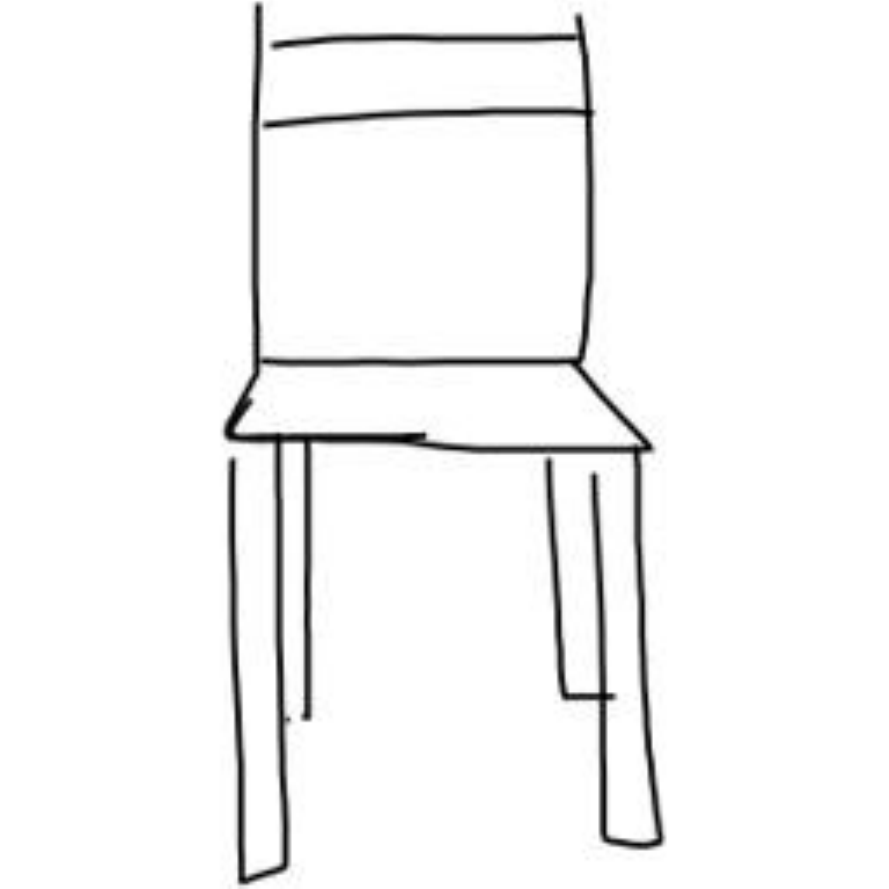}
&\includegraphics[width=0.125\linewidth]{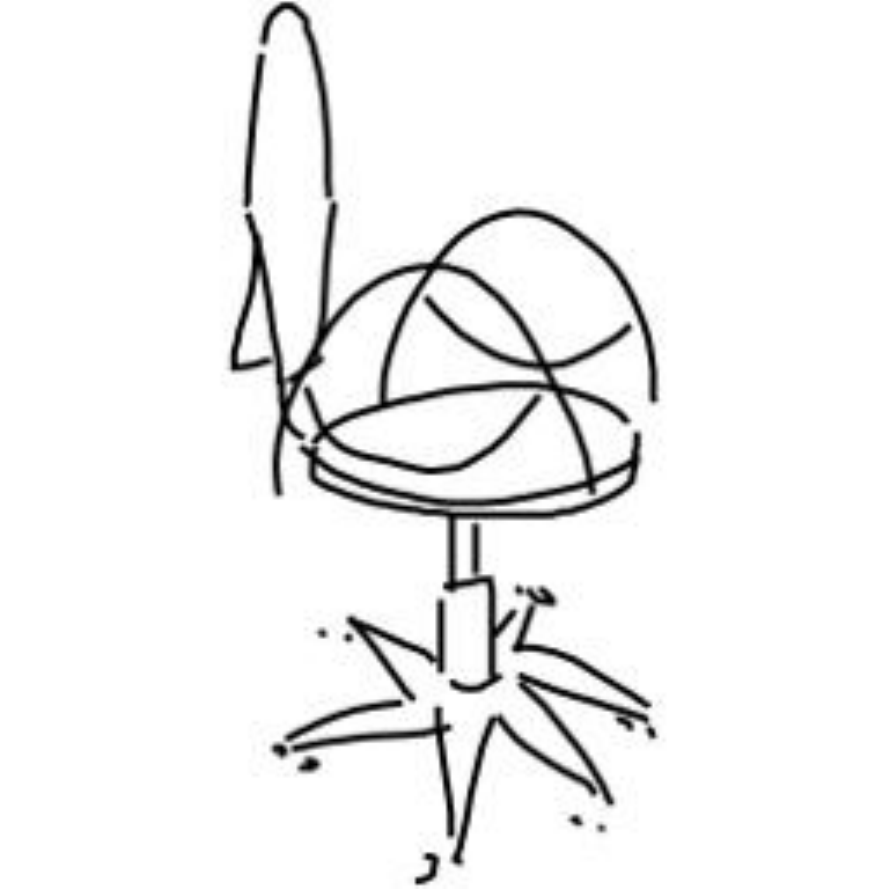}
\\
\includegraphics[trim = 1 1 1 1, clip, width=0.125\linewidth]{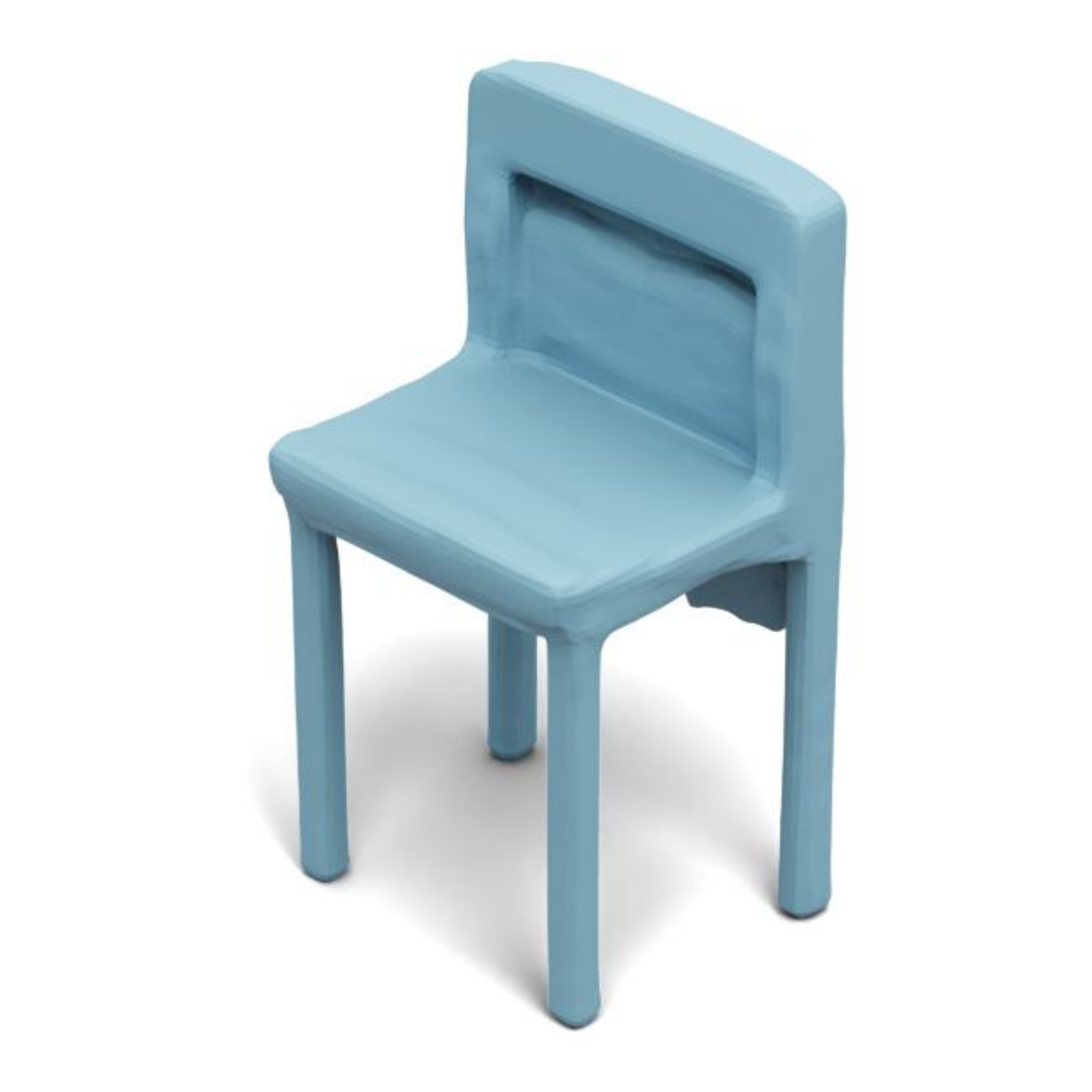}
&\includegraphics[trim = 1 1 1 1, clip, width=0.125\linewidth]{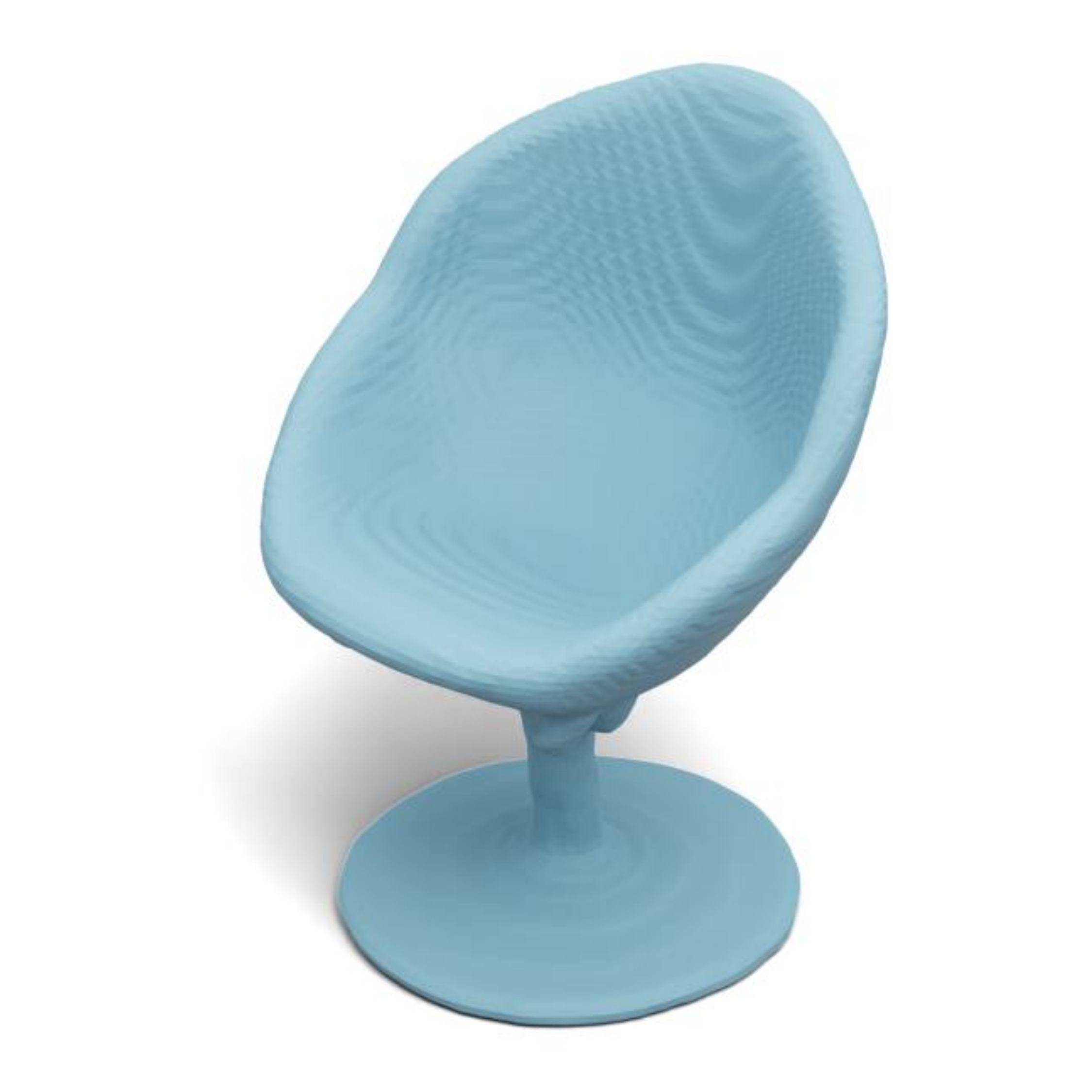}
&\includegraphics[trim = 1 1 1 1, clip, width=0.125\linewidth]{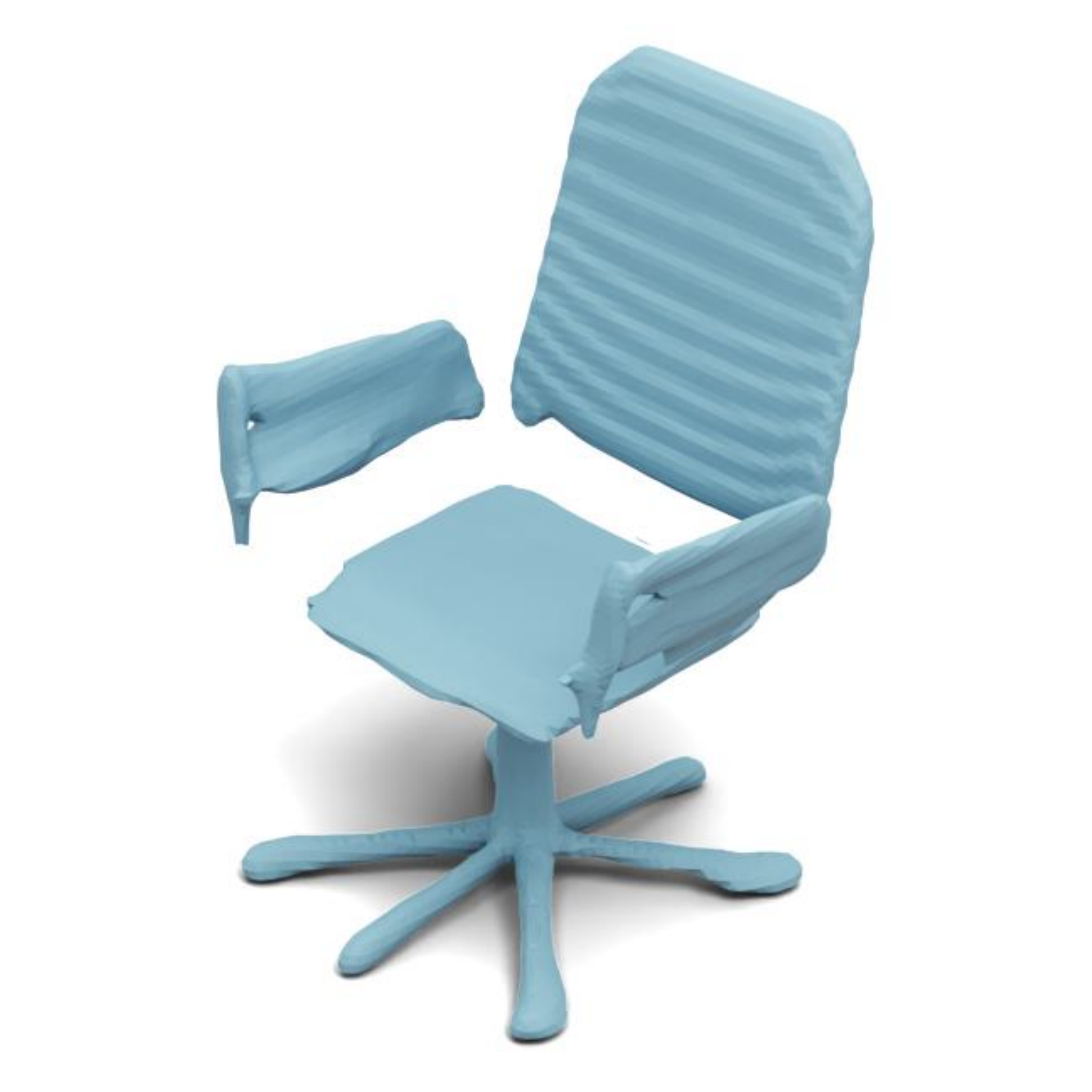}
&\includegraphics[trim = 1 1 1 1, clip, width=0.125\linewidth]{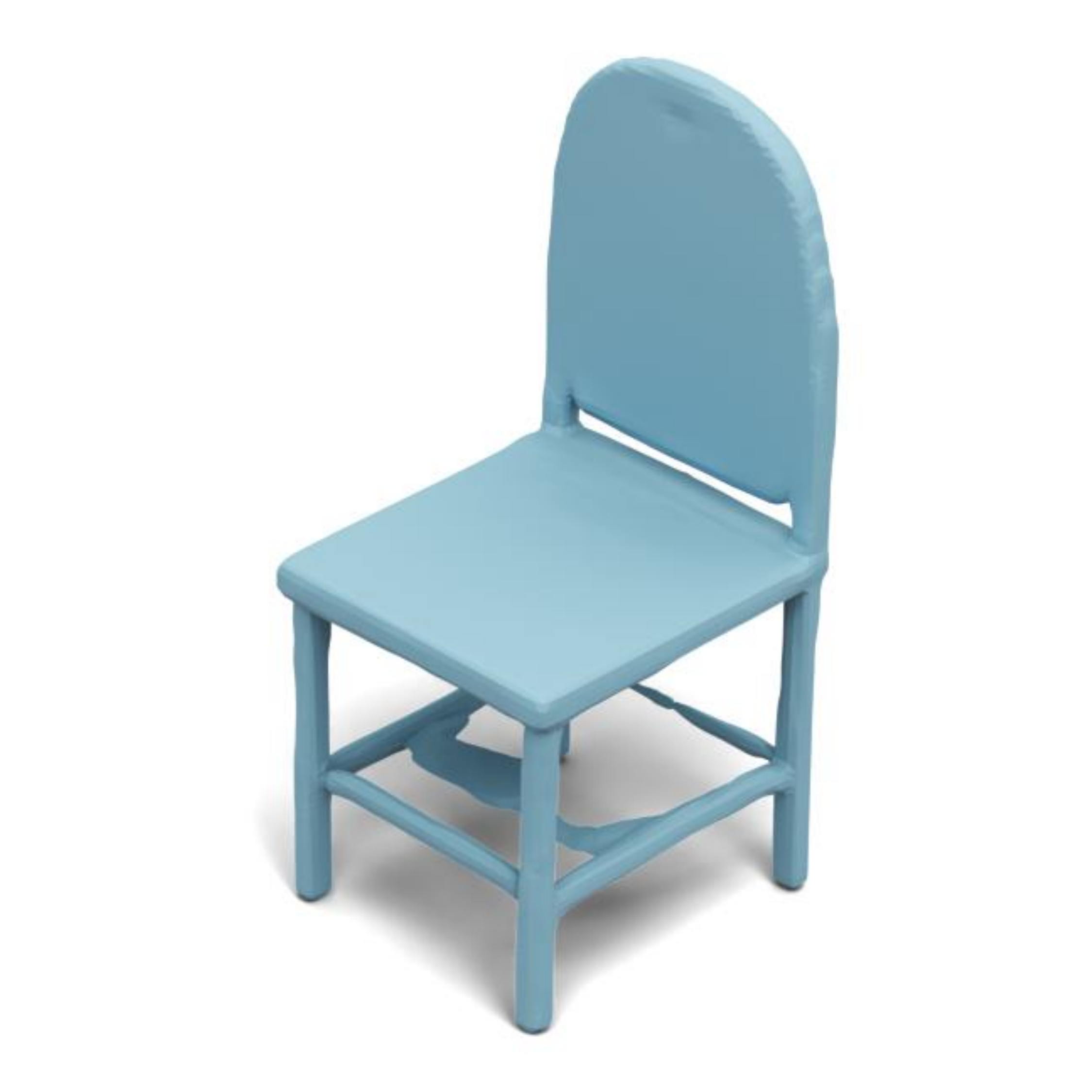}
&\includegraphics[trim = 1 1 1 1, clip, width=0.125\linewidth]{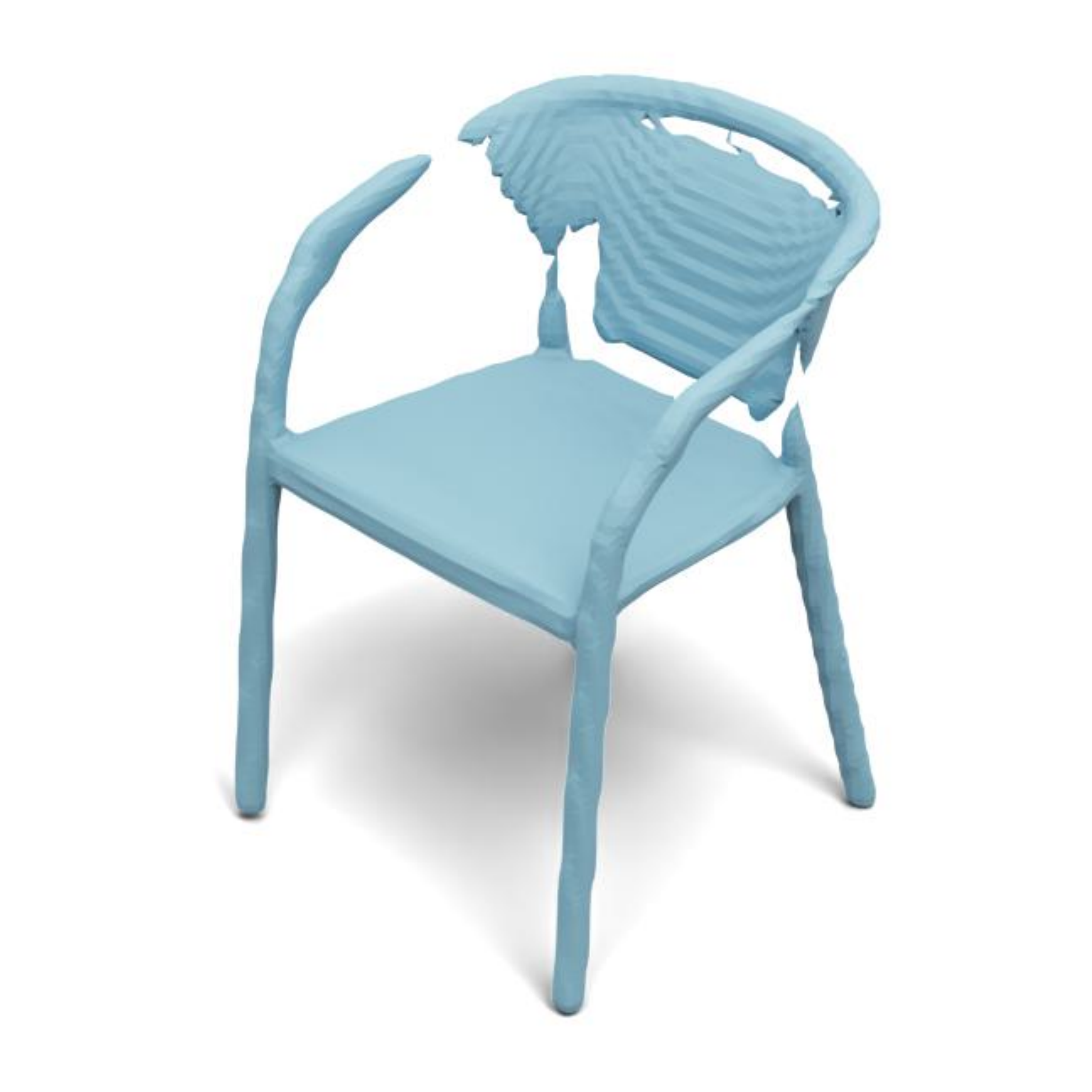}
&\includegraphics[trim = 1 1 1 1, clip, width=0.125\linewidth]{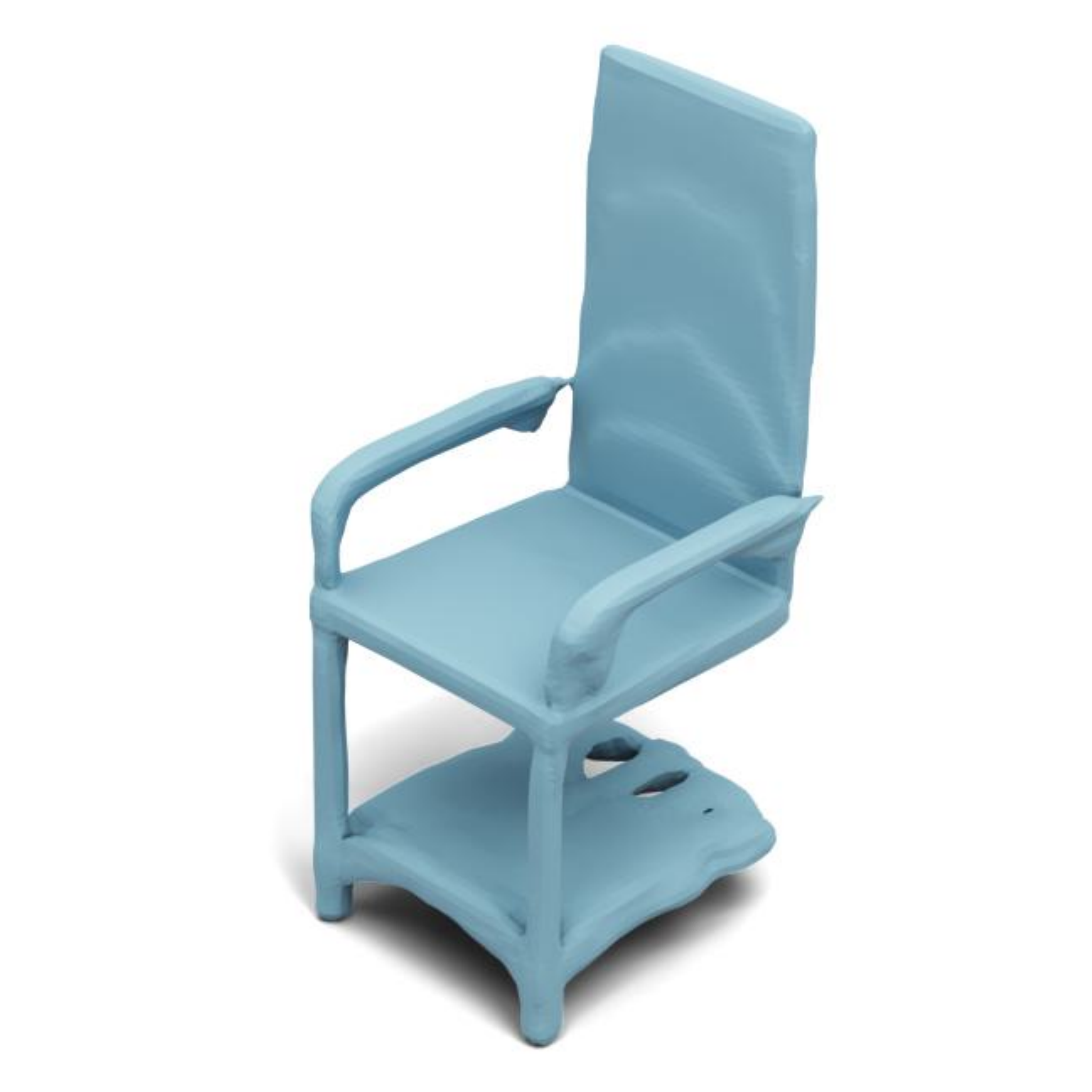}
&\includegraphics[trim = 1 1 1 1, clip, width=0.125\linewidth]{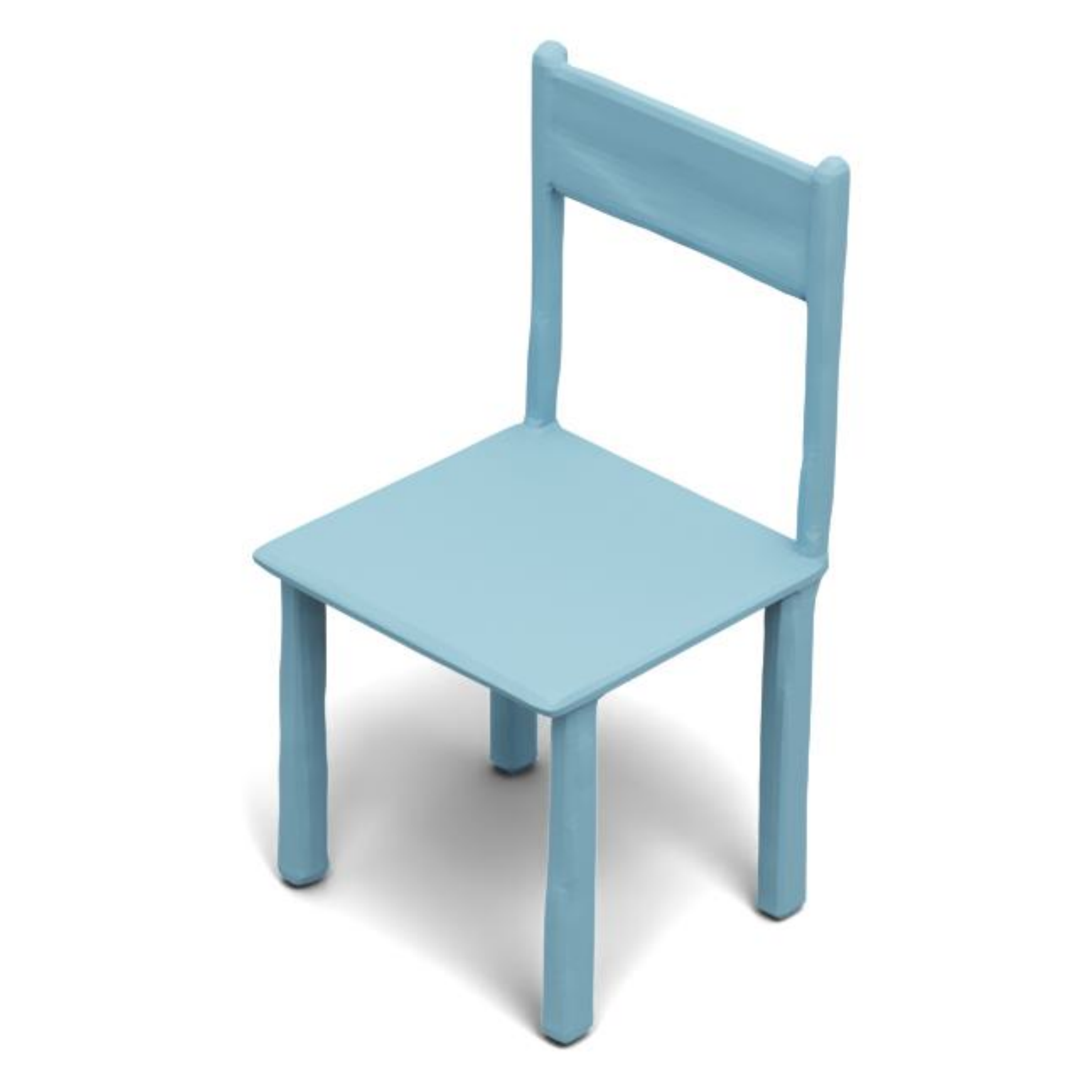}
&\includegraphics[trim = 1 1 1 1, clip, width=0.125\linewidth]{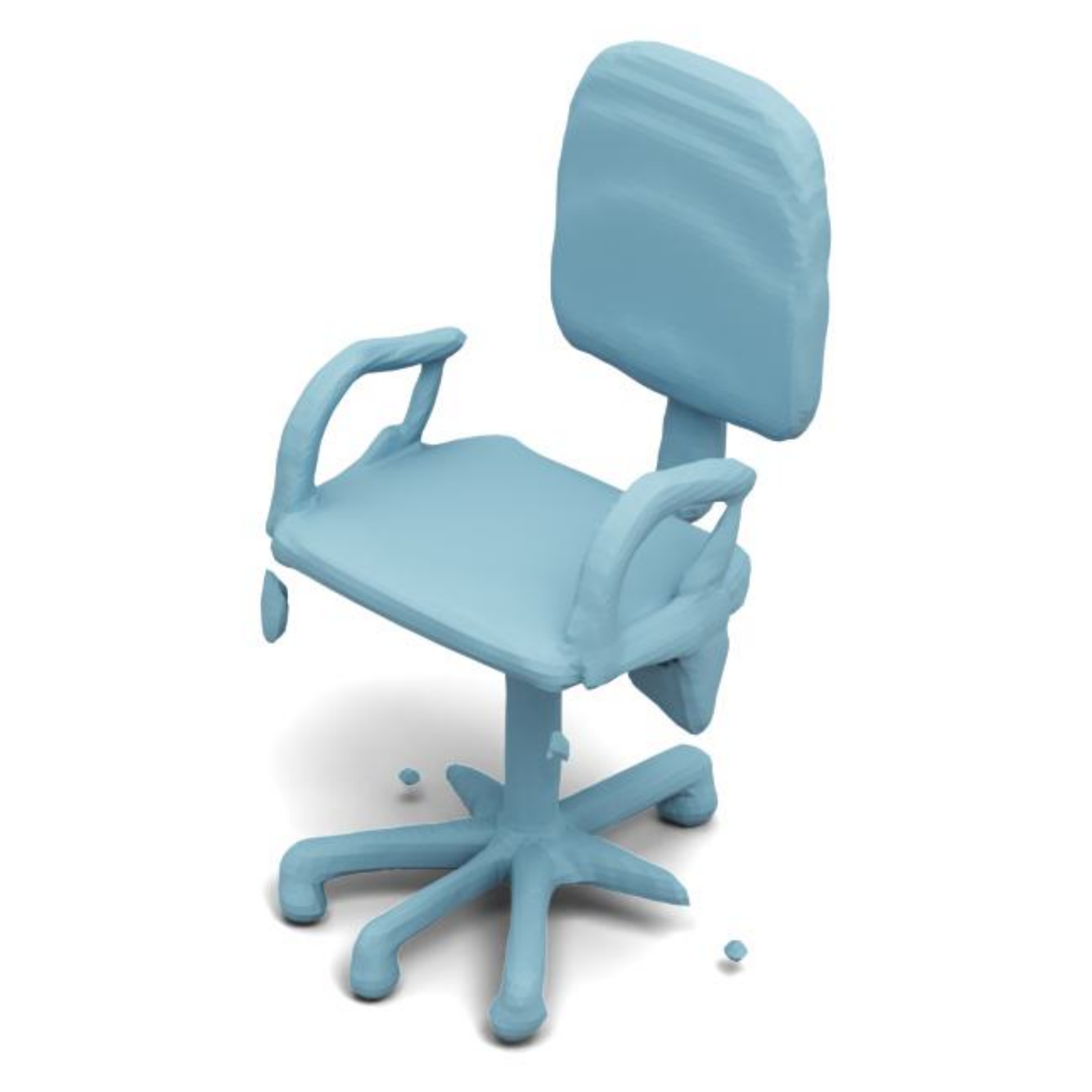}
\\
\includegraphics[width=0.125\linewidth]{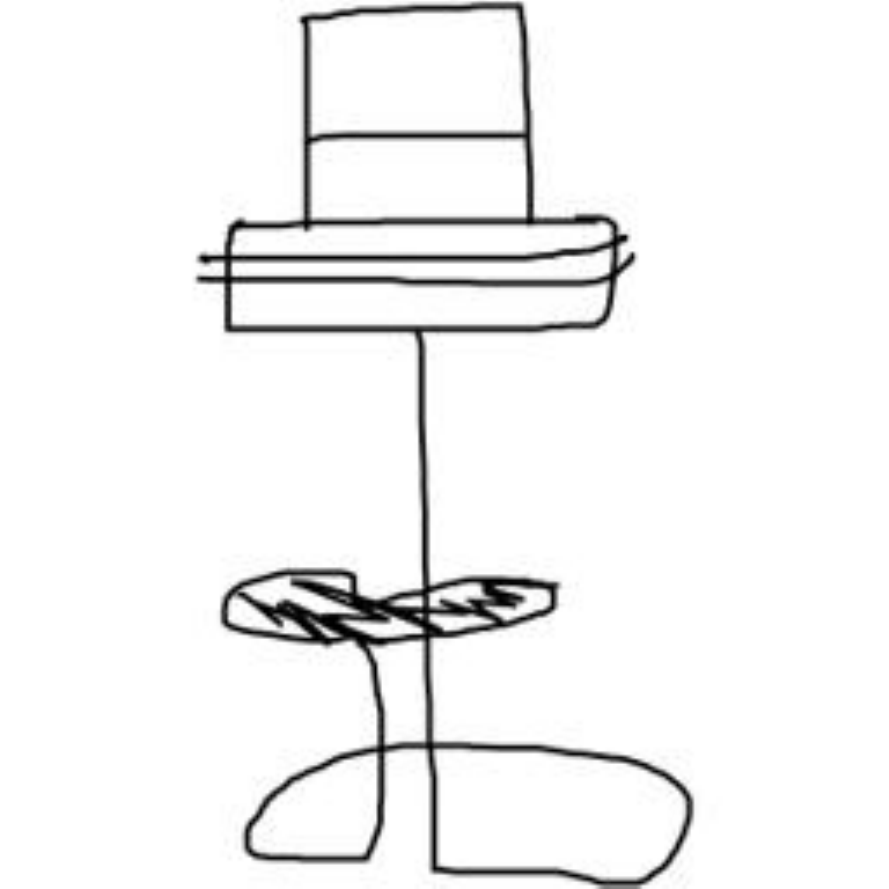}
&\includegraphics[width=0.125\linewidth]{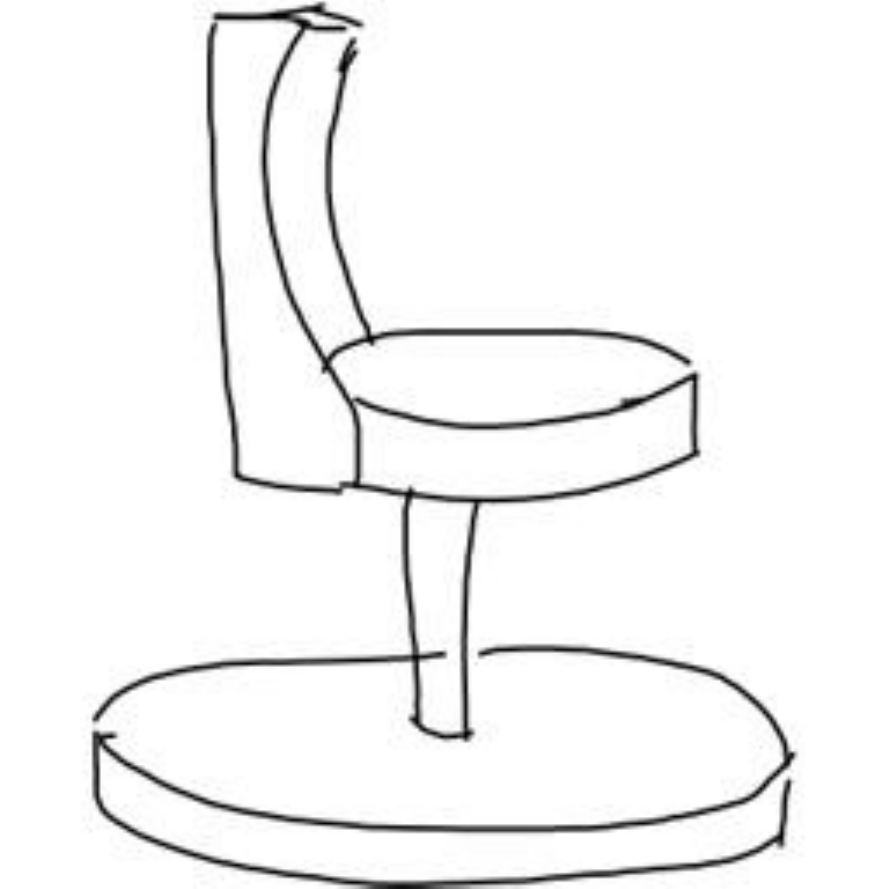}
&\includegraphics[width=0.125\linewidth]{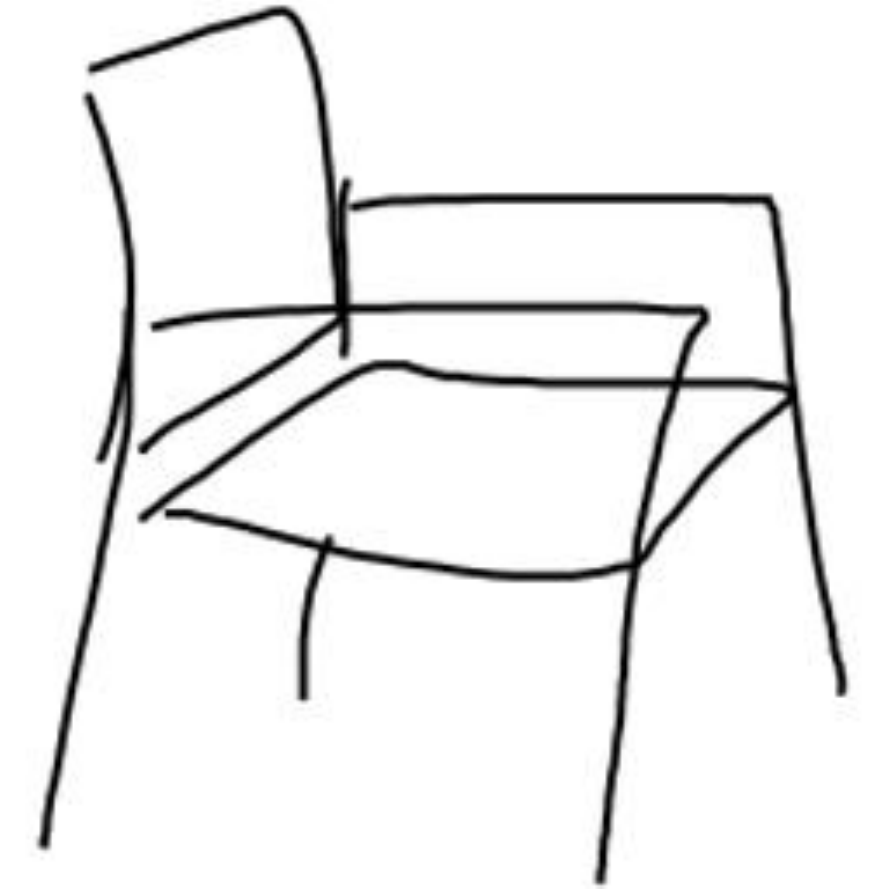}
&\includegraphics[width=0.125\linewidth]{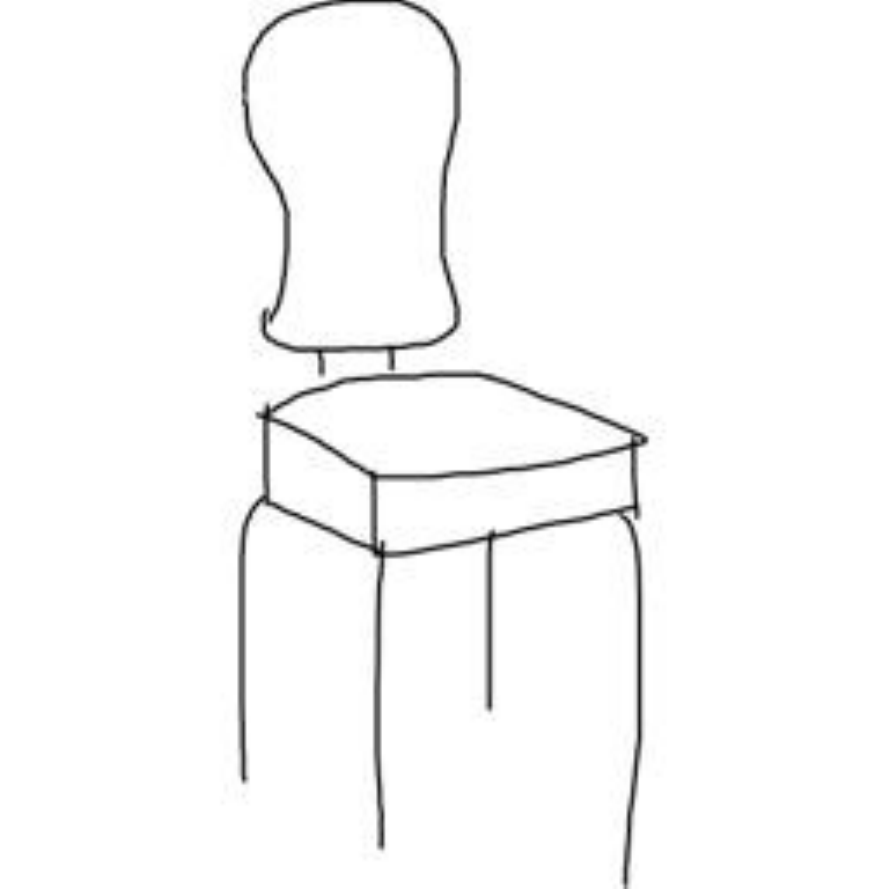}
&\includegraphics[width=0.125\linewidth]{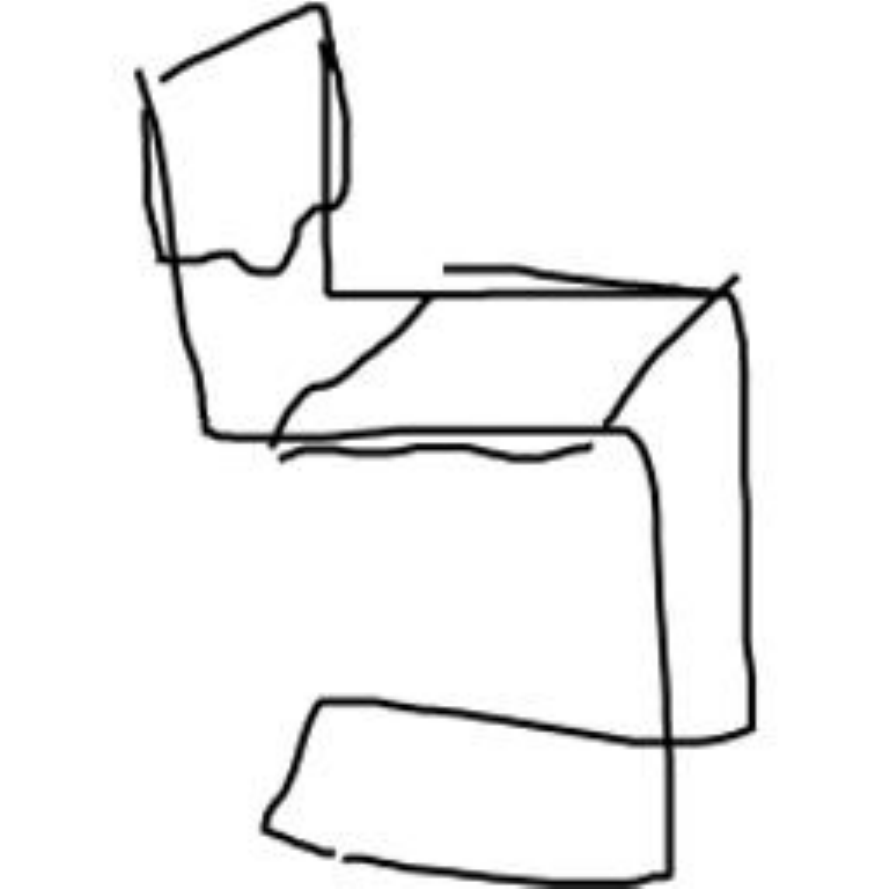}
&\includegraphics[width=0.125\linewidth]{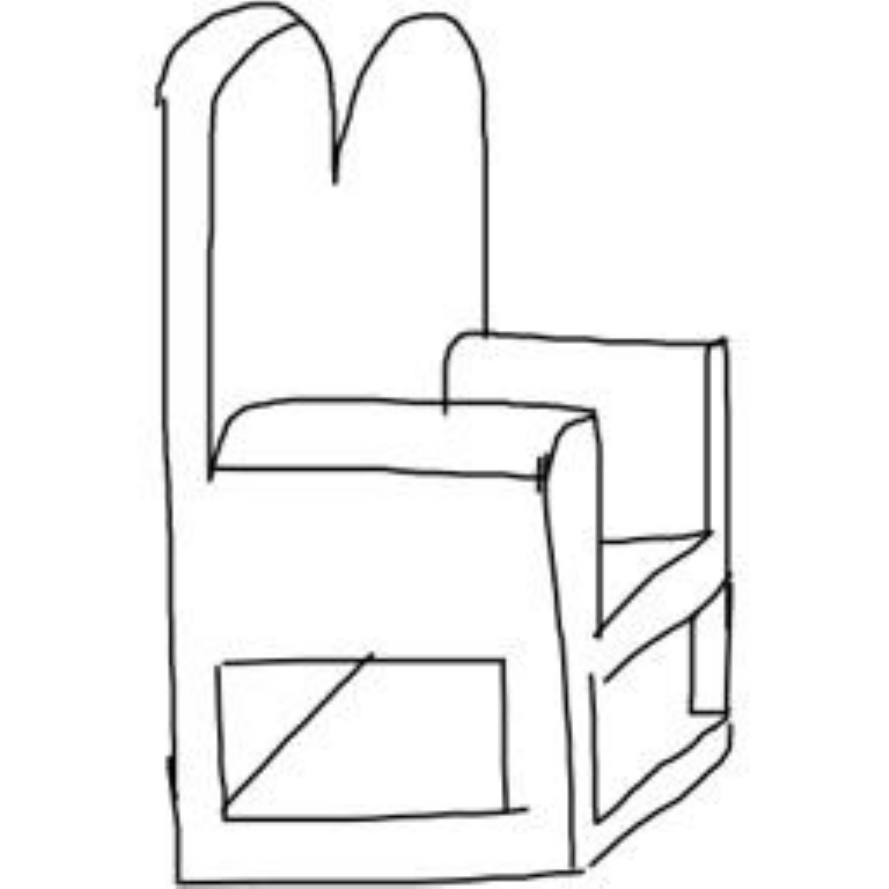}
&\includegraphics[width=0.125\linewidth]{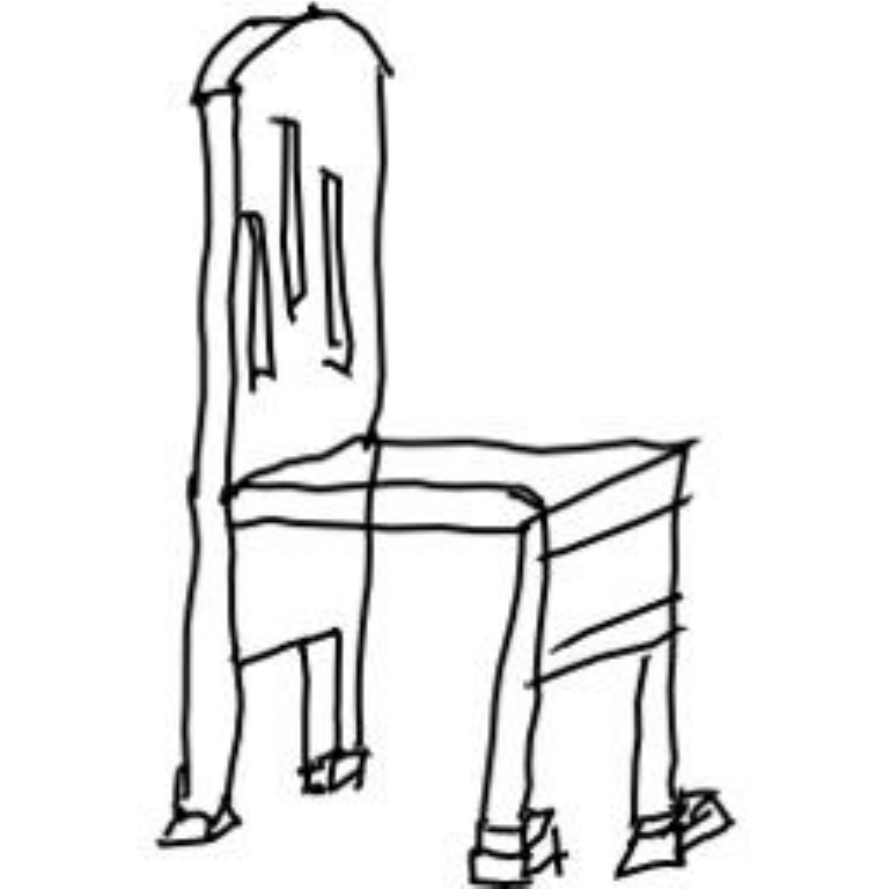}
&\includegraphics[width=0.125\linewidth]{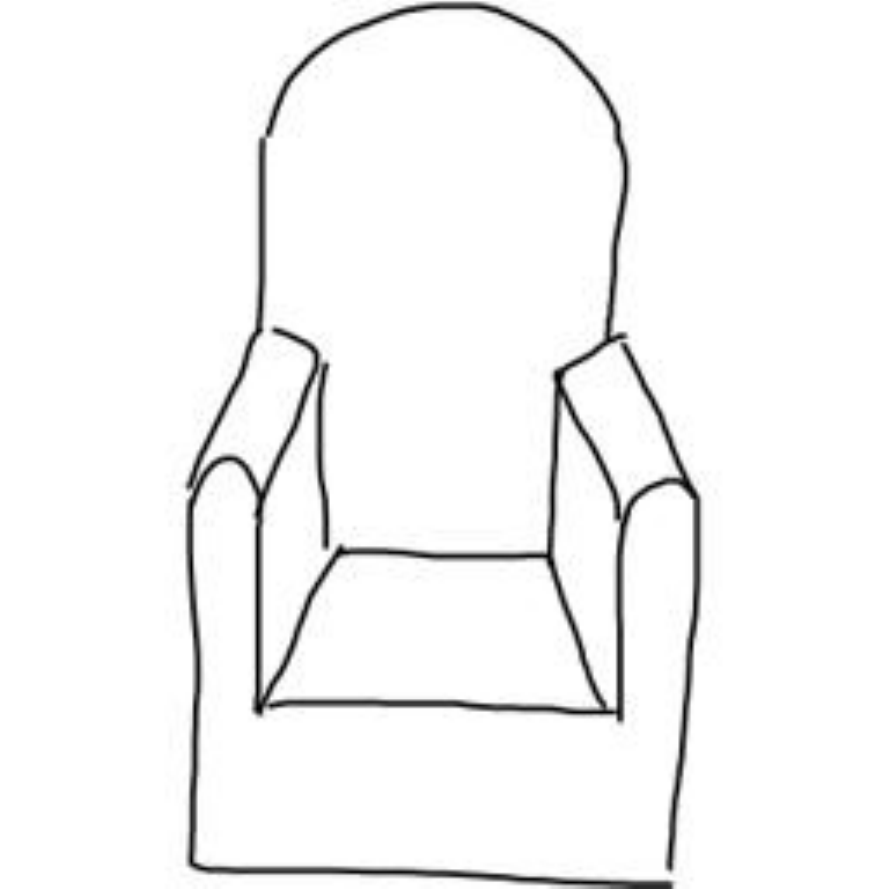}
\\
\includegraphics[trim = 1 1 1 1, clip, width=0.125\linewidth]{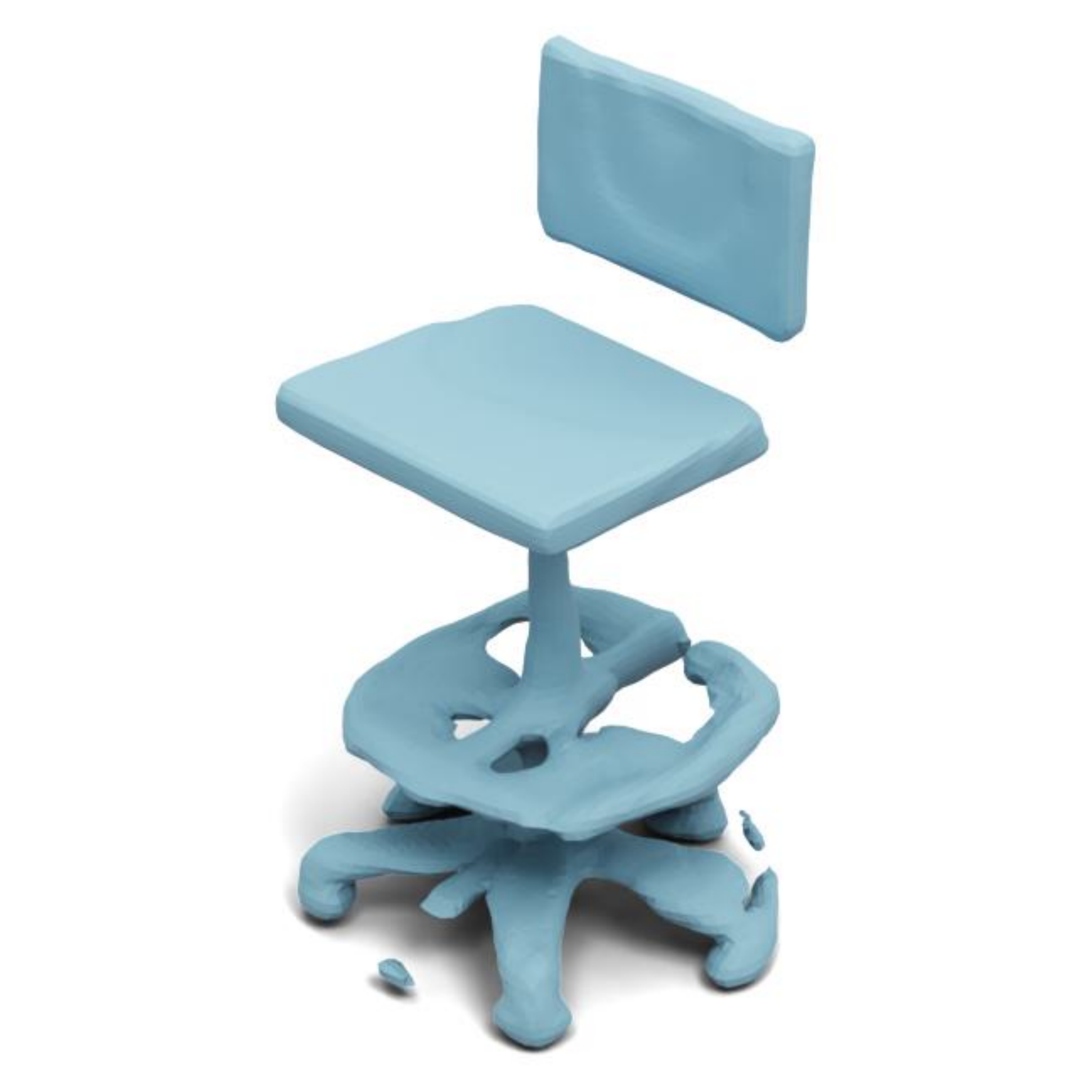}
&\includegraphics[trim = 1 1 1 1, clip, width=0.125\linewidth]{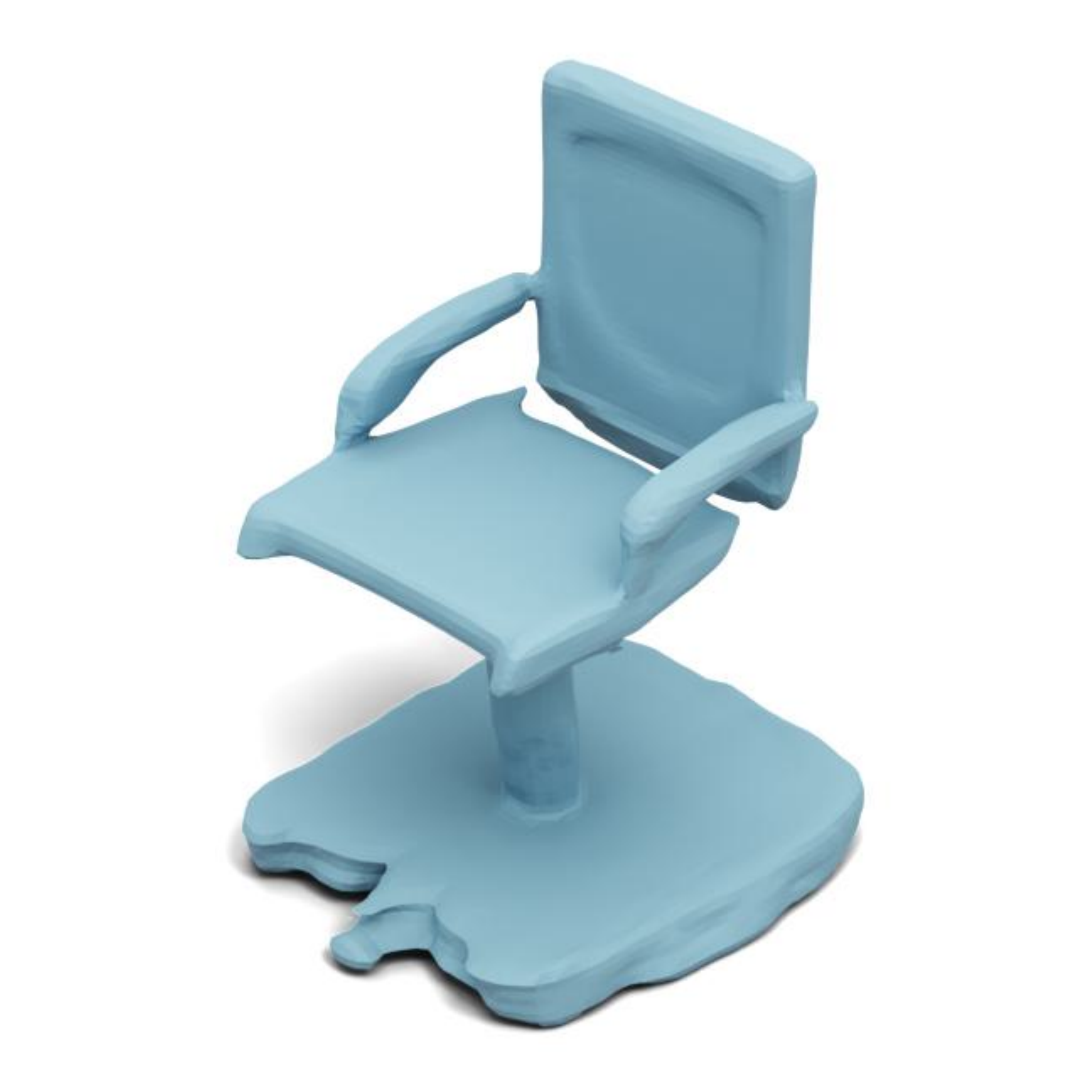}
&\includegraphics[trim = 1 1 1 1, clip, width=0.125\linewidth]{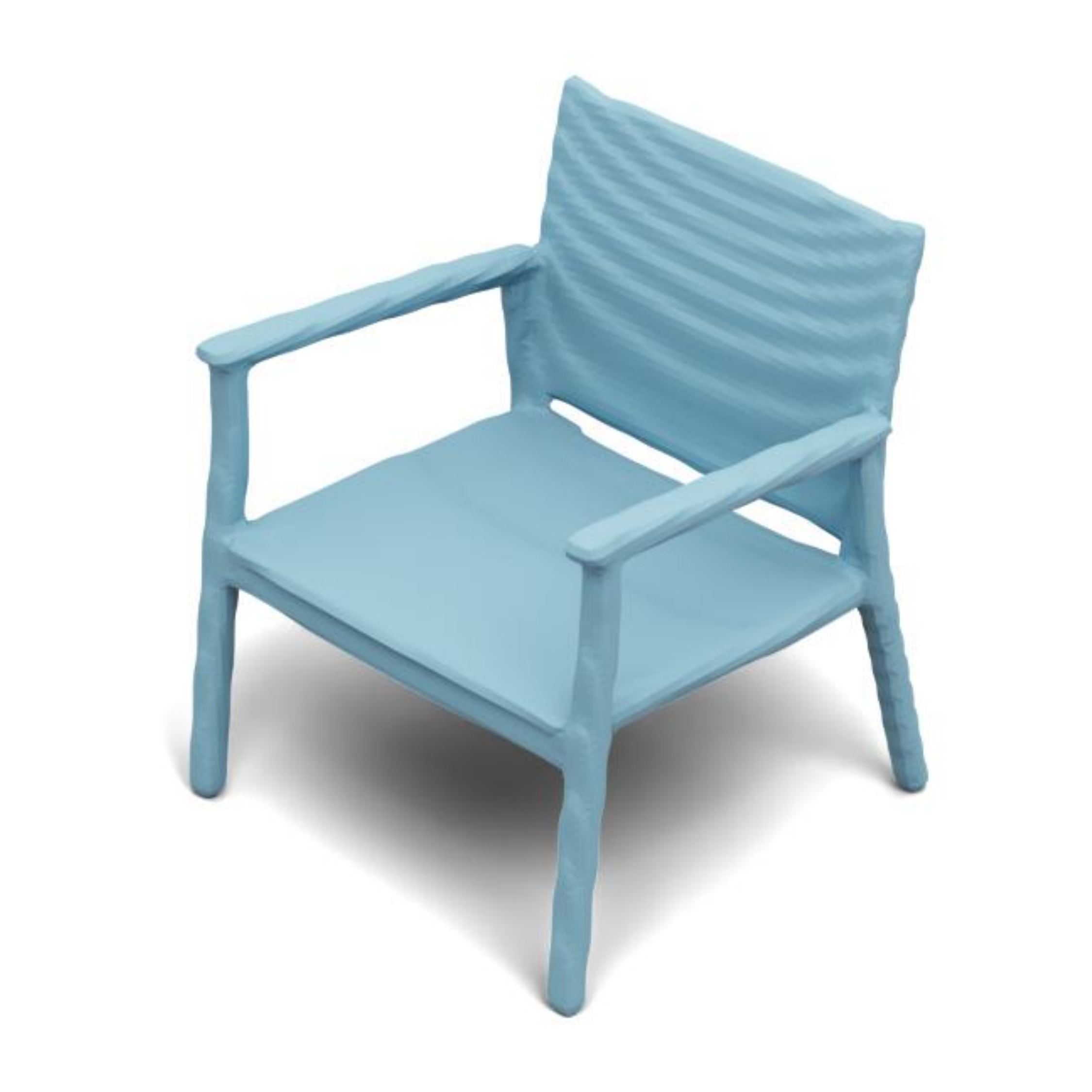}
&\includegraphics[trim = 1 1 1 1, clip, width=0.125\linewidth]{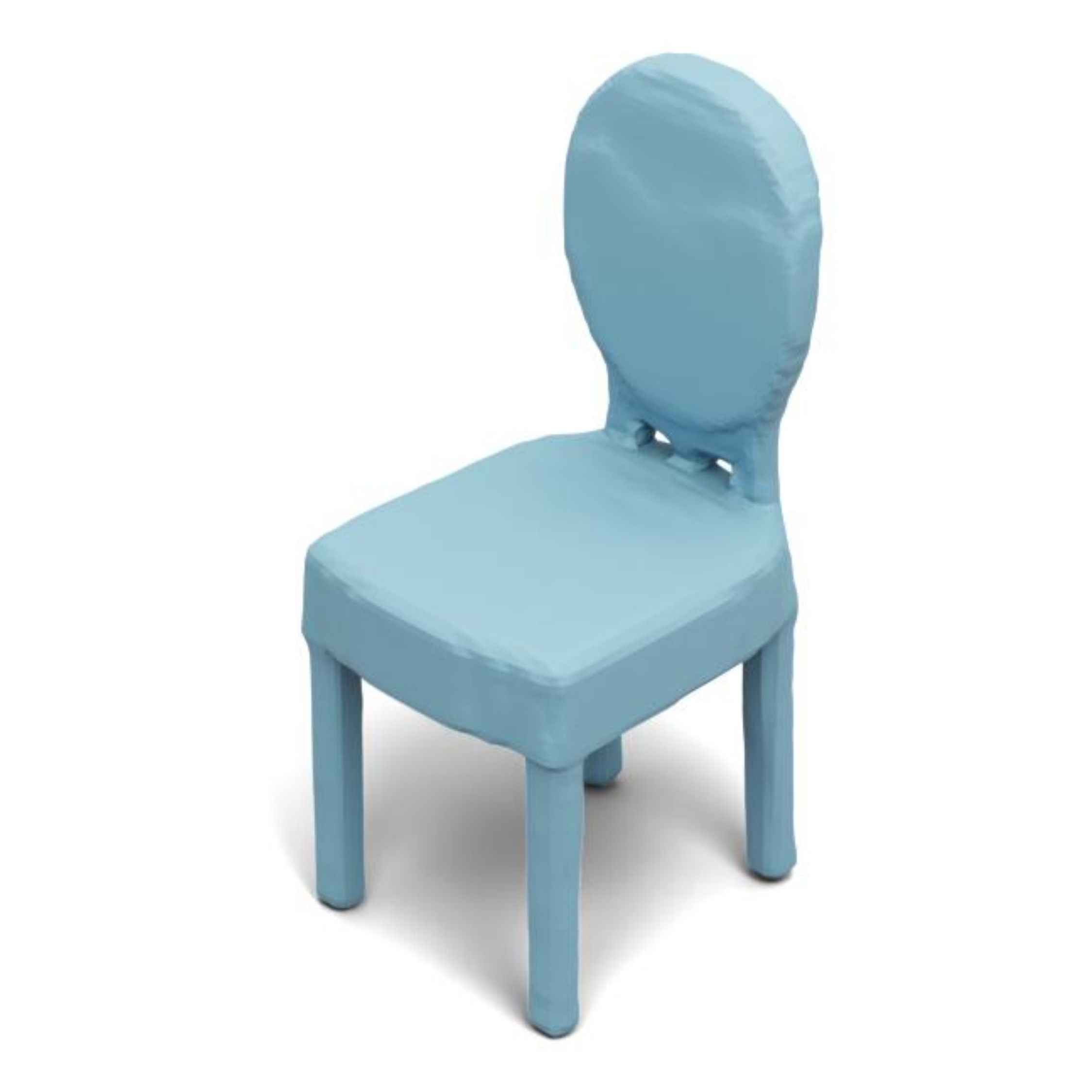}
&\includegraphics[trim = 1 1 1 1, clip, width=0.125\linewidth]{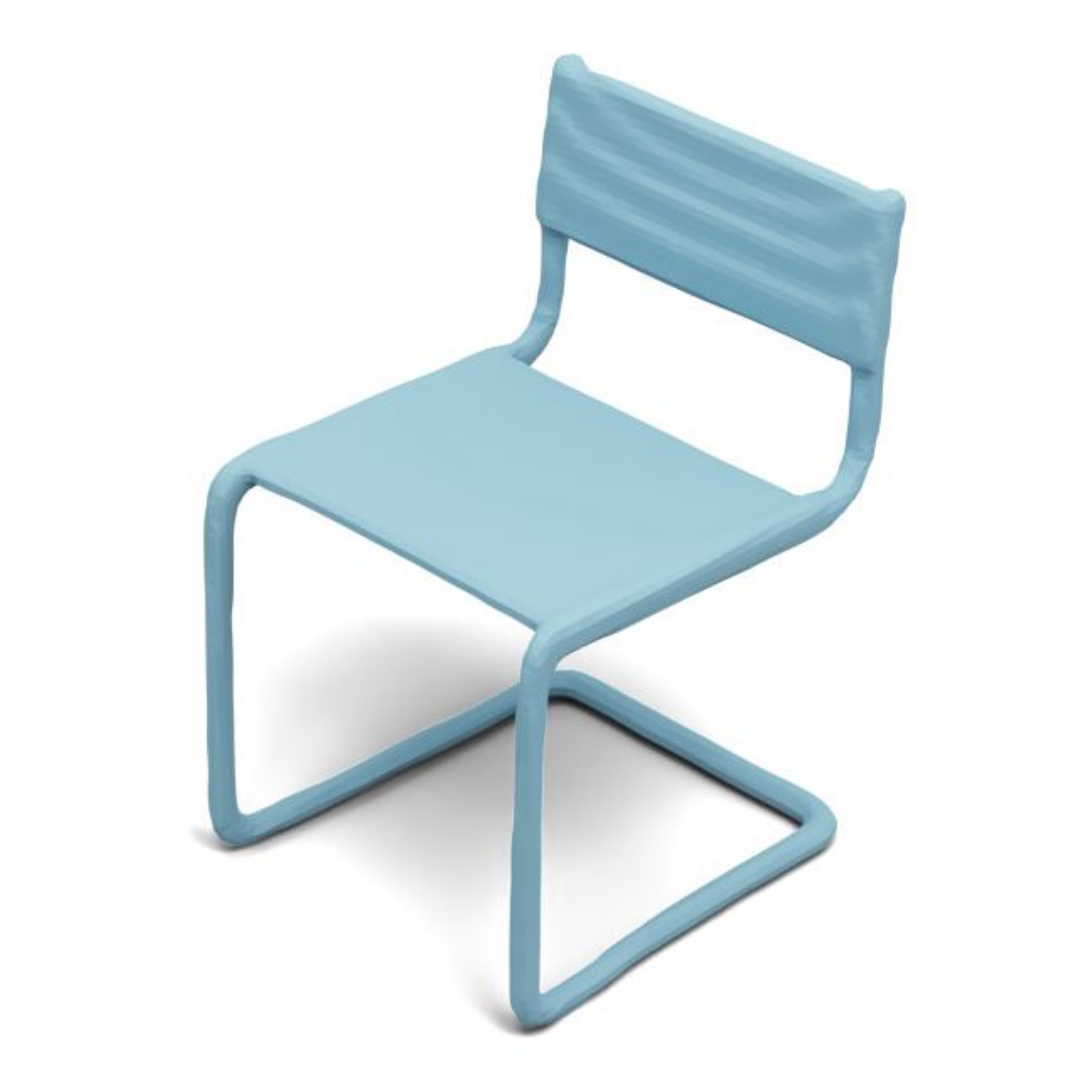}
&\includegraphics[trim = 1 1 1 1, clip, width=0.125\linewidth]{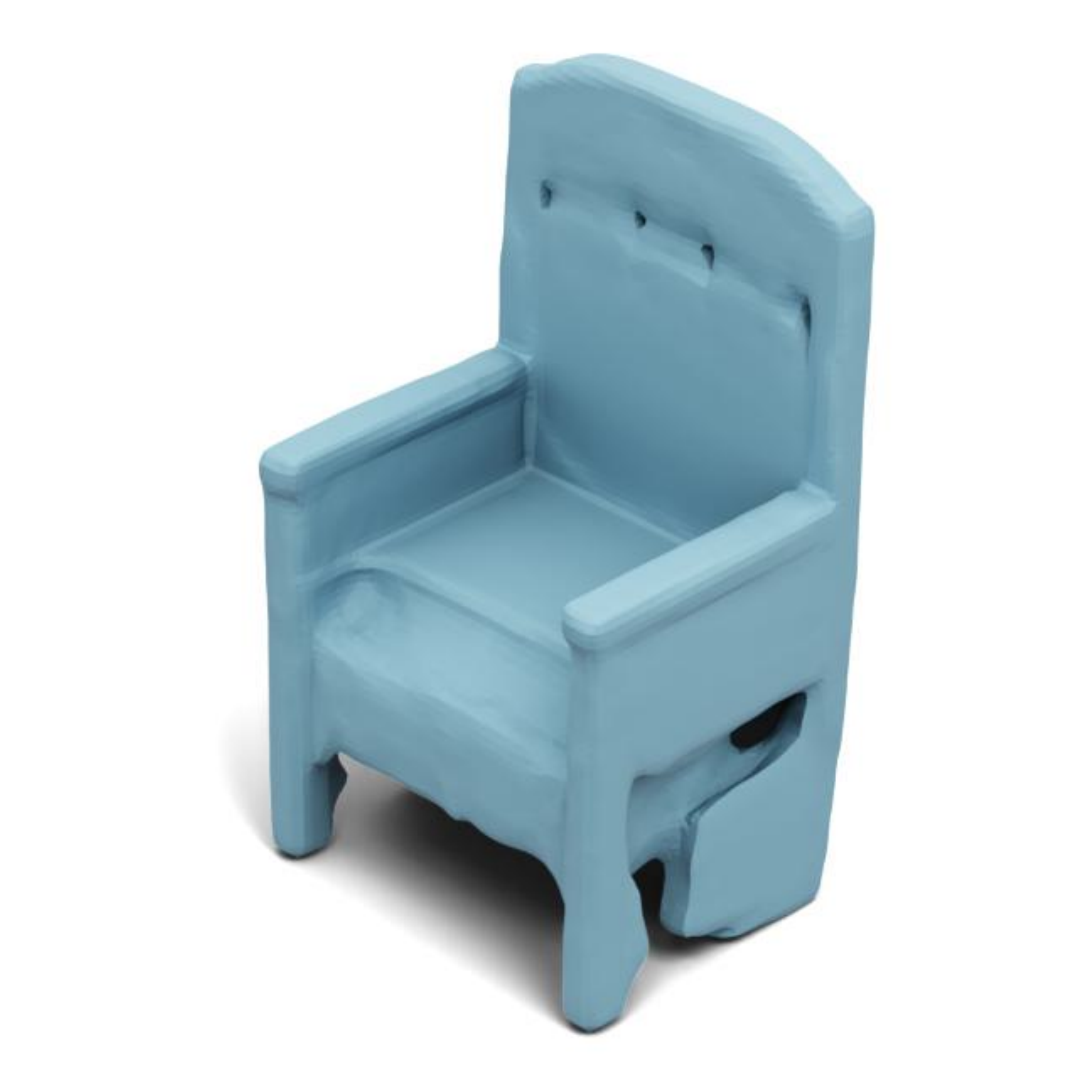}
&\includegraphics[trim = 1 1 1 1, clip, width=0.125\linewidth]{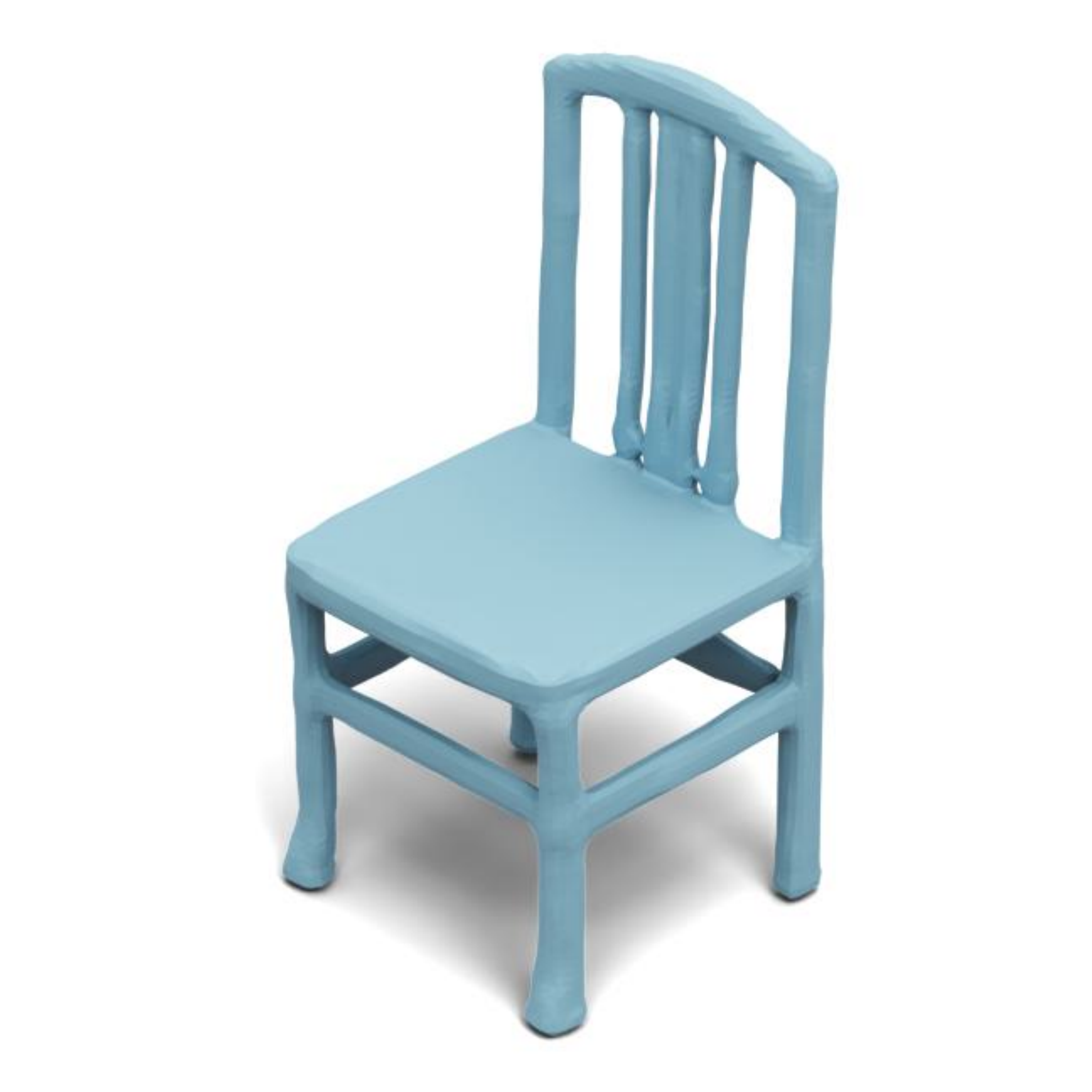}
&\includegraphics[trim = 1 1 1 1, clip, width=0.125\linewidth]{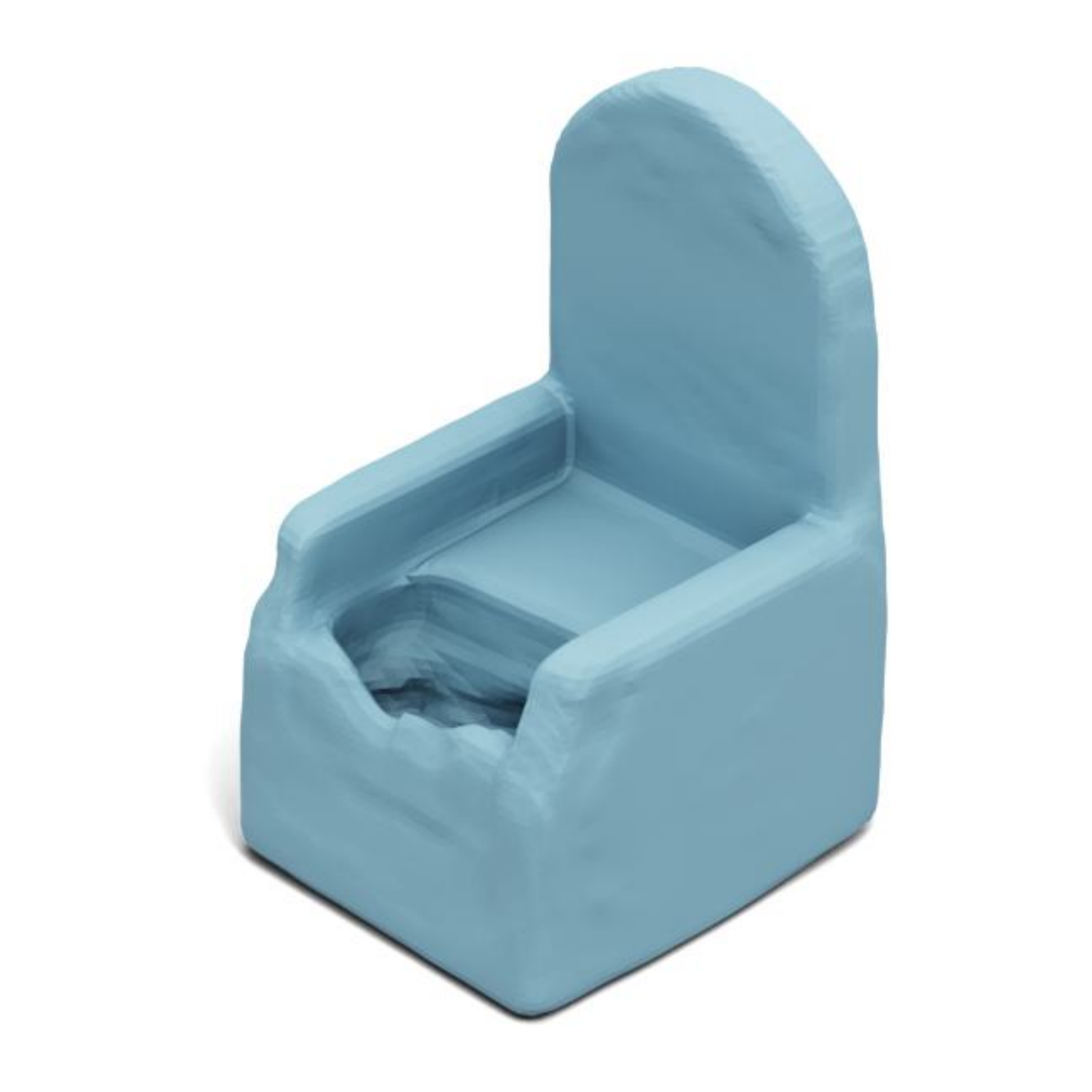}
\\
\includegraphics[width=0.125\linewidth]{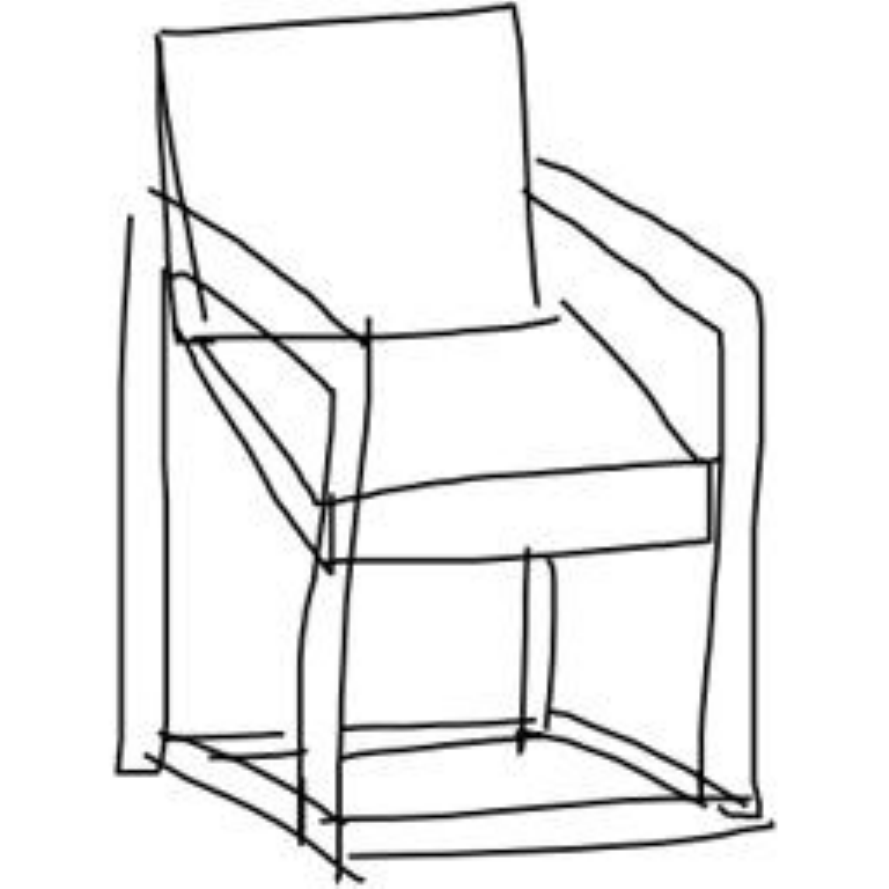}
&\includegraphics[width=0.125\linewidth]{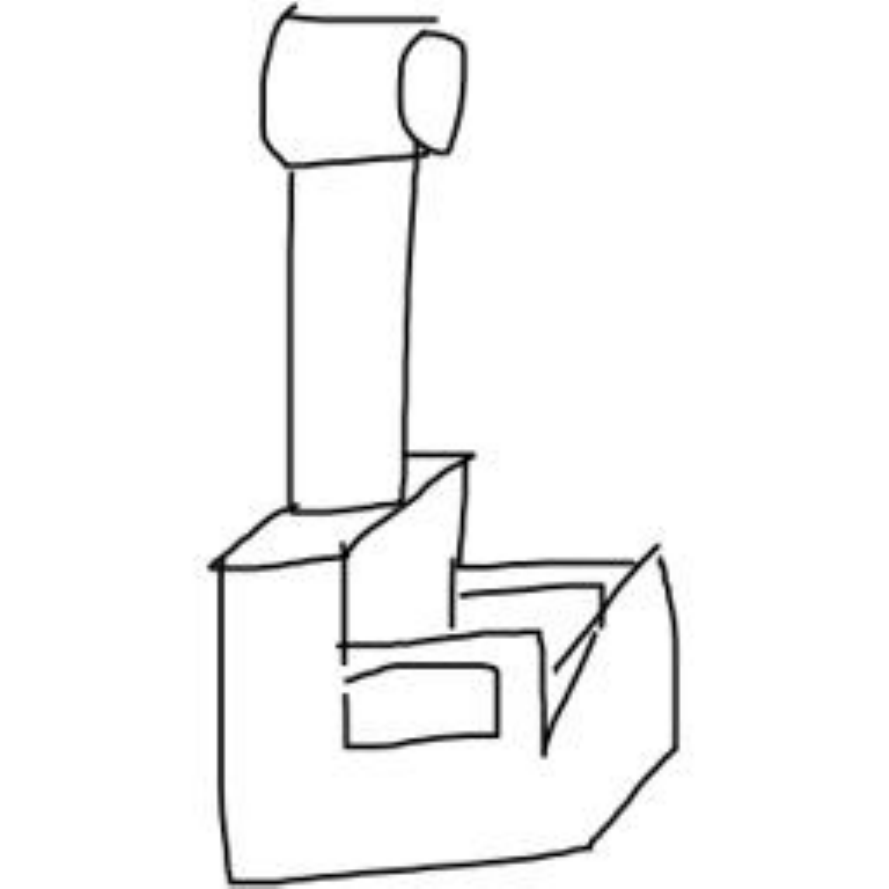}
&\includegraphics[width=0.125\linewidth]{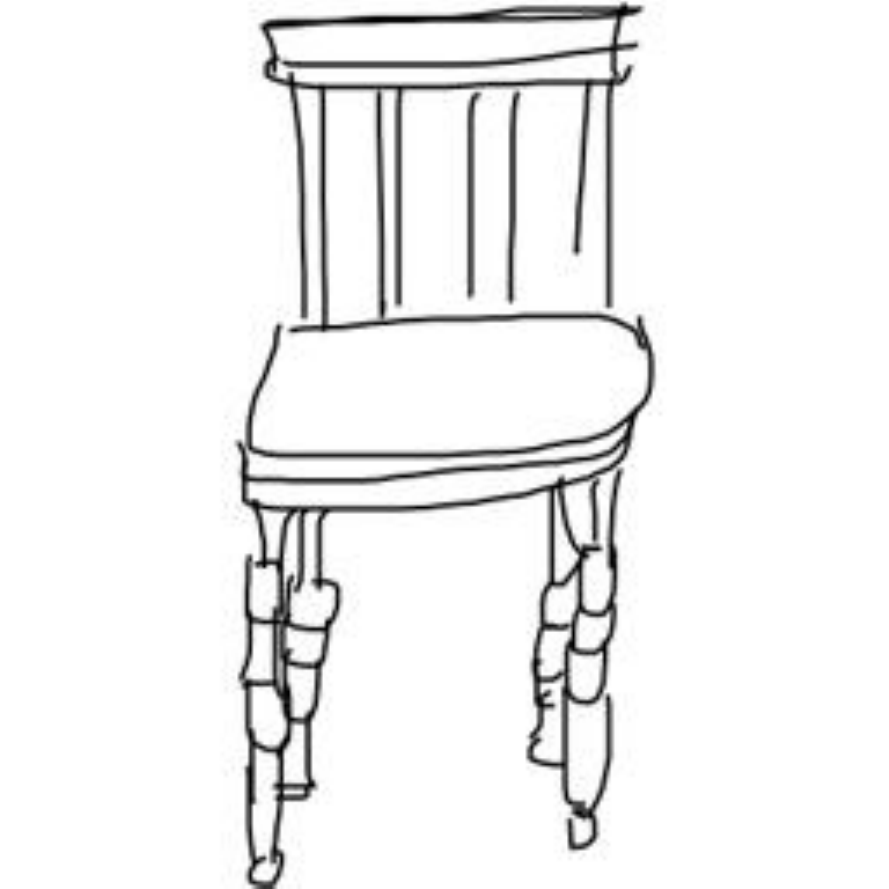}
&\includegraphics[width=0.125\linewidth]{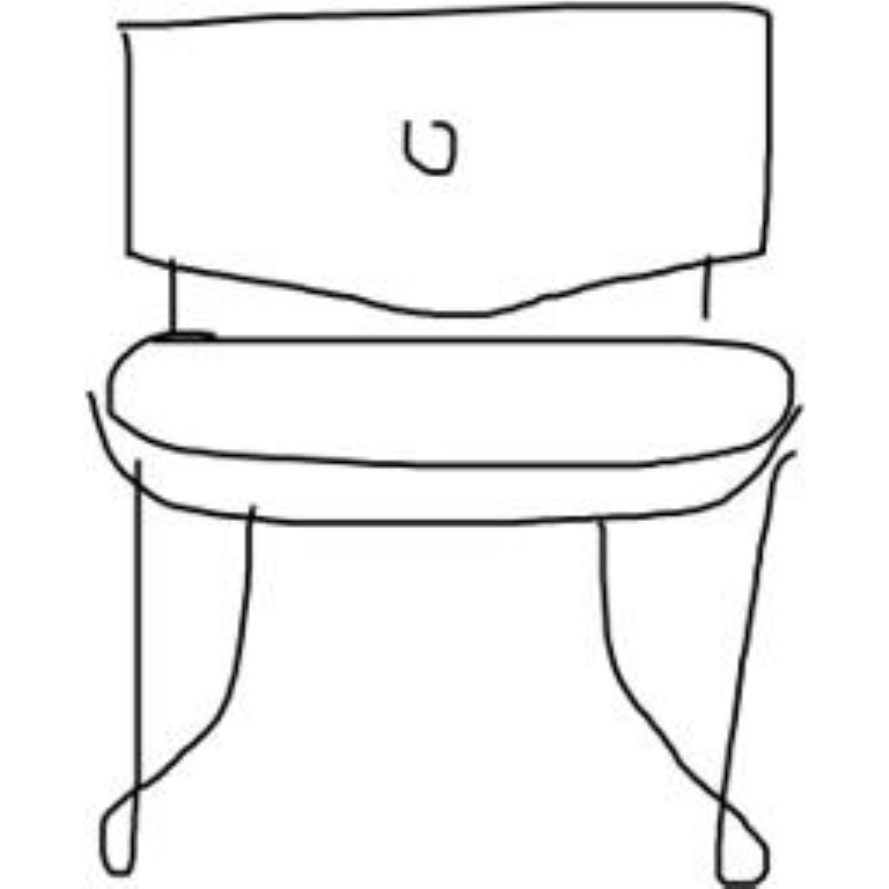}
&\includegraphics[width=0.125\linewidth]{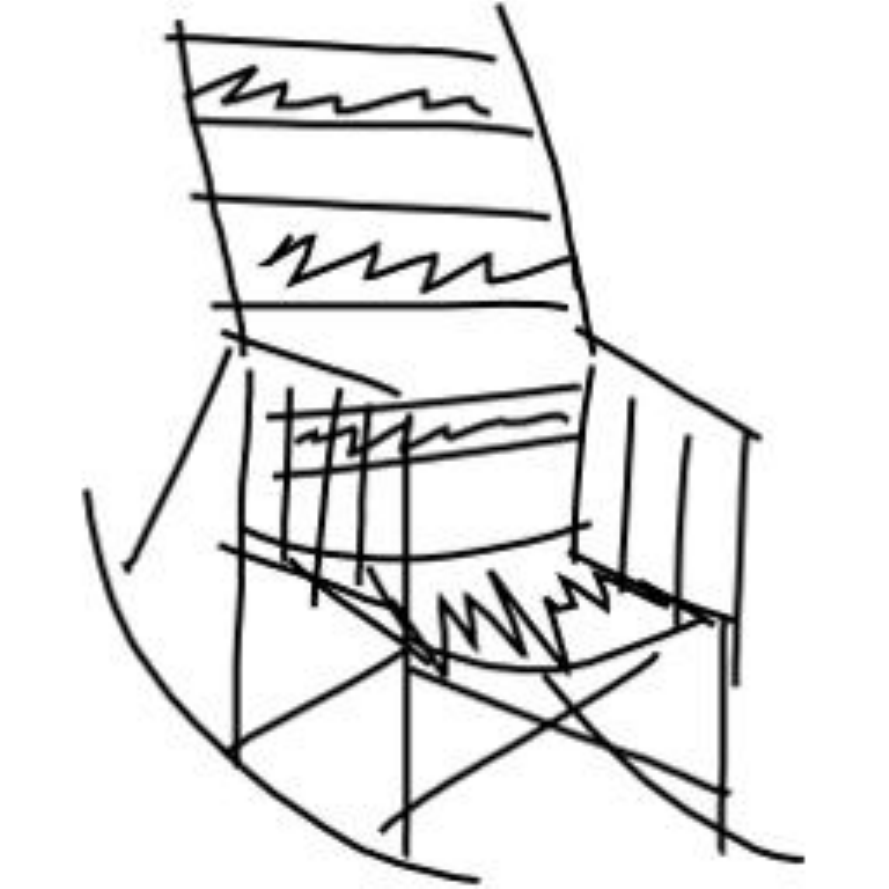}
&\includegraphics[width=0.125\linewidth]{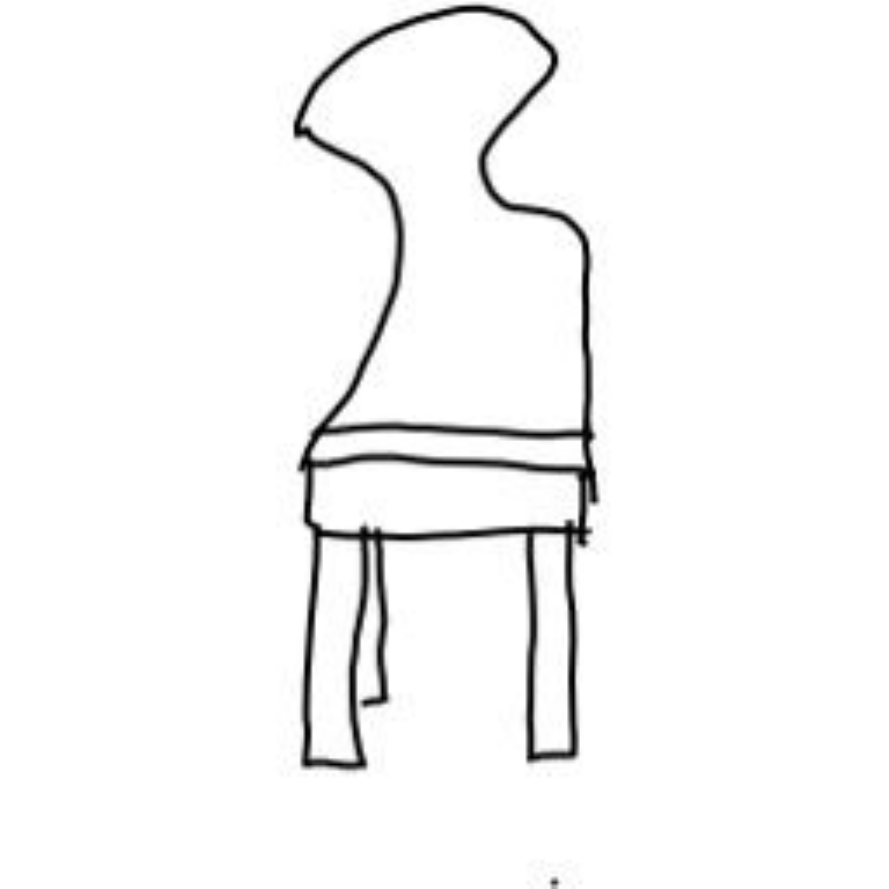}
&\includegraphics[width=0.125\linewidth]{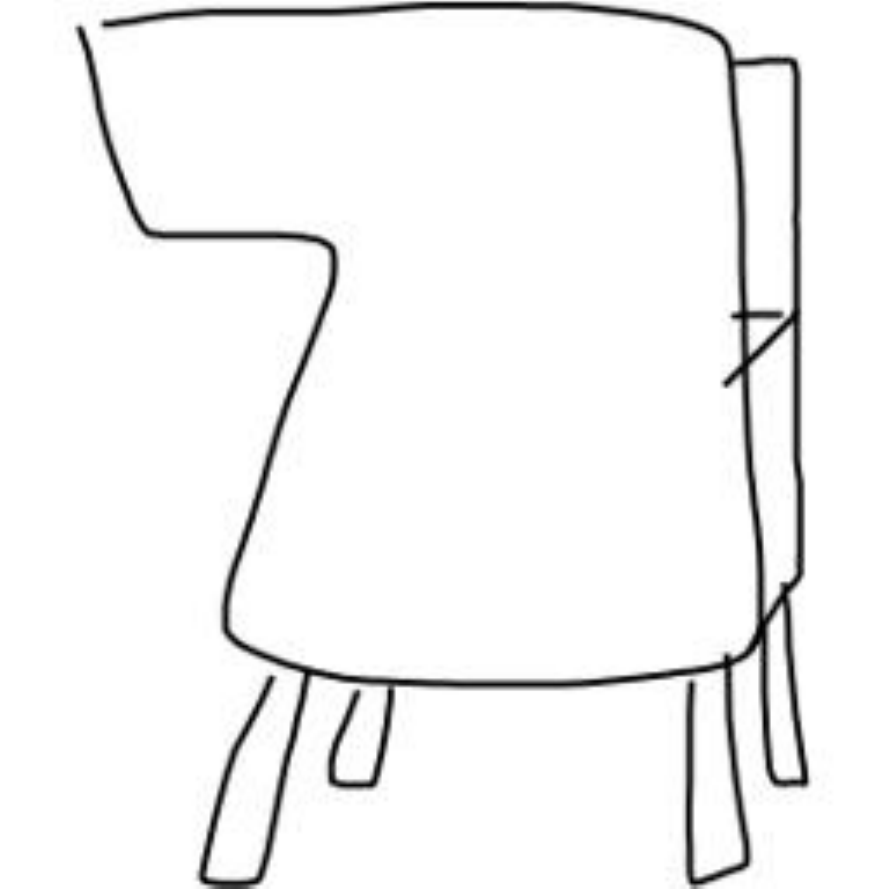}
&\includegraphics[width=0.125\linewidth]{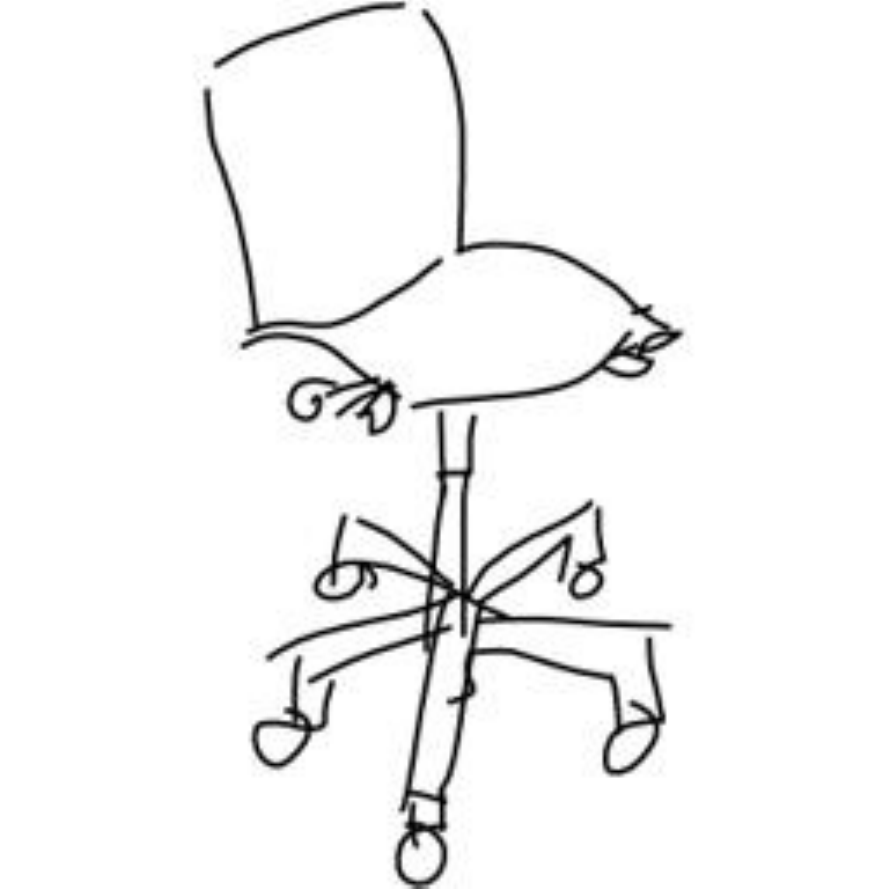}
\\
\includegraphics[trim = 1 1 1 1, clip, width=0.125\linewidth]{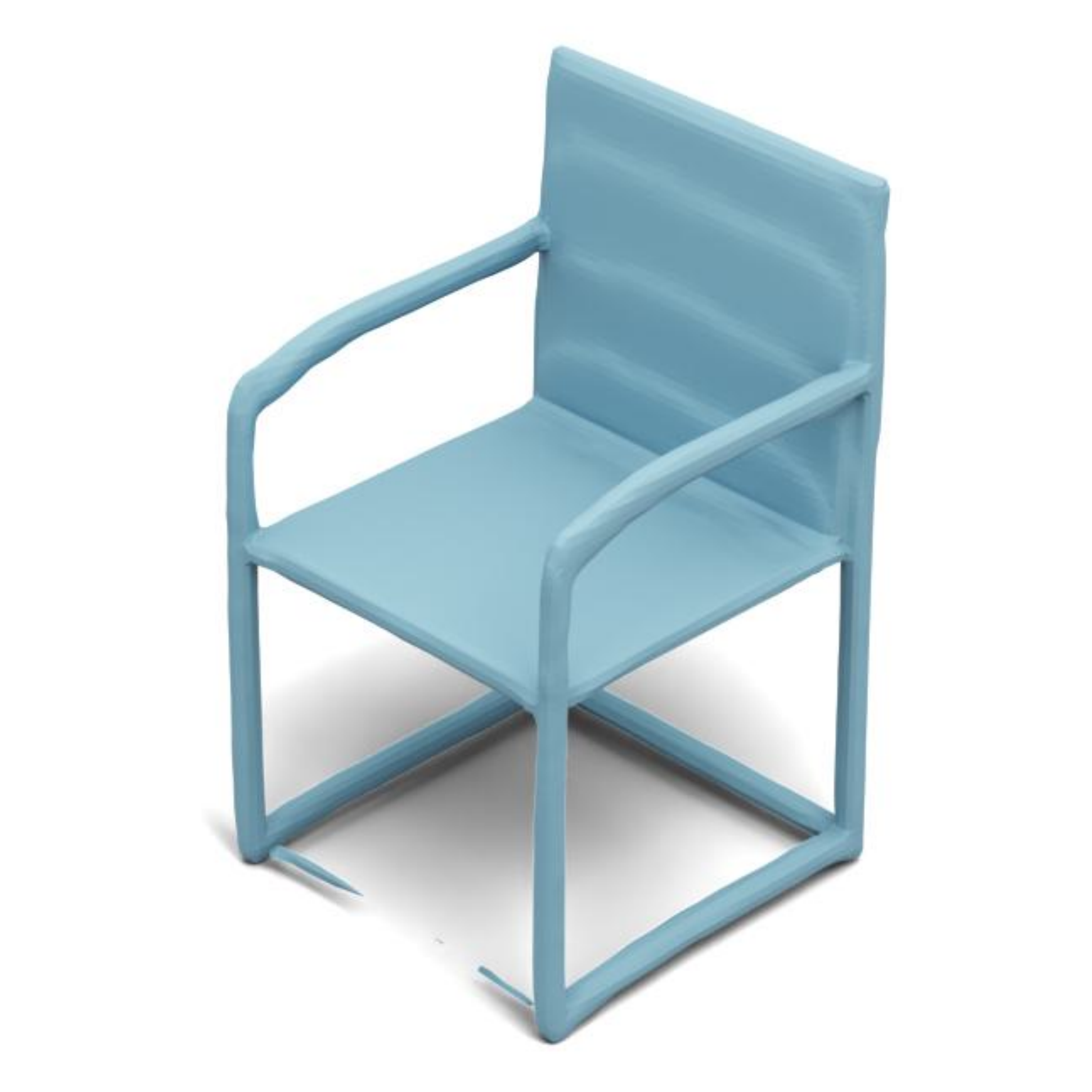}
&\includegraphics[trim = 1 1 1 1, clip, width=0.125\linewidth]{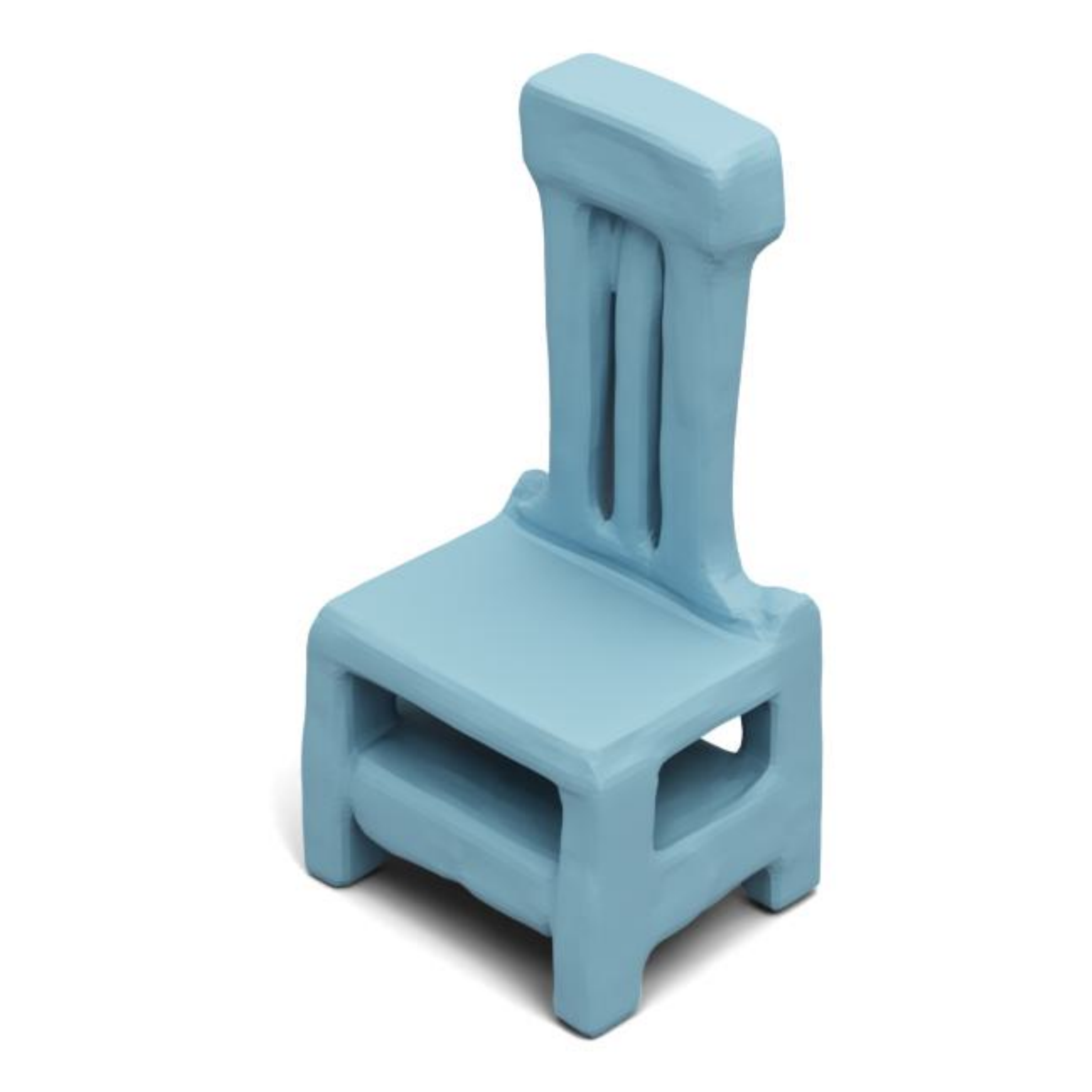}
&\includegraphics[trim = 1 1 1 1, clip, width=0.125\linewidth]{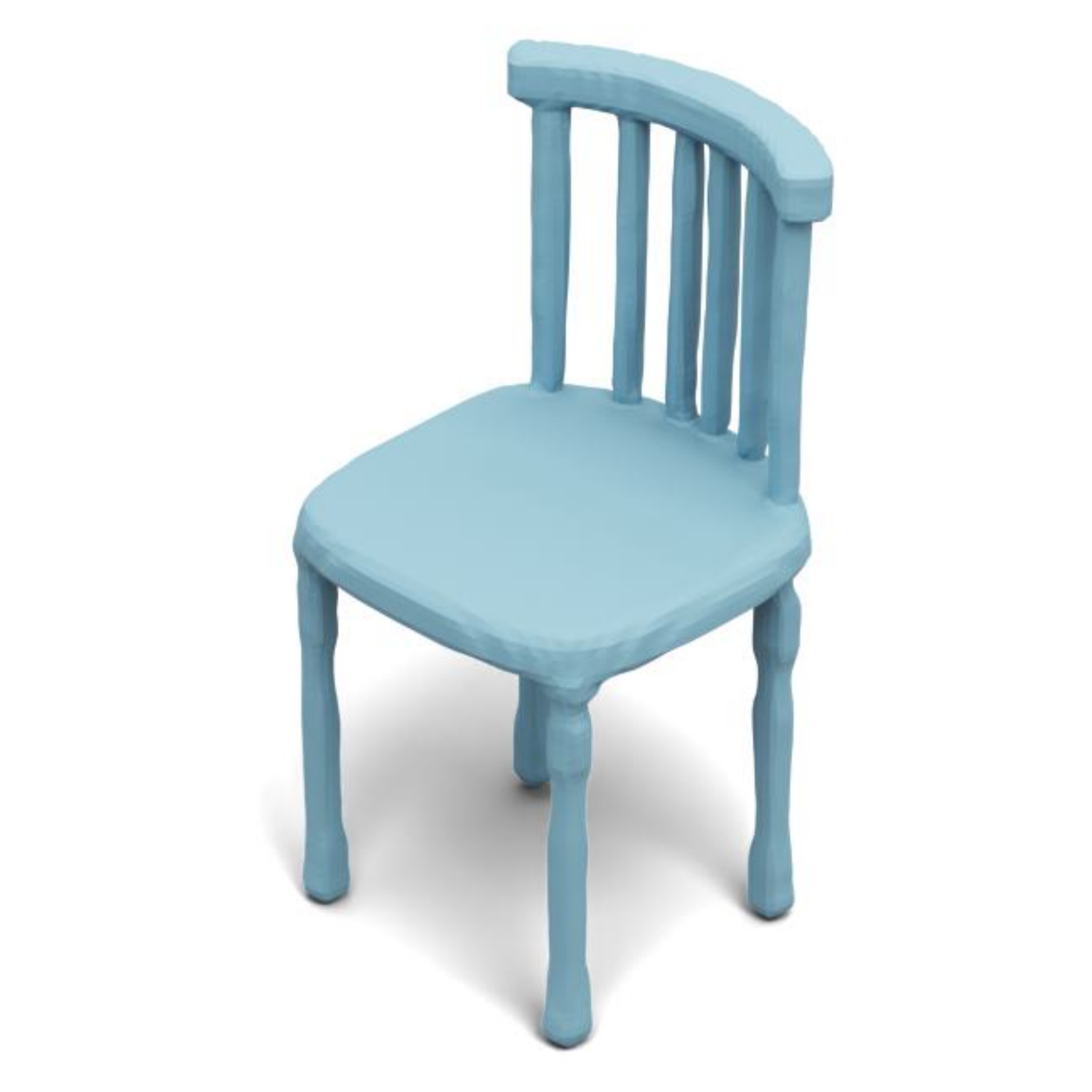}
&\includegraphics[trim = 1 1 1 1, clip, width=0.125\linewidth]{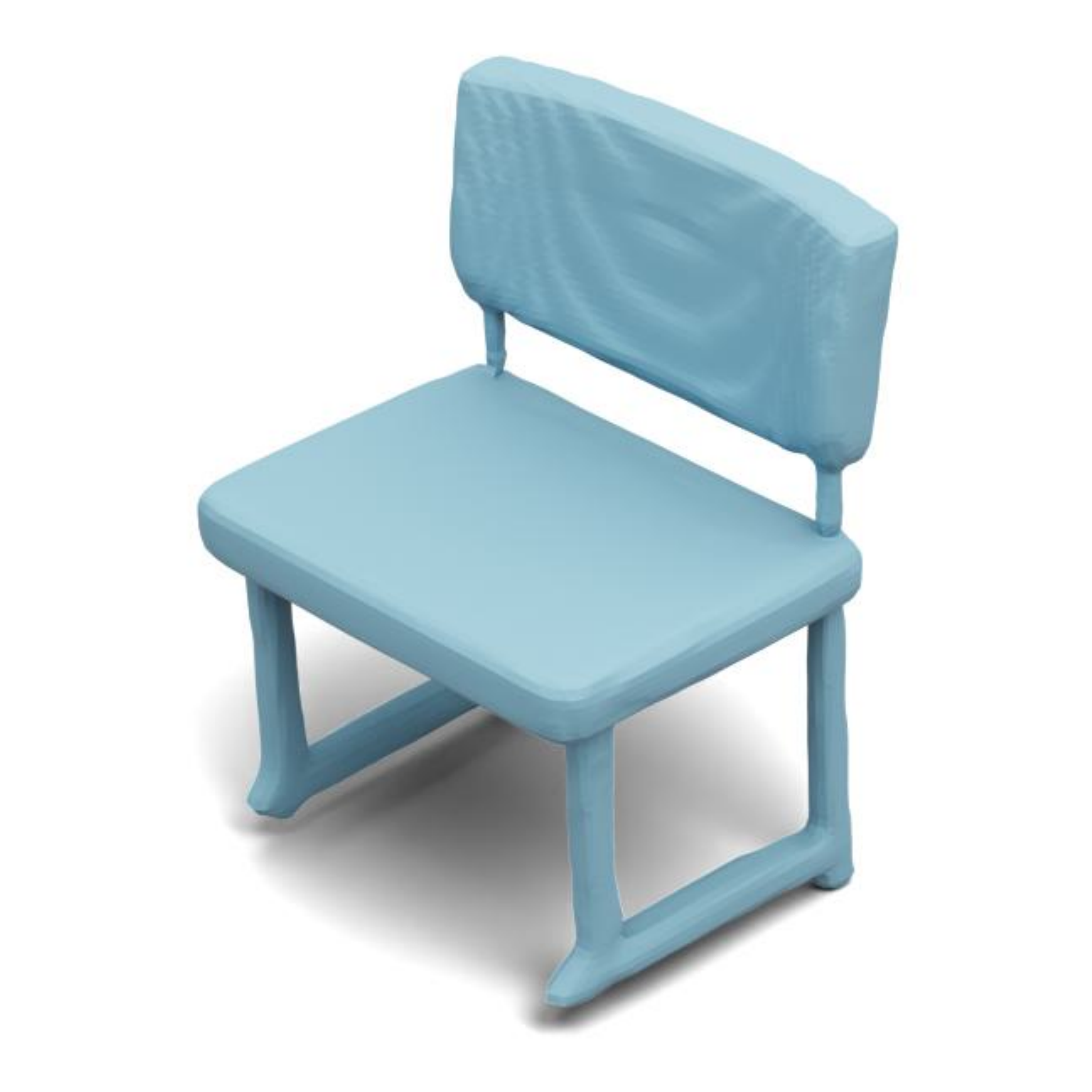}
&\includegraphics[trim = 1 1 1 1, clip, width=0.125\linewidth]{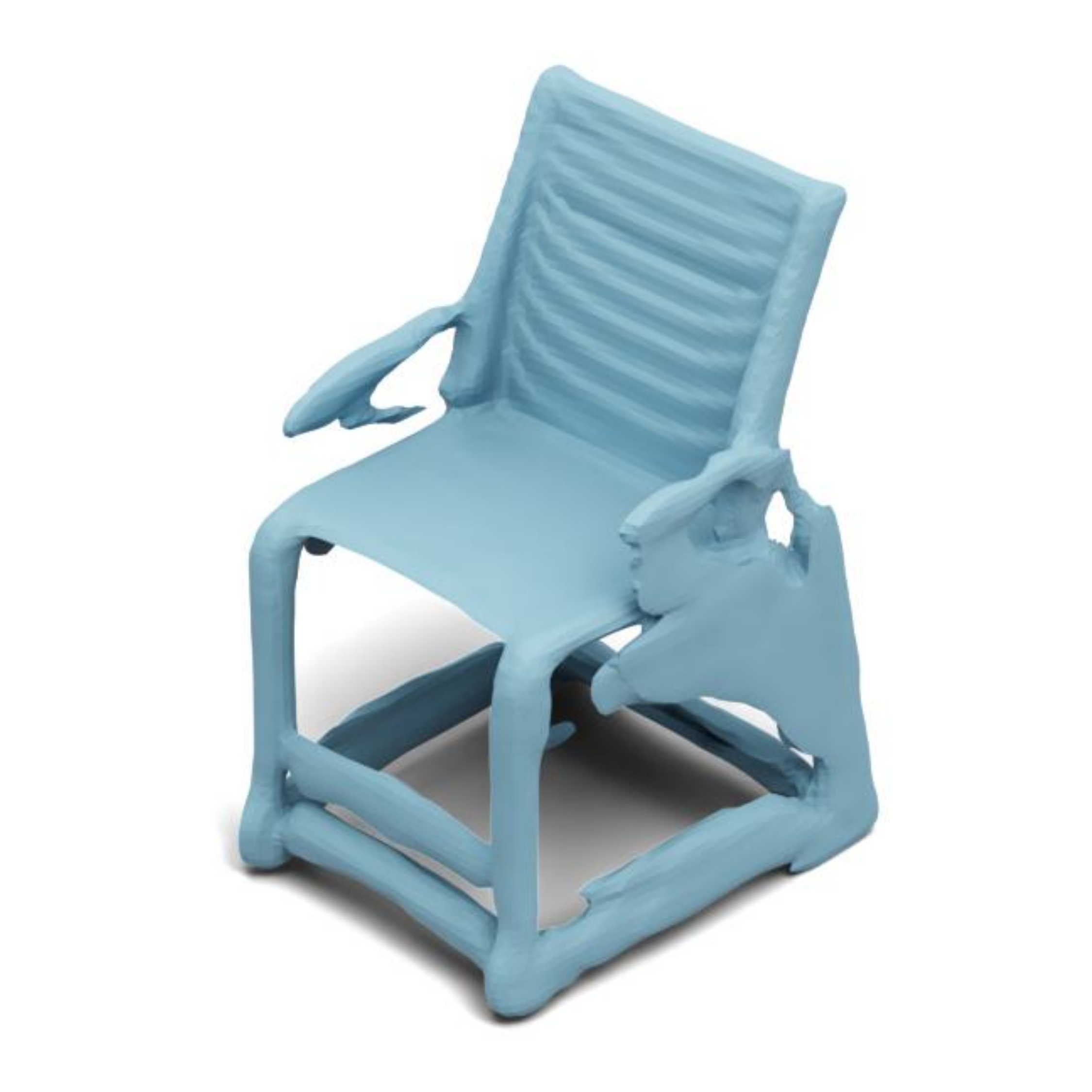}
&\includegraphics[trim = 1 1 1 1, clip, width=0.125\linewidth]{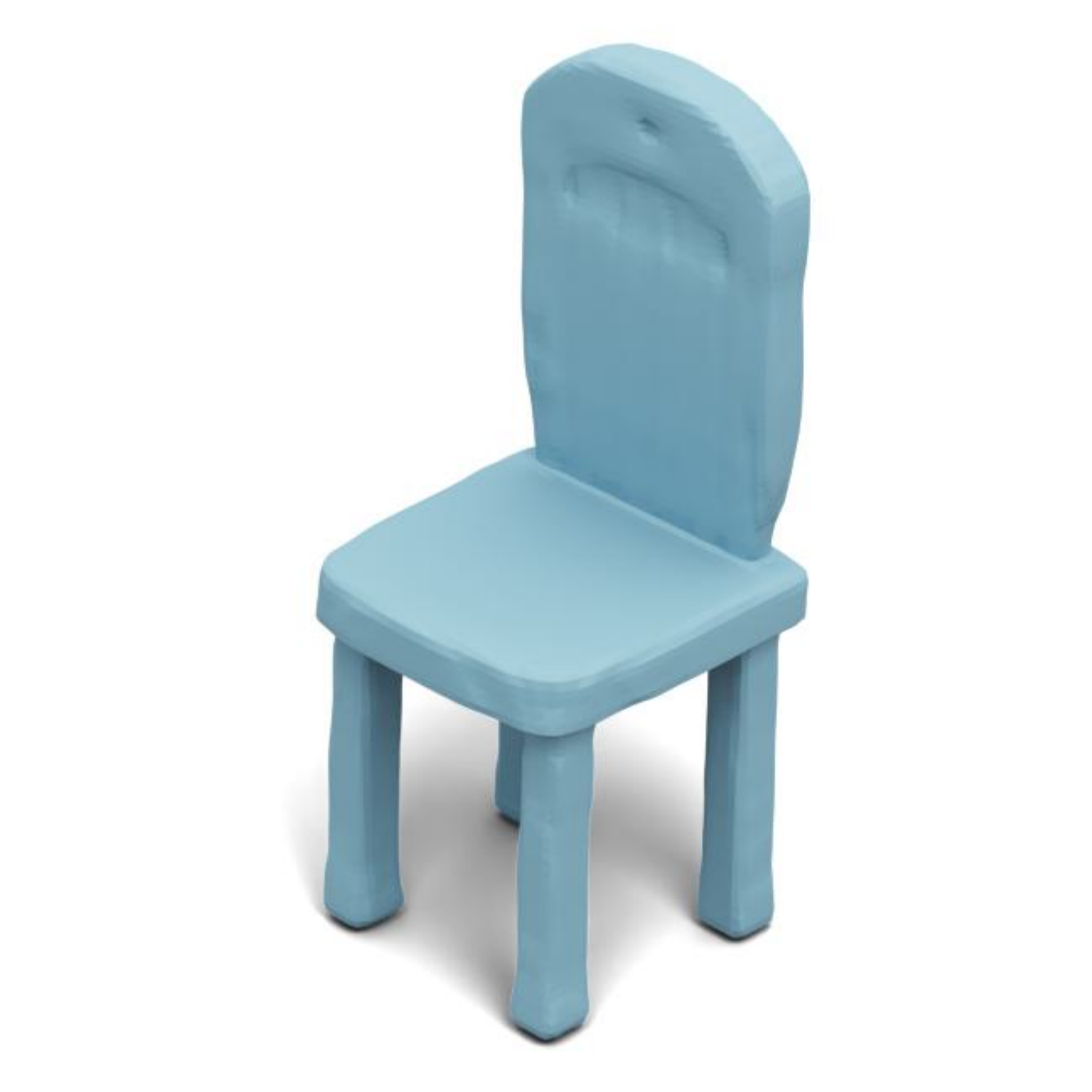}
&\includegraphics[trim = 1 1 1 1, clip, width=0.125\linewidth]{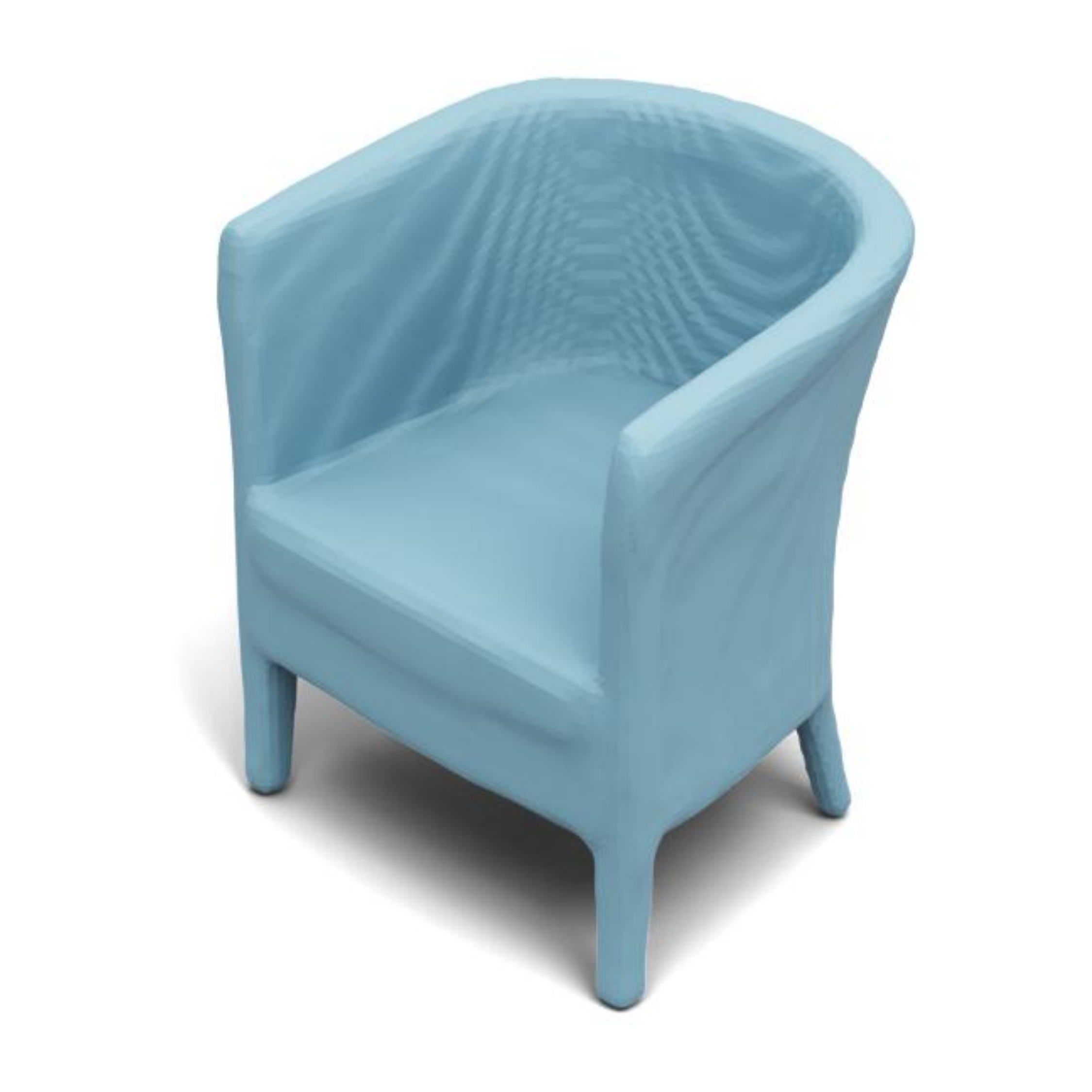}
&\includegraphics[trim = 1 1 1 1, clip, width=0.125\linewidth]{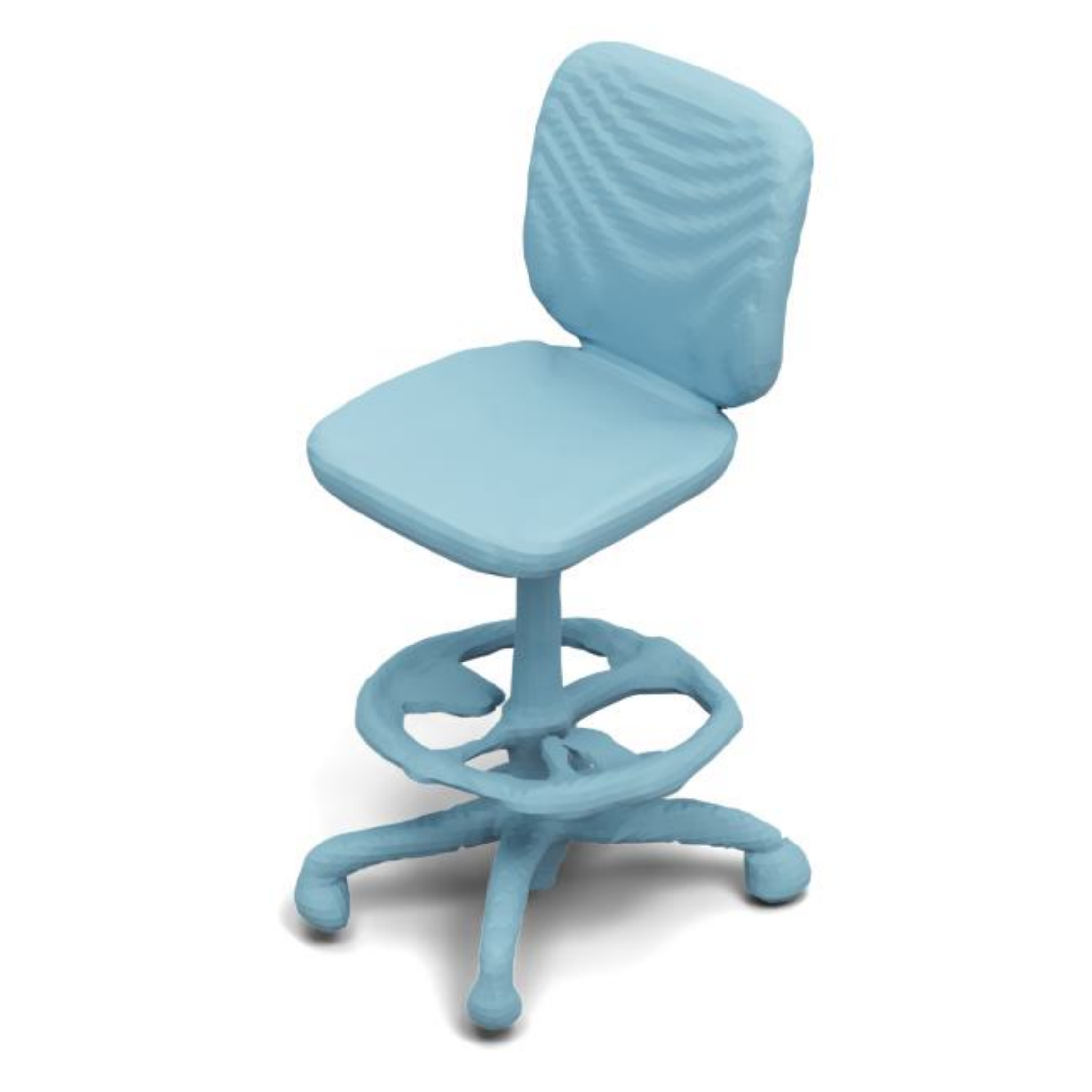}
\\
\end{tabular}
    \caption{We randomly sample sketches from the AmateurSketch dataset and showcase the results of our method.}
	\label{fig:additionalExp1}
\end{figure*}

\begin{figure*}[t]
	\centering
	\small
	\setlength{\tabcolsep}{1pt}
 \begin{tabular}{cccccccc}
\includegraphics[width=0.125\linewidth]{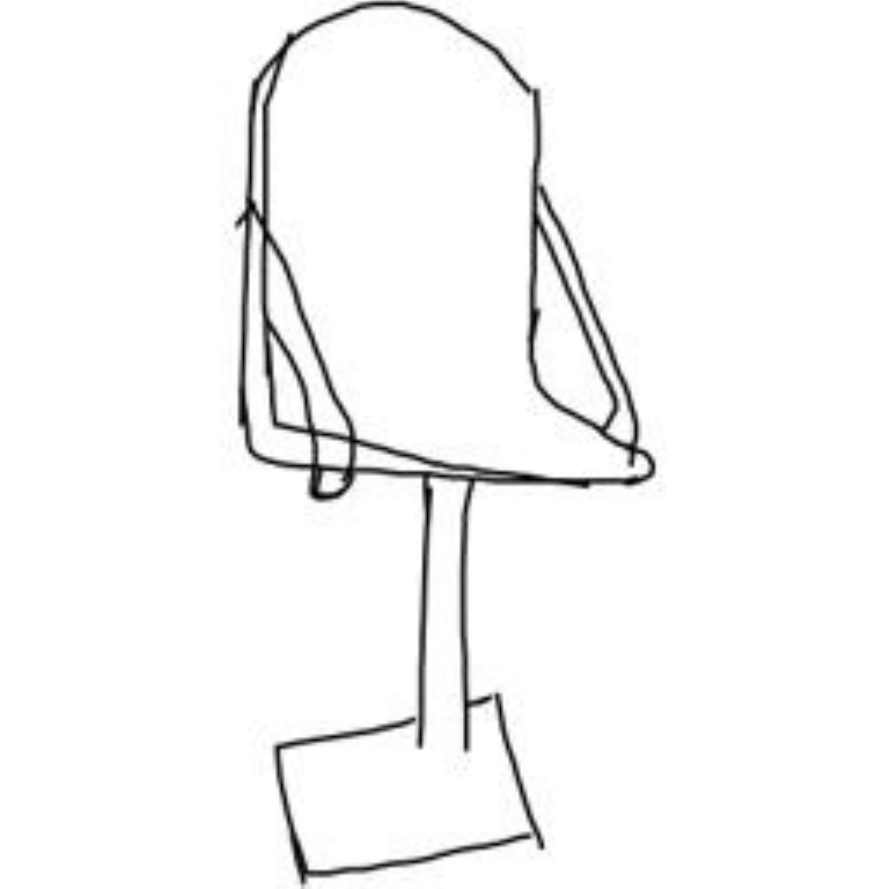}
&\includegraphics[width=0.125\linewidth]{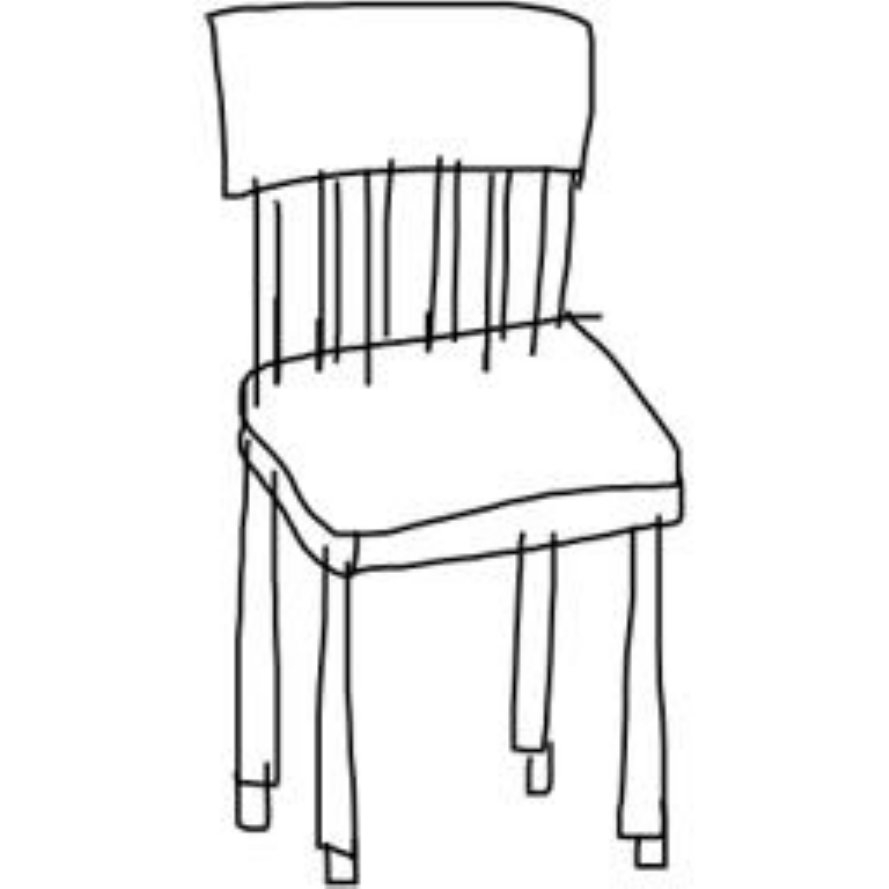}
&\includegraphics[width=0.125\linewidth]{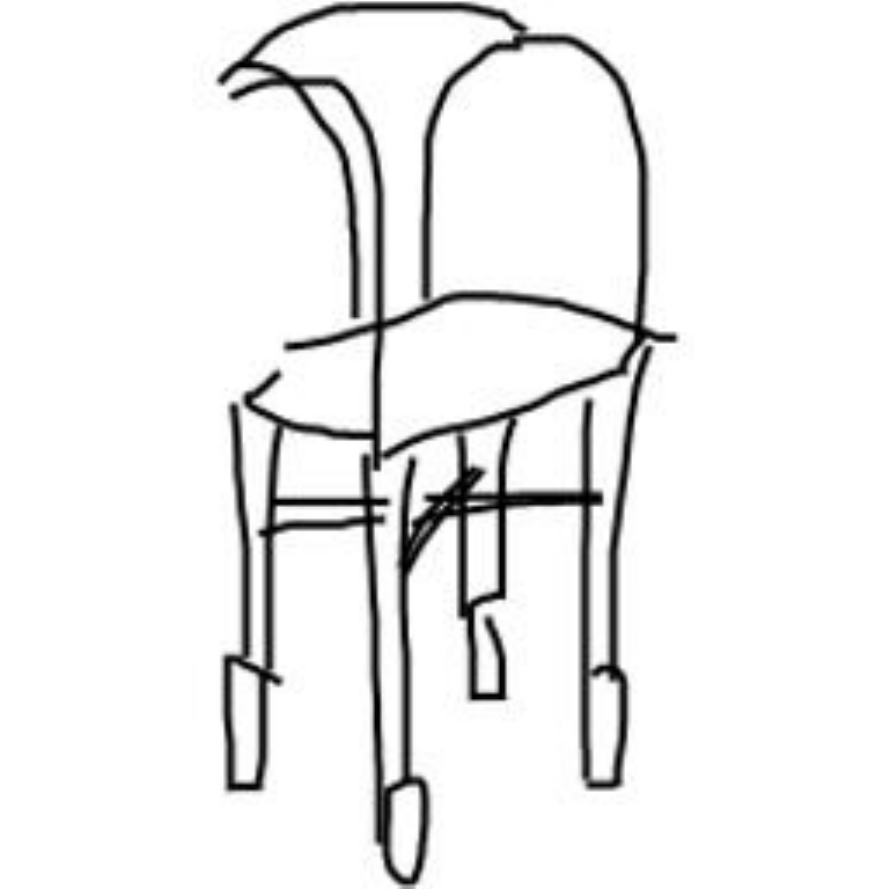}
&\includegraphics[width=0.125\linewidth]{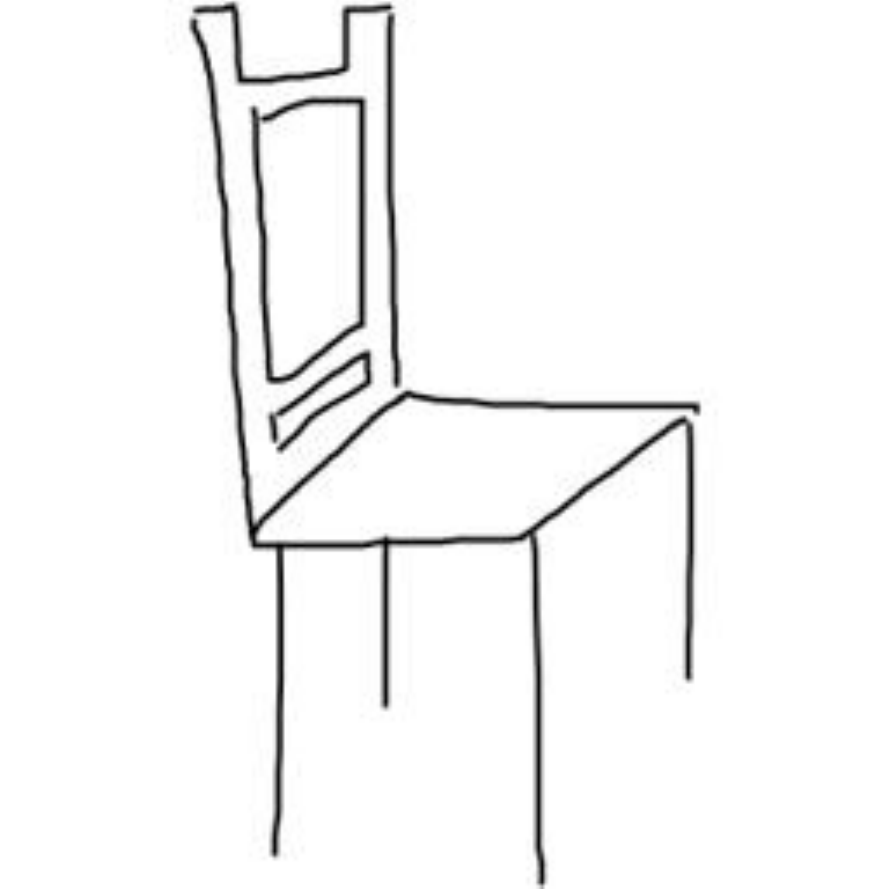}
&\includegraphics[width=0.125\linewidth]{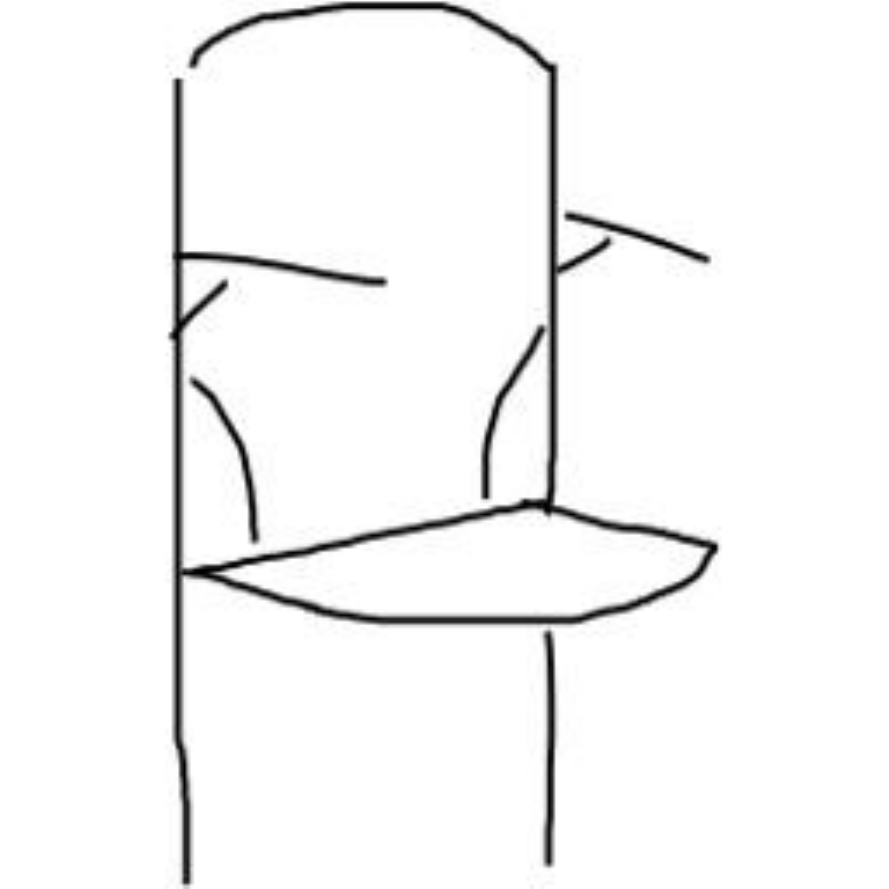}
&\includegraphics[width=0.125\linewidth]{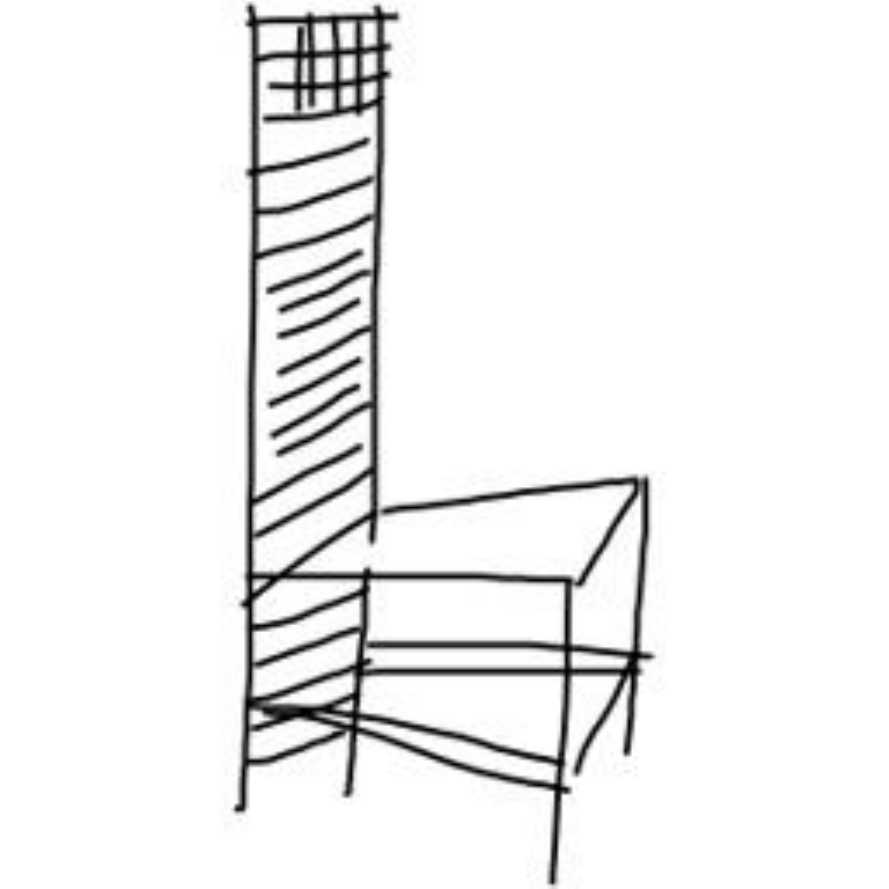}
&\includegraphics[width=0.125\linewidth]{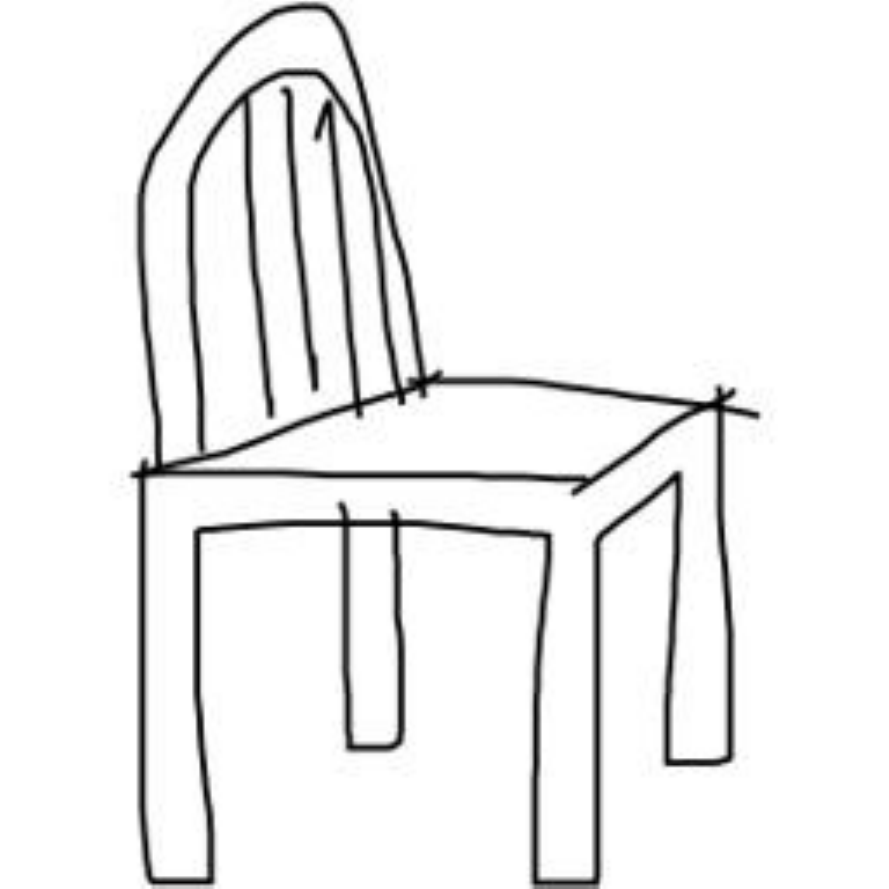}
&\includegraphics[width=0.125\linewidth]{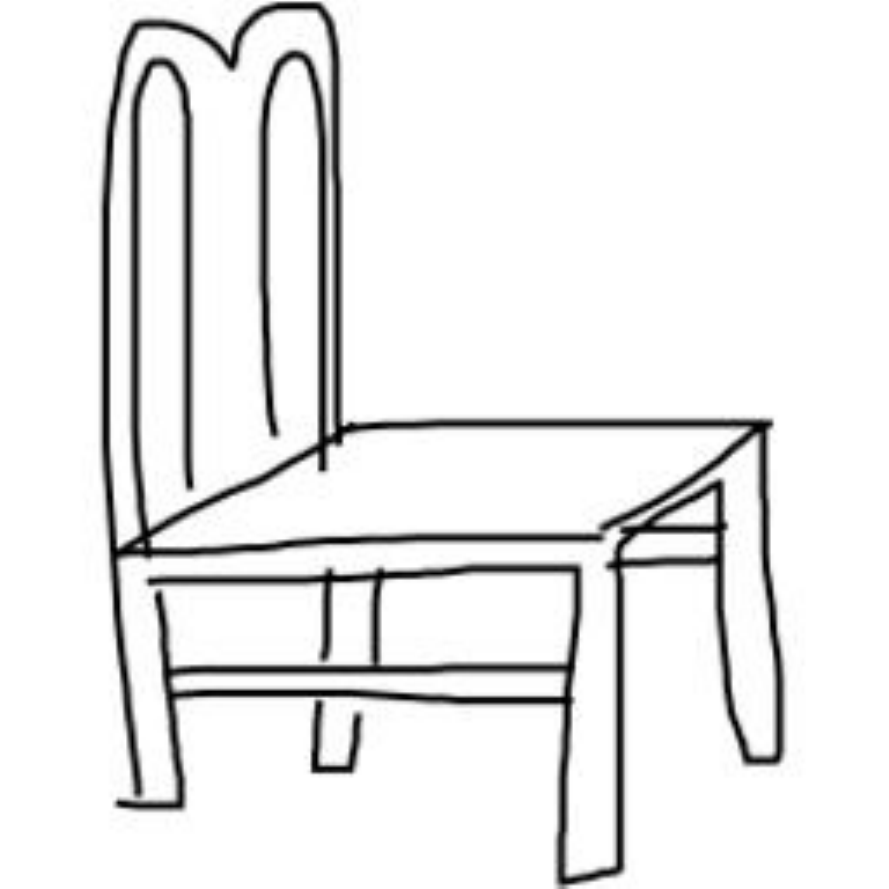}
\\
\includegraphics[trim = 1 1 1 1, clip, width=0.125\linewidth]{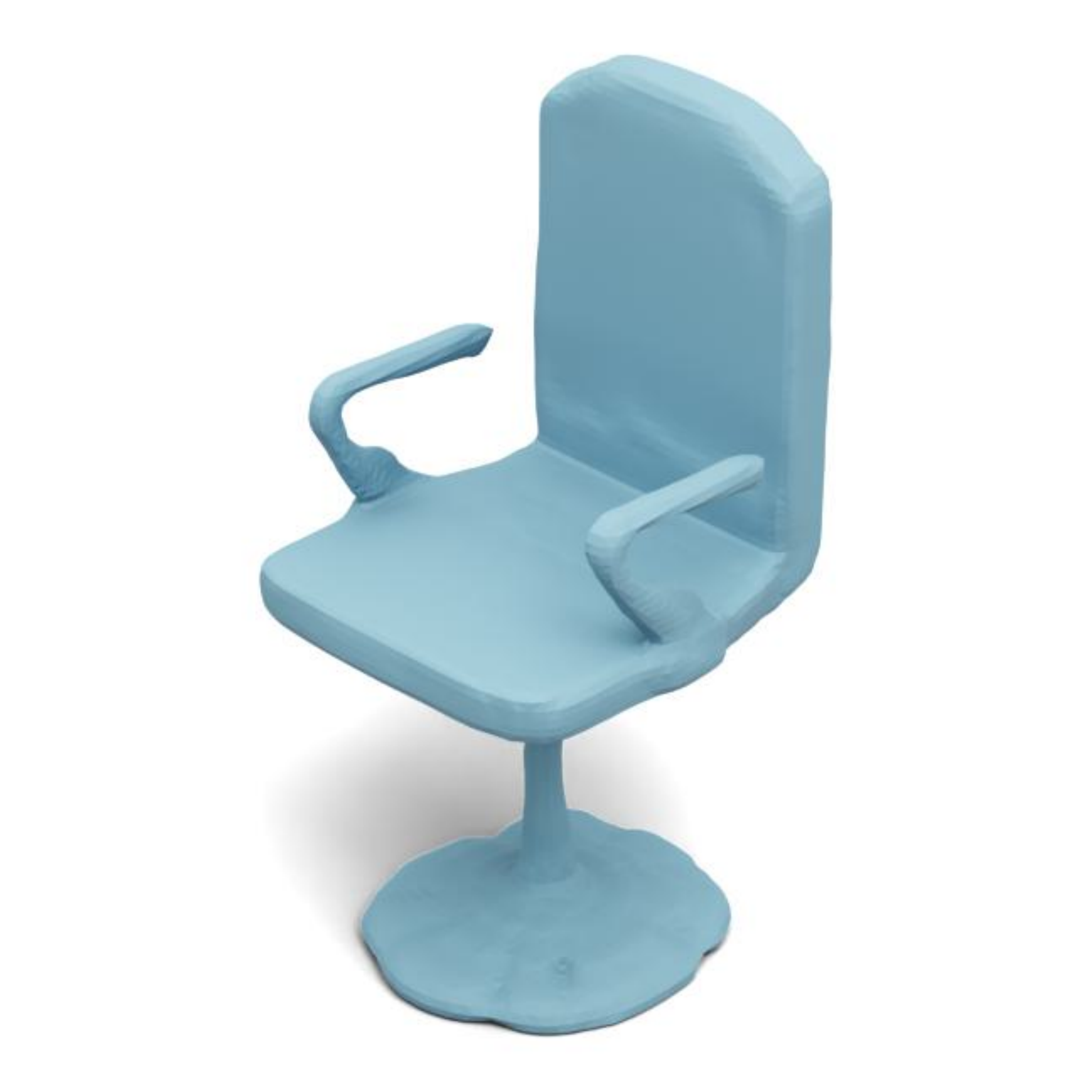}
&\includegraphics[trim = 1 1 1 1, clip, width=0.125\linewidth]{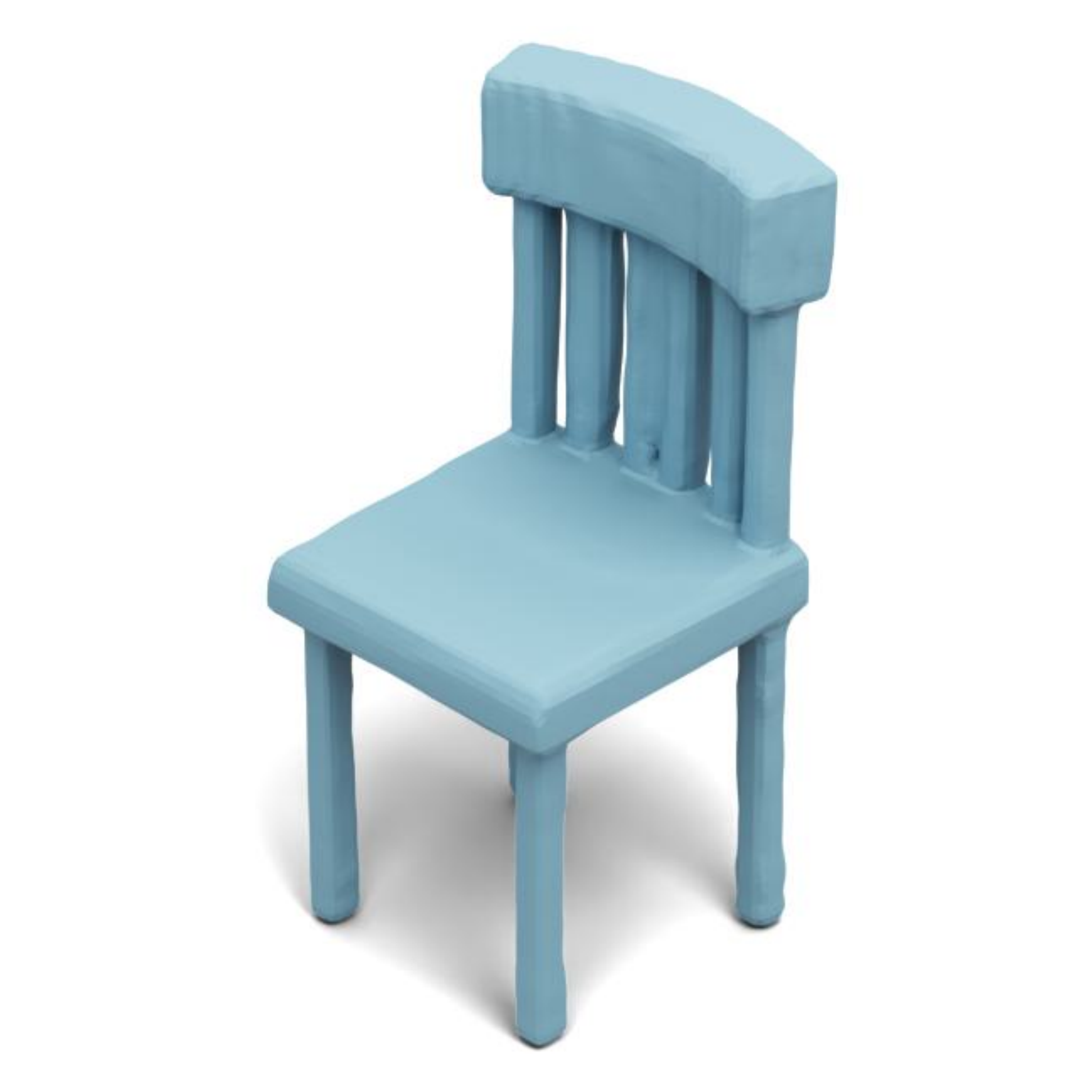}
&\includegraphics[trim = 1 1 1 1, clip, width=0.125\linewidth]{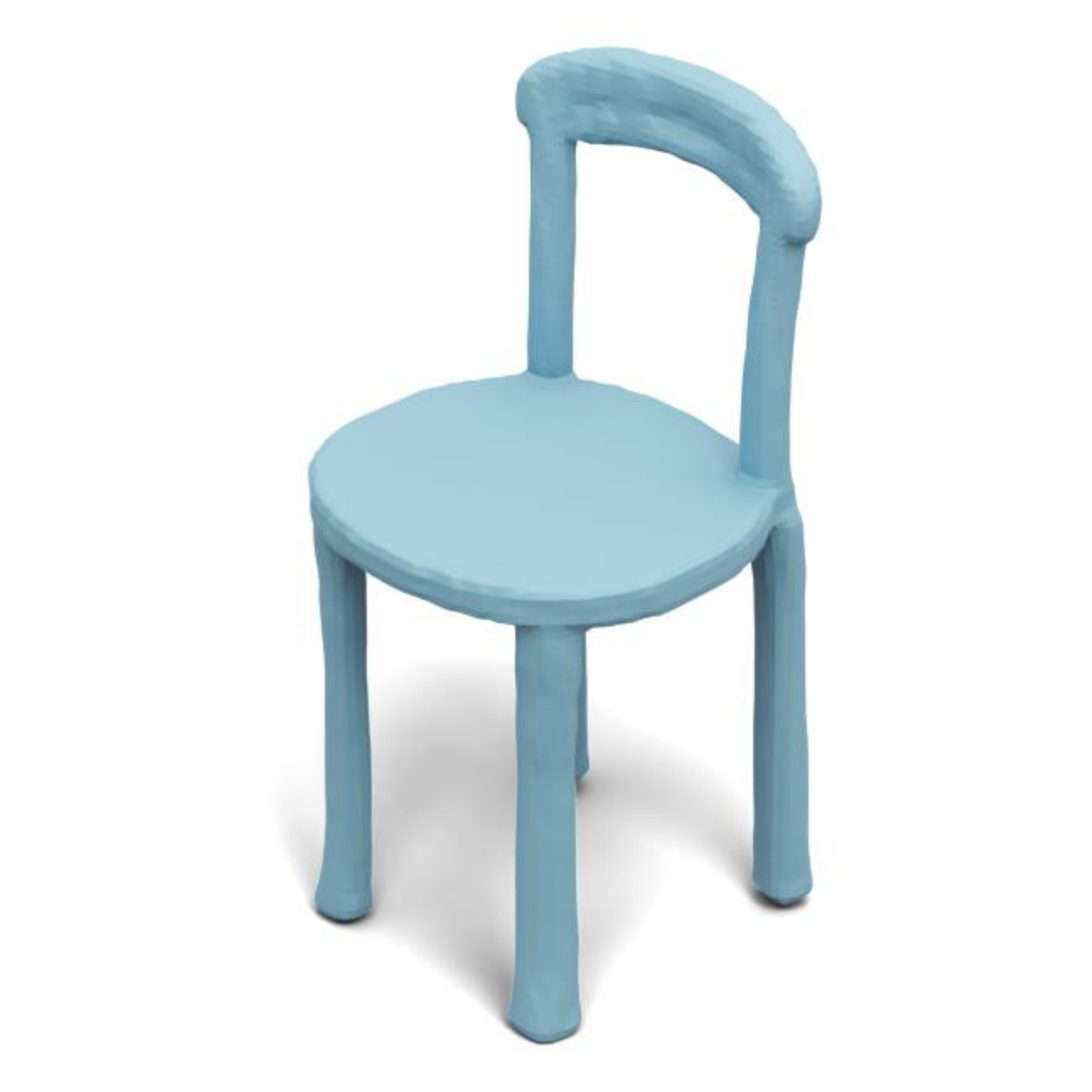}
&\includegraphics[trim = 1 1 1 1, clip, width=0.125\linewidth]{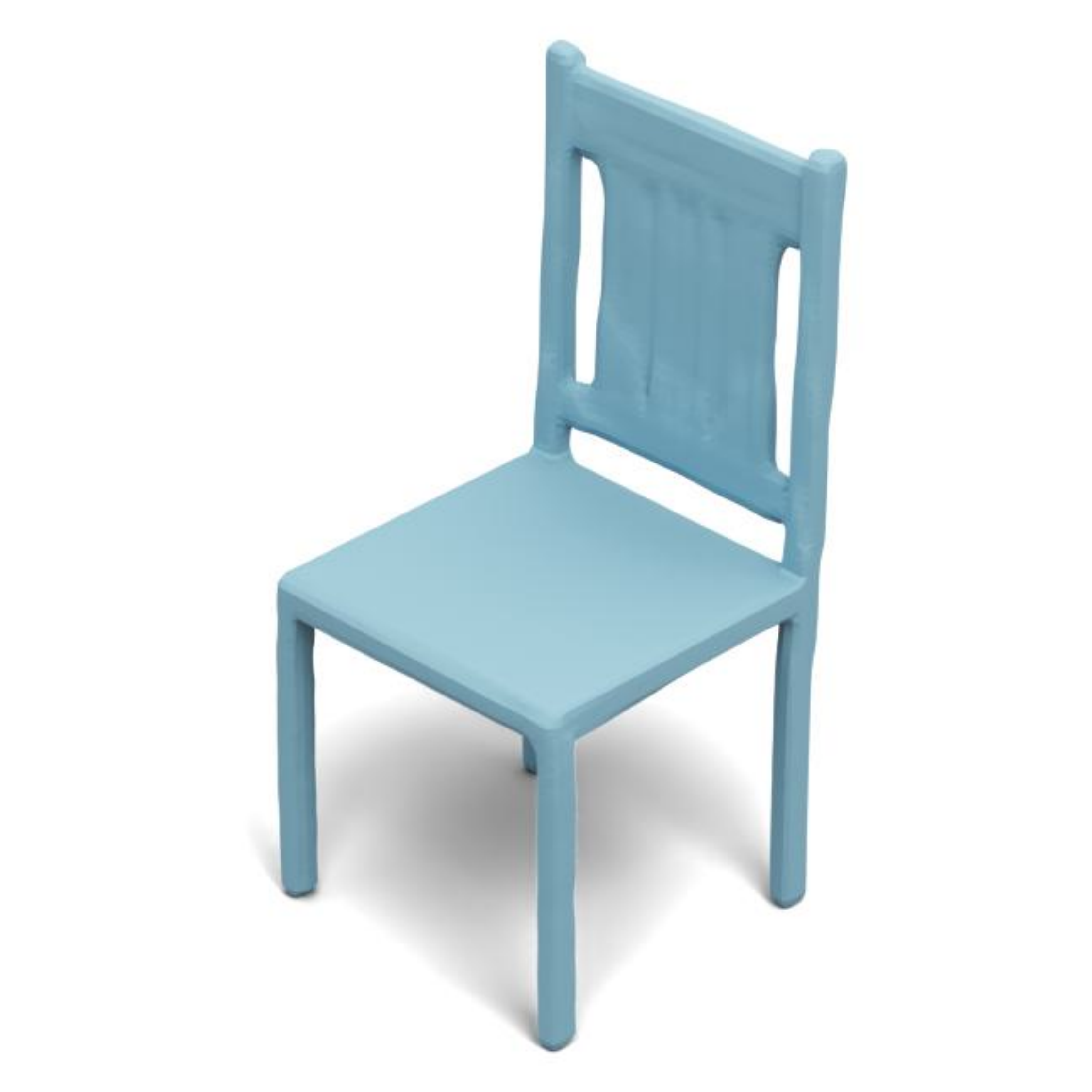}
&\includegraphics[trim = 1 1 1 1, clip, width=0.125\linewidth]{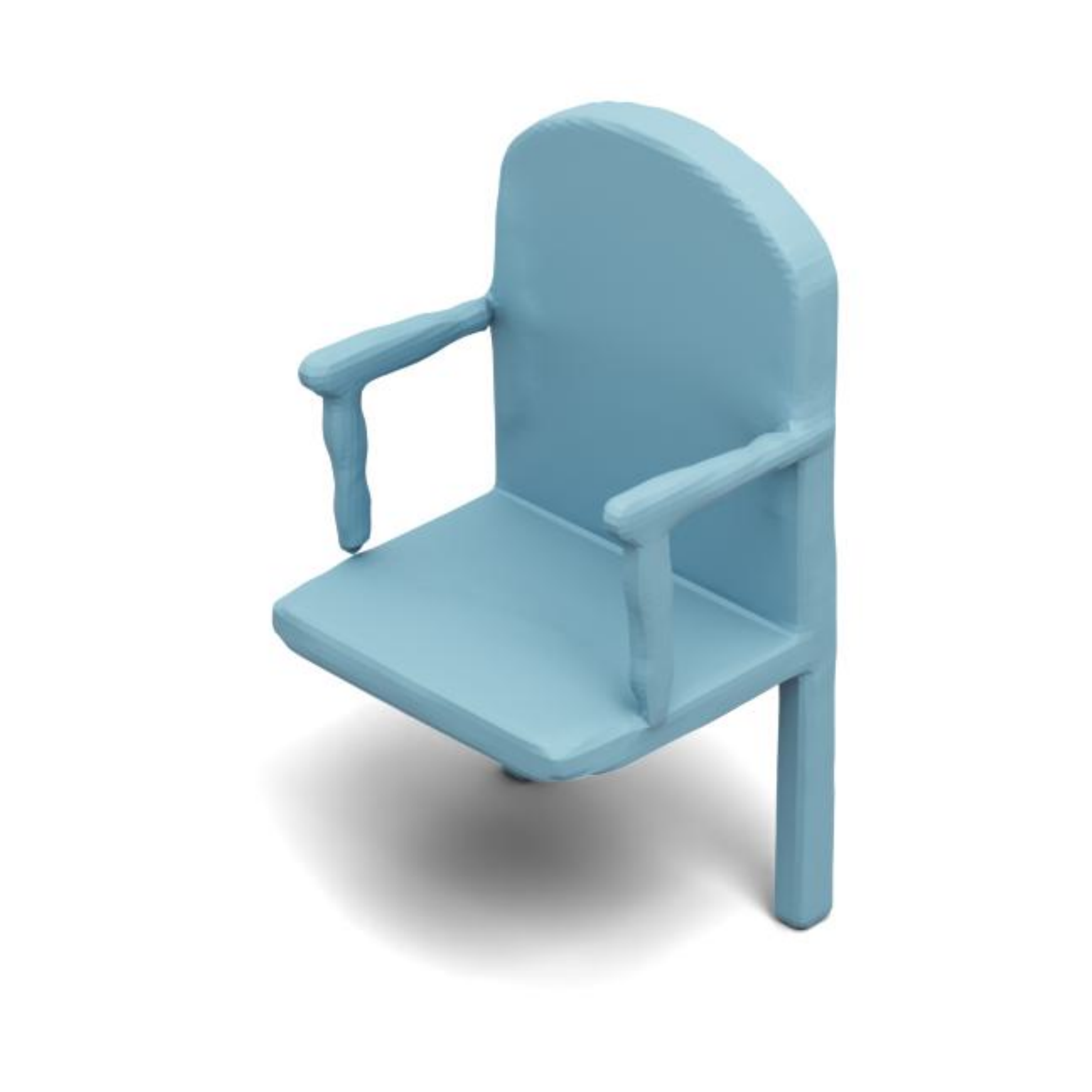}
&\includegraphics[trim = 1 1 1 1, clip, width=0.125\linewidth]{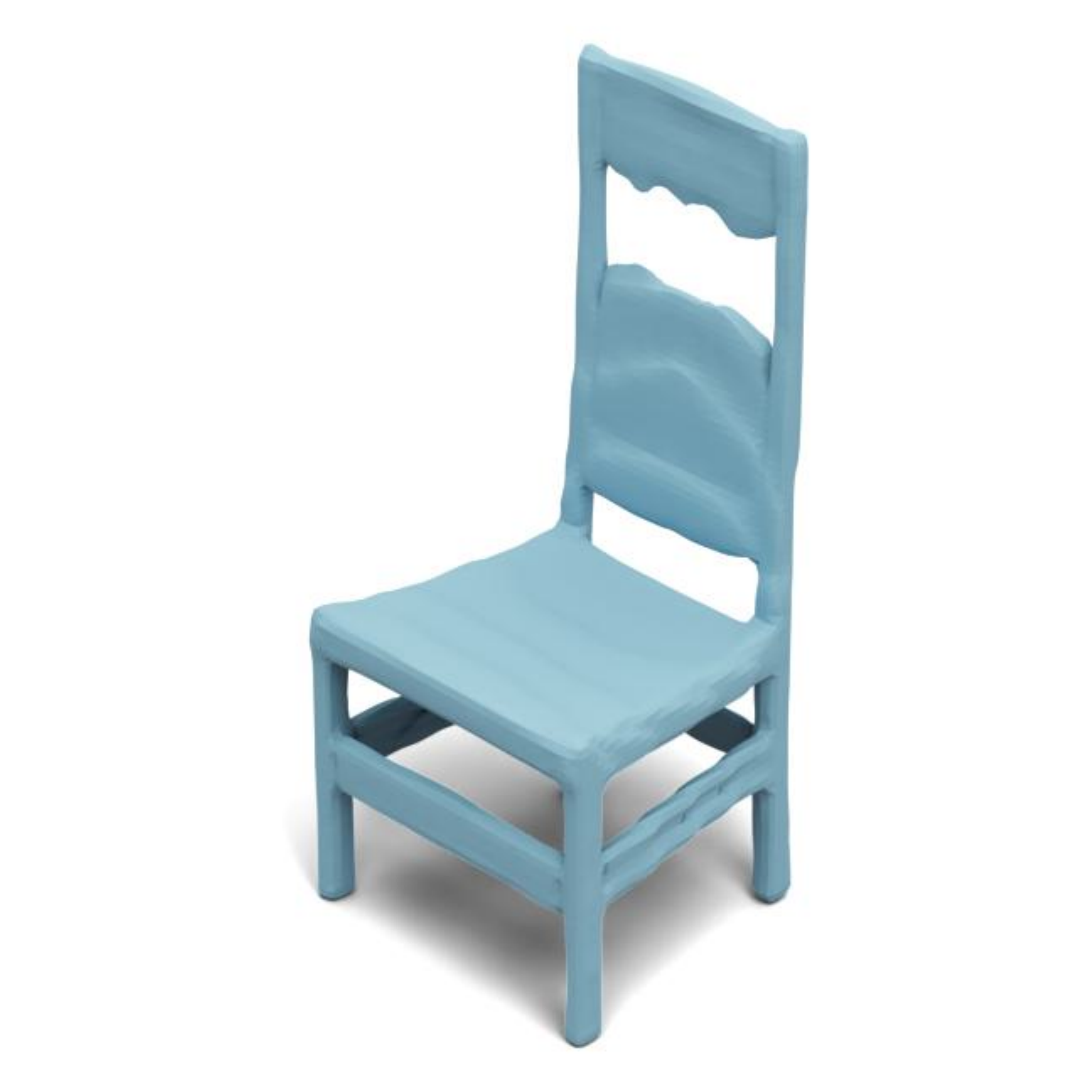}
&\includegraphics[trim = 1 1 1 1, clip, width=0.125\linewidth]{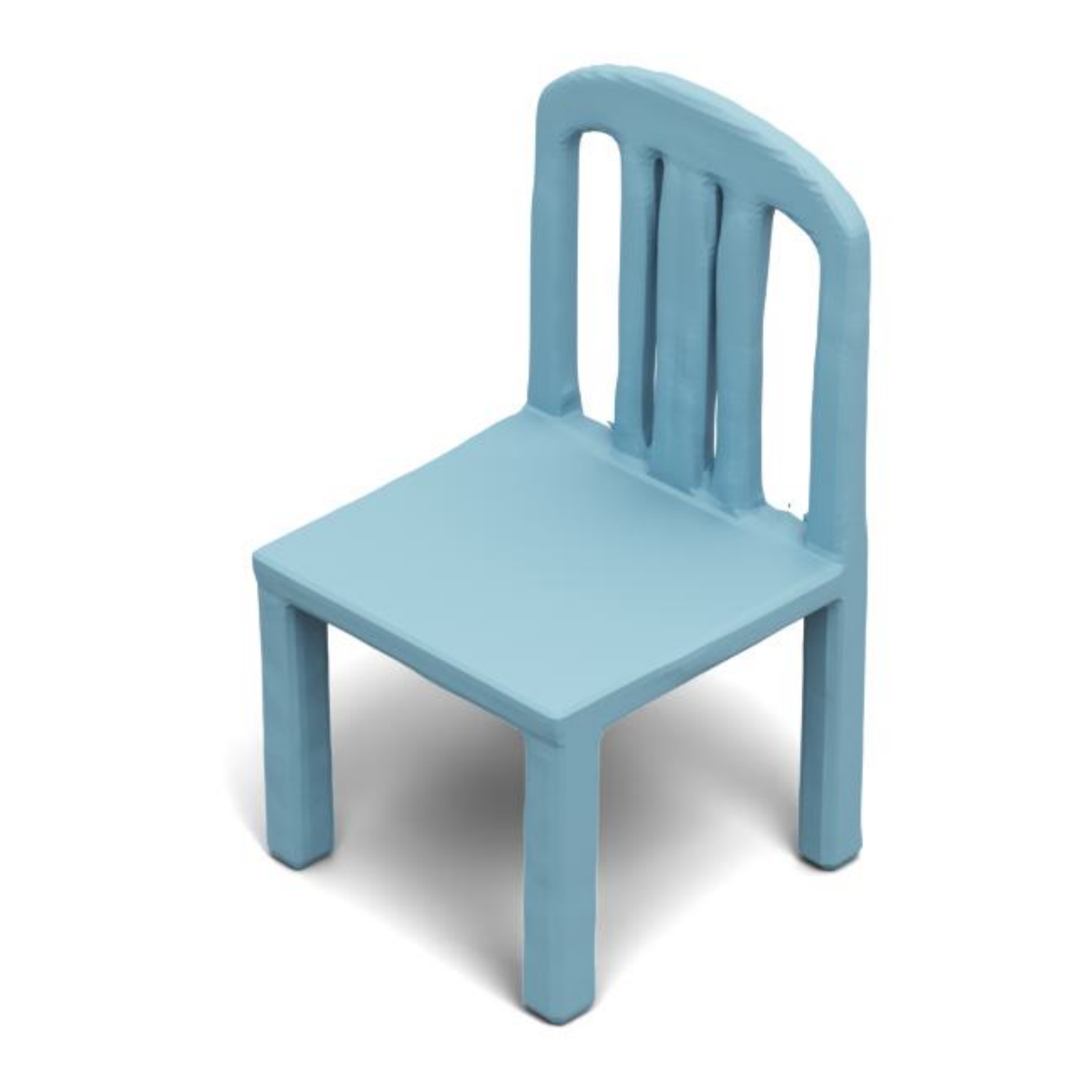}
&\includegraphics[trim = 1 1 1 1, clip, width=0.125\linewidth]{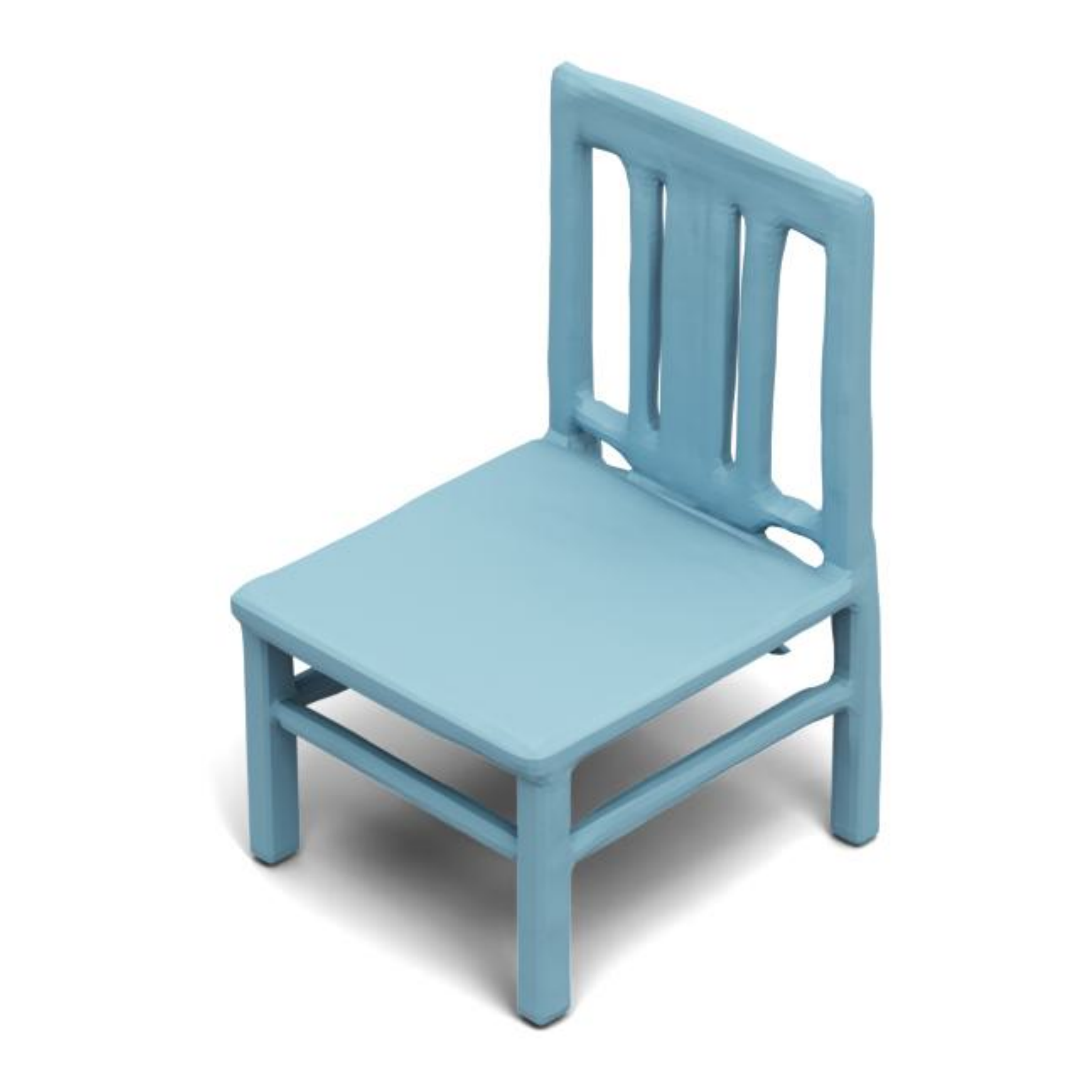}
\\
\includegraphics[width=0.125\linewidth]{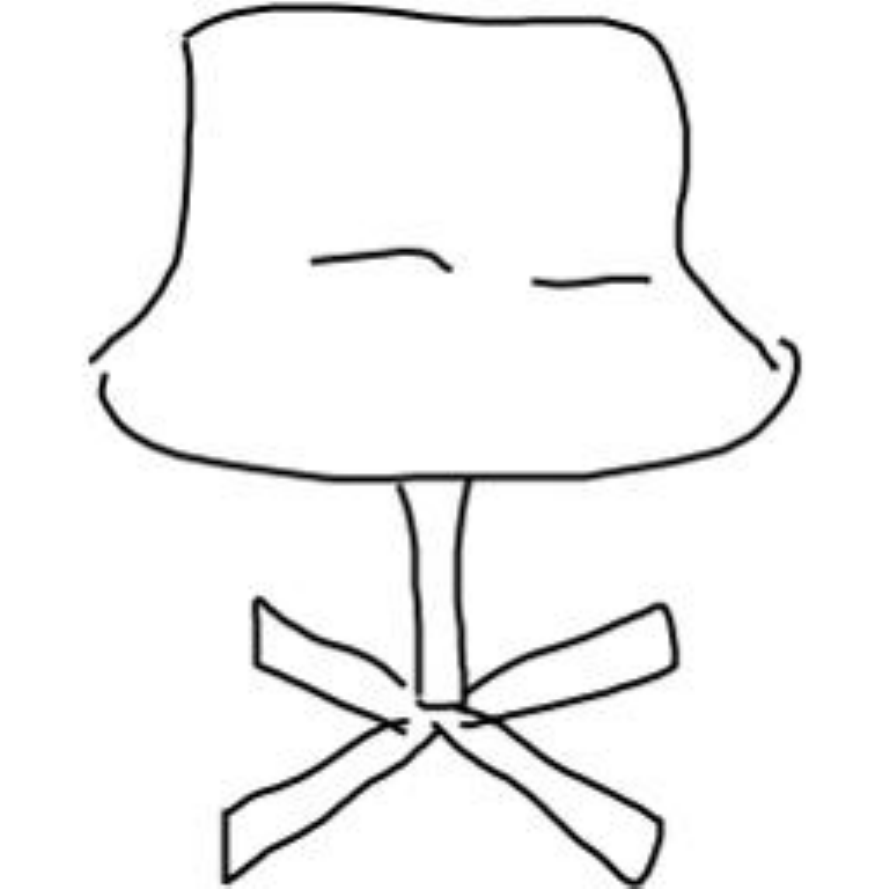}
&\includegraphics[width=0.125\linewidth]{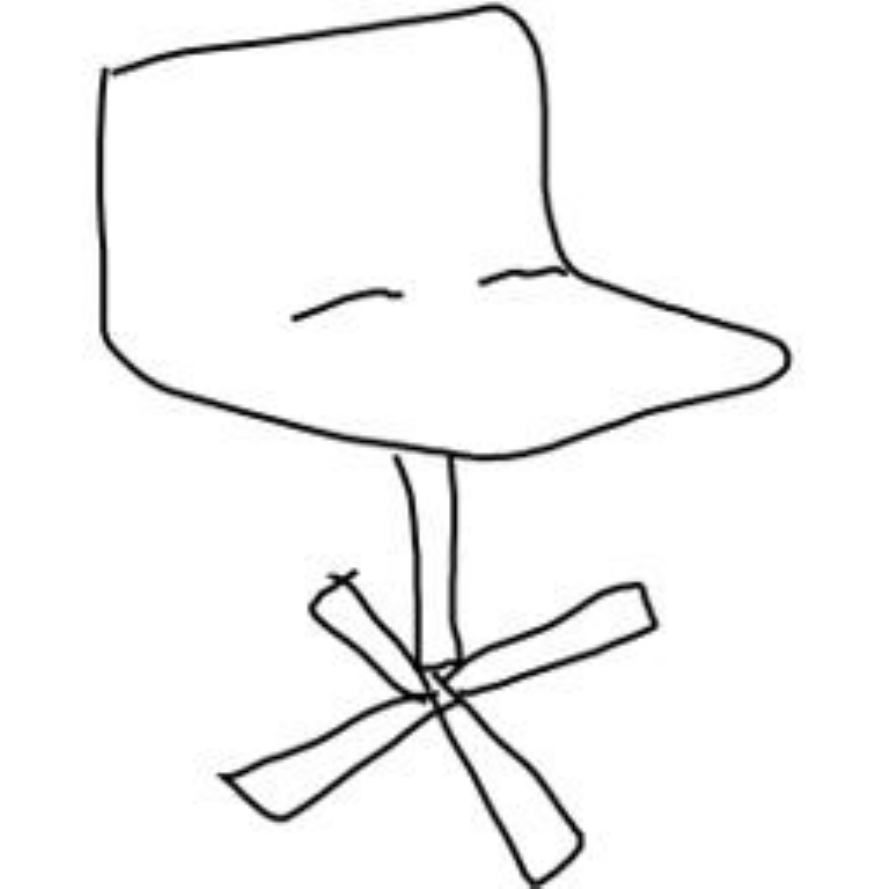}
&\includegraphics[width=0.125\linewidth]{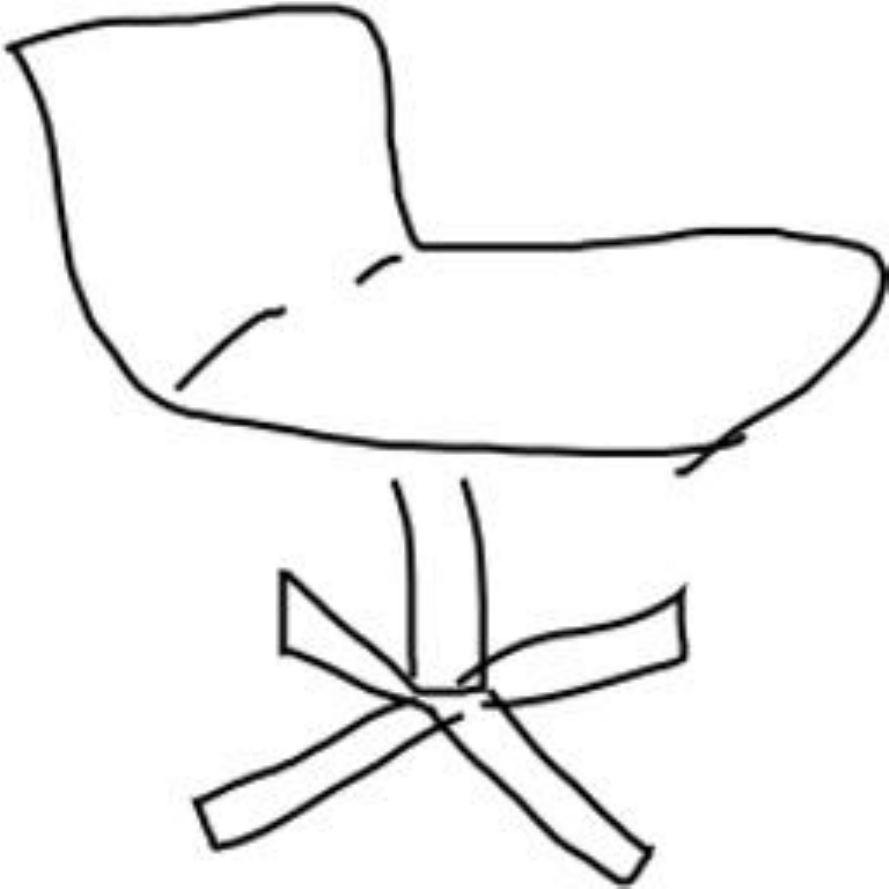}
&\includegraphics[width=0.125\linewidth]{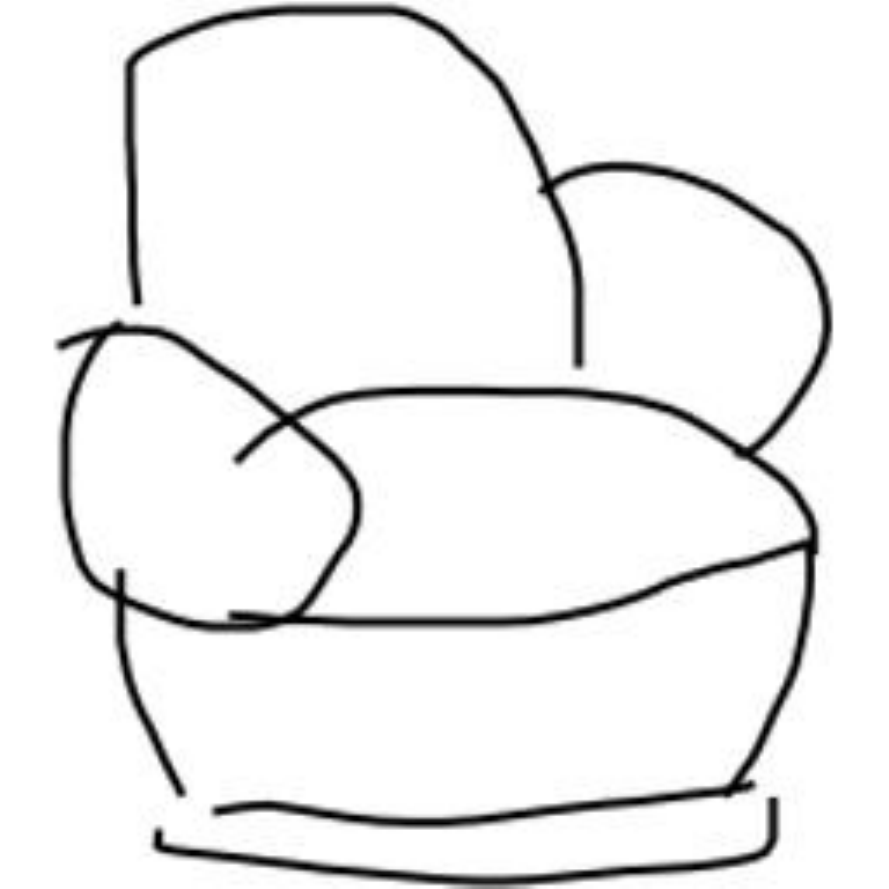}
&\includegraphics[width=0.125\linewidth]{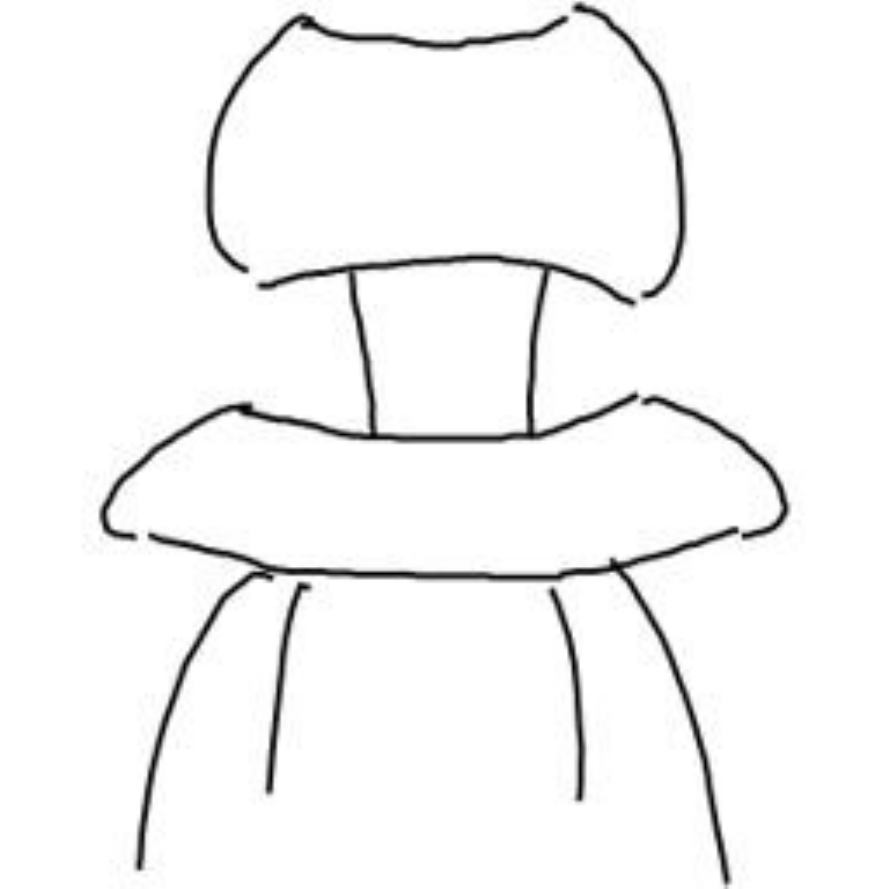}
&\includegraphics[width=0.125\linewidth]{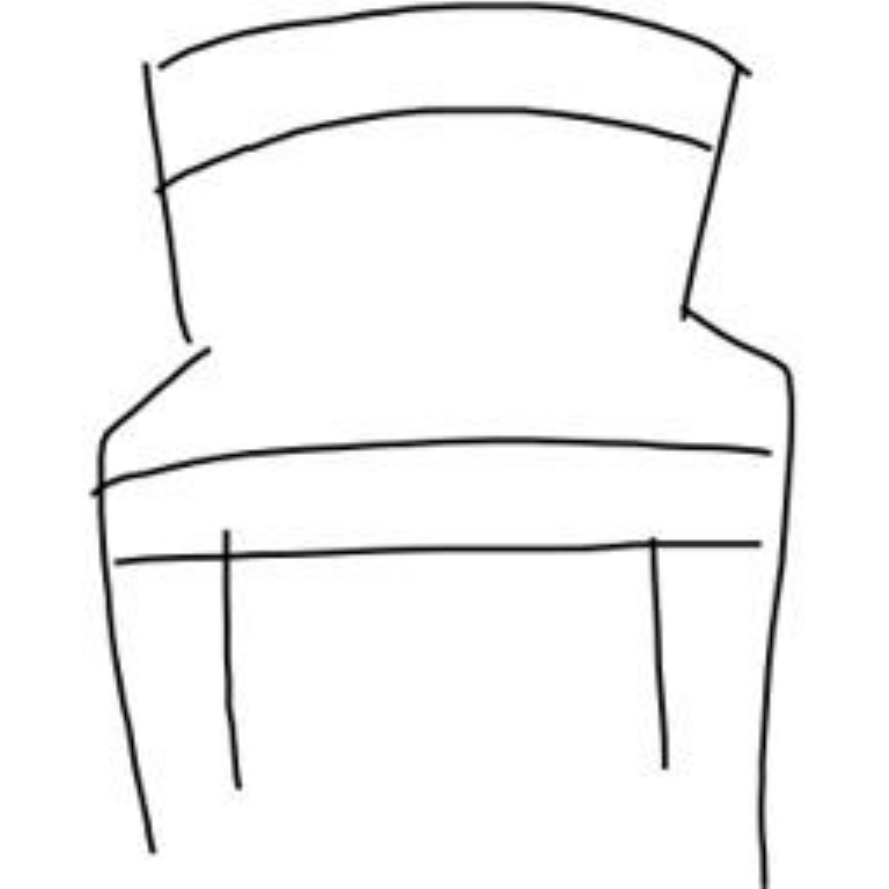}
&\includegraphics[width=0.125\linewidth]{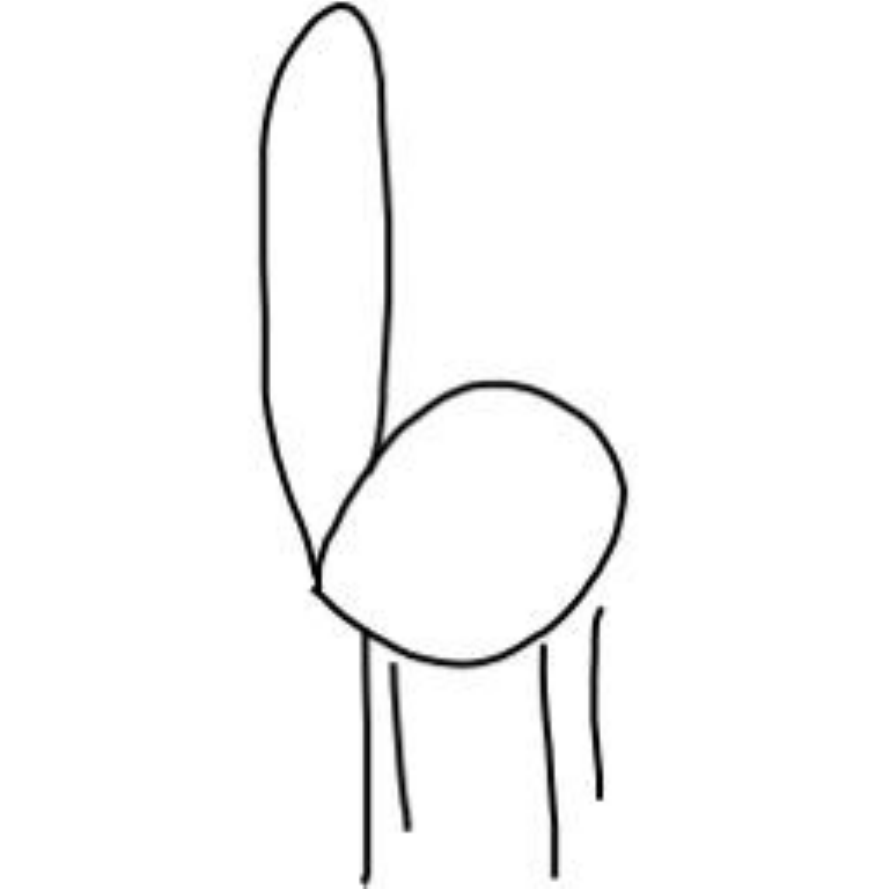}
&\includegraphics[width=0.125\linewidth]{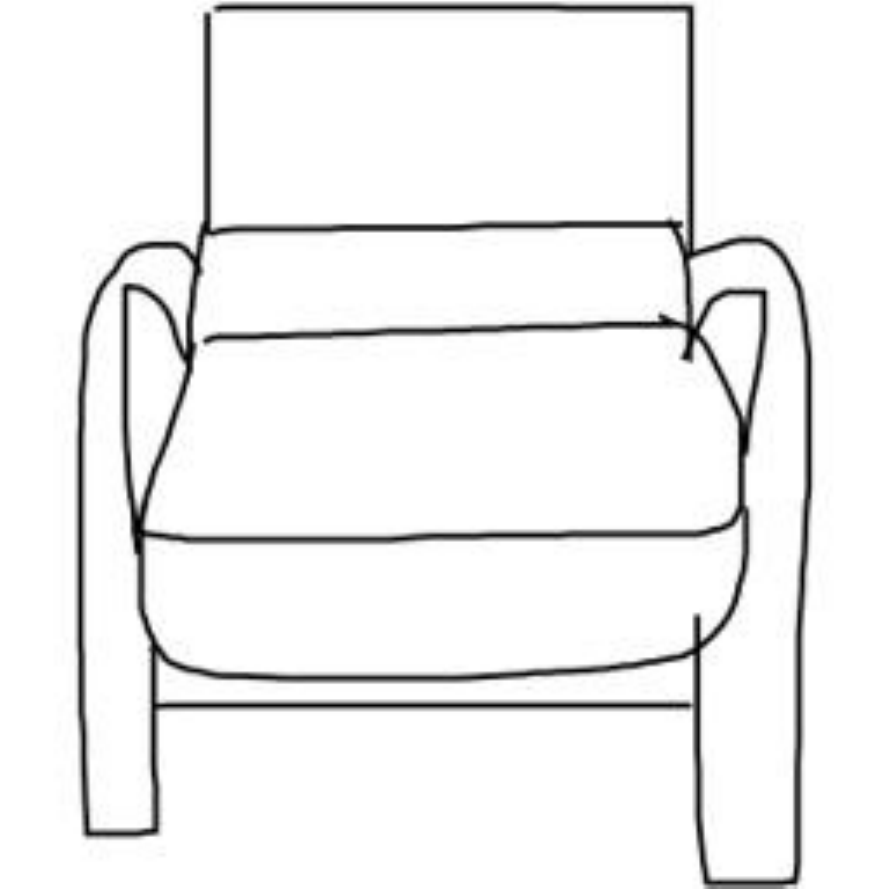}
\\
\includegraphics[trim = 1 1 1 1, clip, width=0.125\linewidth]{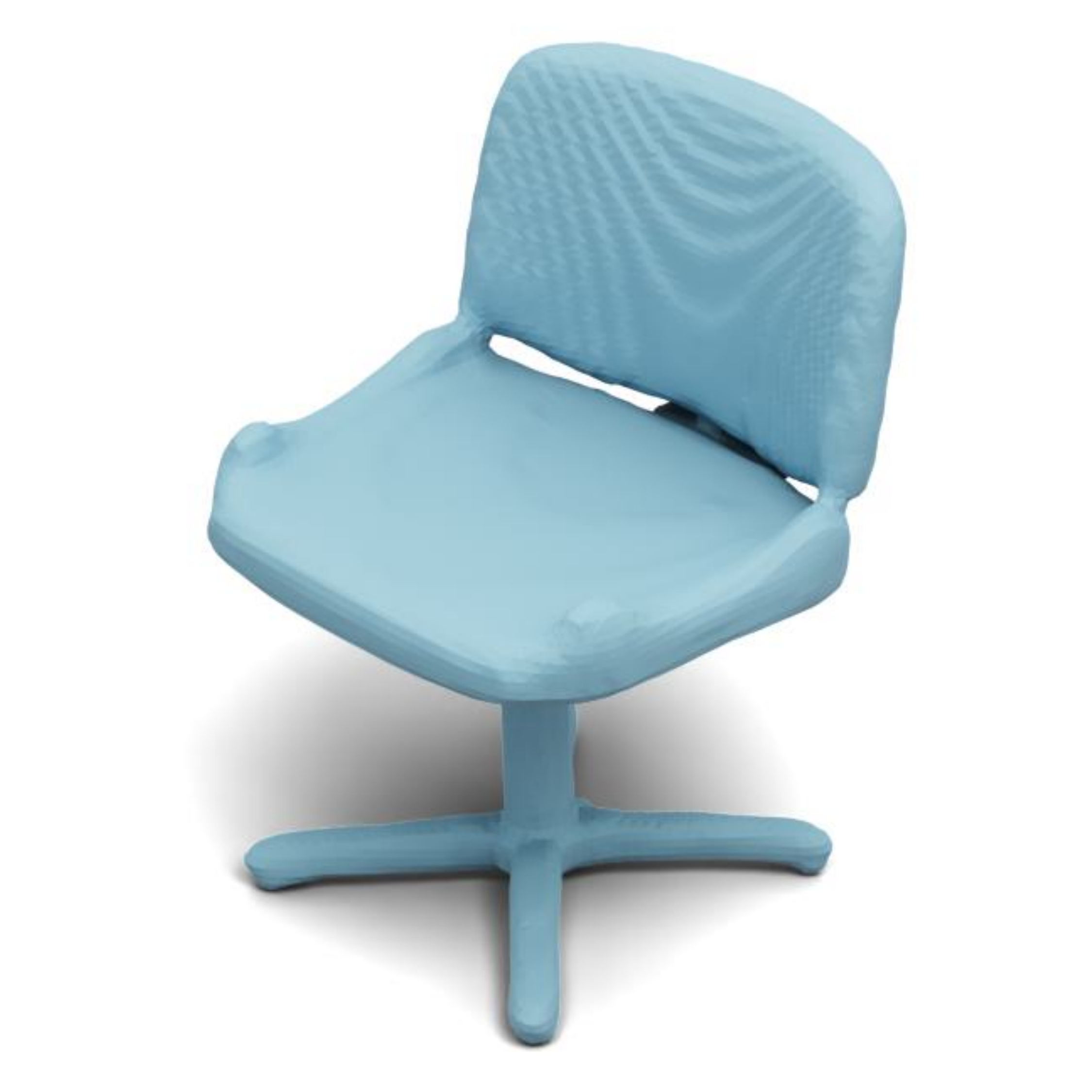}
&\includegraphics[trim = 1 1 1 1, clip, width=0.125\linewidth]{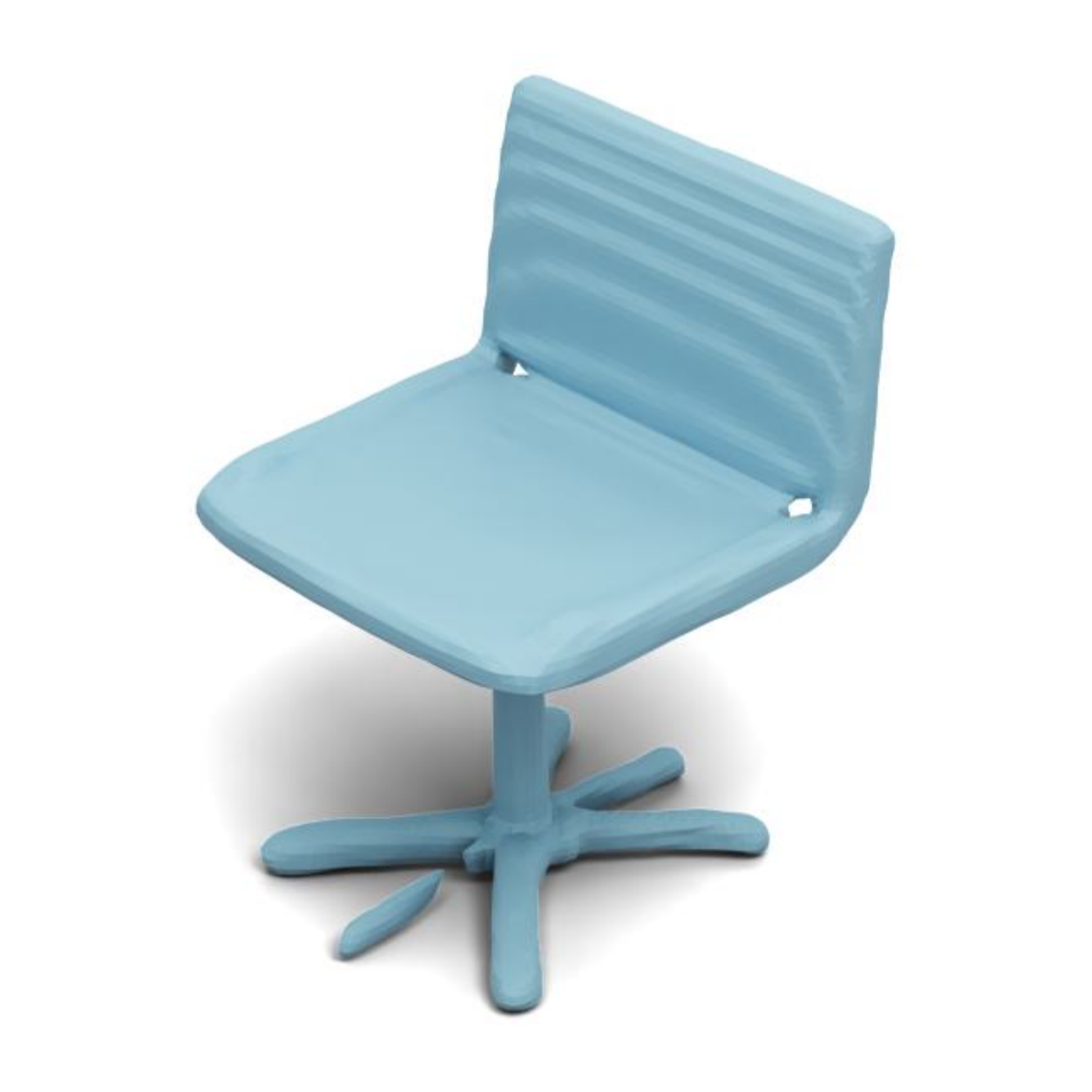}
&\includegraphics[trim = 1 1 1 1, clip, width=0.125\linewidth]{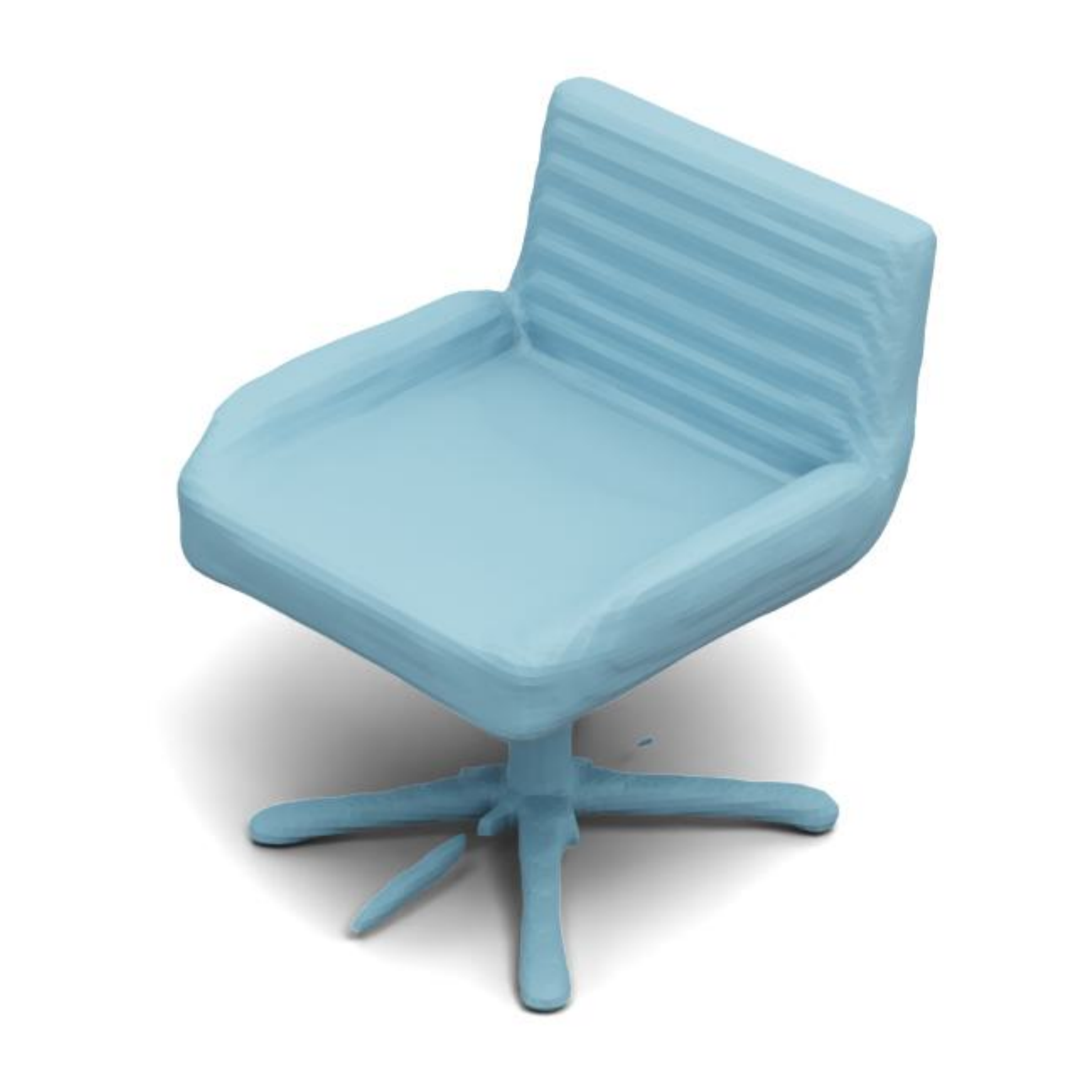}
&\includegraphics[trim = 1 1 1 1, clip, width=0.125\linewidth]{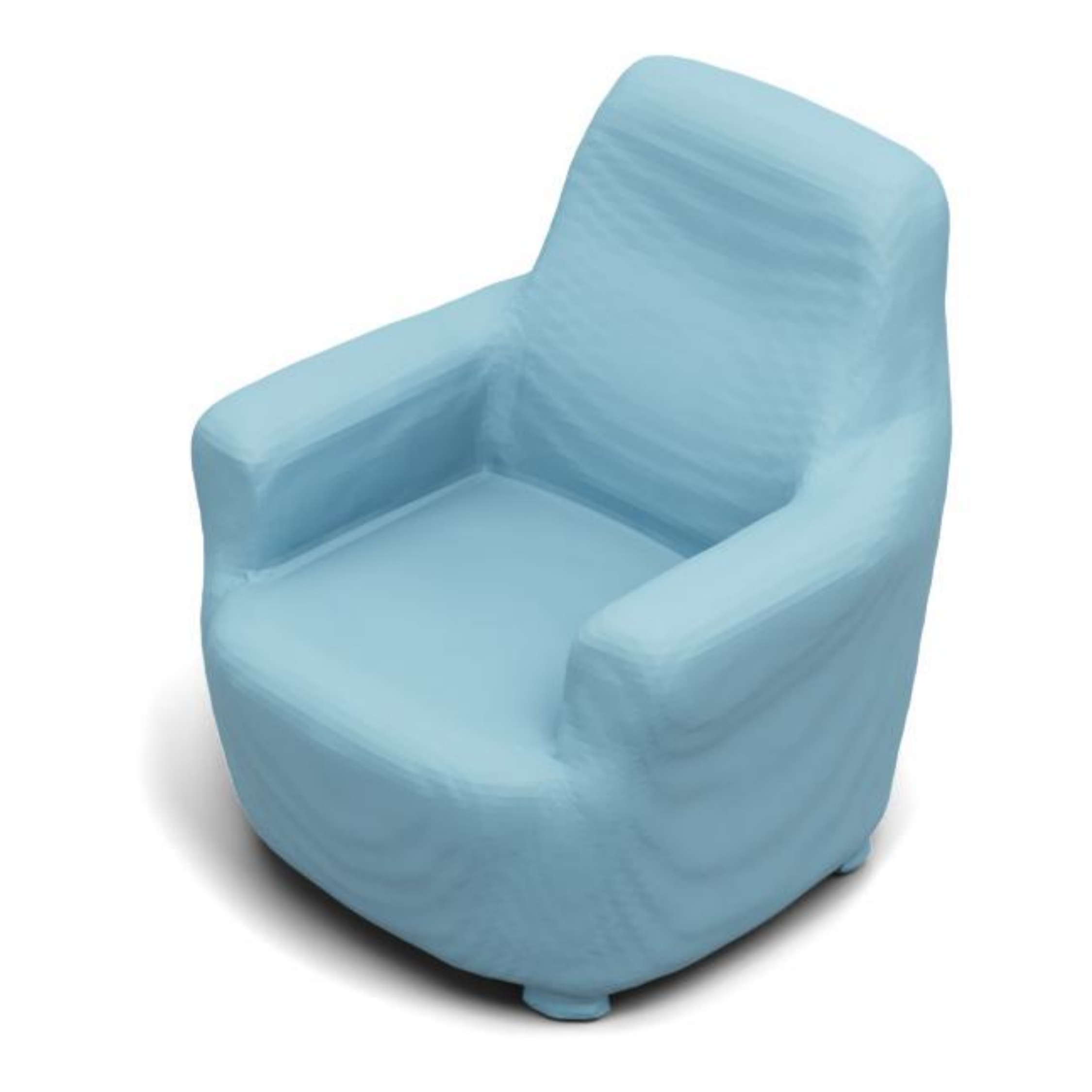}
&\includegraphics[trim = 1 1 1 1, clip, width=0.125\linewidth]{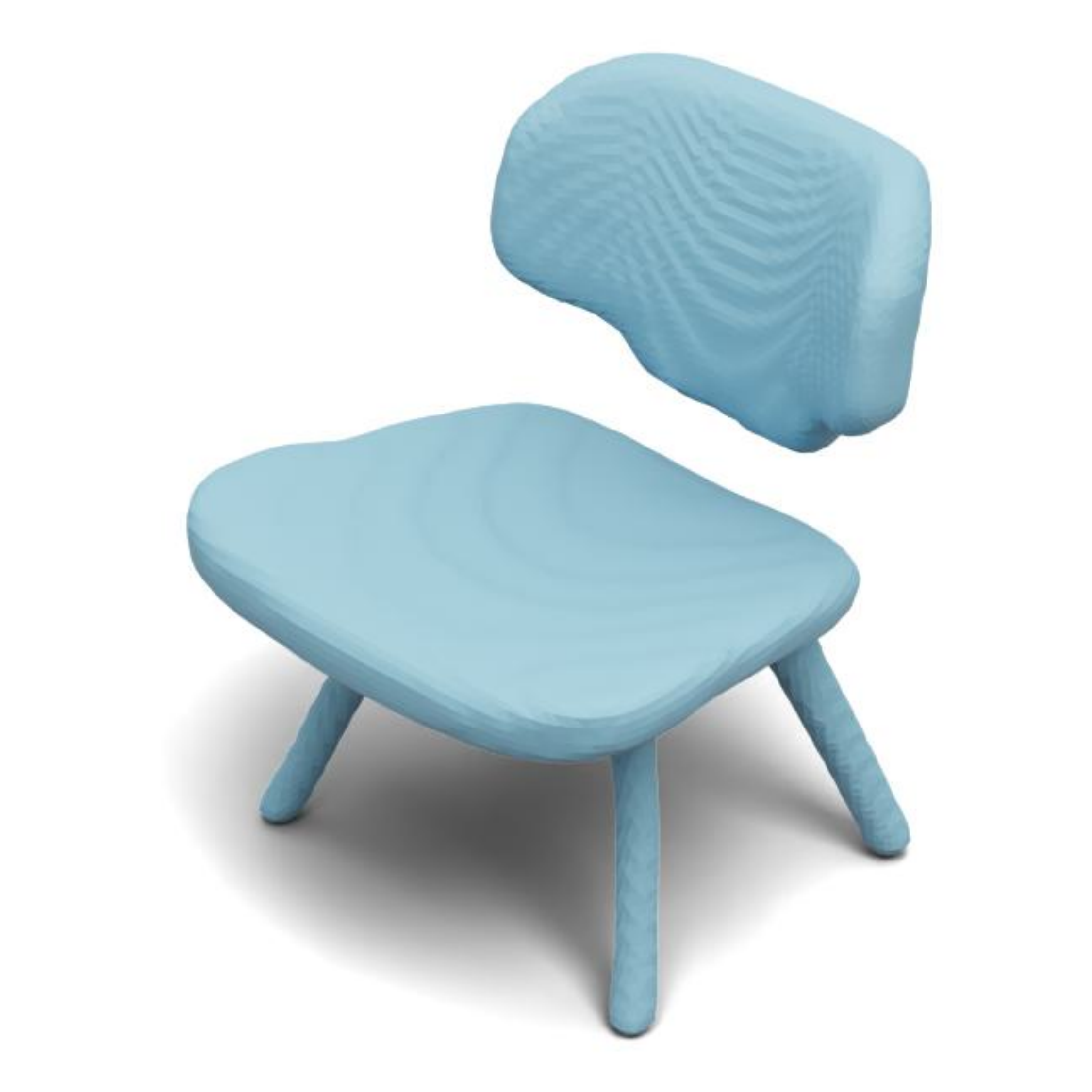}
&\includegraphics[trim = 1 1 1 1, clip, width=0.125\linewidth]{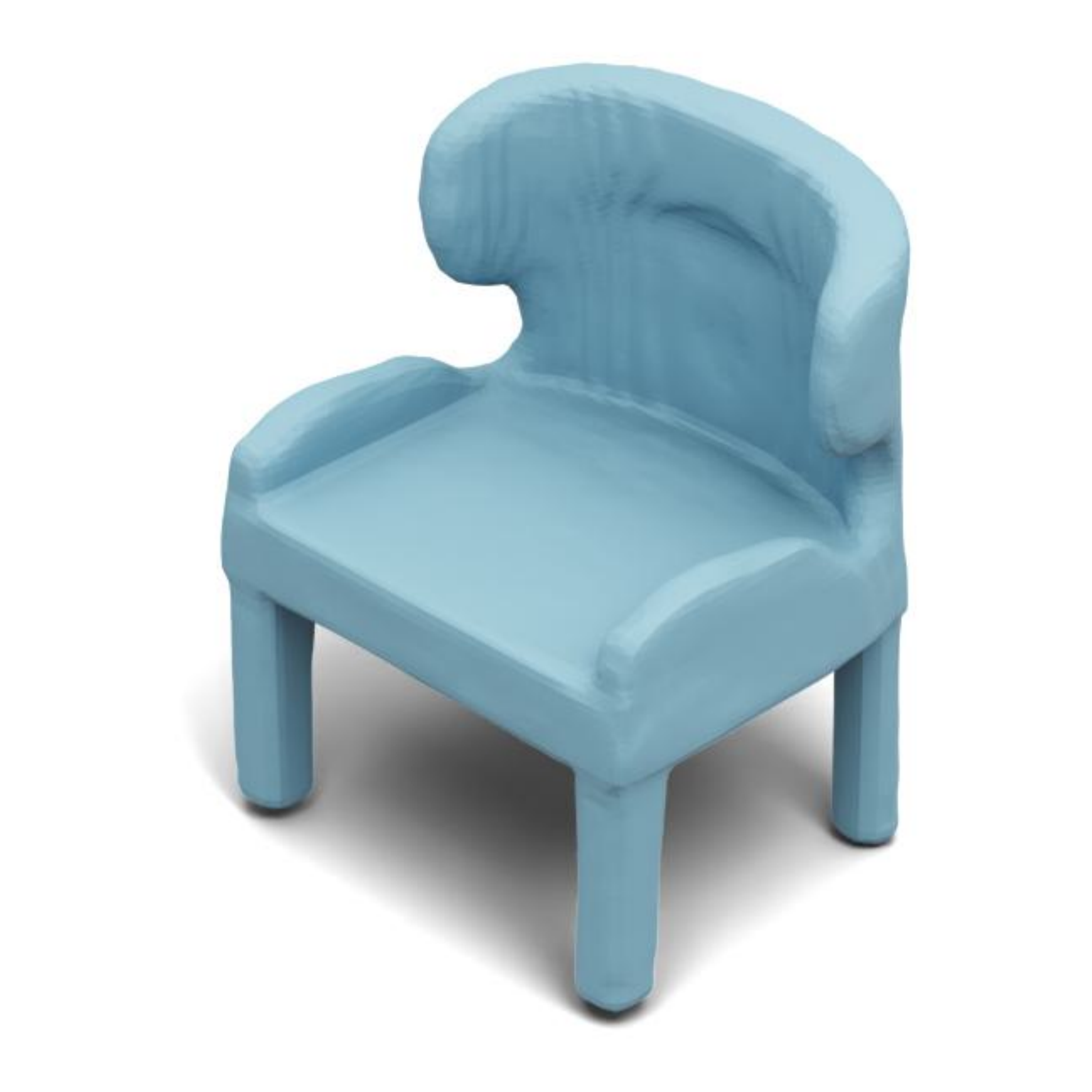}
&\includegraphics[trim = 1 1 1 1, clip, width=0.125\linewidth]{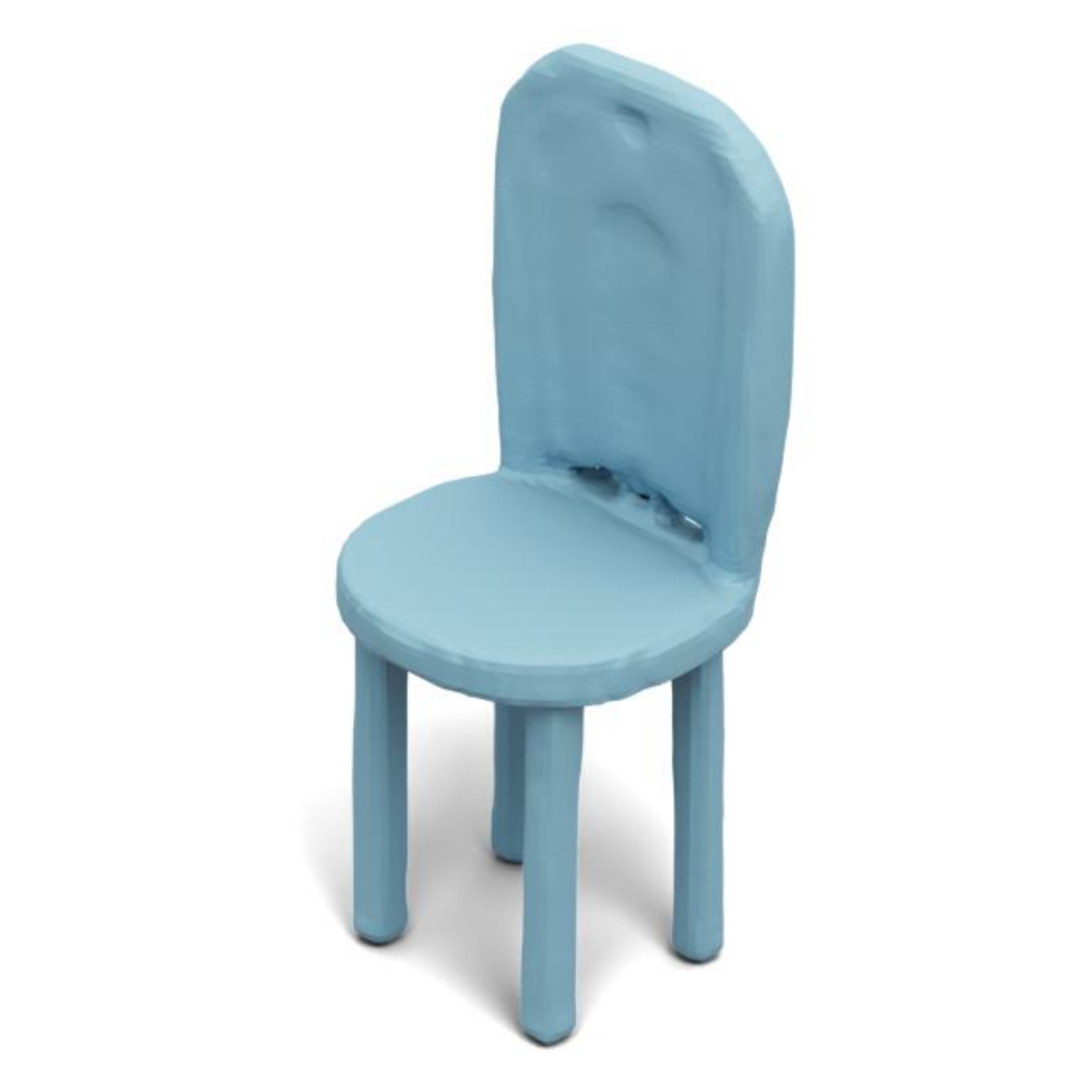}
&\includegraphics[trim = 1 1 1 1, clip, width=0.125\linewidth]{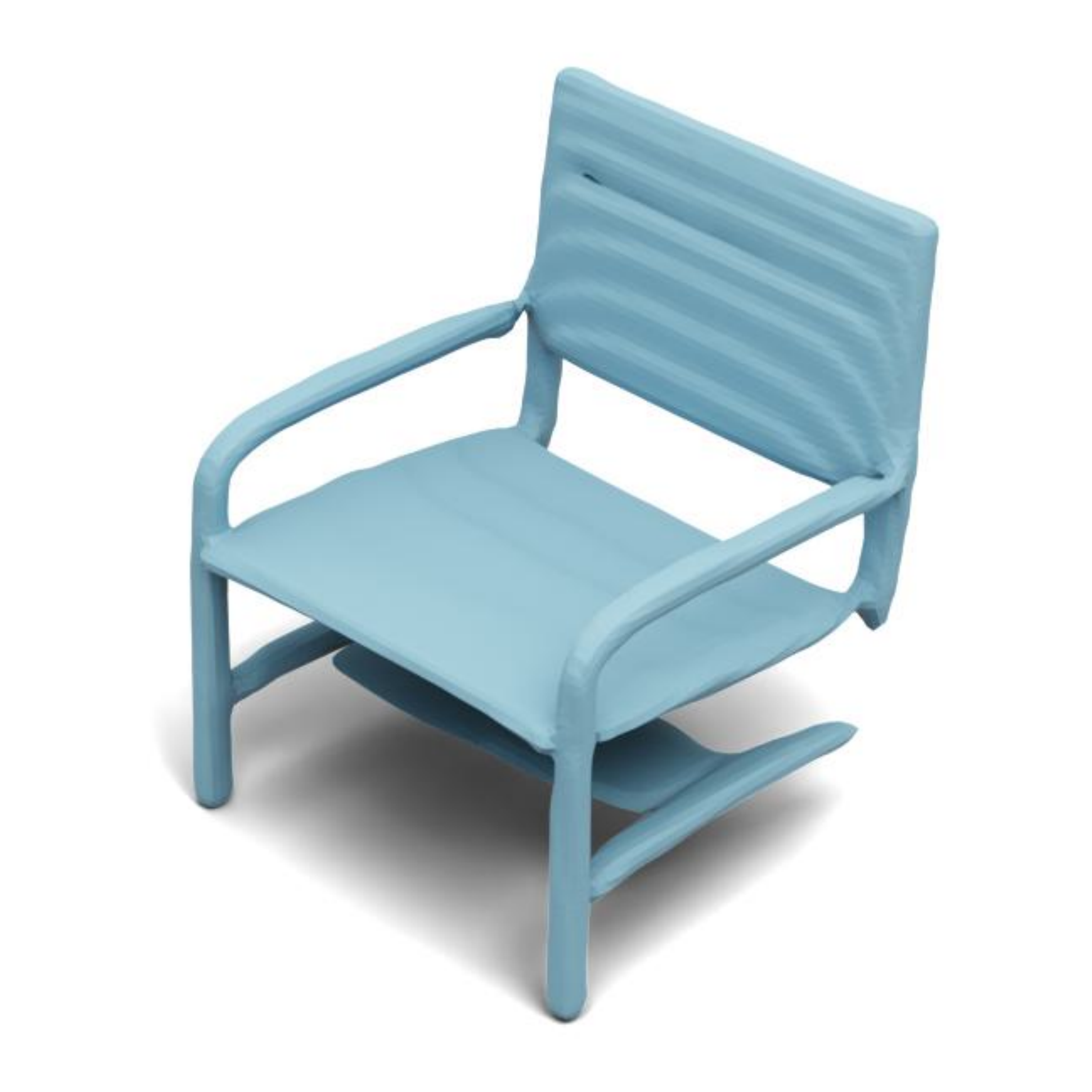}
\\
\includegraphics[width=0.125\linewidth]{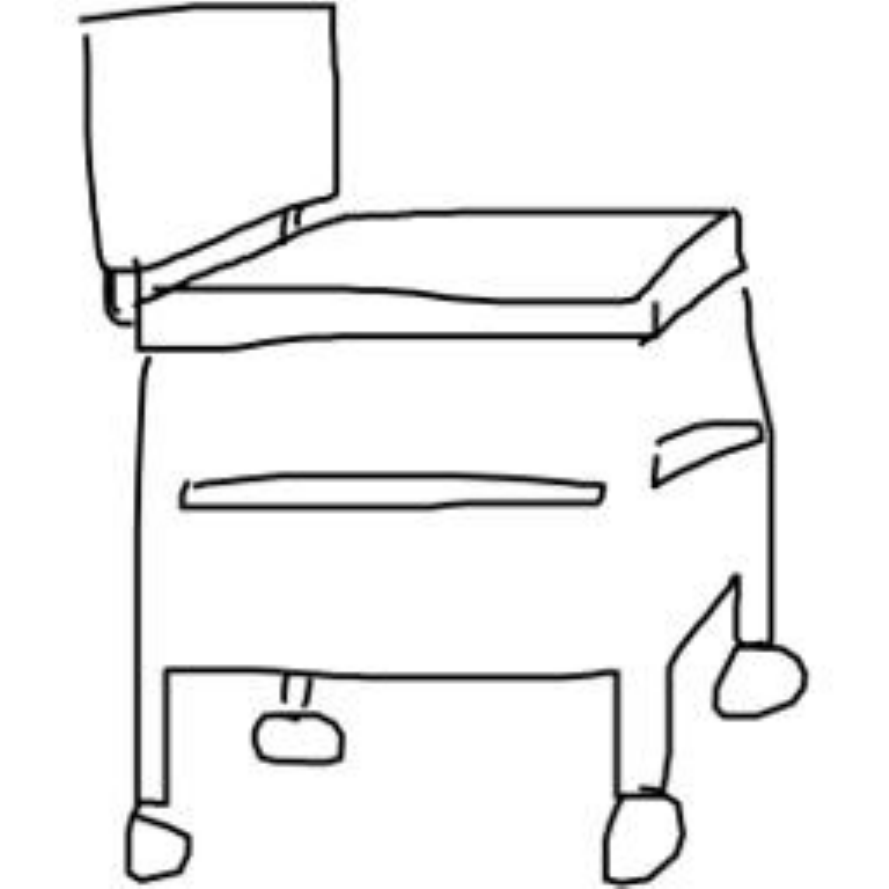}
&\includegraphics[width=0.125\linewidth]{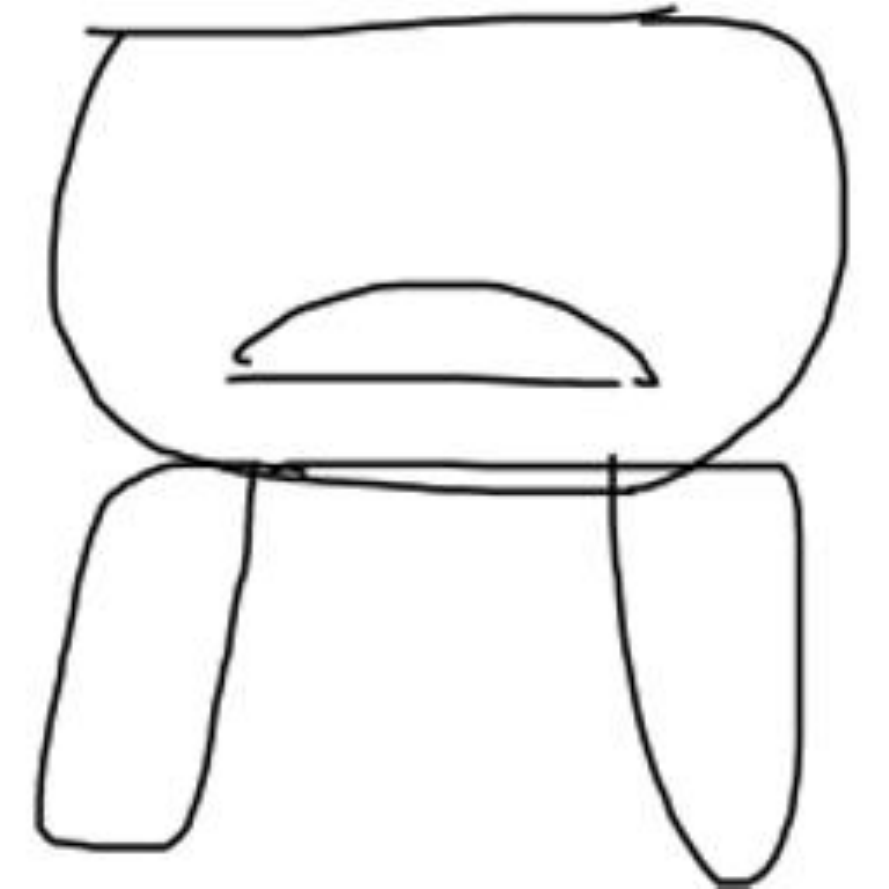}
&\includegraphics[width=0.125\linewidth]{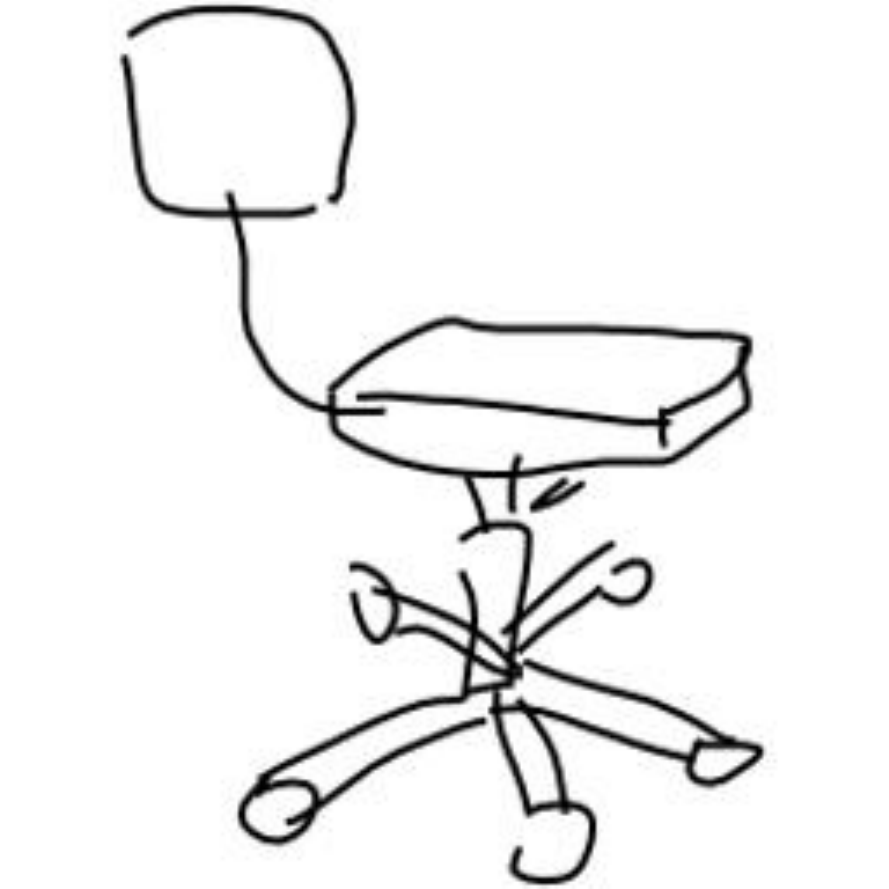}
&\includegraphics[width=0.125\linewidth]{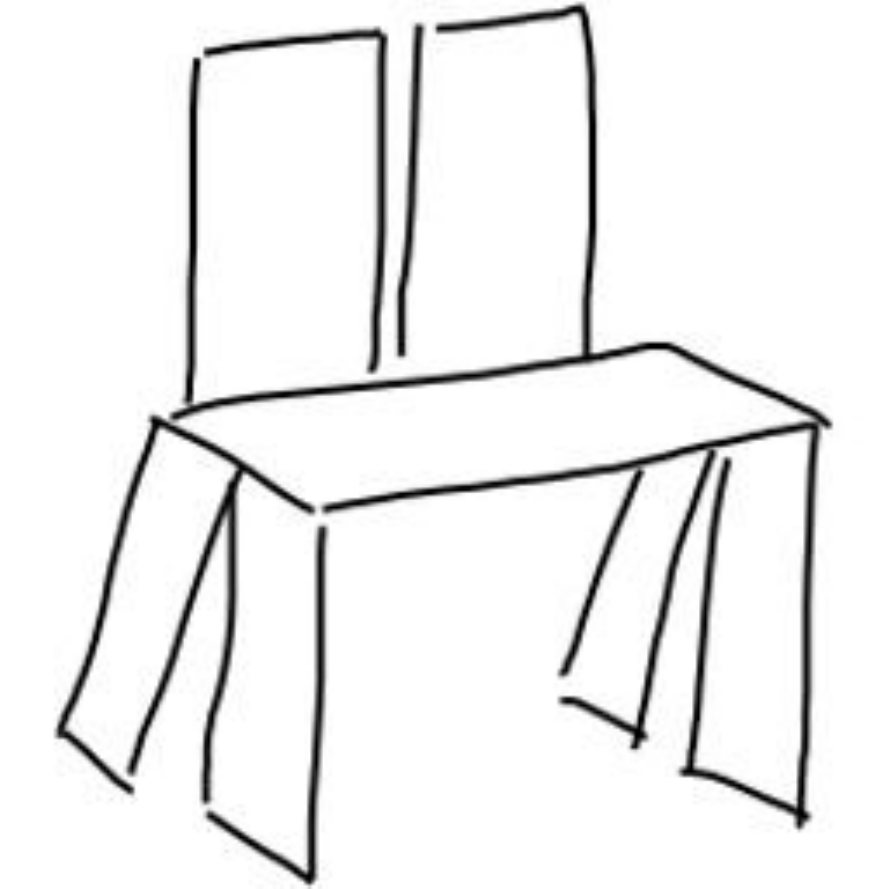}
&\includegraphics[width=0.125\linewidth]{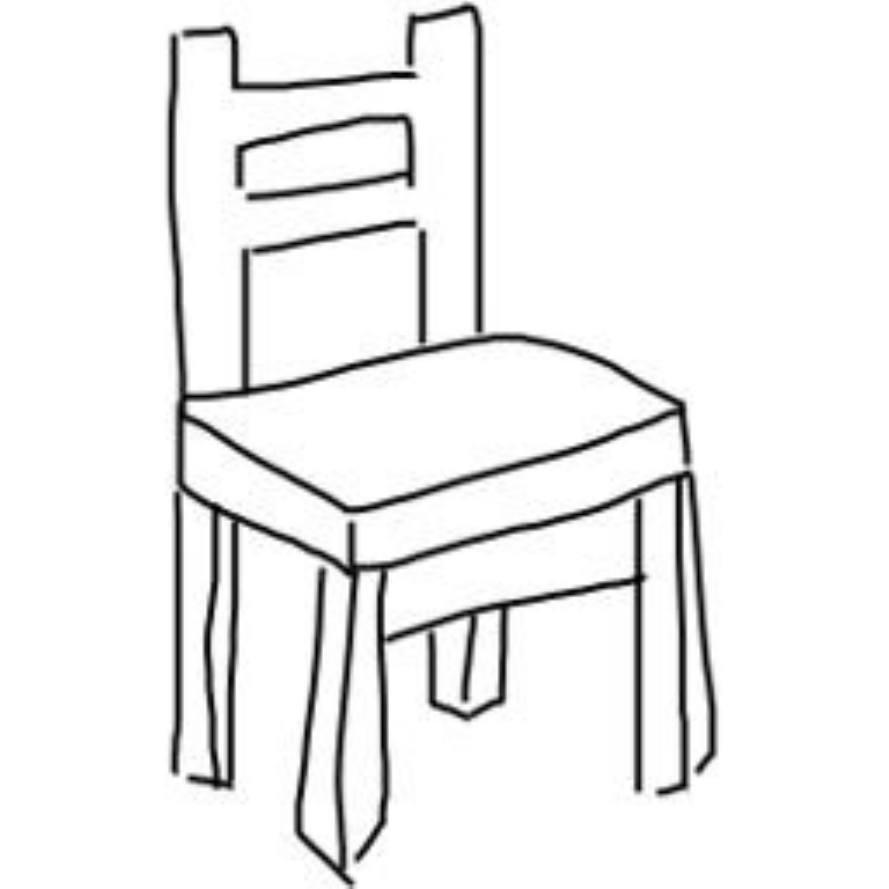}
&\includegraphics[width=0.125\linewidth]{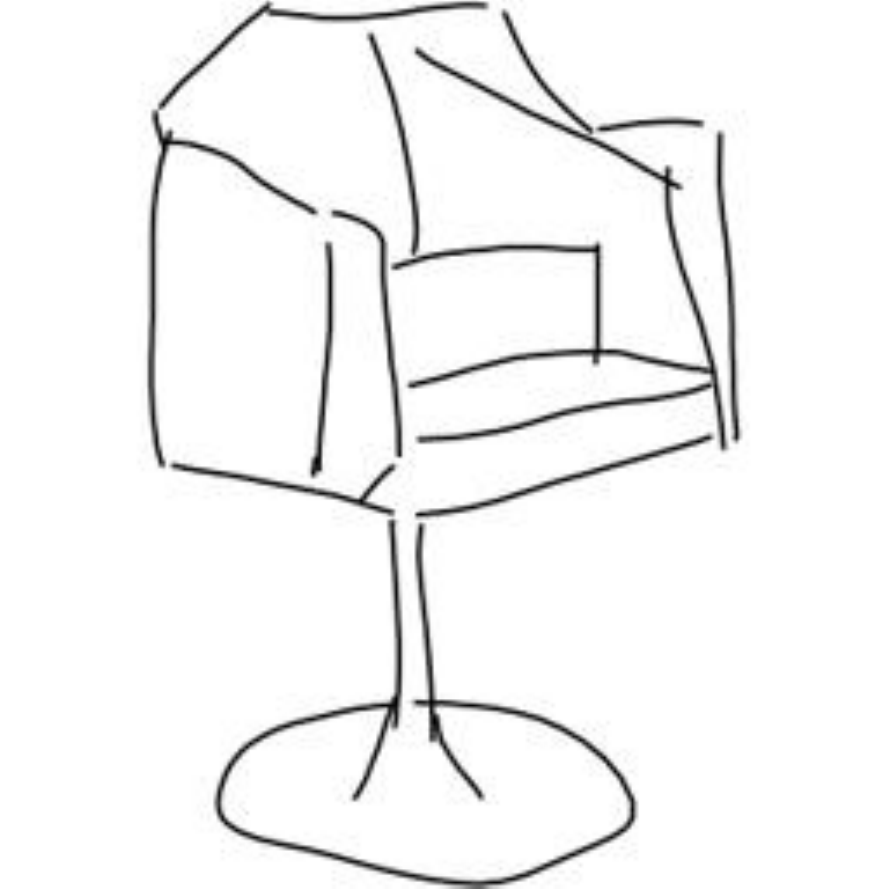}
&\includegraphics[width=0.125\linewidth]{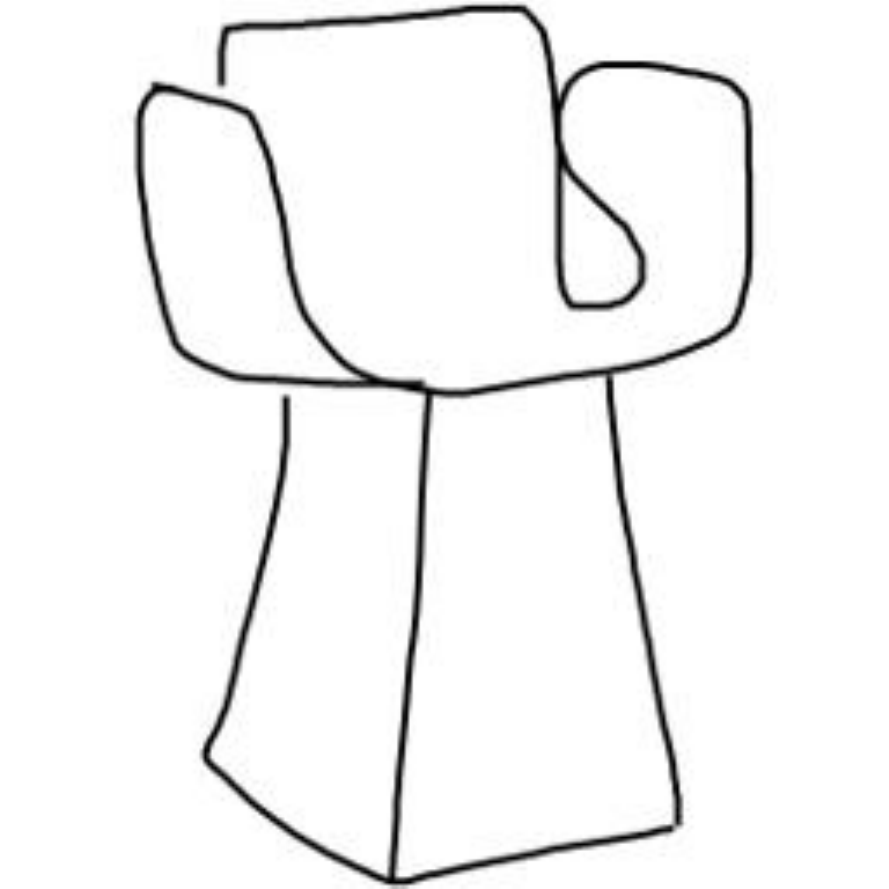}
&\includegraphics[width=0.125\linewidth]{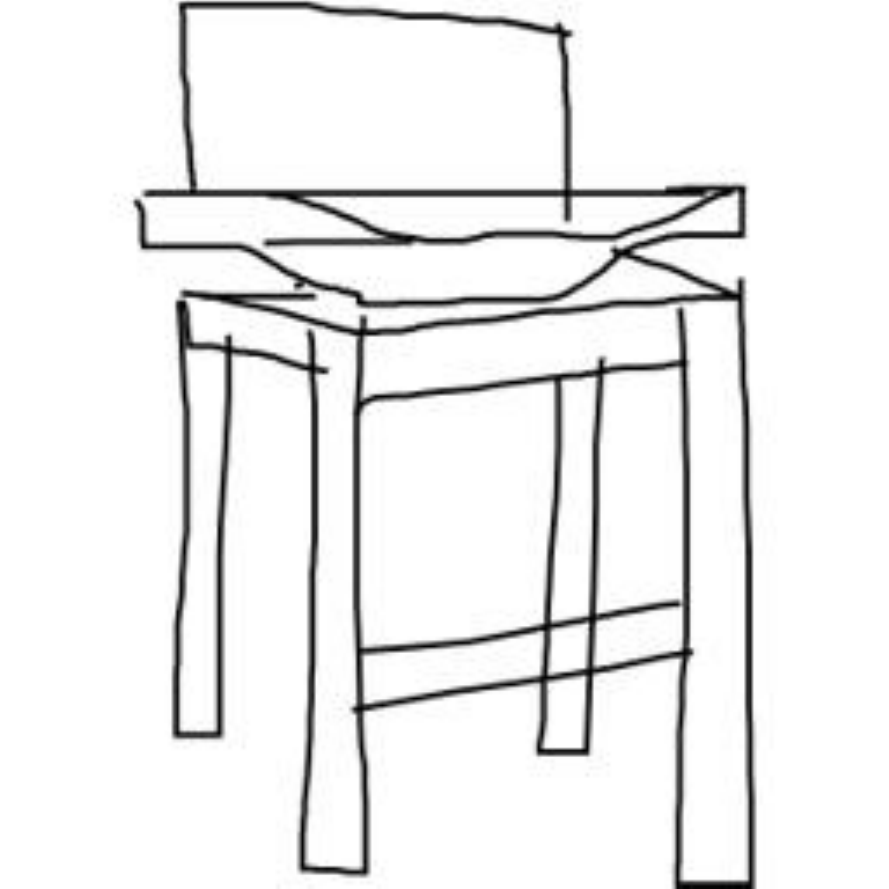}
\\
\includegraphics[trim = 1 1 1 1, clip, width=0.125\linewidth]{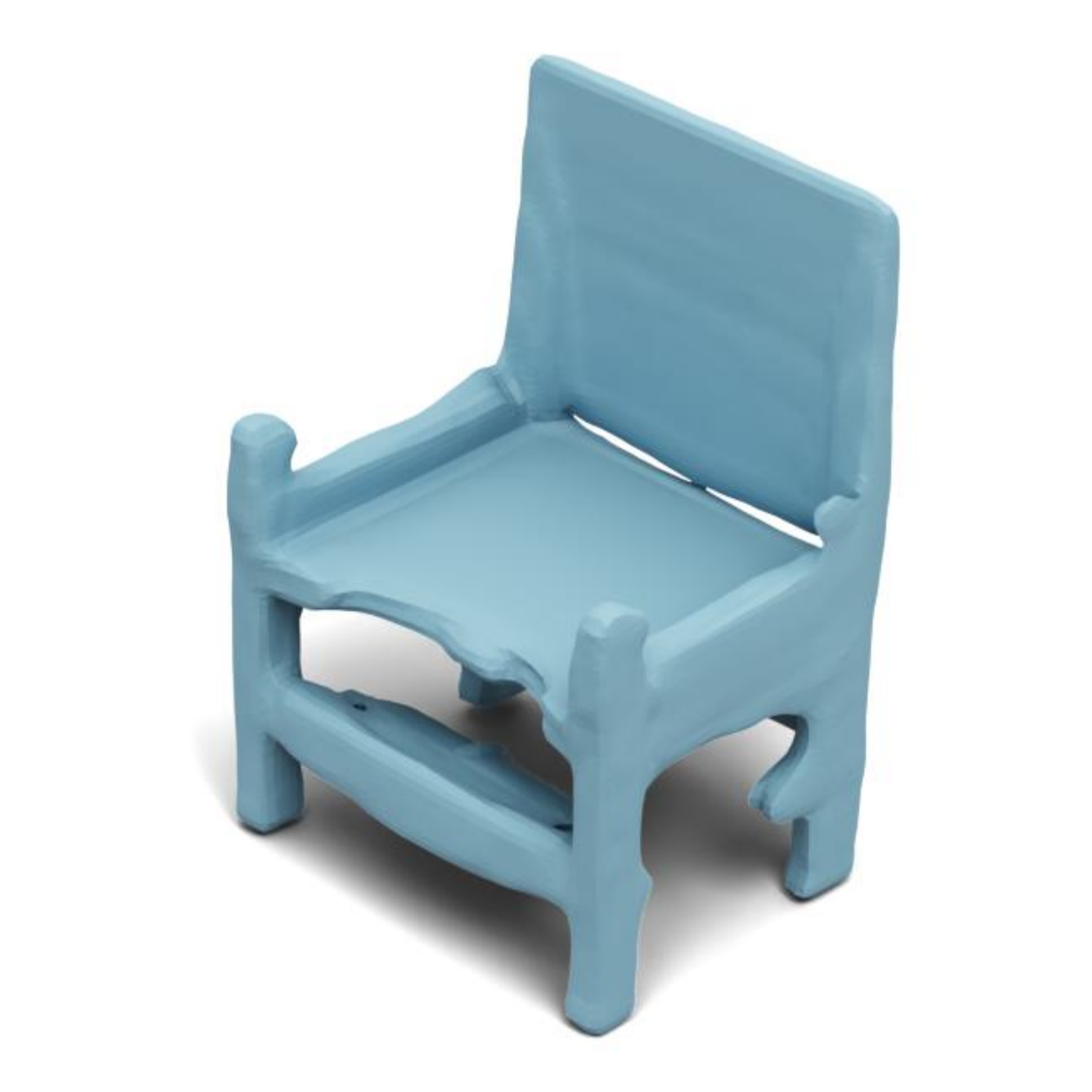}
&\includegraphics[trim = 1 1 1 1, clip, width=0.125\linewidth]{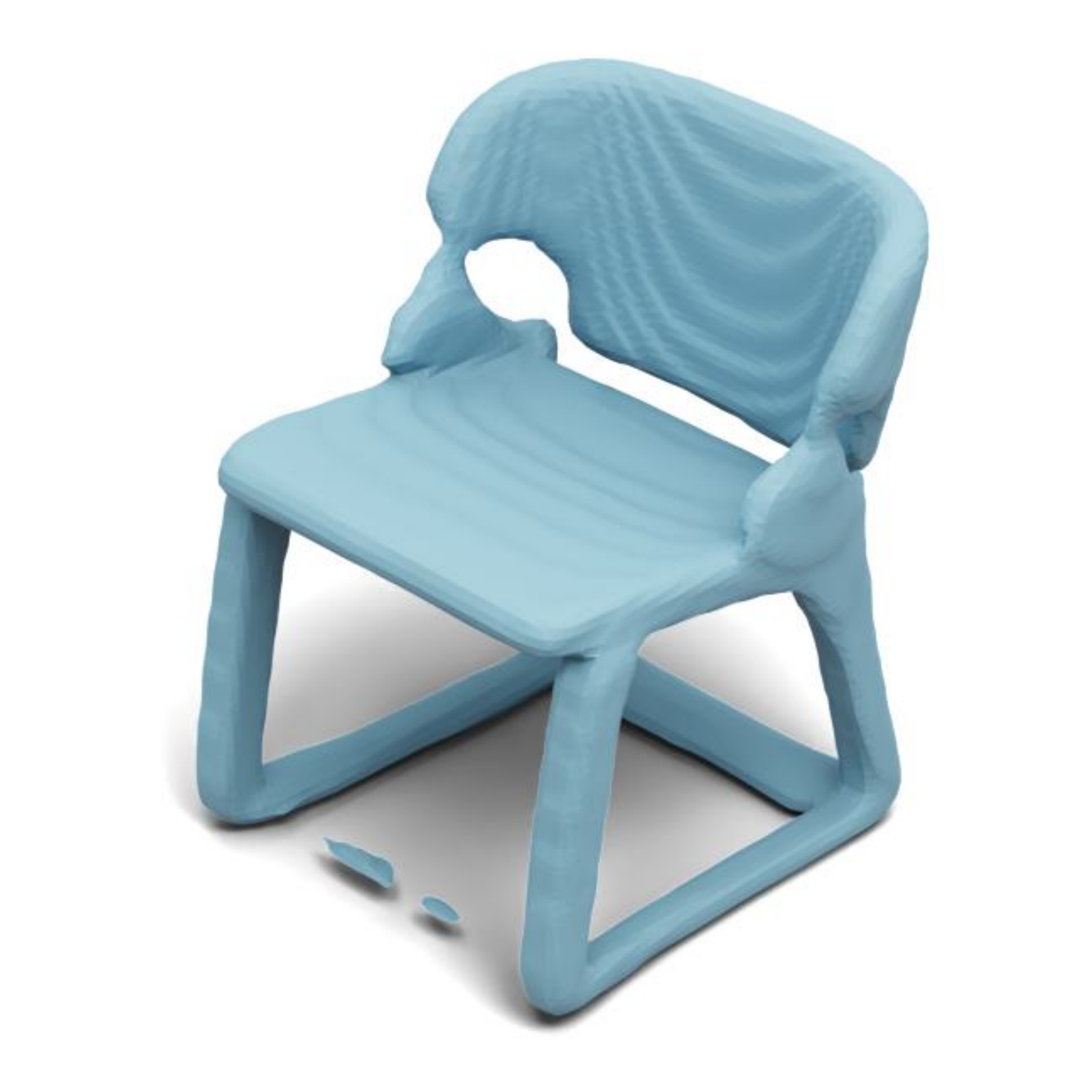}
&\includegraphics[trim = 1 1 1 1, clip, width=0.125\linewidth]{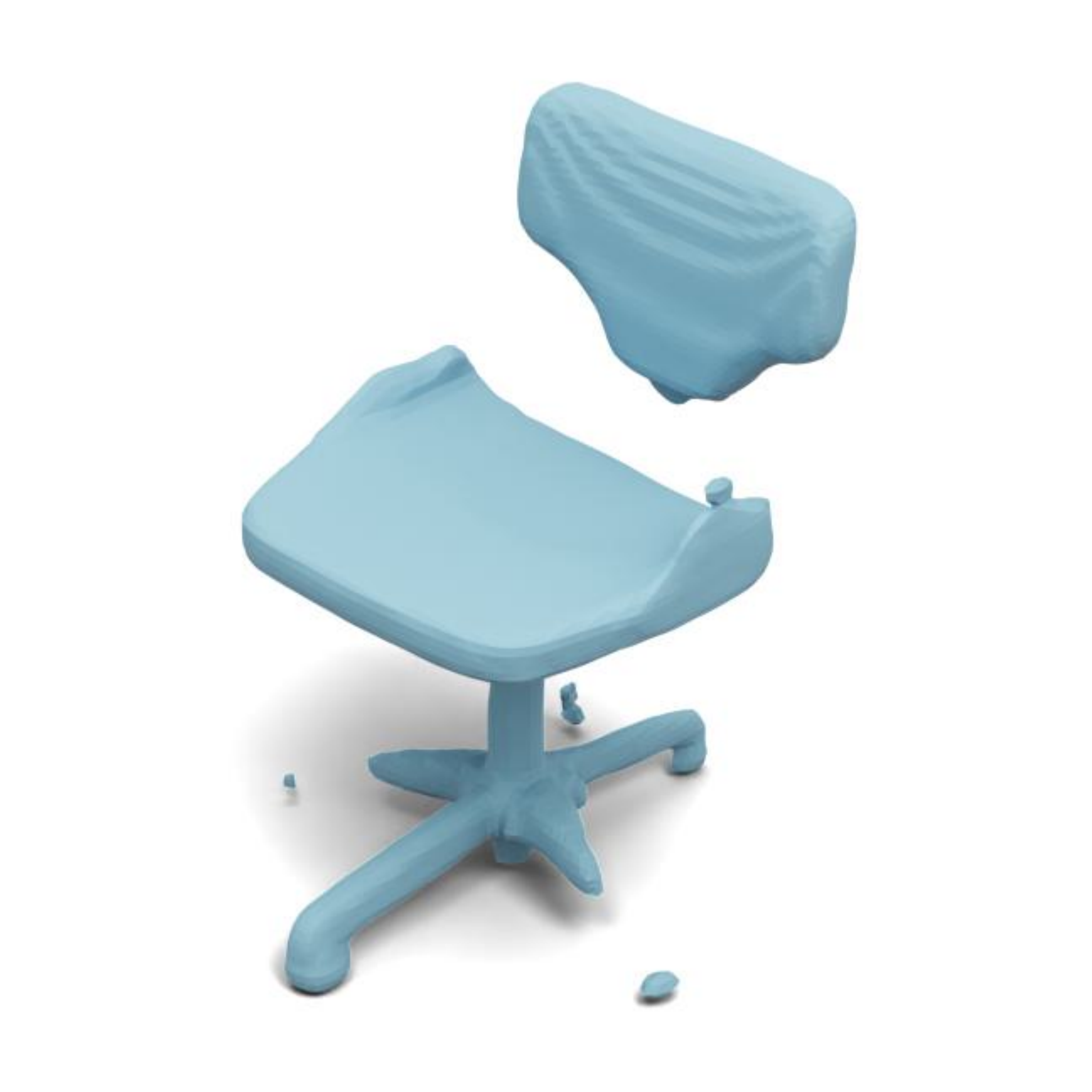}
&\includegraphics[trim = 1 1 1 1, clip, width=0.125\linewidth]{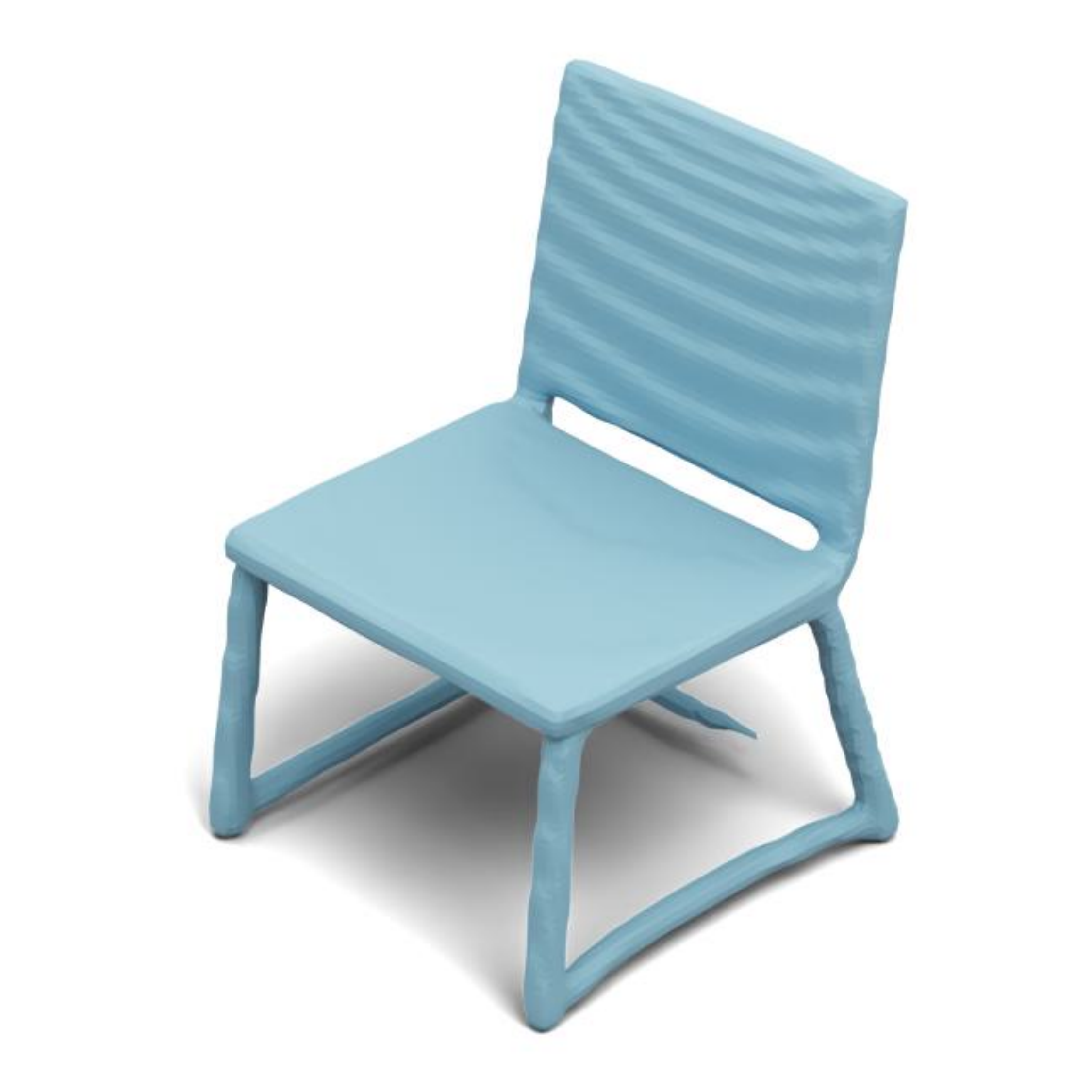}
&\includegraphics[trim = 1 1 1 1, clip, width=0.125\linewidth]{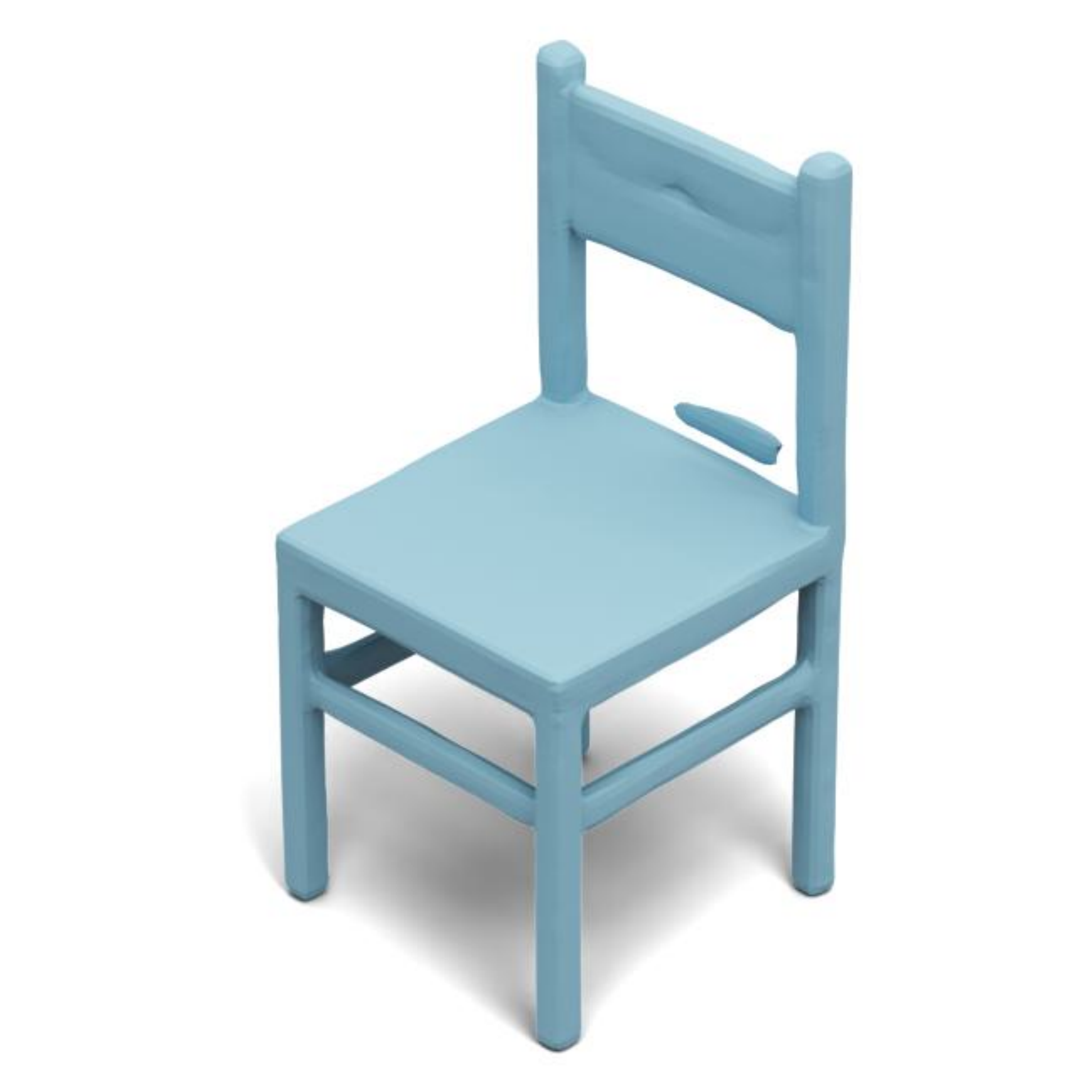}
&\includegraphics[trim = 1 1 1 1, clip, width=0.125\linewidth]{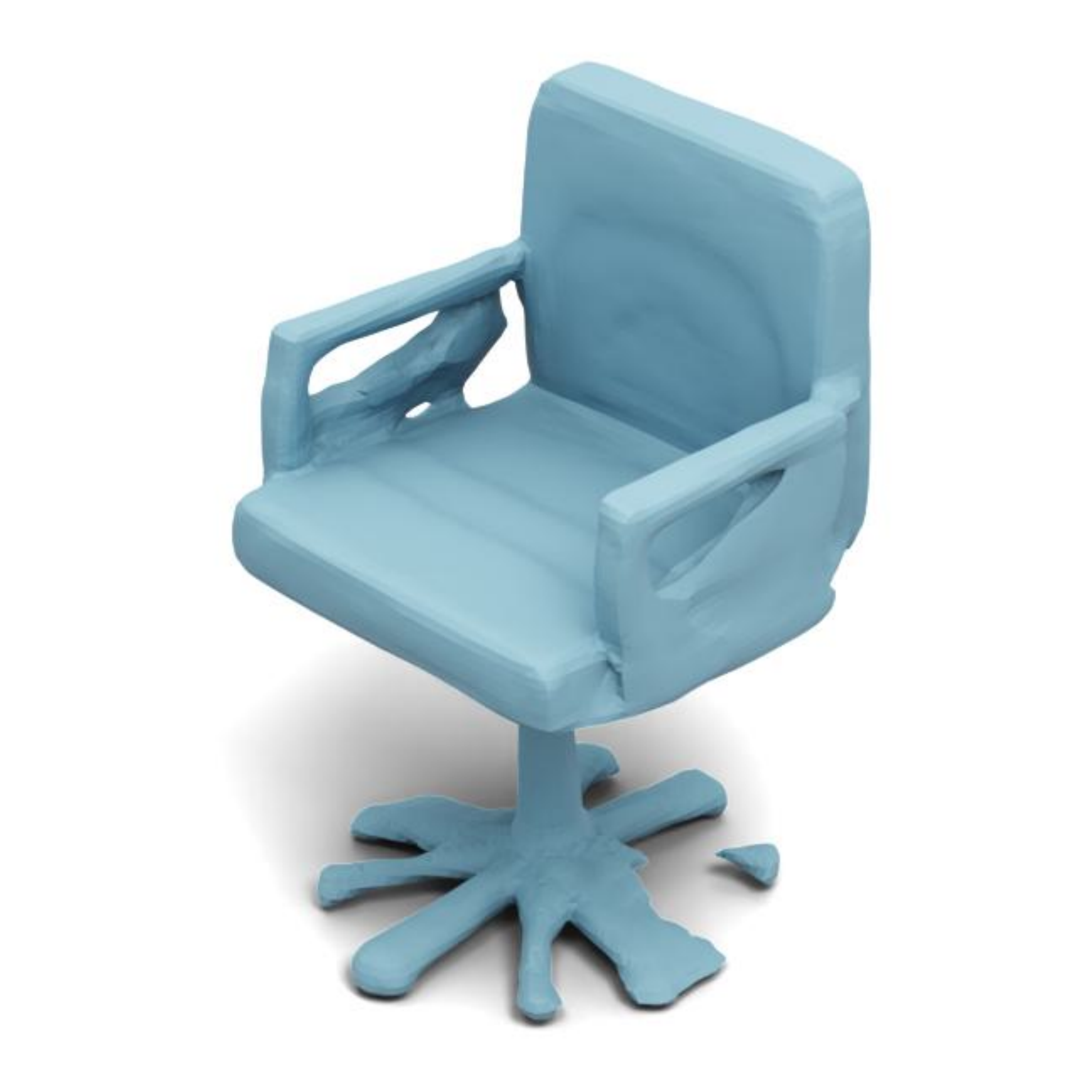}
&\includegraphics[trim = 1 1 1 1, clip, width=0.125\linewidth]{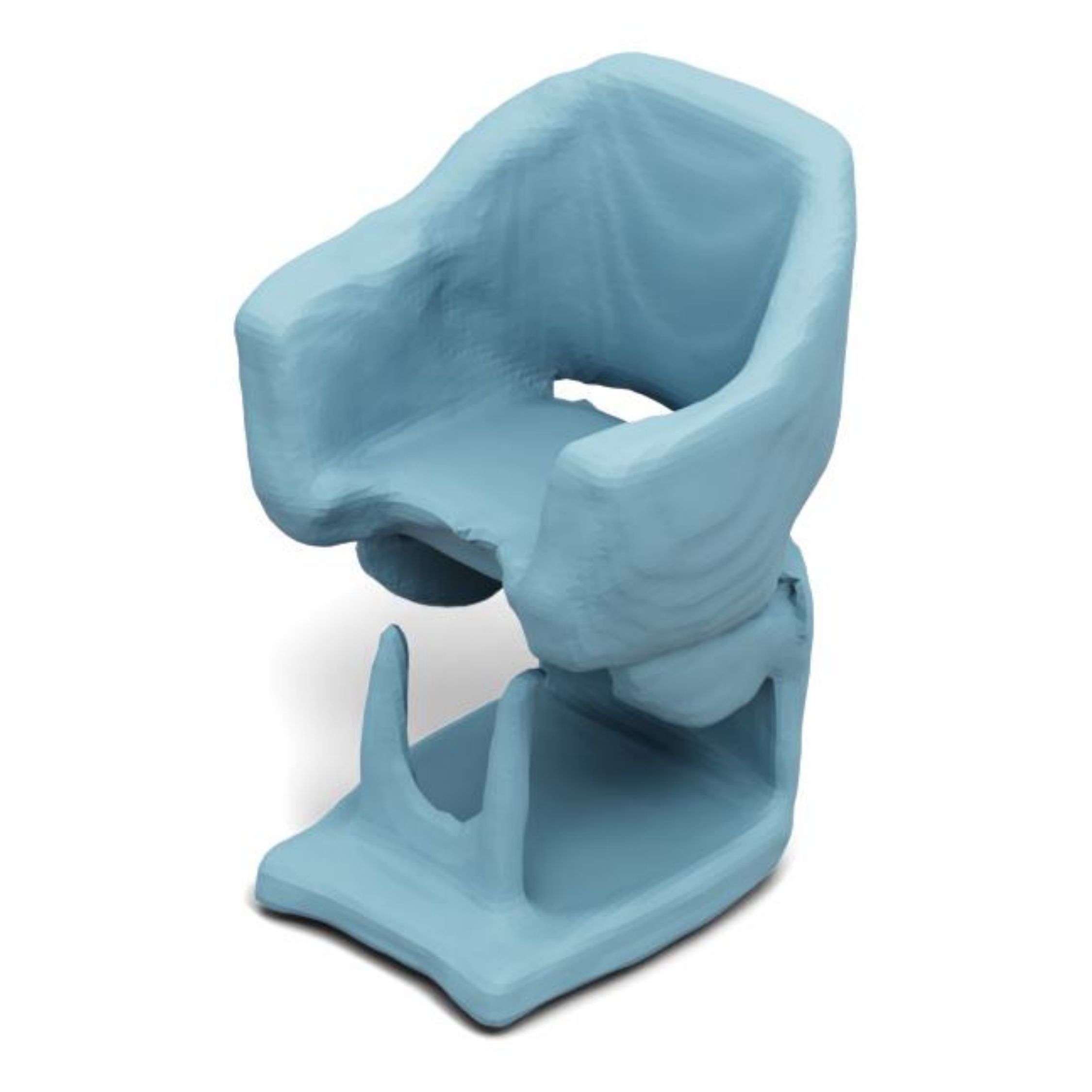}
&\includegraphics[trim = 1 1 1 1, clip, width=0.125\linewidth]{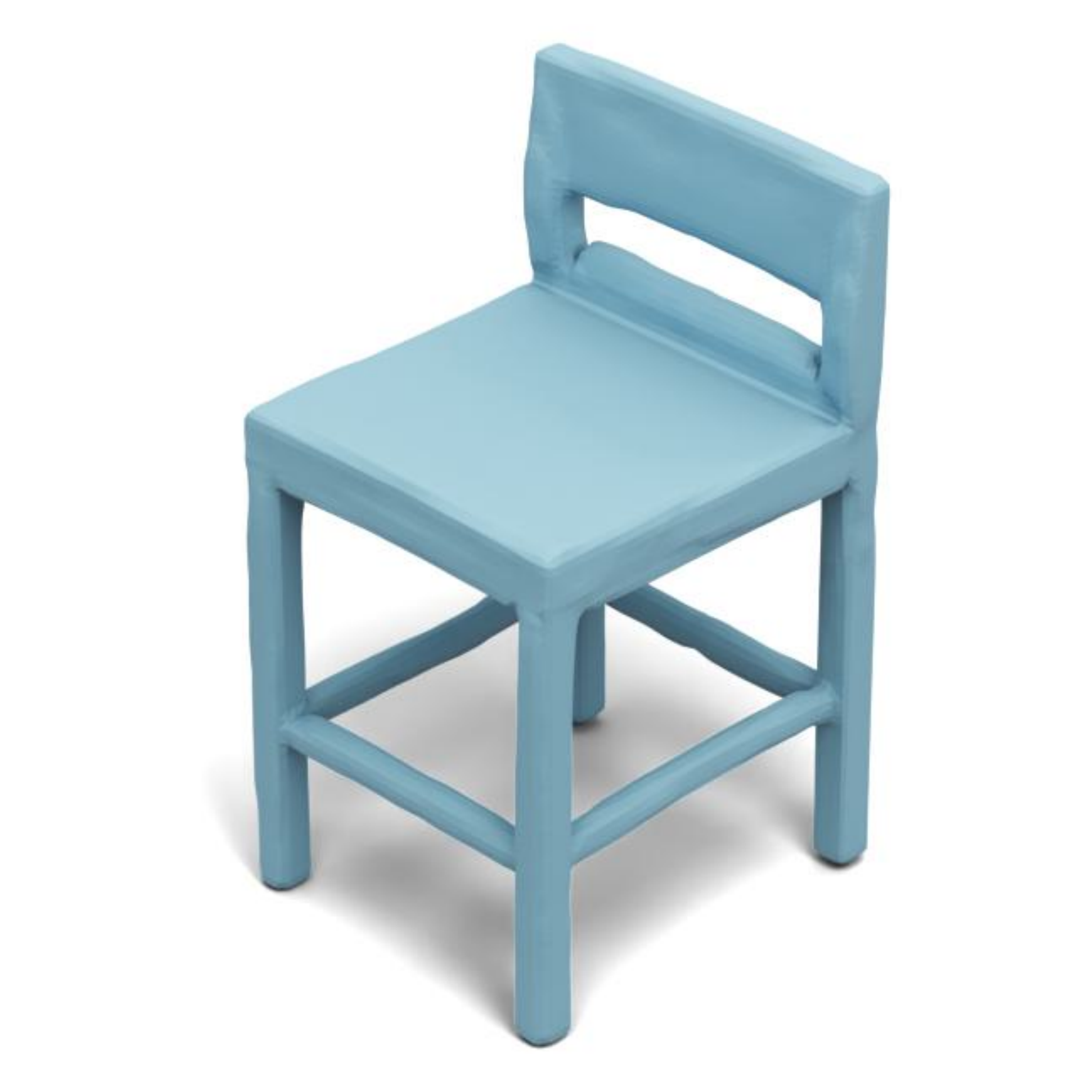}
\\
\includegraphics[width=0.125\linewidth]{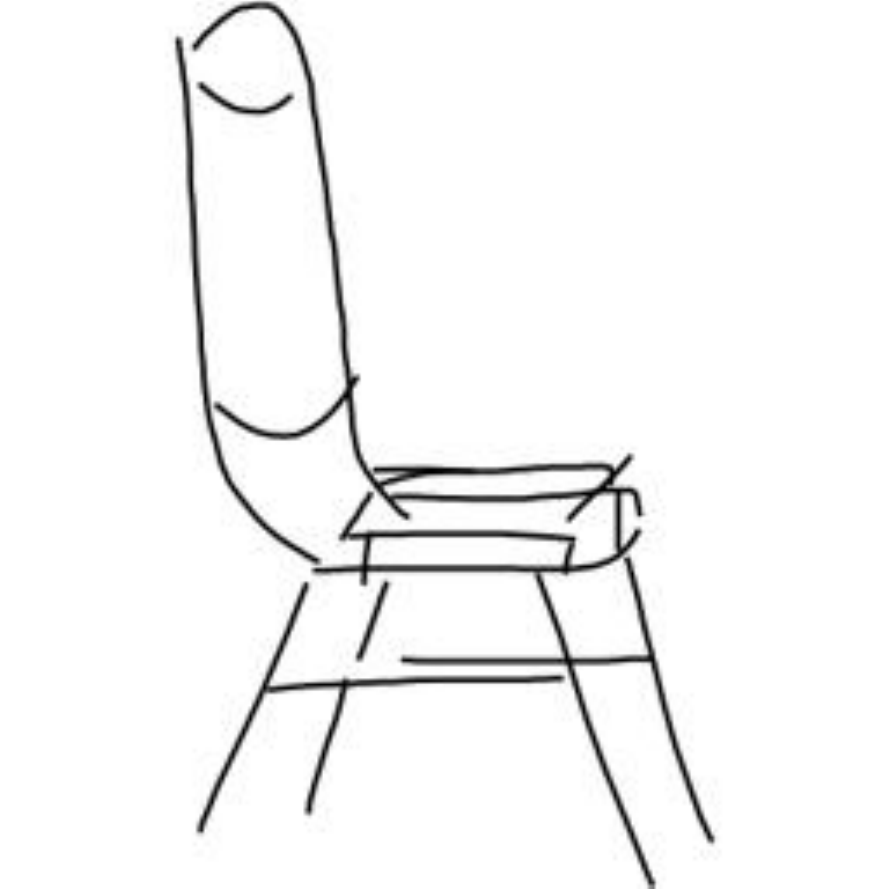}
&\includegraphics[width=0.125\linewidth]{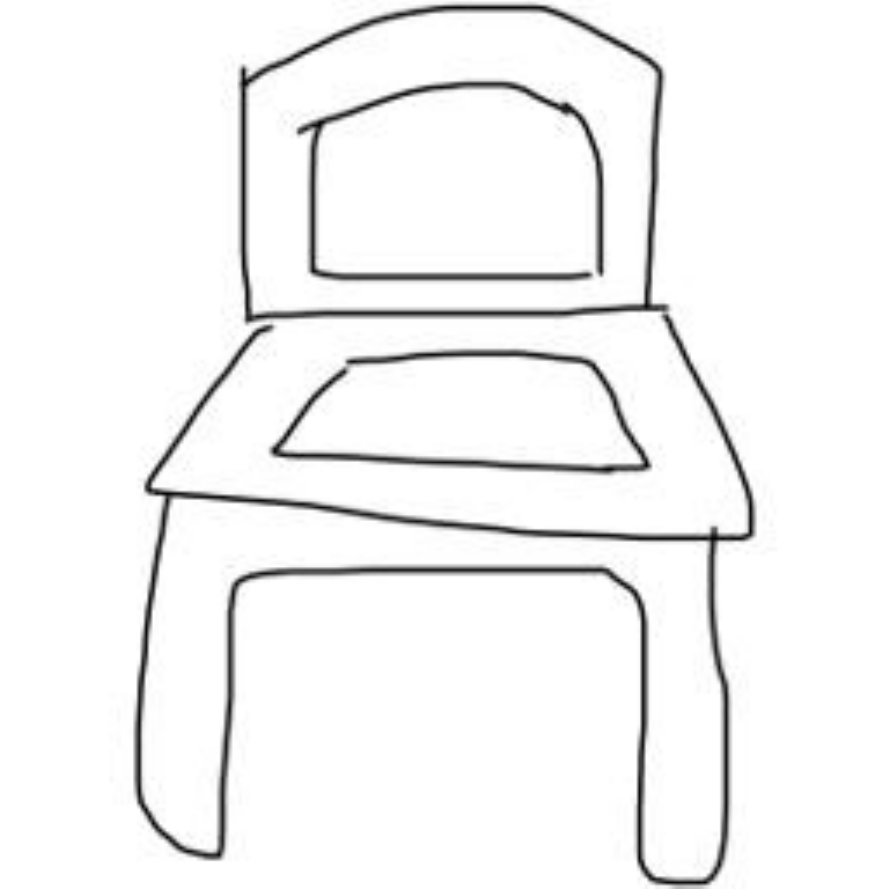}
&\includegraphics[width=0.125\linewidth]{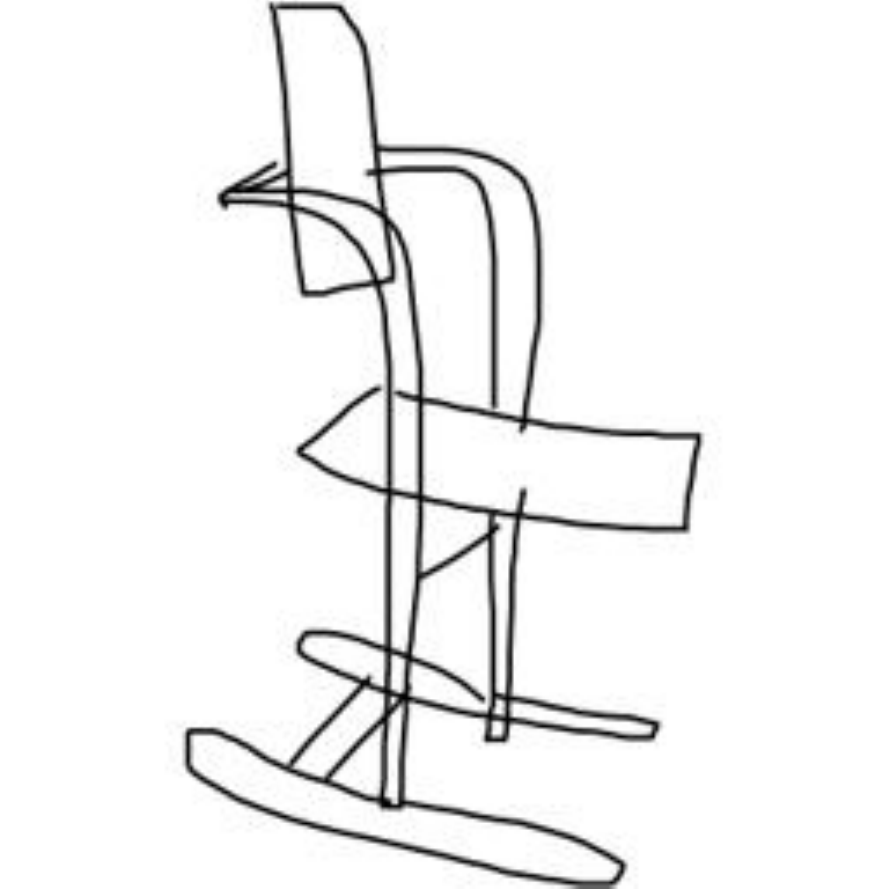}
&\includegraphics[width=0.125\linewidth]{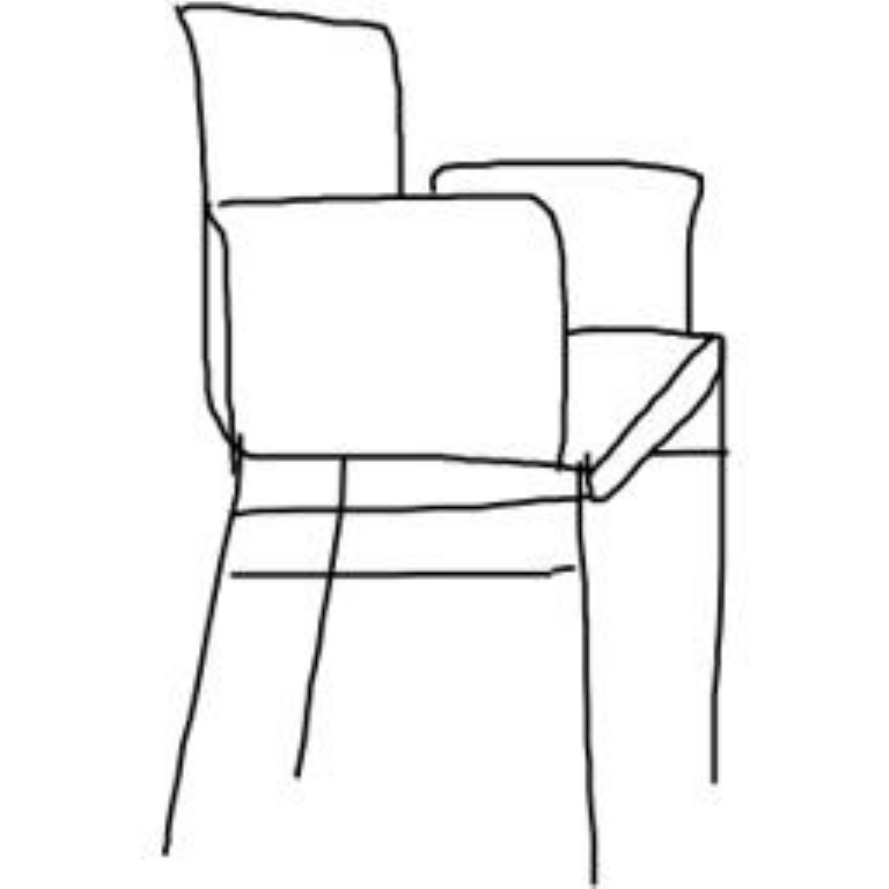}
&\includegraphics[width=0.125\linewidth]{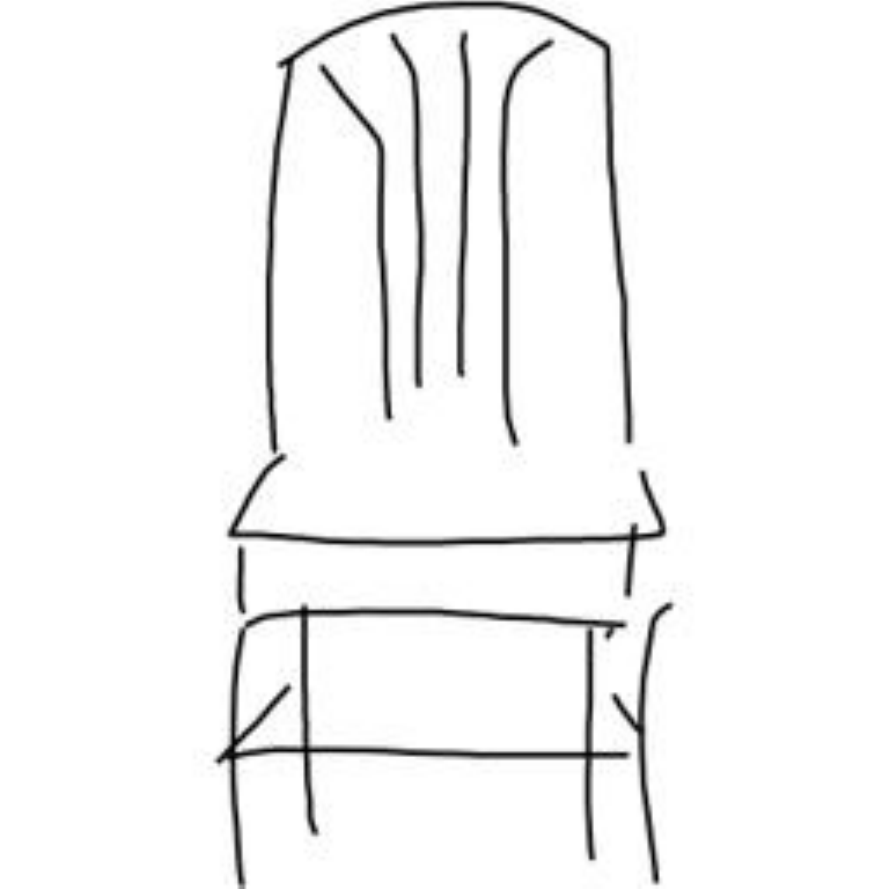}
&\includegraphics[width=0.125\linewidth]{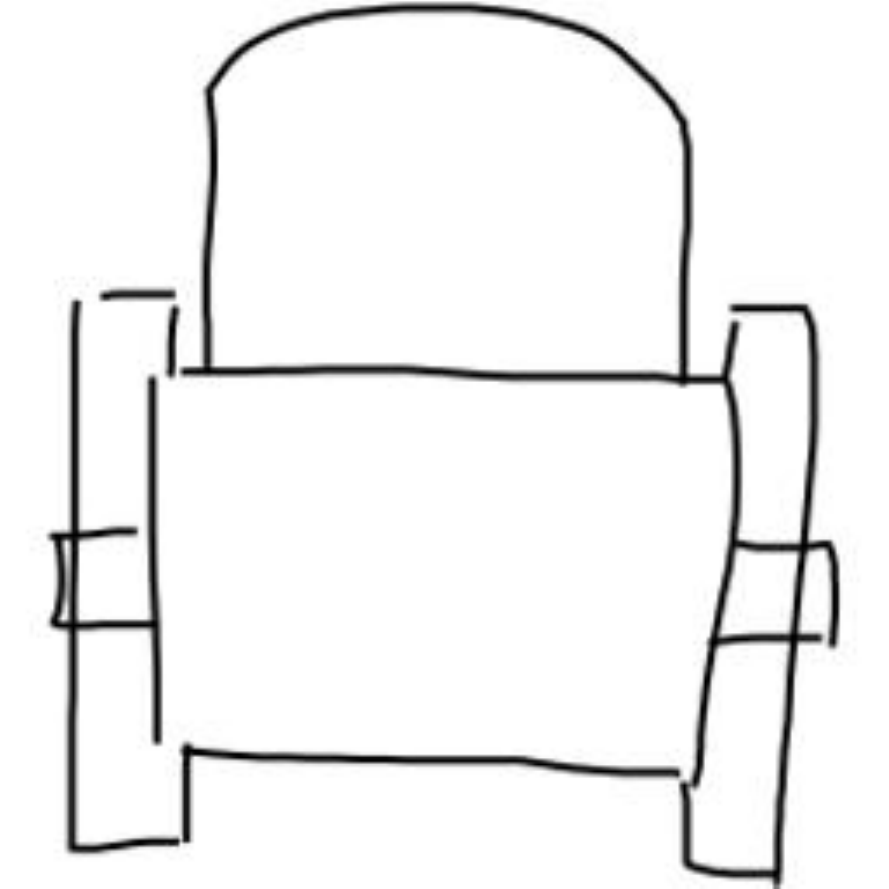}
&\includegraphics[width=0.125\linewidth]{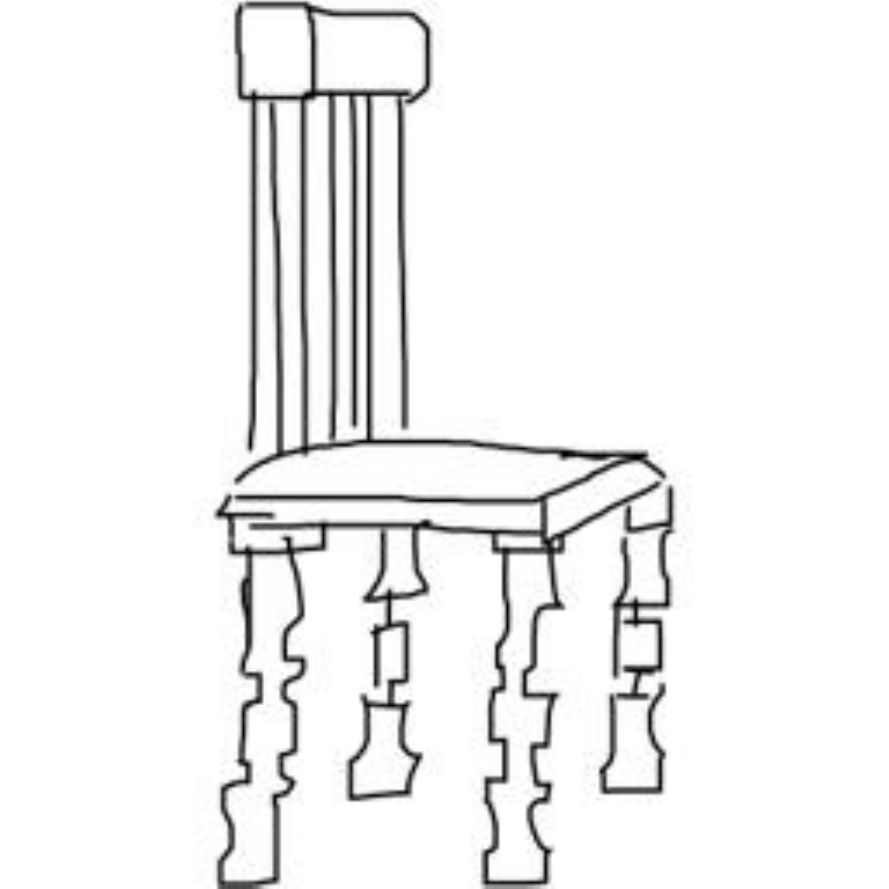}
&\includegraphics[width=0.125\linewidth]{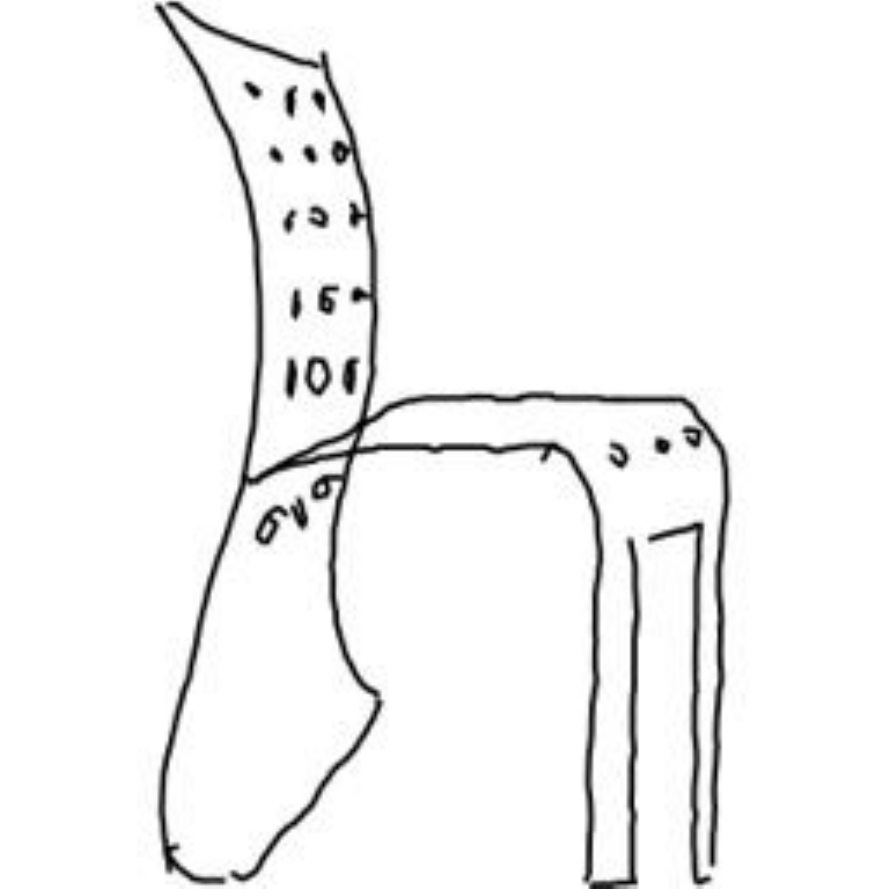}
\\
\includegraphics[trim = 1 1 1 1, clip, width=0.125\linewidth]{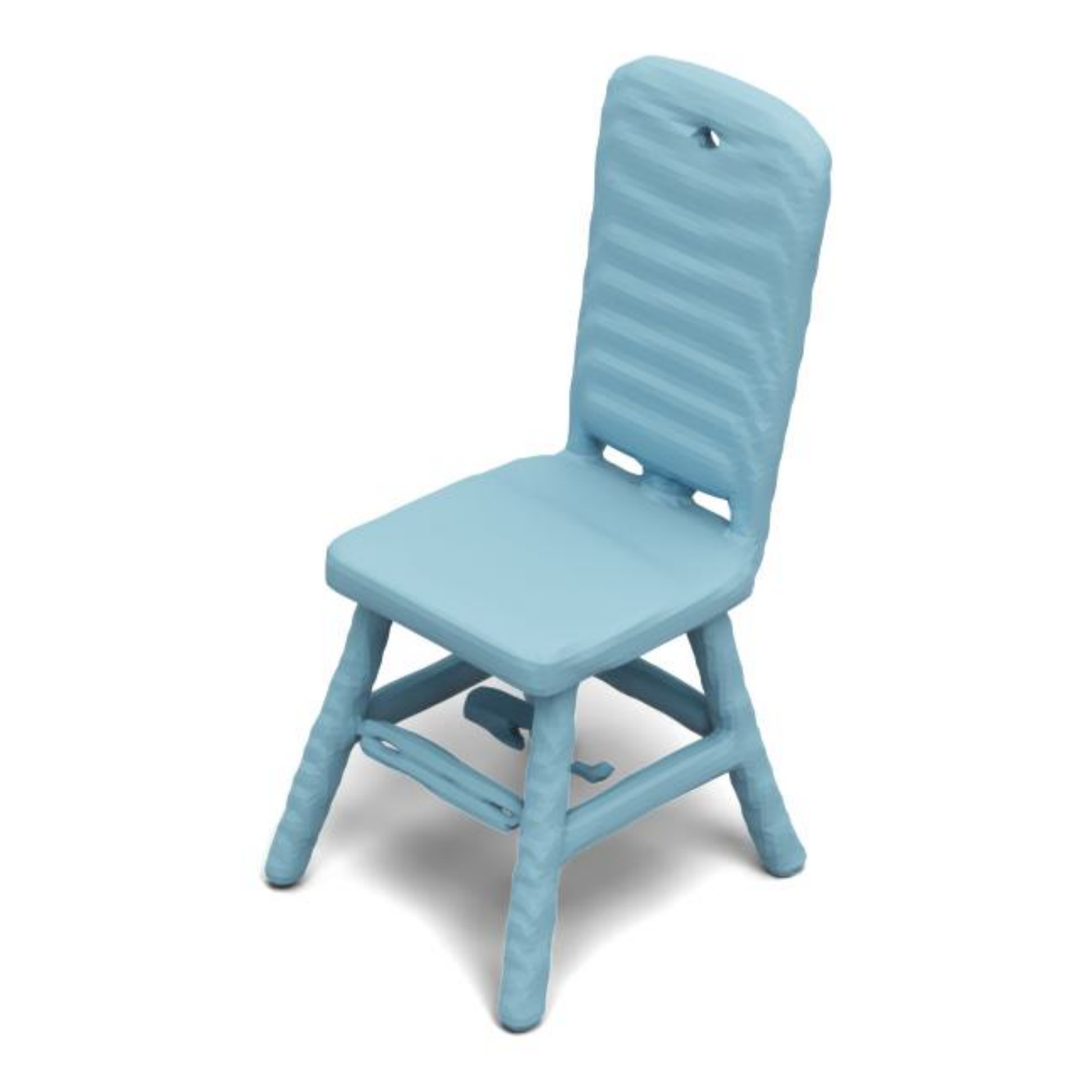}
&\includegraphics[trim = 1 1 1 1, clip, width=0.125\linewidth]{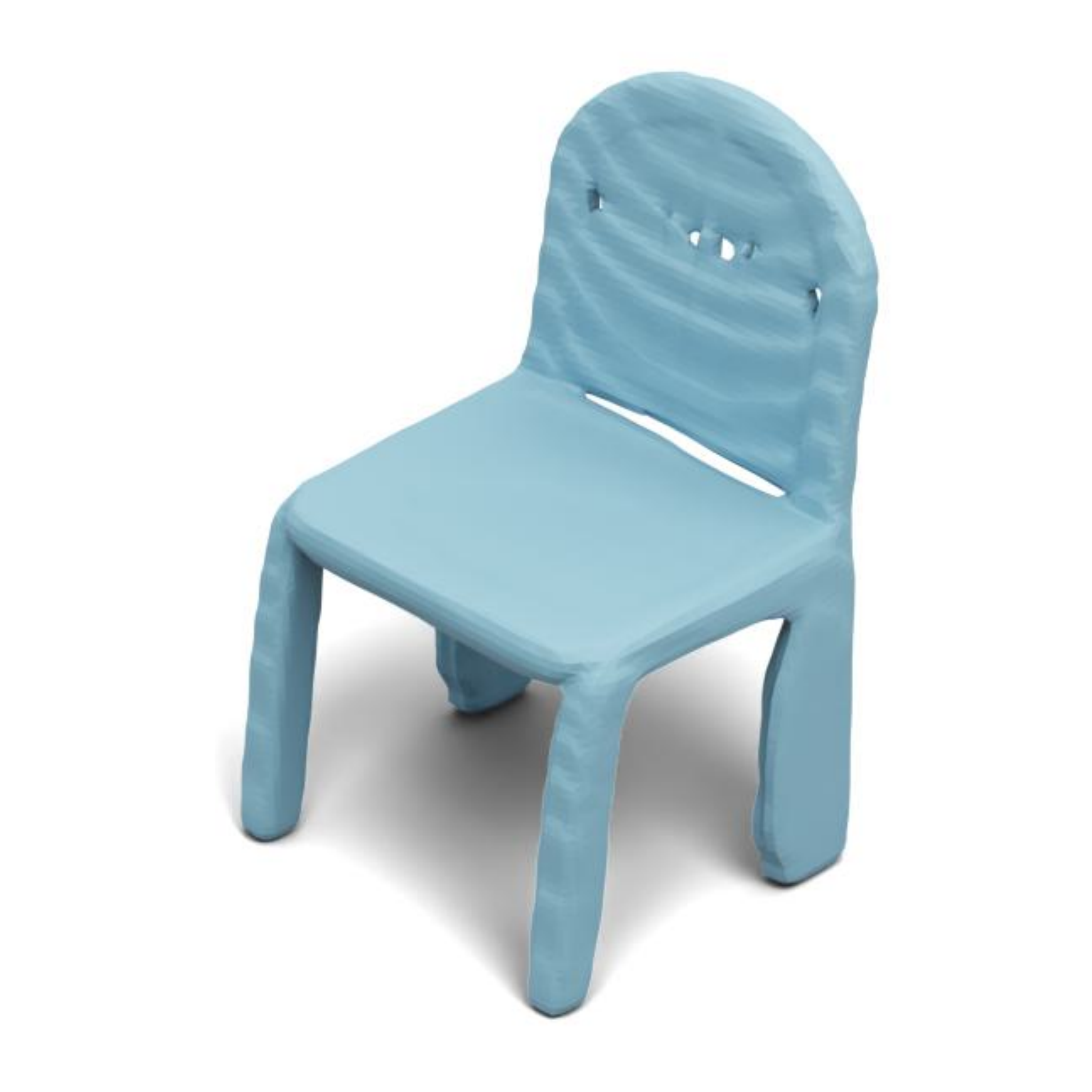}
&\includegraphics[trim = 1 1 1 1, clip, width=0.125\linewidth]{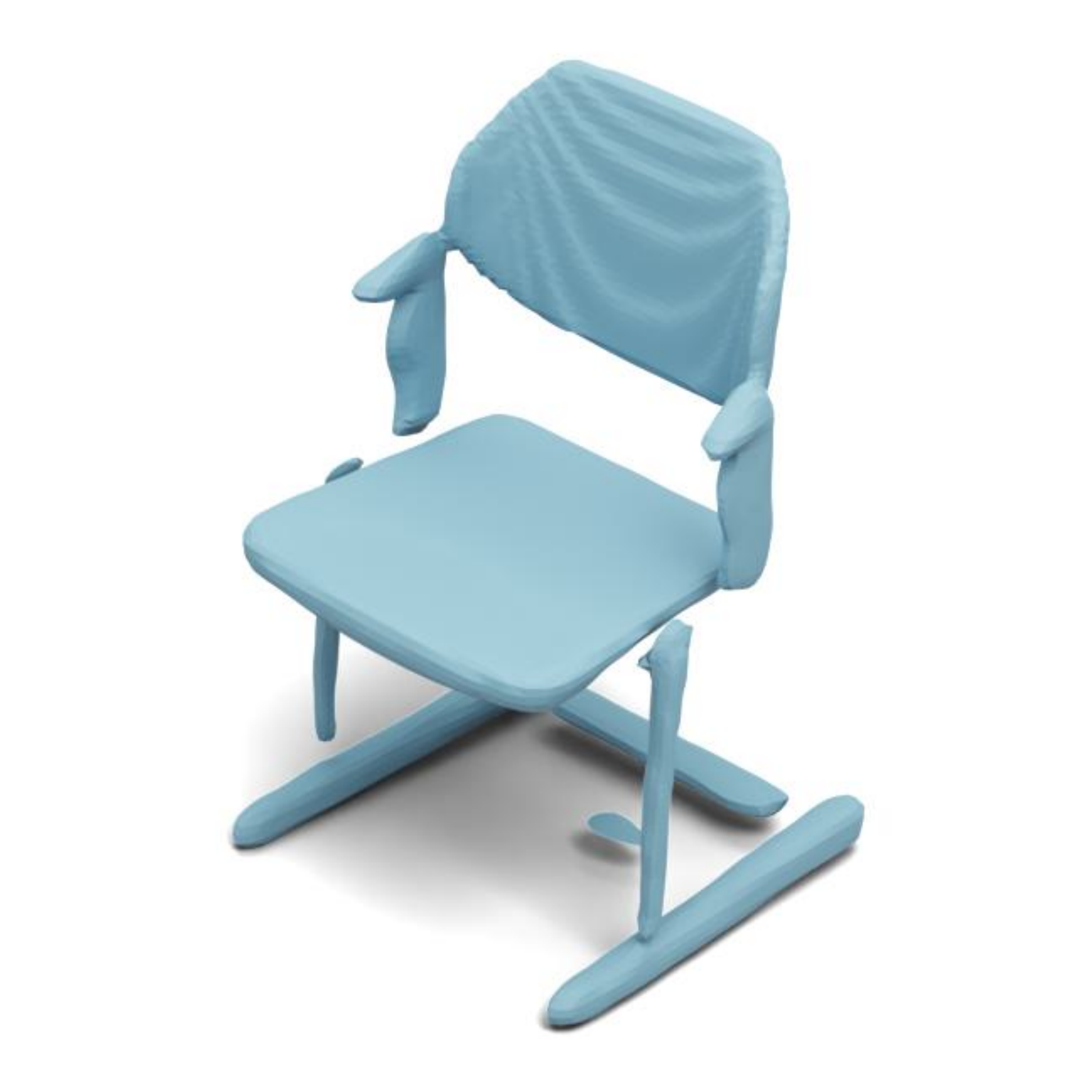}
&\includegraphics[trim = 1 1 1 1, clip, width=0.125\linewidth]{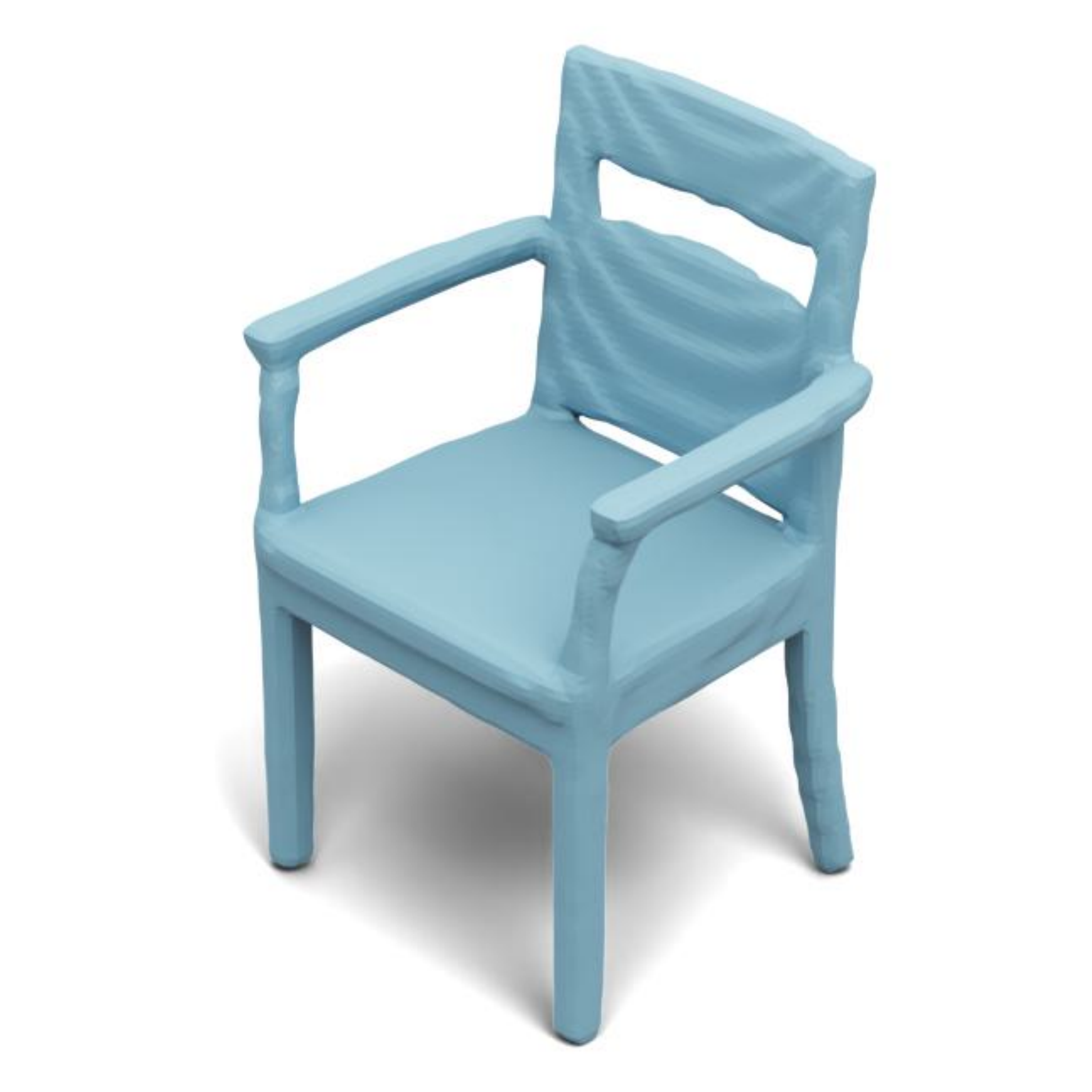}
&\includegraphics[trim = 1 1 1 1, clip, width=0.125\linewidth]{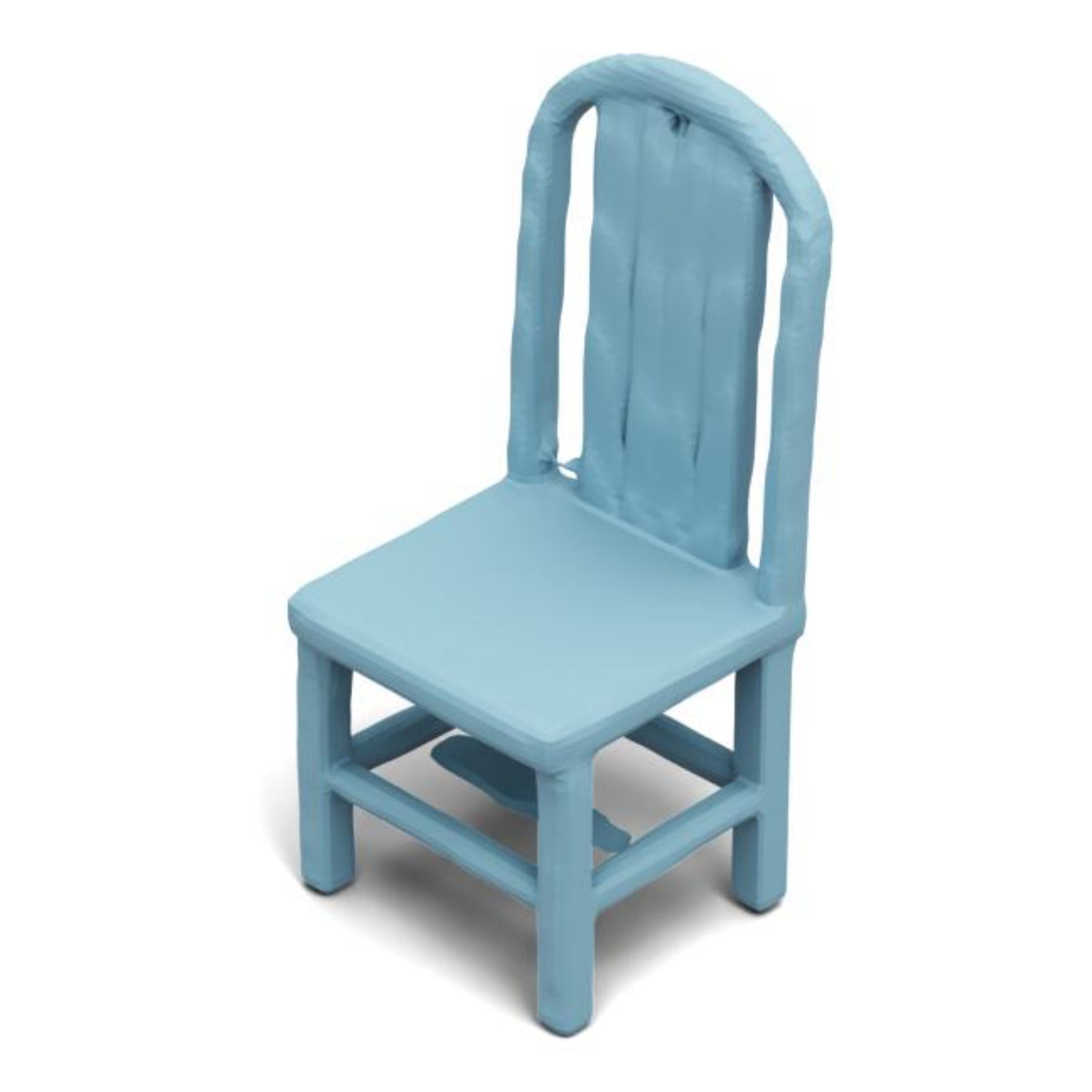}
&\includegraphics[trim = 1 1 1 1, clip, width=0.125\linewidth]{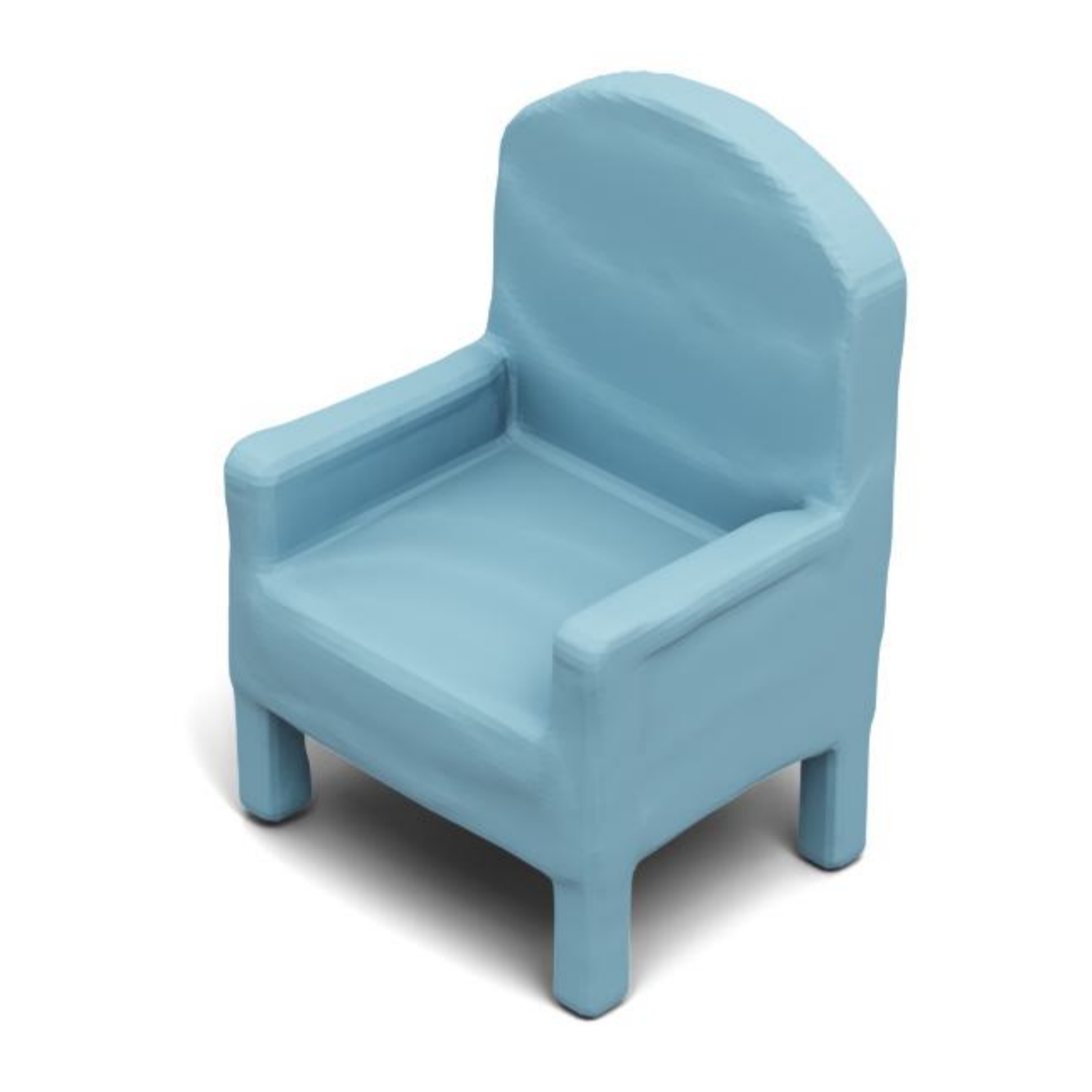}
&\includegraphics[trim = 1 1 1 1, clip, width=0.125\linewidth]{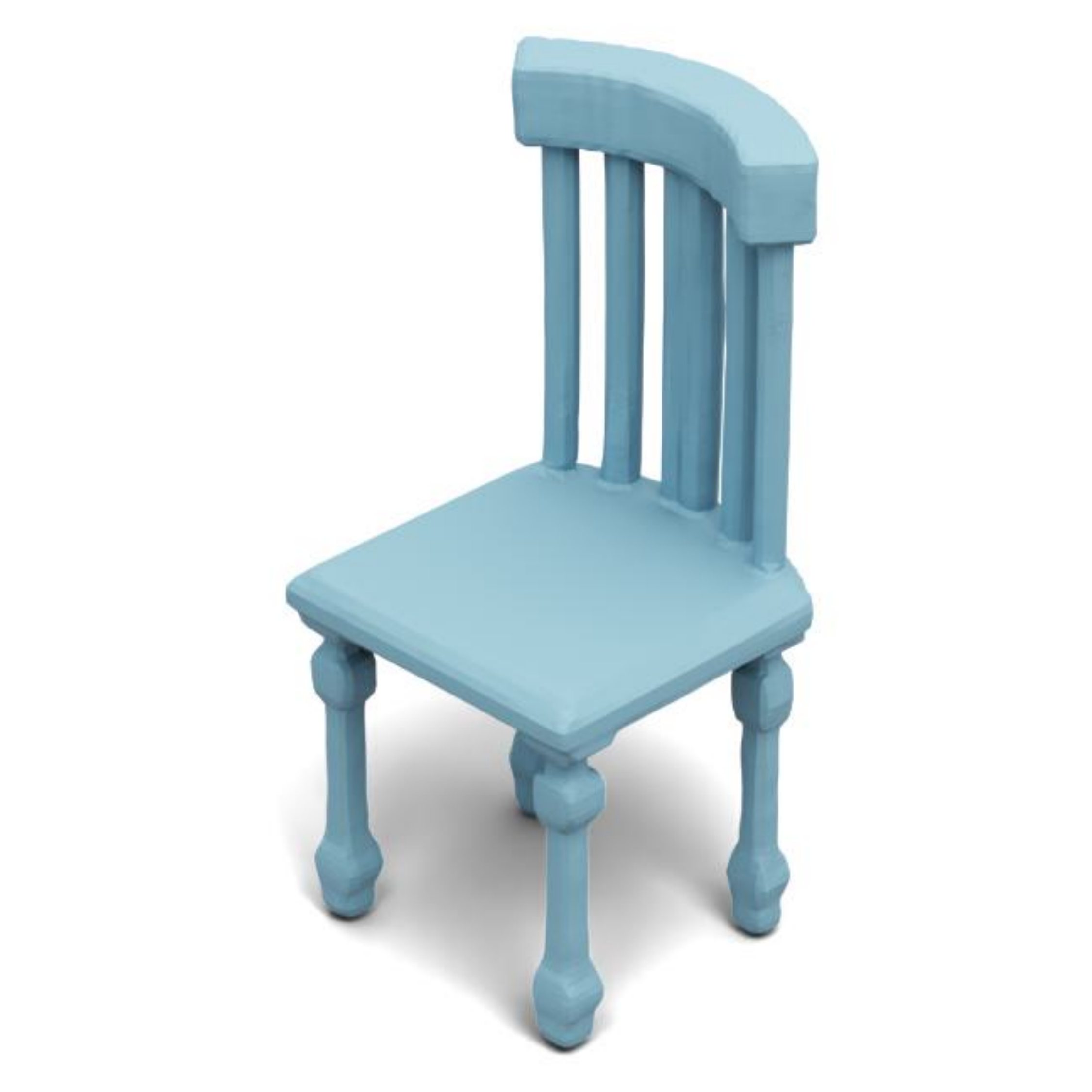}
&\includegraphics[trim = 1 1 1 1, clip, width=0.125\linewidth]{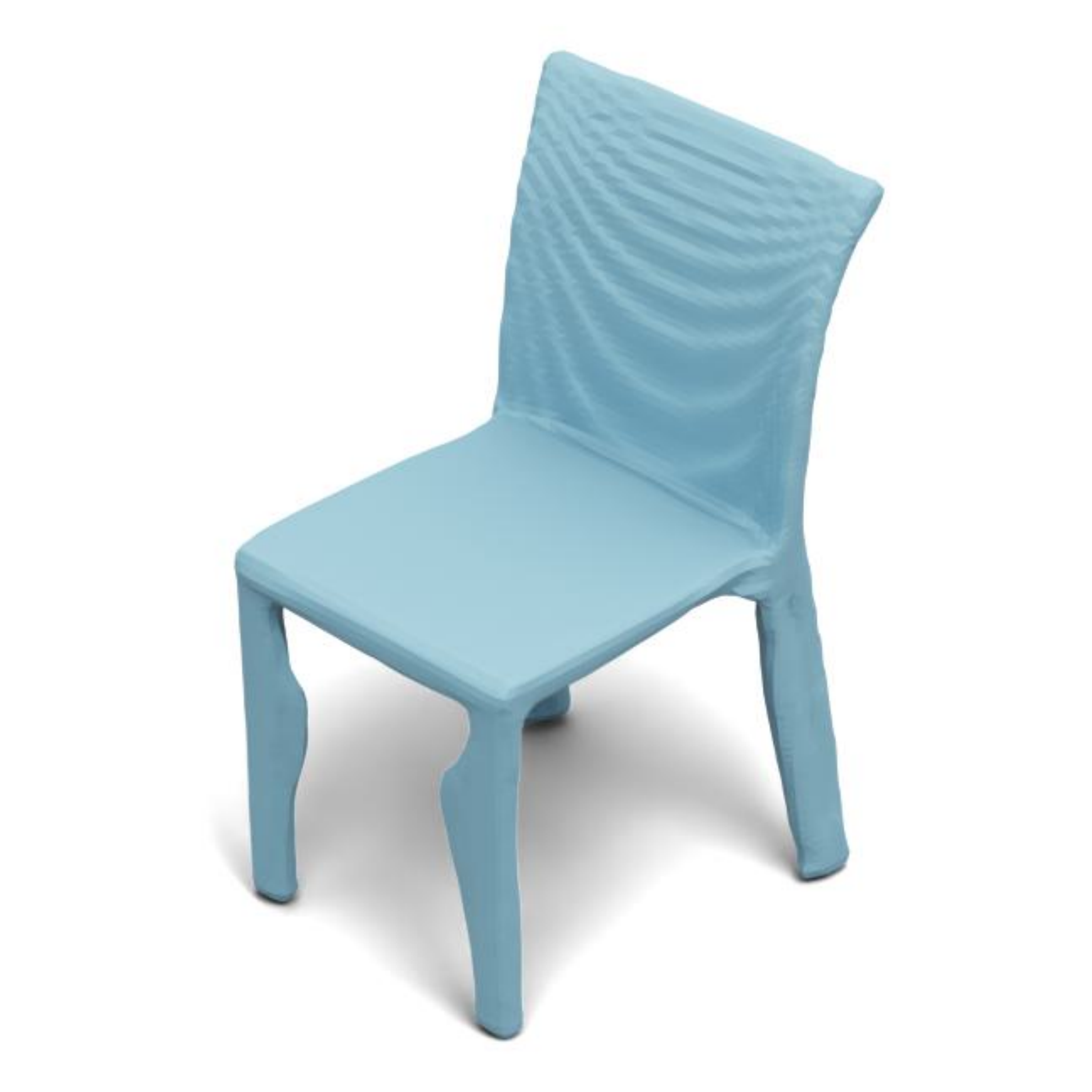}
\\

\end{tabular}
    \caption{We randomly sample sketches from the AmateurSketch dataset and showcase the results of our method.}
	\label{fig:additionalExp2}
\end{figure*}

\begin{figure*}[t]
	\centering
	\small
	\setlength{\tabcolsep}{1pt}
 \begin{tabular}{cccccccc}
\includegraphics[width=0.125\linewidth]{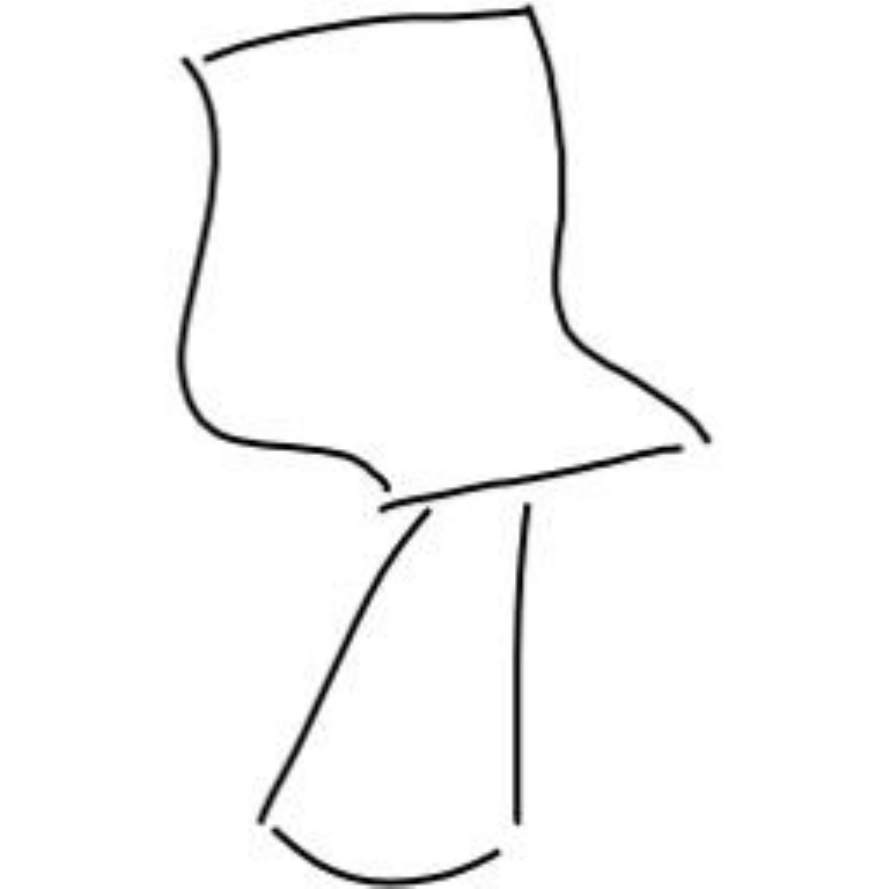}
&\includegraphics[width=0.125\linewidth]{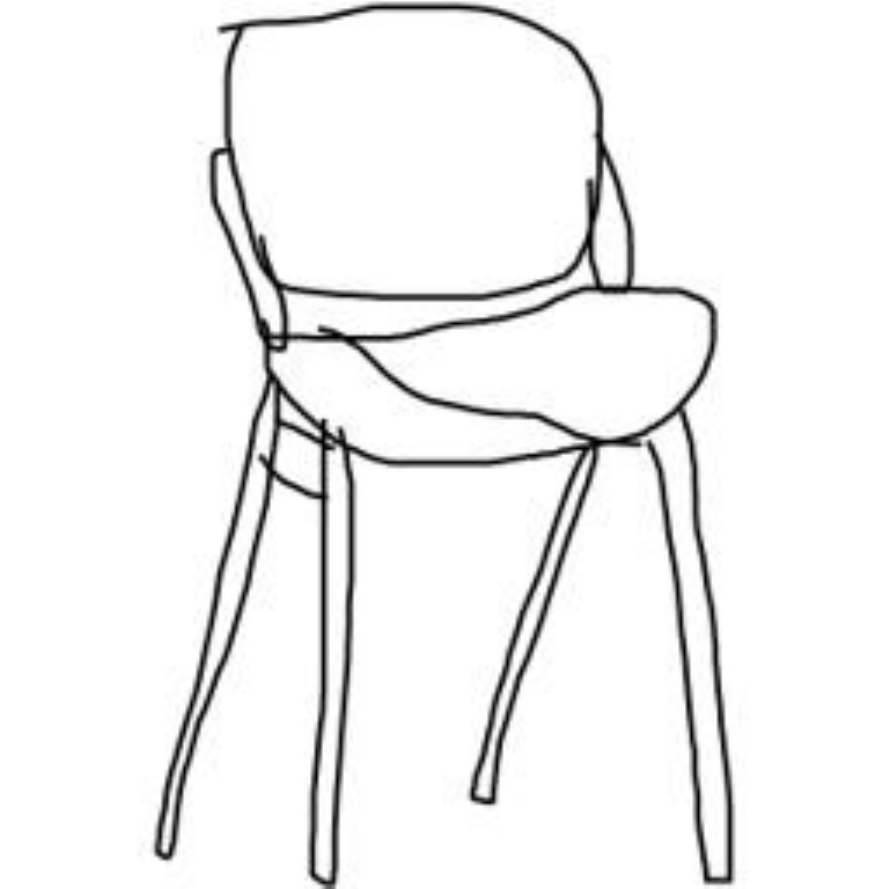}
&\includegraphics[width=0.125\linewidth]{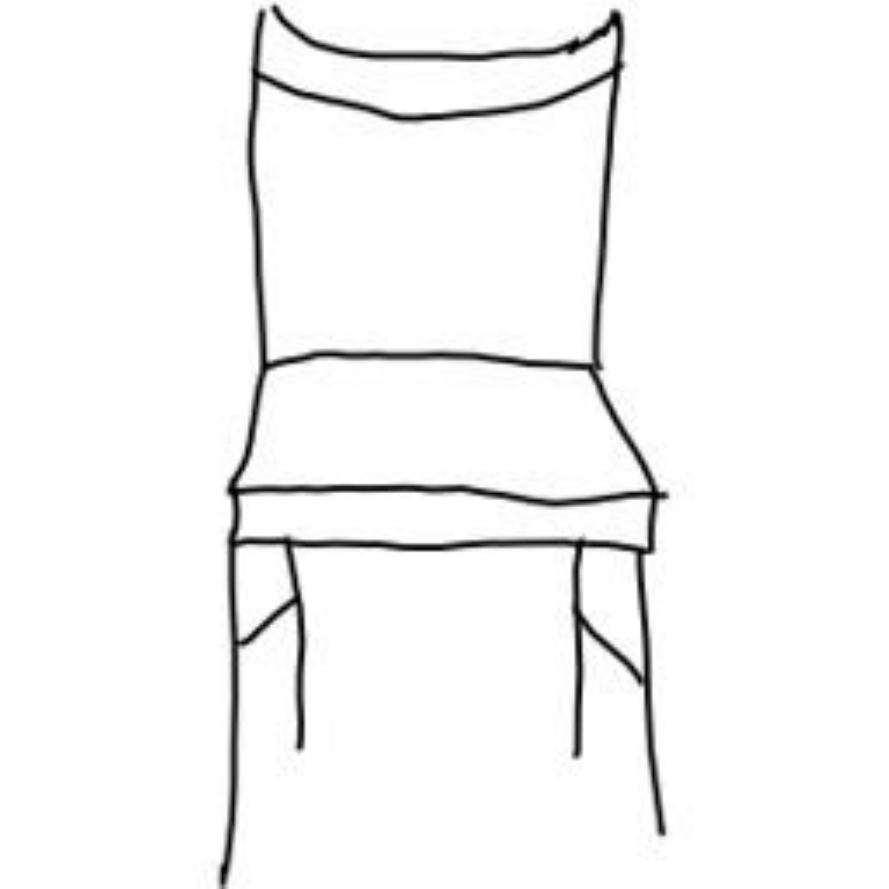}
&\includegraphics[width=0.125\linewidth]{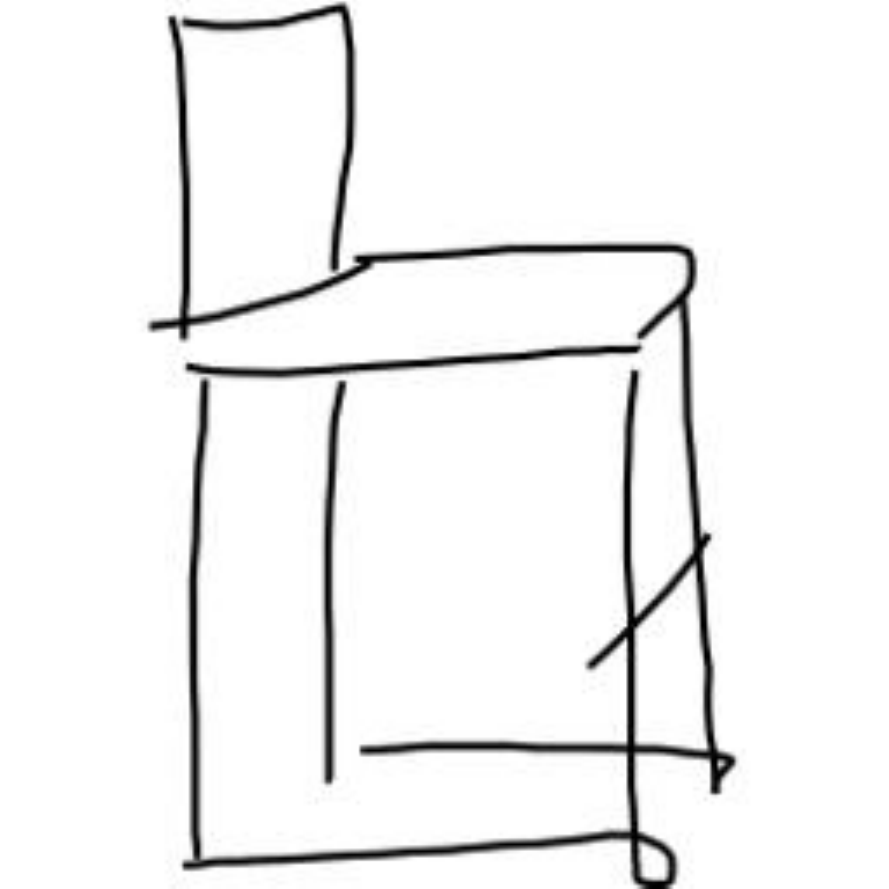}
&\includegraphics[width=0.125\linewidth]{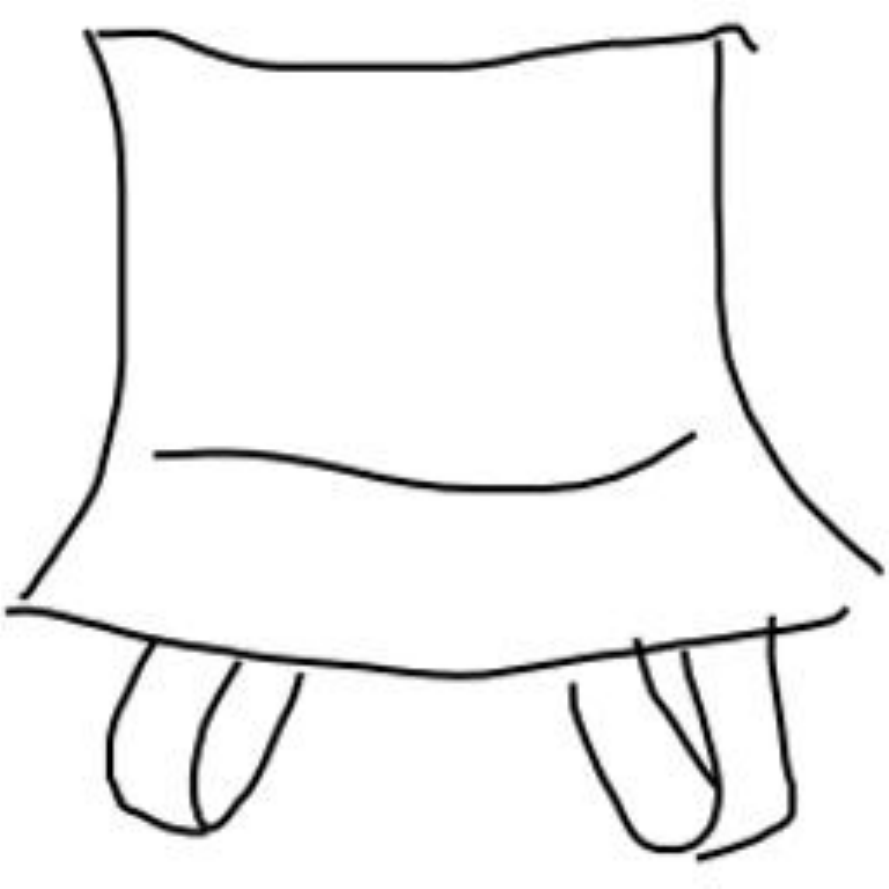}
&\includegraphics[width=0.125\linewidth]{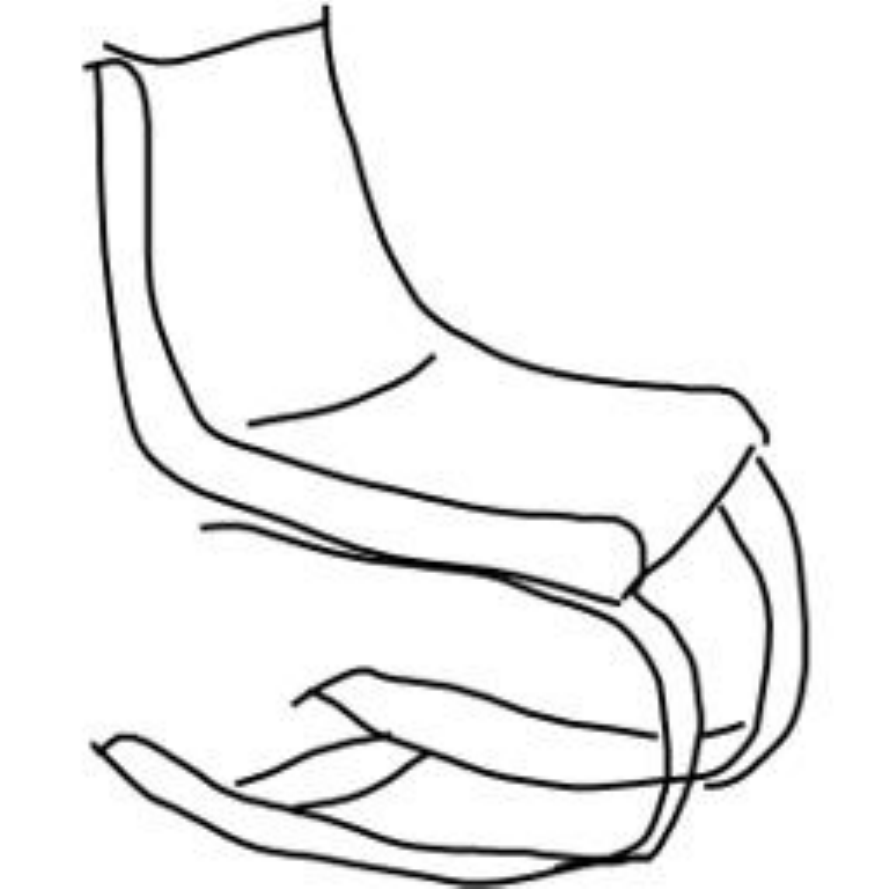}
&\includegraphics[width=0.125\linewidth]{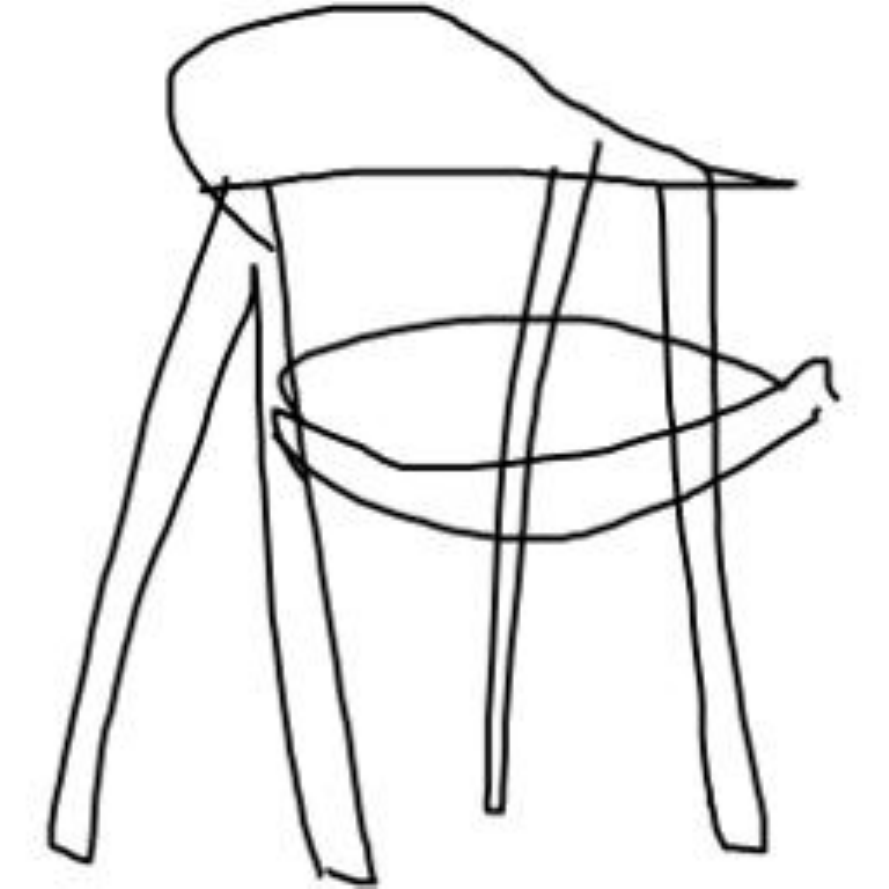}
&\includegraphics[width=0.125\linewidth]{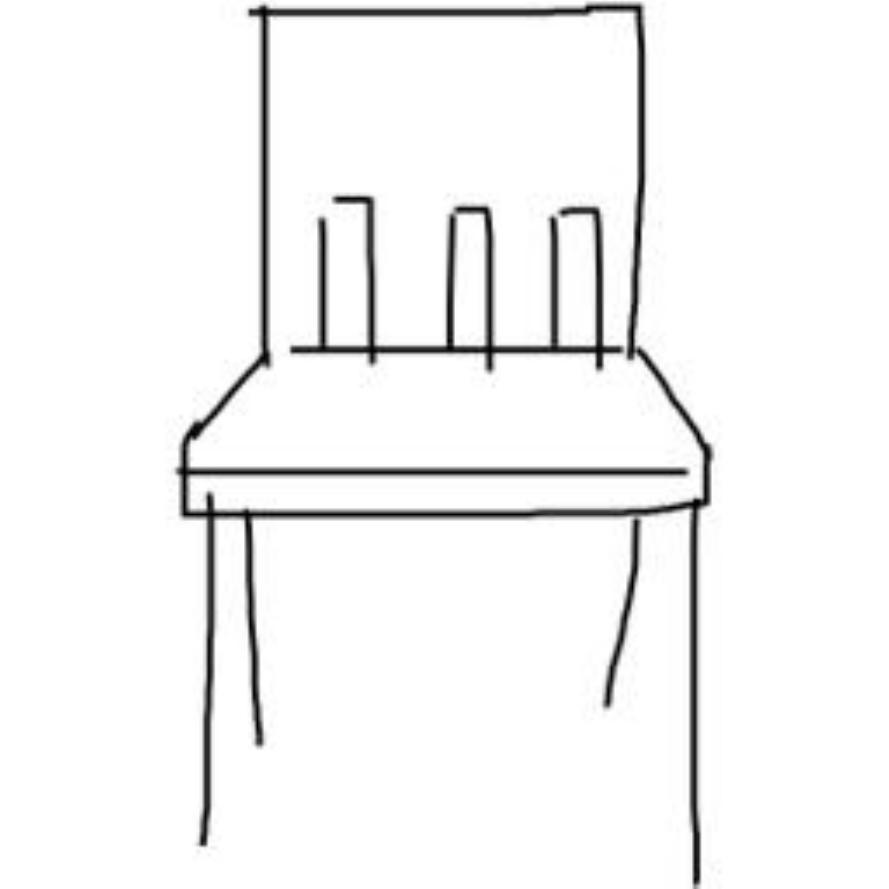}
\\
\includegraphics[trim = 1 1 1 1, clip, width=0.125\linewidth]{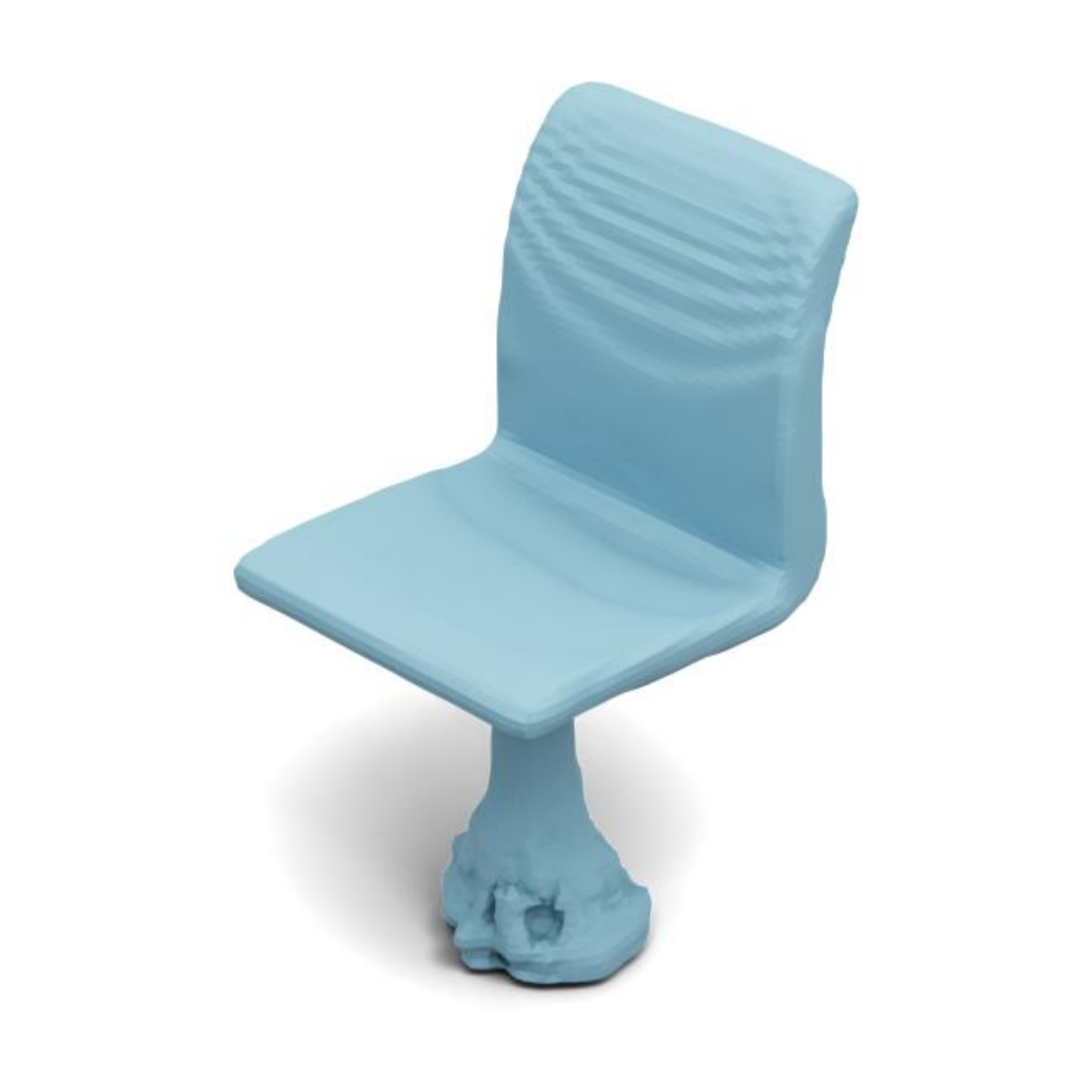}
&\includegraphics[trim = 1 1 1 1, clip, width=0.125\linewidth]{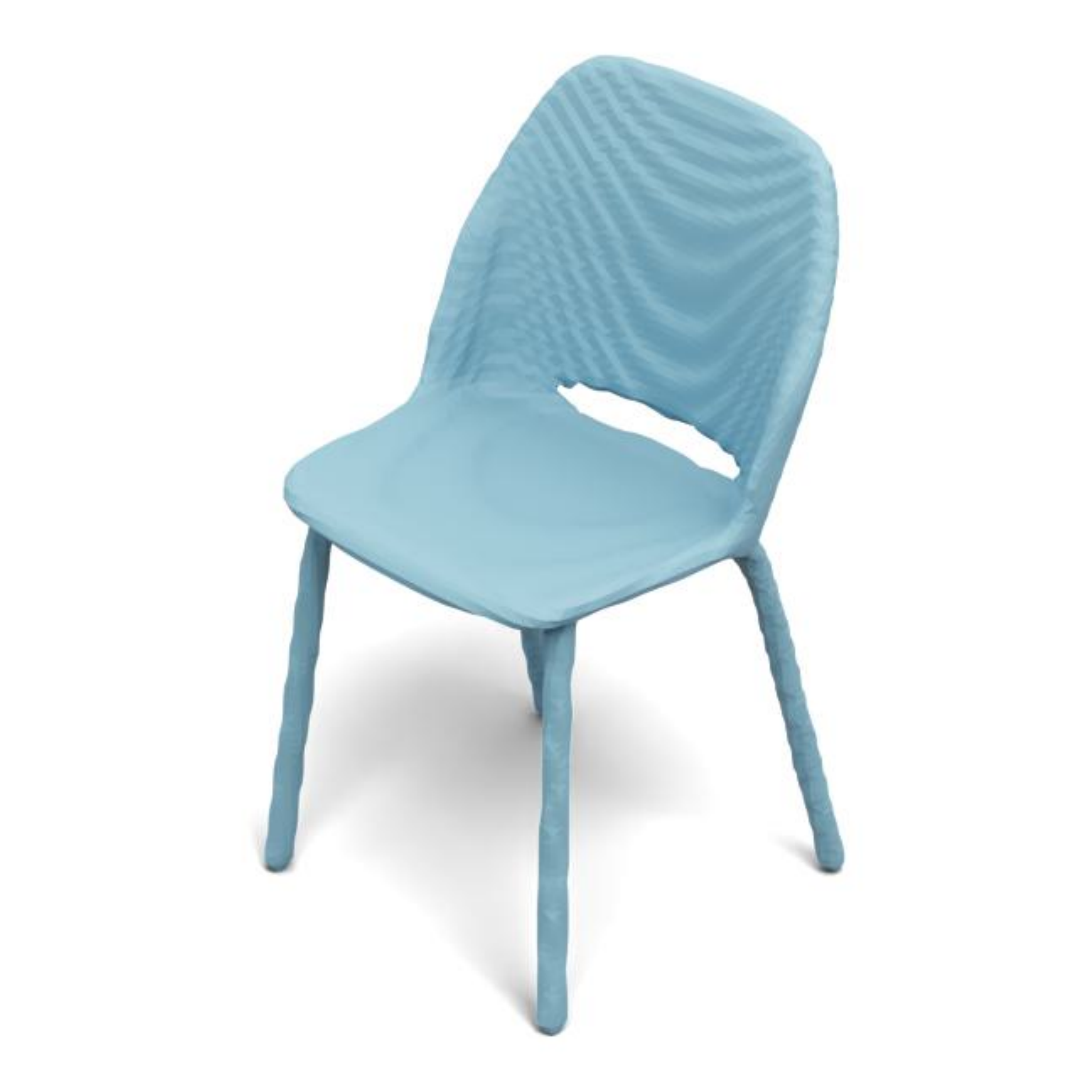}
&\includegraphics[trim = 1 1 1 1, clip, width=0.125\linewidth]{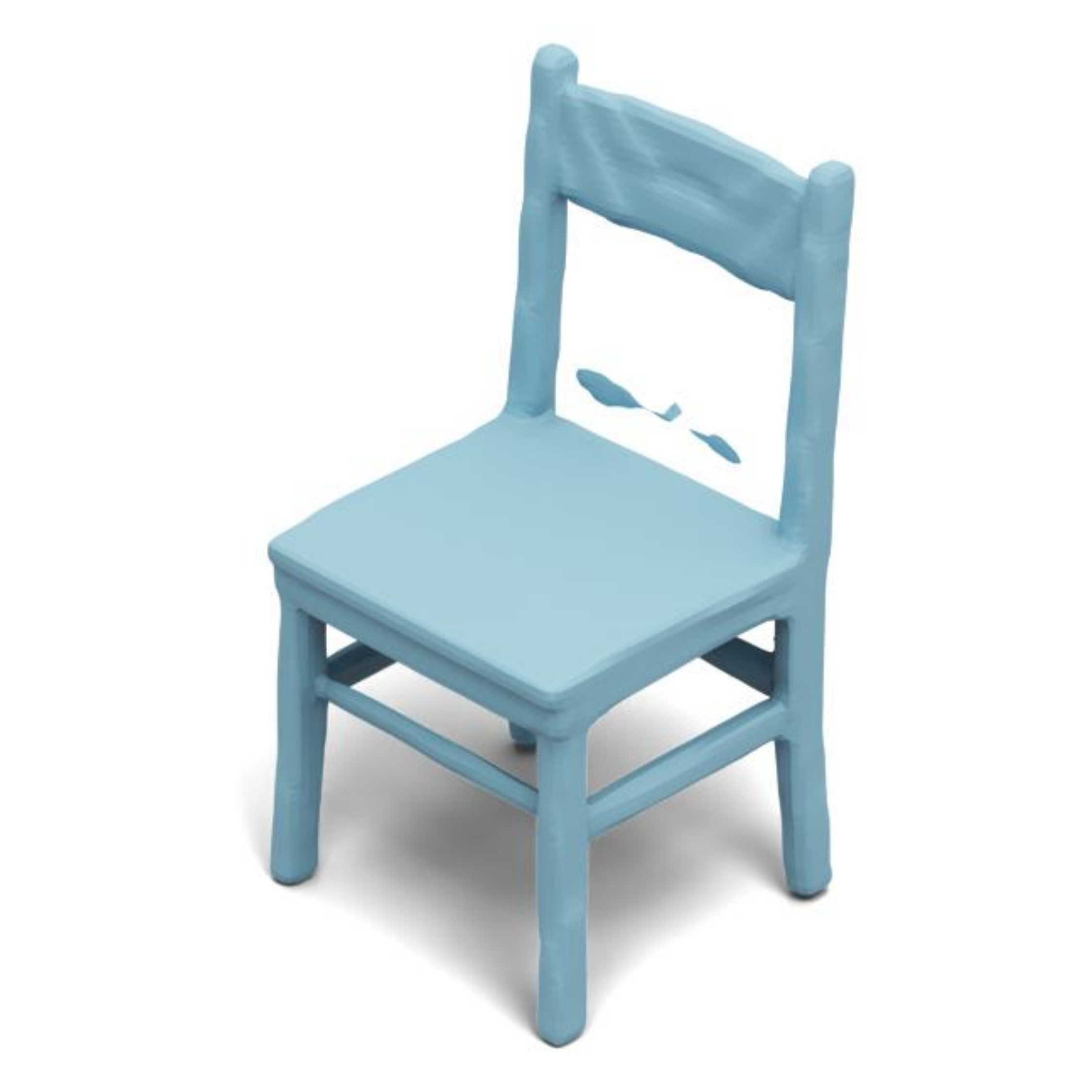}
&\includegraphics[trim = 1 1 1 1, clip, width=0.125\linewidth]{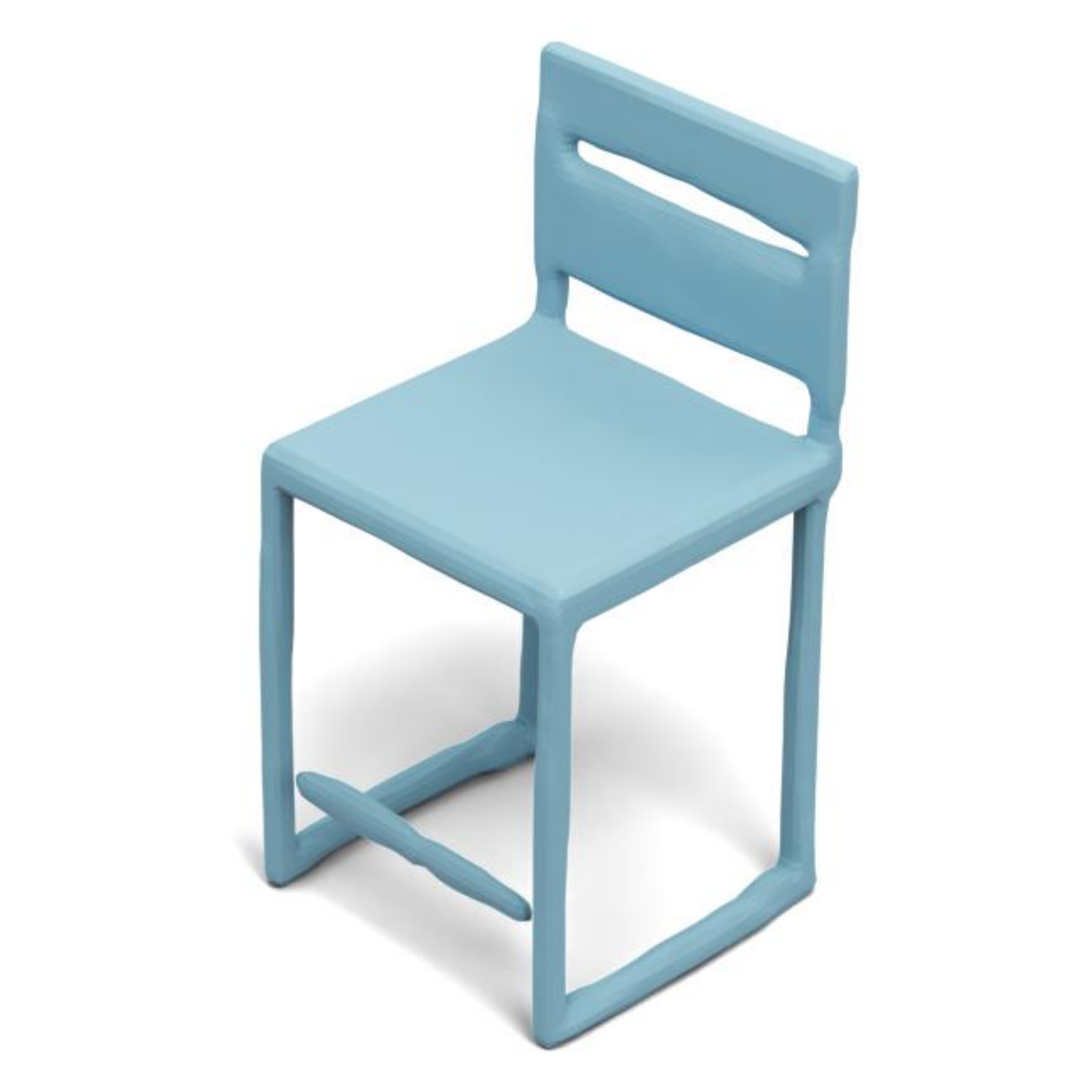}
&\includegraphics[trim = 1 1 1 1, clip, width=0.125\linewidth]{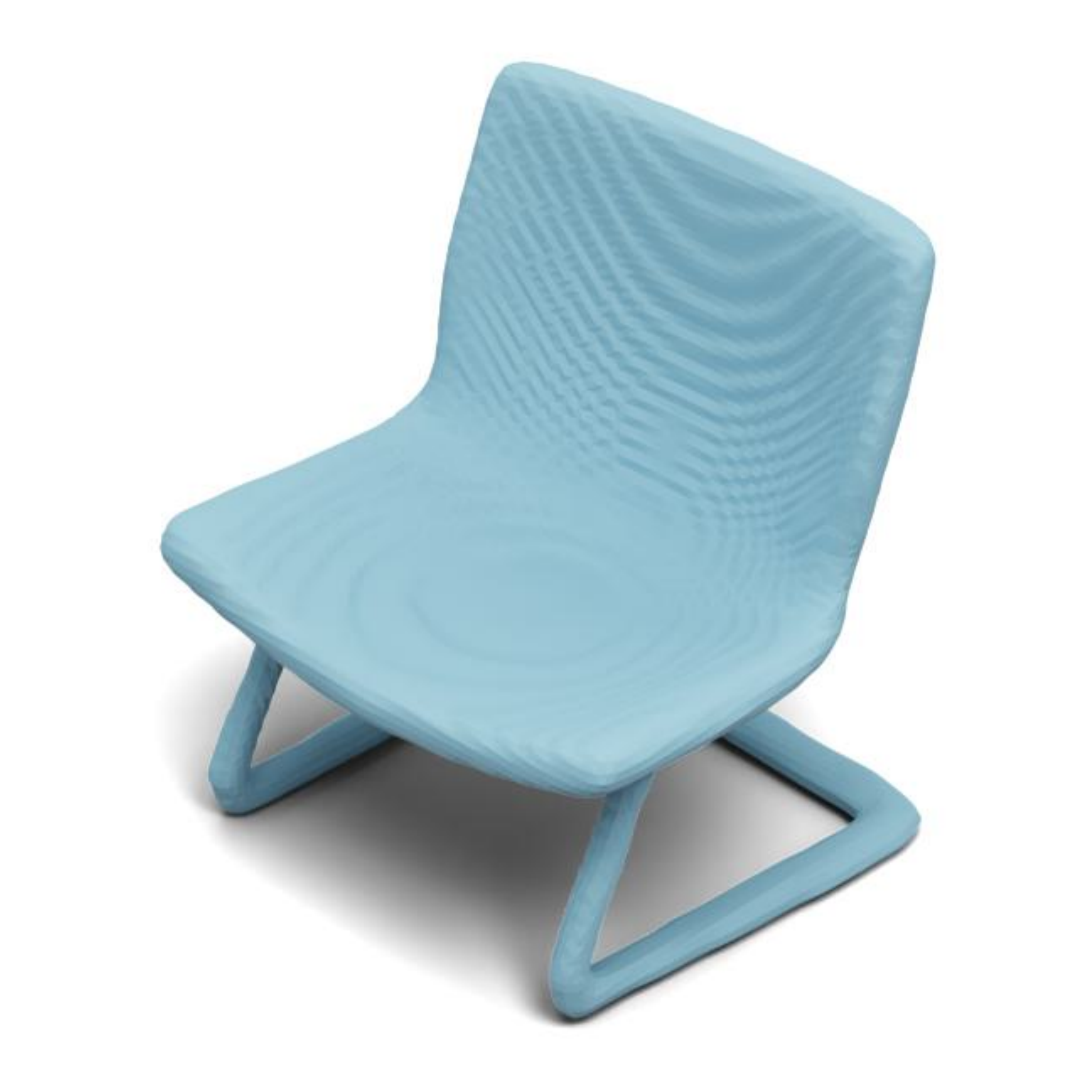}
&\includegraphics[trim = 1 1 1 1, clip, width=0.125\linewidth]{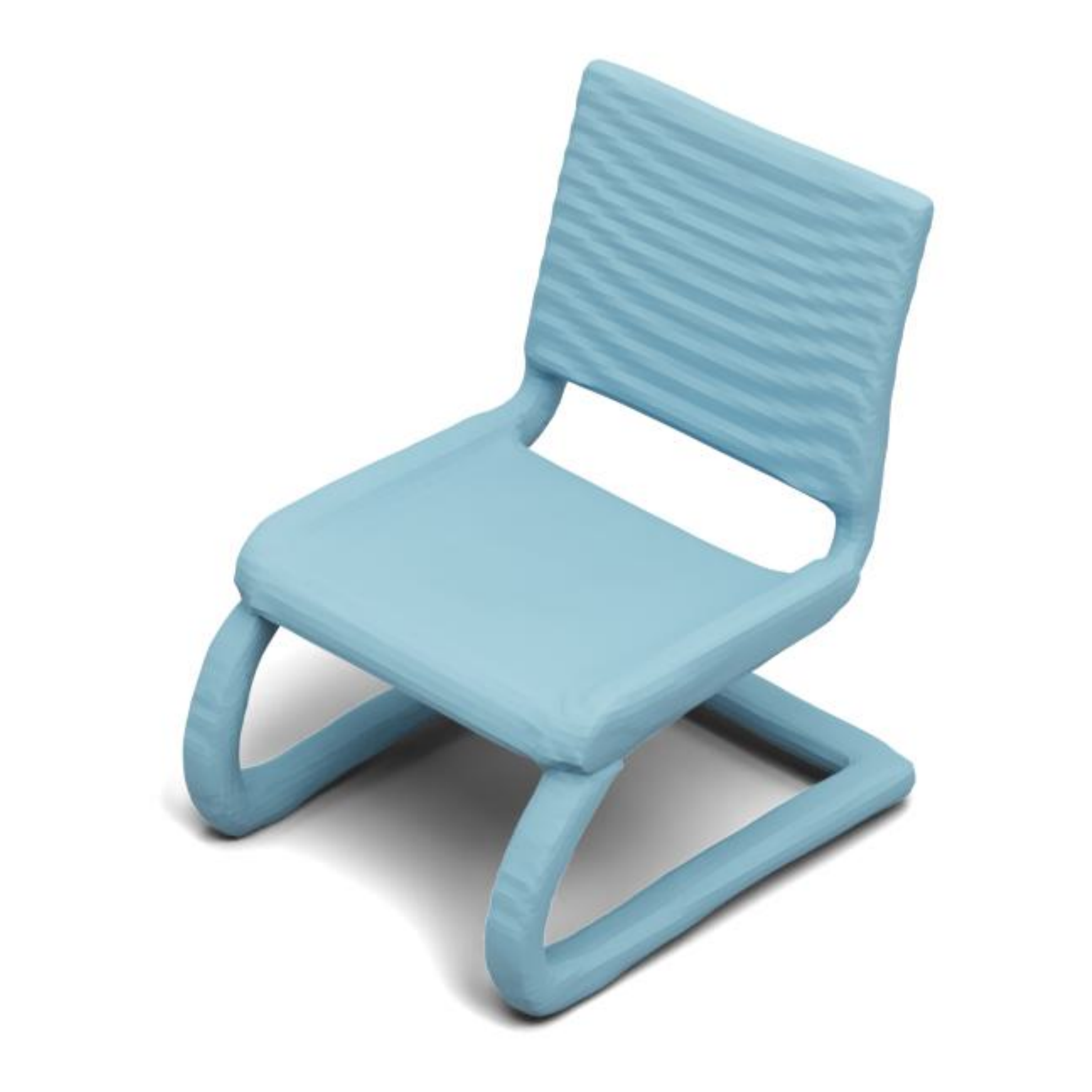}
&\includegraphics[trim = 1 1 1 1, clip, width=0.125\linewidth]{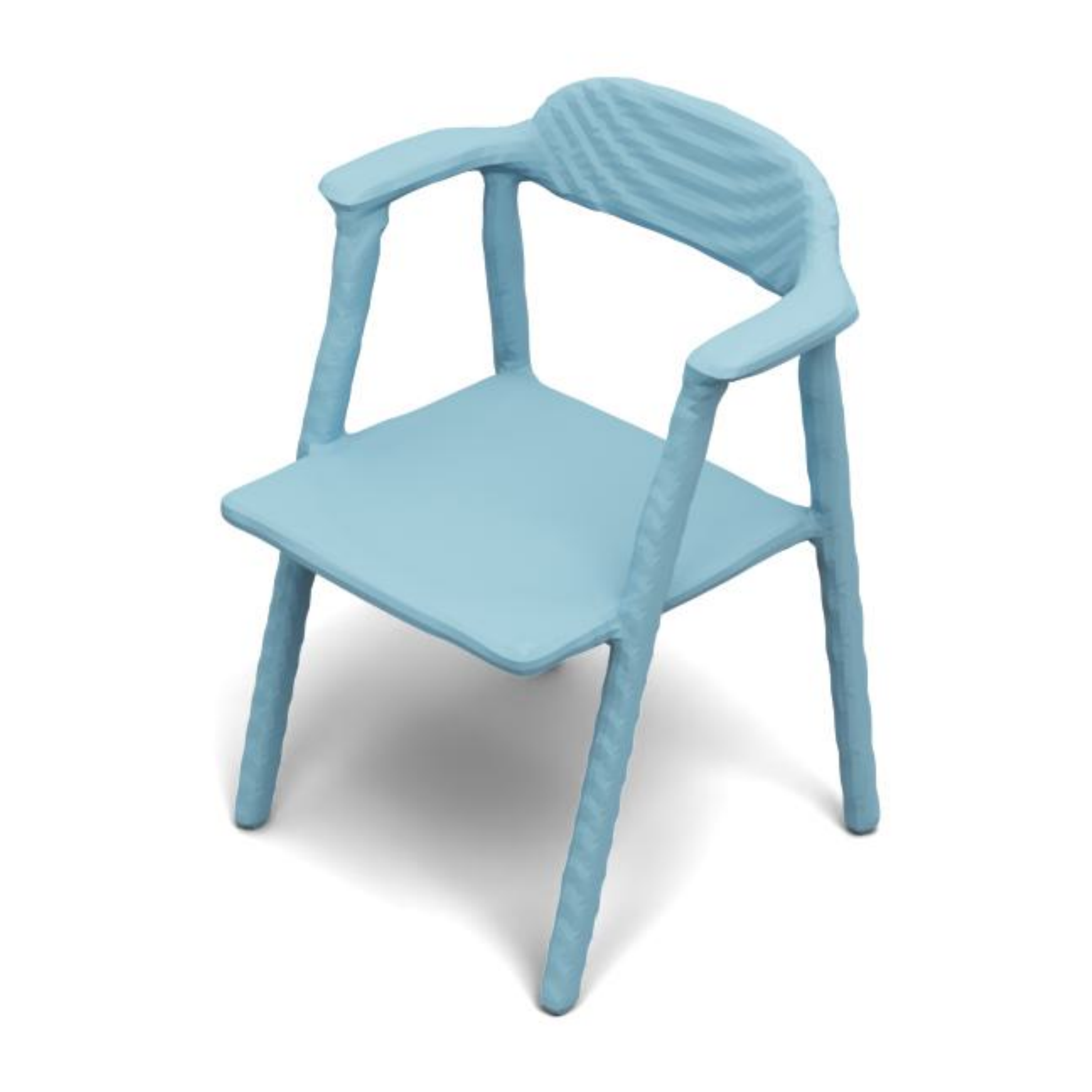}
&\includegraphics[trim = 1 1 1 1, clip, width=0.125\linewidth]{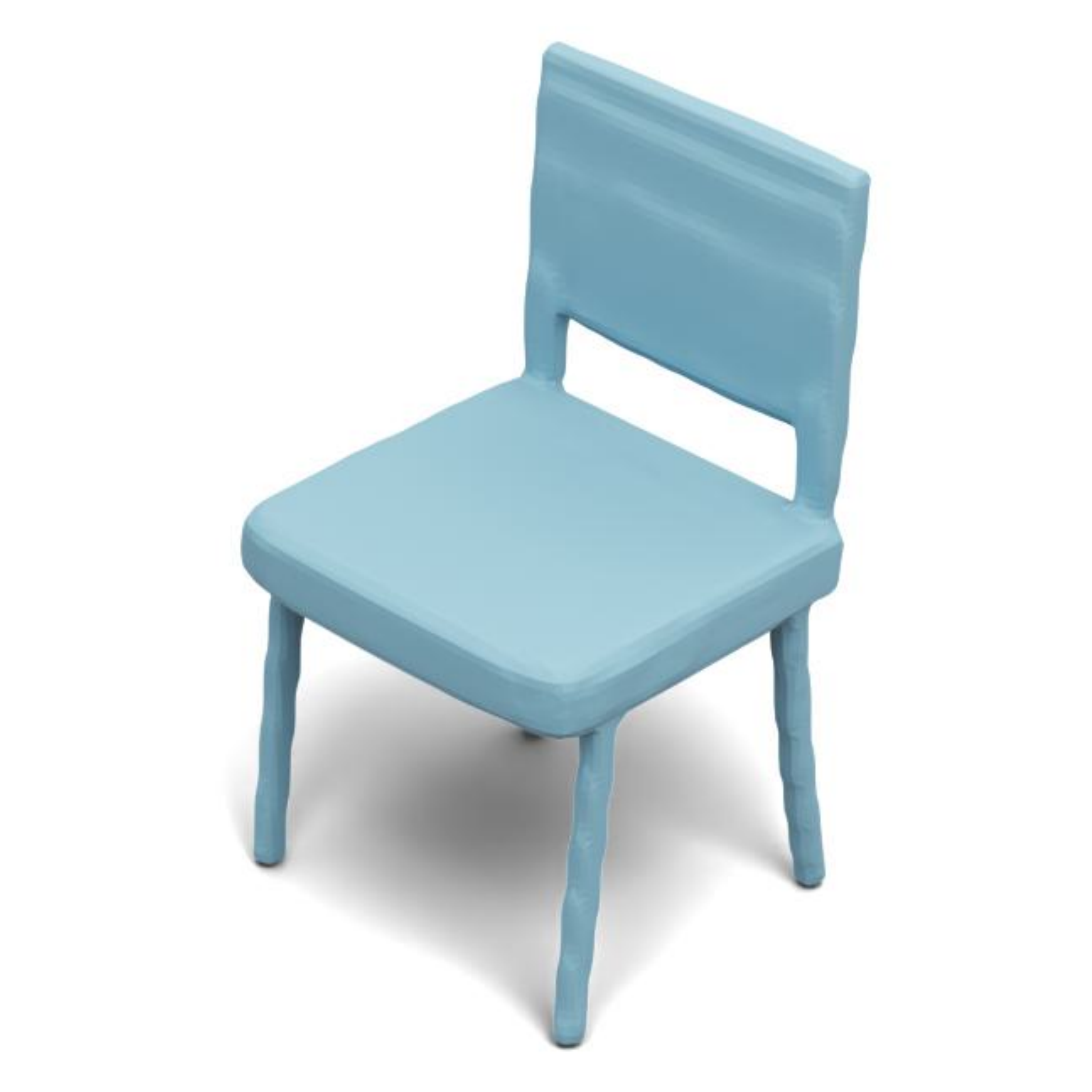}
\\
\includegraphics[width=0.125\linewidth]{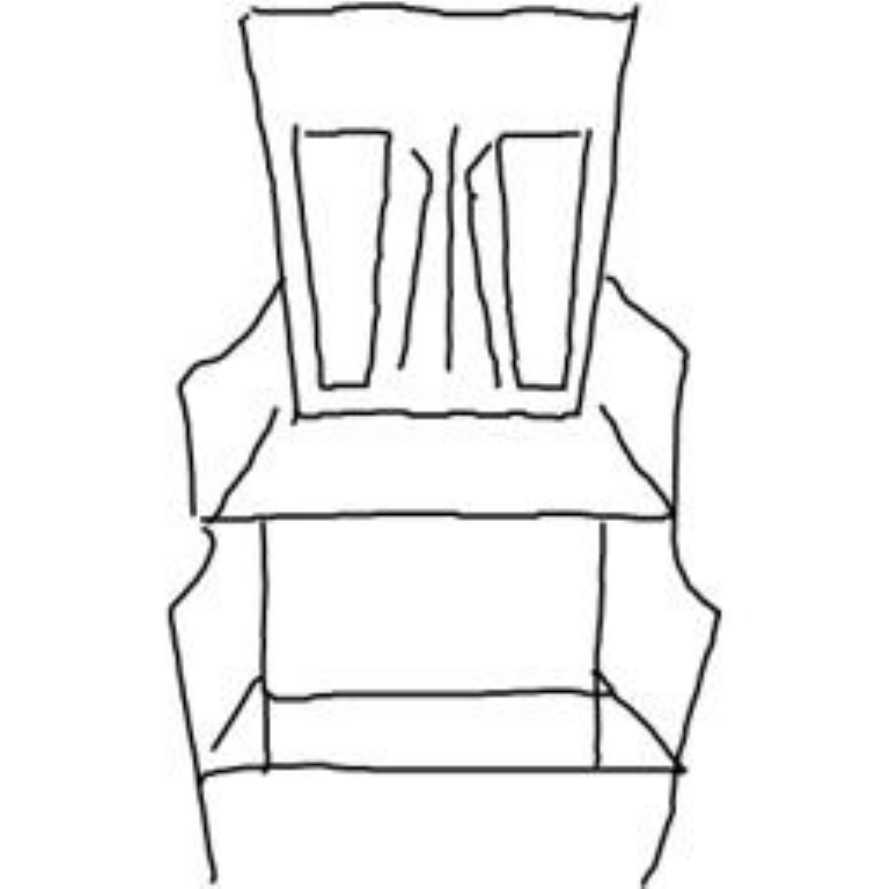}
&\includegraphics[width=0.125\linewidth]{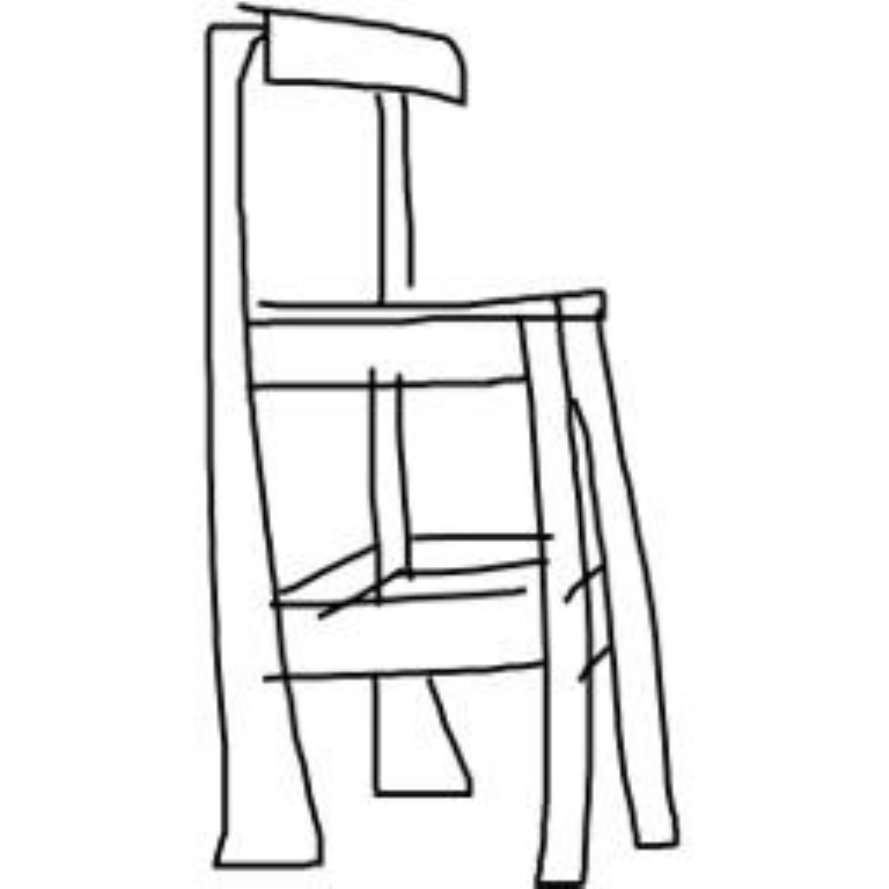}
&\includegraphics[width=0.125\linewidth]{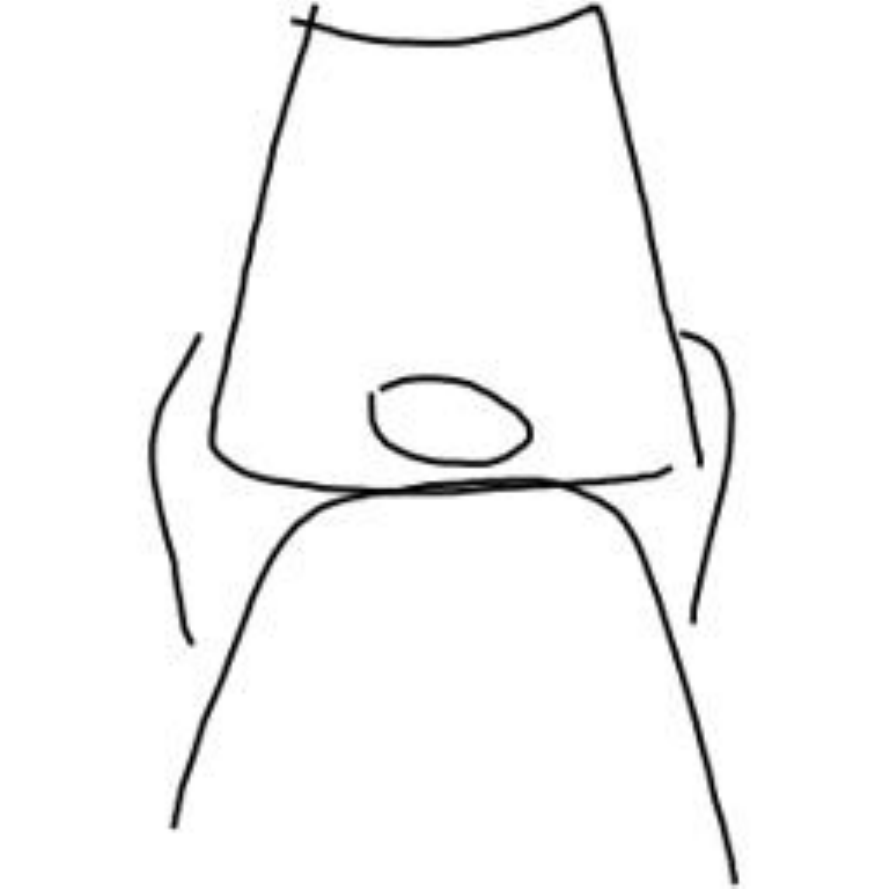}
&\includegraphics[width=0.125\linewidth]{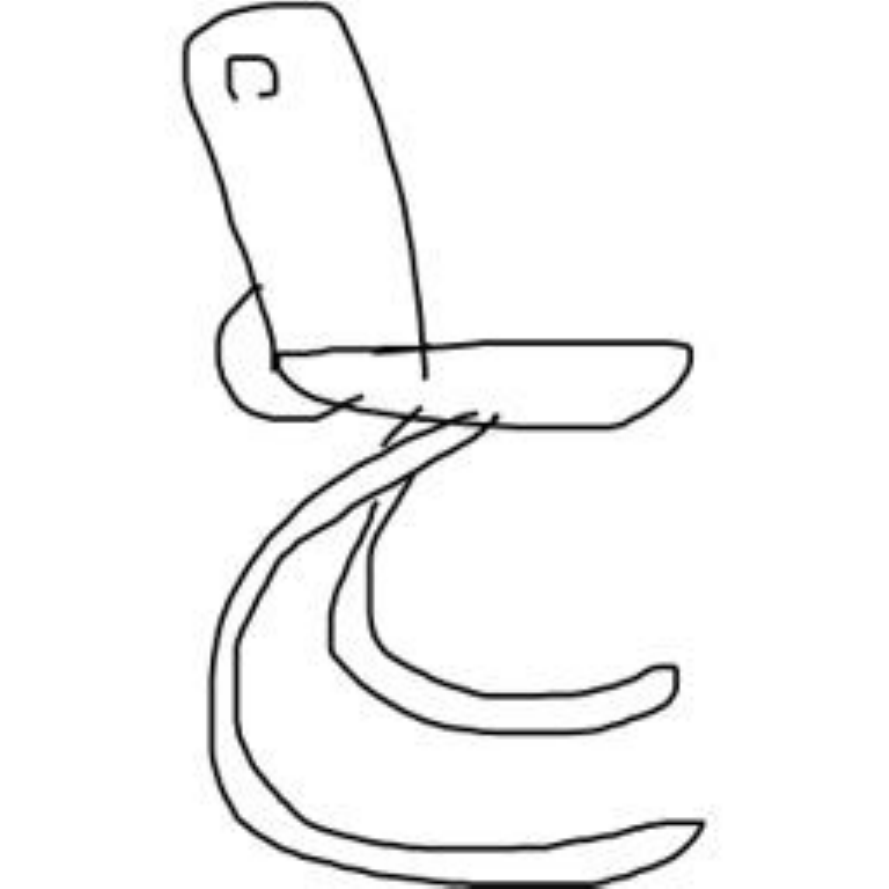}
&\includegraphics[width=0.125\linewidth]{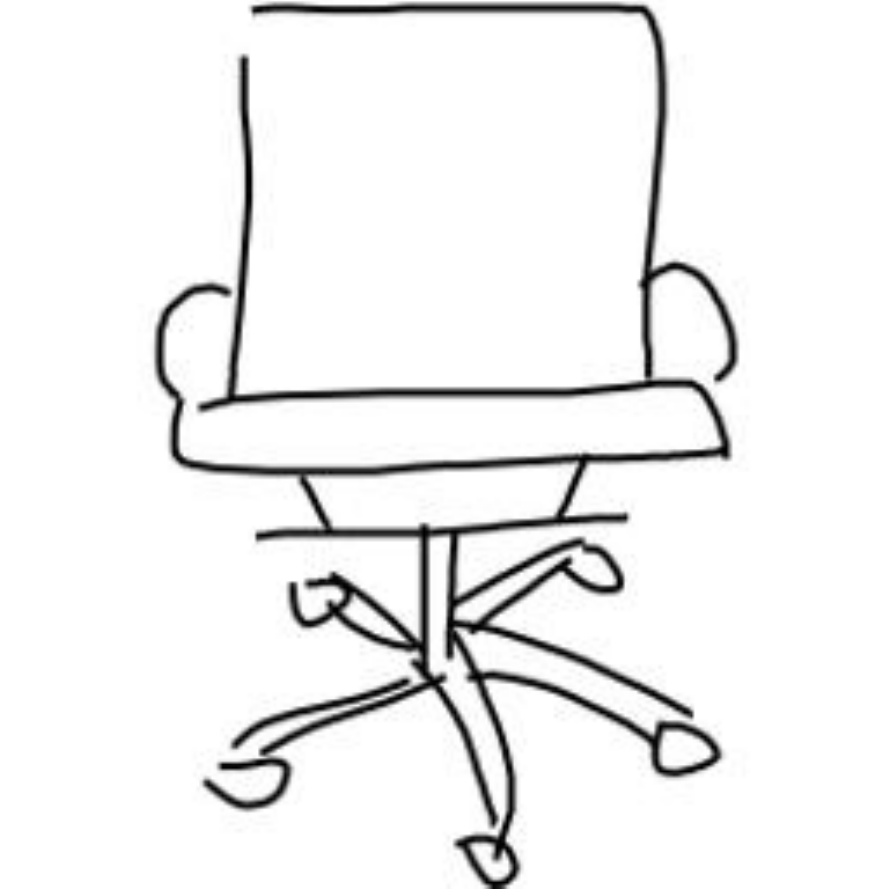}
&\includegraphics[width=0.125\linewidth]{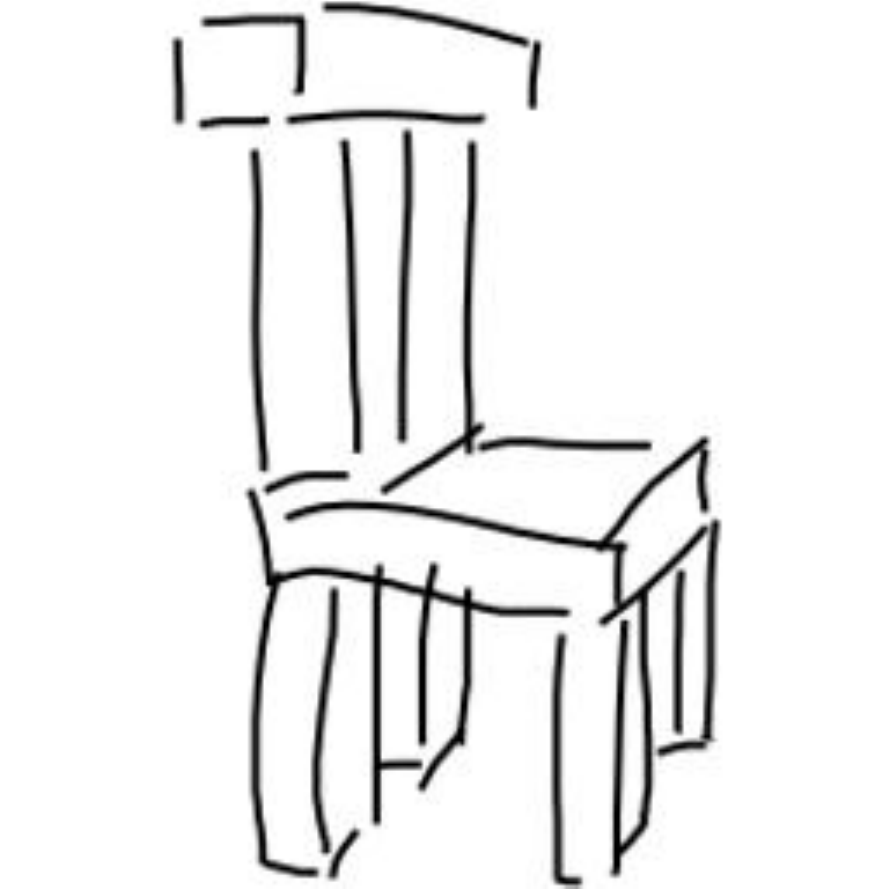}
&\includegraphics[width=0.125\linewidth]{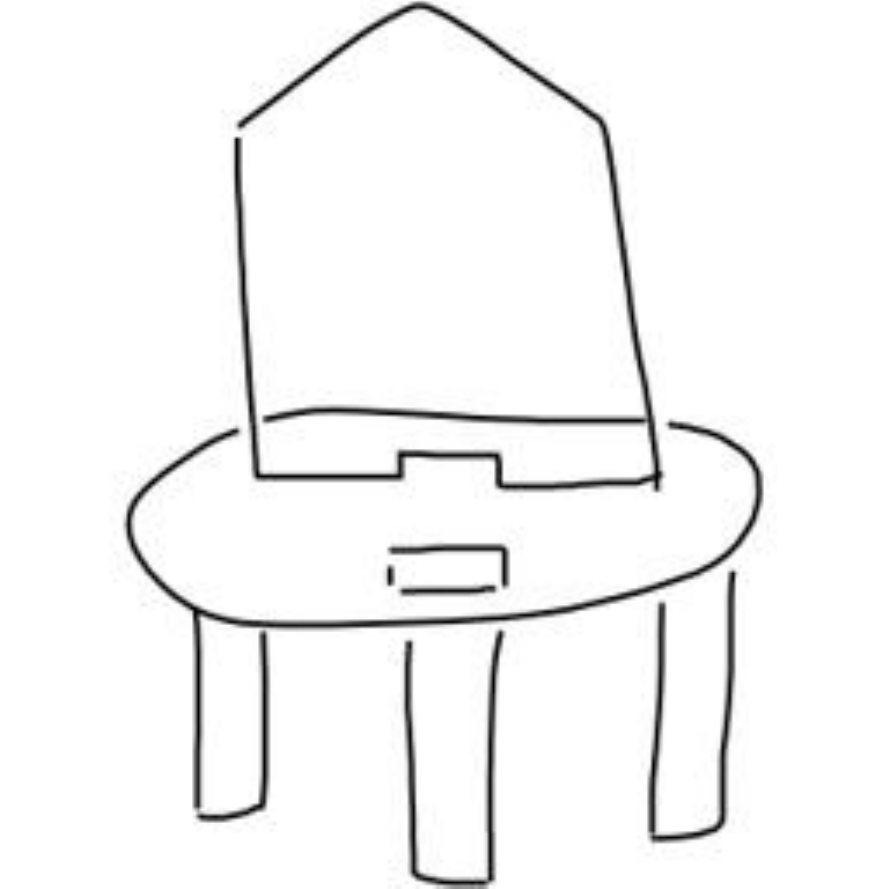}
&\includegraphics[width=0.125\linewidth]{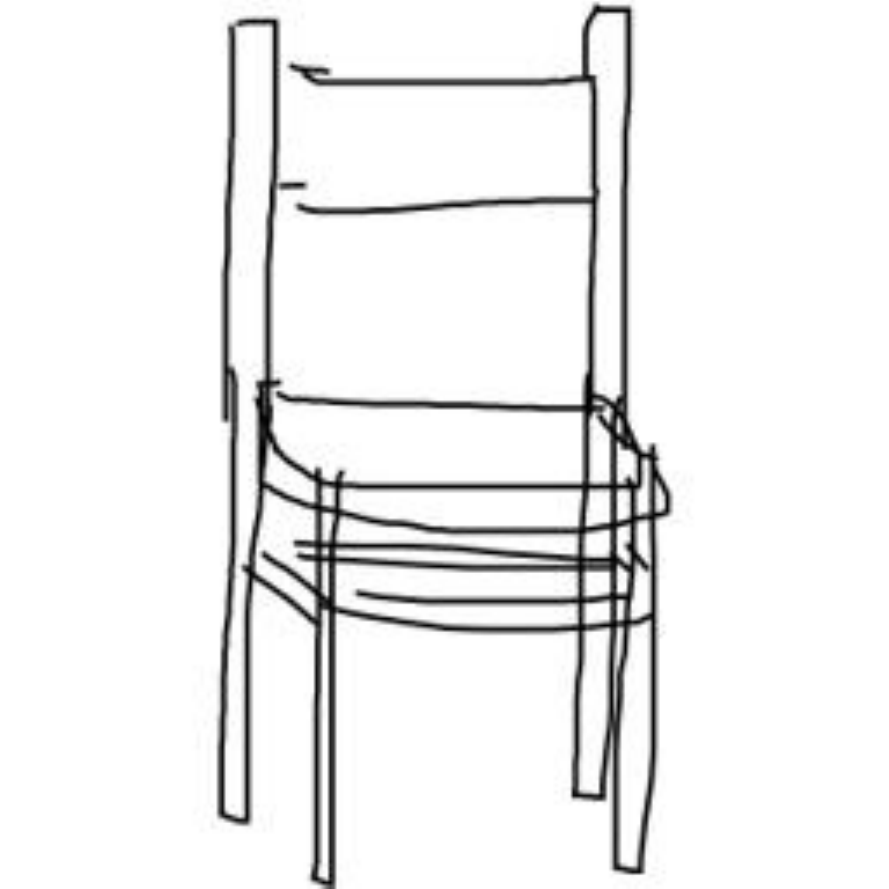}
\\
\includegraphics[trim = 1 1 1 1, clip, width=0.125\linewidth]{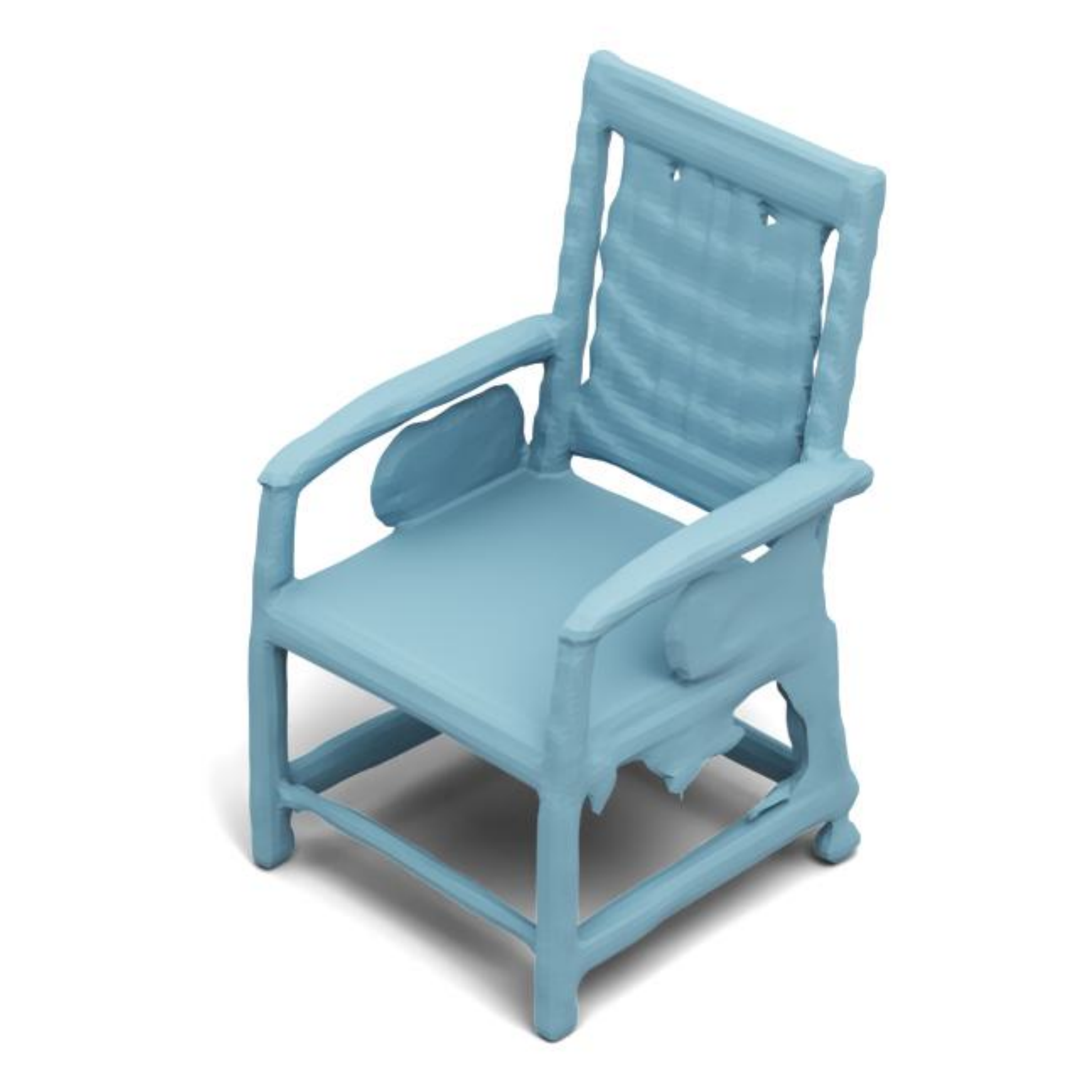}
&\includegraphics[trim = 1 1 1 1, clip, width=0.125\linewidth]{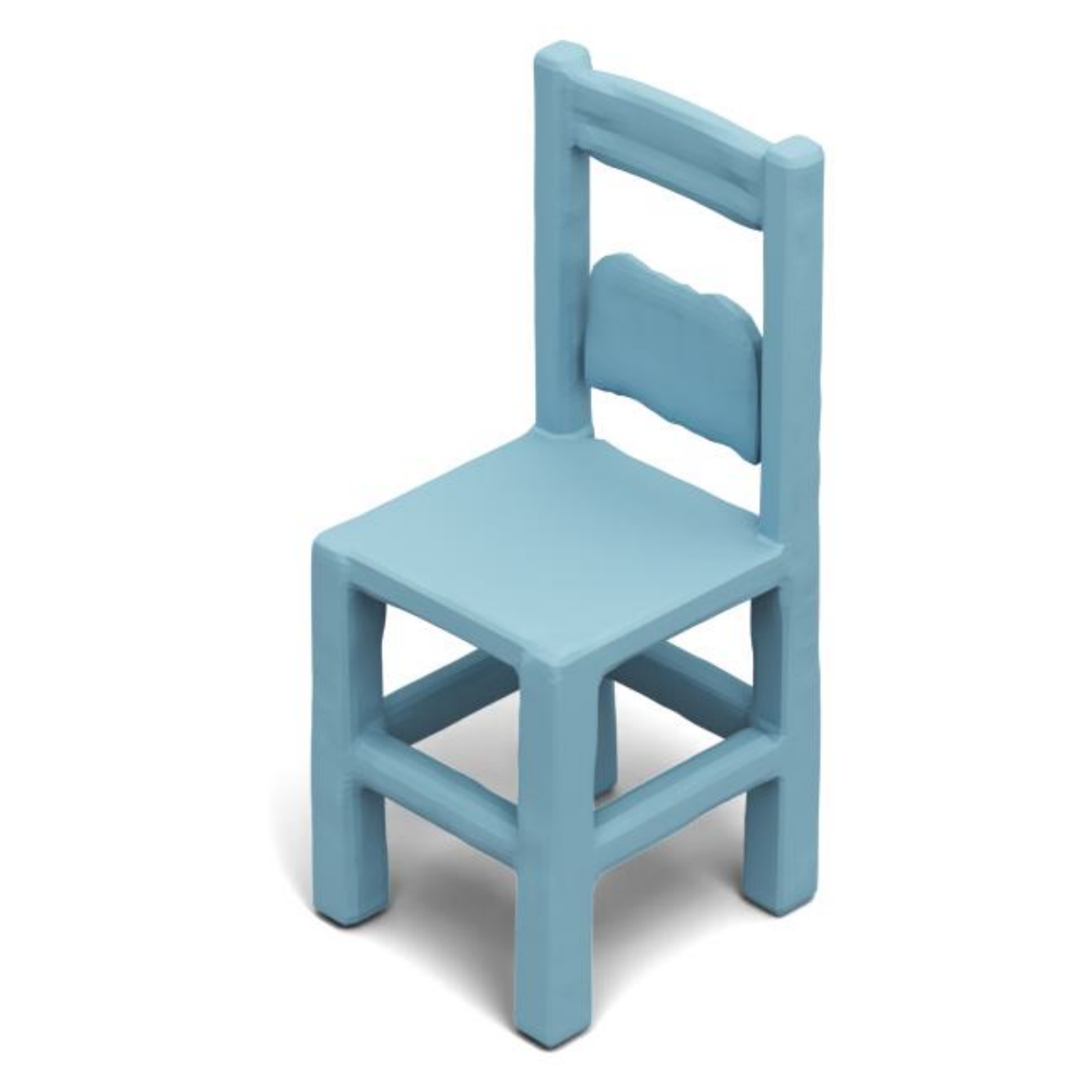}
&\includegraphics[trim = 1 1 1 1, clip, width=0.125\linewidth]{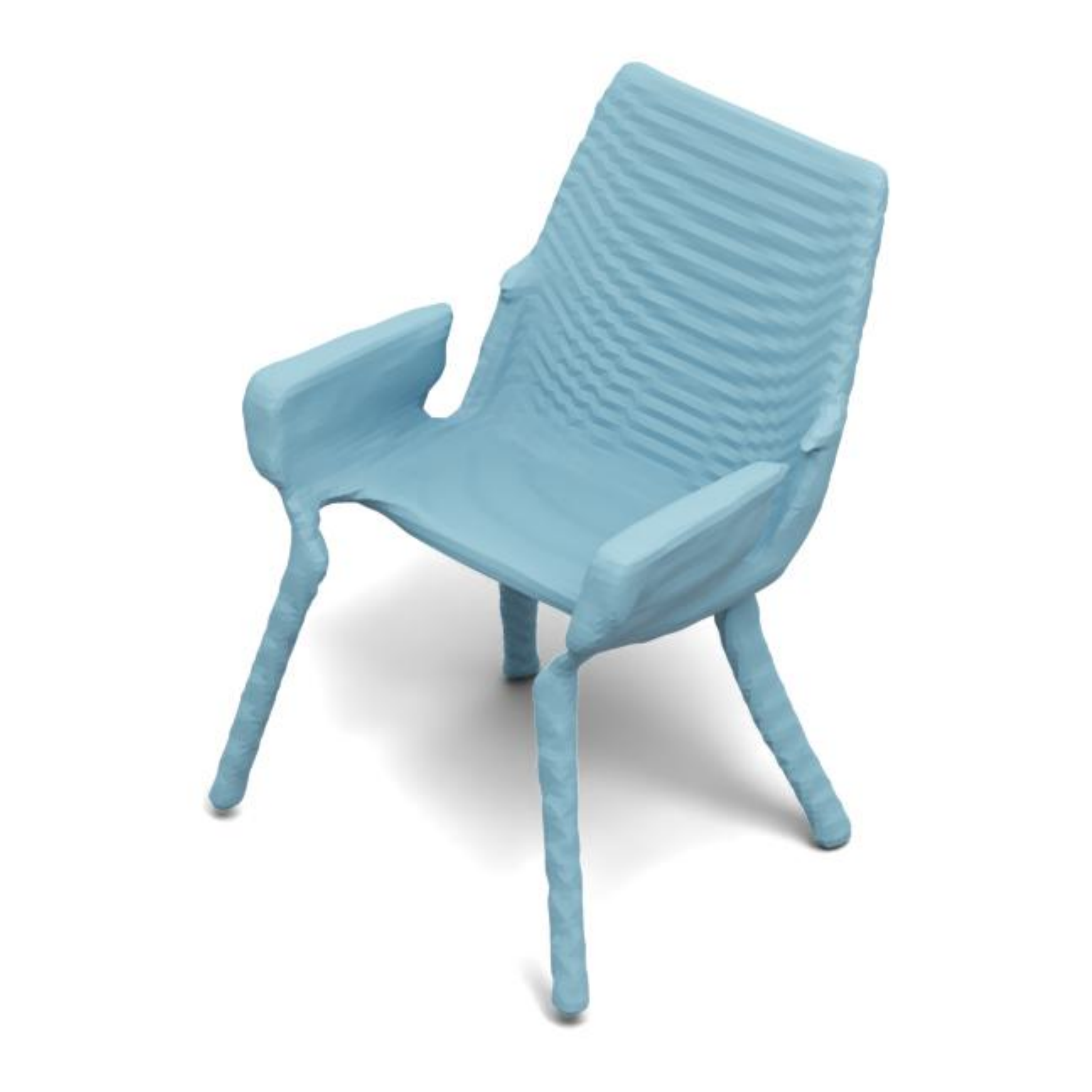}
&\includegraphics[trim = 1 1 1 1, clip, width=0.125\linewidth]{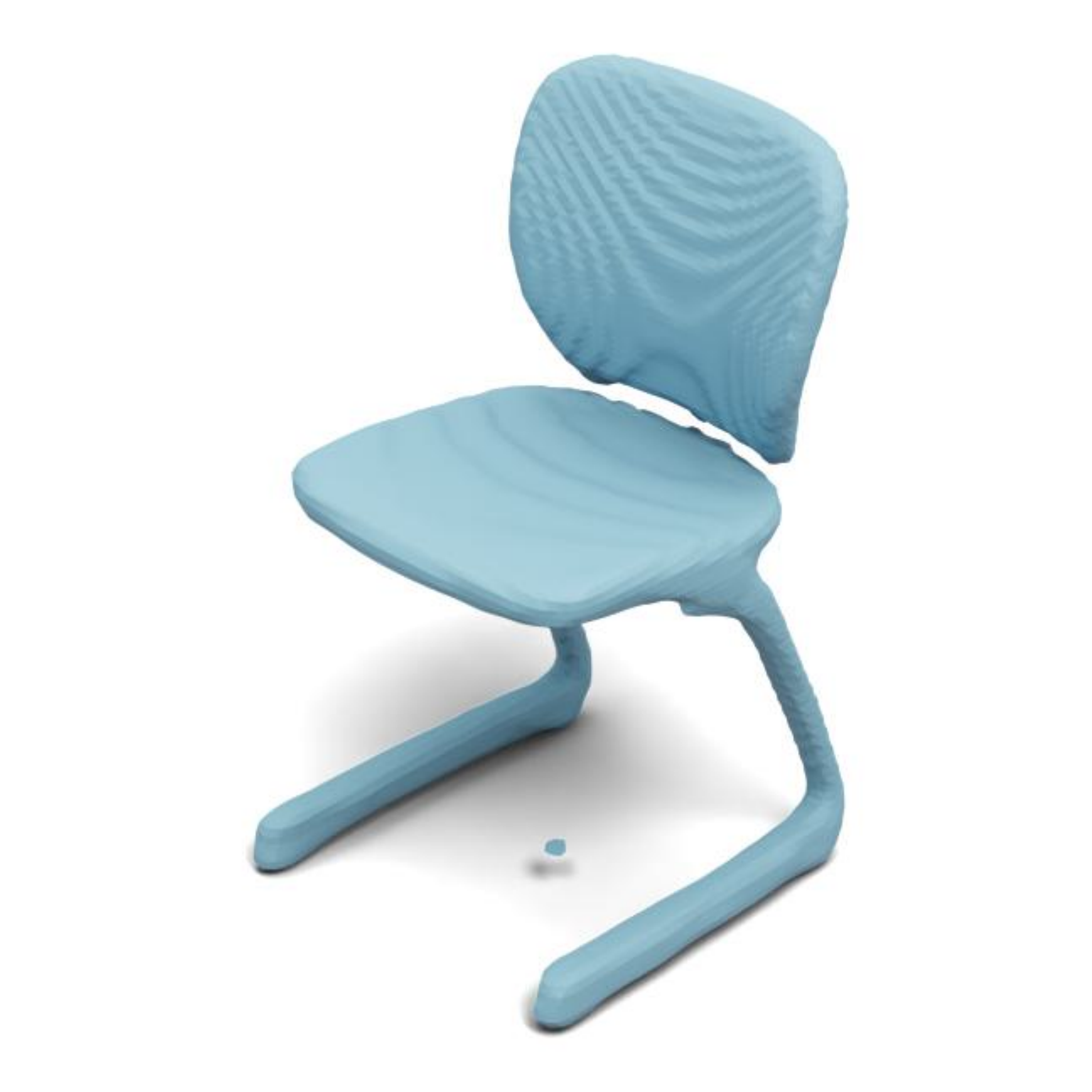}
&\includegraphics[trim = 1 1 1 1, clip, width=0.125\linewidth]{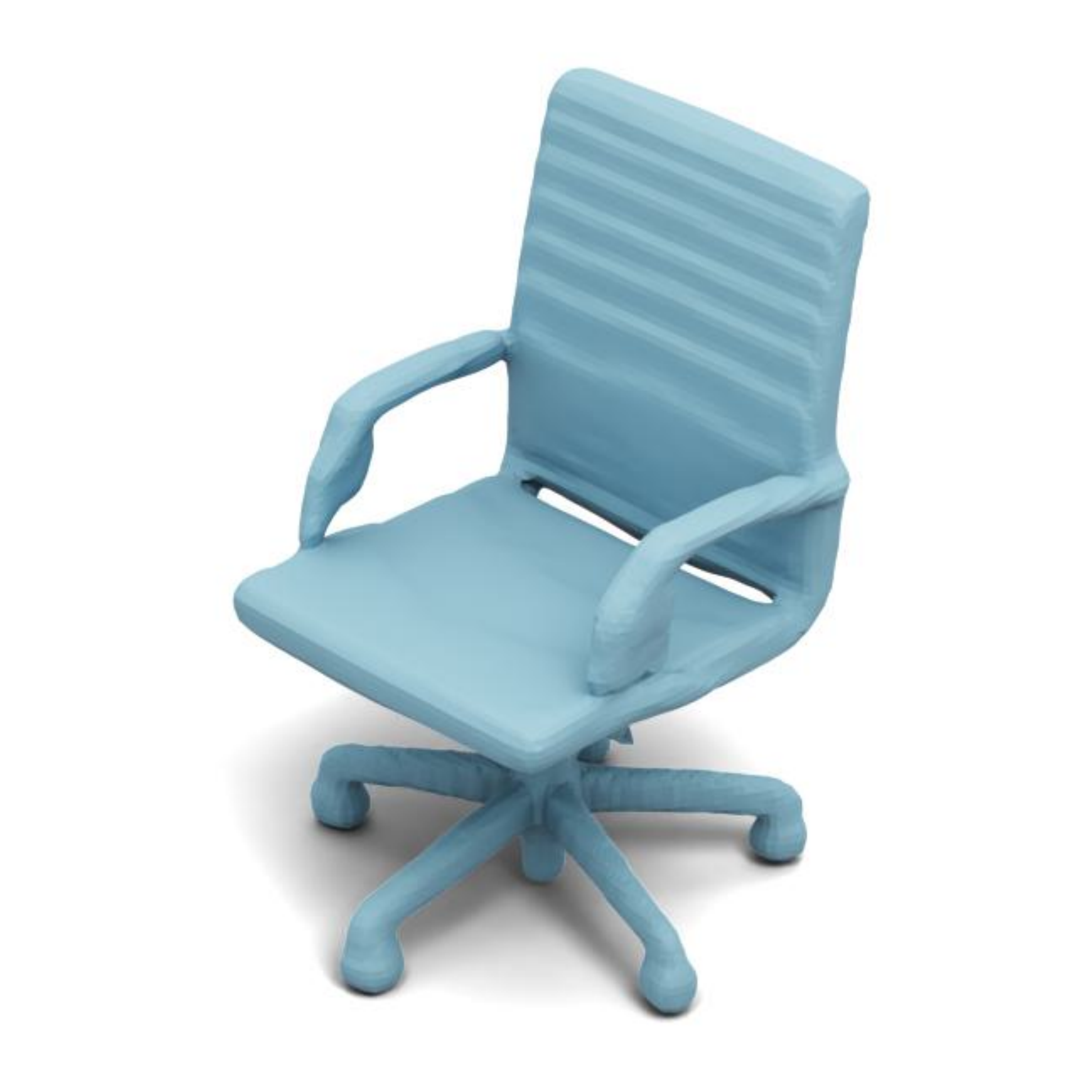}
&\includegraphics[trim = 1 1 1 1, clip, width=0.125\linewidth]{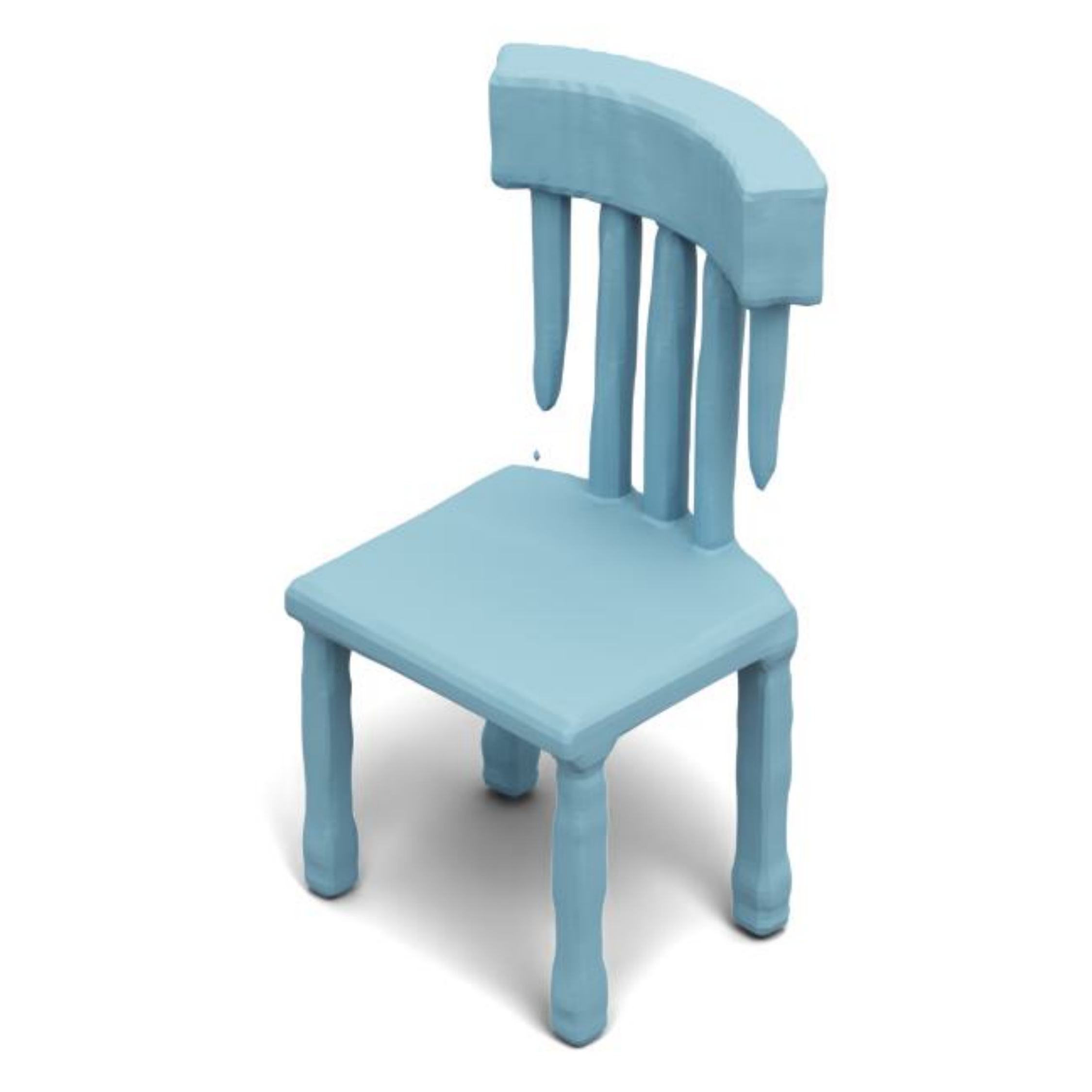}
&\includegraphics[trim = 1 1 1 1, clip, width=0.125\linewidth]{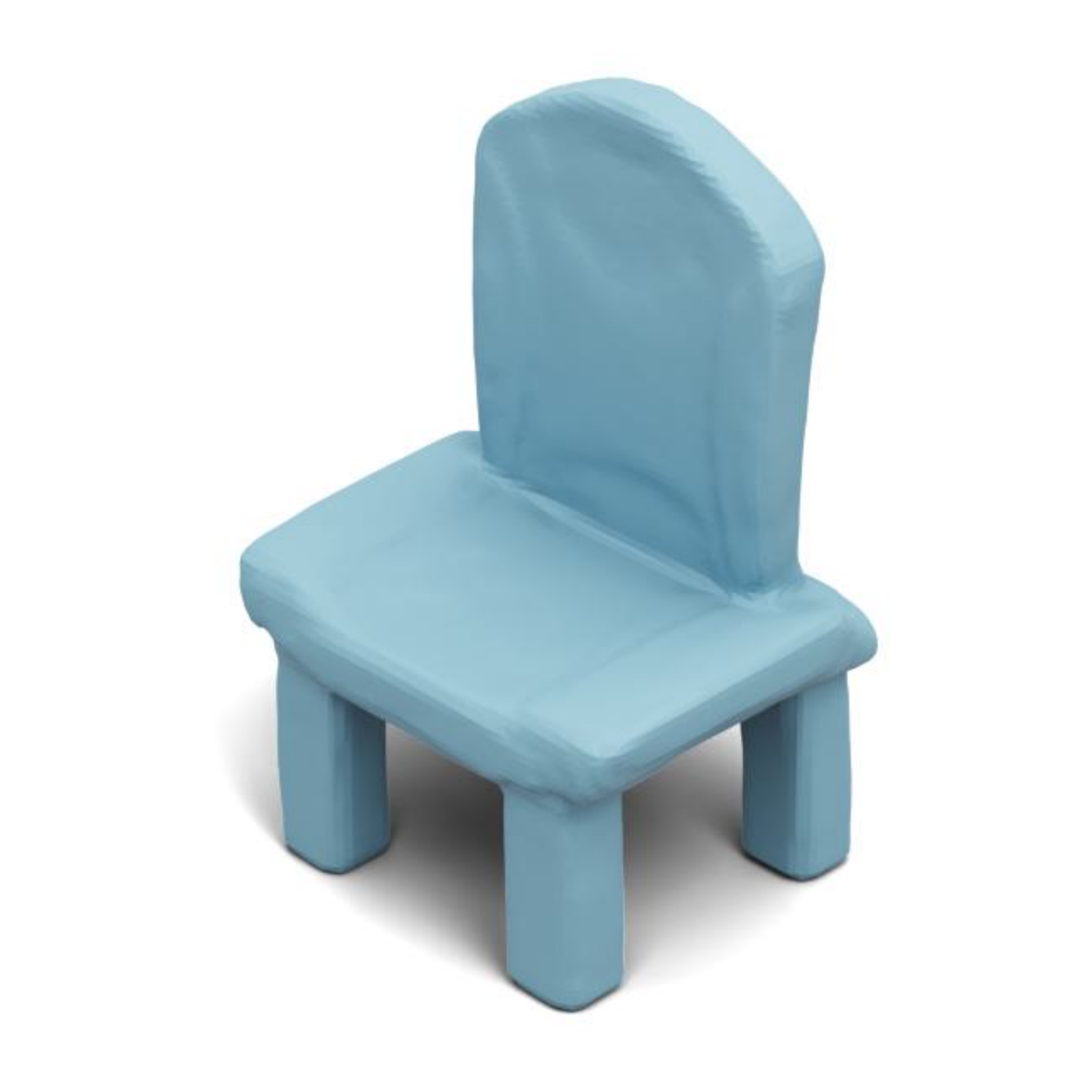}
&\includegraphics[trim = 1 1 1 1, clip, width=0.125\linewidth]{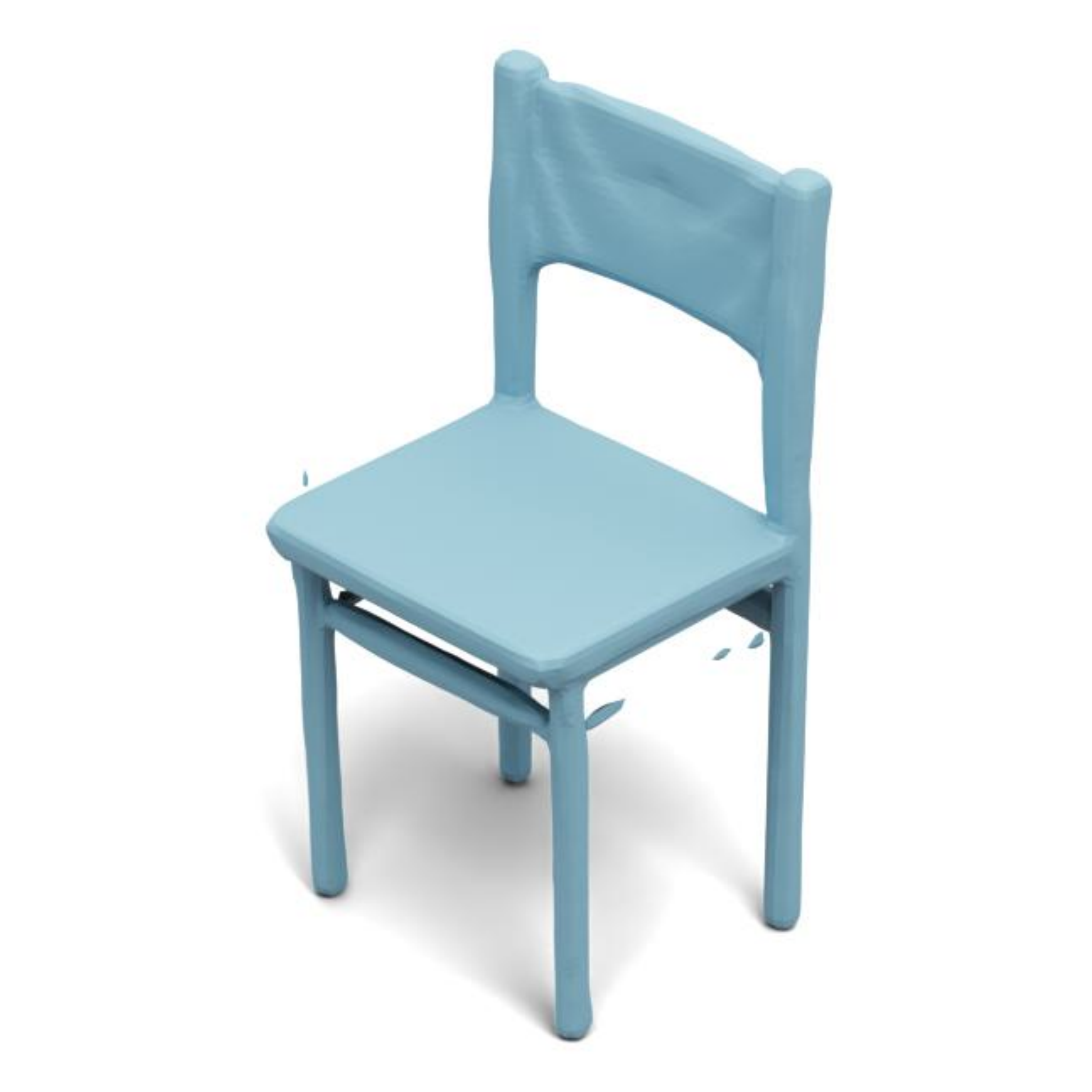}
\\
\includegraphics[width=0.125\linewidth]{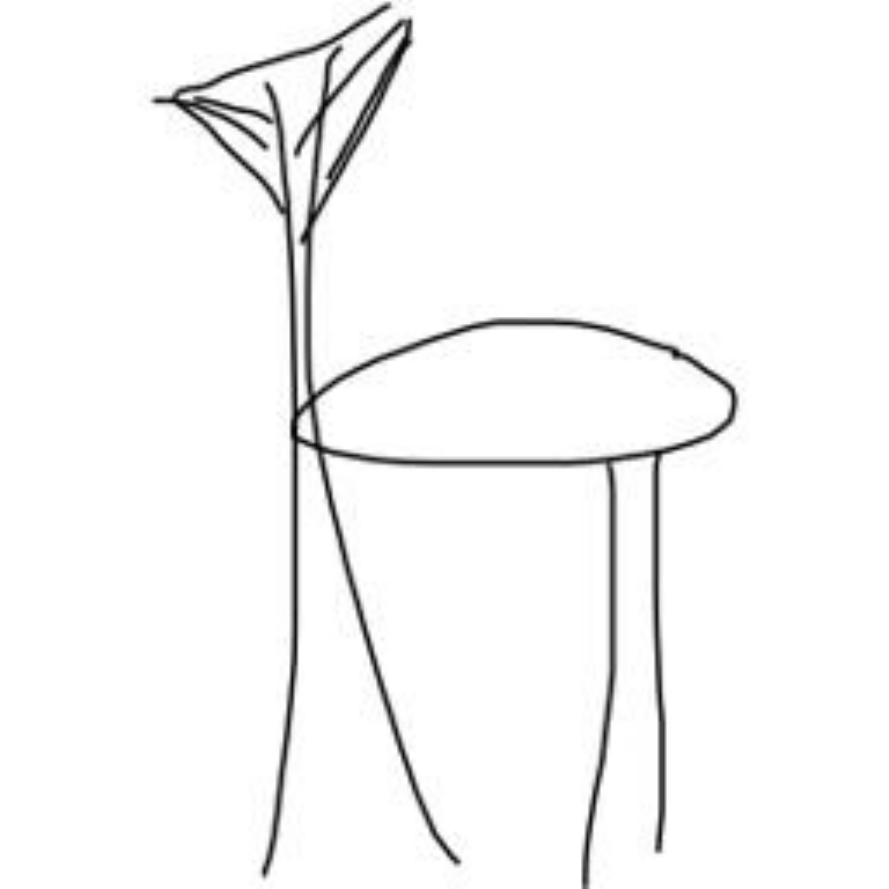}
&\includegraphics[width=0.125\linewidth]{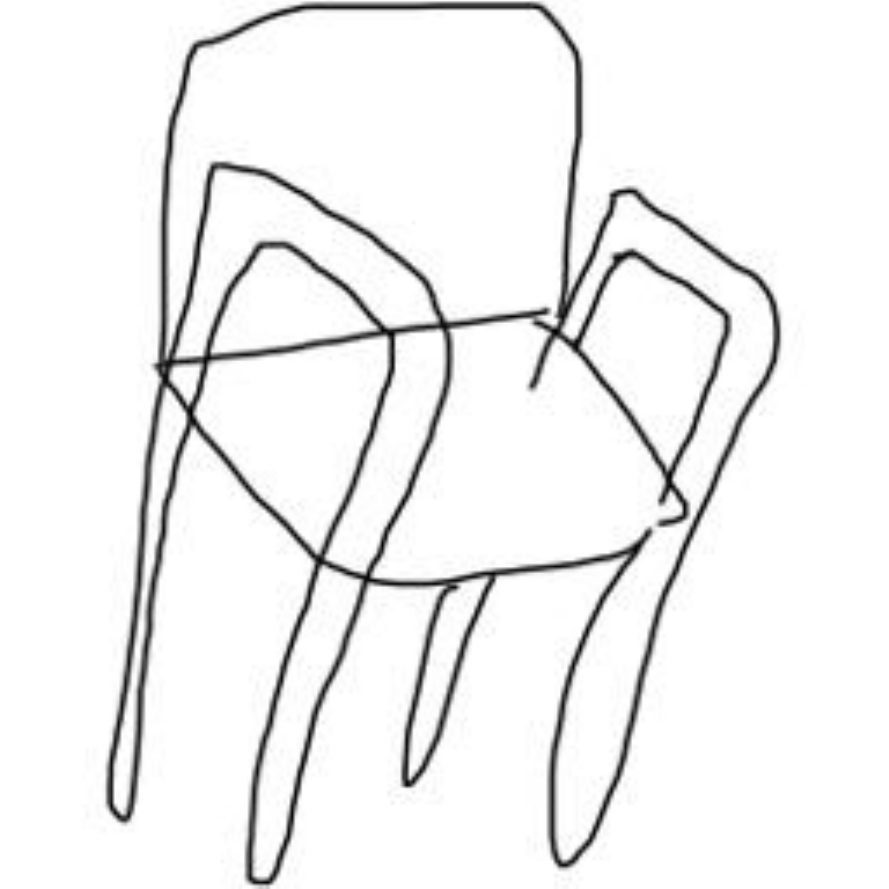}
&\includegraphics[width=0.125\linewidth]{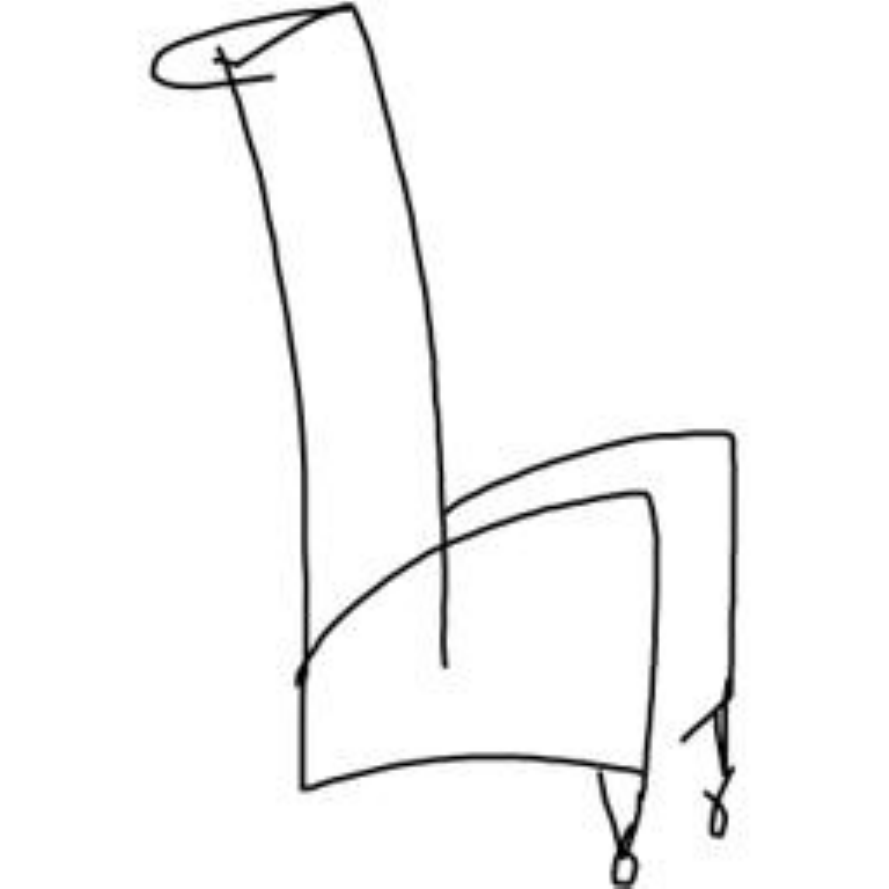}
&\includegraphics[width=0.125\linewidth]{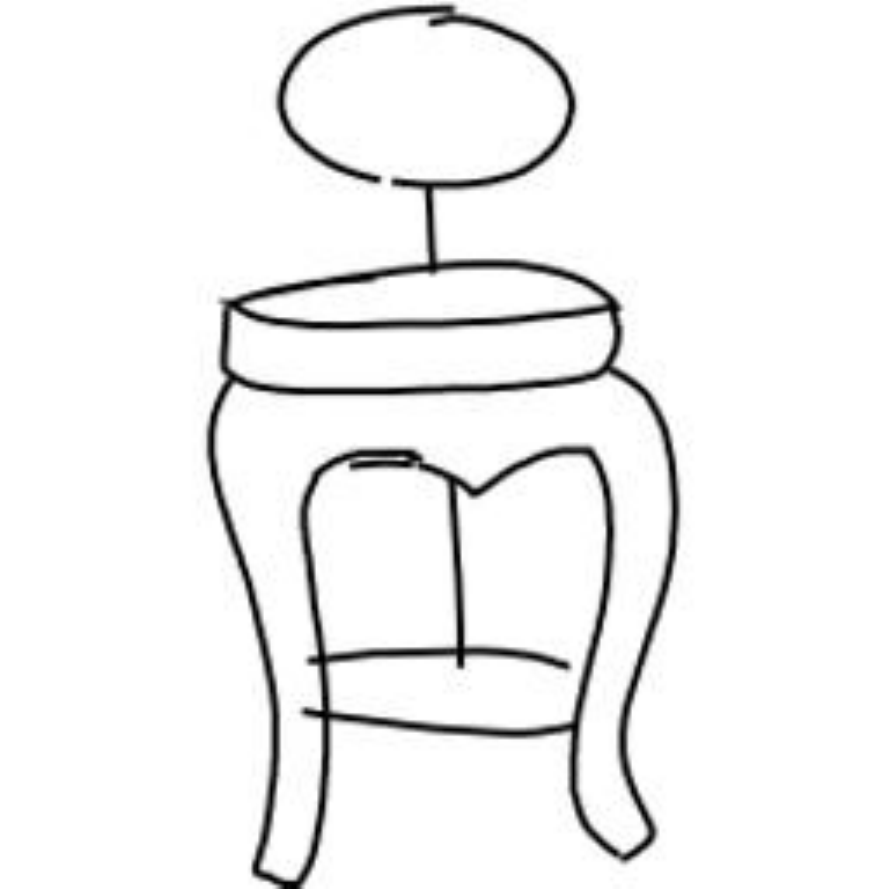}
&\includegraphics[width=0.125\linewidth]{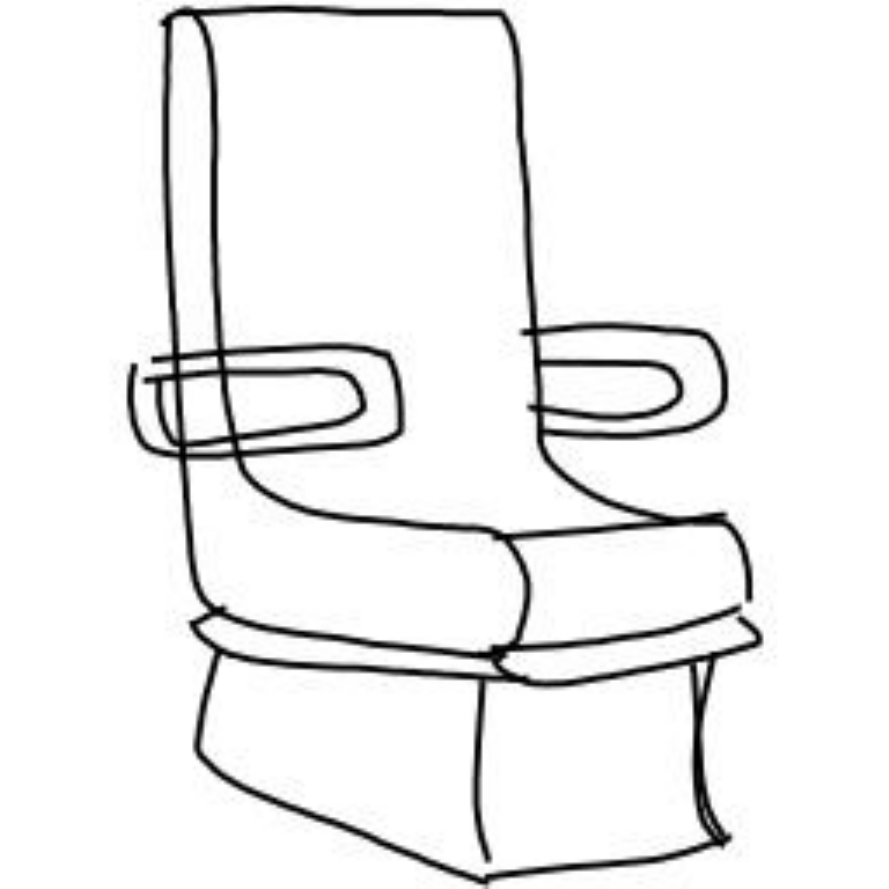}
&\includegraphics[width=0.125\linewidth]{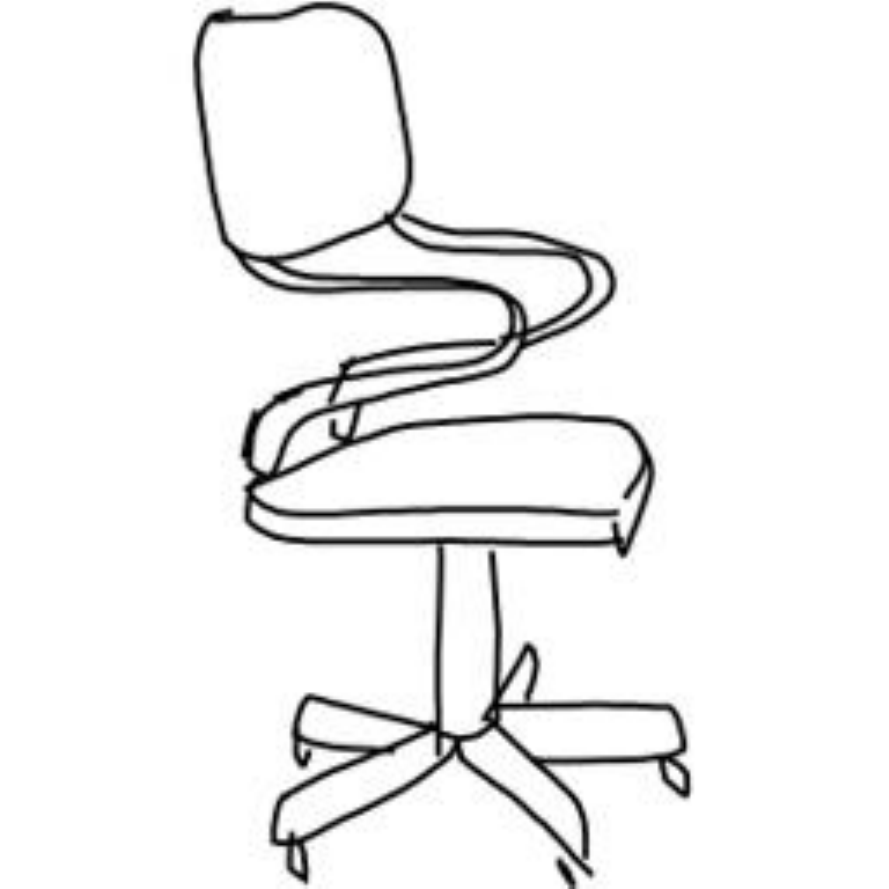}
&\includegraphics[width=0.125\linewidth]{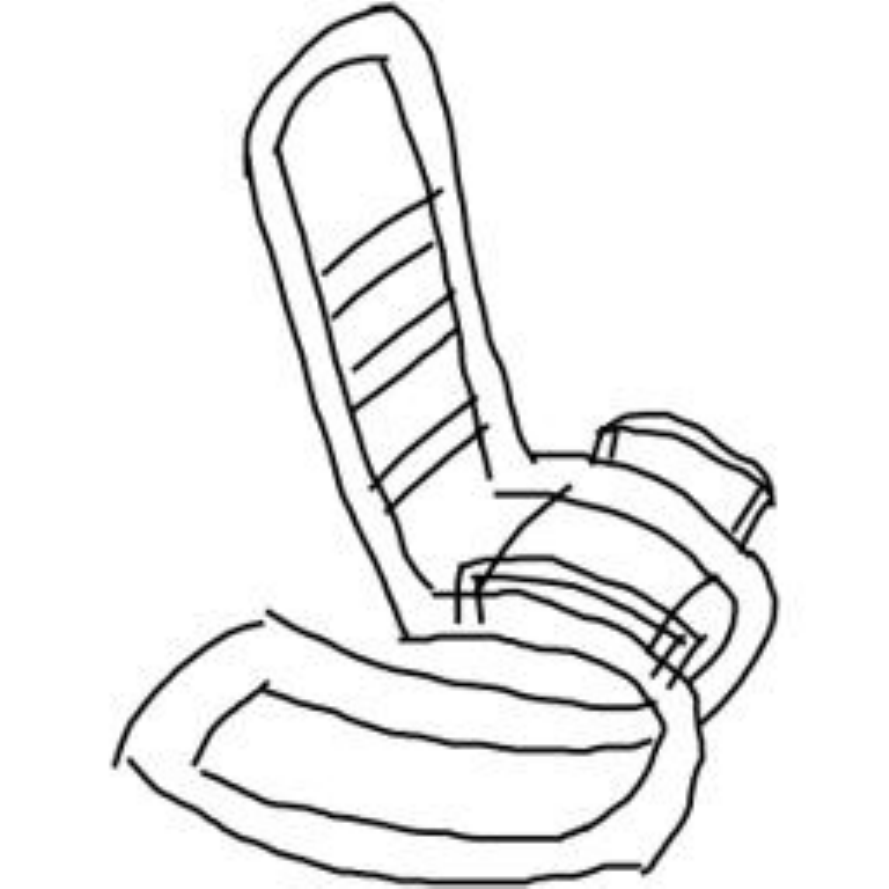}
&\includegraphics[width=0.125\linewidth]{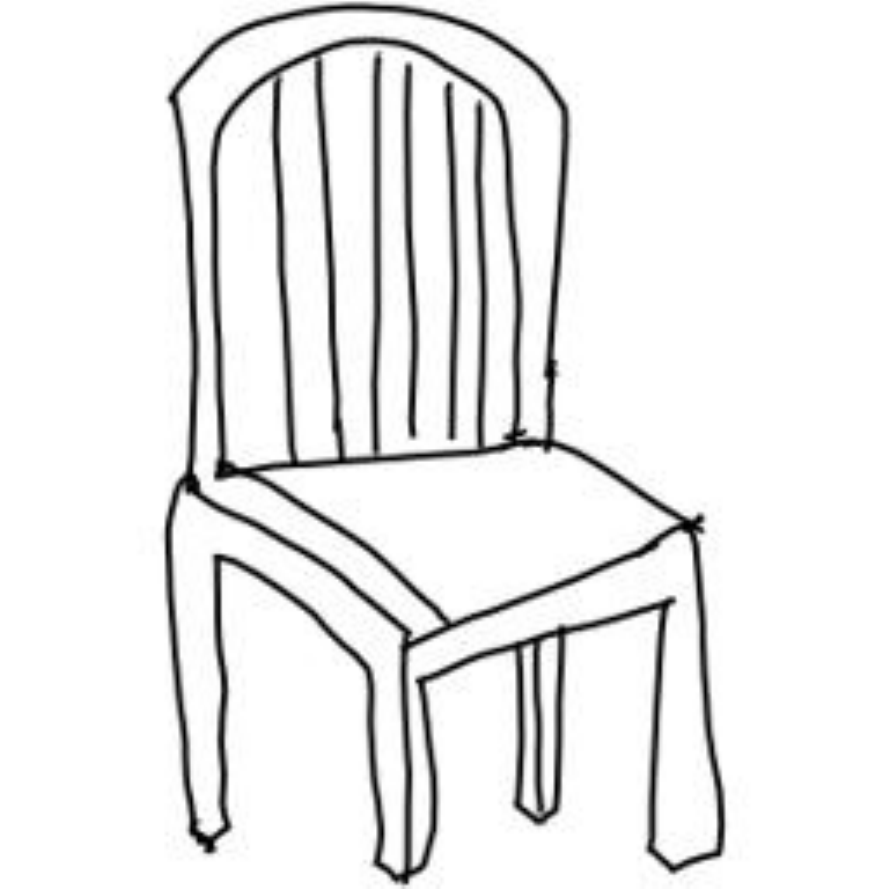}
\\
\includegraphics[trim = 1 1 1 1, clip, width=0.125\linewidth]{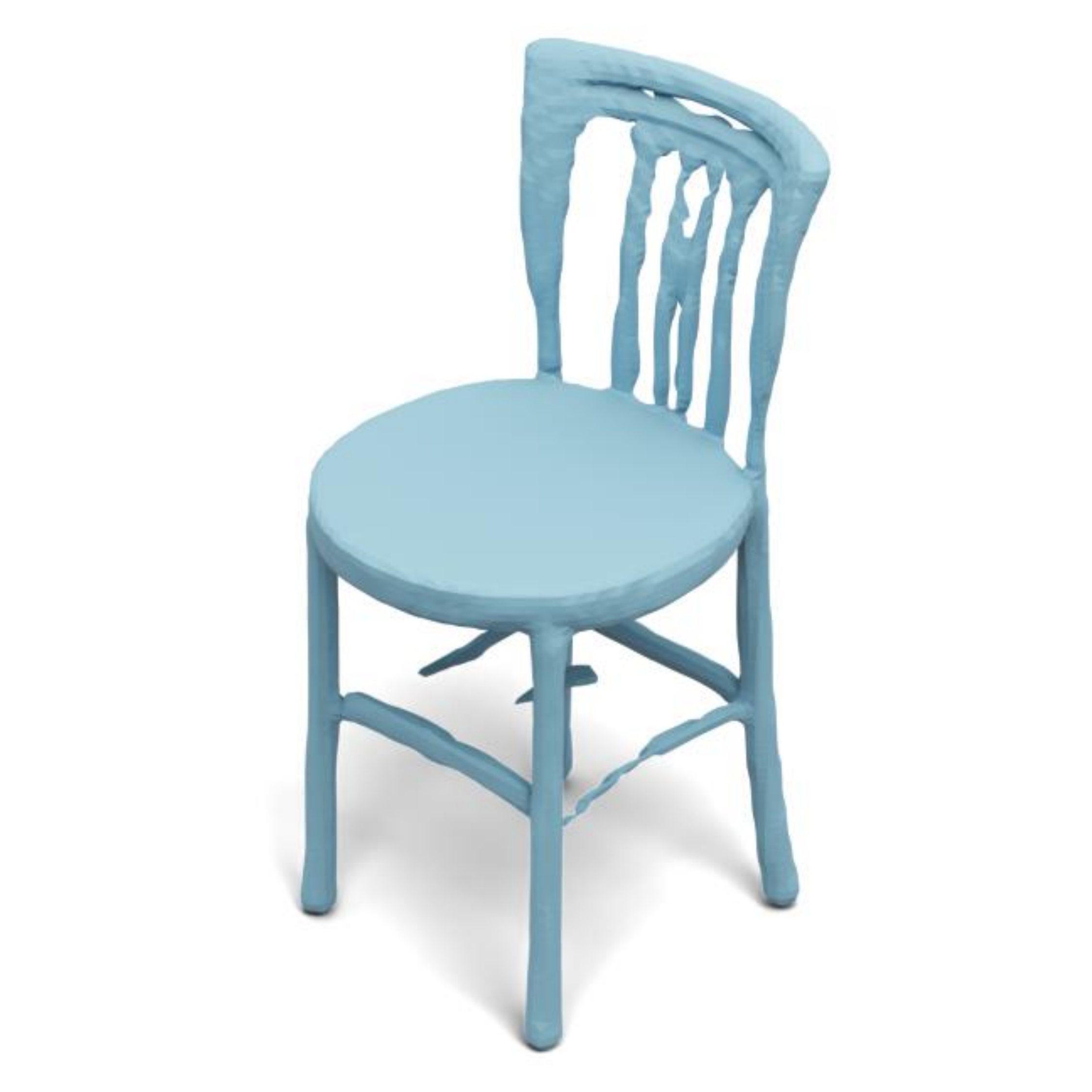}
&\includegraphics[trim = 1 1 1 1, clip, width=0.125\linewidth]{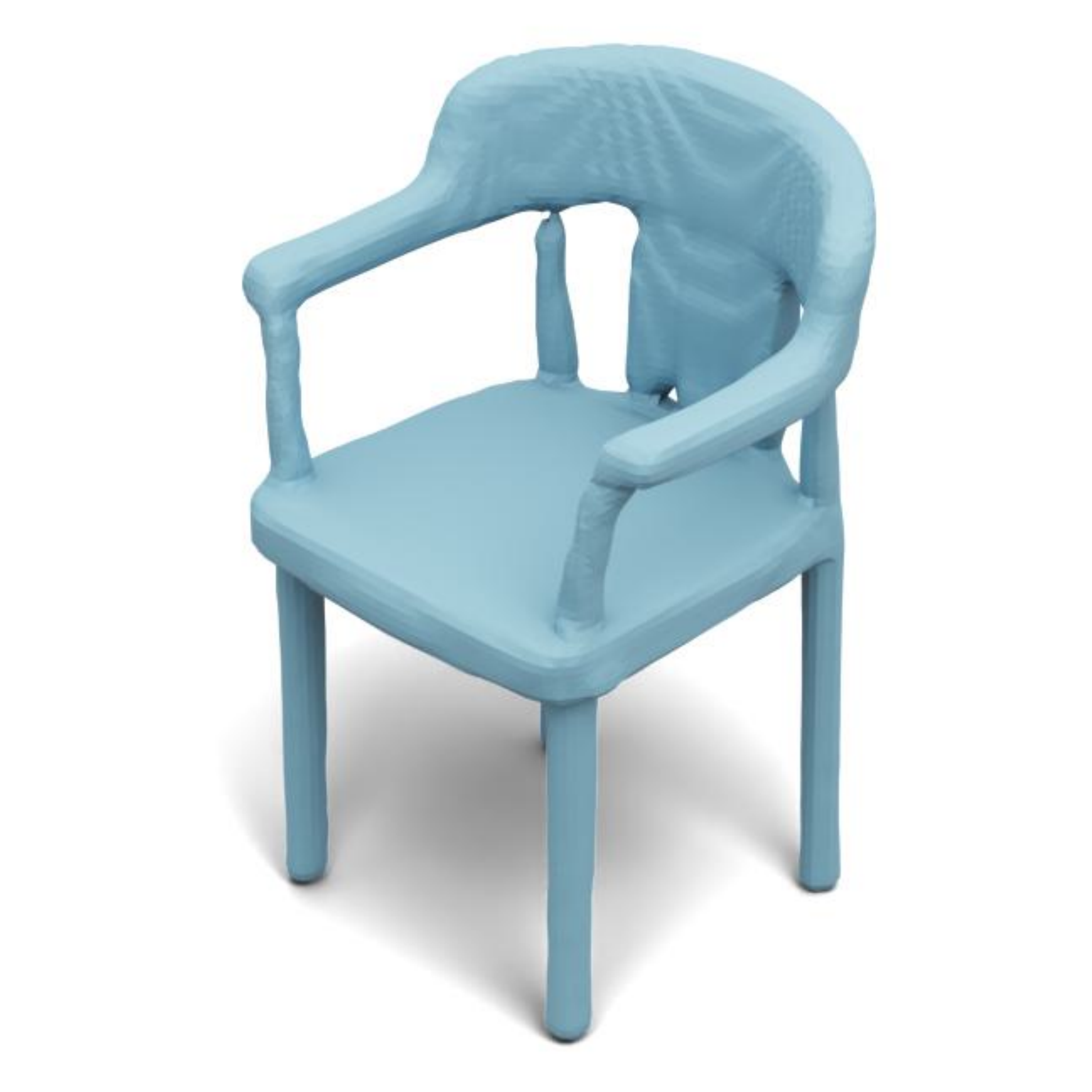}
&\includegraphics[trim = 1 1 1 1, clip, width=0.125\linewidth]{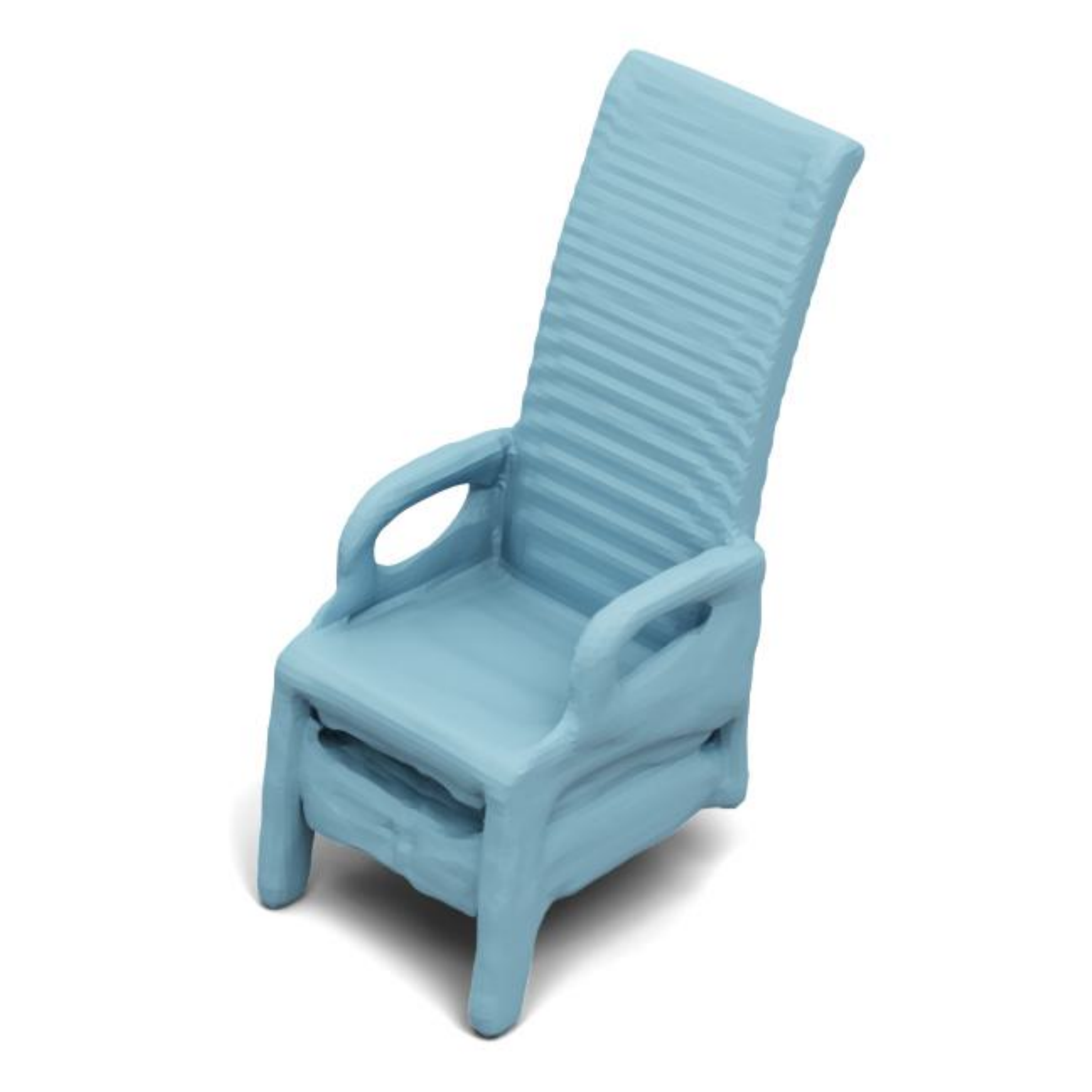}
&\includegraphics[trim = 1 1 1 1, clip, width=0.125\linewidth]{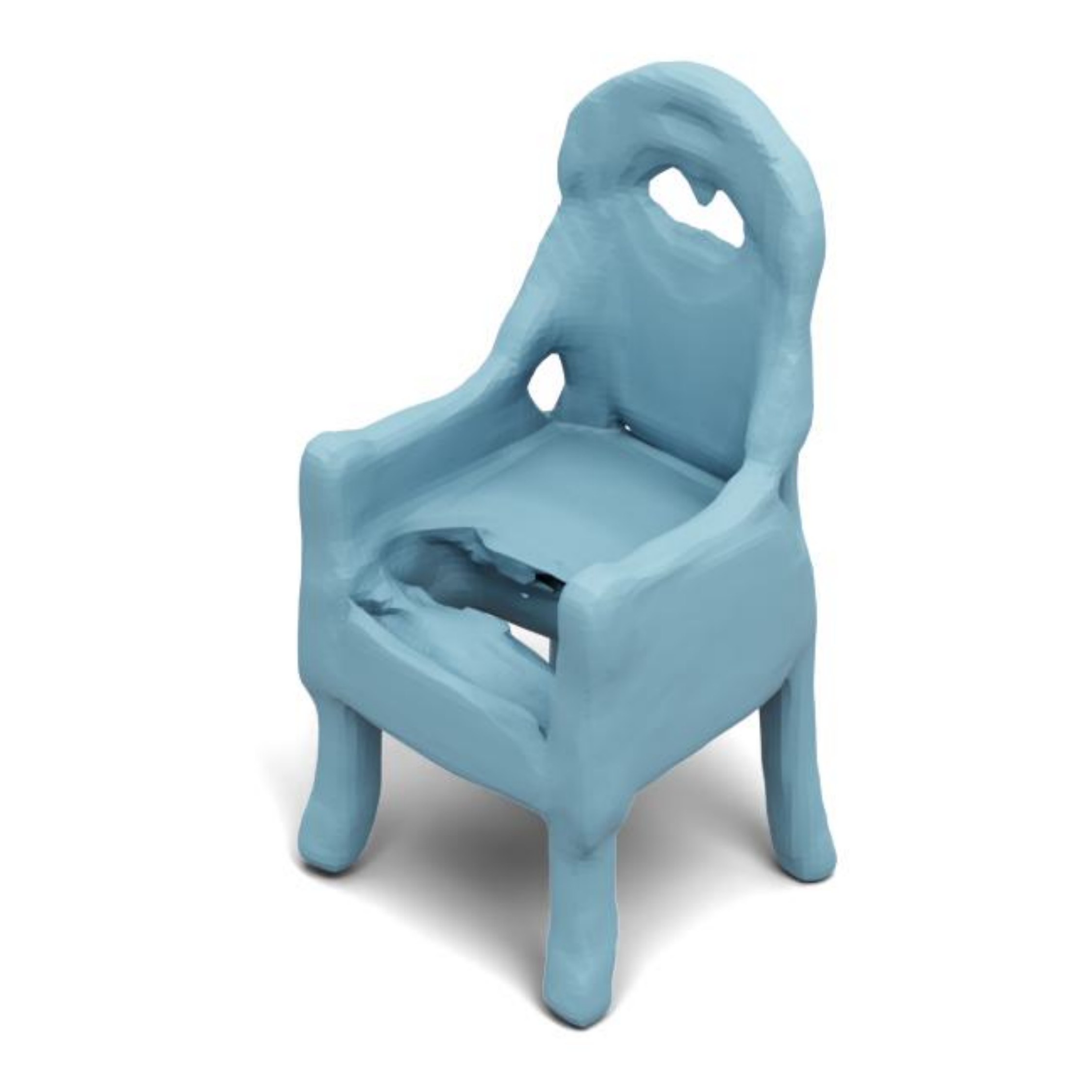}
&\includegraphics[trim = 1 1 1 1, clip, width=0.125\linewidth]{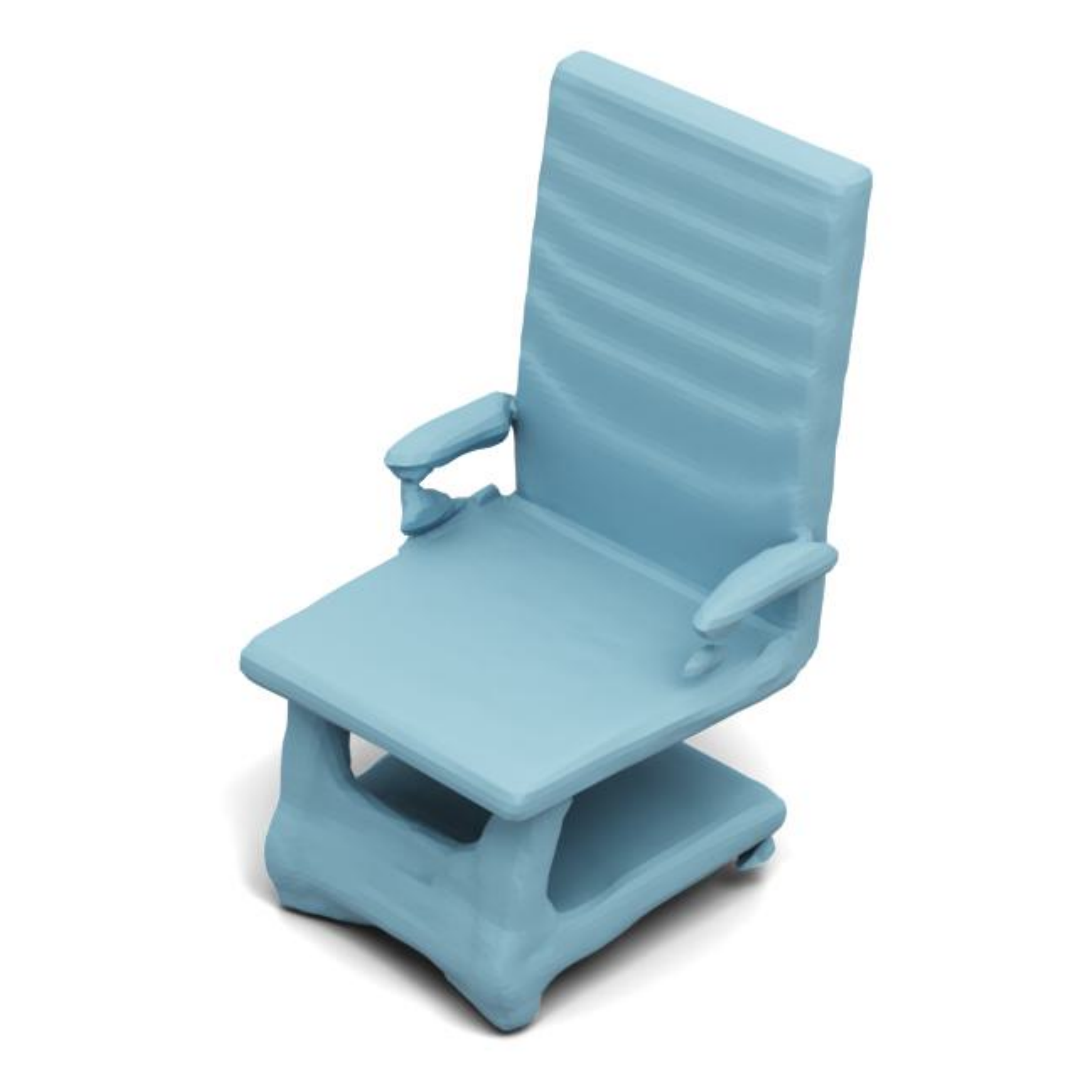}
&\includegraphics[trim = 1 1 1 1, clip, width=0.125\linewidth]{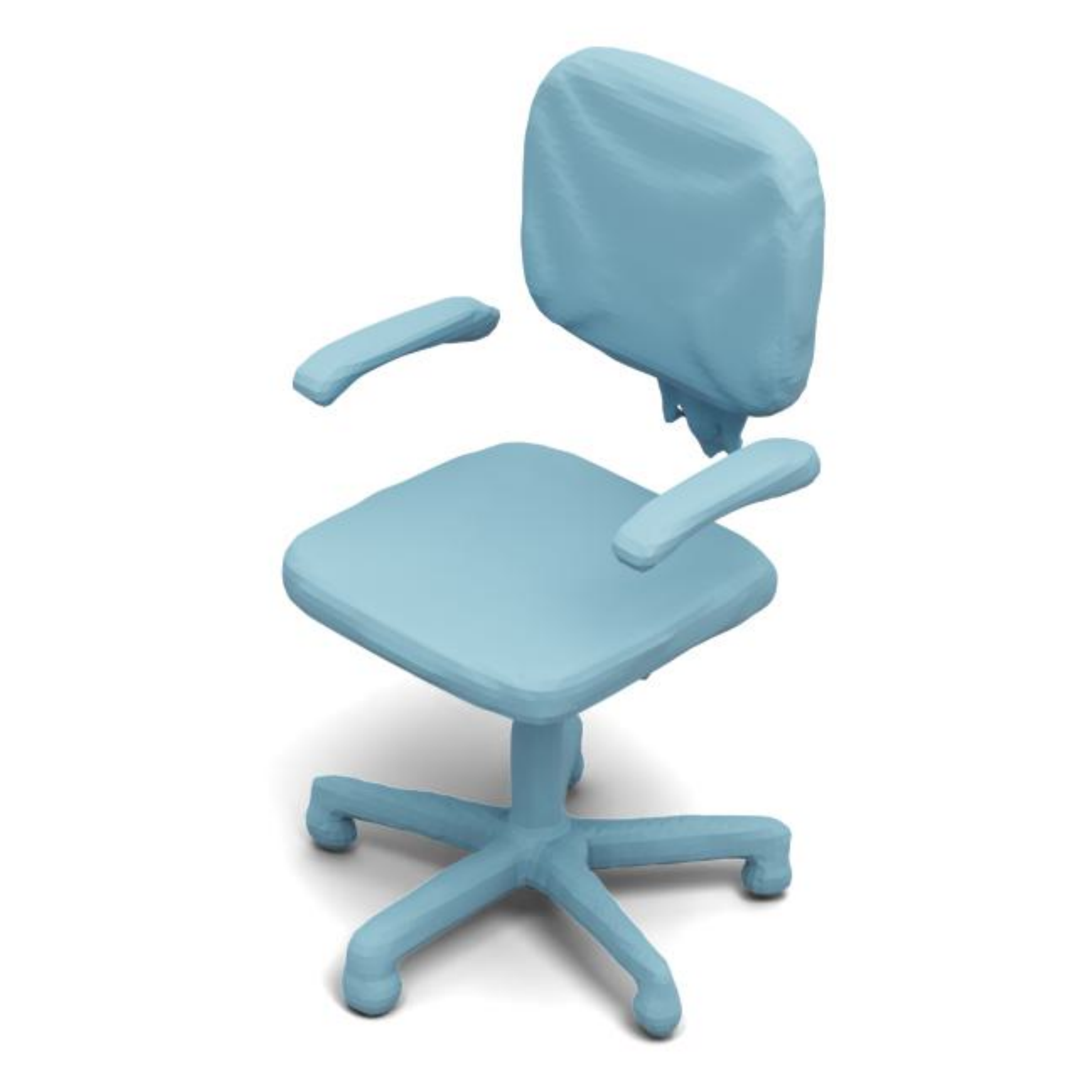}
&\includegraphics[trim = 1 1 1 1, clip, width=0.125\linewidth]{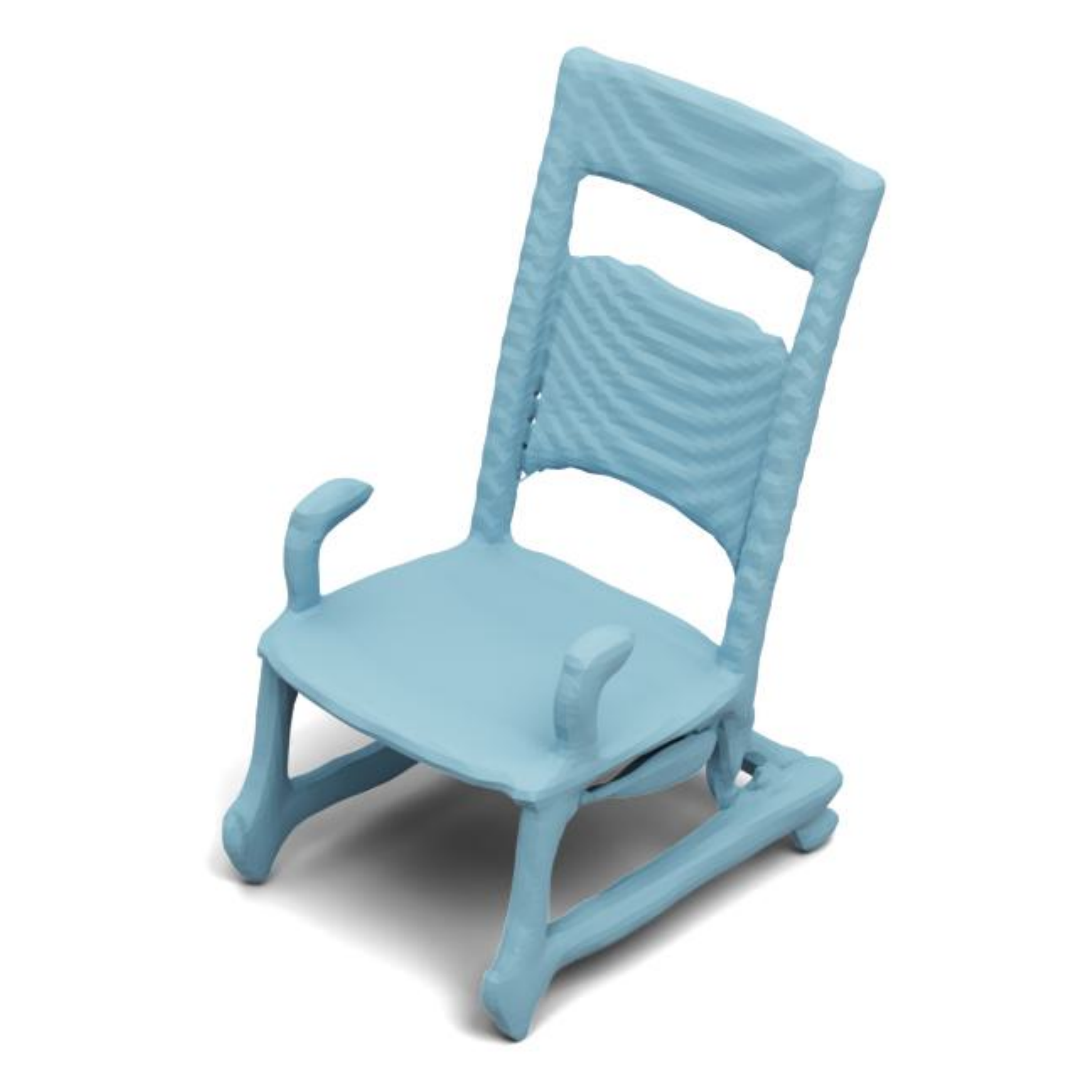}
&\includegraphics[trim = 1 1 1 1, clip, width=0.125\linewidth]{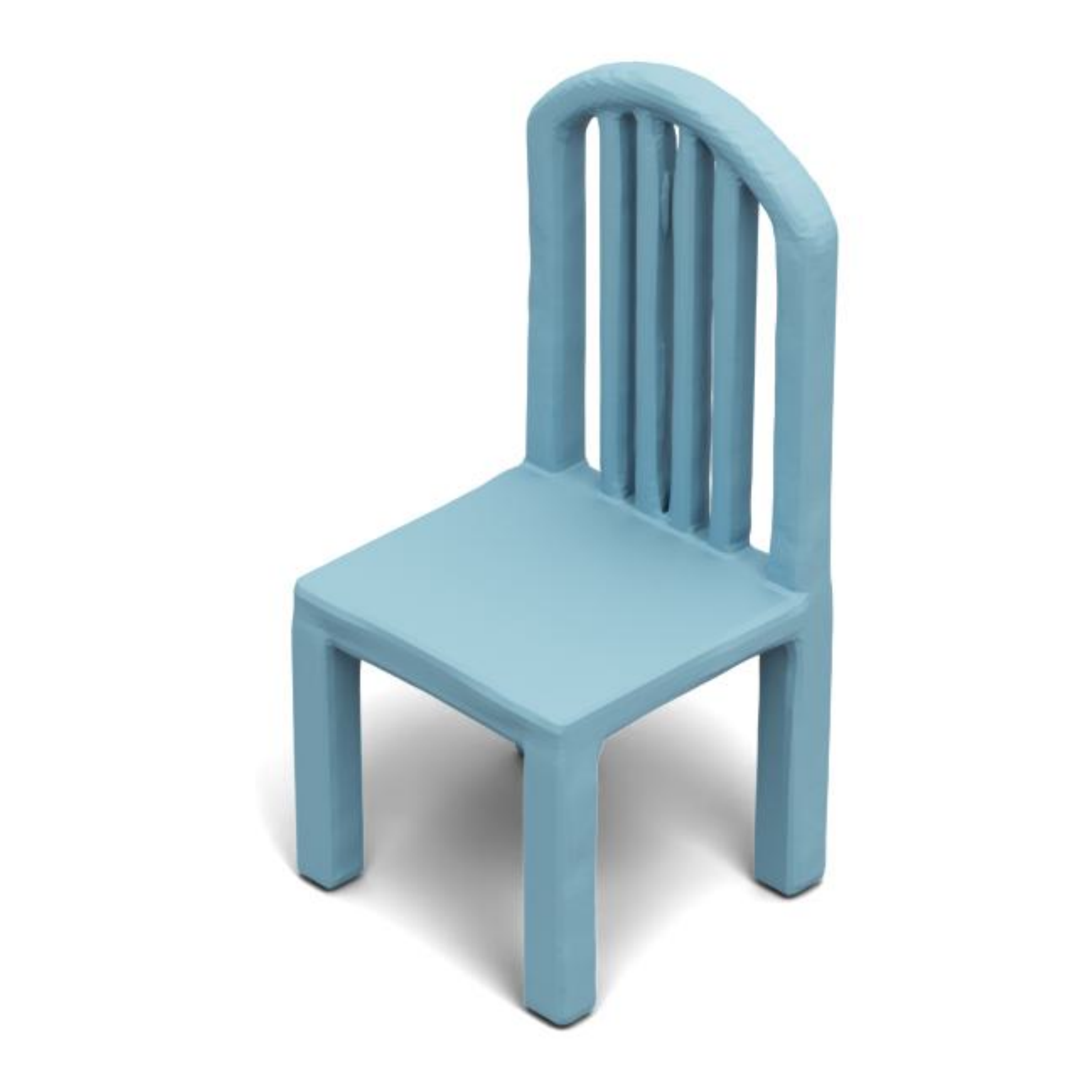}
\\
\includegraphics[width=0.125\linewidth]{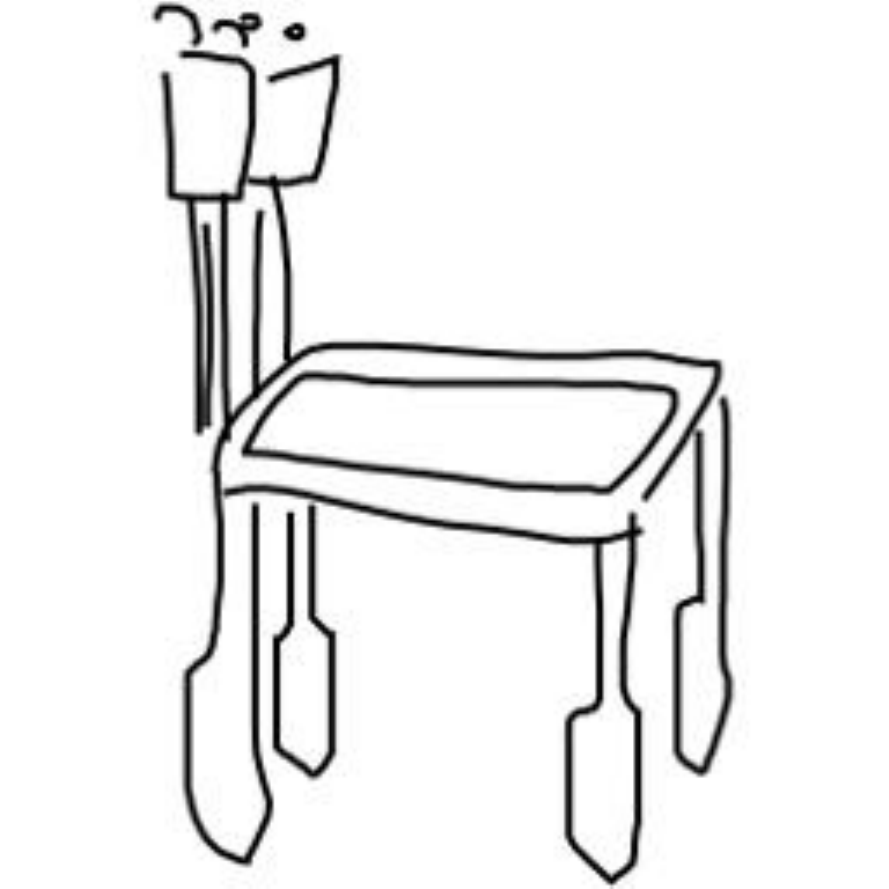}
&\includegraphics[width=0.125\linewidth]{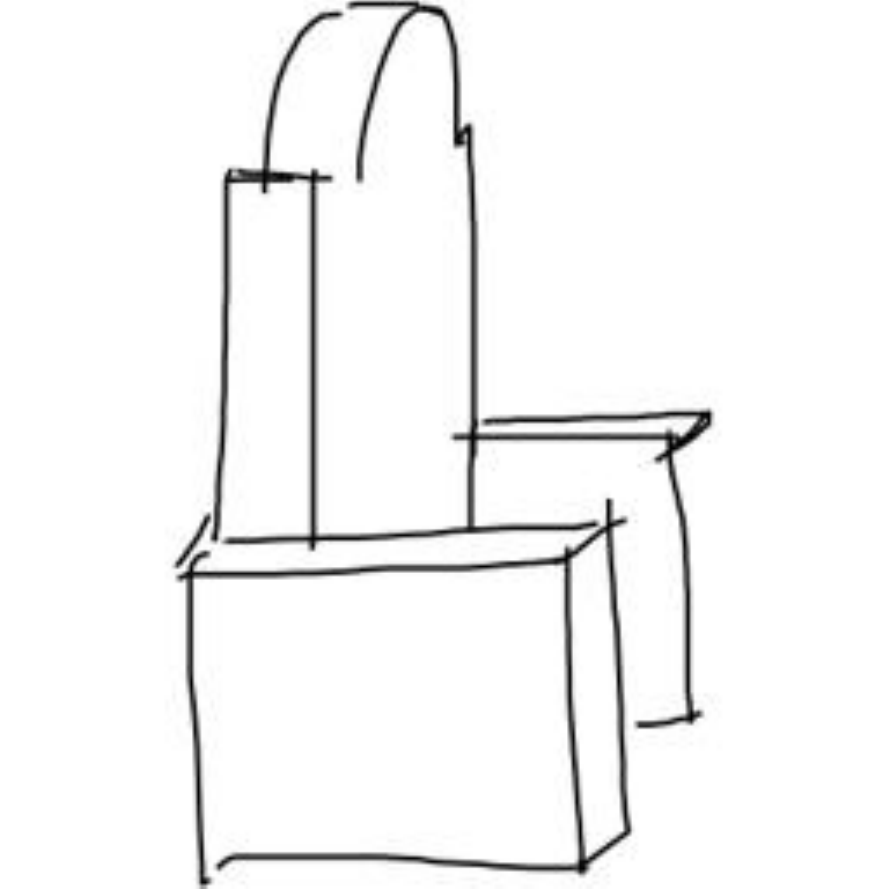}
&\includegraphics[width=0.125\linewidth]{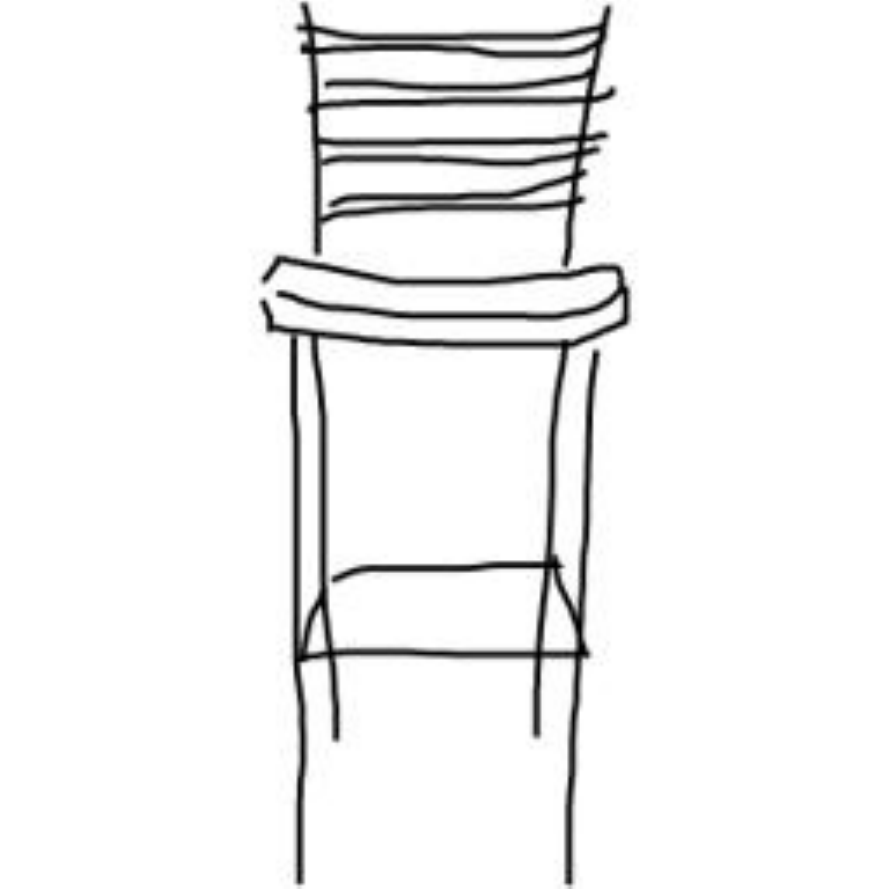}
&\includegraphics[width=0.125\linewidth]{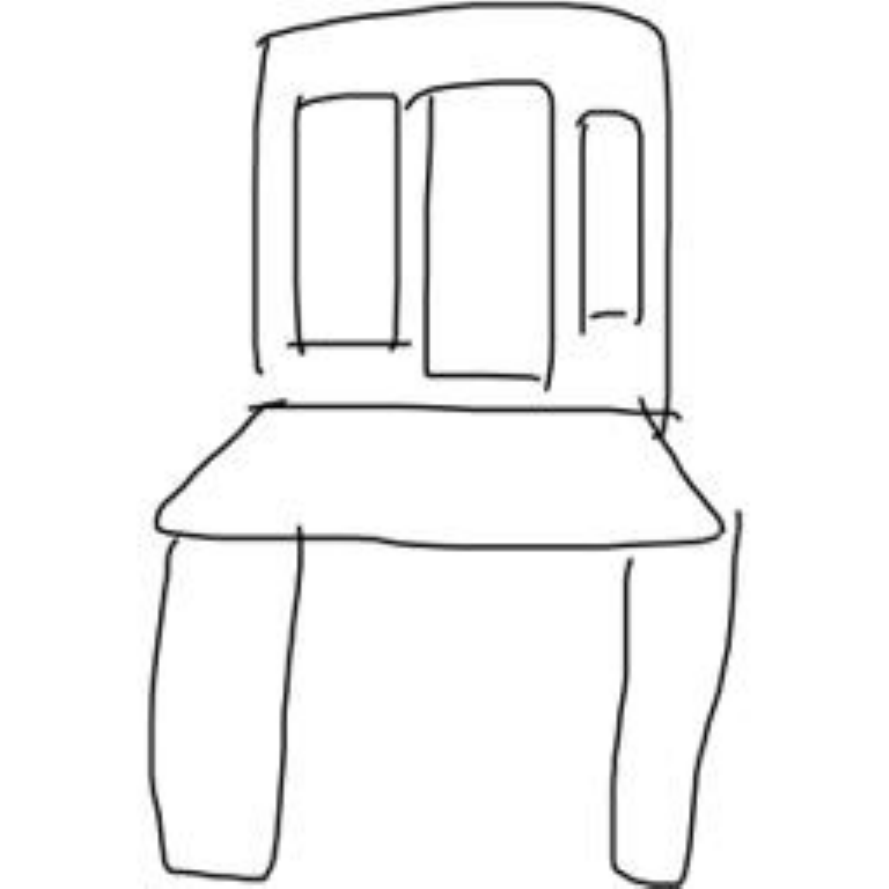}
&\includegraphics[width=0.125\linewidth]{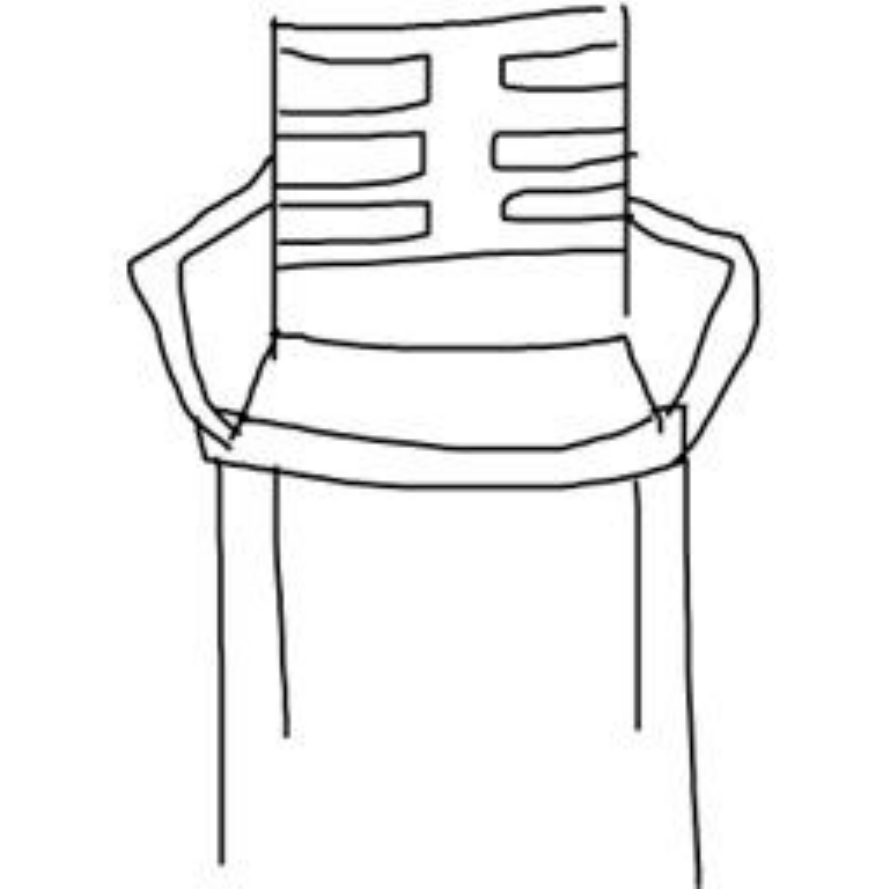}
&\includegraphics[width=0.125\linewidth]{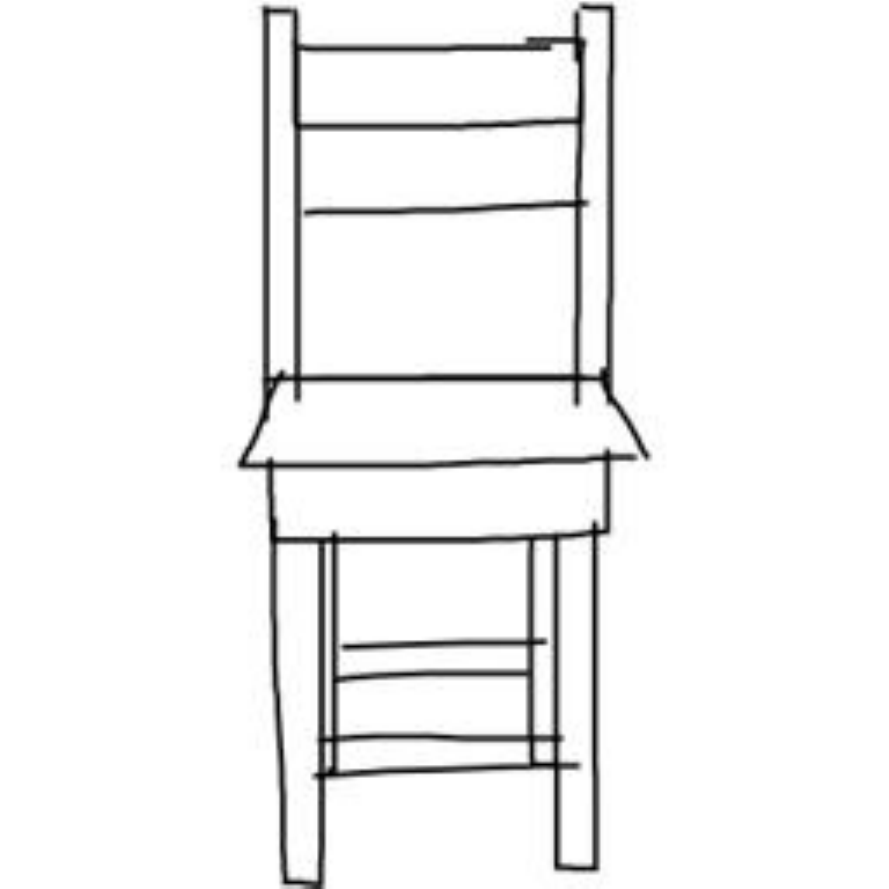}
&\includegraphics[width=0.125\linewidth]{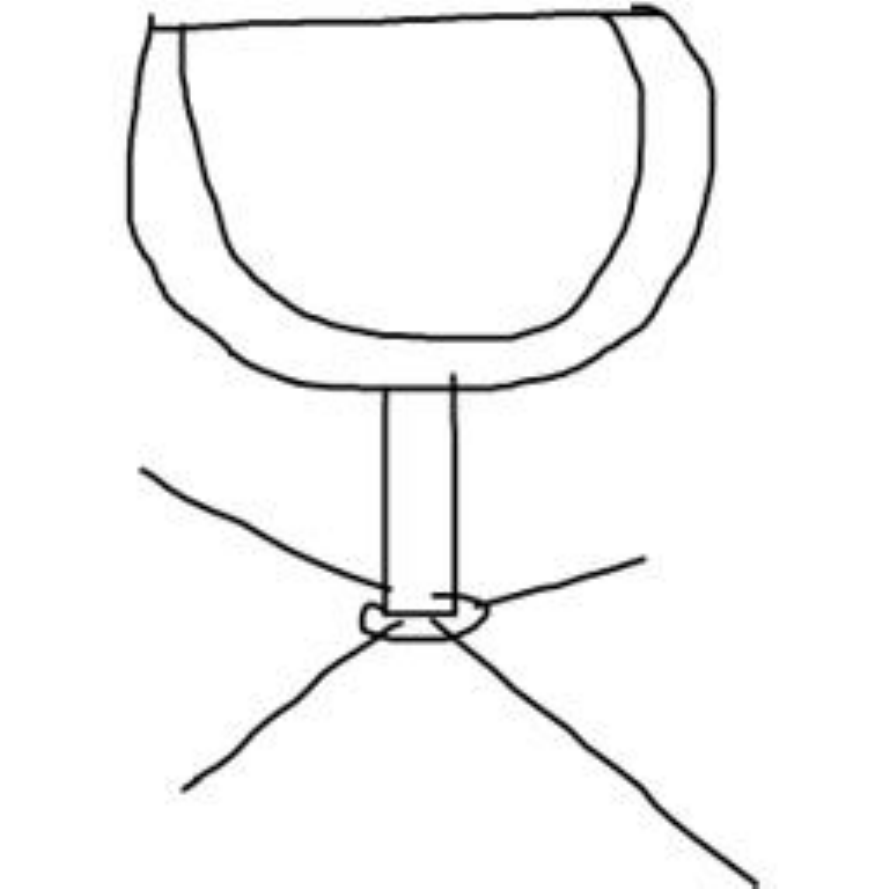}
&\includegraphics[width=0.125\linewidth]{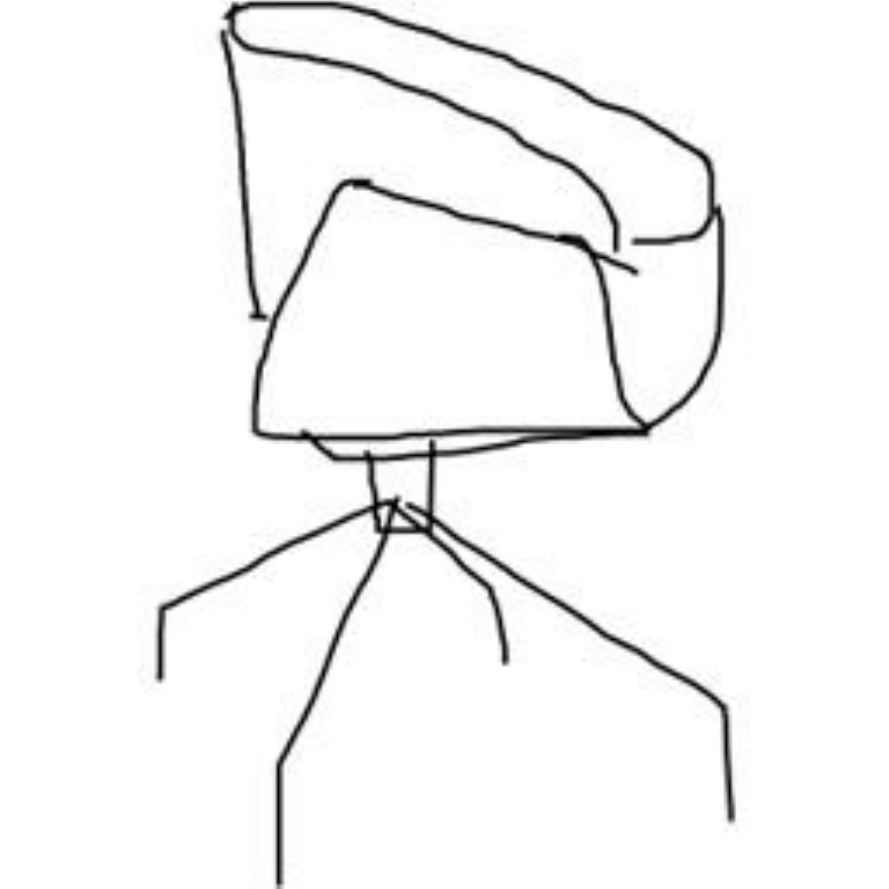}
\\
\includegraphics[trim = 1 1 1 1, clip, width=0.125\linewidth]{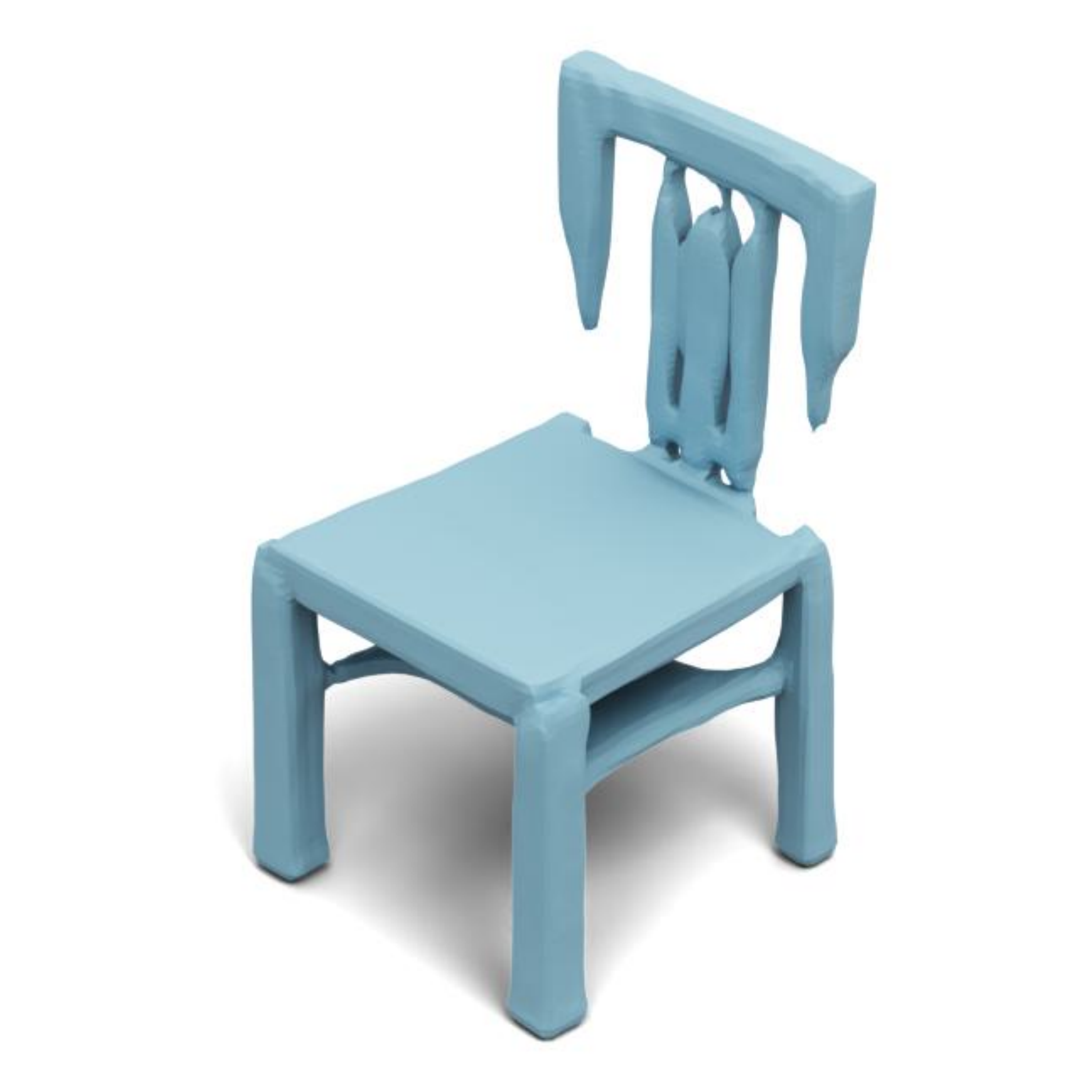}
&\includegraphics[trim = 1 1 1 1, clip, width=0.125\linewidth]{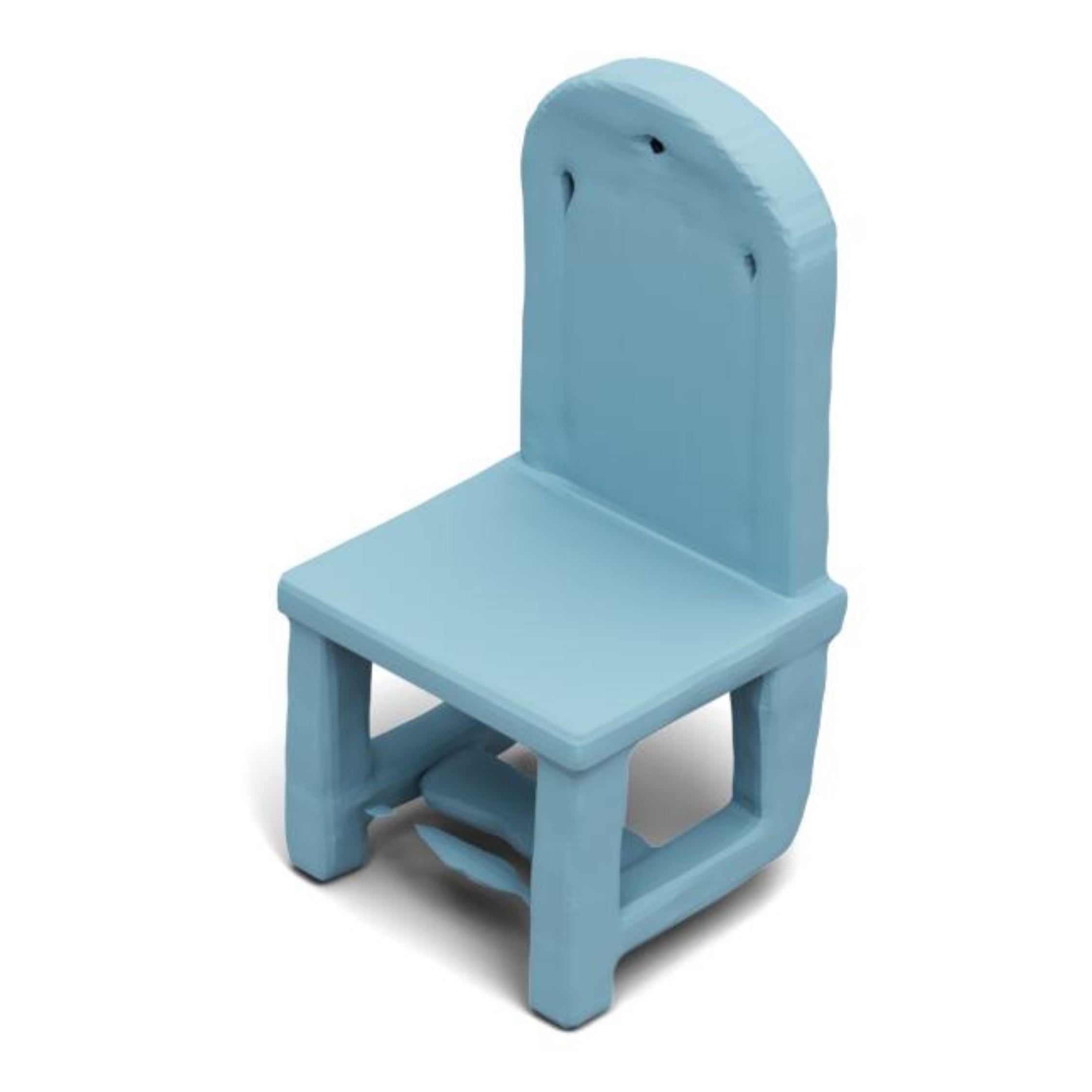}
&\includegraphics[trim = 1 1 1 1, clip, width=0.125\linewidth]{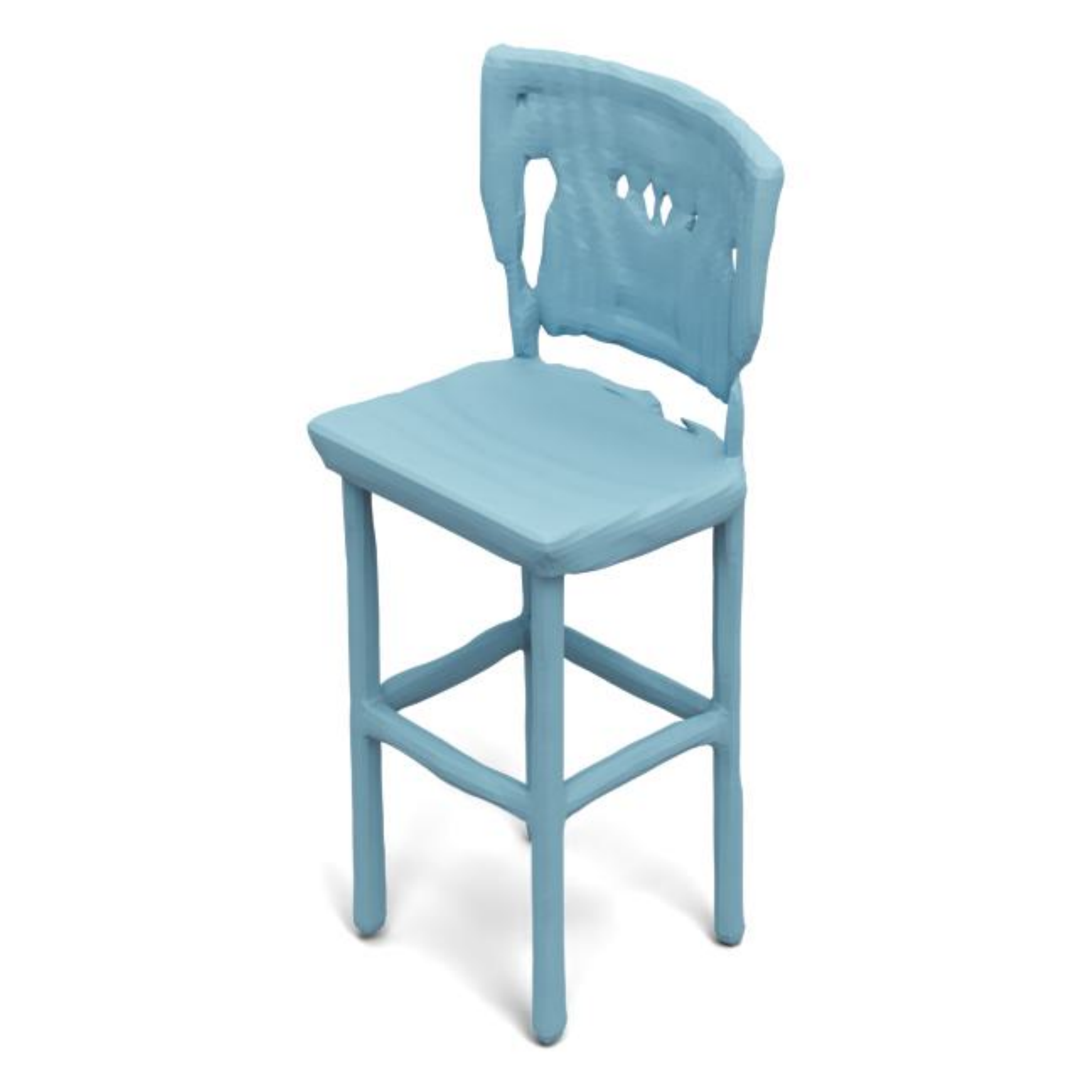}
&\includegraphics[trim = 1 1 1 1, clip, width=0.125\linewidth]{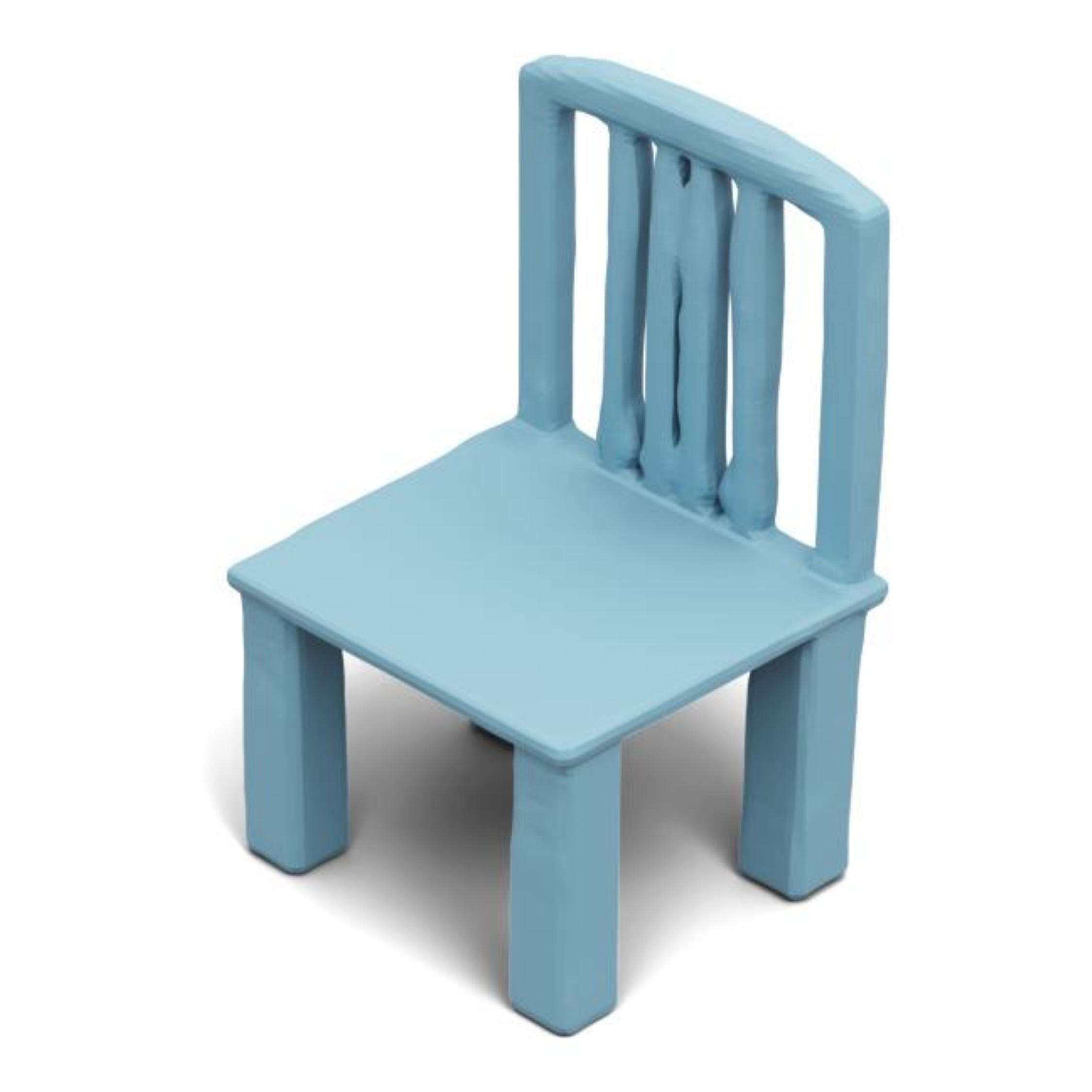}
&\includegraphics[trim = 1 1 1 1, clip, width=0.125\linewidth]{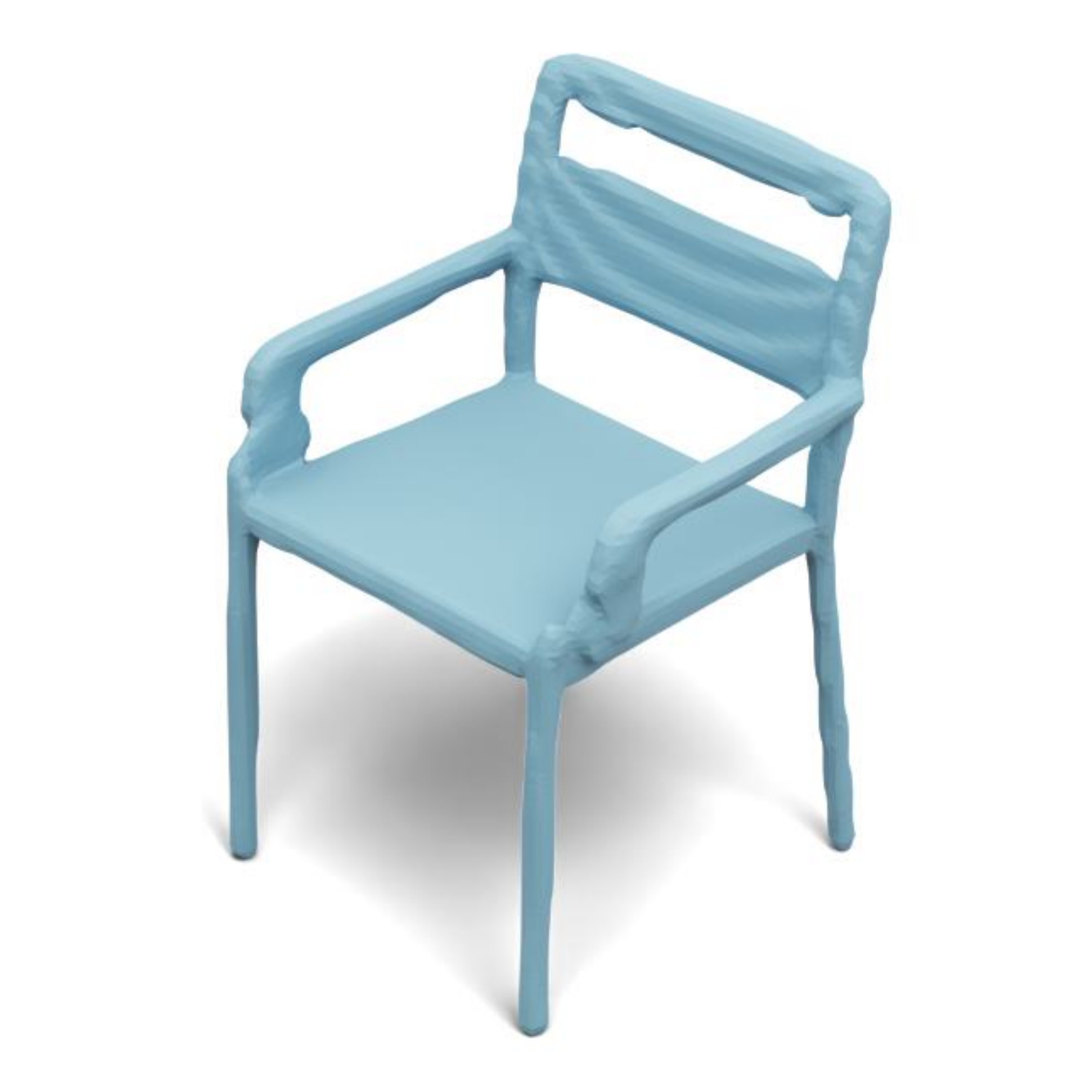}
&\includegraphics[trim = 1 1 1 1, clip, width=0.125\linewidth]{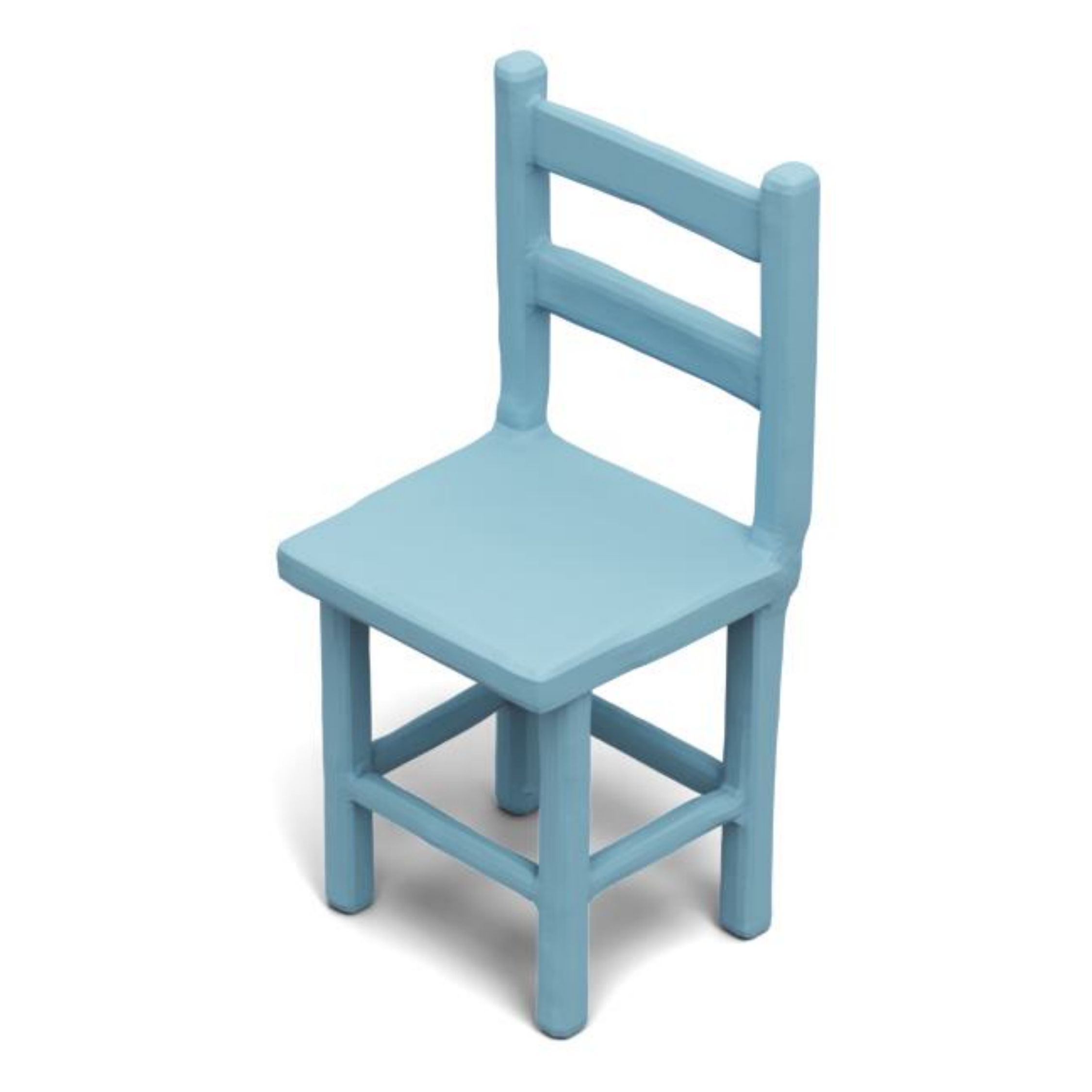}
&\includegraphics[trim = 1 1 1 1, clip, width=0.125\linewidth]{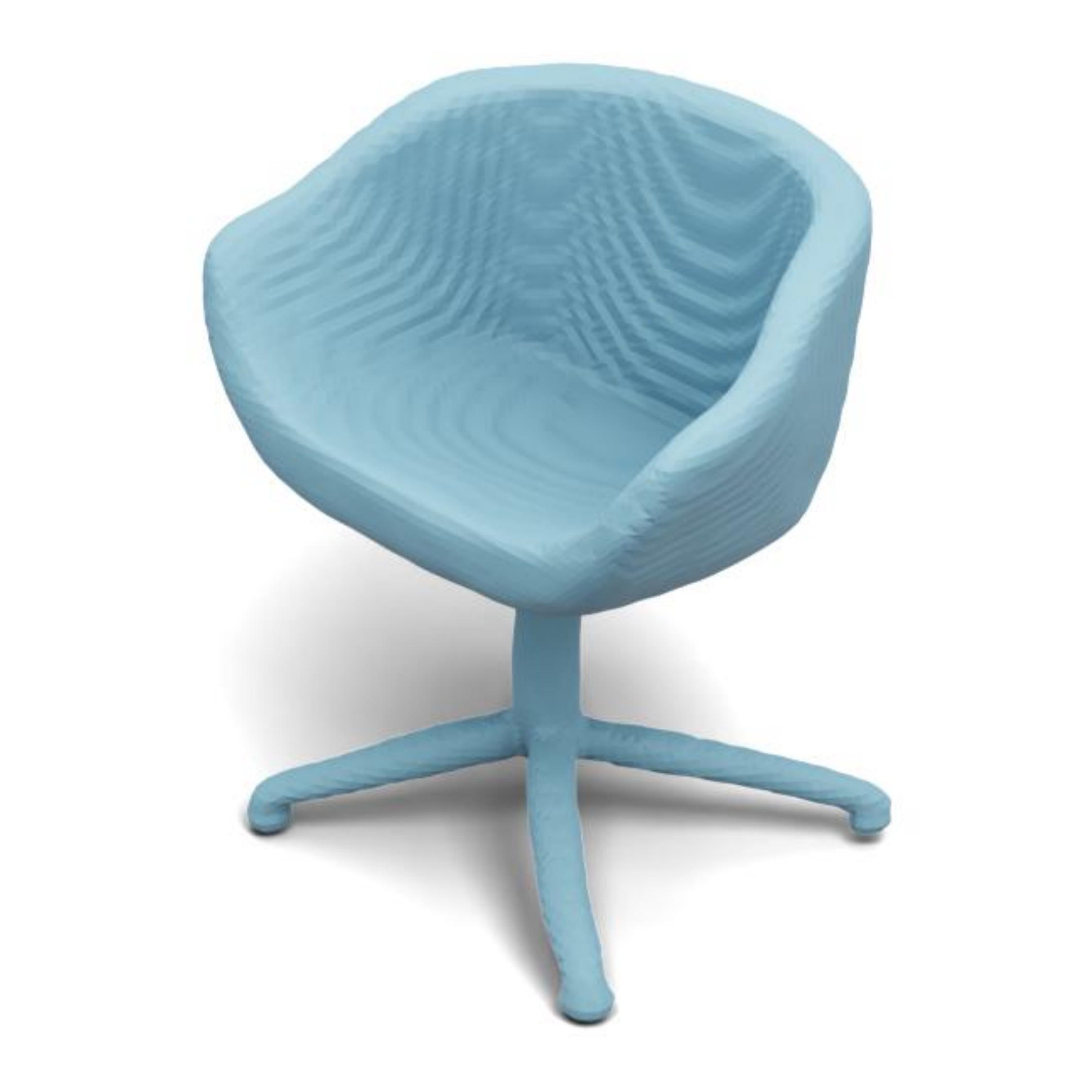}
&\includegraphics[trim = 1 1 1 1, clip, width=0.125\linewidth]{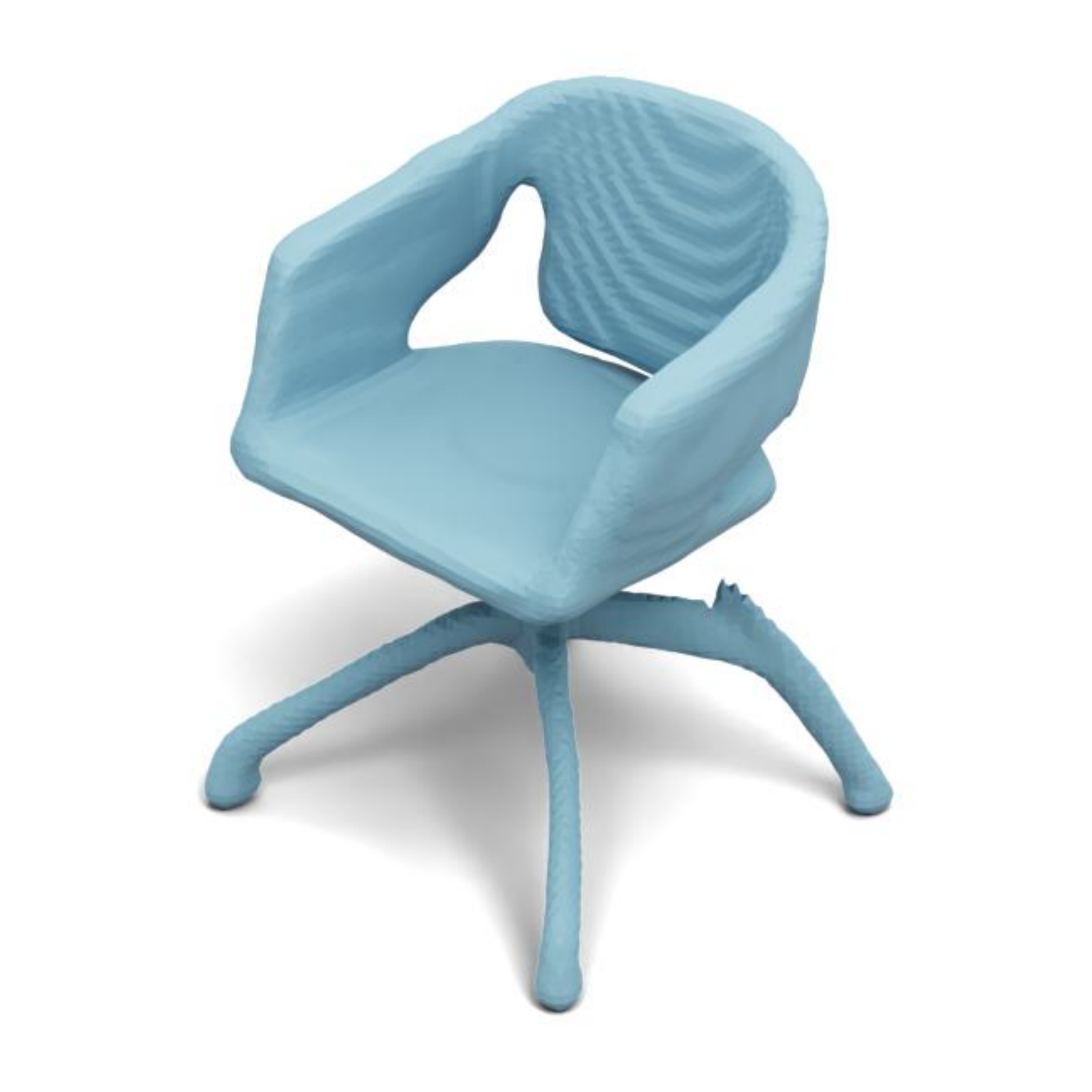}
\\

\end{tabular}
    \caption{We randomly sample sketches from the AmateurSketch dataset and showcase the results of our method.}
	\label{fig:additionalExp3}
\end{figure*}

\begin{figure*}[t]
	\centering
	\small
	\setlength{\tabcolsep}{1pt}
 \begin{tabular}{cccccccc}
\includegraphics[width=0.125\linewidth]{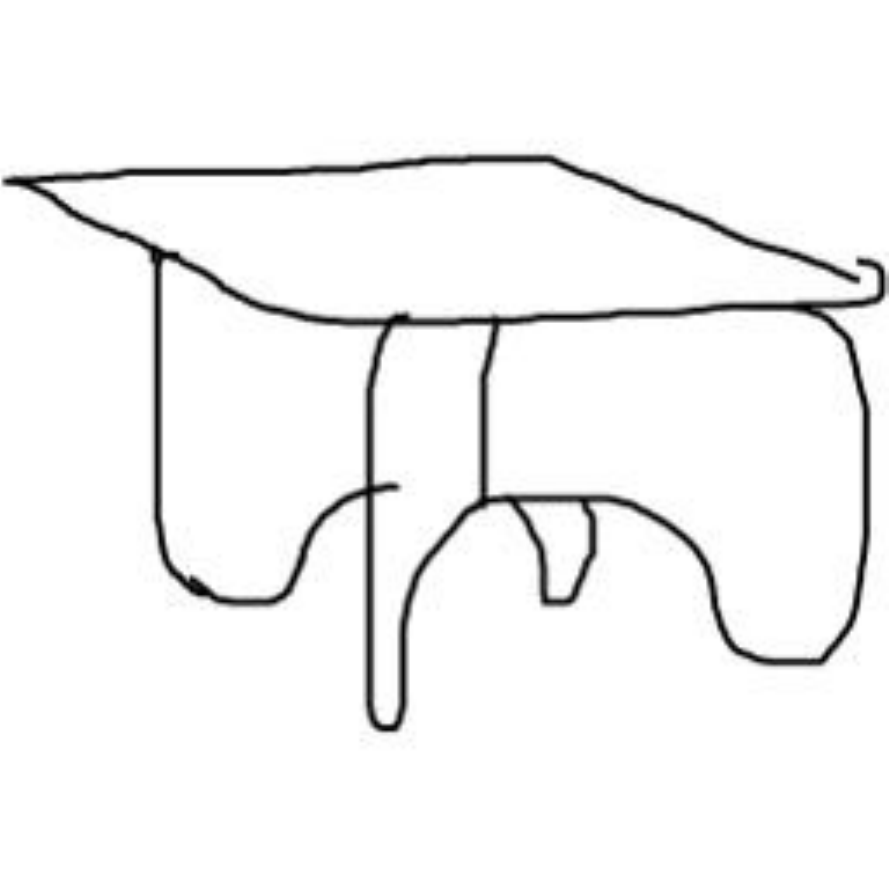}
&\includegraphics[width=0.125\linewidth]{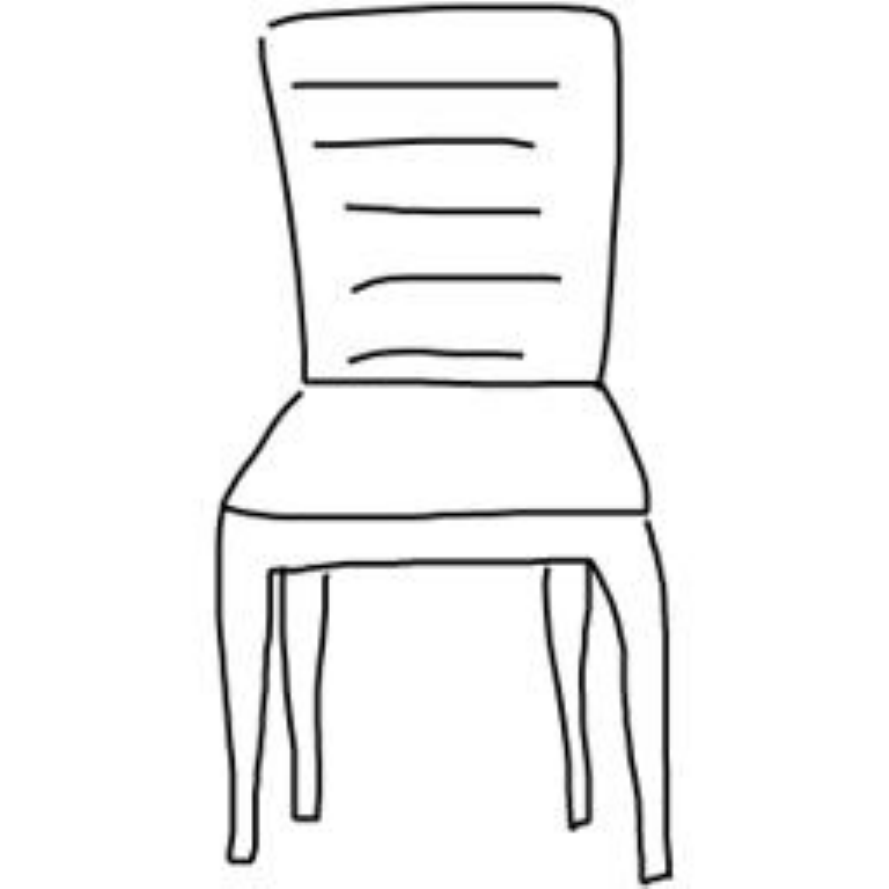}
&\includegraphics[width=0.125\linewidth]{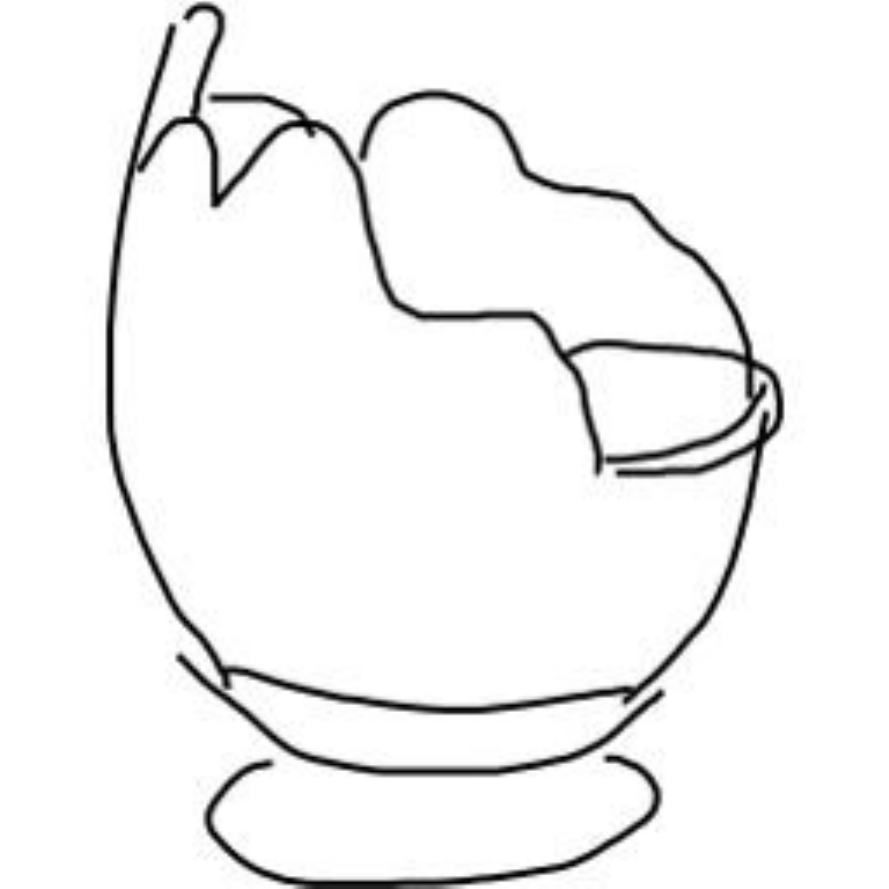}
&\includegraphics[width=0.125\linewidth]{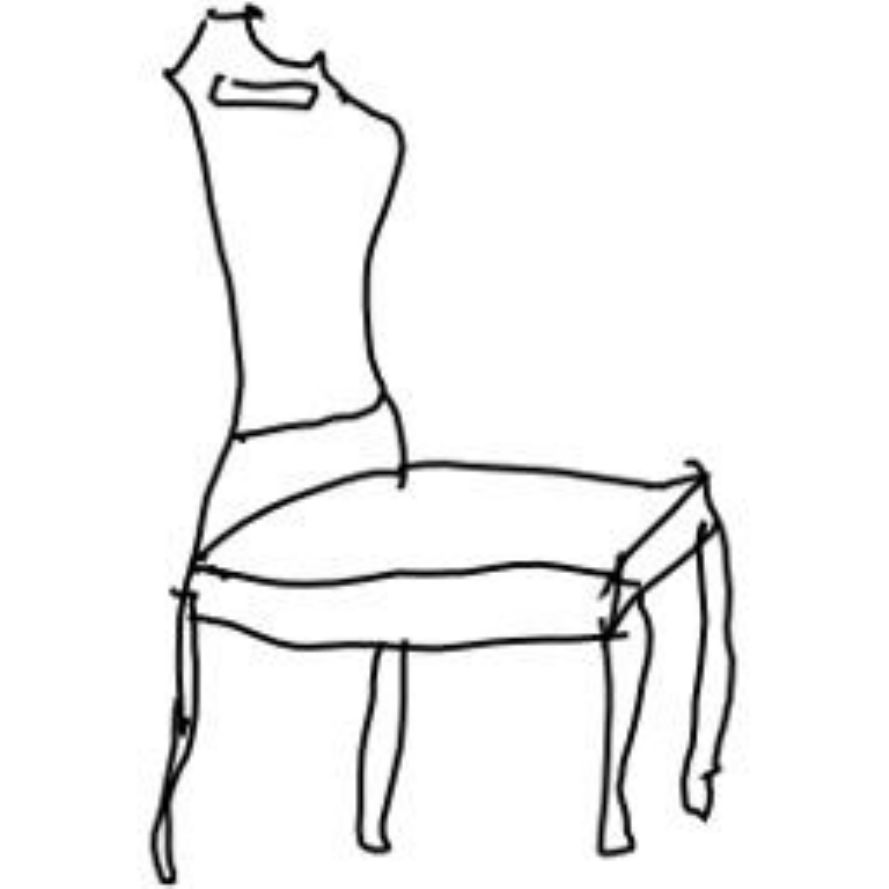}
&\includegraphics[width=0.125\linewidth]{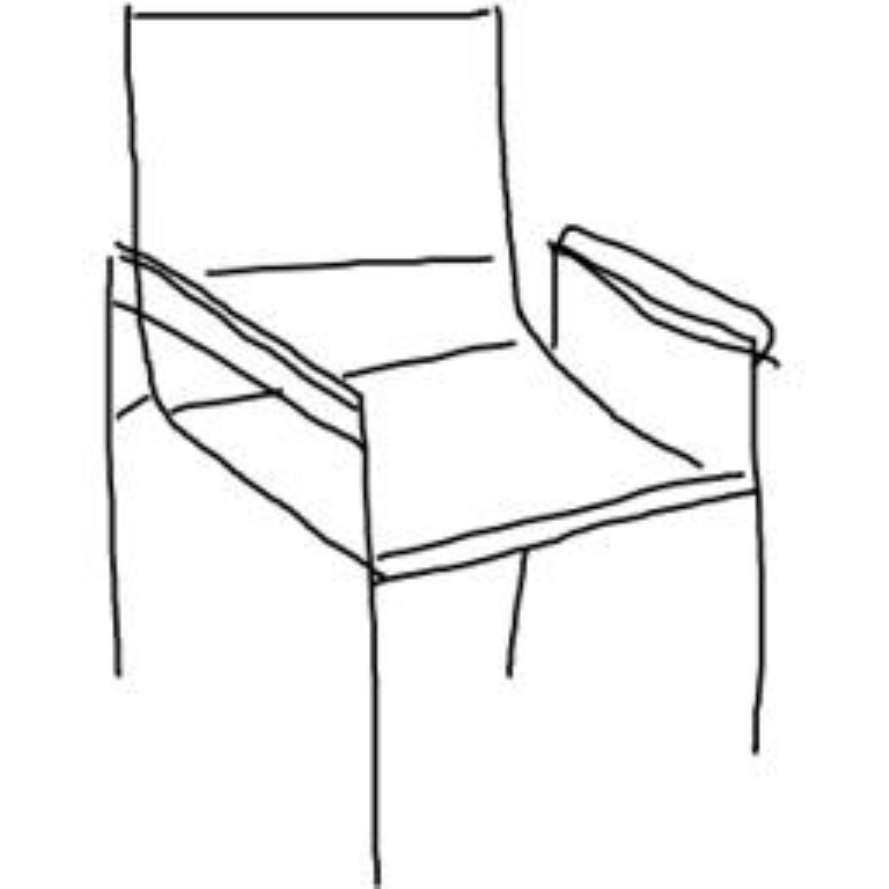}
&\includegraphics[width=0.125\linewidth]{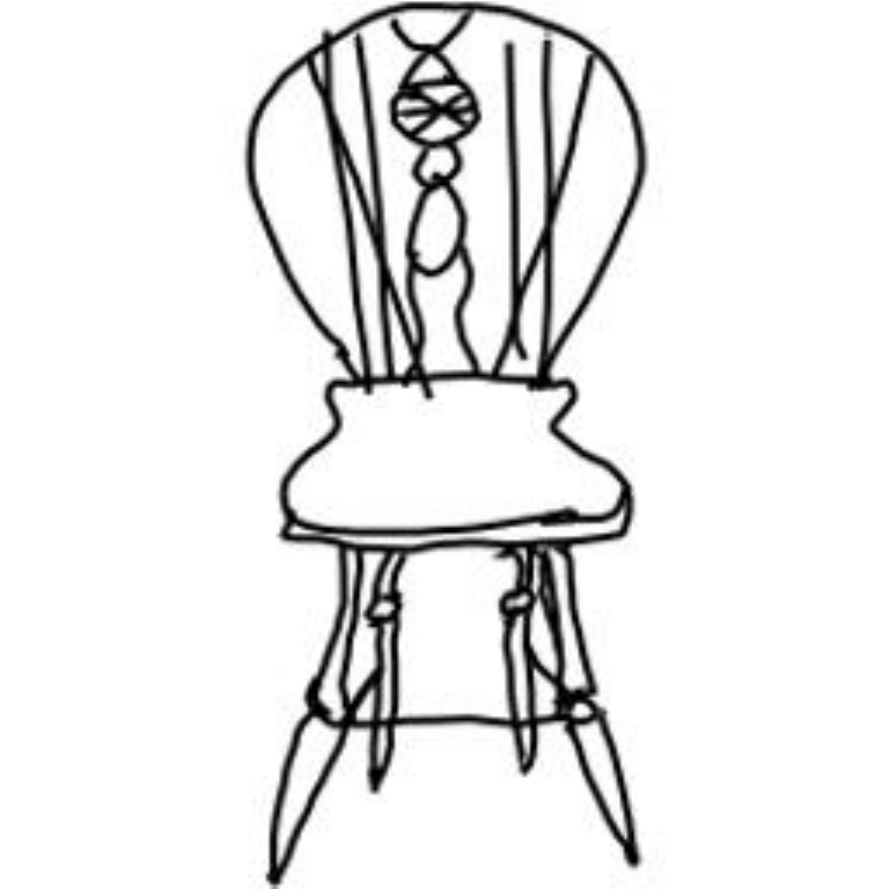}
&\includegraphics[width=0.125\linewidth]{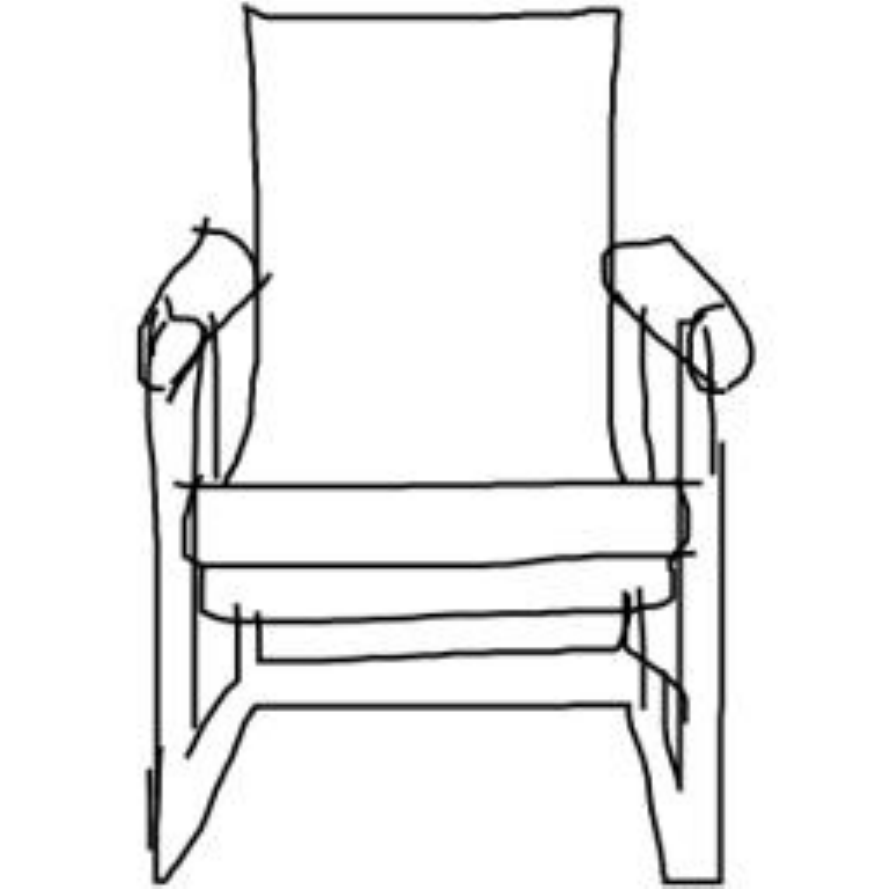}
&\includegraphics[width=0.125\linewidth]{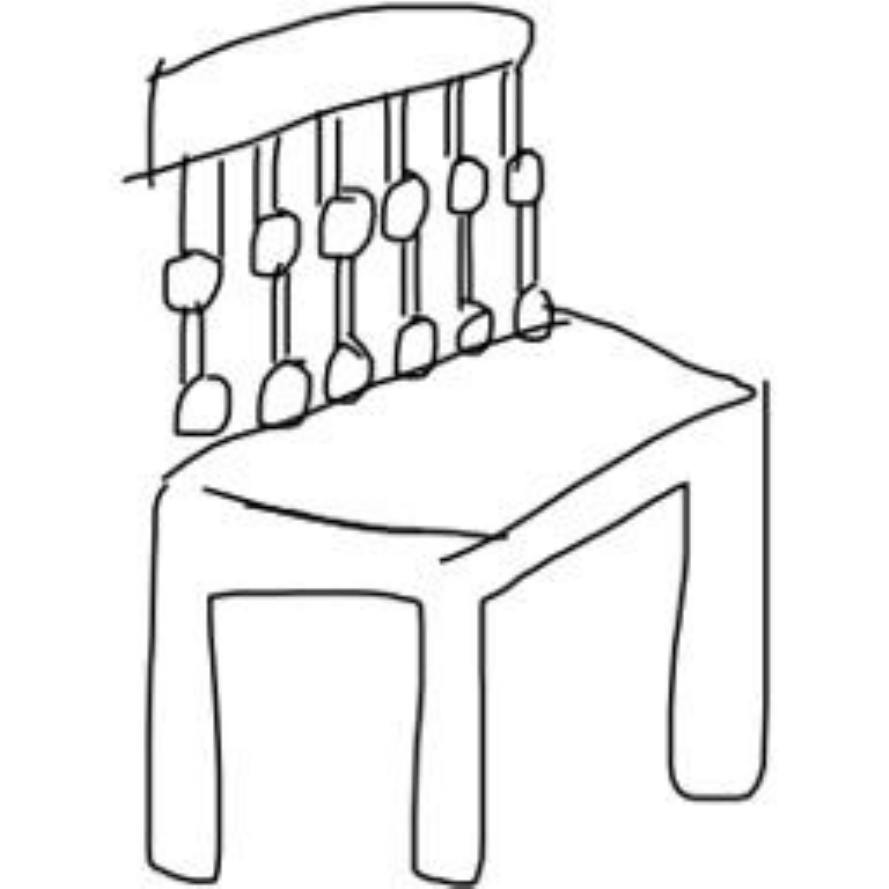}
\\
\includegraphics[trim = 1 1 1 1, clip, width=0.125\linewidth]{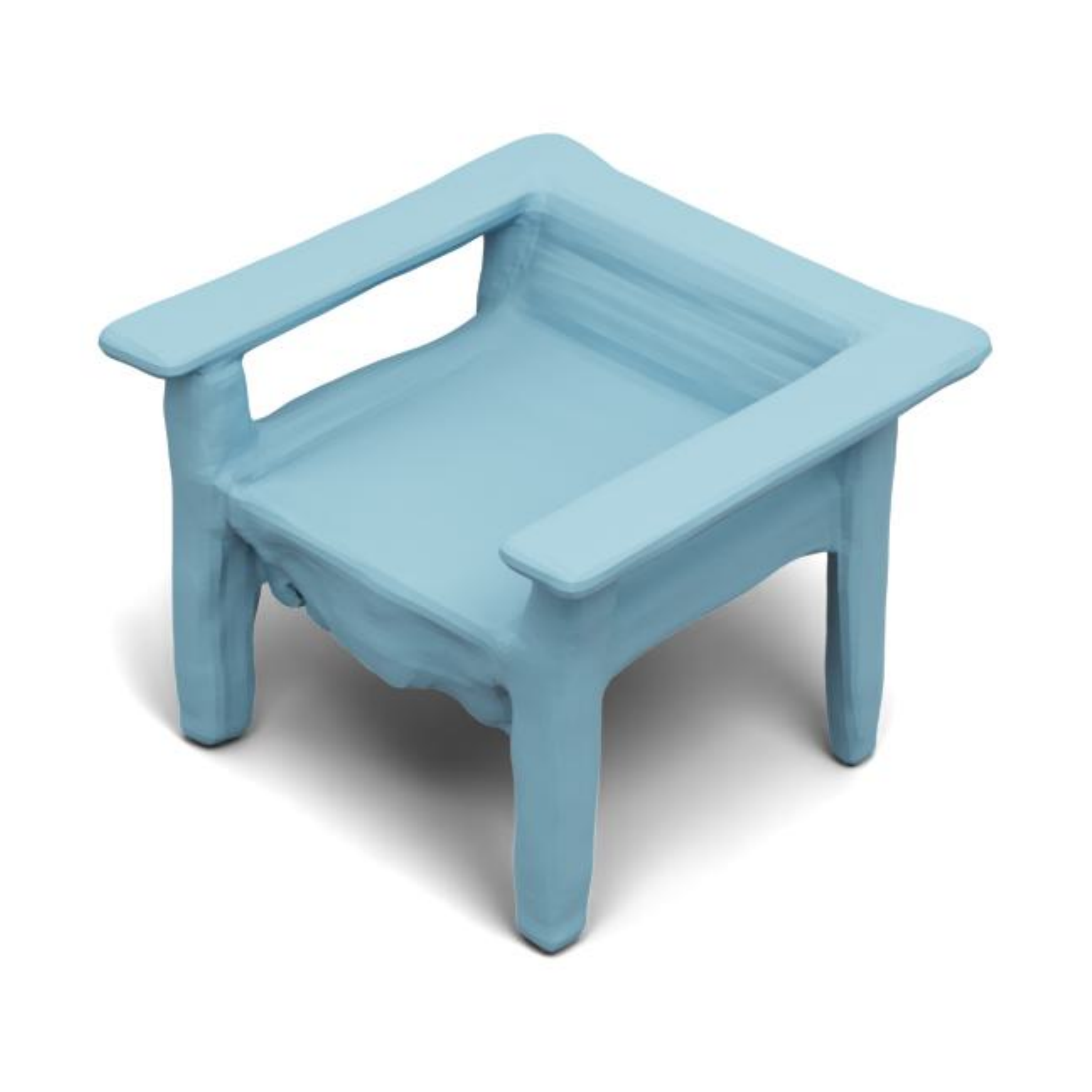}
&\includegraphics[trim = 1 1 1 1, clip, width=0.125\linewidth]{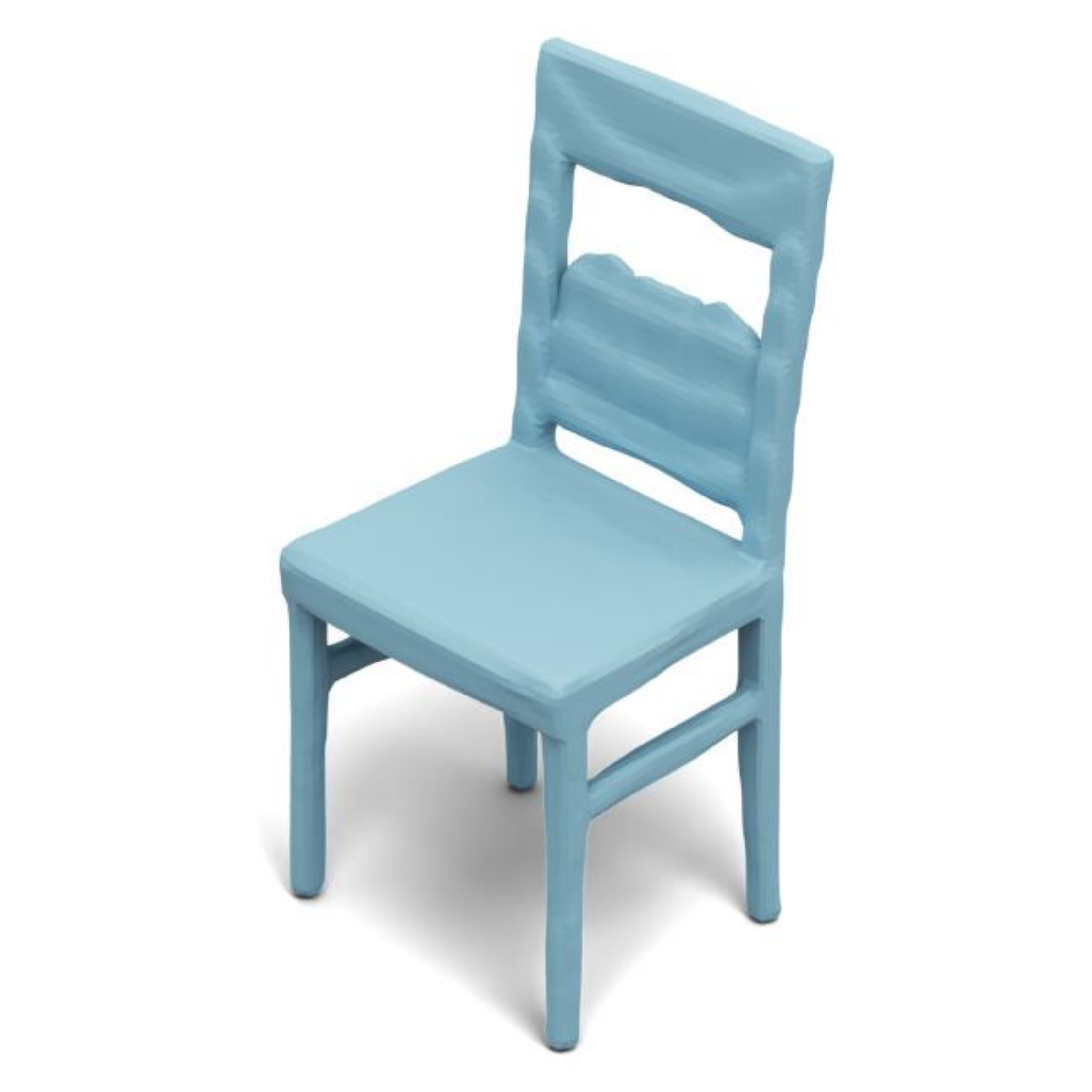}
&\includegraphics[trim = 1 1 1 1, clip, width=0.125\linewidth]{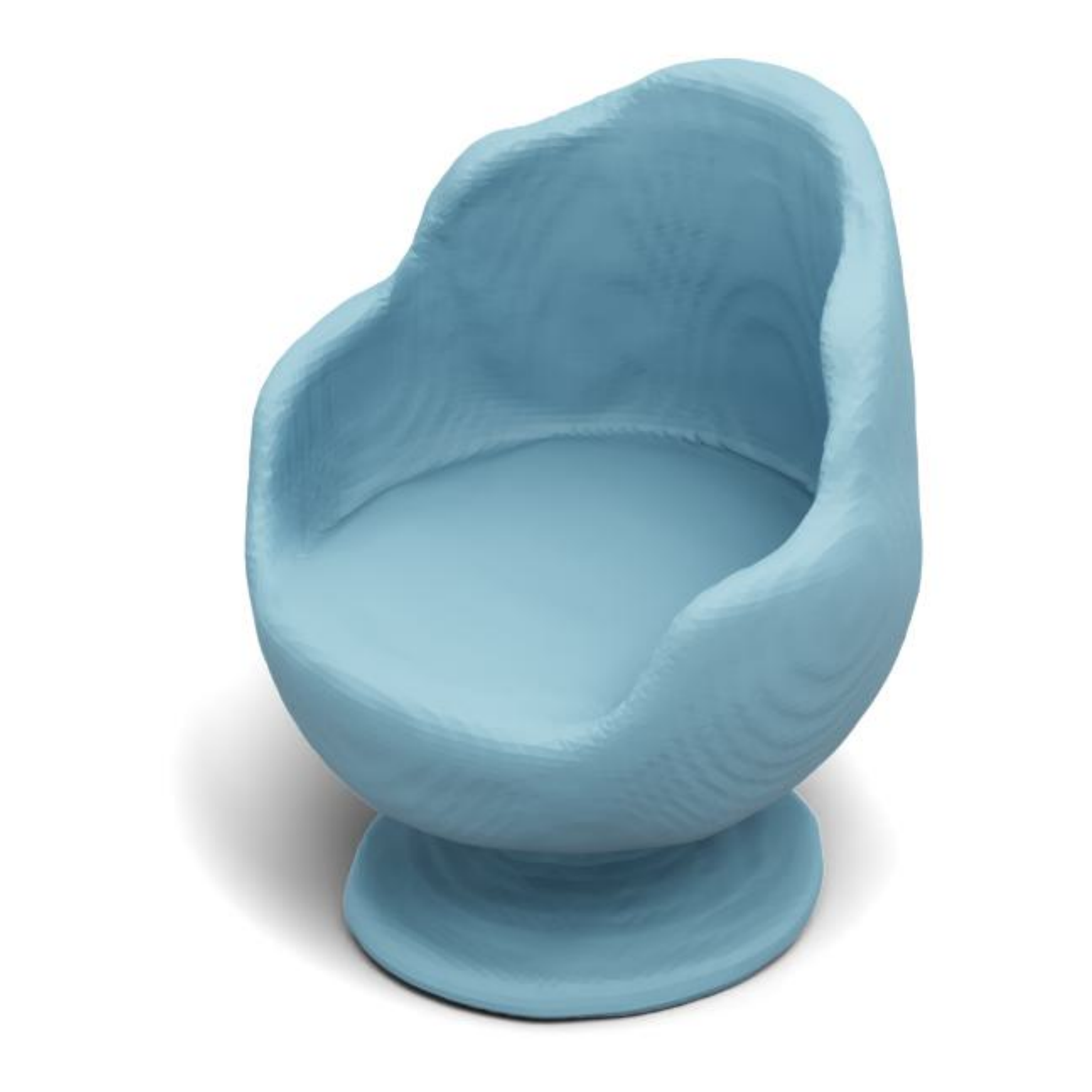}
&\includegraphics[trim = 1 1 1 1, clip, width=0.125\linewidth]{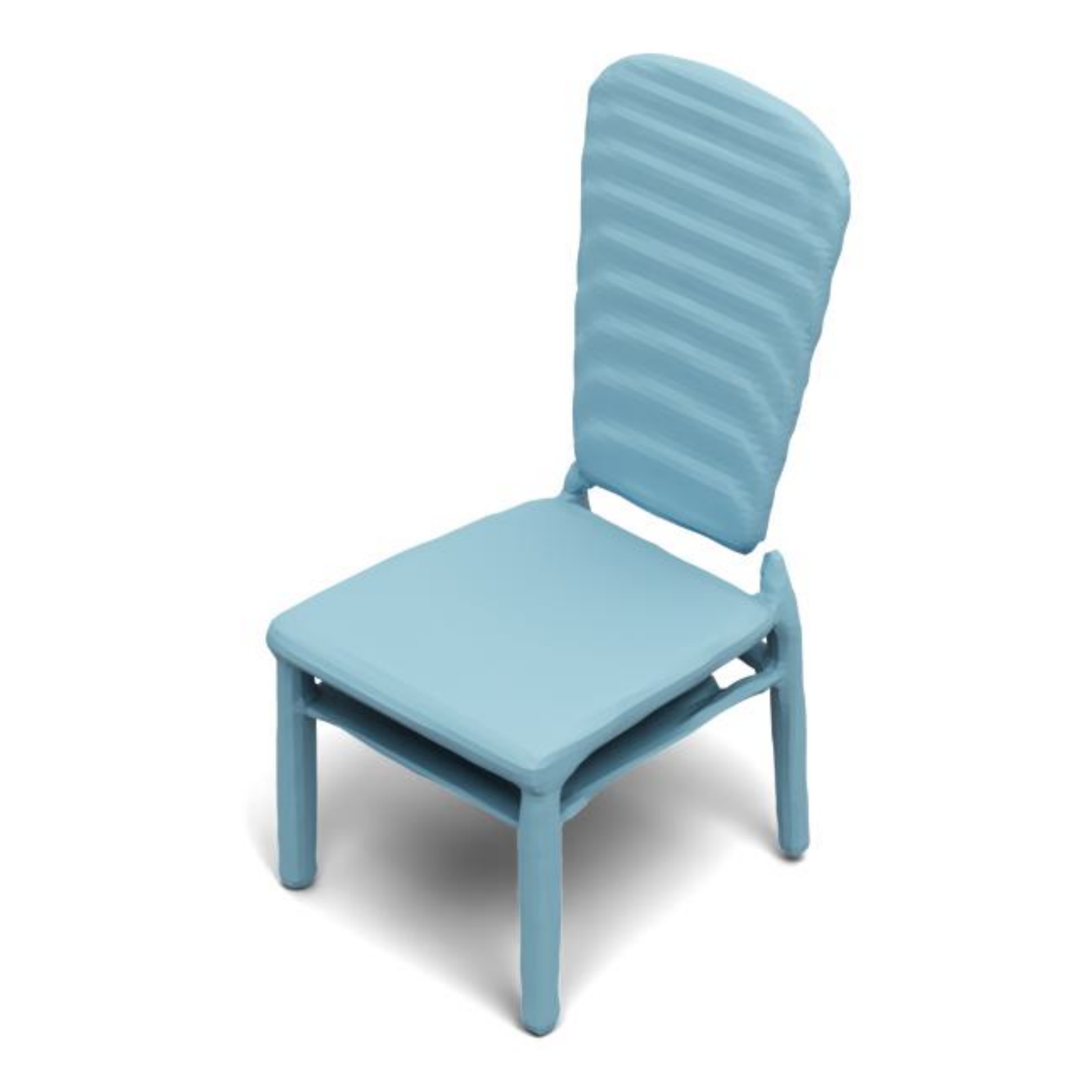}
&\includegraphics[trim = 1 1 1 1, clip, width=0.125\linewidth]{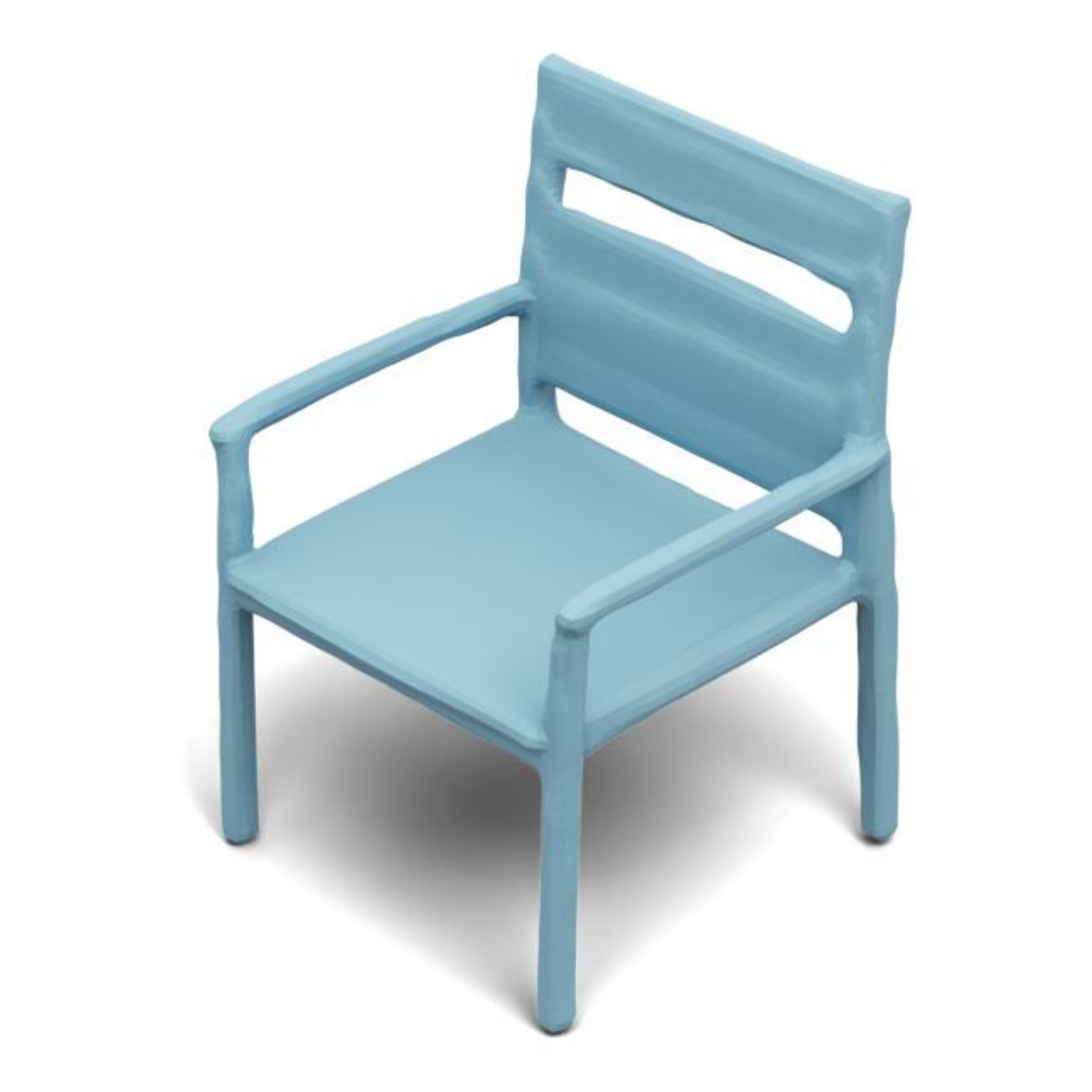}
&\includegraphics[trim = 1 1 1 1, clip, width=0.125\linewidth]{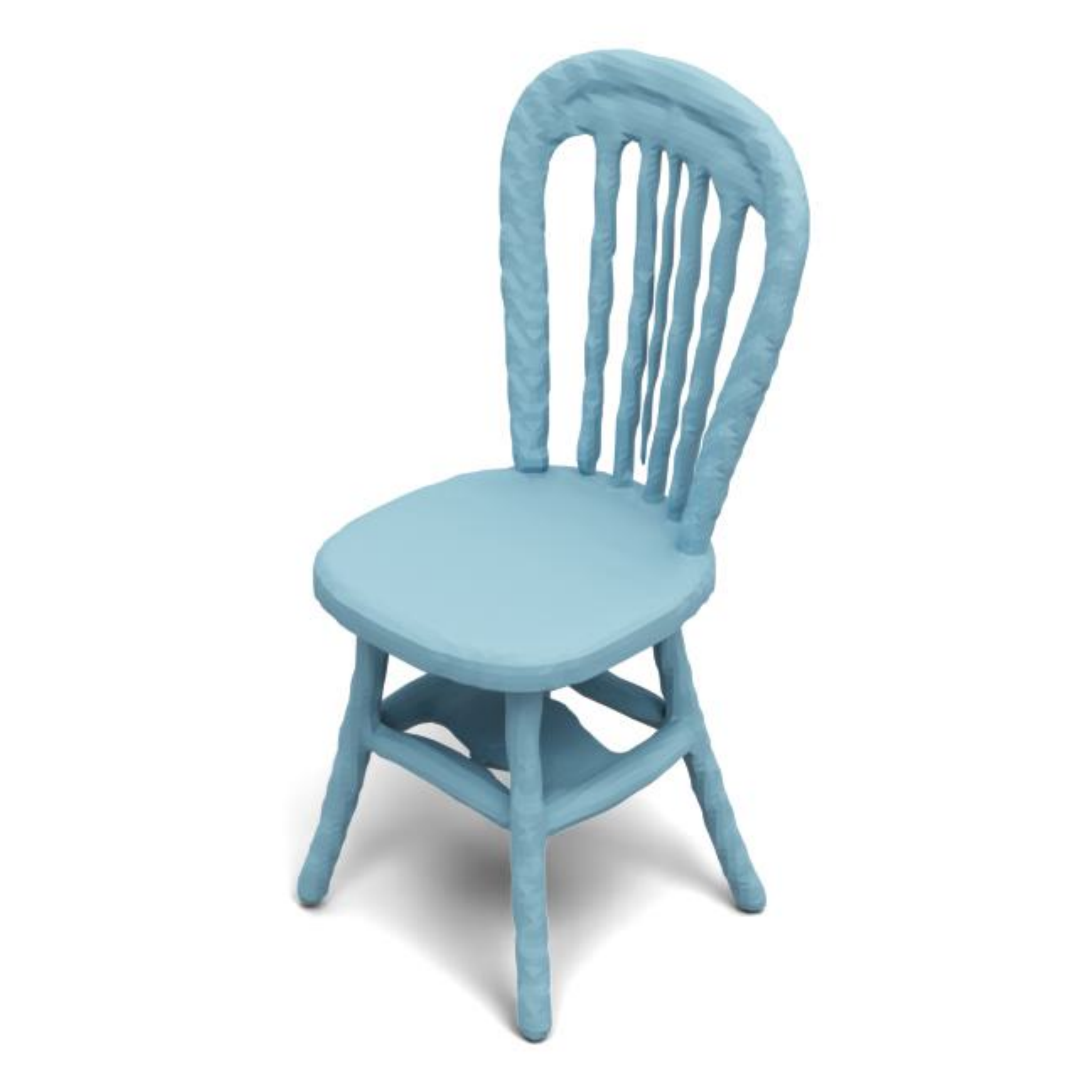}
&\includegraphics[trim = 1 1 1 1, clip, width=0.125\linewidth]{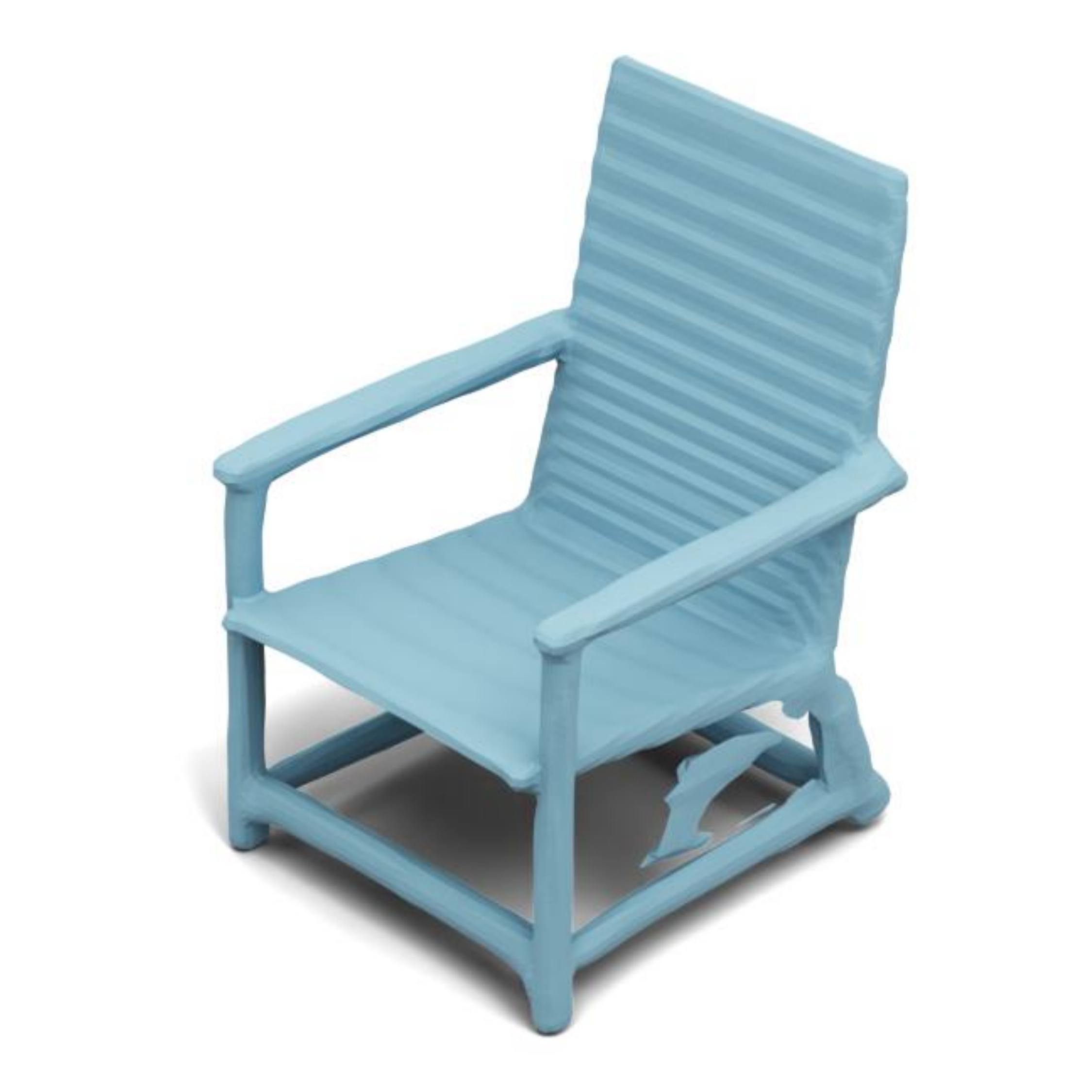}
&\includegraphics[trim = 1 1 1 1, clip, width=0.125\linewidth]{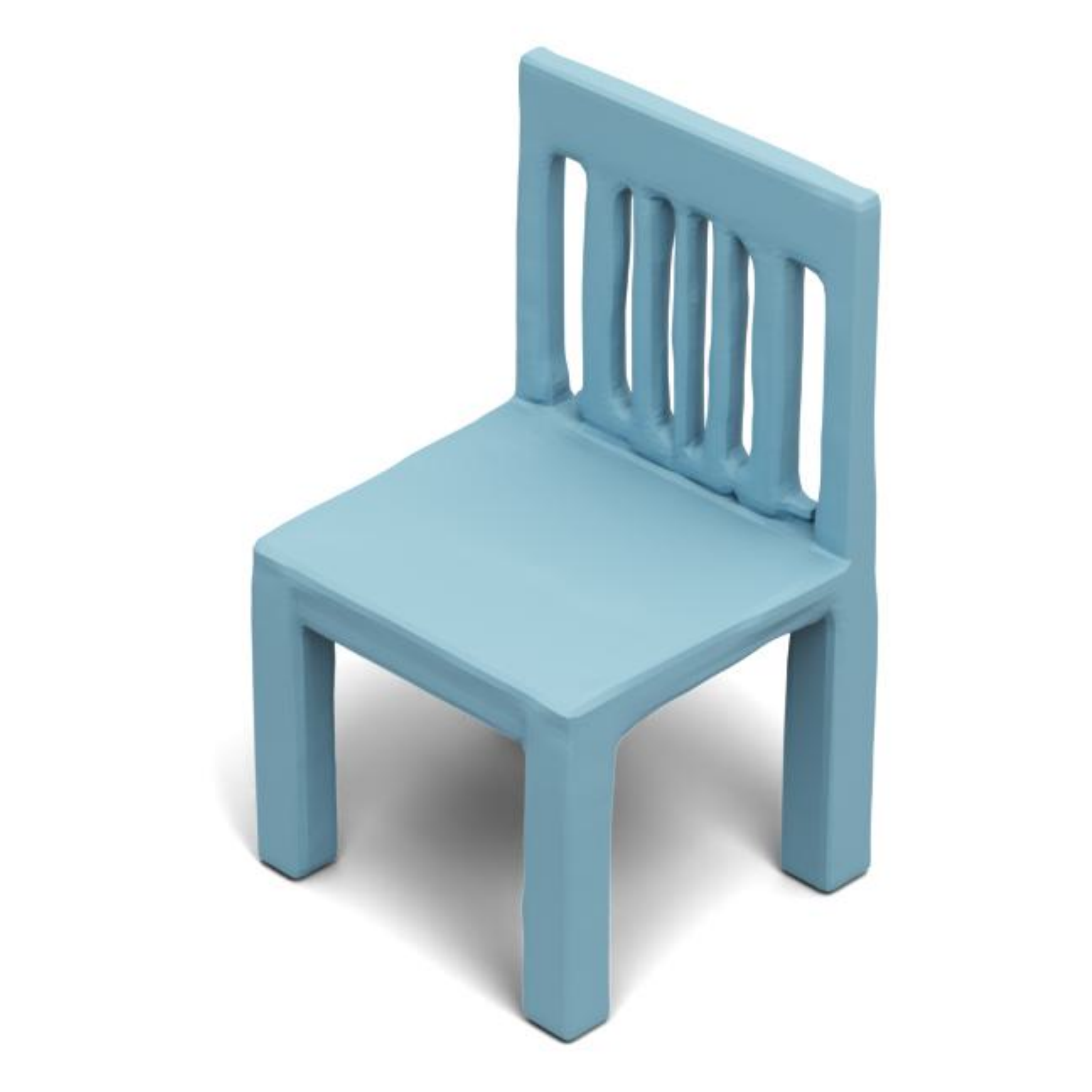}
\\
\includegraphics[width=0.125\linewidth]{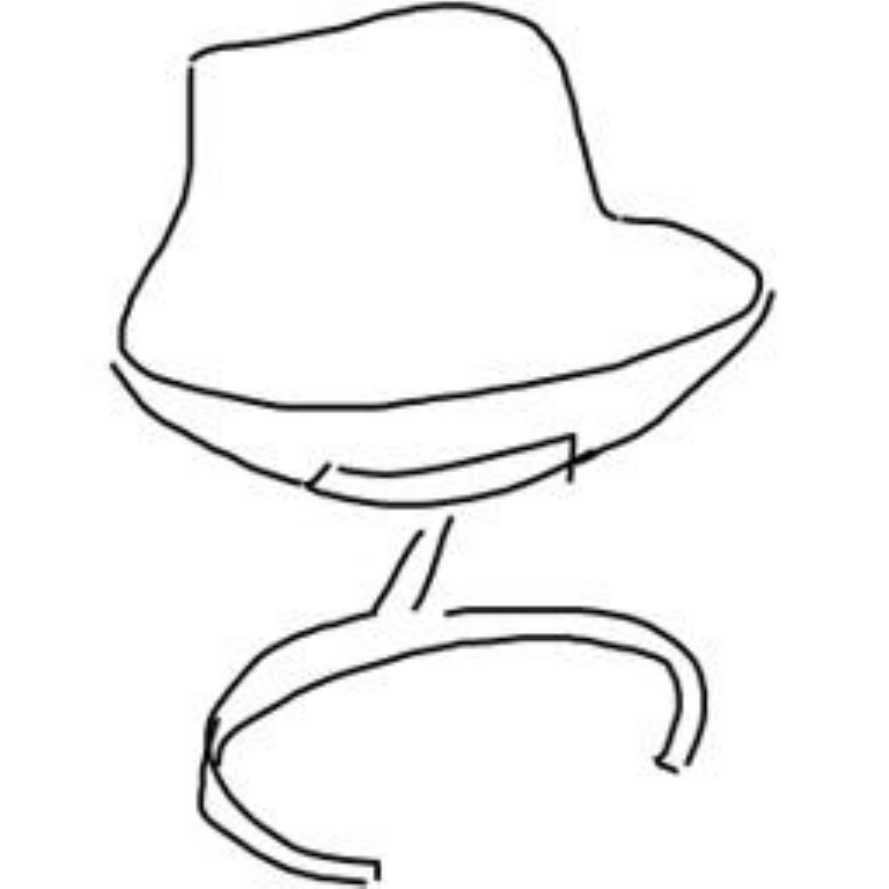}
&\includegraphics[width=0.125\linewidth]{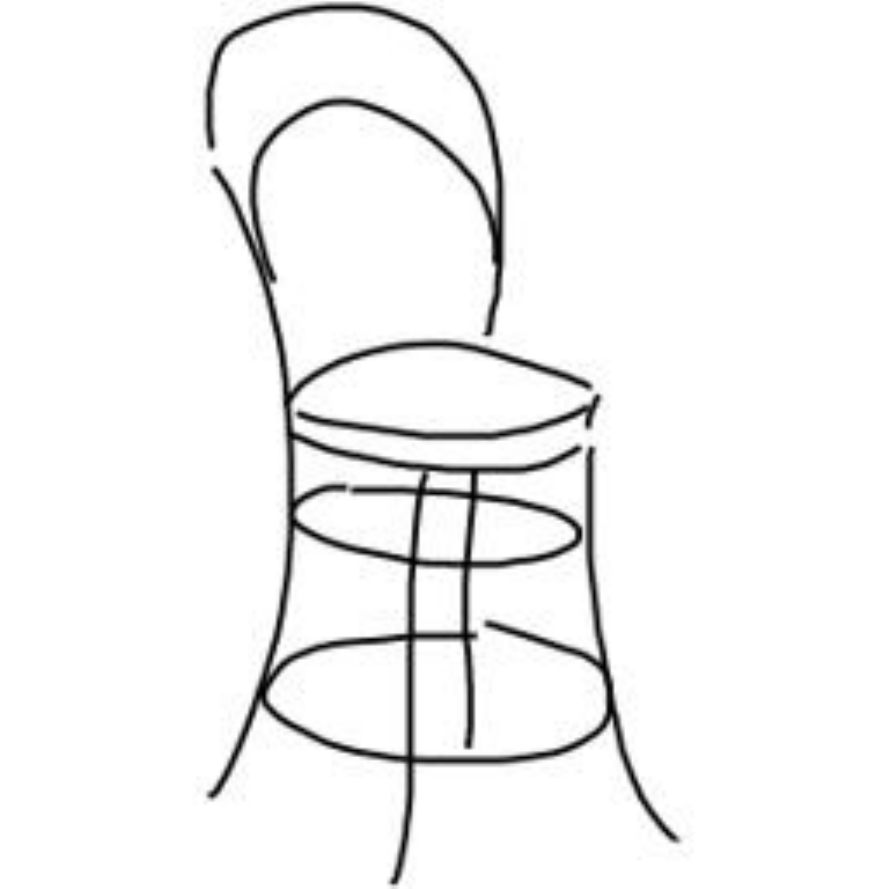}
&\includegraphics[width=0.125\linewidth]{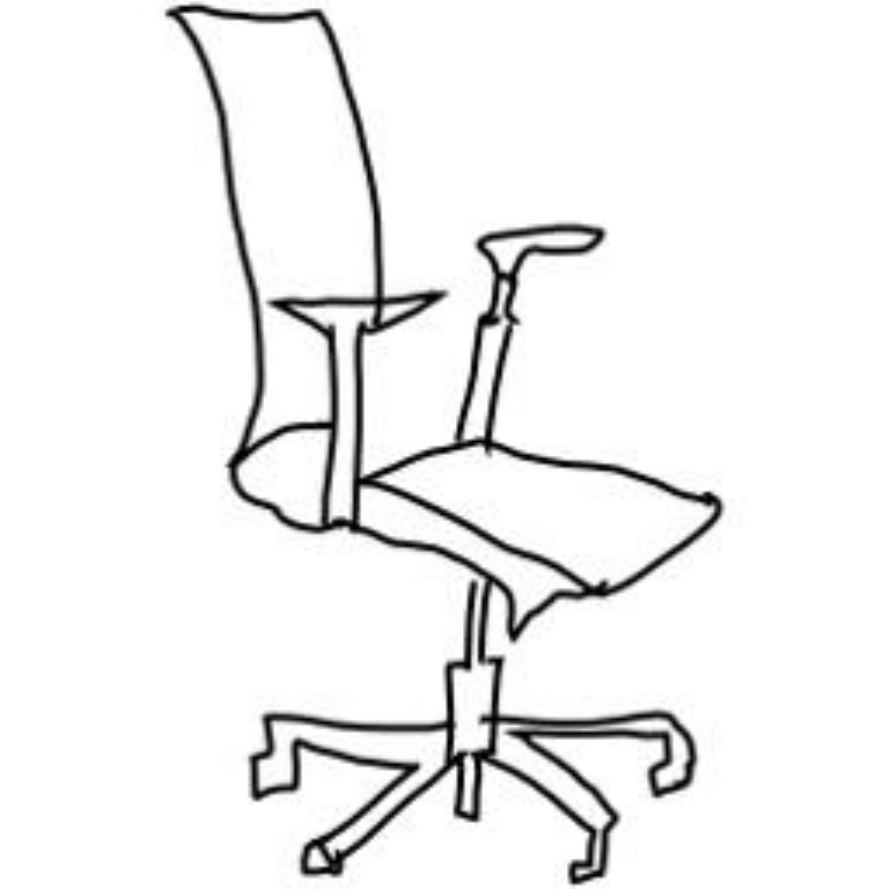}
&\includegraphics[width=0.125\linewidth]{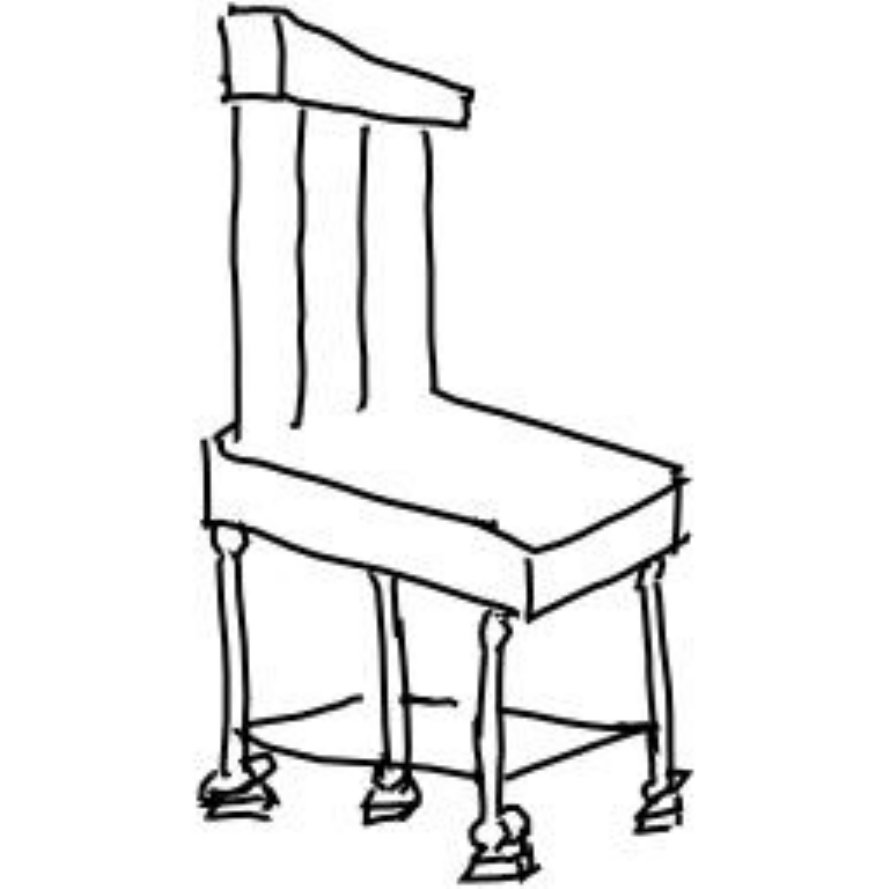}
&\includegraphics[width=0.125\linewidth]{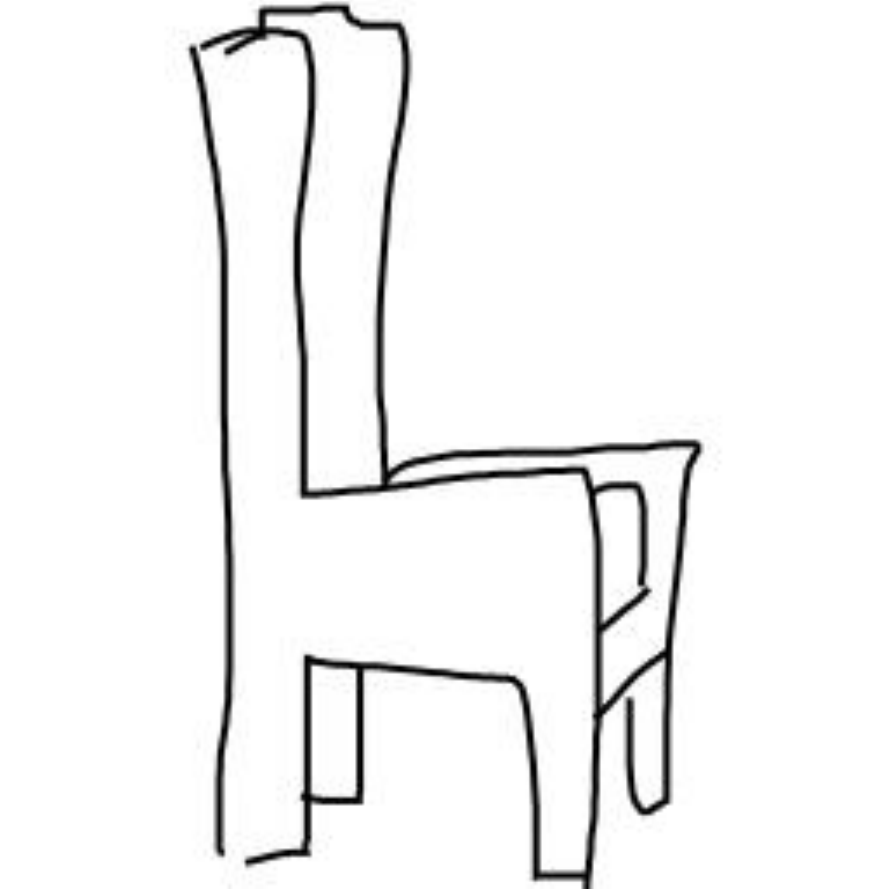}
&\includegraphics[width=0.125\linewidth]{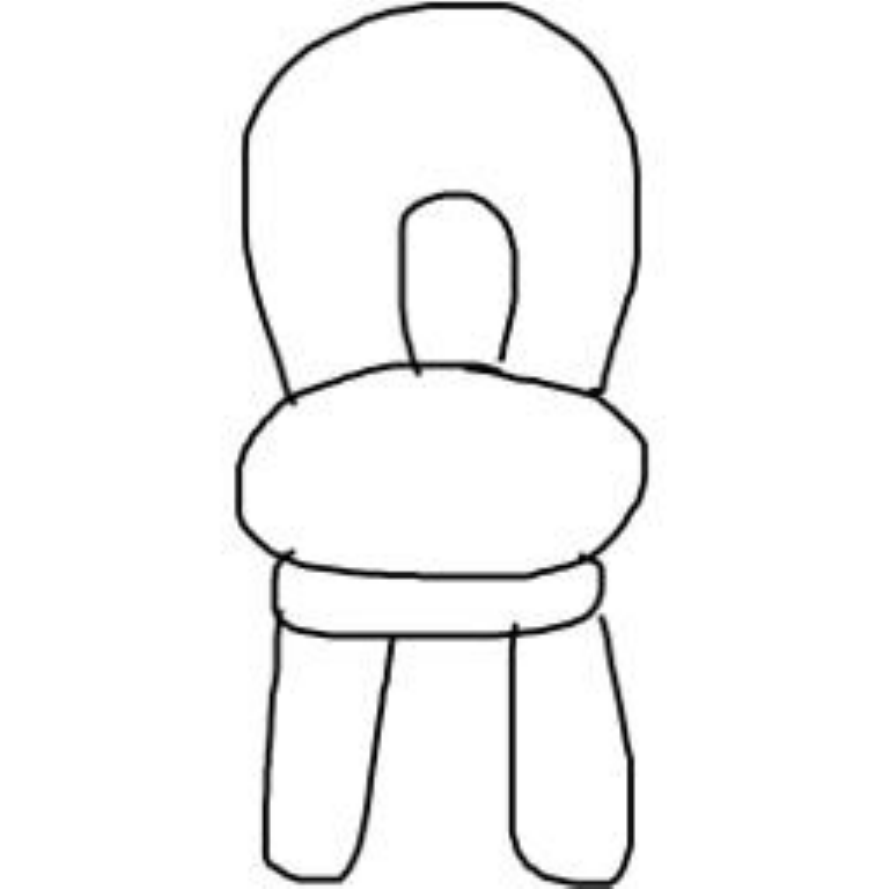}
&\includegraphics[width=0.125\linewidth]{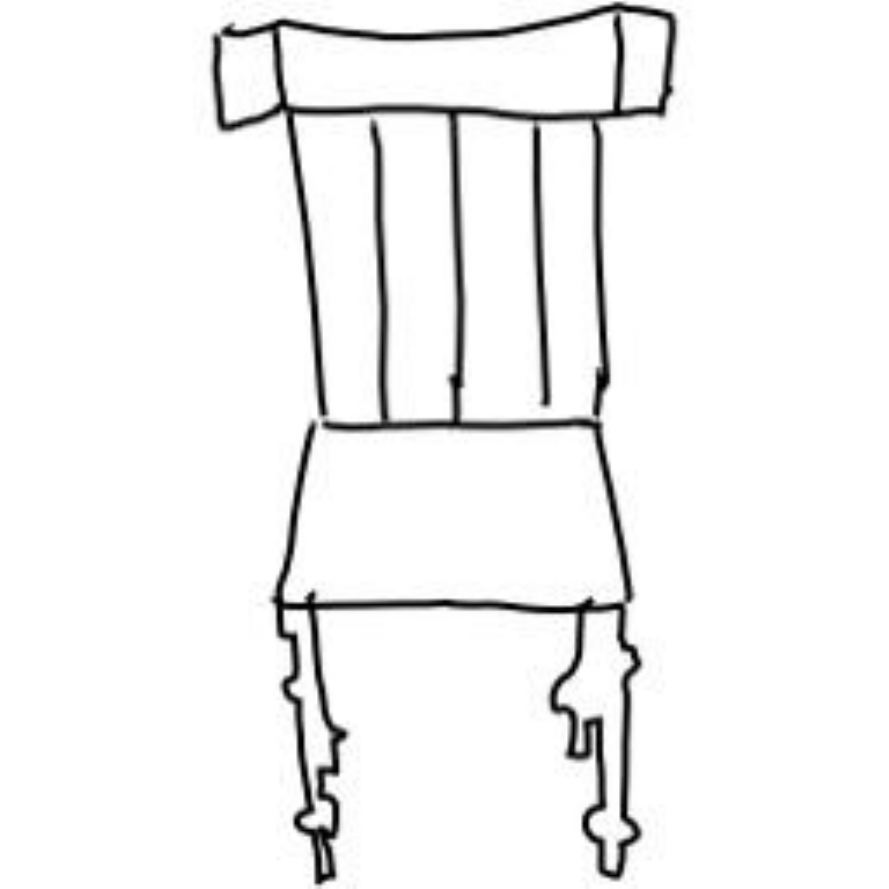}
&\includegraphics[width=0.125\linewidth]{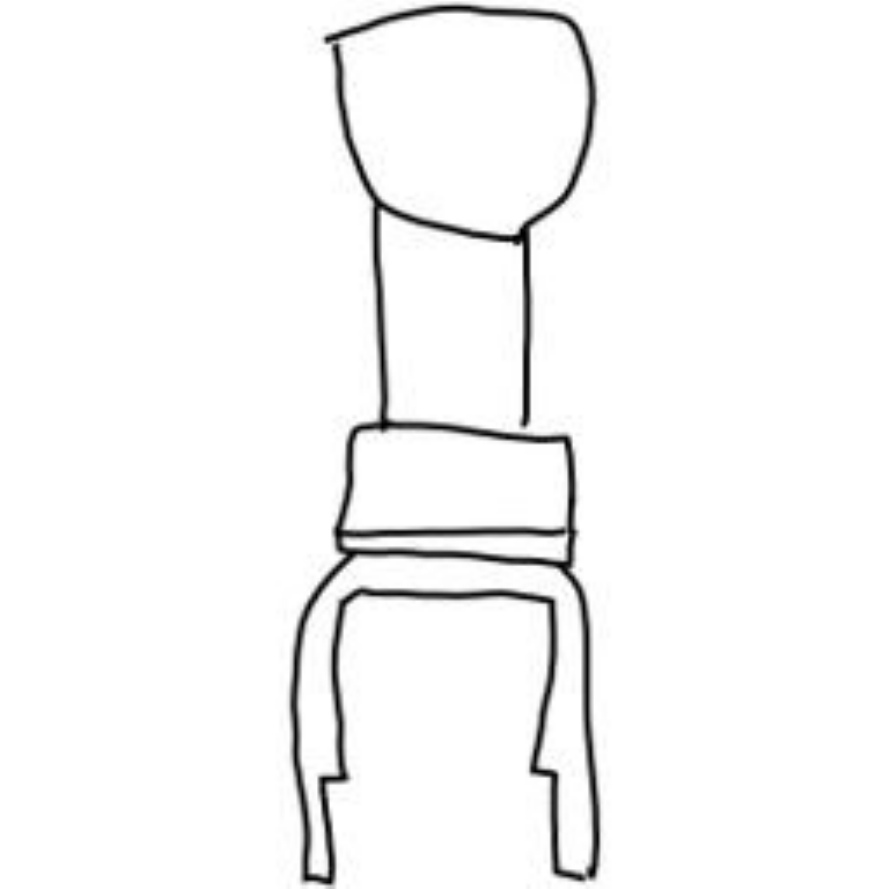}
\\
\includegraphics[trim = 1 1 1 1, clip, width=0.125\linewidth]{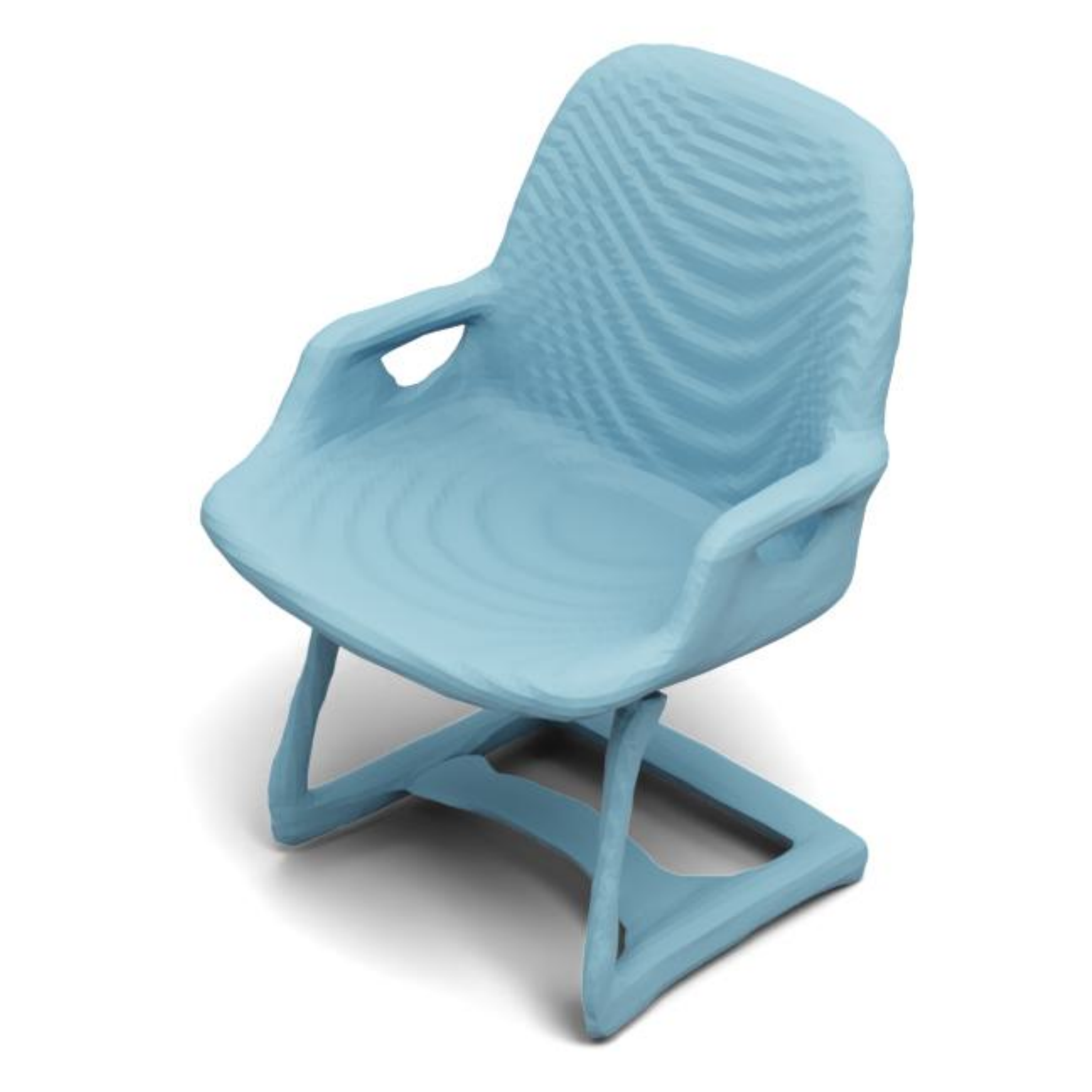}
&\includegraphics[trim = 1 1 1 1, clip, width=0.125\linewidth]{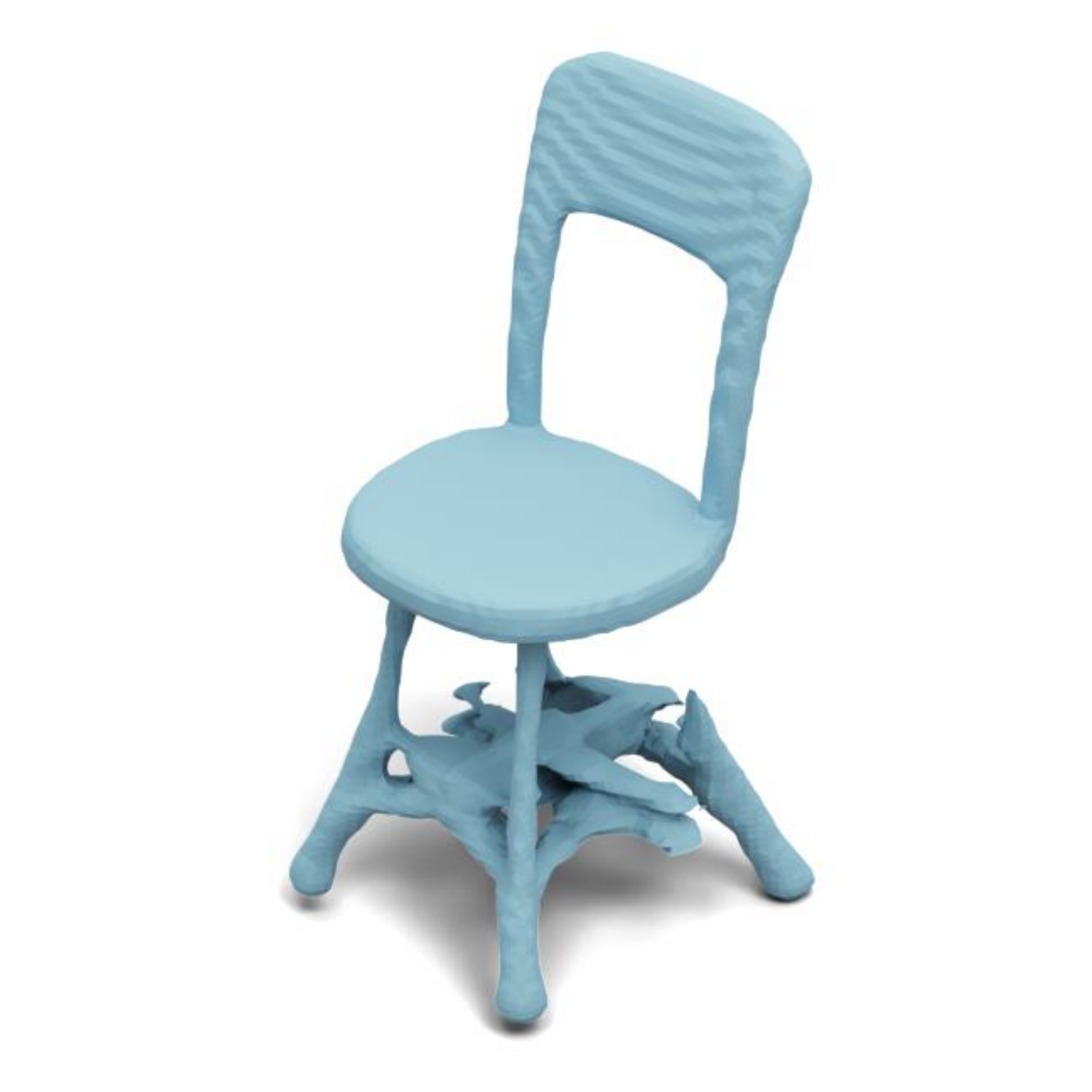}
&\includegraphics[trim = 1 1 1 1, clip, width=0.125\linewidth]{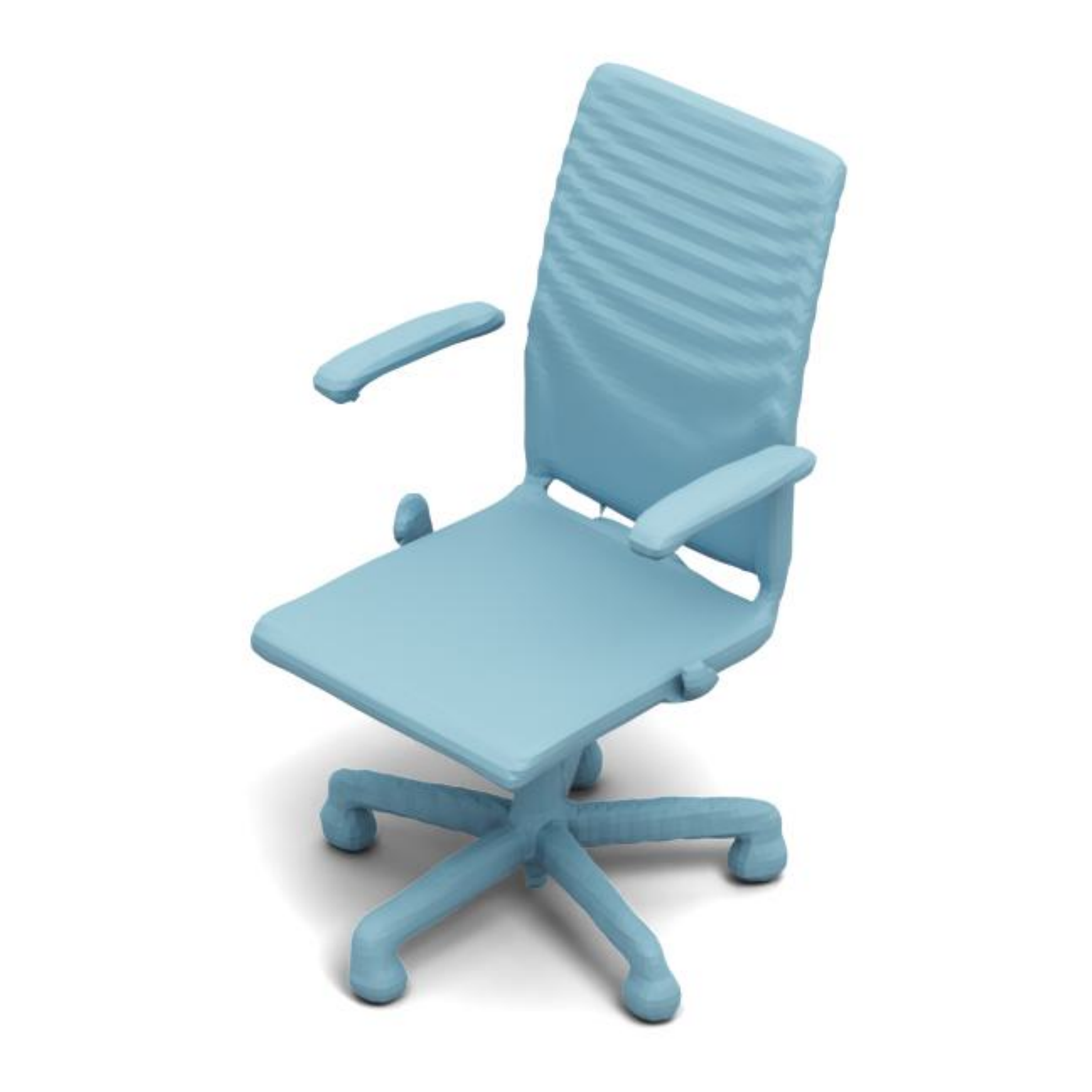}
&\includegraphics[trim = 1 1 1 1, clip, width=0.125\linewidth]{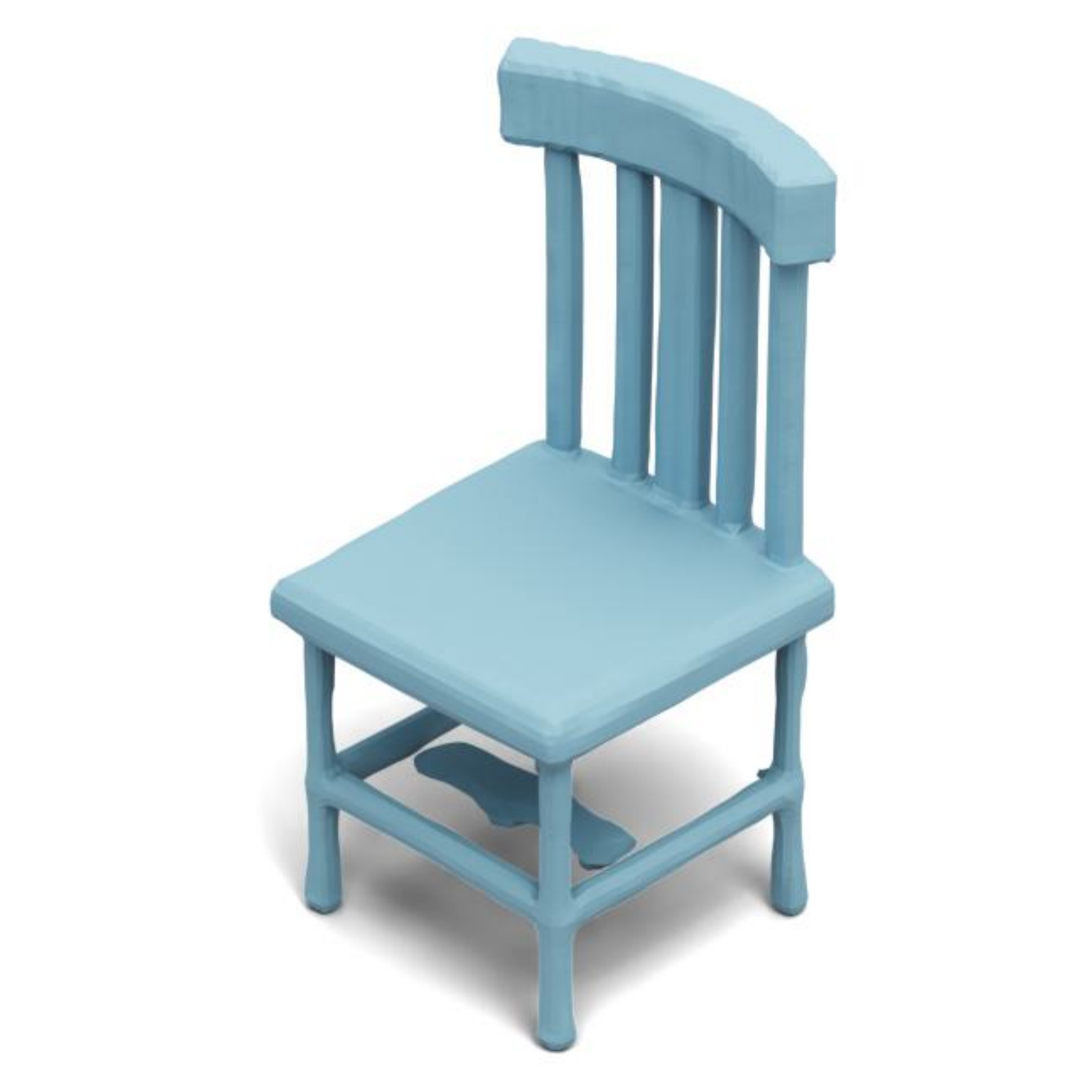}
&\includegraphics[trim = 1 1 1 1, clip, width=0.125\linewidth]{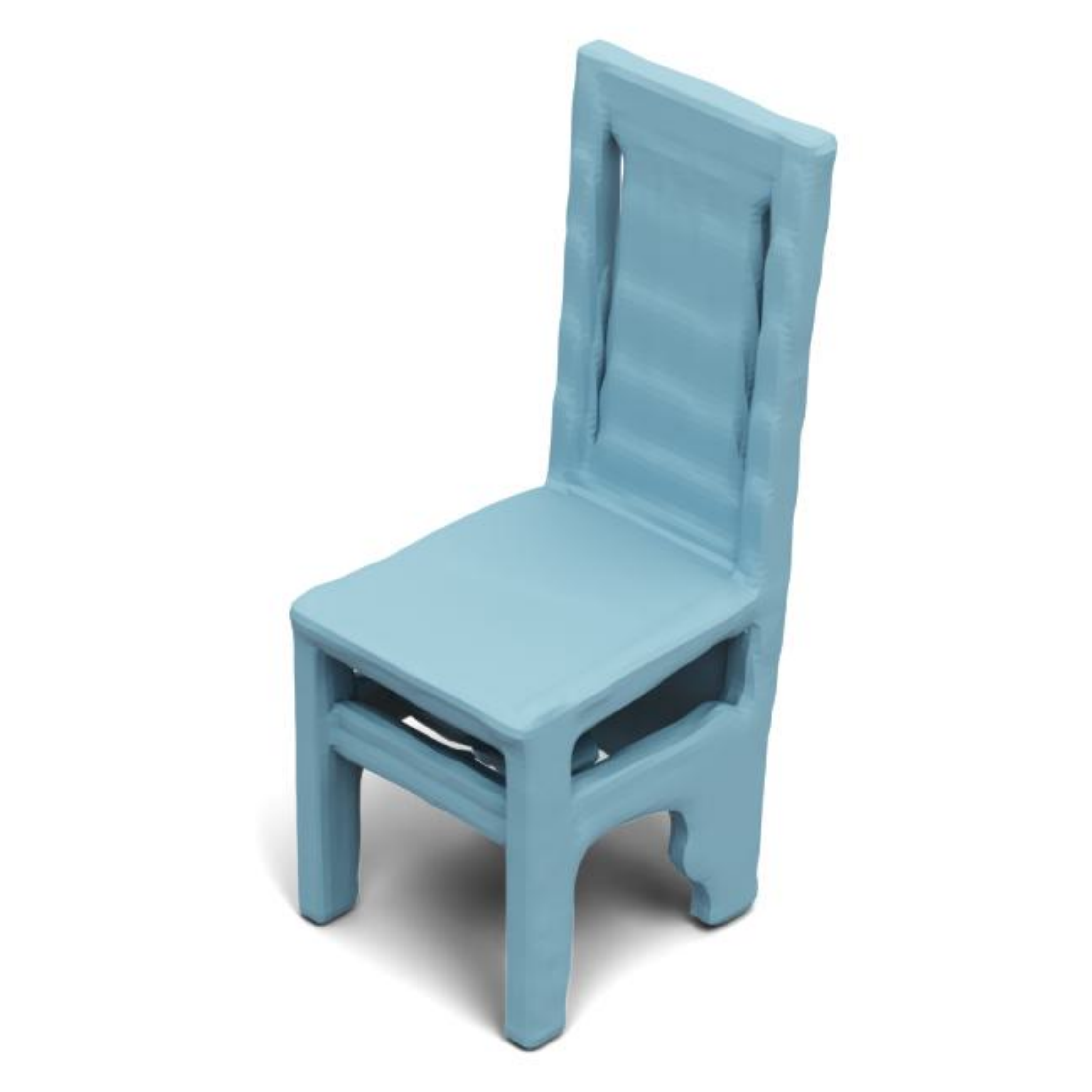}
&\includegraphics[trim = 1 1 1 1, clip, width=0.125\linewidth]{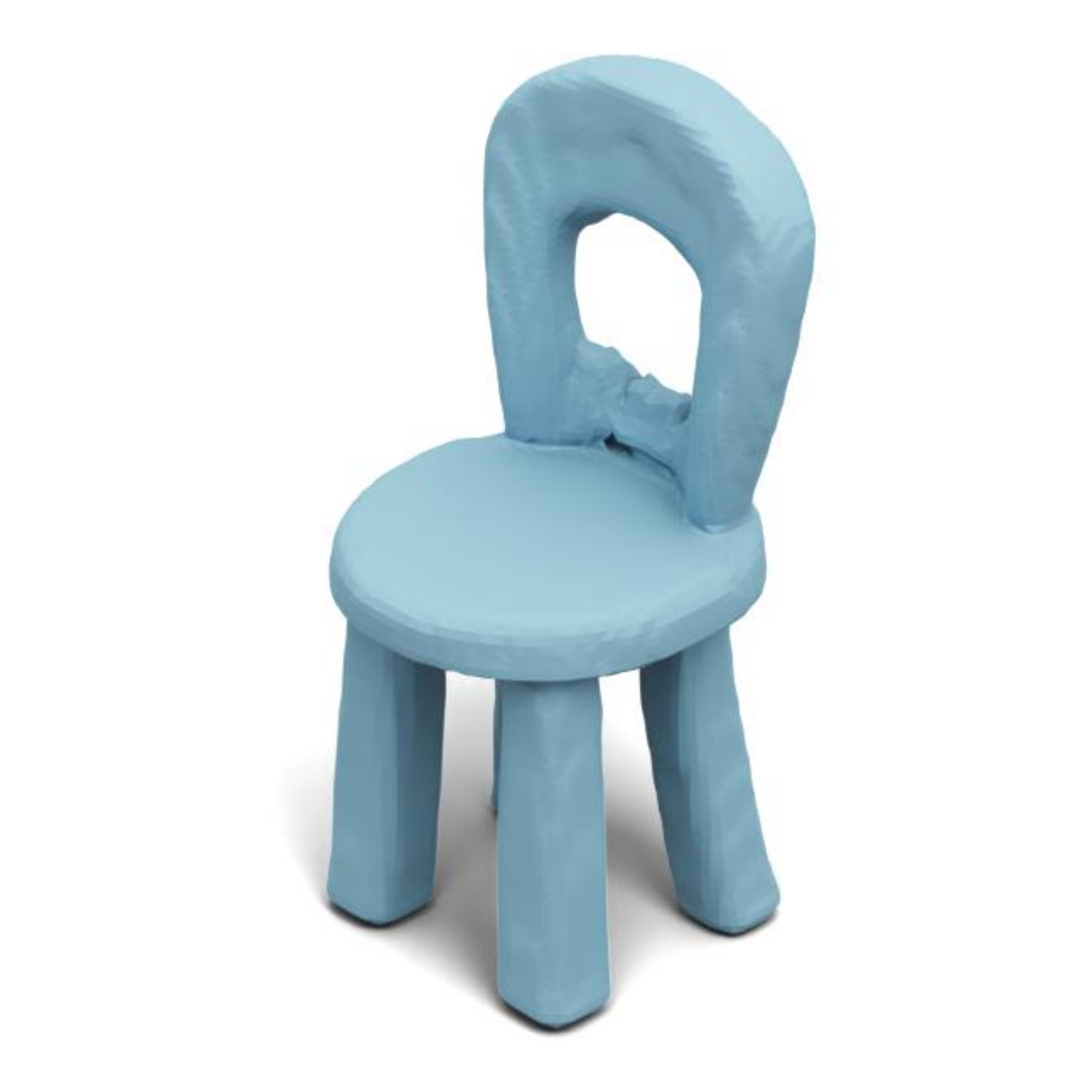}
&\includegraphics[trim = 1 1 1 1, clip, width=0.125\linewidth]{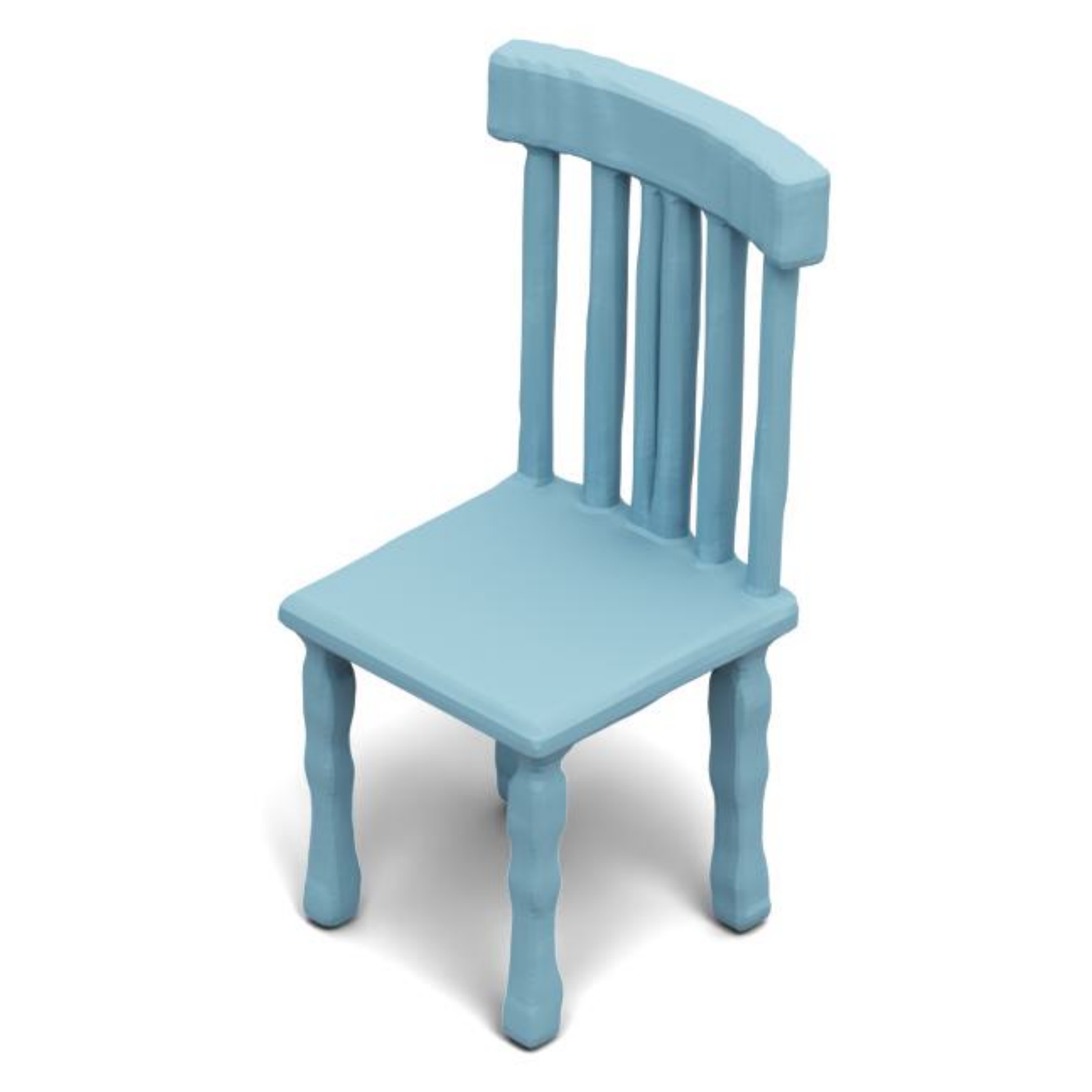}
&\includegraphics[trim = 1 1 1 1, clip, width=0.125\linewidth]{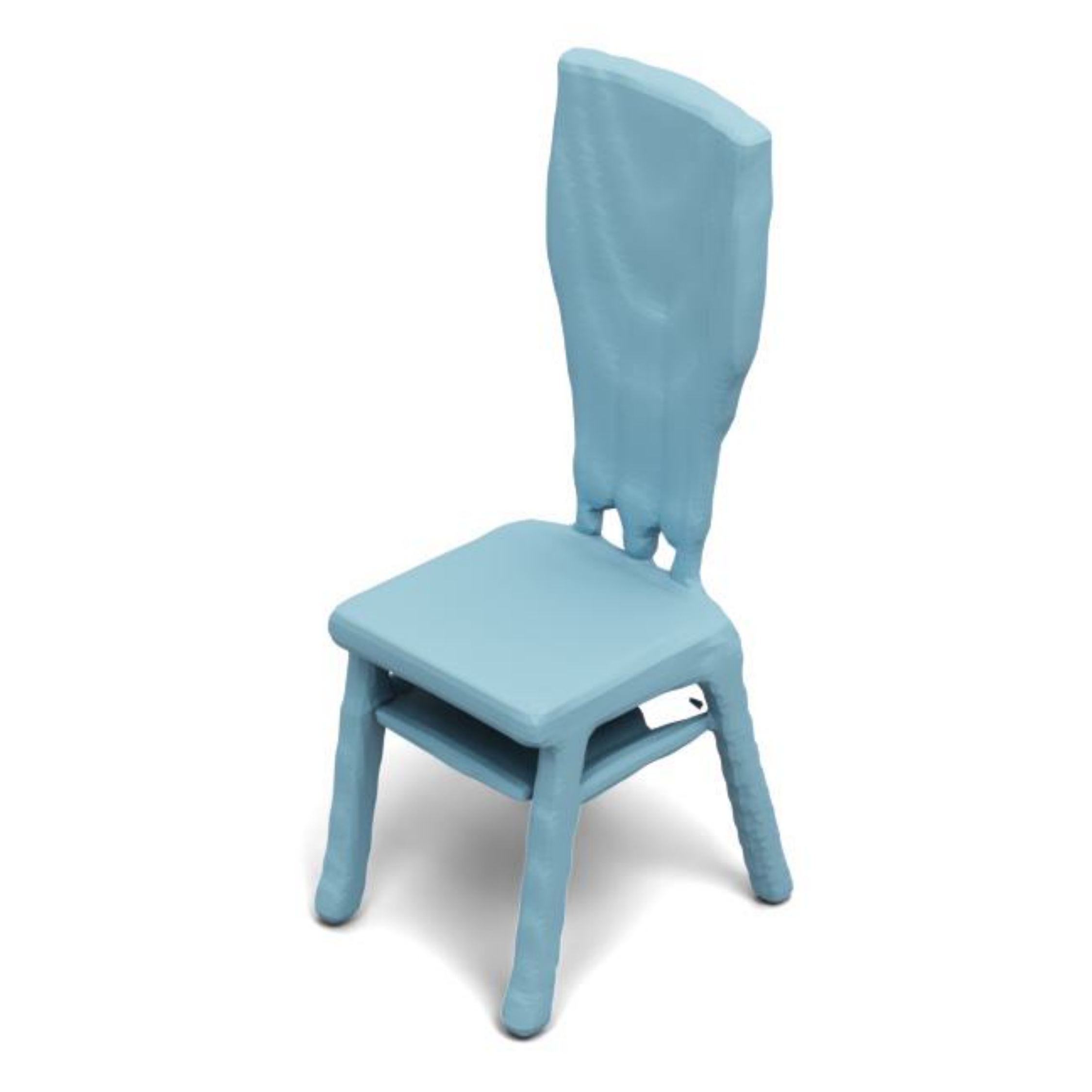}
\\
\includegraphics[width=0.125\linewidth]{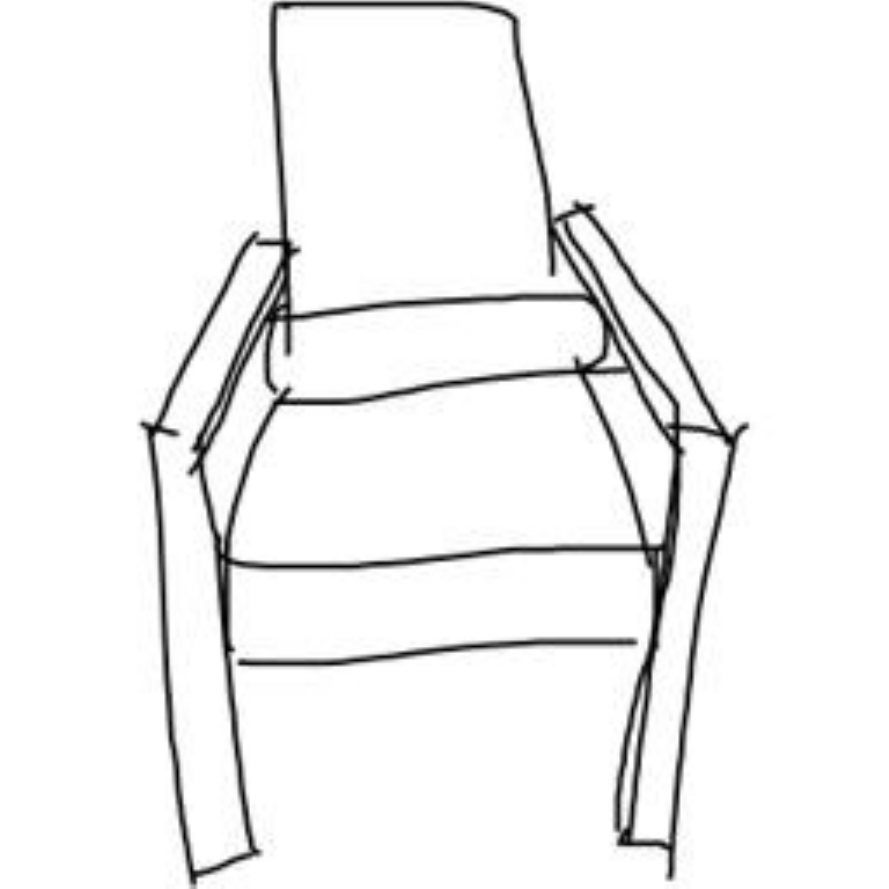}
&\includegraphics[width=0.125\linewidth]{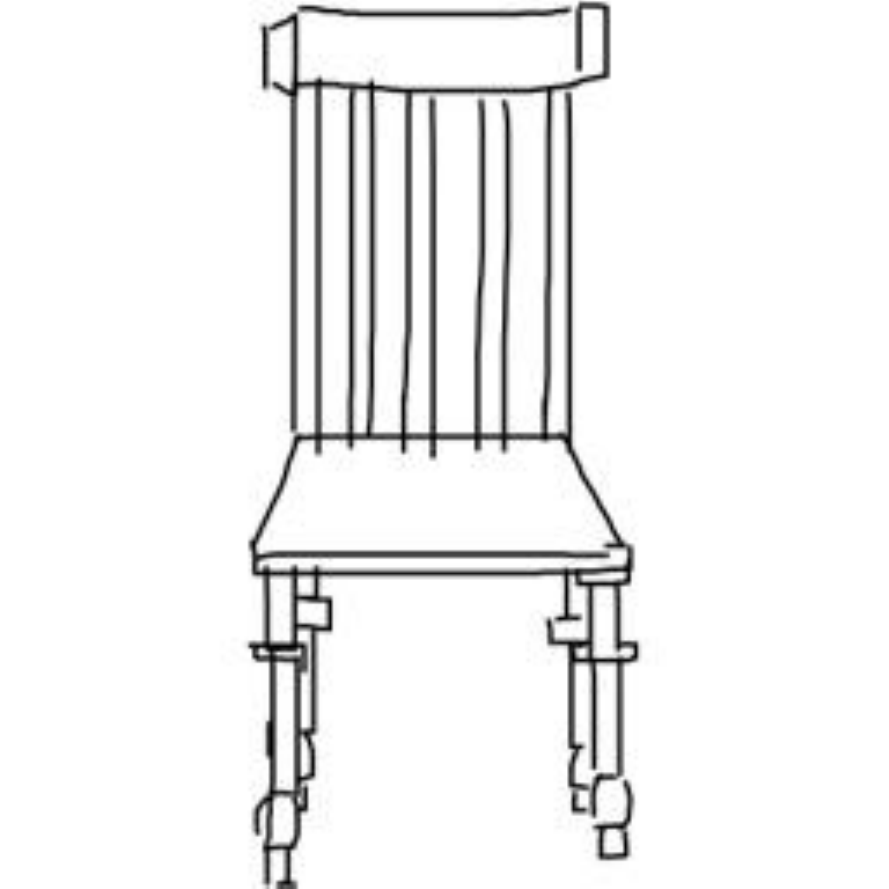}
&\includegraphics[width=0.125\linewidth]{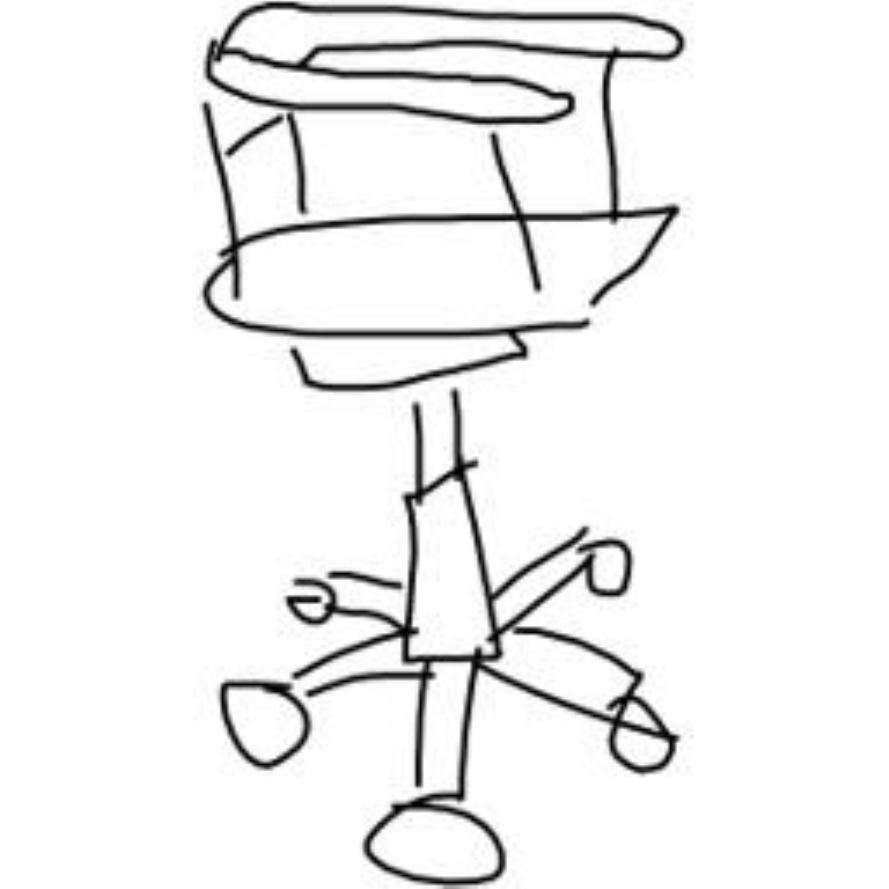}
&\includegraphics[width=0.125\linewidth]{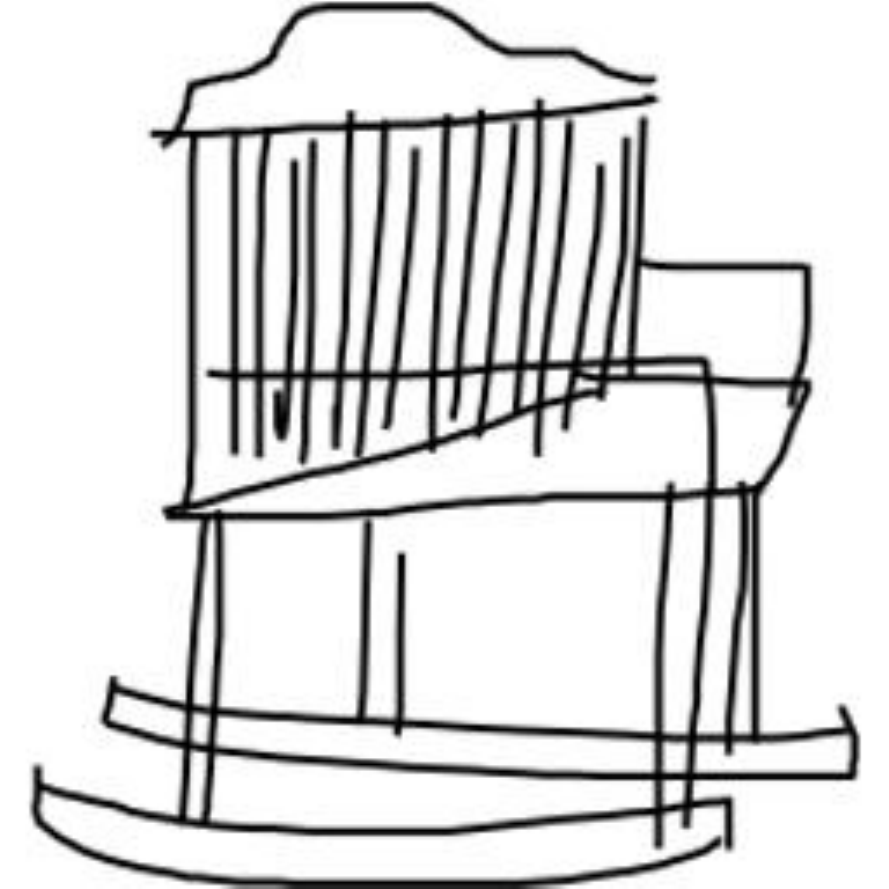}
&\includegraphics[width=0.125\linewidth]{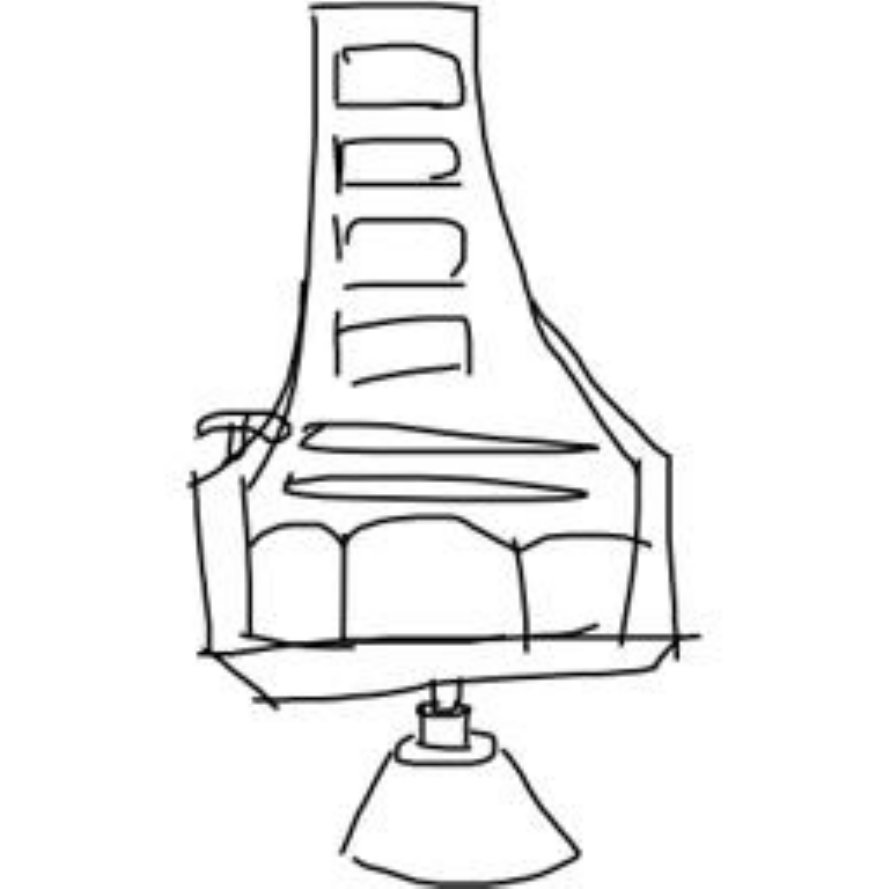}
&\includegraphics[width=0.125\linewidth]{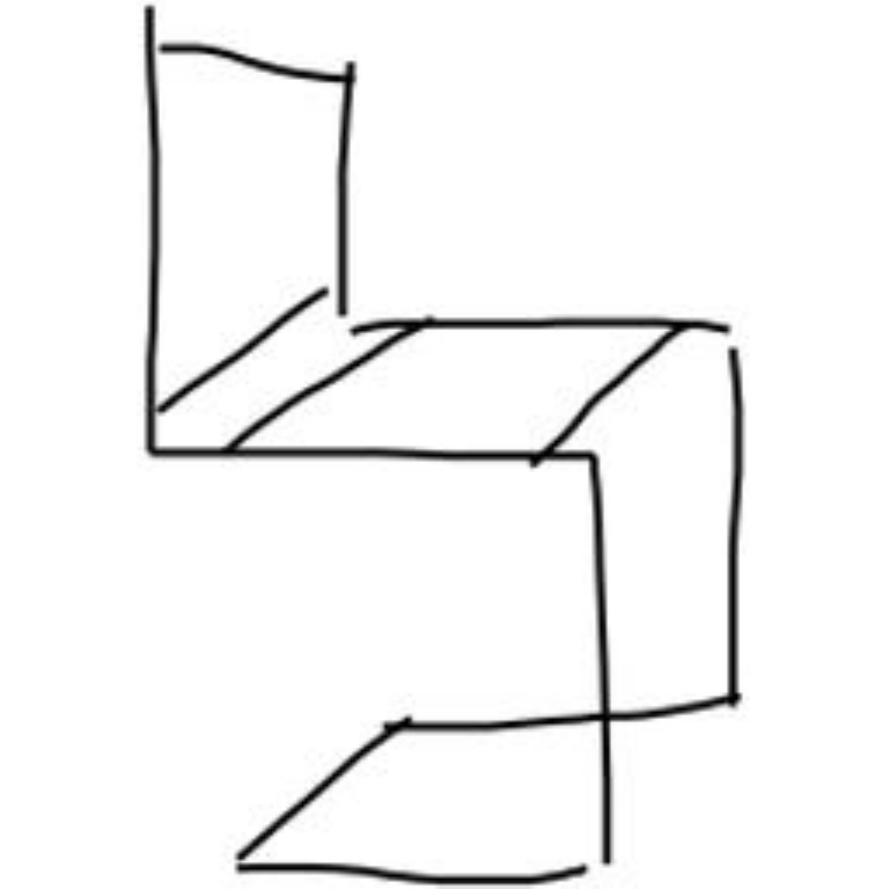}
&\includegraphics[width=0.125\linewidth]{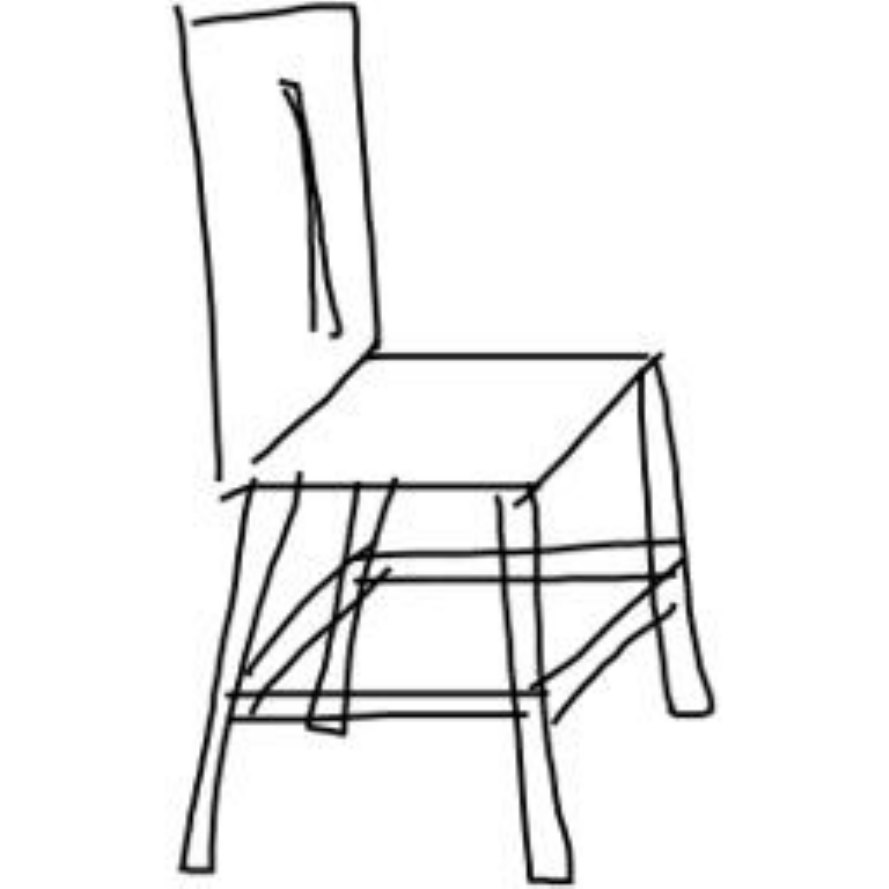}
&\includegraphics[width=0.125\linewidth]{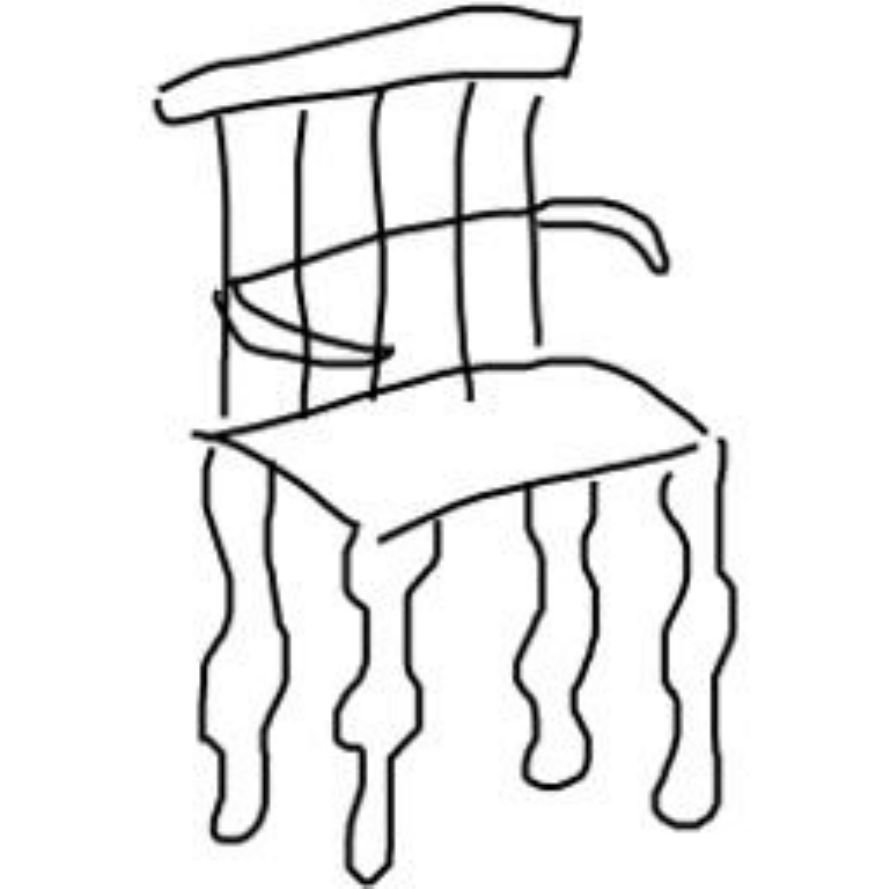}
\\
\includegraphics[trim = 1 1 1 1, clip, width=0.125\linewidth]{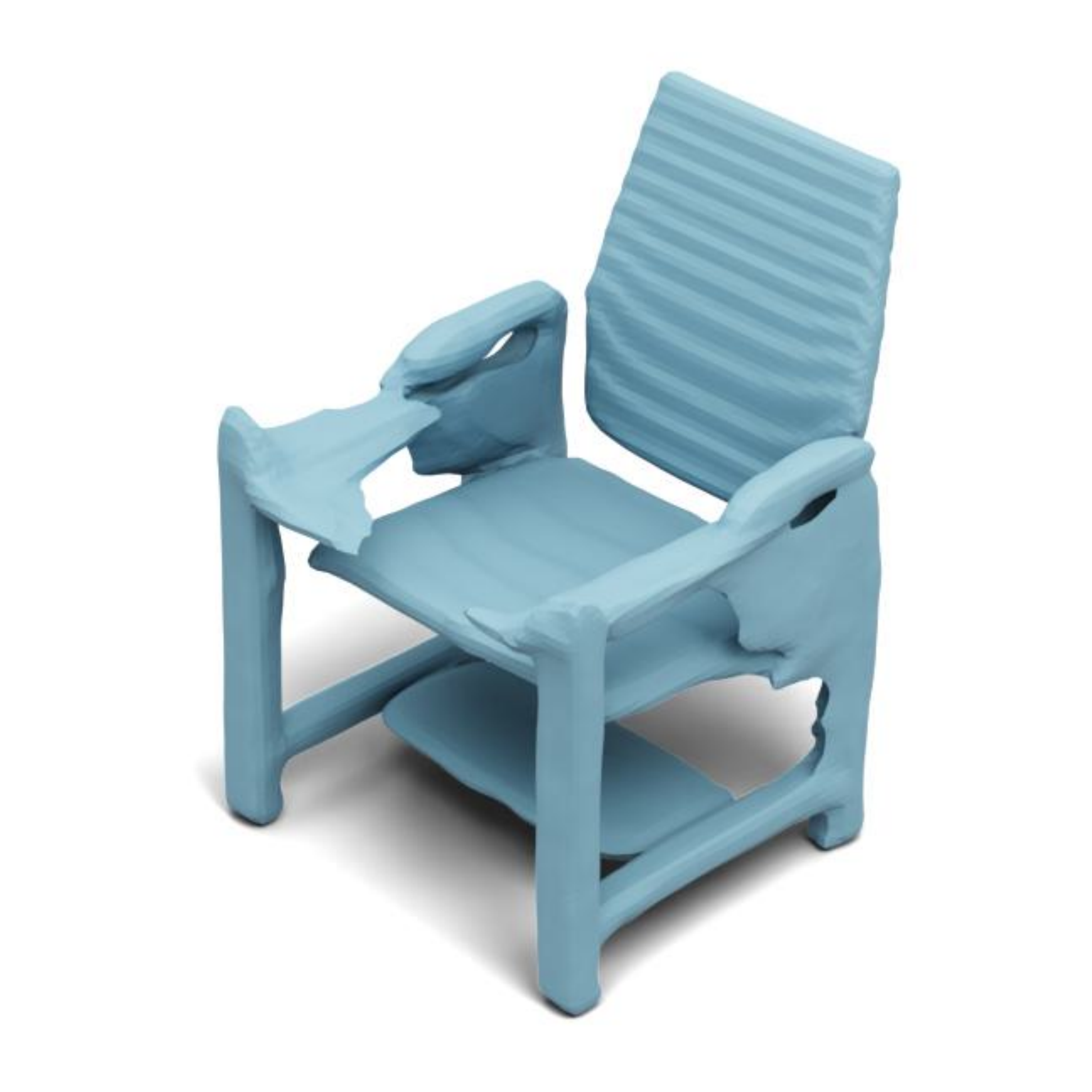}
&\includegraphics[trim = 1 1 1 1, clip, width=0.125\linewidth]{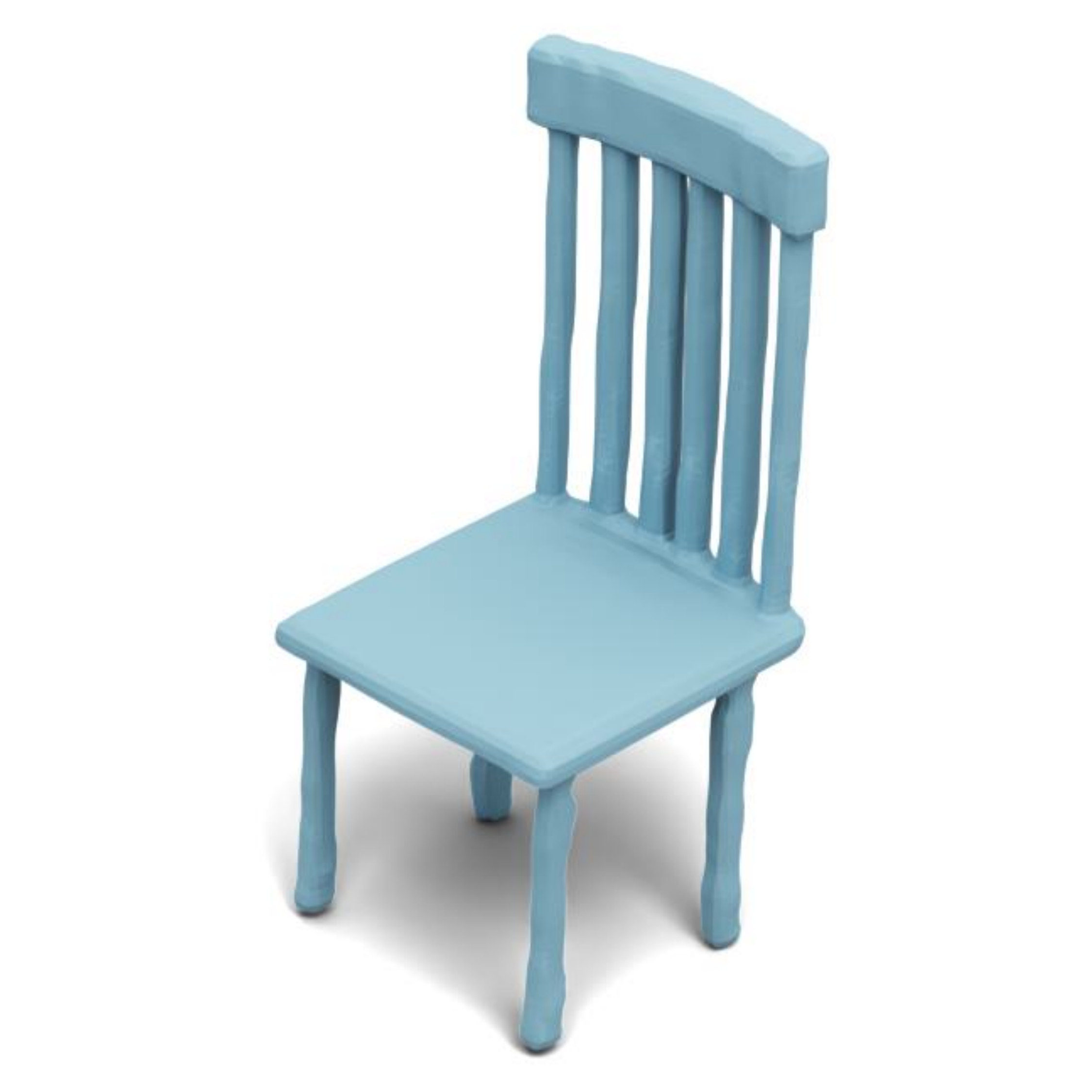}
&\includegraphics[trim = 1 1 1 1, clip, width=0.125\linewidth]{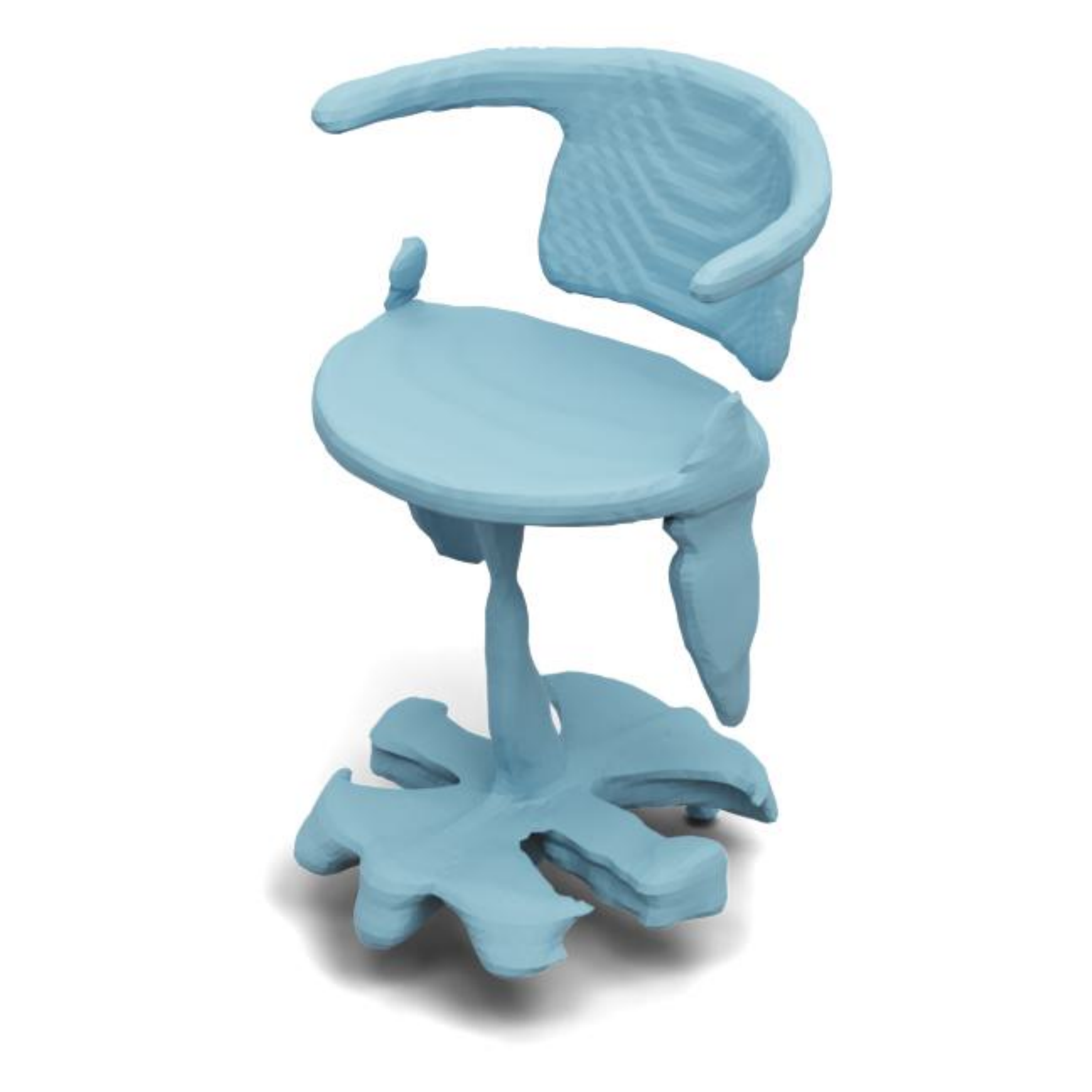}
&\includegraphics[trim = 1 1 1 1, clip, width=0.125\linewidth]{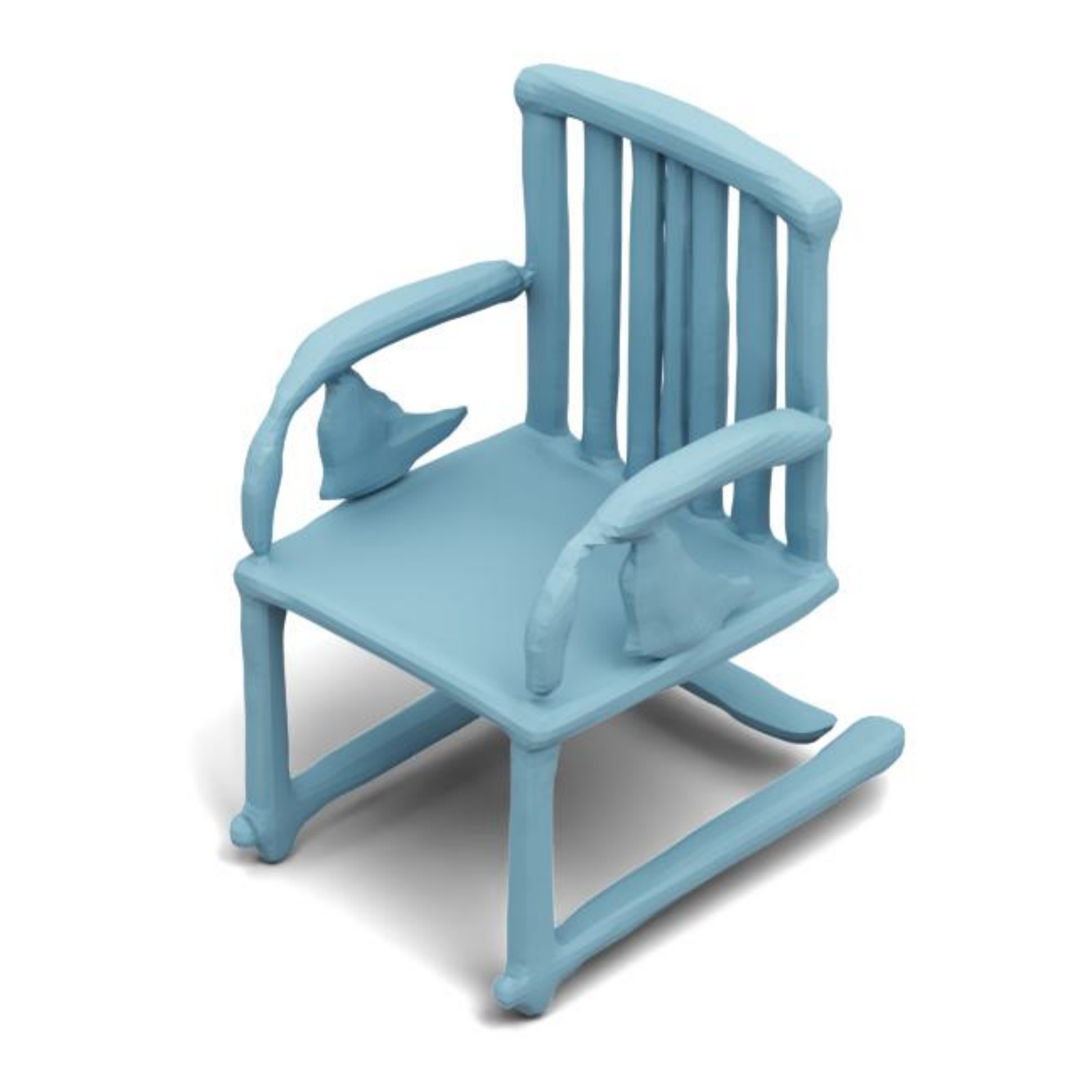}
&\includegraphics[trim = 1 1 1 1, clip, width=0.125\linewidth]{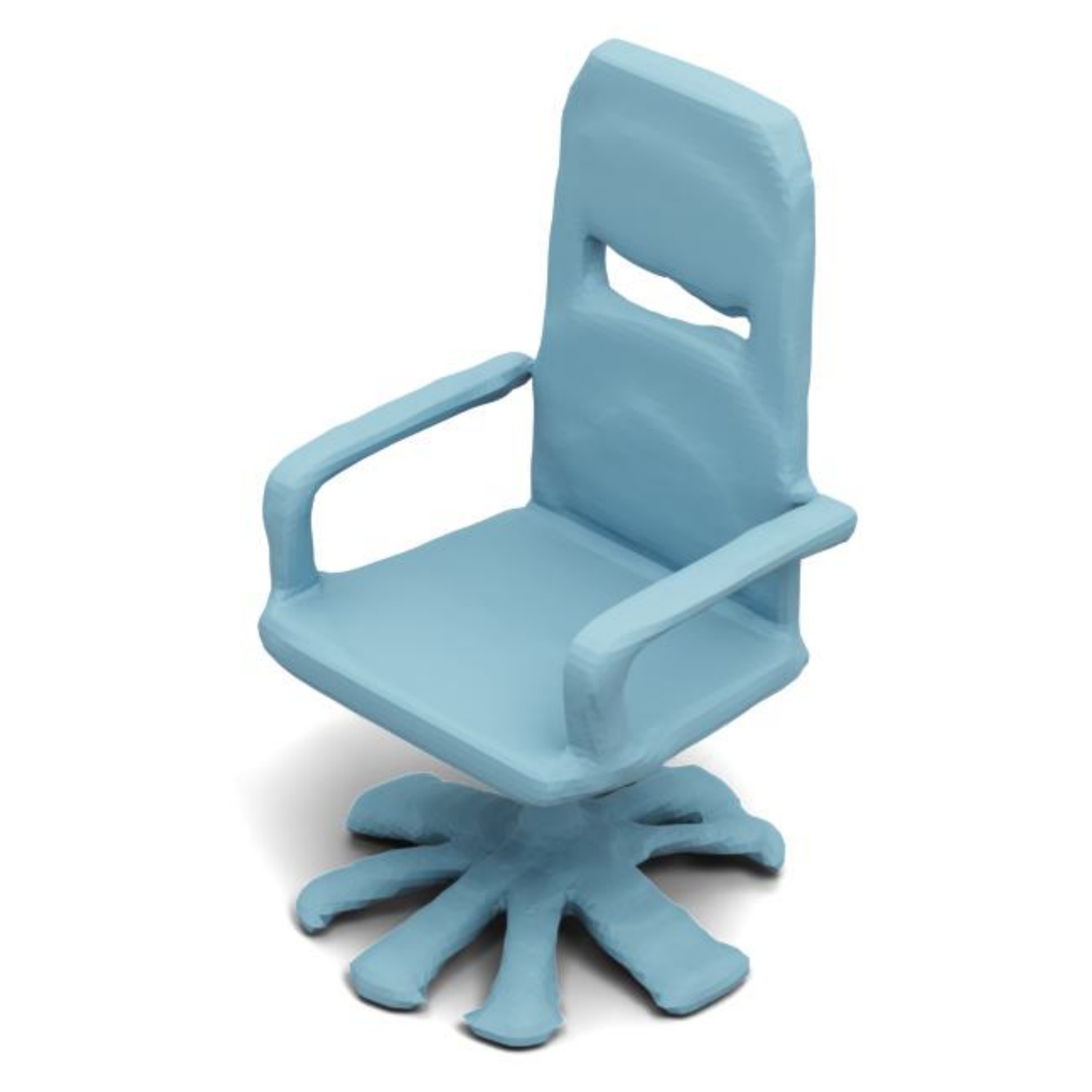}
&\includegraphics[trim = 1 1 1 1, clip, width=0.125\linewidth]{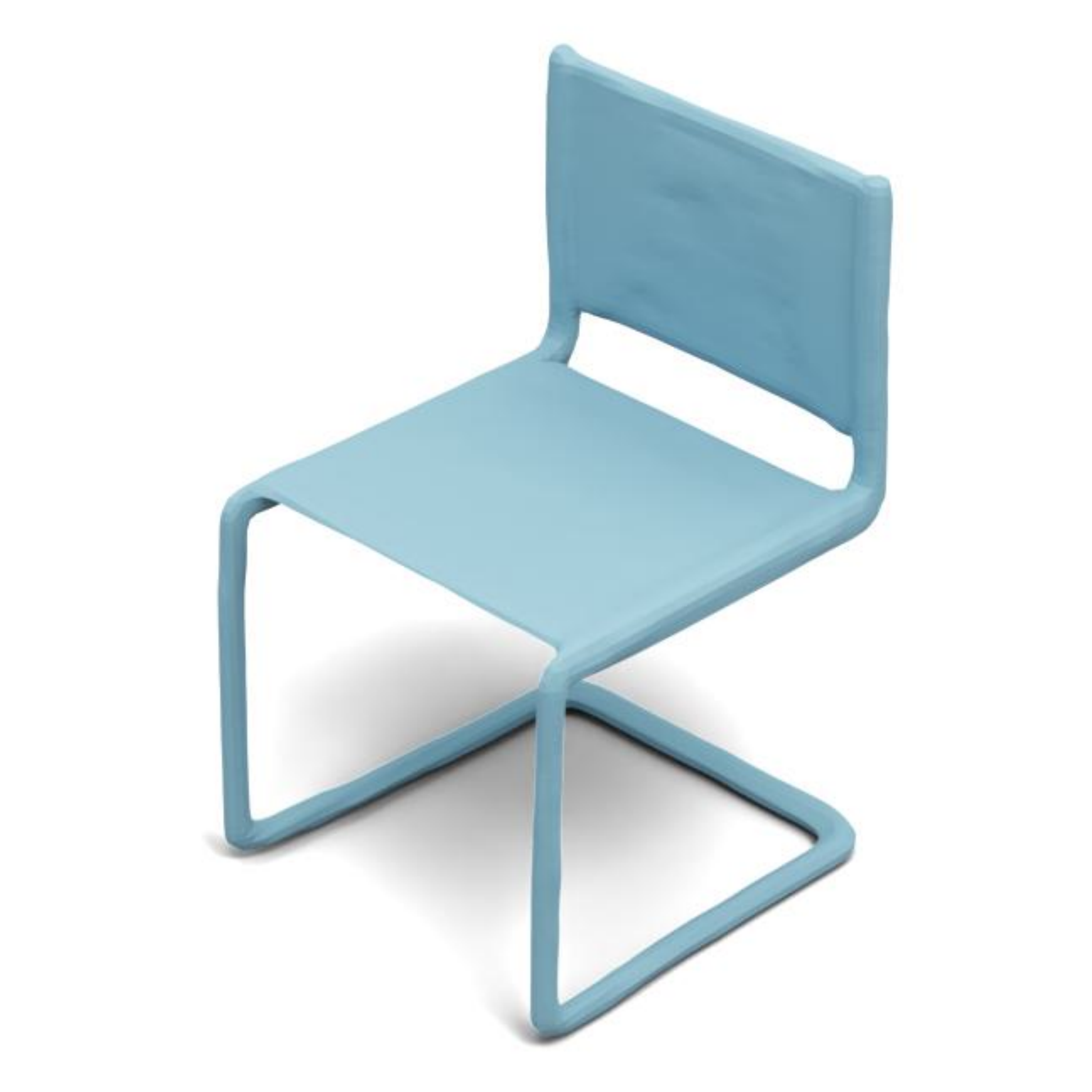}
&\includegraphics[trim = 1 1 1 1, clip, width=0.125\linewidth]{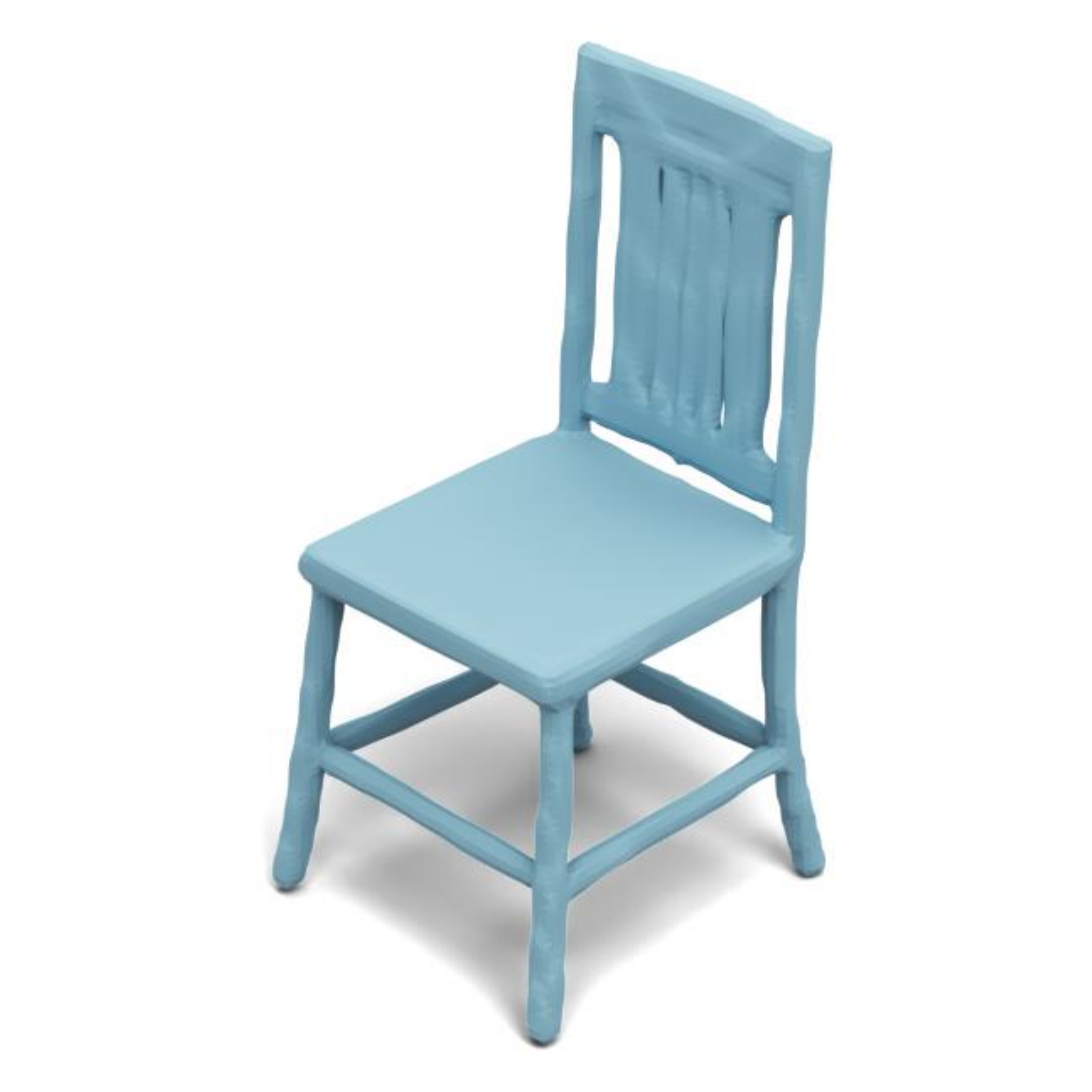}
&\includegraphics[trim = 1 1 1 1, clip, width=0.125\linewidth]{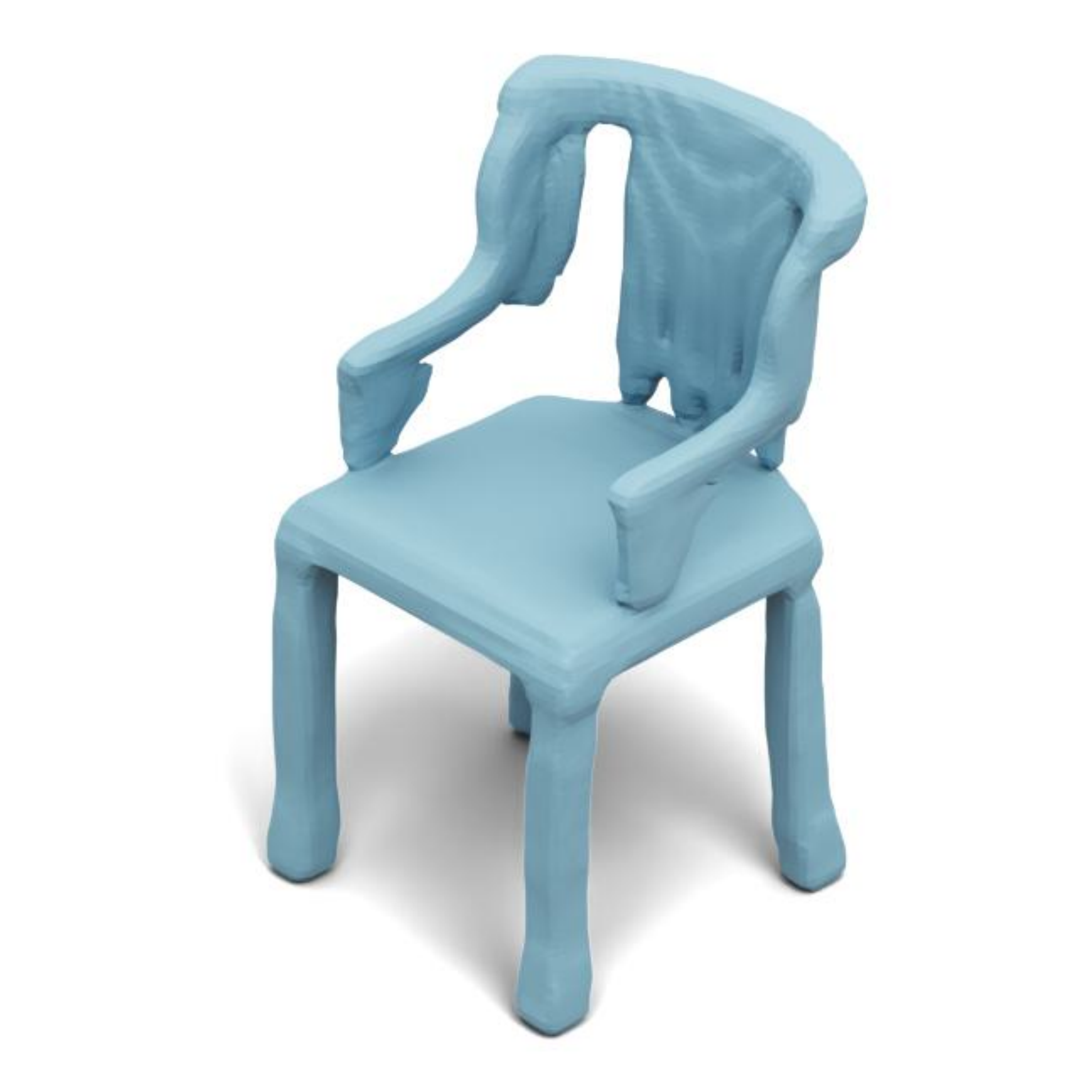}
\\
\includegraphics[width=0.125\linewidth]{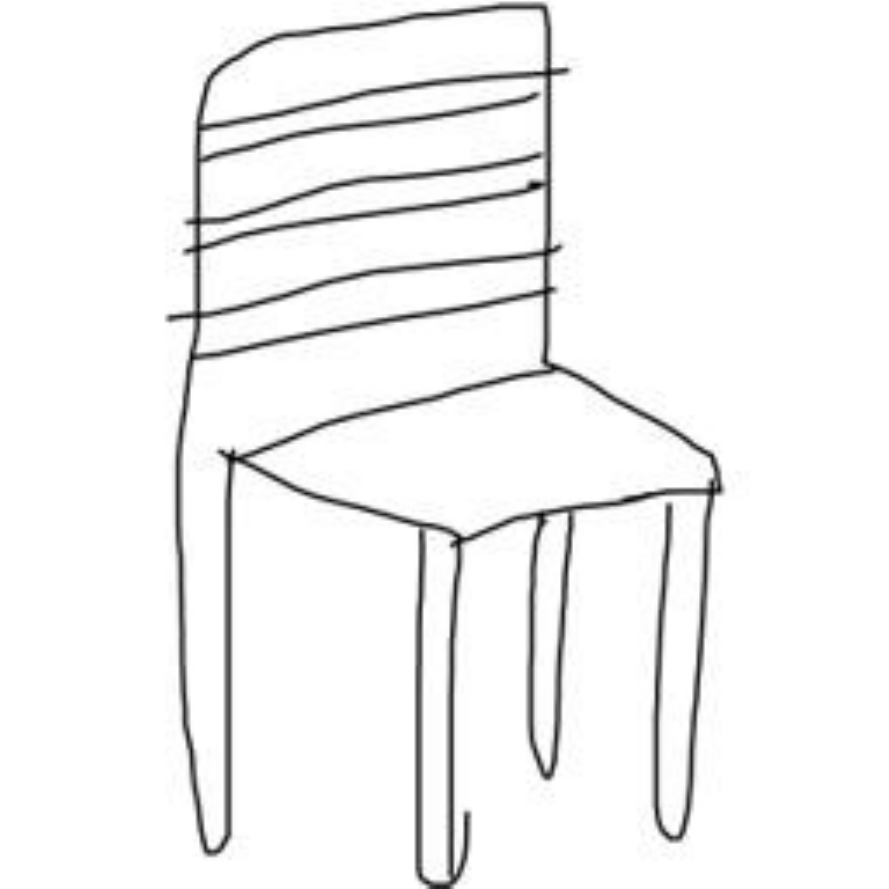}
&\includegraphics[width=0.125\linewidth]{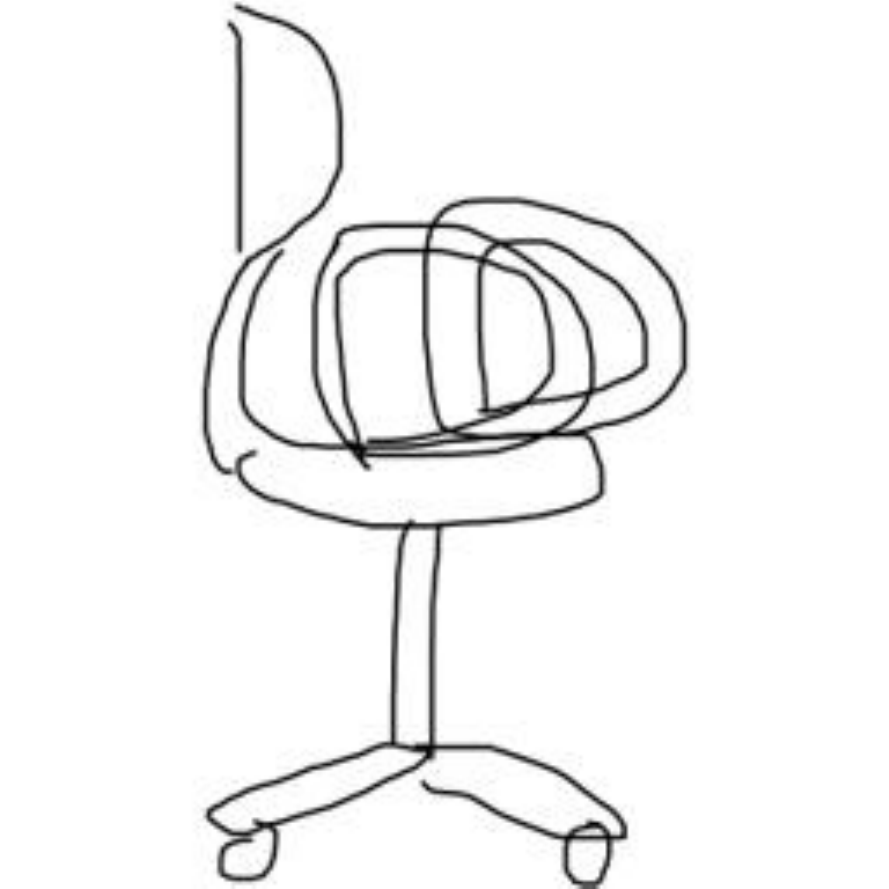}
&\includegraphics[width=0.125\linewidth]{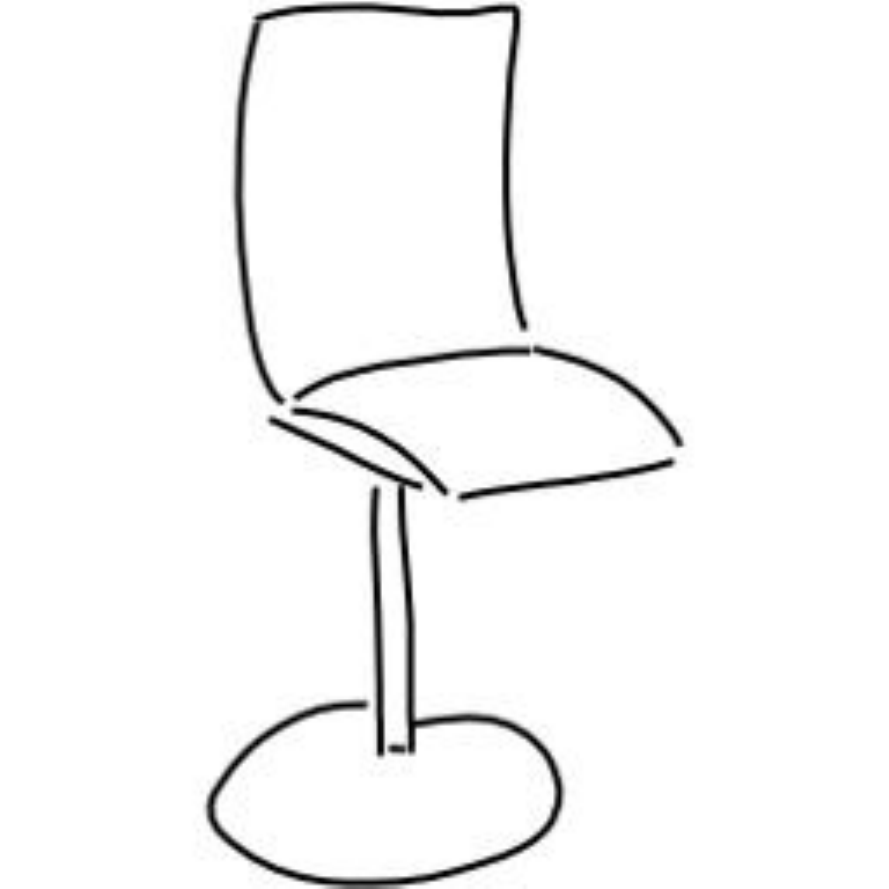}
&\includegraphics[width=0.125\linewidth]{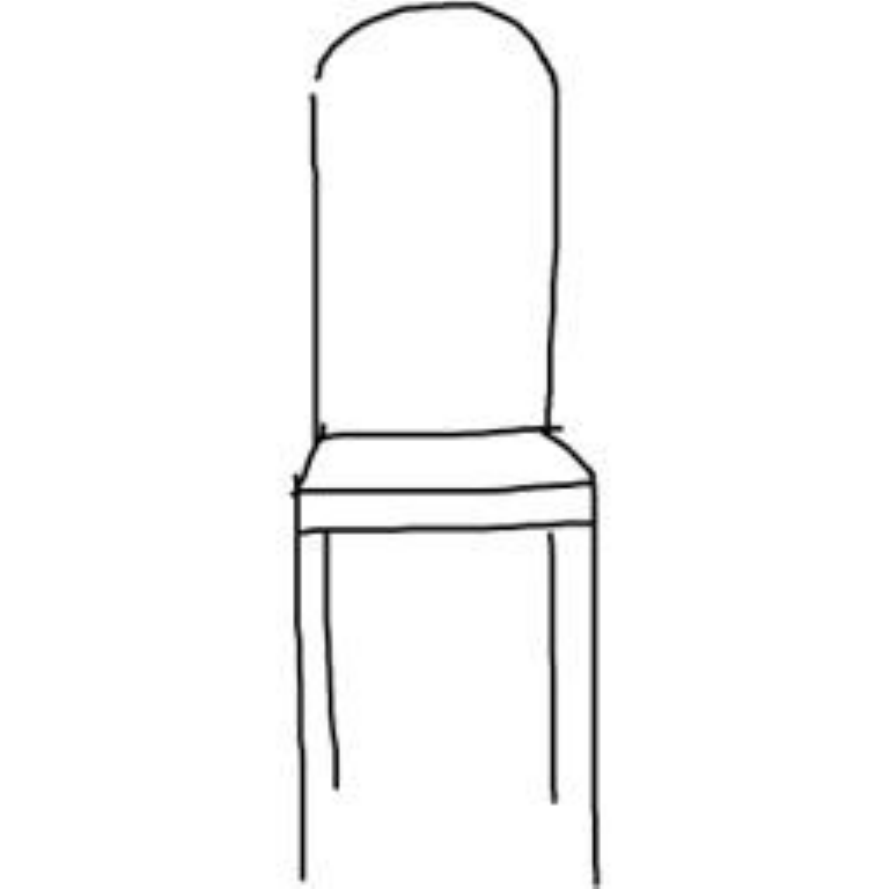}
&\includegraphics[width=0.125\linewidth]{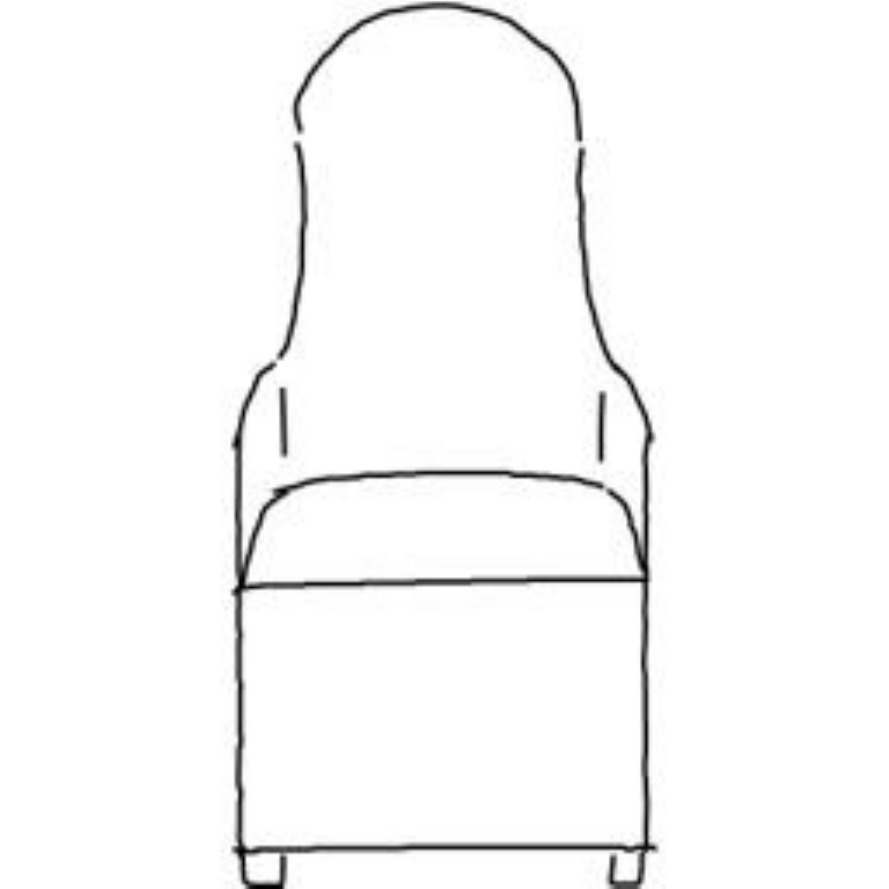}
&\includegraphics[width=0.125\linewidth]{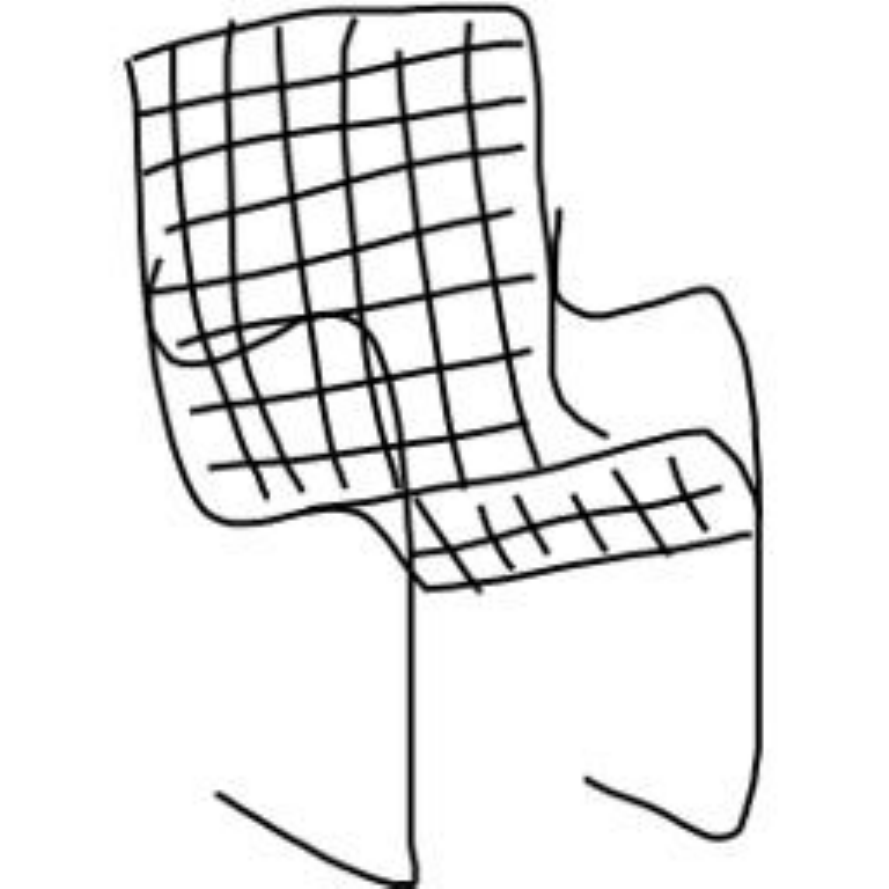}
&\includegraphics[width=0.125\linewidth]{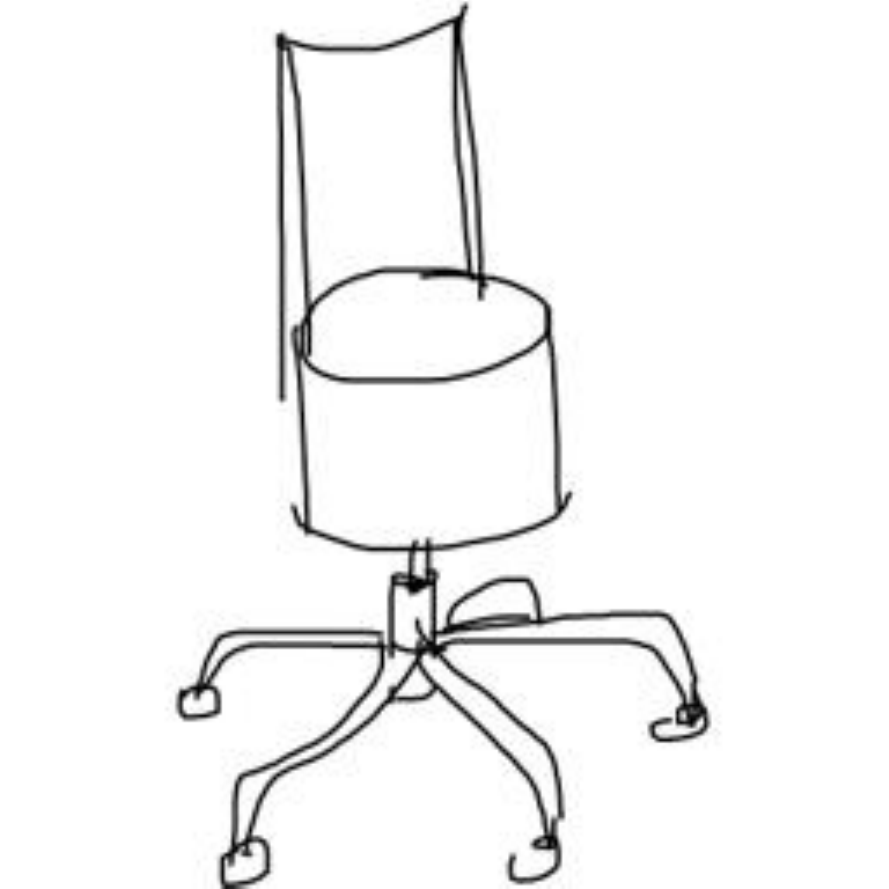}
&\includegraphics[width=0.125\linewidth]{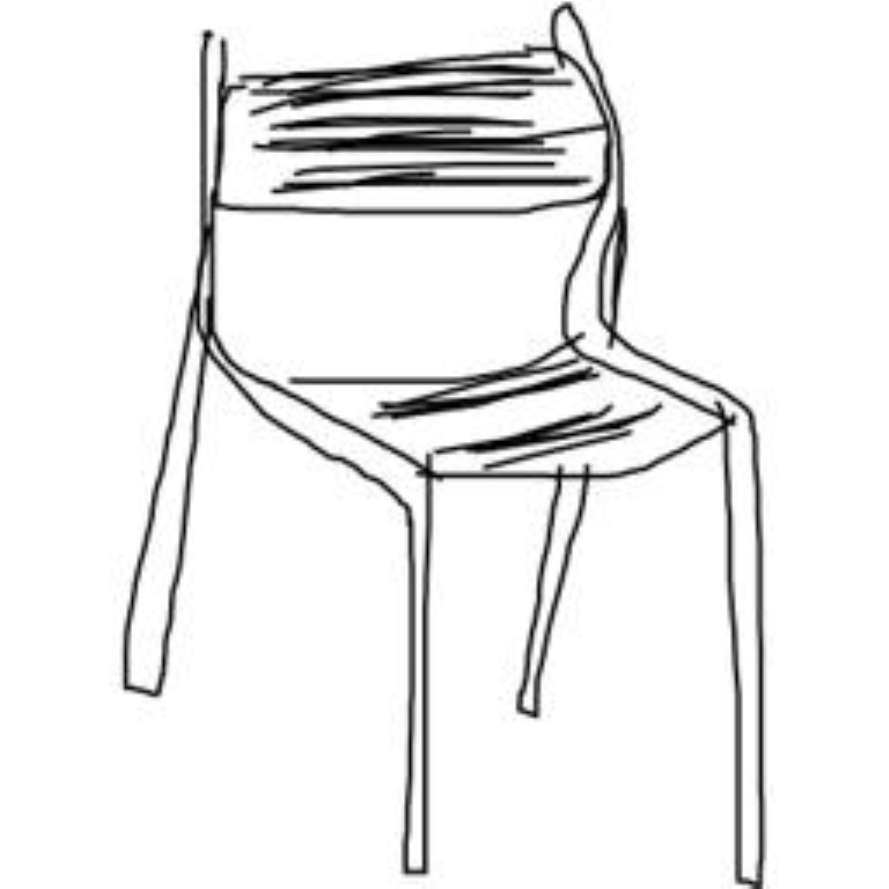}
\\
\includegraphics[trim = 1 1 1 1, clip, width=0.125\linewidth]{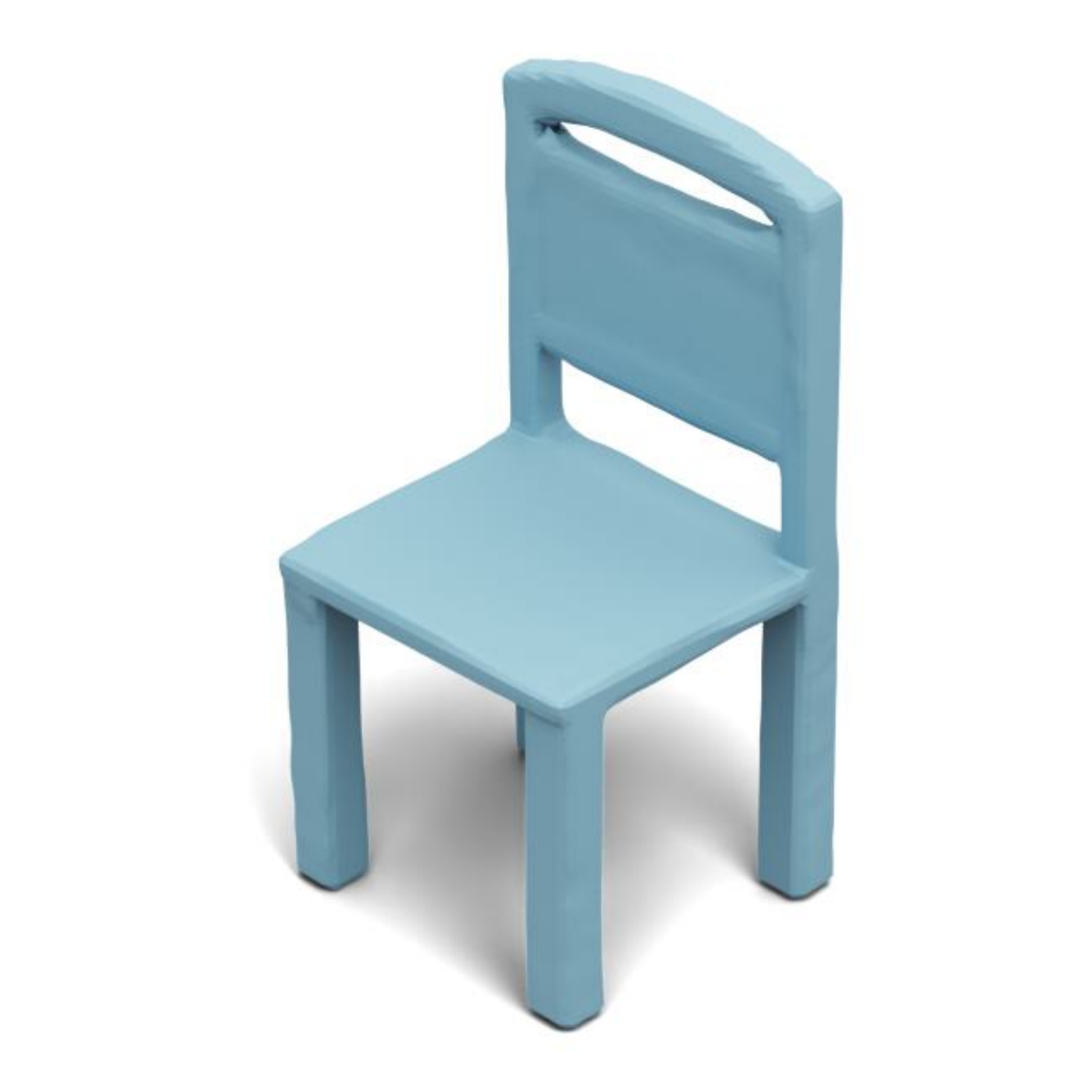}
&\includegraphics[trim = 1 1 1 1, clip, width=0.125\linewidth]{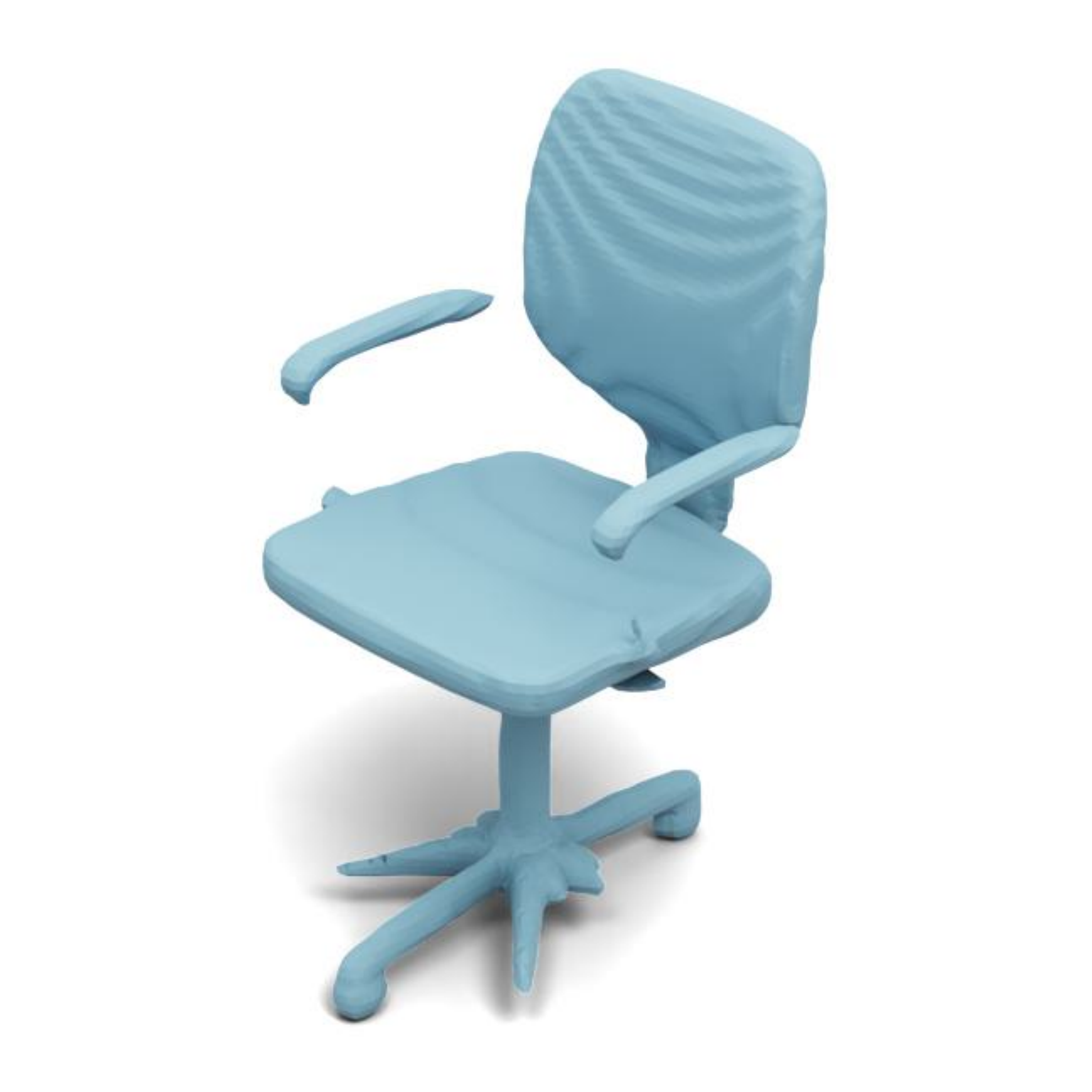}
&\includegraphics[trim = 1 1 1 1, clip, width=0.125\linewidth]{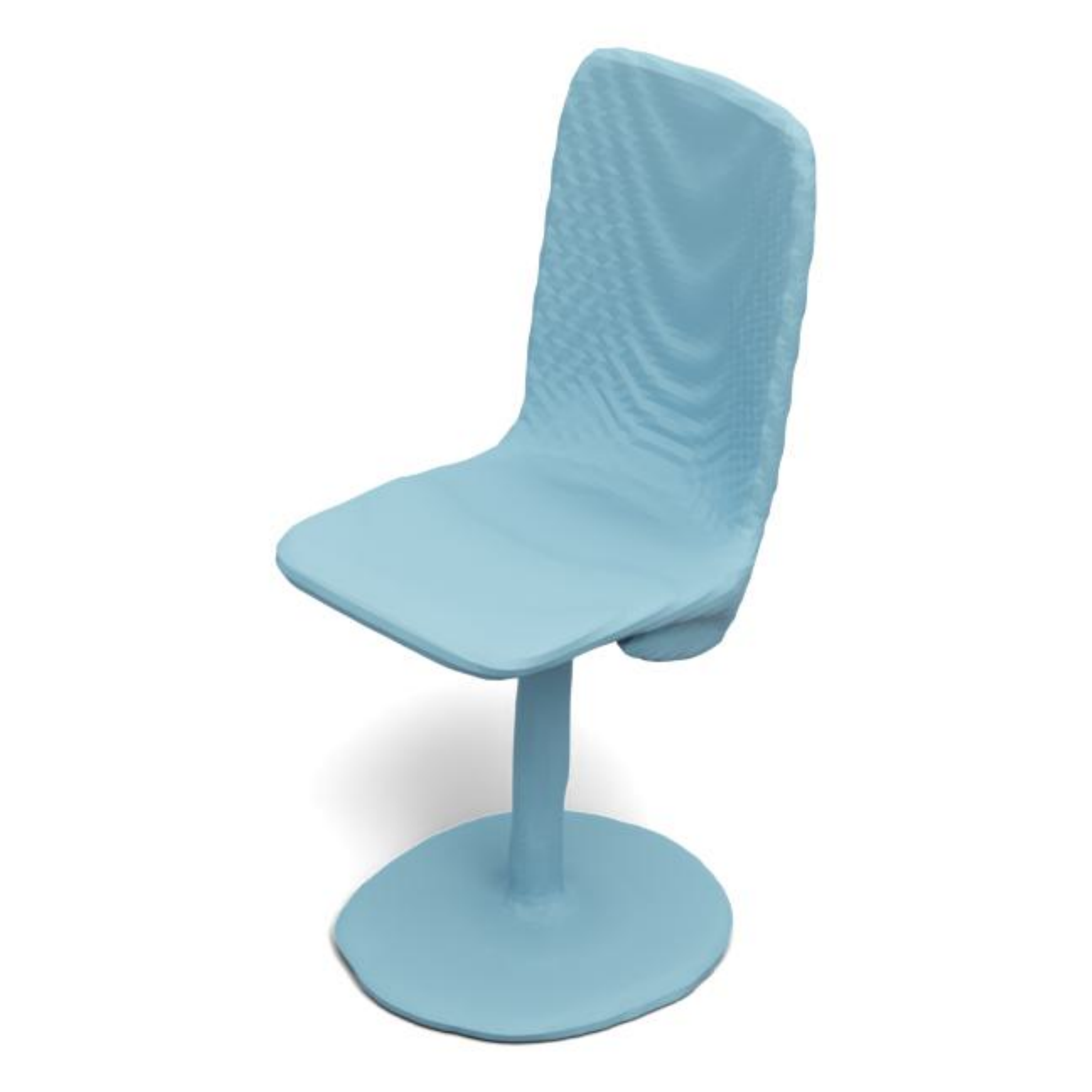}
&\includegraphics[trim = 1 1 1 1, clip, width=0.125\linewidth]{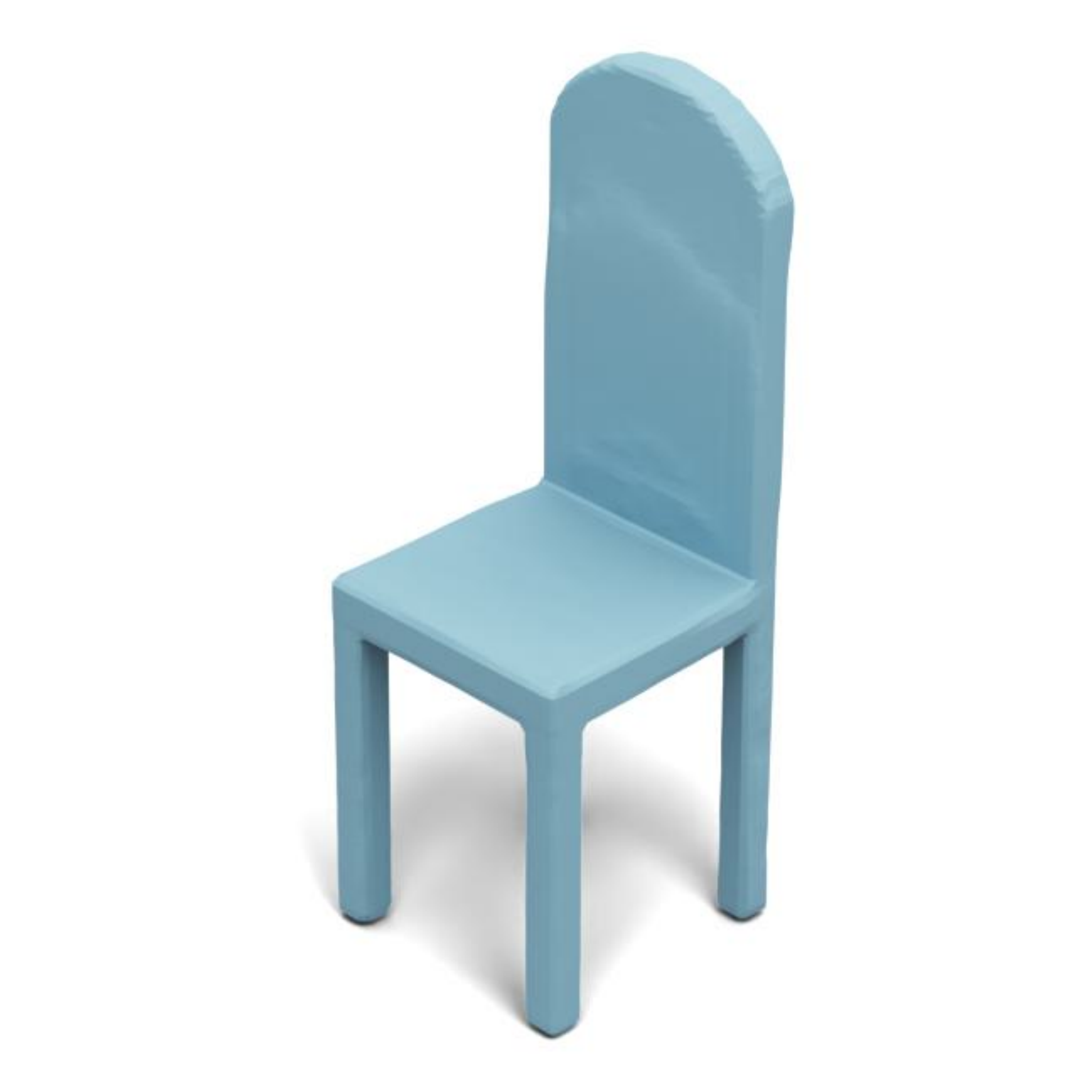}
&\includegraphics[trim = 1 1 1 1, clip, width=0.125\linewidth]{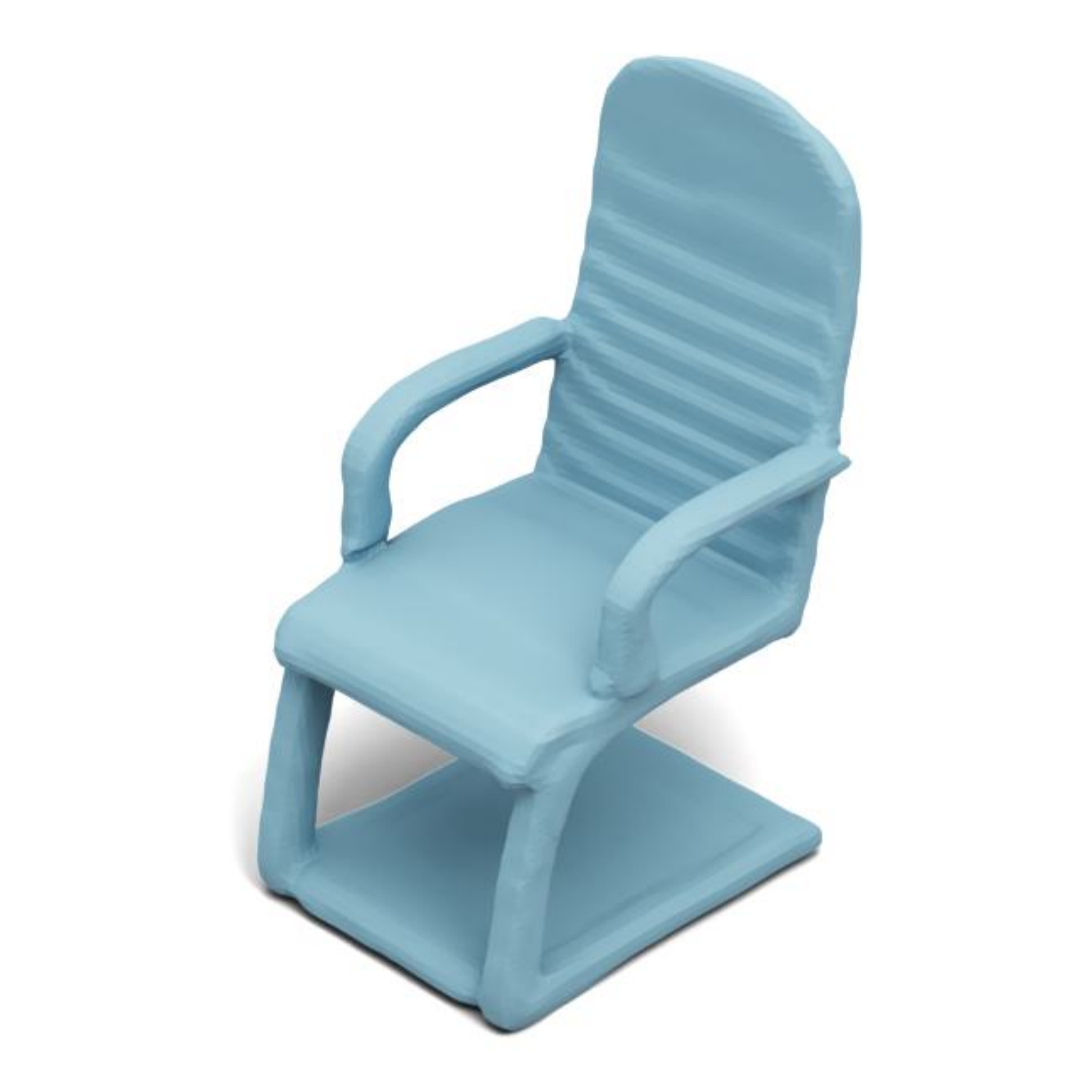}
&\includegraphics[trim = 1 1 1 1, clip, width=0.125\linewidth]{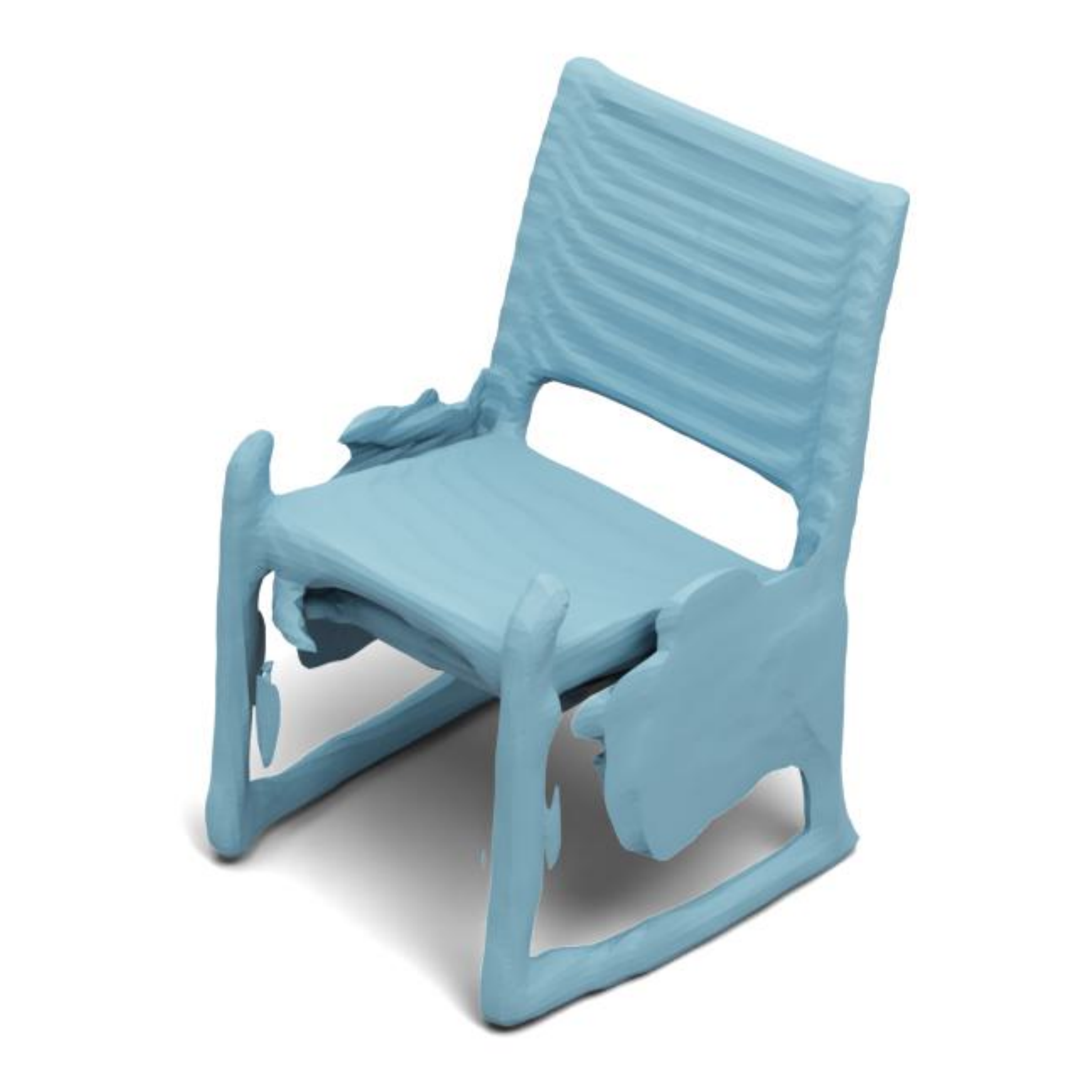}
&\includegraphics[trim = 1 1 1 1, clip, width=0.125\linewidth]{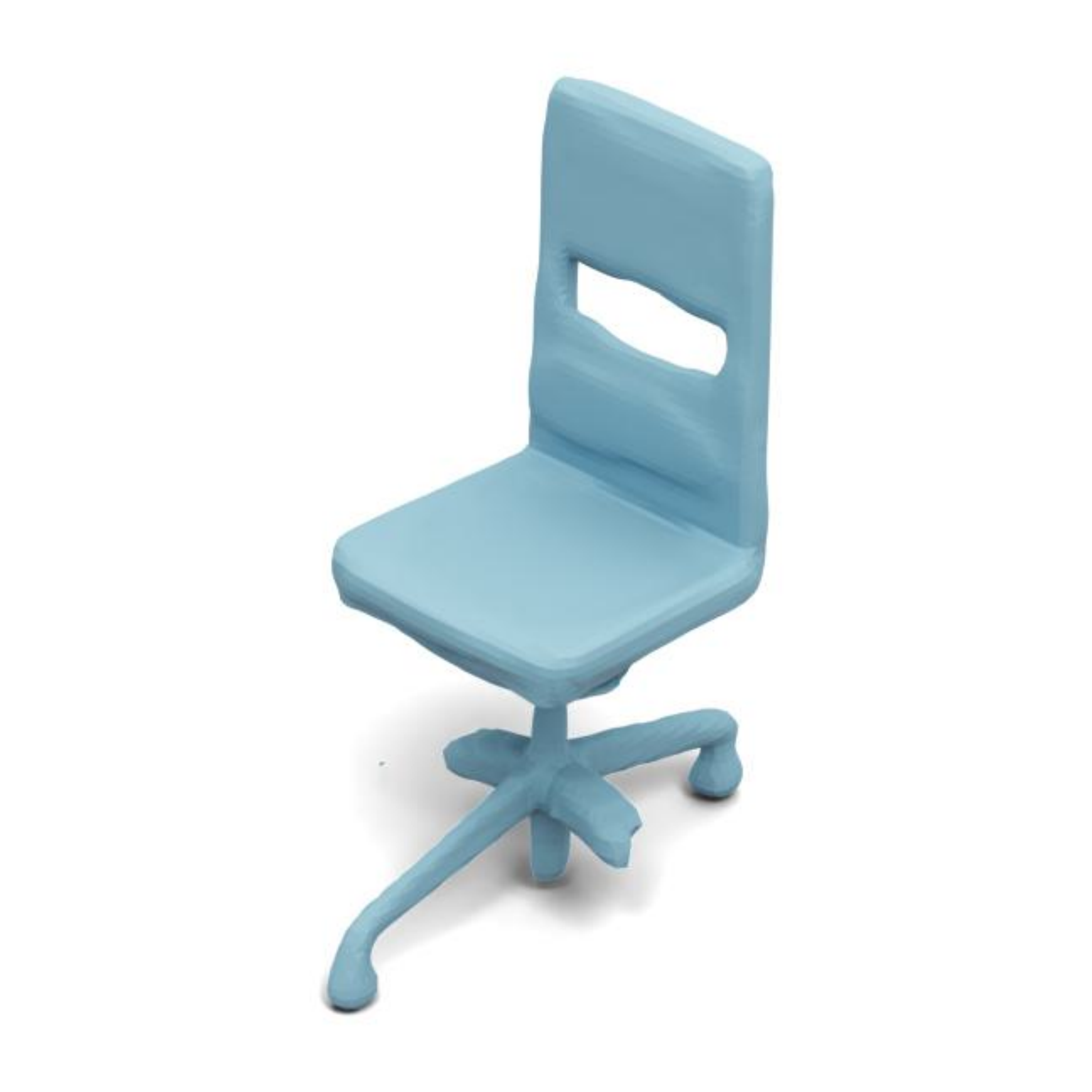}
&\includegraphics[trim = 1 1 1 1, clip, width=0.125\linewidth]{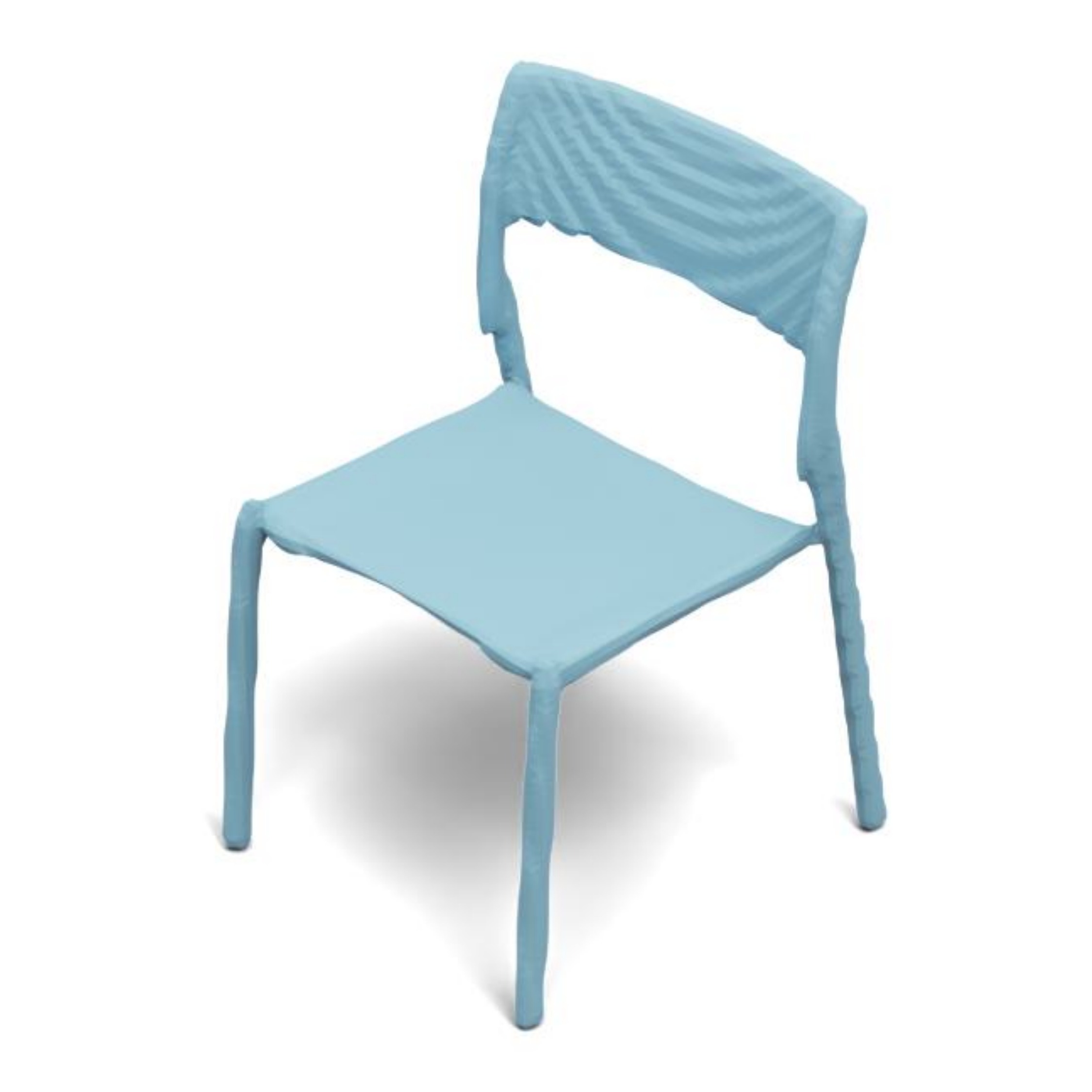}
\\
\end{tabular}
    \caption{We randomly sample sketches from the AmateurSketch dataset and showcase the results of our method.}
	\label{fig:additionalExp4}
\end{figure*}

In addition to the quantitative and qualitative evaluations, we also provide further visual results. We \emph{randomly} sample 128 sketches from the AmateurSketch dataset and present the result of \ourmethod{} in \figref{fig:additionalExp1}, \figref{fig:additionalExp2}, \figref{fig:additionalExp3}, and \figref{fig:additionalExp4}.

\bibliographystyle{eg-alpha-doi} 
\bibliography{references_supplementary}    

\newpage

\newpage

\newpage
